\documentclass[aps,prd,secnumarabic,twocolumn,nofootinbib,superscriptaddress,floatfix]{revtex4}

\usepackage{graphics}
\usepackage{color}
\usepackage{graphicx}
\usepackage{amssymb}
\usepackage{amsmath}
\usepackage{epstopdf}
\usepackage{xspace}

\usepackage{mathbbol}
\usepackage{wasysym} 
\usepackage{bbm}     
 
\usepackage{rotating}
\usepackage{dcolumn}
\usepackage{bm}
\usepackage{longtable}
\usepackage{multirow}
\usepackage{enumitem}

\usepackage{ifpdf}
\ifpdf
\usepackage[pdftex]{hyperref}
\else
\usepackage[hypertex]{hyperref}
\fi

\hypersetup{
 pdftitle={},%
 pdfauthor={},%
 pdfsubject={},%
 pdfkeywords={},%
 pdfstartview={},%
 bookmarksopen=true, breaklinks=true, debug=true,
 colorlinks=true, linkcolor=blue, citecolor=red, urlcolor=red,
 hyperfigures=true
}

\def\figwid{3.30in}

\def\APS{The American Physical Society}
\def\PLB{Elsevier}
\def\NPA{Elsevier}
\def\SPV{Springer-Verlag}
\def\figPermX#1#2#3{From \cite{#1} with kind permission, copyright ({#2}) {#3}}
\def\figPerm#1#2#3#4{{#1} from \cite{#2} with kind permission, copyright ({#3}) {#4}}

\def\figPermXAPS#1#2{\figPermX{#1}{#2}{\APS}}
\def\figPermXSPV#1#2{\figPermX{#1}{#2}{\SPV}}
\def\figPermXPLB#1#2{\figPermX{#1}{#2}{\PLB}}
\def\figPermXNPA#1#2{\figPermX{#1}{#2}{\NPA}}

\def\AfigPerm#1#2#3{\figPerm{Adapted}{#1}{#2}{#3}}
\def\AfigPermAPS#1#2{\AfigPerm{#1}{#2}{\APS}}
\def\AfigPermPLB#1#2{\AfigPerm{#1}{#2}{\PLB}}
\def\AfigPermSPV#1#2{\AfigPerm{#1}{#2}{\SPV}}

\def\figPerms#1#2#3#4{{#1} from \cite{#2} with kind permission,
                      copyrights ({#3}), respectively, {#4}}

\def\figPermsM#1#2#3#4{{#1} from \cite{#2} with kind permission,
                      copyrights ({#3}), {#4}, respectively}

\def\AfigPerms#1#2#3{\figPerms{Adapted}{#1}{#2}{#3}}
\def\AfigPermsAPS#1#2{\AfigPerms{#1}{#2}{\APS}}

\def\subthreesection#1{\vskip0.4cm  \noindent{\it \underline{#1}}\hfill\par\vskip0.1cm}
\def\subfoursection#1{\vskip0.2cm  \hskip0.3cm{\it {$\bullet~$}{#1}}\hfill\par\noindent\vskip0.1cm}

\def\FThing#1#2{#1~\ref{#2}}
\def\FThings#1#2{\FThing{#1s}{#2}}

\def\Thing#1#2{#1.~\ref{#2}}
\def\Things#1#2{\Thing{#1s}{#2}}

\def\Sec#1{\Thing{Sect}{#1}}
\def\Secs#1{\Things{Sect}{#1}}

\def\Section#1{\FThing{Section}{#1}}

\def\Fig#1{\Thing{Fig}{#1}}
\def\Figs#1{\Things{Fig}{#1}}

\def\Figure#1{\FThing{Figure}{#1}}
\def\Figures#1{\FThings{Figure}{#1}}

\def\Tab#1{\FThing{Table}{#1}}
\def\Tabs#1{\FThings{Table}{#1}}

\def\Eq#1{Eq.~(\ref{#1})}
\def\Eqs#1{Eqs.~(\ref{#1})}

\def\beq{\begin{equation}}
\def\eeq{\end{equation}}
\def\beqa{\begin{eqnarray}}
\def\eeqa{\end{eqnarray}}
\def\non{\nonumber}

\newcommand{\etc}   {{\it etc.}}
\newcommand{\ie}   {{\it i.e.,}~}
\newcommand{\eg}   {{\it e.g.,}~}
\newcommand{\etal} {{\it {et al.}}}
\newcommand{\ev}   {~\ensuremath{{\mathrm{eV}}      }}
\newcommand{\tev}  {~\ensuremath{{\mathrm{TeV}}     }}
\newcommand{\kev}  {~\ensuremath{{\mathrm{keV}}     }}
\newcommand{\gev}  {~\ensuremath{{\mathrm{GeV}}     }}
\newcommand{\mev}  {~\ensuremath{{\mathrm{MeV}}     }}
\newcommand{\gevc} {~\ensuremath{{\mathrm{GeV}/c}   }}

\newcommand{\GeV}  {\gev}

\newcommand{\evcc}  {\ev}

\newcommand{\mevcc} {\mev}
\newcommand{\gevcc} {\gev}

\newcommand{\ecm}  {\ensuremath{E_{\mathrm{c.m.}} }}

\newcommand{\brat}   {\ensuremath{ \mathcal{B}}}
\newcommand{\gee}  {\ensuremath{ \Gamma_{\mathrm{ee}} }}

\newcommand{\antid}{\ensuremath{\bar{\rm d}}}
\newcommand{\deut}{\ensuremath{{\rm d}}}

\newcommand{\epem}   {\ensuremath{ e^+e^- }}
\newcommand{\Dze}   {\ensuremath{ D^0 }}
\newcommand{\Dpl}  {\ensuremath{ D^+ }}
\newcommand{\Dmi}   {\ensuremath{ D^- }}
\newcommand{\Dsp}  {\ensuremath{ D_s^+ }}
\newcommand{\Dsm}  {\ensuremath{ D_s^- }}
\newcommand{\Dstp} {\ensuremath{ D^{*+} }}
\newcommand{\Dstn} {\ensuremath{ D^{*0} }}
\newcommand{\Dstm} {\ensuremath{ D^{*-} }}
\newcommand{\Dstsp}  {\ensuremath{ D_s^{*+} }}
\newcommand{\Dstsm}  {\ensuremath{ D_s^{*-} }}

\newcommand{\DDbar}      {\ensuremath{ D   \bar D     }}
\newcommand{\nonDDbar}   {\ensuremath{ {\rm non\text{-}}\DDbar  }}
\newcommand{\DDst}    {\ensuremath{ D   \bar D^* }}
\newcommand{\DstDst}  {\ensuremath{ D^* \bar D^* }}
\newcommand{\DstnDn}  {\ensuremath{\Dstn \bar{\Dze}}}
\newcommand{\DzDz}    {\ensuremath{ \Dze \bar{\Dze} } }
\newcommand{\DpDm}    {\ensuremath{ \Dpl\Dmi }}

\newcommand{\DpDstm}  {\ensuremath{ \Dpl  \Dstm  }}
\newcommand{\DstpDstm}{\ensuremath{ \Dstp \Dstm }}
\newcommand{\DsDs}    {\ensuremath{ \Dsp \Dsm     }}
\newcommand{\DsDsts}  {\ensuremath{ \Dsp \Dstsm }}
\newcommand{\DstsDsts}{\ensuremath{ \Dstsp \Dstsm }}

\newcommand{\DDp}     {\ensuremath{ \Dze \Dmi   \pi^+ }}
\newcommand{\DDstp}   {\ensuremath{ \Dze \Dstm \pi^+ }}

\newcommand{\pipi}    {\ensuremath{\pi^+\pi^- }}
\newcommand{\pppsip}  {\ensuremath{\pipi \psip    }}

\newcommand{\Dtwo}  {\ensuremath{\bar{D}^{*}_2(2460) }}
\newcommand{\Dtch}  {\ensuremath{     D^{*}_2(2460)^+ }}
\newcommand{\Dtnb}  {\ensuremath{\bar{D}^{*}_2(2460)^0 }}

\newcommand{\DDtwo} {\ensuremath{D   \Dtwo }}
\newcommand{\DDtch} {\ensuremath{D^- \Dtch }}
\newcommand{\DDtnb} {\ensuremath{\Dze \Dtnb }}
\newcommand{\lap}   {\ensuremath{\Lambda_c^+ }}
\newcommand{\lam}   {\ensuremath{\Lambda_c^- }}
\newcommand{\lala}  {\ensuremath{\lap\lam }}

\newcommand{\eeDDp}      {\ensuremath{ \epem \to \DDp  }}

\newcommand{\eell}       {\ensuremath{ \epem \to \lala }}

\def\Ups{\ensuremath{\Upsilon}}
\def\ups{\Ups}
\def\Unx#1#2{\ensuremath{\Ups({#1}{#2})}}
\def\UnS#1{\ensuremath{\Unx{#1}{S}}}
\def\UoneD{\ensuremath{\Unx{1}{D}}}
\def\UoneDJ{\ensuremath{\Unx{1}{^3D_J}}}
\def\UoneDO{\ensuremath{\Unx{1}{^3D_1}}}
\def\UoneDT{\ensuremath{\Unx{1}{^3D_2}}}
\def\UoneDH{\ensuremath{\Unx{1}{^3D_3}}}

\newcommand{\Usum}{\ensuremath{\UnS{1} + \UnS{2} + \UnS{3}}}

\newcommand{\jpsi}{\ensuremath{J/\psi}}
\newcommand{\costhk}{$\cos\theta_K$}
\newcommand{\psip}{\ensuremath{\psi(2S)}}
\newcommand{\psit}{\ensuremath{{\psi(3770)}}}
\newcommand{\chicOne}{\ensuremath{\chi_{c1}}}

\newcommand{\ppjp}    {\ensuremath{ \pi^+\pi^-\jpsi }}

\usepackage{relsize}
\def\babar{\mbox{\slshape B\kern-0.1em{\smaller A}\kern-0.1em
    B\kern-0.1em{\smaller A\kern-0.2em R}}}
\newcommand{\DZero}{\rm D\O}

\newcommand{\kpi}{\ensuremath{K\pi^-}}

\newcommand{\psipi}{\ensuremath{\psi\pi^-}}

\newcommand{\psitwospi}{\ensuremath{\psi(2S)\pi^-}}
\newcommand{\mpsitwospi}{\ensuremath{M(\psi(2S)\,\pi^-)}}

\newcommand{\Ksone}{\ensuremath{K^{\ast}(892)}}
\newcommand{\Kstwo}{\ensuremath{K^{\ast}_2(1430)}}

\newcommand{\zBelle}{\ensuremath{Z(4430)^-}}
\newcommand{\zBellep}{\ensuremath{Z(4430)^+}}
\newcommand{\zOne}{\ensuremath{Z_1(4050)^-}}
\newcommand{\zTwo}{\ensuremath{Z_2(4250)^-}}

\newcommand{\Gmumu}{\ensuremath{\Gamma_{\mu\mu}}}
\newcommand{\Gee}{\ensuremath{\Gamma_{ee}}}
\newcommand{\Gtee}{\ensuremath{\tilde{\Gamma}_{ee}}}
\newcommand{\Gtt}{\ensuremath{\Gamma_{\tau\tau}}}
\newcommand{\Btt}{\ensuremath{{\cal B}_{\tau\tau}}}

\newcommand{\Bmumu}{\ensuremath{{\cal B}_{\mu\mu}}}
\newcommand{\Btmumu}{\ensuremath{\tilde{\cal B}_{\mu\mu}}}
\newcommand{\Ghad}{\ensuremath{{\Gamma_{\rm had}}}}
\newcommand{\Gtot}{\ensuremath{{\Gamma_{\rm tot}}}}

\def\dmhf{\Delta m_{\rm hf}}

\def\mchicj{\left<m(1\,^3P_J)\right>}

\def\cost{|\cos\theta_T|}

\def\hqm{m}

\newcommand{\piz}{{\ensuremath{{\pi^0}}}}
\newcommand{\dipiz}{{\ensuremath{{\piz\piz}}}}
\newcommand{\dipi}{{\ensuremath{{\pi^+\pi^-}}}}
\newcommand{\dimu}{{\ensuremath{{\mu^+\mu^-}}}}
\newcommand{\dilep}{{\ensuremath{{\ell^+\ell^-}}}}

\newcommand{\rag}{\rangle}
\newcommand{\lag}{\langle}
\def\diq{\lag\bar{q}q\rag}

\def\lamQ{\Lambda_{\rm QCD}}
\def\MSbar{\overline{\rm MS}}
\def\LMSb{\Lambda_{\MSbar}}
\def\als{{\alpha_{\rm s}}}
\def\simg{{\ \lower-1.2pt\vbox{\hbox{\rlap{$>$}\lower6pt\vbox{\hbox{$\sim$}}}}\ }}

\newcommand{\clg}[1]{\ensuremath{{\cal #1}}}

\newcommand{\BoAf}{\ensuremath{\clg{B}/\clg{A}}}
\newcommand{\ReBA}{\ensuremath{\Re(\clg{B}/\clg{A})}}
\newcommand{\ImBA}{\ensuremath{\Im(\clg{B}/\clg{A})}}

\newcommand{\CoAf}{\ensuremath{\clg{C}/\clg{A}}}

\newcommand{\upsDec}[2]{\ensuremath{\Upsilon({#1}S) \rightarrow \Upsilon({#2}S) \pi \pi}\xspace}

\def\emrm{{{\rm em}}}

\newcommand{\sLj}[3]{{\,\,}^{#1}#2_{#3}}
\newcommand{\lhad}{{\rm LH}}

\newcommand{\Brat}{\ensuremath{{\clg{B}}}}

\newcommand{\etab}{\ensuremath{{\eta_b(1S)}}}
\newcommand{\etac}{\ensuremath{{\eta_c(1S)}}}
\newcommand{\etacp}{\ensuremath{{\eta_c(2S)}}}
\newcommand{\etabp}{\ensuremath{{\eta_b(2S)}}}

\newcommand{\etap}{\ensuremath{{\eta\,'}}}

\newcommand{\gpiz}{{\gamma\piz}}
\newcommand{\geta}{{\gamma\eta}}
\newcommand{\gepr}{{\gamma\etap}}

\newcommand{\jpgpiz}{{\jpsi\to\gpiz}}
\newcommand{\ppgpiz}{{\psip\to\gpiz}}
\newcommand{\ptgpiz}{{\psit\to\gpiz}}

\newcommand{\myspace}{{\ \ \ }}

\newcommand{\ttgeta}{{\myspace\to\geta}}

\newcommand{\ttgepr}{{\myspace\to\gepr}}

\newcommand{\sigabsj}{\ensuremath{\sigma_{\rm abs}^{\jpsi}}}

\newcommand{\sqrtsNN}{\mbox{$\sqrt{s_{_{NN}}}$}}
\newcommand{\sqrts}{\mbox{$\sqrt{s}$}}
\newcommand{\dAu}{d+Au}
\newcommand{\dCu}{d+Cu}
\newcommand{\pAu}{$p$+Au}
\newcommand{\pCu}{$p$+Cu}
\newcommand{\AuAu}{Au+Au}
\newcommand{\CuCu}{Cu+Cu}
\newcommand{\ppcoll}{$pp$}
\newcommand{\AAcoll}{$AA$}
\newcommand{\Ncoll}{\ensuremath{N_{\rm coll}}}
\newcommand{\RAA}{\mbox{$R_{AA}$}}
\newcommand{\RCP}{\mbox{$R_{CP}$}}
\newcommand{\RpAu}{\mbox{$R_{p{\rm Au}}$}}
\newcommand{\RdAu}{\mbox{$R_{\rm dAu}$}}
\newcommand{\RpCu}{\mbox{$R_{p{\rm Cu}}$}}

\def\PANDA{{$\overline{\mbox{\sffamily P}}${\sffamily ANDA}}\xspace}%

\def\pbarp{p\bar{p}}
\def\DbarD{\DDbar}
\def\cbarc{c\bar{c}}
\def\pbar{\bar{p}}
\newcommand{\psift}{\psi(4040)}
\newcommand{\psifto}{\psi(4160)}
\newcommand{\psiftf}{\psi(4415)}
\newcommand{\hsubc}{\ensuremath{h_c(1P)}}
\newcommand{\hsubb}{\ensuremath{h_b(1P)}}

\newcommand{\chicJ}{\chi_{cJ}}
\newcommand{\ptrans}{\ensuremath{p_{\rm T}}}
\newcommand{\chic}{$\chi_c$}
\newcommand{\chib}{$\chi_b$}

\newcommand{\polone}{\ensuremath{1\,^1D_2}}
\newcommand{\poltwo}{\ensuremath{1\,^3D_1}}

\newcommand{\kmax}{\ensuremath{k_{\rm max}}}
\newcommand{\kbind}{\ensuremath{k_{\rm b}}}
\newcommand{\Ebind}{\ensuremath{E_{\rm b}}}
\newcommand{\mred}{\ensuremath{m_{\rm r}}}

\begin{document}

\title{
{\rm
{\hbox to\textwidth{\hfill January 26, 2011}}
{\hbox to\textwidth{\hfill ~~}}
{\hbox to\textwidth{{{\hspace{0.5\textwidth}{TUM-EFT 11/10}}
      {\hfill CLNS~10/2066}}}}
{\hbox to\textwidth{{{\hspace{0.5\textwidth}{ANL-HEP-PR-10-44}
      {\hfill Alberta~Thy~11-10}}}}}
{\hbox to\textwidth{{{\hspace{0.5\textwidth}{CP3-10-37}
      {\hfill FZJ-IKP-TH-2010-24}}}}}
{\hbox to\textwidth{{{\hspace{0.5\textwidth}{INT-PUB-10-059 }
      {\hfill }}}}}
\small~\\
~\\
}
\LARGE Heavy quarkonium: progress, puzzles, and opportunities}

\thispagestyle{empty}

\renewcommand{\thefootnote}{\fnsymbol{footnote}}

\author{N.~Brambilla\footnotemark[1]\footnotemark[2]}
\affiliation{Physik-Department, Technische Universit\"at M\"unchen, James-Franck-Str. 1, 85748 Garching, Germany}

\author{S.~Eidelman\footnotemark[1]\footnotemark[2]}
\affiliation{Budker Institute of Nuclear Physics, Novosibirsk 630090, Russia}
\affiliation{Novosibirsk State University, Novosibirsk 630090, Russia}

\author{B.~K.~Heltsley\footnotemark[1]\footnotemark[2]\footnotemark[3]}
\affiliation{Cornell University, Ithaca, NY 14853, USA}

\author{R.~Vogt\footnotemark[1]\footnotemark[2]}
\affiliation{Physics Division, Lawrence Livermore National Laboratory, Livermore, CA 94551, USA}
\affiliation{Physics Department, University of California at Davis, Davis, CA 95616, USA}

\author{G.~T.~Bodwin\footnotemark[2]}
\affiliation{High Energy Physics Division, Argonne National Laboratory, 9700 South Cass Avenue, Argonne, IL 60439, USA}

\author{E.~Eichten\footnotemark[2]}
\affiliation{Fermi National Accelerator Laboratory, P.O. Box 500, Batavia, IL 60510, USA}

\author{A.~D.~Frawley\footnotemark[2]}
\affiliation{Physics Department, Florida State University, Tallahassee, FL, 32306-4350, USA}

\author{A.~B.~Meyer\footnotemark[2]}
\affiliation{Deutsches Elektronen-Synchrotron DESY, Hamburg, Germany}

\author{R.~E.~Mitchell\footnotemark[2]}
\affiliation{Indiana University, Bloomington, IN 47405, USA}

\author{V.~Papadimitriou\footnotemark[2]}
\affiliation{Fermi National Accelerator Laboratory, P.O. Box 500, Batavia, IL 60510, USA}

\author{P.~Petreczky\footnotemark[2]}
\affiliation{Physics Department, Brookhaven National Laboratory, Upton, NY 11973-5000, USA}

\author{A.~A.~Petrov\footnotemark[2]}
\affiliation{Department of Physics and Astronomy, Wayne State University, Detroit, MI 48201, USA}

\author{P.~Robbe\footnotemark[2]}
\affiliation{Laboratoire de l'Acc\'el\'erateur Lin\'eaire, IN2P3/CNRS and Universit\'e Paris-Sud 11, Centre Scientifique d'Orsay, BP 34, F-91898 Orsay Cedex, France}

\author{A.~Vairo\footnotemark[2]}
\affiliation{Physik-Department, Technische Universit\"at M\"unchen, James-Franck-Str. 1, 85748 Garching, Germany}

\author{A.~Andronic}
\affiliation{GSI Helmholtzzentrum f\"ur Schwerionenforschung, D-64291 Darmstadt, Germany}

\author{R.~Arnaldi}
\affiliation{INFN Sezione di Torino, Via P. Giuria 1, I-10125 Torino, Italy}

\author{P.~Artoisenet}
\affiliation{Department of Physics, The Ohio State University, Columbus, OH 43210, USA}

\author{G.~Bali}
\affiliation{Institut f\"ur Theoretische Physik, Universit\"at Regensburg, 93040 Regensburg, Germany}

\author{A.~Bertolin}
\affiliation{INFN Sezione di Padova, Via Marzolo 8, I-35131 Padova, Italy}

\author{D.~Bettoni}
\affiliation{Universit\`a di Ferrara and INFN Sezione di Ferrara, Via del Paradiso 12, I-44100 Ferrara, Italy}

\author{J.~Brodzicka}
\affiliation{Institute of Nuclear Physics Polish Academy of Sciences, Krak\'ow, Poland}

\author{G.~E.~Bruno}
\affiliation{Universit\`a di Bari and INFN Sezione di Bari, Via Amendola 173, I-70126 Bari, Italy}

\author{A.~Caldwell}
\affiliation{Max Planck Institute for Physics, M\"unchen, Germany}

\author{J.~Catmore}
\affiliation{Department of Physics, Lancaster University, Lancaster, LA1 4YB, UK}

\author{C.-H.~Chang}
\affiliation{CCAST (World Laboratory), P.O.Box 8730, Beijing 100190, China}
\affiliation{Institute of Theoretical Physics, Chinese Academy of Sciences, Beijing 100190, China}

\author{K.-T.~Chao}
\affiliation{Department of Physics, Peking University, Beijing 100871, China}

\author{E.~Chudakov}
\affiliation{Thomas Jefferson National Accelerator Facility, 12000 Jefferson Ave.,  Newport News, VA 23606, USA}

\author{P.~Cortese}
\affiliation{INFN Sezione di Torino, Via P. Giuria 1, I-10125 Torino, Italy}

\author{P.~Crochet}
\affiliation{Clermont Universit\'e, Universit\'e Blaise Pascal, CNRS-IN2P3, LPC, BP 10448, F-63000 Clermont-Ferrand, France}

\author{A.~Drutskoy}
\affiliation{University of Cincinnati, Cincinnati, OH 45221, USA}

\author{U.~Ellwanger}
\affiliation{Laboratoire de Physique Th\'eorique, Unit\'e mixte de Recherche - CNRS - UMR 8627, Universit\'e de Paris-Sud, F-91405 Orsay, France}

\author{P.~Faccioli}
\affiliation{LIP, Av. Elias Garcia 14, 1000-149 Lisbon, Portugal}

\author{A.~Gabareen~Mokhtar}
\affiliation{SLAC National Accelerator Laboratory, Stanford, CA 94309, USA}

\author{X.~Garcia~i~Tormo}
\affiliation{Department of Physics, University of Alberta, Edmonton, Alberta, Canada T6G 2G7}

\author{C.~Hanhart}
\affiliation{Institut f\"{u}r Kernphysik, J\"ulich Center for Hadron Physics, and Institute for Advanced Simulation, Forschungszentrum J\"{u}lich, D--52425 J\"{u}lich, Germany}

\author{F.~A.~Harris}
\affiliation{Department of Physics and Astronomy, University of Hawaii, Honolulu, HI 96822, USA}

\author{D.~M.~Kaplan}
\affiliation{Illinois Institute of Technology, Chicago, IL 60616, USA}

\author{S.~R.~Klein}
\affiliation{Lawrence Berkeley National Laboratory, Berkeley, CA 94720, USA}

\author{H.~Kowalski}
\affiliation{Deutsches Elektronen-Synchrotron DESY, Hamburg, Germany}

\author{J.-P.~Lansberg}
\affiliation{IPNO, Universit\'e Paris-Sud 11, CNRS/IN2P3, Orsay, France}
\affiliation{Centre de Physique Th\'eorique, \'Ecole Polytechnique, CNRS, 91128 Palaiseau, France}

\author{E.~Levichev}
\affiliation{Budker Institute of Nuclear Physics, Novosibirsk 630090, Russia}

\author{V.~Lombardo}
\affiliation{INFN Sezione di Milano, Via Celoria 16, I-20133 Milano, Italy}

\author{C.~Louren\c{c}o}
\affiliation{CERN, CH-1211 Geneva 23, Switzerland}

\author{F.~Maltoni}
\affiliation{Center for Cosmology, Particle Physics and Phenomenology, Universit\'e Catholique de Louvain, B-1348 Louvain-la-Neuve, Belgium}

\author{A.~Mocsy}
\affiliation{Department of Math and Science, Pratt Institute, 200 Willoughby Ave, ARC LL G-35, Brooklyn, NY 11205, USA}

\author{R.~Mussa}
\affiliation{INFN Sezione di Torino, Via P. Giuria 1, I-10125 Torino, Italy}

\author{F.~S.~Navarra}
\affiliation{Instituto de F\'{\i}sica, Universidade de S\~ao Paulo, C.P. 66318, 05315-970 S\~ao Paulo, SP, Brazil}

\author{M.~Negrini}
\affiliation{Universit\`a di Ferrara and INFN Sezione di Ferrara, Via del Paradiso 12, I-44100 Ferrara, Italy}

\author{M.~Nielsen}
\affiliation{Instituto de F\'{\i}sica, Universidade de S\~ao Paulo, C.P. 66318, 05315-970 S\~ao Paulo, SP, Brazil}

\author{S.~L.~Olsen}
\affiliation{Department of Physics \& Astronomy, Seoul National University, Seoul, Korea}

\author{P.~Pakhlov}
\author{G.~Pakhlova}
\affiliation{Institute for Theoretical and Experimental Physics, Moscow 117218, Russia}

\author{K.~Peters}
\affiliation{GSI Helmholtzzentrum f\"ur Schwerionenforschung, D-64291 Darmstadt, Germany}

\author{A.~D.~Polosa}
\affiliation{INFN Sezione di Roma, Piazzale Aldo Moro 2, I-00185 Roma, Italy}

\author{W.~Qian}
\affiliation{Department of Engineering Physics, Tsinghua University, Beijing 100084, China}
\affiliation{Laboratoire de l'Acc\'el\'erateur Lin\'eaire, IN2P3/CNRS and Universit\'e Paris-Sud 11, Centre Scientifique d'Orsay, BP 34, F-91898 Orsay Cedex, France}

\author{J.-W.~Qiu}
\affiliation{Physics Department, Brookhaven National Laboratory, Upton, NY 11973-5000, USA}
\affiliation{C.N.~Yang Institute for Theoretical Physics, Stony Brook University, Stony Brook, NY 11794-3840, USA}

\author{G.~Rong}
\affiliation{Institute of High Energy Physics, Chinese Academy of Sciences, Beijing 100049, China}

\author{M.~A.~Sanchis-Lozano}
\affiliation{Instituto de F\'\i sica Corpuscular (IFIC) and Departamento de F\'\i sica Te\'orica, Centro Mixto Universitat de Valencia-CSIC, Doctor Moliner 50, E-46100 Burjassot,  Valencia, Spain}

\author{E.~Scomparin}
\affiliation{INFN Sezione di Torino, Via P. Giuria 1, I-10125 Torino, Italy}

\author{P.~Senger}
\affiliation{GSI Helmholtzzentrum f\"ur Schwerionenforschung, D-64291 Darmstadt, Germany}

\author{F.~Simon}
\affiliation{Max Planck Institute for Physics, M\"unchen, Germany}
\affiliation{Excellence Cluster `Universe', Technische Universit\"at M\"unchen, Garching, Germany}

\author{S.~Stracka}
\affiliation{INFN Sezione di Milano, Via Celoria 16, I-20133 Milano, Italy}
\affiliation{Dipartimento di Fisica, Universit\`a di Milano, I-20133 Milano, Italy}

\author{Y.~Sumino}
\affiliation{Department of Physics, Tohoku University, Sendai, 980-8578 Japan}

\author{M.~Voloshin}
\affiliation{William I. Fine Theoretical Physics Institute, School of Physics and Astronomy, University of Minnesota, 116 Church Street SE, Minneapolis, MN 55455, USA}

\author{C.~Weiss}
\affiliation{Thomas Jefferson National Accelerator Facility, 12000 Jefferson Ave.,  Newport News, VA 23606, USA}

\author{H.~K.~W\"ohri}
\affiliation{LIP, Av. Elias Garcia 14, 1000-149 Lisbon, Portugal}

\author{C.-Z.~Yuan}
\affiliation{Institute of High Energy Physics, Chinese Academy of Sciences, Beijing 100049, China}

\date{ }


\date{January 26, 2011}
\maketitle

\footnotetext[1]{Editors}
\footnotetext[2]{Section coordinators}
\footnotetext[3]{Corresponding author:
\href{mailto:bkh2@cornell.edu}{\it bkh2@cornell.edu} }

{\small
\noindent 
{\bf Abstract} A golden age for heavy quarkonium physics dawned a decade
ago, initiated by the confluence of exciting advances in 
quantum chromodynamics (QCD) and an explosion of related experimental
activity. The early years of this period
were chronicled in the Quarkonium Working Group (QWG) CERN Yellow
Report (YR) in 2004, which presented a comprehensive review
of the status of the field at that time and provided specific
recommendations for further progress. However, the broad spectrum
of subsequent breakthroughs, surprises, and continuing 
puzzles could only be partially anticipated. Since the release of
the YR, the BESII program concluded only to give birth to BESIII;
the $B$-factories and CLEO-c flourished;
quarkonium production and polarization 
measurements at HERA and the Tevatron matured; 
and heavy-ion collisions at RHIC have opened a window
on the deconfinement regime.
All these experiments leave legacies of quality, precision, and unsolved
mysteries for quarkonium physics,
and therefore beg for continuing investigations at BESIII,
the LHC, RHIC, FAIR, the 
Super Flavor and/or Tau-Charm factories, JLab, the ILC, and beyond.
The list of newly-found conventional states 
expanded to include $\hsubc$, $\chi_{c2}(2P)$, $B_c^+$, and \etab.
In addition, the
unexpected and still-fascinating $X(3872)$ has been joined by
more than a dozen other charmonium- and bottomonium-like ``$XYZ$'' states
that appear to lie outside the quark model.
Many of these still need experimental confirmation. 
The plethora of new states unleashed a flood of theoretical
investigations into new forms of matter such as quark-gluon hybrids, 
mesonic molecules, and tetraquarks. Measurements of the spectroscopy, 
decays, production, and in-medium behavior of $c\bar{c}$, $b\bar{b}$,
and $b\bar{c}$ bound states have been shown to validate some
theoretical approaches to QCD
and highlight lack of quantitative success for others.
Lattice QCD has grown from a tool with computational possibilities to
an industrial-strength effort 
now dependent more on insight and innovation than pure computational power. 
New effective field theories for the description of quarkonium in 
different regimes have been developed and brought to a high 
degree of sophistication, 
thus enabling precise and solid theoretical predictions.
Many expected decays and transitions have either been
measured with precision or for the first time, but the 
confusing patterns of decays, both above and below
open-flavor thresholds, endure and have deepened.
The intriguing details of quarkonium
suppression in heavy-ion collisions that have emerged from RHIC
have elevated the importance of separating hot- and cold-nuclear-matter
effects in quark-gluon plasma studies.
This review systematically addresses all these matters
and concludes by prioritizing directions for 
ongoing and future efforts.}


\renewcommand{\thefootnote}{\arabic{footnote}}

{\small \tableofcontents}

\clearpage
\section*{Dedication}

We dedicate this review to the memories of three
friends of the Quarkonium Working Group. 
Richard Galik, building on his leadership role
in quarkonium physics at CLEO, supported and guided our efforts 
from the start. His relentless focus, unyielding 
objectivity, and insistent collegiality continue 
to inspire those who knew him.
Beate Naroska pioneered quarkonium measurements at HERA
and enthusiastically advocated for the 2007 QWG meeting at DESY
as a member of the local organizing committee.
A visionary scientist and dedicated teacher, she made 
invaluable and enduring contributions to our field.
Andrzej Zieminski had a longstanding devotion
to the study of quarkonium production in hadron collisions.
He worked energetically to ensure the success of the 
inaugural QWG meeting in 2002 as a convener of the 
QCD Tests Working Group, and continued his sustaining commitment
as liason to the \DZero\ Collaboration.
We remember Rich, Beate, and Andrzej with fondness and
gratitude.

\section{Introduction}

Heavy quarkonium is a multiscale system which can probe all regimes of 
quantum chromodynamics (QCD). At high energies, a perturbative 
expansion in the strong coupling constant $\als(Q^2)$ is possible. 
At low energies, nonperturbative effects dominate. In between, the 
approximations and techniques which work at the extremes may not
succeed, necessitating more complex approaches. Hence heavy 
quarkonium presents an ideal laboratory for testing
the interplay between perturbative and nonperturbative QCD 
within a controlled environment. To do so in a 
systematic manner requires the intersection of many avenues of inquiry:
experiments in both particle physics and nuclear physics are required;
perturbative and lattice QCD 
calculations must be performed in conjunction with one another;
characteristics of confinement and deconfinement in matter
must be confronted; phenomenology should be pursued both within the
Standard Model and beyond it. Above all, experiments must continue to
provide measurements which constrain and challenge all aspects
of QCD, and theory must then guide experiment toward the next 
important observables.

Effective field 
theories\footnote{EFTs such as HQEFT, NRQCD, pNRQCD, SCET, ..., 
are described elsewhere in this article.}
(EFTs) describing quarkonium processes 
have continued to develop and now provide a unifying description 
as well as solid and versatile tools yielding well-defined predictions.
EFTs rely on higher-order perturbative calculations and 
lattice simulations. Progress on both fronts has
improved the reach and precision of EFT-based predictions,
enabling, \eg the
increasingly precise determinations of several fundamental
parameters of the Standard Model (\ie $\als$, $m_c$, and $m_b$).

Several experiments operating during this era, primarily \babar\ 
at SLAC and Belle at KEK), CLEO-III and CLEO-c at CESR, 
CDF and \DZero\ at Fermilab, and BESII and BESIII at IHEP 
have, in effect, operated as quarkonium factories, vastly
increasing the available data on quarkonia spectra and decays.
Over the same period, investigations of quarkonium production
in fixed target experiments at Fermilab and CERN,
HERA-B at DESY, and PHENIX and STAR at RHIC have
vastly increased the knowledge base for cold- and hot-medium studies.
The resulting variety of collision types, energy regimes, 
detector technologies, and analysis techniques has yielded quarkonium-related 
physics programs that are both competitive and complementary. 
Taken together, the experimental programs provide the 
confirmations and refutations of newly observed phenomena 
that are crucial for sustained progress in the field as well as the
breadth and depth necessary for a vibrant quarkonium research environment.

The Quarkonium Working Group (QWG) was formed in 2002 as a dedicated
and distinct effort to advance quarkonium studies by drawing 
sometimes disparate communities together in this common cause.
QWG activities bring experts in theory and experiment 
together to discuss the current status and
progress in all the relevant subfields. 
Subsequent participant interactions are intended to synthesize
a consensus of progress and priorities going forward.
Periodic QWG meetings have been effective in achieving this function.
The exhaustive {\it CERN Yellow Report}~\cite{Brambilla:2004wf},
the first document produced by QWG
detailing the state of quarkonium physics and suggestions
for future efforts, was 
released in 2004 to embody such a synthesis. 
Since that report appeared, 
much has been accomplished in theory and experiment,
warranting an updated review.

This review provides a comprehensive exploration of heavy quarkonium
physics applicable to the landscape of 2010, with particular emphases 
on recent developments and future opportunities. The presentation 
is organized into five broad and frequently overlapping categories: 
\begin{itemize}
\item {\it Spectroscopy} (\Sec{sec:SpecChapter}), which focuses on the 
existence, quantum numbers, masses, and widths of 
heavy quarkonium (or quarkonium-like) bound states;
\item {\sl Decay}  (\Sec{sec:DecChapter}), 
an examination of the patterns and properties of quarkonia
transitions and decays, with special attention given to the decay dynamics and
exclusive final-state branching fractions;
\item {\sl Production}  (\Sec{sec:ProdChapter}), 
the study of heavy quarkonium creation in
$\epem$, $p\bar p$, $\ell p$, $\gamma p$, and $pp$ collisions; 
\item {\sl In medium}  (\Sec{sec:MedChapter}), the investigation of
deconfinement and formation of quark-gluon plasma
in heavy-ion collisions via measurement of quarkonium suppression;
\item {\sl Experimental outlook}  (\Sec{sec:FutChapter}), 
the status and physics reach of new and planned experimental facilities.
\end{itemize}
Below we briefly introduce and motivate each of these sections.

Heavy quarkonium {\sl spectroscopy} examines the tableau of
heavy-quark bound states, thereby providing the starting point for
all further investigations. Which states exist? Why?
What are their masses, widths, and quantum numbers?
Which states should exist but have not yet been observed?
Does QCD fully explain the observed terrain? If not, why?
New experimental and theoretical efforts 
over the last decade have provided some answers to
these questions, while also raising new ones. 
Some long-anticipated states have, at last, been measured
(\eg $\hsubc$, $\etacp$, and $\etab$),
while many unanticipated states (\eg $X(3872)$ and $Y(4260)$) also appeared. 
Does the underestimation of the $\etab$
hyperfine splitting by some QCD calculations indicate faults in application of
theory, inaccuracy of measurements, or the presence of new physics?
Have we observed mesonic molecules? 
Tetraquarks? Quark-gluon hybrids?
How would we know if we had? How many of the new states 
are experimental artifacts? 
Do $X(3872)$ decay patterns comport with those of any conventional
quarkonium? Is $X(3872)$ above or below \DstnDn\ threshold?
Is the $\epem$ hadronic cross section enhancement near
10.86\gev\ simply the \UnS{5}\ resonance or does \UnS{5}\ overlap
with a new $Y_b$ state, as suggested by recent
dipion transition data?
These questions, among many others, animate 
continuing theoretical and experimental spectroscopic investigations.

For states away from threshold, theory provides a description, 
at the level of the binding-energy scale, in the form of an EFT called
pNRQCD. Precise and accurate calculation of the \etab\ hyperfine splitting 
remains a challenge for both perturbative and lattice calculations.
With one exception, no EFT description has yet been constructed nor have the 
appropriate degrees of freedom been clearly identified for
most new states close to threshold. The exception is
$X(3872)$, which displays universal characteristics 
due to its proximity to \DstnDn\ threshold, thus prompting
a plethora of calculations based on a single elegant formalism.
Spectroscopy has advanced from both direct and EFT-formulated 
lattice calculations. In general, however, the threshold regions 
remain troublesome for the lattice as well as EFTs, 
excited-state lattice calculations have been only recently 
pioneered, and the full treatment of bottomonium on the lattice
remains a challenge.

A substantial challenge in the realm of quarkonium {\sl decay} is for
theory to keep pace with the large number of new measurements.
These include increasingly precise measurements of prominent
decay modes (\eg dilepton branching fractions and widths of \jpsi\ and
\Ups, branching fractions for and dynamical descriptions of
dipion transitions from \psip\ and \UnS{n}), 
and first measurements or important refinements
of previously low-statistics results (\eg 
$\jpsi\to 3\gamma$;
$\jpsi\to\gamma\etac$;
$\UnS{4}\to\dipi\UnS{1S,2S,3}$), 
and the burgeoning lists of
exclusive hadronic decay modes (\eg $\etac$ and $\chi_{bJ}$).
Some previously puzzling situations 
(\eg theory-experiment disagreements for higher-order multipoles
in $\psip\to\gamma\chi_{cJ}$, $\chi_{cJ}\to\gamma\jpsi$)
have been resolved by improved measurements while others
(\eg the $\rho\pi$ puzzle, suppressed \psip\ and \UnS{1} decays
to $\gamma\eta$) remain. Has the two-peak dipion
mass structure in $\UnS{3}\to\dipi\UnS{1}$ been explained?
What exactly is the source of the distorted photon lineshape
in $\jpsi\to\gamma\etac$? Does the \psit\ have 
\nonDDbar\ decay modes summing to more than $\sim 1\%$?
Our review of decays details new measurements and 
addresses these and related questions.

For a quarkonium with a small radius, an EFT description of radiative 
magnetic dipole transitions has been recently obtained,
replacing the now-outdated model description; its extension to 
electric dipole transitions and to states with larger radius is 
needed.
Steady improvement in NRQCD inclusive decay-width calculations has taken 
place in higher-order expansions in the 
velocity and strong coupling constant as well as in the 
lattice evaluation of matrix elements. 
Predictions match measurements adequately at the level of ratios 
of decay widths. Further improvements would require the lattice 
calculation or data extraction of the NRQCD matrix elements and 
perturbative resummation of large contributions to the 
NRQCD matching coefficients.
The new data on hadronic transitions and hadronic decays pose interesting 
challenges to the theory.

The pioneering measurements of quarkonium {\sl production} at 
the Tevatron were carried out in the early 1990s.
Soon after, NRQCD factorization became the standard
tool for theoretical calculations. Since then,
the Tevatron, $B$-factories, and HERA have all 
performed important measurements, some of which have
given rise to inconsistencies, puzzles, 
and new challenges for both theory and experiment.
Among these are apparent inconsistencies in 
quarkonium polarization at the Tevatron between
Run~I and Run~II for the \jpsi,
between CDF and \DZero\ for the \Ups, and between experiment
and NRQCD factorization predictions for both.
At least as surprising was the
observation at the $B$-factories that close to 60\% 
of \epem\ collisions 
that contain a \jpsi\
also include a charm meson pair.
Photoproduction measurements at HERA revealed discrepancies with LO NRQCD
factorization predictions. In response
to these and other challenges, the theory of quarkonium production has
progressed rapidly. 

NRQCD factorization is the basis for much of the current theoretical
work on quarkonium production. Factorization in exclusive quarkonium
production has recently been proven to all orders in perturbation theory
for both double-charmonium production in \epem\
annihilation and $B$-meson decays to charmonium and a light meson. 
NRQCD factorization for inclusive quarkonium
production has been shown to be valid at NNLO. 
However, an all-orders
proof of factorization for inclusive quarkonium production remains
elusive. This is a key theoretical issue, as a failure of factorization
at any order in perturbation theory would imply that there are large, non-factorizing
contributions, owing to the presence
of soft-gluon effects.

Corrections to hadroproduction have been calculated at NLO, and, 
in the case of the color-singlet channel, partially at NNLO, even 
though just a few years ago these calculations were thought to be 
barely possible. 
The new calculations show that, because of kinematic enhancements, 
higher-order corrections can be orders of magnitude larger than the 
Born-level contributions.
In the case of double-charmonium production in $e^+e^-$ collisions, 
relativistic and perturbative corrections increased the predicted 
cross sections by more than a factor of four, bringing them into 
agreement with experiment.
New NRQCD factorization calculations of quarkonium 
photoproduction to NLO at HERA have also moved 
predictions into agreement with experiment.
The importance of higher-order corrections has raised the issue of the
convergence of the perturbation series. New methods to address this
issue are on the horizon.

New observables have been proposed that may help us to understand the
mechanisms of quarkonium production. For example, alternative methods
for obtaining information about the polarization of produced quarkonia
have been suggested. The associated production of quarkonia may also be an
important tool in understanding new states. The  production
characteristics of the $X(3872)$ may shed light on its exotic nature.
The improved theoretical landscape will soon be confronted with the
first phase of running at the LHC, where 
charmonium and bottomonium production will be measured with high statistics
in a greatly extended kinematic range.

The study of quarkonium {\sl in medium} has also undergone crucial 
development. The large datasets from heavy-ion collisions 
at RHIC suggest that the quark-gluon plasma is actually more like a liquid 
than a plasma. The suppression of quarkonium production in a 
hot medium was proposed as a clean probe of deconfined matter.
However, the use of quarkonium yields as a diagnostic 
tool of the hot medium has turned out to be quite challenging.
Indeed, quarkonium production was already found to be 
suppressed by cold-nuclear-matter 
effects in proton-nucleus collisions. Such effects require dedicated experimental and 
theoretical attention themselves. In high-energy environments 
such as at heavy-ion colliders, where more than one 
$Q \overline Q$ pair may be produced in a collision,
coalescence of $Q$ and $\overline Q$ can lead to secondary 
quarkonium production, requiring understanding of the transport 
properties of the medium to separate 
primary and secondary quarkonium production.  
The interpretation of in-medium hot matter effects requires 
understanding the $Q \overline Q$ interaction in terms of finite
temperature ($T$) QCD.
The successful Hard Thermal Loop effective theory integrates over the
hardest momenta proportional to $T$ for light quark and gluon observables.
To extend the Hard Thermal Loop theory to heavy quarkonium at finite
temperature, the additional scales introduced by the bound state must be
taken into account.
Recently there has been significant progress in constructing a perturbative 
EFT description of quarkonium at finite $T$, resulting in a clearly defined
potential.  This potential displays characteristics that are considerably different 
from the phenomenological, lattice-inspired description 
used up to now with well-defined phenomenological implications, as we further
discuss.
The higher energy of the heavy-ion collisions at the LHC will expand the study
of quarkonium in media to bottomonia production and suppression.  These studies
will be crucial for arriving at a uniform description of heavy quarkonia in
cold and hot nuclear matter.

Lastly, we turn our attention to a discussion of the {\sl experimental outlook}
in the near term as well as longer-term prospects.  For the LHC, the future is
now, with the first quarkonium data presented this year.  While the preliminary
data are encouraging, the full potential for LHC quarkonium
studies is still to come.  There is a future in low energy 
quarkonium hadroproduction studies as well, including two experiments at 
GSI in Darmstadt, Germany. \PANDA will make precision spectroscopy studies
in $\bar{p} p$ and $\bar p A$ interactions, while the CBM detector will
make fixed-target studies of $pA$ and $AA$ interactions to further the 
understanding of quarkonium production and suppression in high 
baryon-density matter. Quarkonium physics goals at the currently-running 
BESIII, as well as at proposed super flavor and tau-charm factories are 
also discussed. Measurements of quarkonium photoproduction 
offer important insight into the gluon generalized parton distribution 
(GPD) in nuclei,
the role of color correlations, and the color-dipole nature of 
quarkonia undergoing elastic scattering at high energies.
These investigations can be performed at JLab, CERN, and the EIC 
in the medium term at lower energies,
whereas higher energy studies will have to await the ENC, EIC, or LHeC. 
Important top quark measurements with high precision 
can be performed
at a future \epem\ linear collider
(ILC or CLIC) in the region just above $t\bar t$ threshold.
Overall, an extremely active and ambitious future 
lies ahead
for the study of heavy quarkonia 
with new facilities.

\clearpage
\section[Spectroscopy]{Spectroscopy\protect\footnotemark[2]}
\addtocounter{footnote}{1}
\footnotetext[2]{
Contributing authors:
N.~Brambilla$^\dag$, B.~K.~Heltsley$^\dag$, A.~A.~Petrov$^\dag$, 
G.~Bali, S.~Eidelman,
U.~Ellwanger, A.~Gabareen Mokhtar, X.~Garcia~i~Tormo, R.~Mussa, F.~S.~Navarra,
M.~Nielsen, S.~L.~Olsen, P.~Pakhlov, G.~Pakhlova, A.~D.~Polosa,
M.~A.~Sanchis-Lozano, Y.~Sumino, and M.~Voloshin}
\label{sec:SpecChapter}

  Spectroscopy is, in part, bump-hunting in mass spectra.
Of late, progress has occurred mostly at $e^+e^-$ colliding beam 
facilities (BES at BEPC, CLEO at CESR, \babar\ at PEP-II, Belle at KEKB,
KEDR at VEPP-4M), 
but other venues have gotten into the game as well, including E835 
at Fermilab ($\bar{p}p$ gas-jet target) and CDF and \DZero\ at the Tevatron 
$p\bar{p}$ collider. Tevatron searches target inclusive 
production of a fully reconstructed state, and can best succeed when 
the presence of leptons (\eg $\jpsi\to\mu^+\mu^-$) 
or displaced vertices (\eg $B$-decay)
can suppress backgrounds and when there is no need to 
observe photons. The main strength 
of $e^+e^-$ colliders is the capability to
obtain large datasets at or near
charmonium and/or bottomonium vector state masses
with well-known initial-state quantum 
numbers and kinematics. Modern $e^+e^-$ detectors feature
precision charged particle trackers, electromagnetic calorimeters, 
Cherenkov-radiation imagers or 
time-of-flight taggers, and muon filters, which together
allow measurement of the individual decay remnants:
$\gamma$, $e^\pm$, $\mu^\pm$, $\pi^\pm$, $K^\pm$, $p(\bar{p})$.
These capabilities in $e^+e^-$ collisions are exploited using 
the following techniques:

{\bf \underline{Full event reconstruction}:} 
Datasets taken on-resonance at vector quarkonium masses 
allow full reconstruction of cascade 
transitions involving another state
(\eg $\psip\to\pi^0 h_c$, $h_c\to\gamma\etac$ or 
$Y(4260)\to\ppjp$).

{\bf \underline{Inclusive spectra}:} One or more 
final state particles are selected in each event.
The measured four-momenta are then used to search
directly for mass peaks, or indirectly via missing mass;
\ie the mass recoiling against the particle(s) selected.
Two examples
are an inclusive photon or \piz\ momentum spectrum to identify
transitions (\eg $\UnS{3}\to\gamma\eta_b$ or
$\psip\to\piz \hsubc$), which typically have small signals
on very large backgrounds. In the continuum reaction
$e^+e^-\to X\jpsi$ with $\jpsi\to\ell^+\ell^-$
(double-charmonium production),
the unmeasured particle(s) $X$ can be identified via
peaks in the missing mass spectrum.

{\bf \underline{Energy scan}:} Scans in $e^+e^-$ center-of-mass energy
($\sqrt{s}$) can map out vector resonances via either inclusive
hadronic-event counting ($R$) and/or exclusive final states
(\eg $\DDst$). This does not use machine time efficiently,
however, because accelerators work best when operated at a single
energy for a long time so tuning can optimize the instantaneous luminosity.
Competing priorities usually limit the duration of such scans.

{\boldmath\bf \underline{$\gamma\gamma$-fusion}:} The process 
$e^+e^-\to e^+e^- \gamma^{(*)}\gamma^{(*)} \to e^+e^- X$ allows searches
for a large range of masses for $X$, but $X$ is restricted to
having spin-0 or 2 and positive $C$-parity (\eg \etacp\
or $\chi_{c2}(2P)$). The outgoing $e^+e^-$ tend to escape the detector 
at small angles, leaving $X$ with very small momentum transverse to 
the beamline.

{\bf\underline{ISR}:} Initial-state radiation~(ISR) allows access to all
vector states with masses below the $\sqrt{s}$ value of the $e^+e^-$
collision. The effective luminosity per unit of radiative
photon energy (or the mass recoiling against it) is well-known,
allowing for well-normalized exposures at all masses.
The ISR photon
tends to escape the detector at small angles, leaving the recoiling
state with small momentum transverse to the beam
but large momentum along it. While the rate for ISR
production at any fixed $\sqrt{s}$ is small per unit luminosity,
factory-sized datasets at \babar\ and Belle make this
a viable tool (\eg $e^+e^-\to\gamma Y(4260)$, 
$e^+e^-\to\gamma \DDst$), and the simultaneous
exposure of all allowed masses recoiling against
the ISR photon contrasts with the discrete points 
available from a direct $e^+e^-$ scan.

{\boldmath\bf\underline{$B$-decays}:} 
Large $B$-factory datasets at the $\UnS{4}$
make it possible to utilize 
two-body kinematics
to search for exclusive decays of $B$-mesons (\eg $B\to K Z_i^+)$.

\begin{figure}[tb]
   \includegraphics*[width=\figwid]{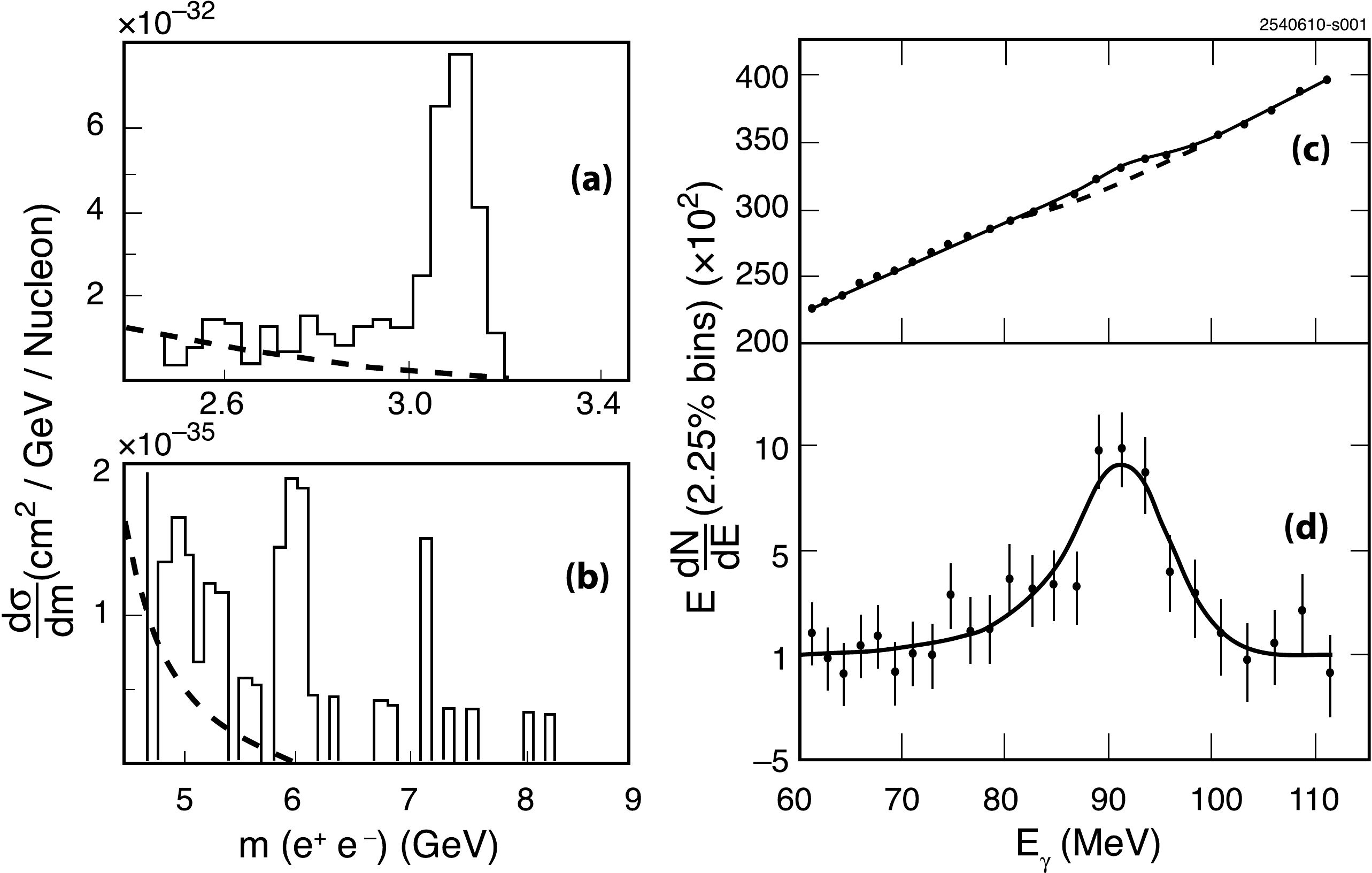}
   \caption{(a),~(b)~Dielectron mass distributions from inclusively
            selected proton-on-beryllium collisions~\cite{Hom:1976cv},
            in which {\it dashed curves} indicate estimated backgrounds.
            (c)~Inclusive photon energy distribution in 
            $\psip$ decays~\cite{Edwards:1981mq} {\it (points with error bars)},
            showing a fit {\it (solid line)} to a signal and a smooth
            background {\it (dashed line)}. 
            Part~(d) is the same as (c)
            but with the smooth backgrounds
            subtracted. Both peaks were later refuted.
            \AfigPermsAPS{Hom:1976cv,Edwards:1981mq}{1976, 1982}
            }
\label{fig:Spec_false}
\end{figure}

It is worth emphasizing here that the key tenet
of experimental science is that discoveries must
be reproducible and {\it verified by independent parties}
as a prerequisite for general acceptance. This is
no small point in a period such as the present
when the world has been bombarded with
more than one new state found per year.
It is worth pondering spectra which initially were thought to
indicate new states in the heavy quarkonium mass region 
but were later proven otherwise by independent measurements.
Such occurrences happen to the best of experiments
and most often can be attributed to fluctuations and/or 
extremely subtle systematic effects, not overt analysis blunders. 
\Figure{fig:Spec_false} highlights two
such examples. \Figures{fig:Spec_false}(a) and (b)
show dielectron mass distributions observed~\cite{Hom:1976cv} inclusively
in 400\gev\ proton collisions on beryllium. A $\jpsi$ peak
is clearly seen in (a) while there is an apparent peak near 6~GeV
in (b). The authors estimated a 2\%\ probability
that the 6~GeV peak was due to background fluctuations,
explicitly cautioned that confirmation was needed, and
suggested that the name ``$\Upsilon$'' be given to
this or any subsequently confirmed high-mass dilepton
peak. The 6~GeV phenomenon was not confirmed several months 
later in a dimuon version~\cite{Snyder:1976ie} of the same experiment.
The same authors discovered the true $\Upsilon(1S)$ shortly 
thereafter~\cite{Herb:1977ek}.
\Figures{fig:Spec_false}(c) and (d) show an inclusively selected 
photon-energy distribution in $\psip$ decays~\cite{Edwards:1981mq}.
The size of the peak near 91\mev\ 
represents a branching fraction of 0.8\% and has
statistical significance of $>6\sigma$.
The peak position corresponds to a mass recoiling
against the photon of 3594$\pm$5\mevcc\ and a width $\Gamma$$<$8\mevcc.
The result stood as an $\etacp$ candidate for twenty
years. It was finally refuted~\cite{Athar:2004dn} 
as having $\cal B$$<$0.2\% at 90\%~CL (confidence level).
Incidents such as these (and many others) have led
many experiments to adopt more stringent criteria
and procedures for establishing signals. These
include requiring a threshold of ``5$\sigma$'' 
statistical significance for claiming ``observation'', 
allowing systematic variations to reduce the reported significance, 
tuning of selection criteria on small subsamples of data not
used in the signal search, and the intentional obscuring
of signal regions until cuts are frozen (so-called ``blind'' analysis).
{\it However, every potential signal deserves independent 
confirmation or refutation.}

This section will first focus on recent measurements:
What is the current status of each state?
How were the measurements performed?
Which need verification? Which are in conflict?
Then the theoretical issues will be addressed.

\subsection{Conventional vectors above open-flavor threshold }
\label{sec:SpecExp_ConVec}

Here we describe recent measurements relevant to the determinations
of mass, width, and open-charm content of the four known
vector charmonia above open-charm threshold.
These states were first observed thirty years ago in \epem\ 
annihilation as enhancements in the total hadronic cross 
section~\cite{Siegrist:1976br,Rapidis:1977cv,Bacino:1977uh,Brandelik:1978ei,Abrams:1979cx}.
No update of their parameters was made until 2005, when a combined 
fit to the Crystal Ball~\cite{Osterheld:1986hw} and BES~\cite{Bai:2001ct} 
$R$-measurements was performed by Seth~\cite{Seth:2005ny}.
Even more recently, BES~\cite{Ablikim:2007gd} reported
new parameter values for the $\psi$ resonances.
A plethora of open charm cross section measurements
has become available and is discussed in what follows.
Finally, recent studies of resonant structures just above
the open-bottom threshold are described.

\subsubsection{Vectors decaying to open charm}
\label{sec:SpecExp_ConVecCharm}

\begin{figure}[b]
\includegraphics*[width=\figwid]{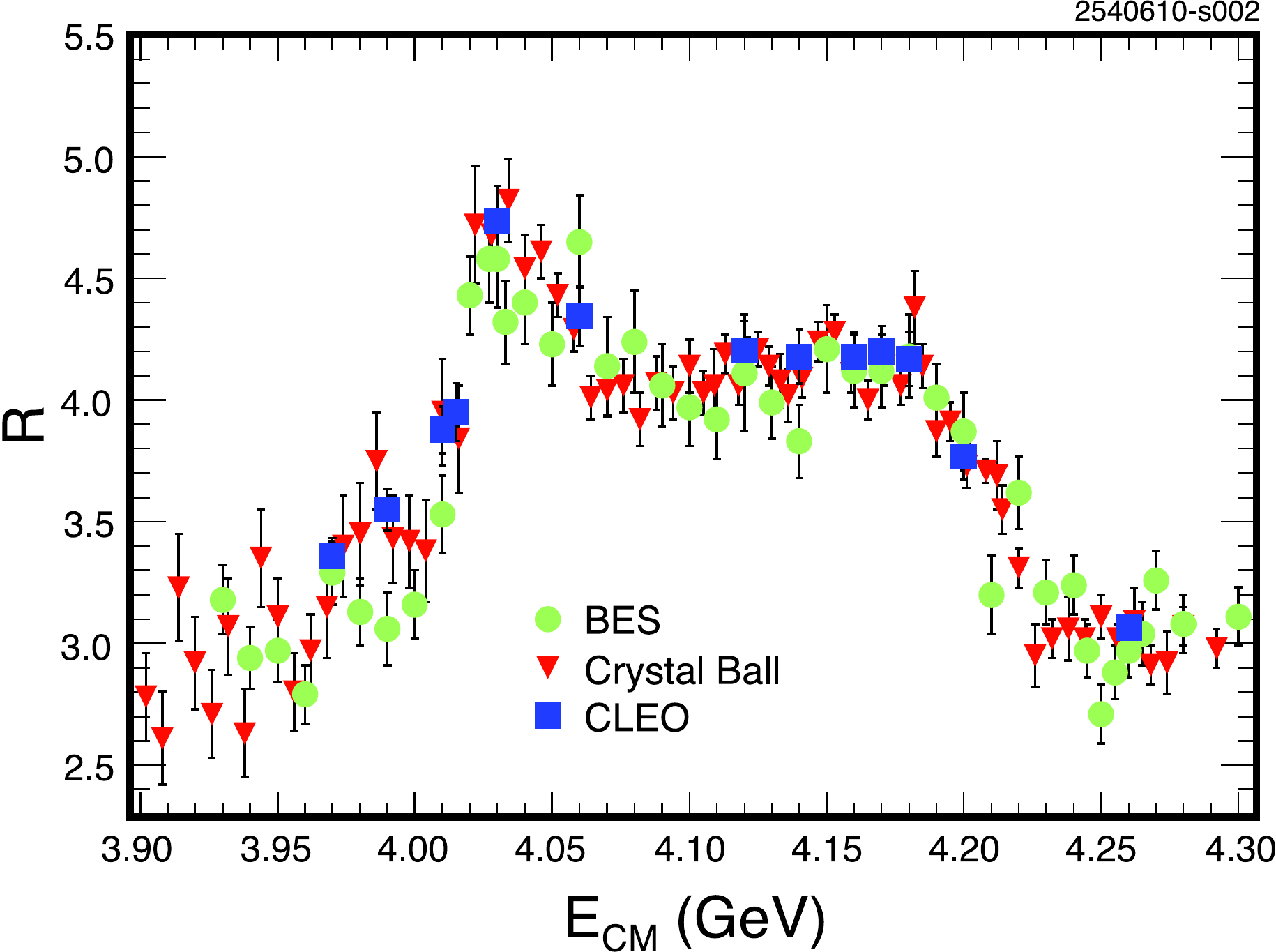}
\caption{Measurements of $R$, including radiative corrections,
in the open charm region. From 
Crystal Ball~\cite{Osterheld:1986hw},
BES~\cite{Bai:2001ct},
and CLEO~\cite{CroninHennessy:2008yi}.
\AfigPermAPS{CroninHennessy:2008yi}{2009}
 }
\label{fig:Spec_RCLEO}
\end{figure}

The total cross section for hadron production in \epem\ annihilation is
usually parametrized in terms of the ratio $R$, defined as
\beq
R=\frac{\sigma(\epem\to\mbox{hadrons})}{\sigma(\epem\to\mu^+\mu^-)}\,, 
\eeq
where the
denominator is the lowest-order QED cross section,
\beq
\sigma(\epem\to\mu^+\mu^-)=\frac{4\pi\alpha^2}{3s}\,.
\eeq
Away from flavor thresholds, measured $R$ values
are consistent with the three-color quark model predictions
plus terms governed by QCD and the running of $\als(Q^2)$.
Resonant states in the vicinity of flavor thresholds can
be studied with fits of measured $R$ distributions.
As a part of a study of open charm cross sections
in the region from 3.97-4.26\gev, CLEO~\cite{CroninHennessy:2008yi}
published radiatively corrected $R$-values
as shown in \Fig{fig:Spec_RCLEO}. These are in good agreement with earlier
measurements~\cite{Osterheld:1986hw,Bai:2001ct}, which are also shown,
demonstrating that in this energy range $R$ values are reasonably 
well-vetted experimentally.

\begin{figure}[tbp]
\includegraphics*[width=\figwid]{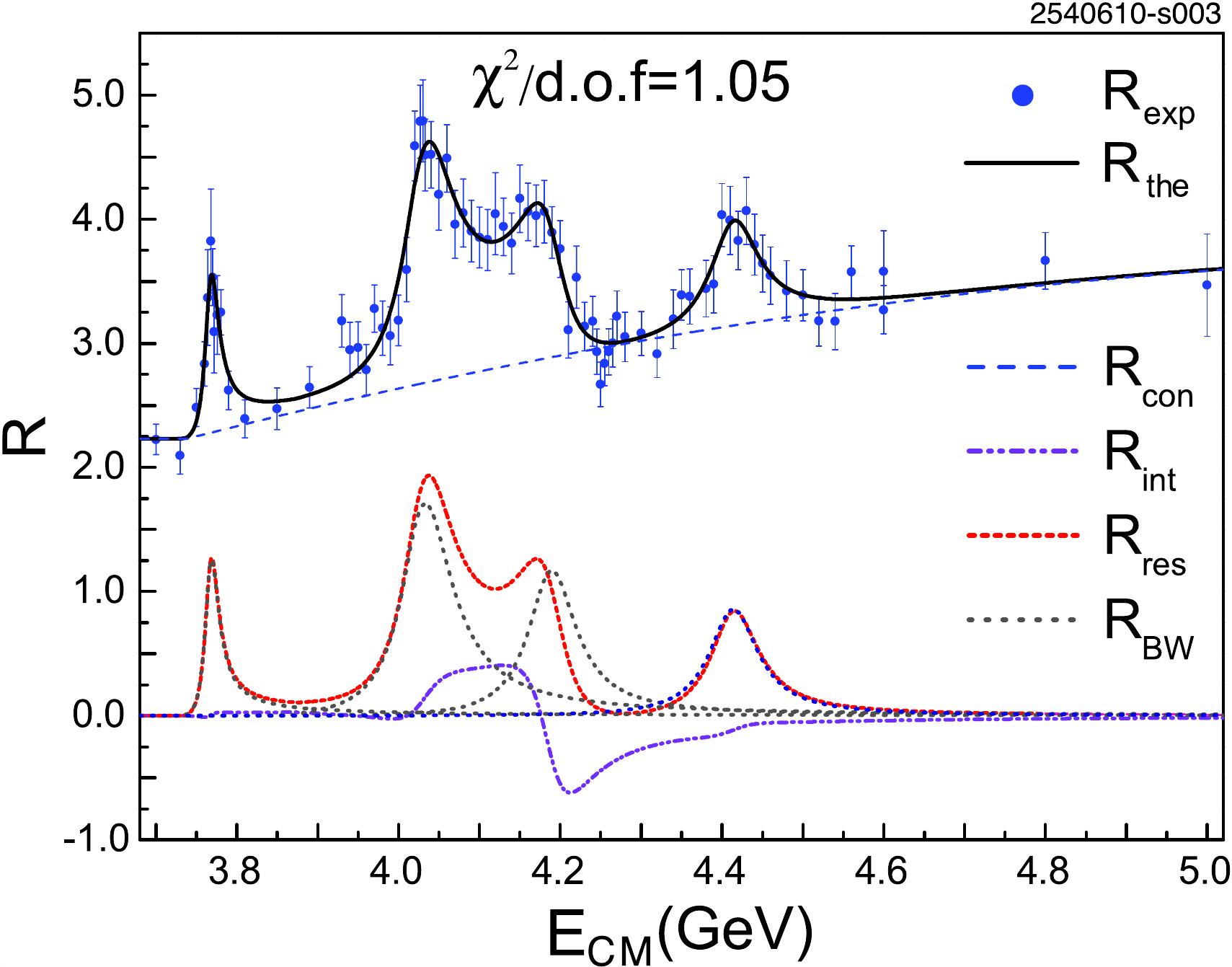}
\caption{From BES~\cite{Ablikim:2007gd}, measured $R$ values from data
  {\it (dots with error bars)} and {\it curves} showing
  the best fit {\it (solid)} and the contributions of its components: 
  continuum background {\it (long dashed)},
  individual resonance {\it (dotted)},
  interference {\it (dash-dot-dot)}, and
  the summation of the non\-background curves {\it (short dashed)}.
  \AfigPermPLB{Ablikim:2007gd}{2008}
}
\label{fig:Spec_RBES}
\end{figure}

The extraction of resonance parameters from such $R$ measurements, however,
has evolved in complexity, causing systematic movement in some of the
parameters over time. The latest BES~~\cite{Ablikim:2007gd} 
fit to their $R$-scan 
data is more sophisticated than previous efforts 
and includes the effects of interference and relative phases,
as shown in \Fig{fig:Spec_RBES} and \Tab{tab:Spec_Rparam}.
To take into account interference, BES relied on model
predictions for branching fractions of $\psi$ states into all possible
two-body charm meson final states. 
Thus the measured parameters from this fit still include some 
model uncertainties which are difficult to estimate. Other systematic
uncertainties are estimated using alternative choices and combinations
of Breit-Wigner forms, energy dependence of the full width, and
continuum charm background. It was found that the results are
sensitive to the form of the energy-dependent total width but are not
sensitive to the form of background. 

  In a separate analysis,
BES~\cite{Ablikim:2008zzc} fit their $R$ data from 3.65-3.90\gev,
finding a 7$\sigma$ preference for two interfering lineshapes
peaked near 3763 and 3781\mevcc\ relative to a single such 
shape for the $\psi(3770)$,
although other sources for the observed distortion of a pure
$D$-wave Breit-Wigner are possible
(see also \Sec{sec:Decay_nonDDbar}).
A very recent preliminary analysis of KEDR~\cite{TodyshevICHEP:2010} 
\epem\ scan data near the \psit\ applies an extended vector dominance model
and includes interference with the tail of the \psip\ resonance,
concluding that the latter interference causes a significant shift
upward in the fitted peak of the \psit\ as compared to
most previous fits, including those of BES. The KEDR measurements
are not consistent with the two-peak distortion seen by BES.

\begin{table}[t]
\caption{The resonance parameters of the high-mass charmonia from the
  BES global fit~\cite{Ablikim:2007gd} together with the values from
  PDG04~\cite{Eidelman:2004wy}, Seth~\cite{Seth:2005ny}, and 
PDG08~\cite{Amsler:2008zzb}
}
\label{tab:Spec_Rparam}
\setlength{\tabcolsep}{0.20pc}
\begin{center}
\begin{tabular}{ccccc}
\hline\hline
\rule[10pt]{-1mm}{0mm}
Resonance & $m$ (MeV) & $\Gamma_{\rm tot}$ (MeV) & $\delta$ ($^\circ$) &
Ref. \\
\hline
\rule[10pt]{-1mm}{0mm}
             & 3769.9$\pm$2.5 & 23.6$\pm$2.7 &            & PDG04~\cite{Eidelman:2004wy} \\
$\psi(3770)$ & 3771.1$\pm$2.4 & 23.0$\pm$2.7 &            & Seth~\cite{Seth:2005ny} \\
             & 3772.0$\pm$1.9 & 30.4$\pm$8.5 & 0          & BES~\cite{Ablikim:2007gd} \\
             & 3772.92$\pm$0.35 & 27.3$\pm$1.0 &          & PDG08~\cite{Amsler:2008zzb} \\
\hline
\rule[10pt]{-1mm}{0mm}
             & 4040$\pm$1     &  52$\pm$10   &            & PDG04~\cite{Eidelman:2004wy} \\
$\psi(4040)$ & 4039$\pm$1.0   &  80$\pm$10   &            & Seth~\cite{Seth:2005ny}  \\
             & 4039.6$\pm$4.3 & 84.5$\pm$12.3& 130$\pm$46 & BES~\cite{Ablikim:2007gd} \\
\hline
\rule[10pt]{-1mm}{0mm}
             & 4159$\pm$20    & 78$\pm$20    &            & PDG04~\cite{Eidelman:2004wy} \\ 
$\psi(4160)$ & 4153$\pm$3     & 103$\pm$8    &            & Seth~\cite{Seth:2005ny}  \\
             & 4191.7$\pm$6.5 & 71.8$\pm$12.3& 293$\pm$57 & BES~\cite{Ablikim:2007gd} \\
\hline
\rule[10pt]{-1mm}{0mm}
             & 4415$\pm$6     & 43$\pm$15    &            & PDG04~\cite{Eidelman:2004wy} \\
$\psi(4415)$ & 4421$\pm$4     & 62$\pm$20    &            & Seth~\cite{Seth:2005ny}  \\
             & 4415.1$\pm$7.9 & 71.5$\pm$19.0& 234$\pm$88 & BES~\cite{Ablikim:2007gd} \\
\hline\hline
\end{tabular}
\end{center}
\end{table}

For determination of the resonance parameters in the open charm
region, inclusive hadronic cross section measurements appear not to supply
enough information to determine the relative strength of different
decay channels. More data and more reliable physical models appear to be 
needed in order to make further progress. The PDG-supplied parameters 
in \Tab{tab:Spec_Rparam} bypass these issues and provide parameters
under the simplest of assumptions, which may or may not
turn out to be correct.

\begin{figure}[t]
\includegraphics*[width=\figwid]{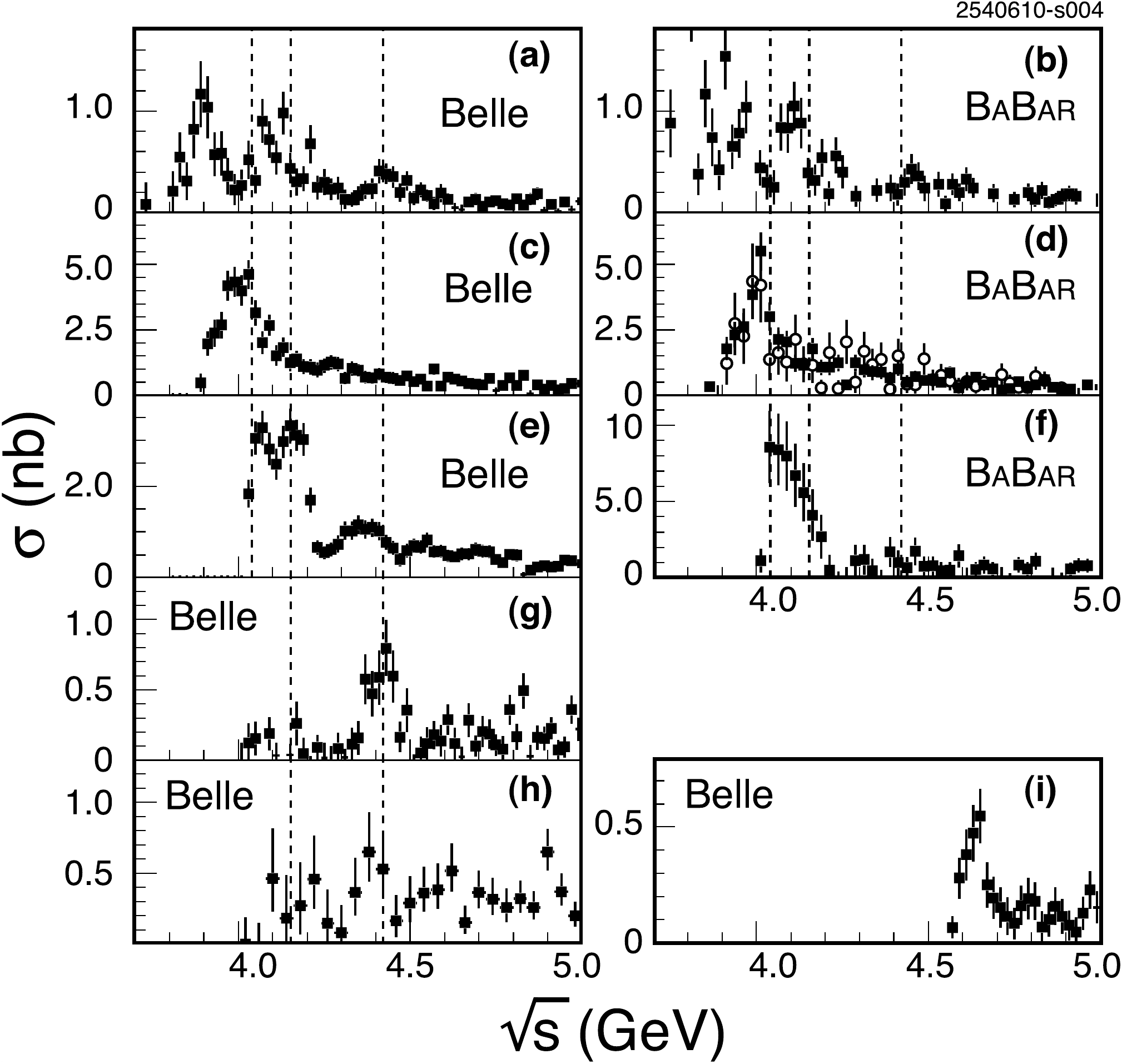}
\caption{Measured \epem\ exclusive open-charm meson- or baryon-pair 
  cross sections for
  $\sqrt{s}=3.7-5.0\gevcc$ 
  from Belle and \babar, showing
  (a)~\DDbar~\cite{Pakhlova:2008zza};
  (b)~\DDbar~\cite{Aubert:2006mi}; 
  (c)~\DpDstm~\cite{Abe:2006fj}; 
  (d)~\DDst\ for $D=\Dze$ {\it (solid squares)} 
  and $D=\Dpl$ {\it (open circles)}~\cite{Aubert:2009xs}; 
  (e)~\DstpDstm~\cite{Abe:2006fj}; 
  (f)~\DstDst~\cite{Aubert:2009xs}; 
  (g)~\DDp~\cite{Pakhlova:2007fq};
  (h)~\DDstp~\cite{Pakhlova:2009jv}; 
  (i)~\lala~\cite{Pakhlova:2008vn}. 
  {\it Vertical dashed lines} indicate $\psi$ masses
  in the region.
 \AfigPermsAPS{Pakhlova:2008zza,Aubert:2006mi,Abe:2006fj,Aubert:2009xs,Pakhlova:2007fq,Pakhlova:2009jv,Pakhlova:2008vn}{2008,2007,2008,2009,2008,2007,2008}
}
\label{fig:Spec_ExclXSec_BelleBaBar}
\end{figure}

Detailed studies of the open-charm content in the charmonium
region were not undertaken until large data\-sets
were obtained by CLEO at discrete energy points and
by the $B$-factory experiments using radiative
returns to obtain a continuous exposure of the mass region. 
The picture that has emerged is complex due to the many
thresholds in the region, nine of which 
are two-body final states using
allowed pairs of \Dze, \Dpl, \Dstn, \Dstp, \Dsp, and \Dstsp.
Moreover, distinguishing genuine two-body from
``multibody'' decays (\eg \DDst\ from \DDp) poses
a challenge. Experimentally, the data are consistent 
where measurements overlap; significant discrepancies with some
predictions mean that theoretical work remains.

\begin{figure}[t]
\includegraphics*[width=\figwid]{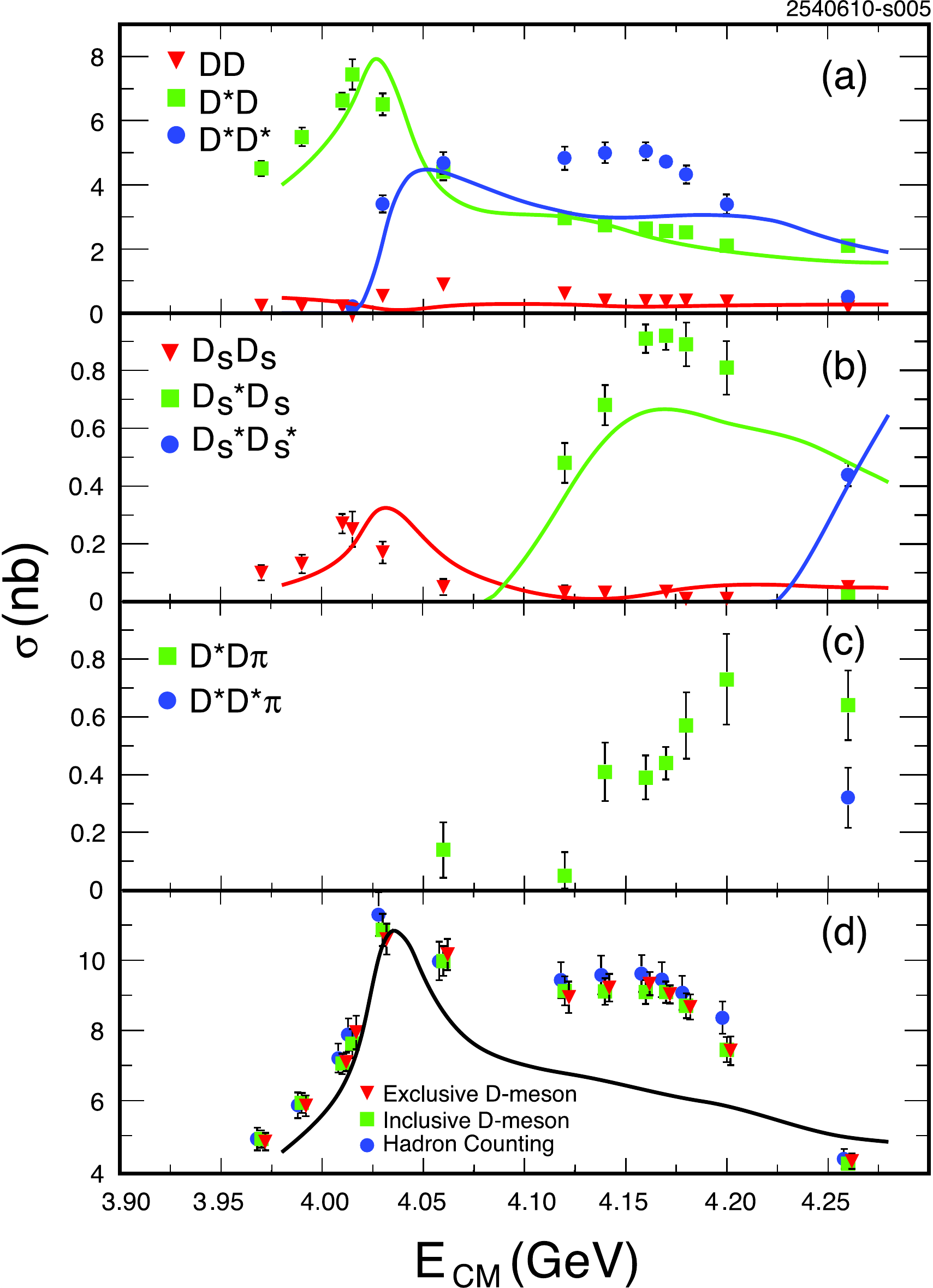}
\caption{From CLEO~\cite{CroninHennessy:2008yi},
cross sections, without radiative corrections, 
for $e^+e^-$ annihilation into:
(a)-(c)~the exclusive open-charm meson-pairs shown;
and (d)~for two methods of
open-charm counting added onto the extrapolated
$uds$ cross section compared to an all-flavor
hadron-counting cross section.
{\it Solid lines} show updated potential model
predictions~\cite{Eichten:1979ms}. 
\AfigPermAPS{CroninHennessy:2008yi}{2009} }
\label{fig:Spec_ExclXSec_CLEO}
\end{figure}

Exclusive \epem\ cross sections for hadronic final states containing
charm mesons were studied by several groups.
In the $\sqrt{s}=3.7-5\gev$ energy region,
Belle~\cite{Pakhlova:2008vn,Abe:2006fj,Pakhlova:2007fq,Pakhlova:2008zza,Pakhlova:2009jv,YuanICHEP:2010},
and
\babar~\cite{Aubert:2006mi,Aubert:2009xs,:2010vb} used initial-state radiation
(ISR) to reach the charmonium region, and 
CLEO used its large data sample taken at the  
$\psi(3770)$ peak~\cite{Dobbs:2007zt} 
and its scan over
$\sqrt{s}=3.97-4.26\gev$~\cite{CroninHennessy:2008yi}. 
Some of these results can be seen in 
\Tab{tab:SpecExp_TwoBodyRatios}
and \Figs{fig:Spec_ExclXSec_BelleBaBar}
and \ref{fig:Spec_ExclXSec_CLEO}.
Measurements of the neutral to charged \DDbar\ cross section ratio 
at the $\psi(3770)$ peak show consistency but the PDG08~\cite{Amsler:2008zzb}
world average, $1.260\pm 0.021$, is dominated
by the CLEO~\cite{Dobbs:2007zt} value.
The \DDbar\ cross sections across the entire charm energy range from
Belle~\cite{Pakhlova:2008zza} and
\babar~\cite{Aubert:2006mi}
appear in \Fig{fig:Spec_ExclXSec_BelleBaBar} and are consistent
with one another.
Both observe a structure in the ISR \DDbar\ cross section
(\Figs{fig:Spec_ExclXSec_BelleBaBar}(a) and (b)),
known as $G(3900)$,
which must be taken into account to describe both the
\DDbar\ cross section and $R$
in the region between $\psi(3770)$ and $\psi(4040)$.
The $G(3900)$ is not considered to be a specific
$c\bar{c}$ bound state, as it is qualitatively consistent
with a prediction from a coupled-channel model~\cite{Eichten:1979ms}.
The \DpDstm\ cross sections from Belle~\cite{Abe:2006fj}
and \babar~\cite{Aubert:2009xs} exhibit a single broad peak near threshold
whereas \DstpDstm\ results~\cite{Abe:2006fj,Aubert:2009xs} 
feature several local maxima and minima across this energy range.
The \eell\ cross section measured by Belle~\cite{Pakhlova:2008vn},
shown in \Fig{fig:Spec_ExclXSec_BelleBaBar}(i), exhibits a substantial
enhancement just above threshold near 4.6\gev (addressed below).

\babar~\cite{Aubert:2009xs,:2010vb}
performed unbinned maximum likelihood fits to the \DDbar, \DDst,
\DstDst, and $D_s^{(*)+}D_s^{(*)-}$ spectra. The expected $\psi$ signals were
parametrized by $P$-wave relativistic Breit-Wigner (RBW) functions
with their parameters fixed to the PDG08
values~\cite{Amsler:2008zzb}. 
An interference between the resonances and
the non\-resonant contributions was required in the fit. 
The computed ratios of the branching
fractions for the $\psi$ resonances to nonstrange
open-charm meson pairs and the quark model predictions
are presented in \Tab{tab:SpecExp_TwoBodyRatios}. The \babar\ results 
deviate from some of the theoretical expectations. 
The \babar~\cite{:2010vb} cross sections for $D_s^{(*)+}D_s^{(*)-}$ production
show evidence for $\psi(4040)$ in $D_s^+D_s^-$
and $\psi(4160)$ in $D_s^{*+}D_s^-$, and are consistent
with the CLEO~\cite{CroninHennessy:2008yi} results where they overlap.

The \eeDDp\ cross section measured by
Belle~\cite{Pakhlova:2007fq} 
is shown in \Fig{fig:Spec_ExclXSec_BelleBaBar}(g)
and exhibits an unambiguous $\psi(4415)$ signal. 
A study of the resonant structure 
shows clear signals for the \Dtnb\ and
\Dtch\ mesons and constructive interference between the neutral
\DDtnb\ and the charged \DDtch\ decay amplitudes.
Belle performed a likelihood fit to the \DDtwo\ mass distribution with a 
$\psi(4415)$ signal parametrized by an $S$-wave RBW function. The
significance of the signal is $\sim$10$\sigma$ and the peak mass
and total width are in good agreement
with the PDG06~\cite{Yao:2006px} values and the BES fit
results~\cite{Ablikim:2007gd}. 
The branching fraction for $\psi(4415)\to\DDtwo\to\DDbar\pi^+$ 
was found to be between 10\% and 20\%, depending
on the $\psi(4415)$ parametrization.
The fraction of \DDtwo$\to\DDbar\pi^+$ final states
composed of non\-resonant \DDp\  was found to be $<22\%$.
Similarly, the \DDstp\ content of $\psi(4415)$, shown in
\Fig{fig:Spec_ExclXSec_BelleBaBar}(h), has
been determined by Belle~\cite{Pakhlova:2009jv};
a marginal signal is found ($3.1\sigma$), and
its branching fraction was limited to $<10.6\%$.
Belle~\cite{YuanICHEP:2010} has also reported
a preliminary spectrum of $\epem\to D_s^{(*)+}D_s^{(*)-}$
cross sections from $\sqrt{s}=3.8$-5\gev\ using ISR from a data sample of 
967~fb$^{-1}$ in 40\mev\ bins; the values
are consistent with but higher-statistics and more finely binned 
than those of \babar~\cite{:2010vb}.

The CLEO exclusive cross sections~\cite{CroninHennessy:2008yi}
in the top three frames of \Fig{fig:Spec_ExclXSec_CLEO} are not directly
comparable to those from \babar\ and Belle as they 
are not radiatively corrected, but generally seem
to reflect consistency.
The updated potential model predictions of 
Eichten~\cite{CroninHennessy:2008yi,Eichten:1979ms}
shown in \Fig{fig:Spec_ExclXSec_CLEO} 
fail to describe many features of the data.
The CLEO total cross section determinations, 
shown in \Fig{fig:Spec_ExclXSec_CLEO}(d),
reveal that, within the measurement accuracy of 5-10\%, 
two- and three-body modes with open charm saturate the yield of all
multihadronic events above the extrapolated $uds$ contribution.

\begin{table}[t]
\caption{From \babar~\cite{Aubert:2009xs}, ratios of branching fractions for the 
  $\psi(4040)$, $\psi(4160)$ and $\psi(4415)$ resonances. 
  The first error is statistical, the second systematic. 
  Theoretical expectations are from models denoted
  $^3P_0$~\cite{Barnes:2005pb}, 
  $C^3$~\cite{Eichten:2005ga},
  and $\rho K \rho$~\cite{Swanson:2006st}
}
\label{tab:SpecExp_TwoBodyRatios}
\setlength{\tabcolsep}{0.30pc}
\begin{center}
\begin{tabular}{crcccc}
\hline\hline
\rule[10pt]{-1mm}{0mm}
State & Ratio\ \ \ \ \ & Measured & $^3P_0$ & $C^3$ & $\rho K \rho$ \\
\hline
\rule[10pt]{-1mm}{0mm}
$\psi(4040)$ & $\DDbar/\DDst$     & 0.24$\pm$0.05$\pm$0.12 & 0.003 &      & 0.14 \\
     & $\DstDst/\DDst$ & 0.18$\pm$0.14$\pm$0.03 & 1.0   &      & 0.29 \\
\hline
\rule[10pt]{-1mm}{0mm}
$\psi(4160)$ & $\DDbar/\DstDst$   & 0.02$\pm$0.03$\pm$0.02 & 0.46  & 0.08 &      \\
     & $\DDst/\DstDst$ & 0.34$\pm$0.14$\pm$0.05 & 0.011 & 0.16 &      \\
\hline
\rule[10pt]{-1mm}{0mm}
$\psi(4415)$ & $\DDbar/\DstDst$   & 0.14$\pm$0.12$\pm$0.03 & 0.025 &      &      \\
     & $\DDst/\DstDst$ & 0.17$\pm$0.25$\pm$0.03 & 0.14  &      &      \\
\hline\hline
\end{tabular}
\end{center}
\end{table}

\subsubsection{Vectors decaying to open bottom}
\label{sec:SpecExp_ConVecBeauty}

\begin{figure}[t]
   \begin{center}
     \includegraphics[width=\figwid]{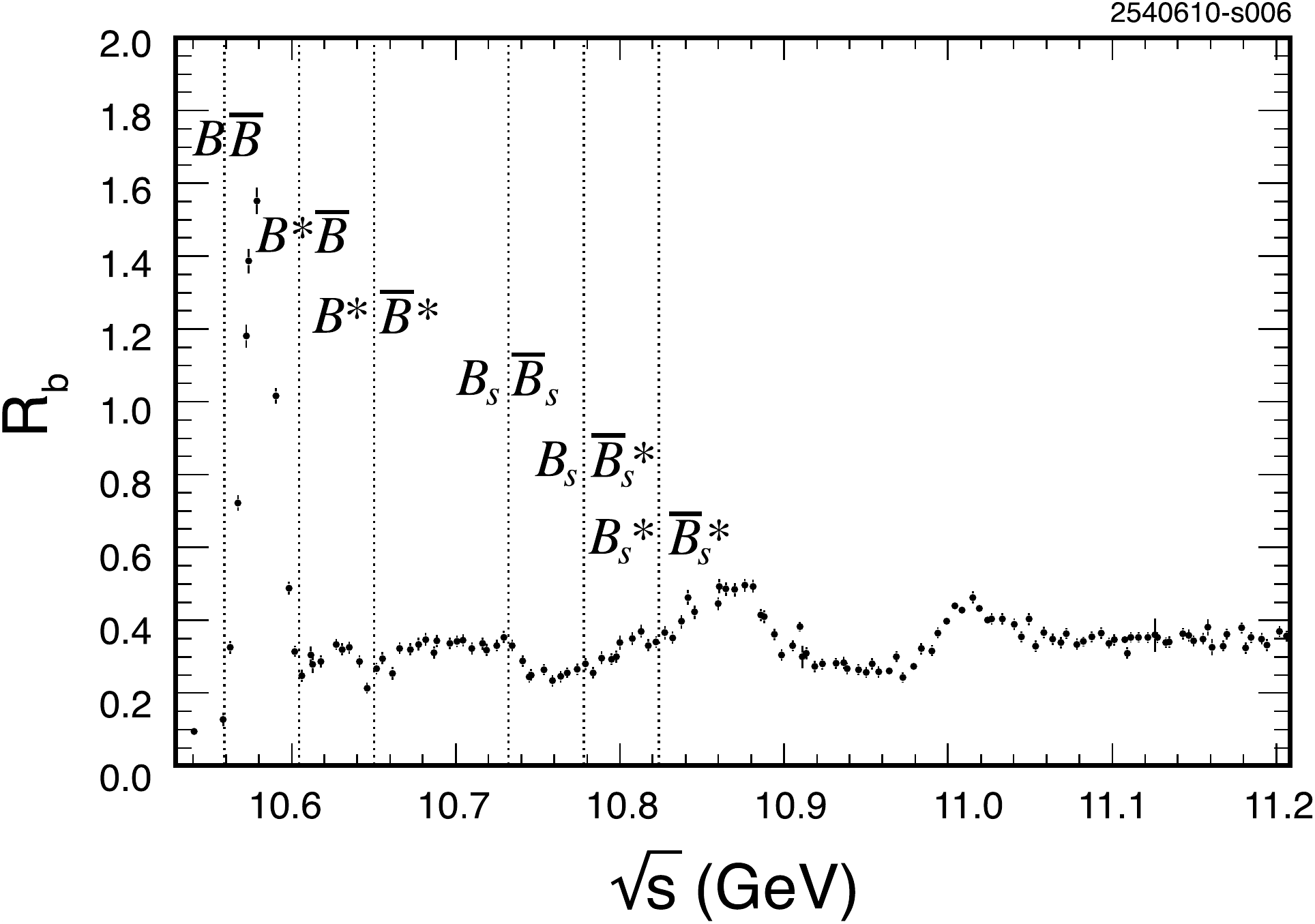}
      \caption{From \babar~\cite{Aubert:2008hx}, measured values of
               the hadronic cross section attributable to
               $b$-flavored
               states, normalized to the point muon pair cross section,
               from a scan of the center-of-mass energy region just below
               the \UnS{4}\ to above the \UnS{6}. {\it Dashed vertical lines}
               indicate the various $B\bar{B}$ mass thresholds.
               \AfigPermAPS{Aubert:2008hx}{2009}}
    \label{fig:Spec_5S6Sscan_babar}
   \end{center}
\end{figure}

The current generation of $B$-factories have scanned the energy range above 
open bottom threshold.
\babar~\cite{Aubert:2008hx} performed a comprehensive 
low-luminosity (25~pb$^{-1}$ per 
point), high-granularity ($\approx5\mev$ steps) scan between 
10.54 and 11.2~GeV, 
followed by an eight-point scan, 0.6~fb$^{-1}$ total,
in the proximity of the \UnS{6}\ peak.
Belle~\cite{Chen:2008pu} acquired $\approx 30$~pb$^{-1}$ for just nine points
over 10.80-11.02~GeV, as well as 8.1~fb$^{-1}$ spread over seven additional points
more focused on the \UnS{5}\ peak.
The \babar\ scan is shown in \Fig{fig:Spec_5S6Sscan_babar}.
Both scans suggest instead that the simple Breit-Wigner 
parametrization, previously used to
model the peaks observed in the CLEO \cite{Besson:1984bd}
and CUSB \cite{Lovelock:1985nb} scans, is 
not adequate for the description of the complex dynamics in the proximity
of the $B^{(*)}\bar{B}^{(*)}$ and $B_s^{(*)}\bar{B_s}^{(*)}$ thresholds.
Data points on $R_b=\sigma(b\bar{b})/\sigma(\mu\mu)$ are better modeled 
assuming a flat $b\bar b$ continuum contribution which interferes constructively
with the $5S$ and $6S$ Breit-Wigner resonances, and a second flat contribution
which adds incoherently. 
Such fits strongly alter the PDG results on the $5S$ and $6S$ peaks, as shown in 
\Tab{tab:Spec_5S6Sfits}.
Strong qualitative agreement is observed between the experimental behavior 
of the $R_b$ ratio and the theory predictions 
based on the coupled-channel approach~\cite{Tornqvist:1984fx}.

\begin{table}[b]
    \caption{New \babar\ and Belle results on masses and widths of the
             $\Upsilon(5S)$ and $\Upsilon(6S)$ resonances, compared to PDG averages.
             The angle $\phi$ parametrizes the phase of interfering continuum }
\label{tab:Spec_5S6Sfits}
\setlength{\tabcolsep}{0.55pc}
\begin{center}
\begin{tabular}{ccccl}
\hline
\hline
\rule[10pt]{-1mm}{0mm}
$\Upsilon$ & $m$~(MeV) & $\Gamma$~(MeV) & $\phi$~(rad) &Ref.\\
\hline
\rule[10pt]{-1mm}{0mm}
5S  & $10876\pm 2$ &  $ 43\pm 4 $     &    $2.11\pm 0.12$ & \babar~\cite{Aubert:2008hx} \\[0.7mm]
    & $10879\pm 3$ &  $ 46^{+9}_{-7}$ &    $2.33^{+0.26}_{-0.24}$ & Belle~\cite{Chen:2008pu} \\[0.7mm]
    & $10865\pm 8$ & $110\pm 13$  & - & PDG08~\cite{Amsler:2008zzb}\\[0.7mm]
\hline
\rule[10pt]{-1mm}{0mm}
6S  & $10996\pm 2$ &  $37 \pm 3 $ &  $0.12\pm 0.07$ & \babar~\cite{Aubert:2008hx}\\[0.7mm]
    & $11019\pm 8$ & $79 \pm 16$  & - & PDG08~\cite{Amsler:2008zzb}\\[0.7mm]
\hline
\hline
\end{tabular}
\end{center}
\end{table}

Additional insight can be provided by the exclusive decomposition of the
two-body (\ie $B\bar{B},B\bar{B}^*,B^*\bar{B}^*$) and many-body decay modes.
Results from $e^+e^-$ collisions have been given by 
Belle~\cite{Drutskoy:2010an} using a
data sample of 
23.6~fb$^{-1}$ acquired at the $\Upsilon$(5S). 
Charged $B$-mesons were reconstructed in two decay channels, $K^\pm J/\psi$ and 
$D^0\pi^\pm$ (with $J/\psi\to l^+l^-$ and 
$D^0 \to K\pi,K\pi\pi\pi$). 
Neutral $B$ mesons were reconstructed in $K^{*0} J/\psi$ and $D^\pm\pi^\mp $, 
with $D^\pm\to K^\pm \pi^\pm \pi^\mp$.
The $B^*$ mesons were reconstructed via their radiative transition.
Belle observes a large fraction (about 16.4\% of the total $b\bar{b}$ pairs)
from 3- and 4-body decay modes, \ie  
$B^{(*)}\bar{B}^{(*)}\pi,B^{(*)}\bar{B}^{(*)}\pi\pi$.
A significant fraction of these events can actually be expected from 
ISR production of $\Upsilon(4S)$.
Theory predictions on multibody decays at $\Upsilon(5S)$ range from 
0.03\%~\cite{Simonov:2008cr} to 0.3\%~\cite{Lellouch:1992bq}.

\begin{table*}[hbt]
\caption{New {\it conventional} states in the $c\bar{c}$, $b\bar{c}$, and
$b\bar{b}$ regions, ordered by mass. Masses $m$ and widths $\Gamma$ represent
the weighted averages from the listed sources.
Quoted uncertainties reflect quadrature summation from individual experiments.
In the Process column, the decay mode of the new state 
claimed is indicated in parentheses.
Ellipses (...) indicate inclusively selected event topologies;
\ie additional particles not required by the Experiments to be present.
For each Experiment a citation is given, as well as
the statistical significance in number of
standard deviations (\#$\sigma$), or ``(np)'' for ``not provided''. 
The Year column gives the date of first measurement cited, which
is the first with significance of $>5\sigma$. 
The Status column indicates that the state 
has been observed 
by at most one ({\color{red} NC!}-needs confirmation) or
at least two independent experiments with significance of $>$5$\sigma$ (OK).
The state labelled $\chi_{c2}(2P)$ has previously been called $Z(3930)$
} 

\setlength{\tabcolsep}{0.30pc}
\begin{center}
\begin{tabular}{lccclccc}
\hline\hline
\rule[10pt]{-1mm}{0mm}
 State & $m$~(MeV) & $\Gamma$~(MeV) & $J^{PC}$ & \ \ \ \ Process~(mode) & 
     Experiment~(\#$\sigma$) & Year & Status \\[0.7mm]
\hline
\rule[10pt]{-1mm}{0mm}
$\hsubc$ & $3525.45\pm 0.15$ & 0.73$\pm$0.53 & $1^{+-}$ &
    $\psi(2S)\to\pi^0 (\gamma\etac)$ & 
    {\color{red} CLEO}~\cite{Rubin:2005px,Rosner:2005ry,Dobbs:2008ec}~(13.2) &
    2004 & OK \\ [0.7mm]
& & ($<$1.44) & &    $\psi(2S)\to\pi^0 (\gamma ...)$ & 
    {\color{red} CLEO}~\cite{Rubin:2005px,Rosner:2005ry,Dobbs:2008ec}~(10.0), BES~\cite{Ablikim:2010rc}~(18.6) & & \\ [0.7mm]
& & & & $p\bar p\to (\gamma \eta_{c})\to (\gamma\gamma\gamma)$ &
    E835~\cite{Andreotti:2005vu}~(3.1)& &\\[0.7mm]
& & & &    $\psi(2S)\to\pi^0 (...)$ & 
    BESIII~\cite{Ablikim:2010rc}~(9.5) & & \\ [1.79mm]
$\etacp$ & $3637\pm4$ &14$\pm$7 &$0^{-+}$ &
     $B\to K (K_S^0K^-\pi^+)$ &
     {\color{red} Belle}~\cite{Choi:2002na}~(6.0) & 2002 & OK\\[0.7mm]
& & & & $e^+e^-\to e^+e^-(K_S^0 K^-\pi^+)$ &
     \babar~\cite{Aubert:2003pt}~(4.9), CLEO~\cite{Asner:2003wv}~(6.5), &
     & \\[0.7mm]
& & & &  & Belle~\cite{Nakazawa:2008zz}~(6) & &\\[0.7mm]
& & & & $e^+e^-\to J/\psi\ (...)$ &
     \babar~\cite{Aubert:2005tj}~(np), Belle~\cite{Abe:2007jn}~(8.1)  & &\\[1.79mm]
$\chi_{c2}(2P)$ & $3927.2\pm2.6$ & 24.1$\pm$6.1 & $2^{++}$ &
     $e^+e^-\to e^+e^- (D\bar{D})$ &
     {\color{red} Belle}~\cite{Uehara:2005qd}~(5.3), \babar~\cite{:2010hka}~(5.8) & 2005 & OK\\ [1.79mm]
$B_c^+$ & $6277.1\pm4.1$ & - & $0^-$ & $\bar{p}p\to (\pi^+\jpsi) ...$
        & {\color{red} CDF}~\cite{Aaltonen:2007gv}~(8.0),
          \DZero~\cite{Abazov:2008kv}~(5.2) & 2007 & OK\\[1.79mm]
$\eta_b(1S)$ & $9390.7\pm2.9$  & ? & 0$^{-+}$ &
     $\Unx{3}{S}\to\gamma + (...)$ &
     {\color{red} \babar}~\cite{:2008vj}~(10),
     CLEO~\cite{Bonvicini:2009hs}~(4.0)  & 2008 & {\color{red} NC!}\\[0.7mm]
  & & & & $\Unx{2}{S}\to\gamma + (...)$ &
     \babar~\cite{:2009pz}~(3.0)  & & \\[1.79mm]
\UoneDT\  & $ 10163.8\pm1.4$ & ? & 2$^{--}$ &
     $\UnS{3}\to\gamma\gamma(\gamma\gamma\UnS{1})$ & 
     CLEO~\cite{Bonvicini:2004yj}~(10.2) & 2004 & OK\\[0.7mm]
&&&&     $\UnS{3}\to\gamma\gamma(\pi^+\pi^-\UnS{1})$ & 
     \babar~\cite{Sanchez:2010kz}~(5.8) & & \\[0.7mm]
\hline\hline
\end{tabular}
\end{center}
\label{tab:Spec_ExpSumCon}
\end{table*}

\subsection{Newly found conventional quarkonia}
\label{sec:SpecExpNewCon}

\Tab{tab:Spec_ExpSumCon} lists properties of new
conventional heavy quarkonium states. The $h_c$ is the $^1P_1$ state of 
charmonium, singlet partner of the long-known $\chi_{cJ}$ triplet
$^3P_J$. The $\etacp$ is the first excited state of
the pseudoscalar ground state \etac, lying just
below the mass of its vector counterpart, $\psi(2S)$. The first $B$-meson seen that
contains charm is the $B_c$. The ground state of
bottomonium is the $\eta_b(1S)$. And the \UoneD\  is the
lowest-lying $D$-wave triplet of the $b\bar{b}$ system.
All fit into their respective spectroscopies roughly 
where expected. Their exact masses, production mechanisms, and decay modes
provide guidance to their descriptions within QCD.

\subsubsection{Observation of $\hsubc$}
\label{sec:SpecExp_hc}

\begin{figure}[t]
    \includegraphics[width=\figwid]{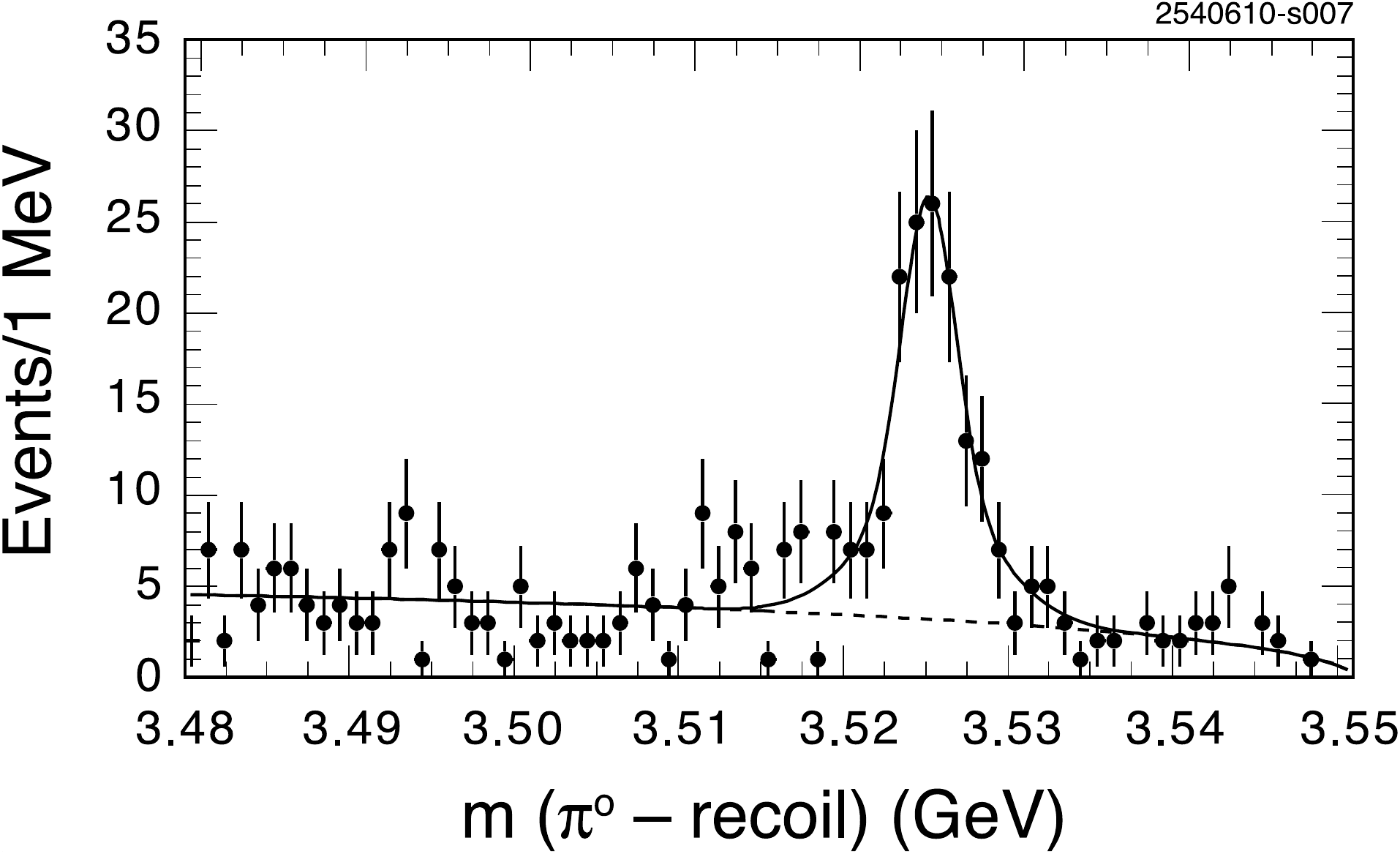}
     \caption{From CLEO~\cite{Dobbs:2008ec}, 
              the mass recoiling against the $\pi^0$ in the
               $\psip\to\pi^0 h_c$, $h_c\to\gamma\etac$
              {\it exclusive} sample
               in which the $\pi^0$, $\gamma$, and $\etac$ are all
               explicitly reconstructed in the detector.
               \AfigPermAPS{Dobbs:2008ec}{2008}}
    \label{fig:Spec_hc_excl}
\end{figure}

  Two experiments reported $\hsubc$ sightings in 2005,
with CLEO~\cite{Rubin:2005px,Rosner:2005ry} 
reporting an observation at $>6\sigma$ in the
isospin-forbidden decay chain $e^+e^-\to\psip\to\pi^0 h_c$,
$h_c\to\gamma\etac$, and E835~\cite{Andreotti:2005vu} 
found $3\sigma$ evidence in 
$p\bar{p}\to h_c$, $h_c\to\gamma\etac$, $\etac\to\gamma\gamma$.
CLEO~\cite{Dobbs:2008ec} 
later updated its measurements with a larger dataset,
refining its mass measurement to a precision of just over 0.2\mevcc,
finding a central value slightly more accurate 
than that of E835, which has an uncertainty of just under 0.3\mevcc.
CLEO utilized two detection methods. The first was a semi-inclusive
selection that required detection of both the transition
$\pi^0$ and radiative photon but only inferred the presence of the $\etac$
through kinematics. The second employed full reconstruction
in fifteen different $\etac$ decay modes, five of them 
previously unseen. The two methods had some statistical
and almost full systematic correlation for the mass 
measurement because both rely on the $\pi^0$ momentum
determination. As the parent $\psip$ has precisely known mass
and is produced nearly at rest by the incoming
$e^+e^-$ pair, the mass of the $\hsubc$ is most accurately
determined by fitting the distribution of the mass
recoiling against the $\pi^0$, as shown for the exclusive 
analysis in \Fig{fig:Spec_hc_excl}.
CLEO's two methods had comparable precision and 
gave consistent masses within their uncorrelated uncertainties.
Statistical uncertainties from the numbers of signal
(background) events in the exclusive (inclusive) analysis
are larger than the systematic errors attributable 
to calorimeter energy resolution.
The E835 measurement relies on knowledge of the initial
center-of-mass energy of the $p\bar p$ for each event
during a scan of the $\hsubc$ mass region as well as upon reconstruction
of all three photons with kinematics consistent with
the production and decay hypothesis. Unlike 
the CLEO result, backgrounds
are negligible. Mass measurement
accuracy was limited equally by statistics
(13 signal events with a standard deviation
in center-of-mass energy of 0.07\mev) 
and systematics of $\bar p$ beam energy stability. 
Using a sample of 106M $\psi(2S)$, in 2010 BESIII~\cite{Ablikim:2010rc} reported a mass
result using the $\pi^0\gamma$ inclusive method,
matching CLEO's precision.
The spin-averaged centroid of the triplet states,
$\mchicj\equiv [m(\chi_{c0}) + 3m(\chi_{c1}) + 5m(\chi_{c2})]/9$,
is expected to be near the $\hsubc$ mass, making
the hyperfine mass splitting,
$\dmhf[\hsubc]\equiv\mchicj - m[\hsubc]$, an important measure of the
spin-spin interaction. 
The $h_c$-related quantities are summarized in \Tab{tab:Spec_hc};
mass measurements are consistent.
It could be a coincidence~\cite{Dobbs:2008ec} 
that $\dmhf[\hsubc]_{\rm exp}\approx 0$, 
the same as the lowest-order perturbative QCD
expectation, 
because the same theoretical assumptions lead to the prediction
\beq
\frac{m(\chi_{c1})-m(\chi_{c0})}{m(\chi_{c2})-m(\chi_{c1})}=\frac{5}{2}\, ,
\eeq
whereas measured masses~\cite{Amsler:2008zzb}
yield a value of 2.1, 20\% smaller than predicted.

\subsubsection{Observation of $\etacp$}
\label{sec:SpecExp_etac2s}

\begin{figure}[t]
    \includegraphics[width=\figwid]{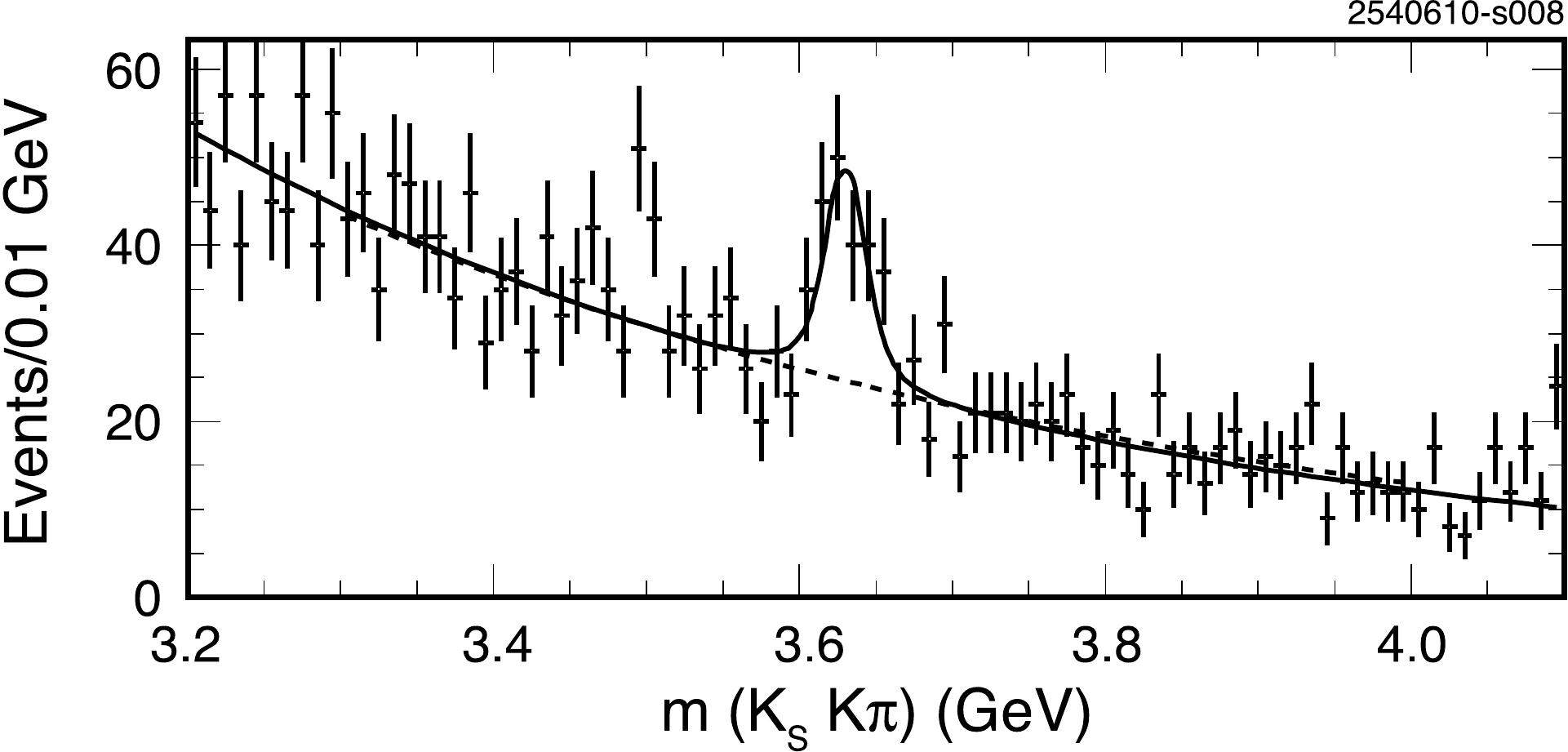}
     \caption{From \babar~\cite{Aubert:2003pt}, 
              the $K\bar{K}\pi$ invariant mass distribution
              from selected $e^+e^-\to e^+e^- K_S^0 K^+\pi^-$ events,
              data {\it (points with error bars)} overlaid with a fit
              {\it (solid line)}
              having two components, a smooth background {\it (dashed
              line)} and an $\etacp$ signal.
              \AfigPermAPS{Aubert:2003pt}{2004}
              }
    \label{fig:Spec_etac2s_Babar}
\end{figure}

  The search for a reproducible $\etacp$ signal has
a long and checkered history. There were hints in
early $e^+e^-\to c\bar{c}$ data for a purported $\etacp$
with mass near 3455\mevcc\ in $\psip\to\gamma\gamma\jpsi$
events~\cite{Whitaker:1976hb} 
and in inclusive radiative $\psip$ decays~\cite{Biddick:1977sv,Tanenbaum:1977eg}.
A possible signal~\cite{Bartel:1978ux} near 3591\mevcc\ was reported
in 1978 in $\psip\to\gamma\gamma\jpsi$. Crystal Ball ruled 
out that result in 1982~\cite{Edwards:1981mq} and
also reported an $\etacp$ signal in 
{\it inclusive} radiative $\psip$ decays
with a mass of 3592$\pm$5\mevcc.
The latter result persisted, in limbo, unconfirmed
and unrefuted, for twenty years, until Belle~\cite{Choi:2002na}
found a signal in $B\to K\etacp$ in the exclusive 
$\etacp\to K_S^0 K^-\pi^+$ decay mode
(a favorite all-charged final state for \etac), 
at 3654$\pm$6$\pm$8\mevcc.
Since then measurements of $\etacp$ in that mass region have been
reported by \babar~\cite{Aubert:2003pt} 
(see \Fig{fig:Spec_etac2s_Babar}),
CLEO~\cite{Asner:2003wv},
and Belle~\cite{Nakazawa:2008zz}
in $\gamma\gamma$-fusion to $K\bar{K}\pi$ final states 
and by \babar~\cite{Aubert:2005tj} and Belle~\cite{Abe:2007jn}
in double charmonium production. 
\begin{table}[t]
    \caption{Width and mass measurements of the $\hsubc$.
             $\left<m(1\,^3P_J)\right>$ and $\dmhf$ are defined
             in the text
             }
    \label{tab:Spec_hc}
\setlength{\tabcolsep}{0.65pc}
\begin{center}
\begin{tabular}{ccc} 
\hline\hline
\rule[10pt]{-1mm}{0mm}
Quantity       & Value~(MeV) & Ref. ($\chi^2$/d.o.f.) \\[0.7mm]
\hline
\rule[10pt]{-1mm}{0mm}
Width & $0.73\pm0.45\pm0.28$ & BES~\cite{Ablikim:2010rc} \\[0.7mm]
      & $<$1.44@90\%~CL      & BES~\cite{Ablikim:2010rc} \\[0.7mm]
\hline
\rule[10pt]{-1mm}{0mm}
Mass & $3525.8\pm0.2\pm0.2$    & E835~\cite{Andreotti:2005vu} \\[0.7mm]
     & $3525.28\pm0.19\pm0.12$ & CLEO~\cite{Dobbs:2008ec} \\[0.7mm]
     & $3525.40\pm0.13\pm0.18$ & BES~\cite{Ablikim:2010rc} \\[0.7mm]
     & $3525.45\pm0.15$        & Avg$^3$ (2.2/2) \\[0.7mm]
\hline
\rule[10pt]{-1mm}{0mm}
$\mchicj$ & $3525.30\pm0.07$ & PDG08~\cite{Amsler:2008zzb}\\[0.7mm]
$\dmhf$   & $-0.15\pm0.17$ & \\[0.7mm]
\hline
\hline
\end{tabular}
\end{center}
\end{table}
   \addtocounter{footnote}{1}
   \footnotetext{A note concerning tables in this section: where the
                 label ``Avg'' is attached to a number, it signifies
                 an inverse-square-error-weighted average of values
                 appearing directly above, for
                 which all statistical and systematic errors
                 were combined in quadrature without accounting
                 for any possible correlations between them.
                 The uncertainty on this average is inflated
                 by the multiplicative factor $S$ 
                 if $S^2\equiv\chi^2$/d.o.f.$>$1}
With this plethora of independent measurements in
three different production mechanisms and two methods
of mass reconstruction (fully reconstructed exclusive decay
to $K\bar{K}\pi$ and missing mass), it might have been
reasonable to expect clarity and cohesion to have emerged.
However, complete experimental unity eludes us because,
while the mass values are all in the same vicinity, 
when averaged they have a PDG $S$-factor of 1.7, 
the factor by which the weighted-average uncertainty
is inflated. The two most precise measurements,
both from $\gamma\gamma\to\etacp$,
disagree by $2.5\sigma$; the two least precise,
both by Belle, disagree by $2.2\sigma$; the
two double-charmonium results disagree by $1.6\sigma$;
and the two \babar\ results disagree by $1.4\sigma$. 
There are no
easily identifiable outliers to discard. The lesson 
here may be that statistics in all the methods
utilized are hard to come by, or that background
shapes are more complicated than assumed, or
that these measurements have been plagued by extraordinary bad luck.
In any case, further exploration 
is clearly merited. CLEO~\cite{:2009vg} 
attempted to find exclusive $\etacp$ decays in radiative
$\psip$ decays, guided by the success of such methods
for \etac~\cite{:2008fb}, but found no clear signals
in its sample of 25M $\psip$.

Just prior to submission of this article, Belle~\cite{NakazawaICHEP2010}
announced a preliminary observation of \etacp, produced in two-photon fusion,
in three new decay modes ($3(\dipi)$, $K^+K^-\,2(\dipi)$, and $K_S^0
K^-\,\dipi\pi^+$). These modes will offer more concrete 
avenues of approach to \etacp\
in order to better measure its properties.

\subsubsection{Observation of $\chi_{c2}(2P)$}
\label{sec:SpecExp_chic22p}

\begin{figure}[t]
    \includegraphics[width=\figwid]{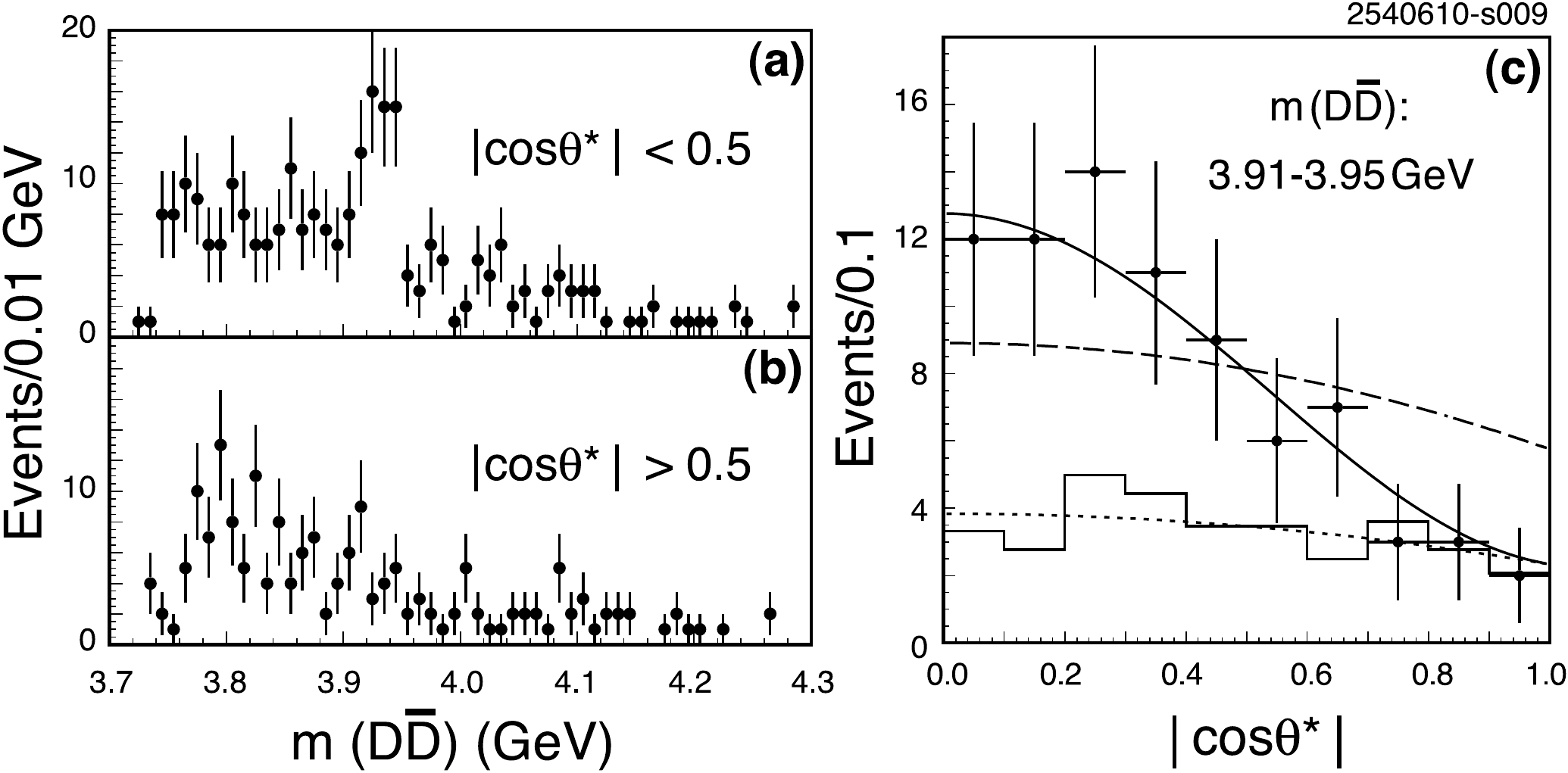}
     \caption{From the Belle~\cite{Uehara:2005qd} observation
              of $\chi_{c2}(2P)\to\DDbar$. 
              (a) and (b)~$m(D\bar{D})$ distributions 
              from selected $e^+e^-\to e^+e^- D\bar{D}$ events,
              for the $|\cos\theta^*|$ regions indicated,
              where $\theta^*$ is the polar angle of a $D$ momentum
              vector in the $\gamma\gamma$ center-of-mass frame.
              Part~(c) shows the corresponding $|\cos\theta^*|$ distributions for the
              $m(D\bar{D})$ region indicated, from data
              {\it (points with error bars)} and background
              {\it (solid line histogram)}. Also shown are
              expected distributions for the spin-2
              (helicity-2) {\it (solid curve)} and spin-zero
              {\it (dashed curve)} hypotheses, both of which
              include background {\it (dotted curve)}.
              \AfigPermAPS{Uehara:2005qd}{2006} }
    \label{fig:Spec_chic22p_Belle}
\end{figure}

In 2005 Belle~\cite{Uehara:2005qd} observed an enhancement
in the \DDbar\ mass spectrum from 
$e^+e^-\to e^+e^- \DDbar$ events
with a statistical significance of $5.3\sigma$.
Properties are shown in \Tab{tab:Spec_chic22P}.
It was initially dubbed the $Z(3930)$, but since has
been widely\footnote{Lattice calculations~\cite{Dudek:2009kk}
suggest that the $\chi_{c2}(2P)$ (\ie the $2\,^3P_2$ $c\bar c$ state)
and the $1\,^3F_2$ state could be quite close in mass,
so that perhaps the $Z(3930)$ is {\it not} the
$2\,^3P_2$ but rather the $1\,^3F_2$.}
(if not universally) accepted as the $\chi_{c2}(2P)$. 
The analysis selects fully reconstructed
$D$ meson pairs with at most one $\pi^0$
and at most six pions/kaons per event, 
using the decays $\Dze\to\ K^-\pi^+$,
$K^-\pi^+\pi^0$, and $K^-\pi^+\pi^+\pi^-$, and
$\Dpl\to K^-\pi^+\pi^+$.  The outgoing
$e^+e^-$ were presumed to exit the detector at small angles.
This $\gamma\gamma$-fusion signature was enforced by requiring small
transverse momentum with respect to the beam direction
in the $e^+e^-$ center-of-mass frame
and restricting the \DDbar\ longitudinal momentum 
to kinematically exclude $e^+e^-\to\gamma\DDbar$.
\Figure{fig:Spec_chic22p_Belle} shows the resulting
\DDbar\ mass and angular distributions; the latter are
consistent with the spin-2, helicity-2 hypothesis
but disagree with spin-0.
\babar~\cite{:2010hka} confirmed the Belle observation
in $\gamma\gamma$-fusion with significance of $5.8\sigma$
and found properties consistent with those from Belle.

\begin{table}[b]
    \caption{Properties of the $\chi_{c2}(2P)$ (originally $Z(3930)$)
             }
    \label{tab:Spec_chic22P}
\setlength{\tabcolsep}{0.80pc}
\begin{center}
\begin{tabular}{ccc} 
\hline\hline
\rule[10pt]{-1mm}{0mm}
Quantity       & Value & Ref. ($\chi^2$/d.o.f.) \\[0.7mm]
\hline
\rule[10pt]{-1mm}{0mm}
Mass (MeV) & 3929$\pm$5$\pm$2 & Belle~\cite{Uehara:2005qd} \\[0.7mm]
     &3926.7$\pm$2.7$\pm$1.1 & \babar~\cite{:2010hka} \\[0.7mm]
     & 3927.2$\pm$2.6 & Avg$^3$ (0.14/1) \\[0.7mm]
\hline\rule[10pt]{-1mm}{0mm}
Width (MeV) & 29$\pm$10$\pm$2 & Belle~\cite{Uehara:2005qd} \\[0.7mm]
     & 21.3$\pm$6.8$\pm$3.6 & \babar~\cite{:2010hka} \\[0.7mm]
     & 24.1$\pm$6.1 & Avg$^3$ (0.37/1) \\ [0.7mm]
\hline\rule[10pt]{-1mm}{0mm}
$\Gamma_{\gamma\gamma}\times {\cal B}(\DDbar)$ & 0.18$\pm$0.05$\pm$0.03 & Belle~\cite{Uehara:2005qd} \\[0.7mm]
  (keV)  & 0.24$\pm$0.05$\pm$0.04 & \babar~\cite{:2010hka} \\[0.7mm]
         & 0.21$\pm$0.04 & Avg$^3$ (0.46/1) \\[0.7mm]
\hline
\hline
\end{tabular}
\end{center}
\end{table}

\subsubsection{Observation of $B_c^+$}
\label{sec:SpecExp_Bc}

\begin{figure}[t]
    \includegraphics[width=\figwid]{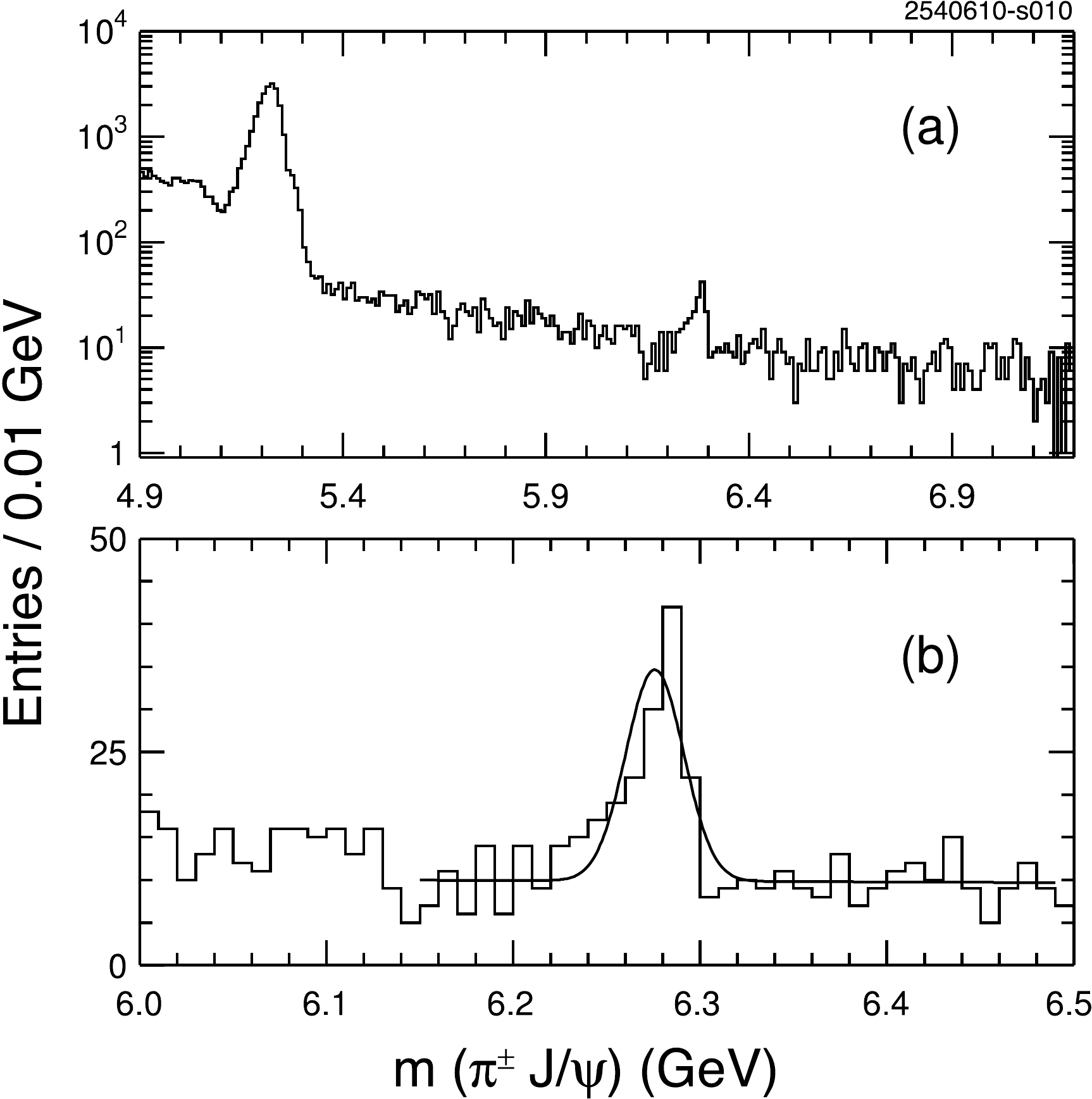}
     \caption{From CDF~\cite{Aaltonen:2007gv},
              (a)~$\jpsi\pi^+$ invariant mass combinations
              from selected $\bar{p}p\to\pi^+\jpsi X$ events.
              The bump near 5.2\gevcc\ is due to
              $B^+\to K^+\jpsi$ decays with a pion mass assignment for
              the kaon. (b)~As in (a), but
              zoomed in on the 6.0-6.5\gevcc\ mass region; 
              the {\it solid curve} indicates the projection of the
              $B_c^+$ maximum likelihood fit to the data.
              \AfigPermAPS{Aaltonen:2007gv}{2008} }
    \label{fig:Spec_B_c_CDFmass}
\end{figure}

Unique among mesons is the $B_c^+$ because it
is the lowest-lying (and only observed) meson composed of
a heavy quark and a heavy antiquark of different flavors.
As such, its mass, lifetime, decay,
and production mechanisms garner attention so as
to constrain and cross-check QCD calculations
similar to those used for other heavy quarkonia. 
Since its mass is well above 6\gevcc, production
of pairs $B_c^+B_c^-$ at the $e^+e^-$ $B$-factories,
which take most of their data near the $\UnS{4}$,
has not been possible.
Although a hint from OPAL~\cite{Ackerstaff:1998zf} 
at LEP and then suggestive evidence
from CDF~\cite{Abe:1998wi} for the existence of $B_c^+$ 
were published in 1998, it was not until a
decade later that two confirming observations
in excess of $5\sigma$ significance were made in Run~II at the
Tevatron. Both mass measurements used the decay chain
$B_c^+\to\jpsi\pi^+$, $\jpsi\to\mu^+\mu^-$, and obtain
for mass and statistical significance the values
\beqa
m(B_c^+) &= 6275.6\pm2.9\pm2.5\mevcc\ &(8\sigma)~{\rm CDF}~\text{\cite{Aaltonen:2007gv}}\non\\
         &= 6300\pm14\pm5\mevcc\      &(5.2\sigma)~\DZero~\text{\cite{Abazov:2008kv}}\, .~~~~~~
\eeqa
The CDF $B_c^+$ mass plot is shown in \Fig{fig:Spec_B_c_CDFmass}. 
Their weighted
average (\Tab{tab:Spec_ExpSumCon}) is about 2$\sigma$ lower
than the lattice QCD prediction~\cite{Allison:2004be} of
6304$\pm$12$^{+18}_{-0}$\mevcc. The only observed decay
modes for $B_c^+$ are $\jpsi\pi^+$ and 
$\jpsi\ell^+\nu_\ell$. 
The semileptonic mode has been
used by both CDF~\cite{Abulencia:2006zu,Nigmanov:2009gu} 
and 
\DZero~\cite{Abazov:2008rba} to measure the $B_c^+$
lifetime. Their results are consistent with each other and have
a weighted average~\cite{Nigmanov:2009gu} of 
0.46$\pm$0.04~ps. (See also \Sec{sec:Prod_Bc} for
discussion of $B_c^+$).

\subsubsection{Observation of $\eta_b(1S)$}
\label{sec:SpecExp_etab}

Nonobservation of the bottomonium ground state 
was an annoying 
thorn in the side of heavy quarkonium spectroscopy
until 2008,
when \babar~\cite{:2008vj,:2009pz} 
succeeded in observing the $\eta_b(1S)$
where previous efforts had failed. 
The hyperfine mass-splitting of singlet-triplet states, 
$\dmhf\equiv m(1^3S_1)-m(1^1S_0)$, 
probes the spin-dependence of bound-state energy levels,
and, once measured, imposes constraints
on theoretical descriptions. 
The $\eta_b$ remained elusive for a variety of
reasons. Branching fractions for transitions
from the $\UnS{n}$ states are small
and no low-multiplicity, high-rate 
``golden'' decay modes analogous to 
$\etac\to K\bar{K}\pi$
appear to exist for $\eta_b$. This left 
inclusive 
$\UnS{n}\to\gamma\eta_b$
as the first line of attack. 

\begin{figure}[t]
  \begin{center}
    \includegraphics[width=\figwid]{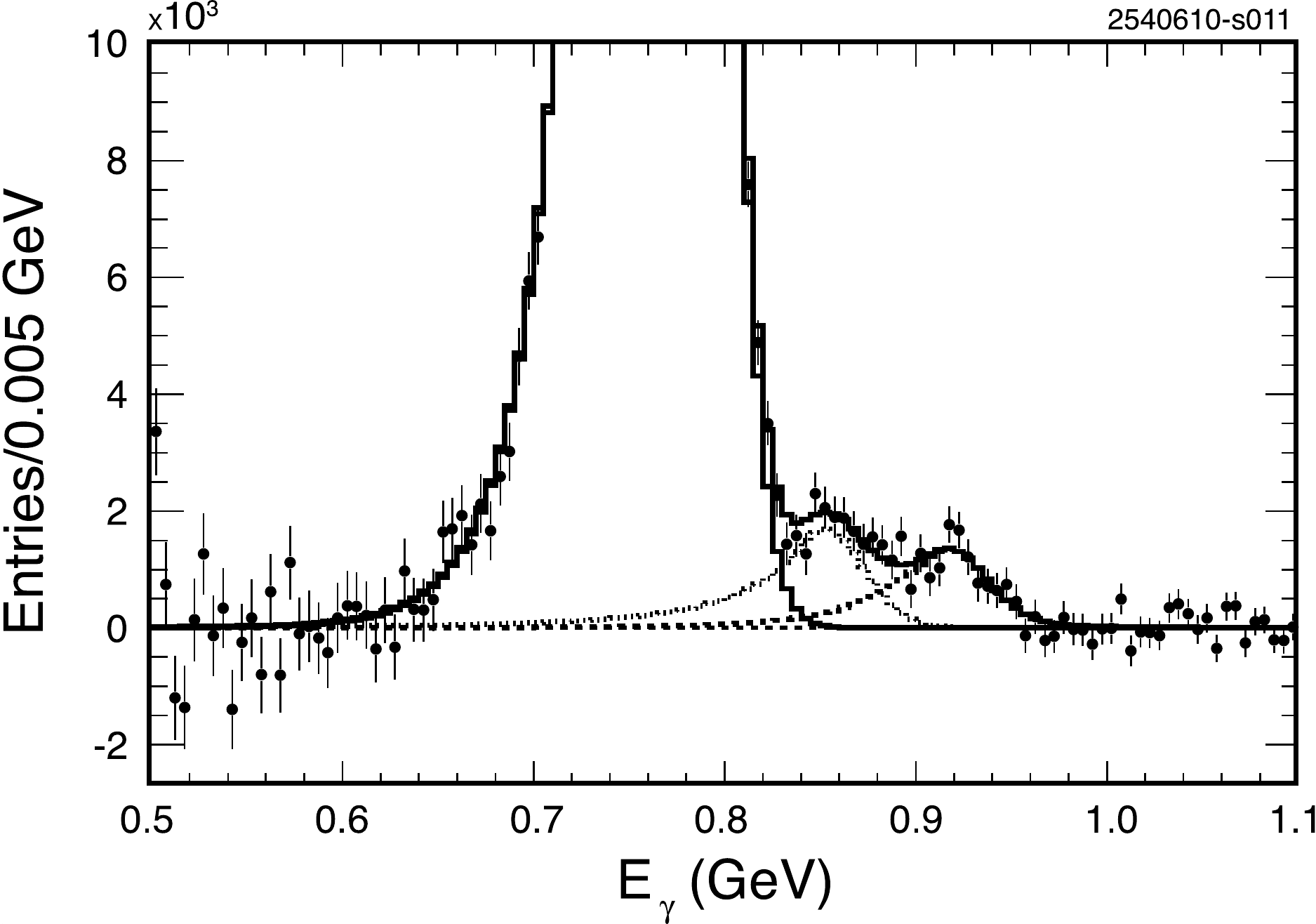}
     \caption{From \babar~\cite{:2008vj}, 
              the inclusive photon energy spectrum in the $e^+e^-$
              center-of-mass frame \UnS{3} data
              after subtraction of the smooth background. 
              The {\it solid curve} shows the best fit, and the 
              {\it peaks} correspond to, from left to right, 
              $\chi_{b1,2}\to\gamma\UnS{1}$, ISR production of
              \UnS{1}, and  $\UnS{3}\to\gamma\eta_b$.
              \AfigPermAPS{:2008vj}{2008} }
    \label{fig:Spec_etab3S_Babar}
   \end{center}
\end{figure}

 \babar's success was mainly due to large data samples 
obtained just prior to shutdown of the experiment.
For the express objective of $\eta_b$-discovery (among others),
\babar\ accumulated 122M \UnS{3}\ and 
100M \UnS{2}\  decays, compared
to CLEO (9M~\UnS{2}\  and 6M~\UnS{3}) and
Belle (11M~\UnS{3}).
Even with such large data samples and
a high-performance cesium iodide crystal
calorimeter, \babar's task was far from
trivial: the expected photon line was
buried under a sea of $\pi^0$-decay photons
even after all photon candidates that combine 
with any other photon to form a $\pi^0$ were vetoed.
The $\eta_b$ photon line was also obscured by two 
other physics processes, each inducing
structure in $E_\gamma$, the photon energy in the
\UnS{n}\ rest frame.
The $\eta_b$ photon line lies in the high energy tail of 
the three Doppler-smeared and merged 
$\chi_{bJ}(nP)\to\gamma\UnS{1}$
peaks and adjacent to that of the radiative return
process, $e^+e^- \to \gamma \UnS{1}$.
\babar\ introduced a method to suppress 
non\-resonant ``continuum'' photons and thereby enhance
experimental signal-squared-to-background ratio ($S^2/B$),
noting that such backgrounds tend to follow initial parton (jet) directions,
whereas the $\eta_b$ decay products will
have direction uncorrelated with that of the transition photon. 
The angle $\theta_T$ was defined to be the angle
between each transition photon candidate
and the thrust axis~\cite{Brandt:1964sa} of the 
rest of the event.
(The thrust axis is the direction that maximizes the 
sum of absolute values of momenta projected upon it,
and, on a statistical basis, follows the axis of two-jet events.)
The thrust angle associated with each candidate radiative 
photon was calculated and
required to satisfy $\cost<0.7$, the criterion
found by \babar\ to maximize $S^2/B$. The analysis
extracted a signal by fitting the $E_\gamma$
distribution to four components:
an empirically determined smooth background,
merged $\chi_{bJ}$ peaks, a 
mono\-chromatic ISR photon line,
and an $\eta_b$ signal. The resulting $E_\gamma$ spectrum
from the \babar~\cite{:2008vj} \UnS{3}\  analysis
is shown in \Fig{fig:Spec_etab3S_Babar},
with an $\eta_b$ signal of 
significance of $>10\sigma$. A few months after this discovery,
\babar~\cite{:2009pz} announced confirmation of their signal with
a nearly identical analysis of their \UnS{2}\  data,
albeit with smaller signficance (3.0$\sigma$).
To avoid bias, these analyses established procedures while ``blind''
to the $\eta_b$ signal region in  $E_\gamma$.

Initially, there was some worry that the 
\babar\ results were in mild conflict with earlier
non\-observation upper limits from a CLEO~\cite{Artuso:2004fp}
analysis, which had as its primary focus a detailed study of
the dipole transitions $\UnS{n}\to\gamma\chi_{bJ}(mP)$.
However, CLEO~\cite{Bonvicini:2009hs} 
later corrected errors and omissions
in that analysis and announced new results
consistent with but less precise than \babar's, including
4$\sigma$ evidence for $\UnS{3}\to\gamma\eta_b$ and
a larger upper limit on ${\cal B}(\UnS{2}\to\gamma\eta_b)$.
In addition to including the initially omitted ISR peak in the
fit to $E_\gamma$ and assuming a more reasonable width,
$\Gamma(\eta_b)=10\mev$, for the signal, 
CLEO exploited an $E_\gamma$ resolution
slightly better than \babar, parametrized the
observed photon lineshape more accurately than before, and
added a new twist to the \babar-inspired thrust-angle
restriction. Instead of simply rejecting a high-background
region of thrust angle, CLEO accumulated three $E_\gamma$ distributions,
one for each of three $\cost$ bins, the middle one being $0.3<\cost<0.7$.
Hence no statistical power was wasted
by throwing any data away, and an
improved $S^2/B$ in the combined $\cost<0.3$ and 0.3-0.7 bins
relative to $<0.7$ was exploited. 
A \babar-like fit to the measured $E_\gamma$ 
distributions in all three 
$\cost$ bins simultaneously
extracted the $\eta_b$ signal.
CLEO left the photon selection criteria from the original
analysis unchanged and quoted a final mass, rate, and significance 
which were each the mean
from an ensemble of fits with reasonable confidence levels, 
not on any arbitrarily chosen individual fit.
The fit ensemble contained many variations,
each specifying a different background parametrization, 
$E_\gamma$ range, and/or logarithmic or linear $E_\gamma$ scale.

\Tabs{tab:Spec_ExpSumCon} and \ref{tab:Spec_etab} 
summarize the experimental $\eta_b$ results, which
together yield
\beq
\dmhf[\etab]_{\rm exp}=69.6\pm2.9\mevcc\,.
\label{eqn:SpecExp_HFSexp}
\eeq
Belle is poised to search for $\eta_b$
using its 11M~\UnS{3}\  events and 
recently augmented 160M~\UnS{2}\  dataset.

  Theoretical predictions for
\etab\ hyperfine splitting are discussed
\Secs{sec:SpecTh_lowestspectra},
\ref{sec:SpecTh_nrqcdlc},
\ref{sec:SpecTh_etab},
\ref{sec:SpecTh_alphasdec}, and
\ref{sec:SpecTh_mixing}.

\begin{table}[t]
    \caption{Measured \etab\ properties. The value quoted
for the weighted average of $\dmhf$ includes all three 
measurements
            }
    \label{tab:Spec_etab}
\setlength{\tabcolsep}{0.10pc}
\begin{center}
\begin{tabular}{cccc} 
\hline\hline
\rule[10pt]{-1mm}{0mm}
Quantity        & $\UnS{2}\to\gamma\eta_b$      &
                  $\UnS{3}\to\gamma\eta_b$      & Ref.~($\chi^2$/d.o.f.) \\[0.7mm]
\hline
\rule[10pt]{-1mm}{0mm}
$E_\gamma$      & $610.5^{+4.5}_{-4.3}\pm1.8$   & 
                  $921.2^{+2.1}_{-2.8}\pm2.4$   & \babar~\cite{:2008vj,:2009pz} \\[0.7mm]
 (MeV)          &  -  & 918.6$\pm6.0\pm 1.8$    & CLEO~\cite{Bonvicini:2009hs}   \\[0.7mm]
\hline
\rule[10pt]{-1mm}{0mm}
$m(\eta_b)$     & $9392.9^{+4.6}_{-4.8}\pm1.8$  &
                  $9388.9^{+3.1}_{-2.3}\pm 2.7$ & \babar~\cite{:2008vj,:2009pz} \\[0.7mm]
   (MeV)        & -   & $9391.8\pm6.6\pm 2.0$   & CLEO~\cite{Bonvicini:2009hs}   \\[0.7mm]
\hline
\rule[10pt]{-1mm}{0mm}
$\dmhf$ &  $67.4^{+4.8}_{-4.5}\pm1.9$   &
                   $71.4^{+2.3}_{-3.1}\pm 2.7$  & \babar~\cite{:2008vj,:2009pz} \\[0.7mm]
   (MeV)        &  - & $68.5\pm6.6\pm 2.0$      &
                   CLEO~\cite{Bonvicini:2009hs}   \\[0.7mm]
&& $69.6\pm2.9$ & Avg$^3$ (0.6/2)\\[0.7mm]
\hline
\rule[10pt]{-1mm}{0mm}
${\cal B}\times 10^4$ & $4.2^{+1.1}_{-1.0} \pm 0.9$   &
                  $4.8\pm 0.5 \pm 1.2$          & \babar~\cite{:2008vj,:2009pz} \\[0.7mm]
                & $<$8.4  & 7.1$\pm 1.8\pm 1.1$ & CLEO~\cite{Bonvicini:2009hs}   \\[0.7mm]
                & $<$5.1  & $<$4.3              & CLEO~\cite{Artuso:2004fp}\\[0.7mm]
\hline
\hline
\end{tabular}
\end{center}
\end{table}

\subsubsection{Search for $\hsubb$}
\label{sec:SpecExp_hb}

A preliminary analysis from \babar~\cite{fulsomICHEP}
describes two searches for $\hsubb$ in a 
sample of 122M \UnS{3} decays. The first
search employs a method similar to the CLEO $\hsubc$
inclusive search (see \Sec{sec:SpecExp_hc}) 
by selecting $\UnS{3}$ decays
with both a soft $\piz$ and a radiative photon,
looking for the decay chain $\UnS{3}\to\piz\hsubb$,
$\hsubb\to\gamma\etab$. With the radiative photon
restricted to the range allowed for the transition
to \etab, the mass recoiling against the soft
\piz\ is plotted and scanned for a peak above
a smooth background. \babar\ sees a $2.7\sigma$
effect at $m(\piz-{\rm recoil})=9903\pm4\pm1$\mev. In a second search,
an upper limit of  
\beqa
\Brat(\UnS{3}\to\dipi\hsubb)<2.5\times 10^{-4}\non\\
 {\rm for}~9.88<m(\hsubb)<9.92\gev{\rm ~at~90\%~CL}
\eeqa
is set.

\subsubsection{Observation of $\UoneDT$}
\label{sec:SpecExp_Ups1D}

\begin{figure}[t]
    \includegraphics[width=\figwid]{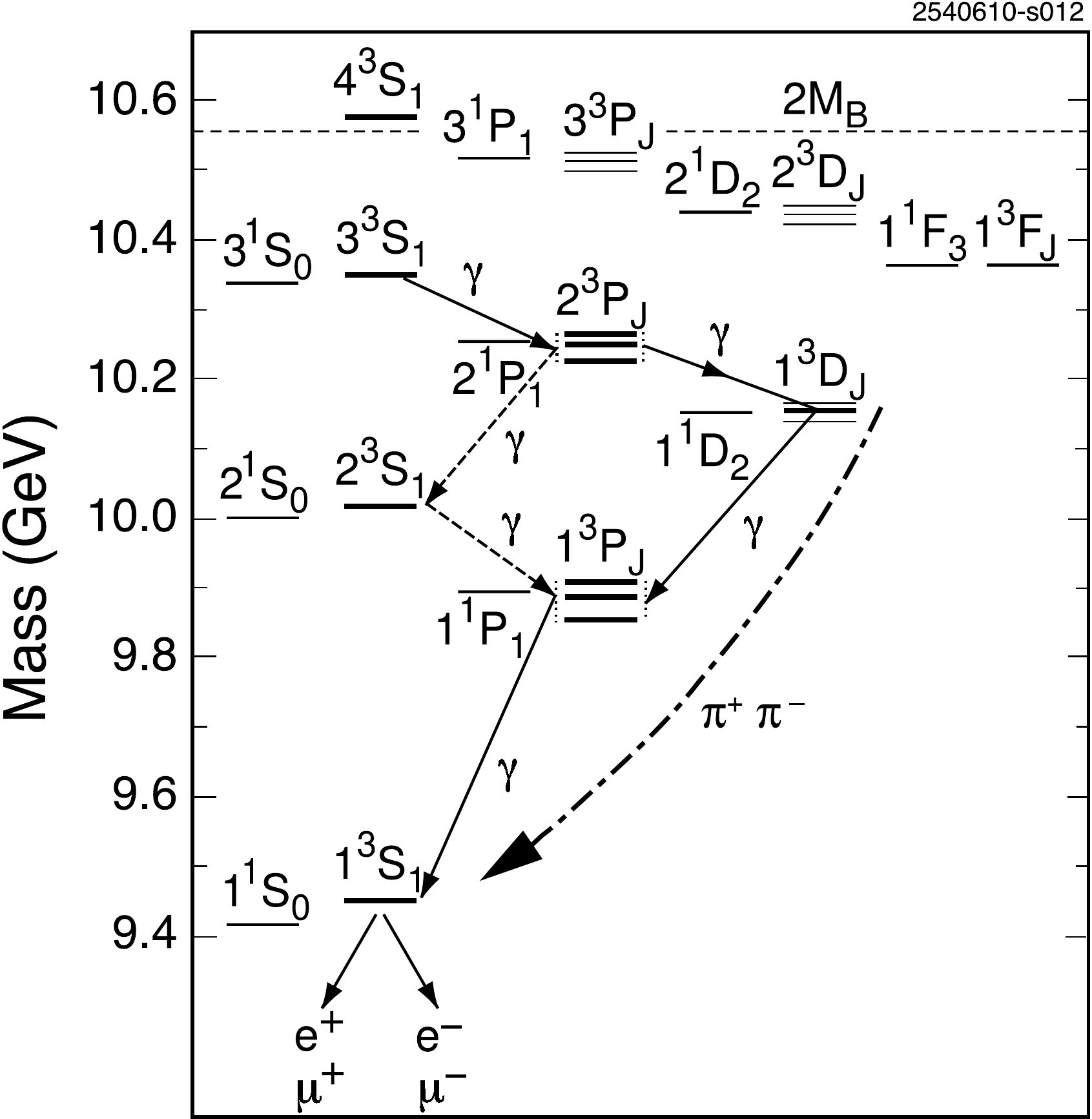}
     \caption{Expected $b\bar{b}$ bound-state 
              mass levels. The four-photon transition
              sequence from the \UnS{3}\  to the $\UnS{1}$ via the
              \UoneD\  states is shown {\it (solid lines)}. An
              alternative route for the four-photon cascade
              via the \UnS{2}\  state is also displayed {\it (dashed lines)}.
              The hadronic dipion transition from \UoneD\ to \UnS{1}
              is indicated by the {\it dot-dash curve}. 
              \AfigPermAPS{Bonvicini:2004yj}{2004} }
    \label{fig:Spec_Ups1D_cascade}
\end{figure}

\begin{figure}[t]
    \includegraphics[width=\figwid]{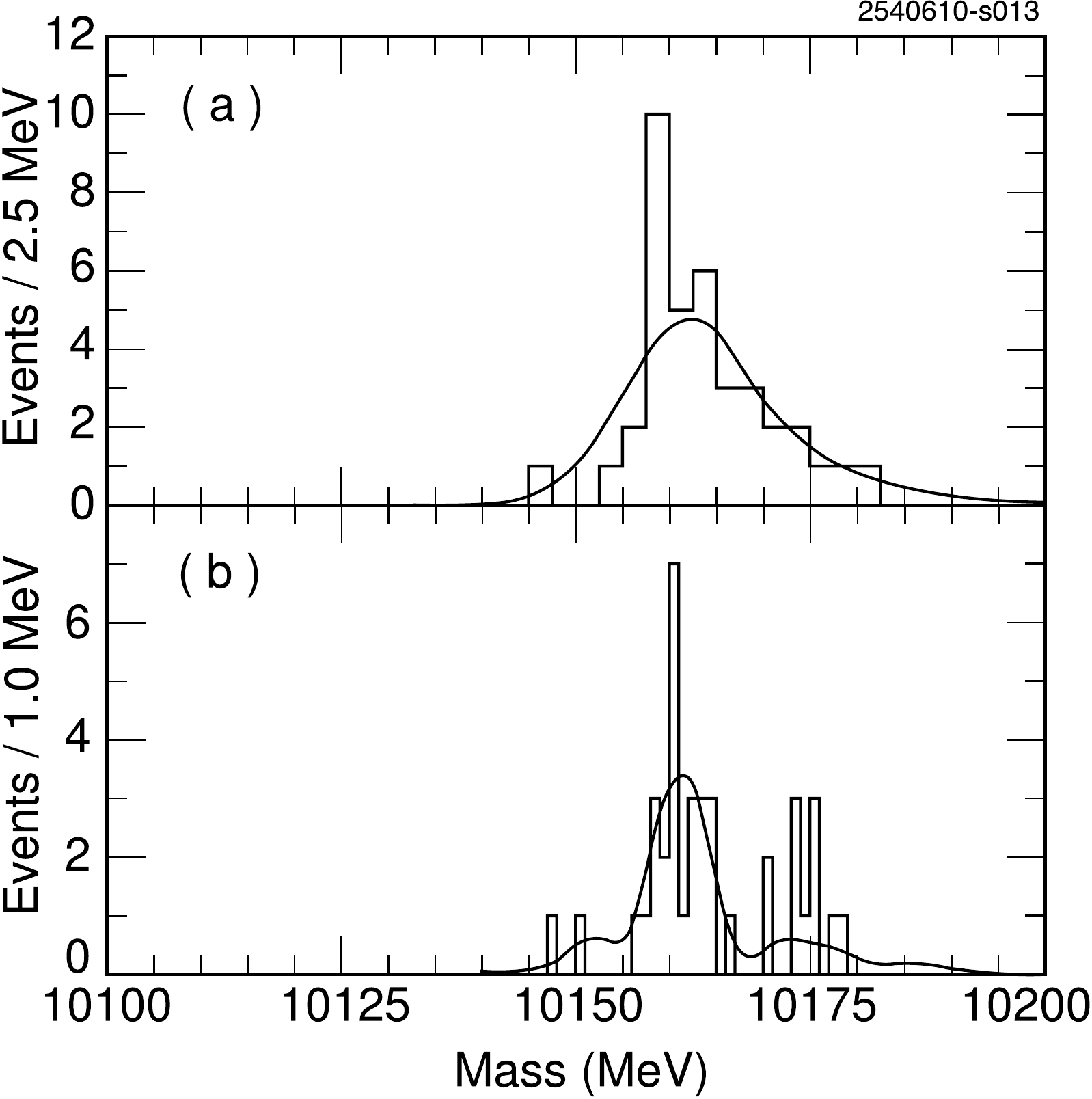}
     \caption{From CLEO~\cite{Bonvicini:2004yj},
     distributions of (a)~mass recoiling against the
     softest two photons, and (b)~mass
     that produces the smallest $\chi^2(1D)$ (see text) per event,
     from $\UnS{3}\to\gamma\gamma\gamma\gamma\ell^+\ell^-$ events, 
     selected to be consistent with a four-photon cascade through $\chi_{bJ}(2P)$,
     \UoneD, and $\chi_{bJ}(1P)$ to $\UnS{1}$.
     The {\it solid line histogram} represents data, and
     the {\it curves} represent the CLEO fits.
     \AfigPermAPS{Bonvicini:2004yj}{2004} }
    \label{fig:Spec_Ups1D_mass}
\end{figure}

CLEO~\cite{Bonvicini:2004yj} made the first of two 
observations of \UoneD, using the four-photon cascade 
shown in \Fig{fig:Spec_Ups1D_cascade}:
\beqa
\UnS{3}\to \gamma\chi_{bJ}(2P)\,,\hspace{1.9in}\non\\ 
\chi_{bJ}(2P)\to \gamma\UoneD\, ,\hspace{1.4in}\non\\
\UoneD\to\gamma\chi_{bJ}(1P)\,,\hspace{0.9in}\non\\ 
\chi_{bJ}(1P)\to\gamma\UnS{1}\,,\hspace{0.4in}\non\\
\UnS{1}\to\ell^+\ell^-\,,~~~~
\eeqa 
where $\ell^\pm\equiv e^\pm$ or $\mu^\pm$.
The largest background source of four soft photons and an $\UnS{1}$
is $\UnS{3}\to\pi^0\pi^0\UnS{1}$, which was suppressed by
vetoing events with two photon-pairings that 
are both consistent with $\pi^0$ masses. The next-most pernicious
background is the quite similar four-photon cascade through \UnS{2}\ 
instead of \UoneD\ 
(also shown in \Fig{fig:Spec_Ups1D_cascade}); 
the softer two photons overlap
the signal photons within the experimental resolution.
This latter background was suppressed by kinematically
constraining each event to a \UoneD\  hypothesis 
with unknown \UoneD\  mass and including all $J$ possibilities
for the intermediate $\chi_{bJ}(nP)$ states,
and then requiring a good fit quality, $\chi^2(1D)$. Masses from the surviving
candidates are shown in \Fig{fig:Spec_Ups1D_mass}, with the mass recoiling
against the softest two photons in (a) and the value obtained
from the minimum $\chi^2(1D)$ combination in (b). 
Both give consistent masses for an  
$\Upsilon(1^3D_2)$, and the latter has an inconclusive
1.9$\sigma$ hint of a second peak 13\mevcc\ above the primary one,
which could be an indication of the corresponding $\Upsilon(1^3D_3)$ state. 
The observed 34.5$\pm$6.4 signal events in the central peak
correspond to a statistical significance of 10$\sigma$,
most of which are attributed to cascades involving the
$\chi_{b1}(nP)$ for both $n$=1 and 2 and to production
of an $\Unx{1}{^3D_2}$. The product branching fraction
for the entire cascade 
was found to be $(2.5\pm0.7)\times10^{-5}$.
Upper limits on other possible decays relative to the
four-photon cascade were also set
 to be $<0.25$ for $\UoneD\to\eta\UnS{1}$ and
$<1.2$ for $\pi^+\pi^-\UoneD$, both at 90\%~CL.

\begin{figure}[!]
    \includegraphics[width=\figwid]{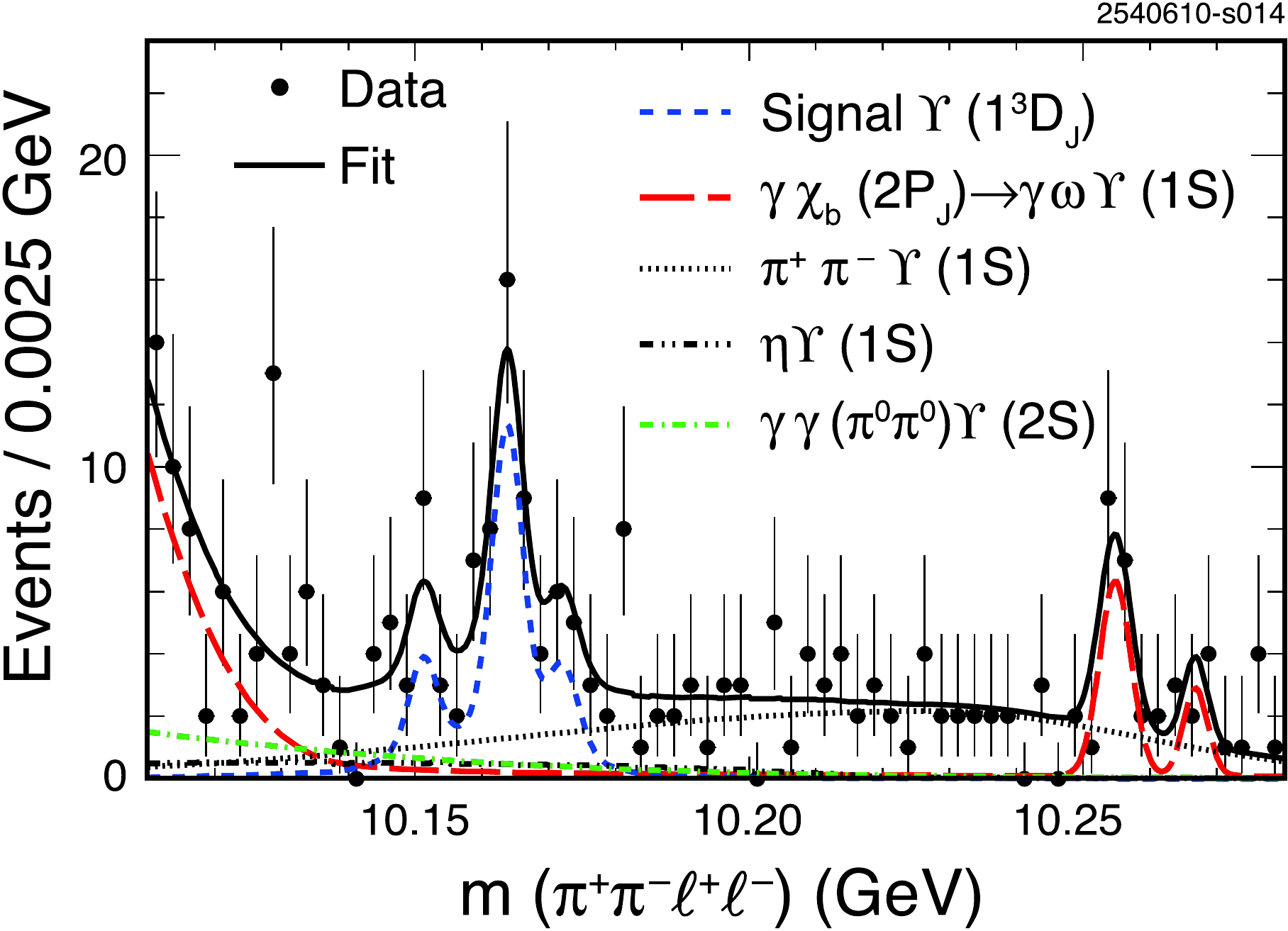}
     \caption{From \babar~\cite{Sanchez:2010kz}, the 
              $\pi^+\pi^-\ell^+\ell^-$ invariant mass   
              restricted to those with dilepton masses near
              that of the $\UnS{1}$. {\it Curves} represent
              $\Unx{1}{^3D_J}$ signals only {\it (short dash)},
              the total fit {\it (solid)} which includes four backgrounds
              {it (others)}. \AfigPermAPS{Sanchez:2010kz}{2010}
     }
    \label{fig:Spec_Ups1D_mass_babar}
\end{figure}

Belle has an \UnS{3}\ dataset slightly larger than CLEO 
and therefore could mount a comparable \UoneD\ search. 
\babar\ has twenty times more \UnS{3} than CLEO, 
and therefore has the
capability to search for other decay chains and to
explore hyperfine mass structure of the allowed \UoneD\ spin states.
While neither Belle nor \babar\ has yet explored the
four-photon cascade, \babar\ has observed~\cite{Sanchez:2010kz}  
$\Unx{1}{^3D_2}$ produced from a two-photon cascade from
$\UnS{3}$ decay as does CLEO, but then undergoing a 
charged dipion transition to the \UnS{1},
which then decays to $\ell^+\ell^-$. The $\pi^+\pi^-\ell^+\ell^-$
invariant mass distribution from such events,
restricted to those with dilepton masses near
that of the $\UnS{1}$, is shown in \Fig{fig:Spec_Ups1D_mass_babar}.
In addition to confirming the $\Unx{1}{^3D_2}$ signal
at 5.8$\sigma$, the $\Unx{1}{^3D_1}$ and $\Unx{1}{^3D_3}$
states are seen at 1.8$\sigma$ and 1.6$\sigma$, respectively.
The $\Unx{1}{^3D_2}$ mass is somewhat larger than
but consistent with the CLEO value,
as shown in \Tab{tab:Spec_U1D_mass}. The \babar\ analysis
also concludes first, that the dipion
invariant mass distribution is in substantially better agreement
with that predicted for a $\Unx{1}{^3D_J}$ than
an $S$ or $P$ state, and second, that angular distributions
of the $\Unx{1}{^3D_2}$ signal events are consistent with
the quantum number assignments of $J=2$ and $P=-1$.

\subsection{New unanticipated states}
\label{sec:SpecExp_Unanticipated}

\subsubsection{$X(3872)$, the enduring exotic}
\label{sec:SpecExpX3872}

The $X(3872)$ occupies a unique niche in the 
menagerie of unanticipated states
listed in \Tab{tab:Spec_ExpSumUnc} as
both the first and the most intriguing. 
At this point it is widely studied, yet
its interpretation demands much more
experimental attention. Its apparent
quantum numbers, mass, and decay patterns 
make it an unlikely conventional charmonium candidate,
and no consensus explanation has been found.

\begin{table}[t]
\caption{Measured mass values of the $\Unx{1}{^3D_2}$ 
}
\label{tab:Spec_U1D_mass}
\setlength{\tabcolsep}{0.15pc}
\begin{center}
\begin{tabular}{ccc}
\hline\hline
\rule[10pt]{-1mm}{0mm}
 Decay & Value~(MeV) & Ref. ($\chi^2$/d.o.f.)\\[0.7mm]
\hline
\rule[10pt]{-1mm}{0mm}
 $\UnS{3}\to\gamma\gamma\gamma\gamma\UnS{1} $ &
       10161.1$\pm$0.6$\pm$1.6 & CLEO~\cite{Bonvicini:2004yj} \\[0.7mm]
 $\UnS{3}\to\gamma\gamma\pi^+\pi^-\UnS{1} $ &
       10164.5$\pm$0.8$\pm$0.5 & \babar~\cite{Sanchez:2010kz} \\[0.7mm]
 Both of above & 10163.8$\pm$1.4 & Avg$^3$ (3.1/1) \\[0.7mm]
\hline\hline
\end{tabular}
\end{center}
\end{table}

\begin{table*}[tbp]
\caption{As in \Tab{tab:Spec_ExpSumCon}, but for new 
{\it unconventional} states in the $c\bar{c}$ and
$b\bar{b}$ regions, ordered by mass. 
For $X(3872)$, the values given are based only upon 
decays to $\pi^+\pi^- J/\psi$. 
$X(3945)$ and $Y(3940)$ have been subsumed under 
$X(3915)$ due to compatible properties. 
The state known as $Z(3930)$ appears as
the $\chi_{c2}(2P)$ in \Tab{tab:Spec_ExpSumCon}.
See also the reviews in 
\cite{Eichten:2007qx,Godfrey:2008nc,Barnes:2009zza,Pakhlova:2010zz} } 
\setlength{\tabcolsep}{0.23pc}
\label{tab:Spec_ExpSumUnc}
\begin{center}
\begin{tabular}{lccclccc}
\hline\hline
\rule[10pt]{-1mm}{0mm}
 State & $m$~(MeV) & $\Gamma$~(MeV) & $J^{PC}$ & \ \ \ \ Process~(mode) & 
     Experiment~(\#$\sigma$) & Year & Status \\[0.7mm]
\hline
\rule[10pt]{-1mm}{0mm}
$X(3872)$& 3871.52$\pm$0.20 & 1.3$\pm$0.6 &
    $1^{++}/2^{-+}$
    & $B\to K (\pi^+\pi^-J/\psi)$ &
    {\color{red} Belle} \cite{Choi:2003ue,Adachi:2008te}~(12.8), 
    \babar~\cite{Aubert:2008gu}~(8.6) & 2003 & OK \\[0.7mm]
& &($<$2.2) & & $p\bar p\to (\pi^+\pi^- J/\psi)+ ...$ &
    CDF~\cite{Acosta:2003zx,Abulencia:2006ma,Aaltonen:2009vj}~(np), \DZero~\cite{Abazov:2004kp}~(5.2) & &\\[0.7mm]
& & &   & $B\to K (\omega J/\psi)$ &
    Belle~\cite{Abe:2005ix}~(4.3),
    \babar~\cite{delAmoSanchez:2010jr}~(4.0) & &\\[0.7mm]
& & & & $B\to K (\DstnDn)$ &
    Belle~\cite{Gokhroo:2006bt,Aushev:2008su}~(6.4), 
    \babar~\cite{Aubert:2007rva}~(4.9) & &\\[0.7mm]
& & & & $B\to K (\gamma J/\psi)$ &
    Belle~\cite{Abe:2005ix}~(4.0), \babar~\cite{Aubert:2006aj,Aubert:2008rn}~(3.6)&&\\[0.7mm]
& & & & $B\to K (\gamma \psi(2S))$ & \babar~\cite{Aubert:2008rn}~(3.5), Belle~\cite{Bhardwaj:2010qwg}~(0.4) & & \\[1.89mm]
$X(3915)$ & $3915.6\pm3.1$ & 28$\pm$10 & $0/2^{?+}$ &
    $B\to K (\omega \jpsi)$ &
    {\color{red} Belle}~\cite{Abe:2004zs}~(8.1),
    \babar~\cite{Aubert:2007vj}~(19) & 2004 & OK\\ [0.7mm]
     & & & & $e^+e^-\to e^+e^- (\omega\jpsi)$ &
    {\color{red} Belle}~\cite{Uehara:2009tx}~(7.7) &&\\[1.89mm]
$X(3940)$ & $3942^{+9}_{-8}$ & $37^{+27}_{-17}$ & $?^{?+}$ &
     $e^+e^-\to J/\psi(\DDst)$ &
     {\color{red} Belle}~\cite{Abe:2007sya}~(6.0) & 2007 & {\color{red} NC!}\\ [0.7mm]
&&&& $e^+e^-\to J/\psi\ (...)$ &
     {\color{red} Belle}~\cite{Abe:2007jn}~(5.0) & \\ [1.89mm]
$G(3900)$ & $3943\pm21$ & 52$\pm$11 & $1^{--}$ &
     $e^+e^-\to\gamma  (\DDbar)$ &
     {\color{red} \babar}~\cite{Aubert:2006mi}~(np),
     Belle~\cite{Pakhlova:2008zza}~(np)
     & 2007 & OK \\ [1.89mm]
$Y(4008)$ & $4008^{+121}_{-\ 49}$ & 226$\pm$97 & $1^{--}$ &
     $e^+e^-\to\gamma  (\pi^+\pi^-J/\psi)$ &
     {\color{red} Belle}~\cite{Belle:2007sj}~(7.4)
     & 2007 & {\color{red} NC!} \\[1.89mm]
$Z_1(4050)^+$ & $4051^{+24}_{-43}$ & $82^{+51}_{-55}$ & ?&
     $ B\to K (\pi^+\chi_{c1}(1P))$ &
     {\color{red} Belle}~\cite{Mizuk:2008me}~(5.0)  & 2008 & {\color{red} NC!}\\[1.89mm]
$Y(4140)$ & $4143.4\pm3.0 $ & $15^{+11}_{-\ 7}$ & $?^{?+}$ &
     $B\to K (\phi J/\psi)$ &
     {\color{red} CDF}~\cite{Aaltonen:2009tz,collaboration:2010aa}~(5.0) & 2009 & {\color{red} NC!}\\[1.89mm]
$X(4160)$ & $4156^{+29}_{-25} $ & $139^{+113}_{-65}$ & $?^{?+}$ &
     $e^+e^-\to\jpsi(\DDst)$ &
     {\color{red} Belle}~\cite{Abe:2007sya}~(5.5) & 2007 & {\color{red} NC!}\\[1.89mm]
$Z_2(4250)^+$ & $4248^{+185}_{-\ 45}$ &
     177$^{+321}_{-\ 72}$ &?&
     $ B\to K (\pi^+\chi_{c1}(1P))$ &
     {\color{red} Belle}~\cite{Mizuk:2008me}~(5.0)  & 2008 &{\color{red} NC!}\\[1.89mm]
$Y(4260)$ & $4263\pm5$ & 108$\pm$14 & $1^{--}$ &
     $e^+e^-\to\gamma  (\pi^+\pi^- J/\psi)$ &
     {\color{red} \babar}~\cite{Aubert:2005rm,Aubert:2008ic}~(8.0) & 2005 & OK \\ [0.7mm]
     &&&&& CLEO~\cite{He:2006kg}~(5.4) & & \\ [0.7mm]
     &&&&& Belle~\cite{Belle:2007sj}~(15) & & \\ [0.7mm]
& & & & $e^+e^-\to (\pi^+\pi^- J/\psi)$ & CLEO~\cite{Coan:2006rv}~(11)& &\\[0.7mm]
& & & & $e^+e^-\to (\pi^0\pi^0 J/\psi)$ & CLEO~\cite{Coan:2006rv}~(5.1) & &\\[1.89mm]
$Y(4274)$ & $4274.4^{+8.4}_{-6.7}$ & $32^{+22}_{-15}$ & $?^{?+}$ &
     $B\to K (\phi J/\psi)$ &
     {\color{red} CDF}~\cite{collaboration:2010aa}~(3.1) & 2010 & {\color{red} NC!}\\[1.89mm]
$X(4350)$ & $4350.6^{+4.6}_{-5.1}$ & $13.3^{+18.4}_{-10.0}$ & 0,2$^{++}$ &
     $e^+e^-\to e^+e^- (\phi\jpsi)$ &
     {\color{red} Belle}~\cite{Shen:2009vs}~(3.2) & 2009 & {\color{red} NC!}\\ [1.89mm]
$Y(4360)$ & $4353\pm11$ & 96$\pm$42 & $1^{--}$ &
     $e^+e^-\to\gamma  (\pi^+\pi^- \psip)$ &
     {\color{red} \babar}~\cite{Aubert:2006ge}~(np),
     Belle~\cite{:2007ea}~(8.0) & 2007 &  OK \\ [1.89mm]
$Z(4430)^+$ & $4443^{+24}_{-18}$ & $107^{+113}_{-\ 71}$ & ?&
     $B\to K (\pi^+\psi(2S))$ &
     {\color{red} Belle}~\cite{Choi:2007wga,Mizuk:2009da}~(6.4)
     & 2007 & {\color{red} NC!}\\[1.89mm]
$X(4630)$ & $4634^{+\ 9}_{-11}$ & $92^{+41}_{-32}$ & $1^{--}$ &
     $e^+e^-\to\gamma (\lala)$ &
     {\color{red} Belle}~\cite{Pakhlova:2008vn}~(8.2)  & 2007 & {\color{red} NC!}\\ [1.89mm]
$Y(4660)$ & 4664$\pm$12 & 48$\pm$15 & $1^{--}$ &
     $e^+e^-\to\gamma (\pi^+\pi^- \psi(2S))$ &
     {\color{red} Belle}~\cite{:2007ea}~(5.8)  & 2007 & {\color{red} NC!}\\ [1.89mm]
$Y_b(10888)$ & 10888.4$\pm$3.0 & 30.7$^{+8.9}_{-7.7}$ & $1^{--}$ &
      $e^+e^-\to(\pi^+\pi^- \Upsilon(nS))$ &
      {\color{red} Belle}~\cite{Chen:2008pu,Abe:2007tk}~(3.2)& 2010 & {\color{red} NC!}\\[0.7mm]
\hline\hline
\end{tabular}
\end{center}
\end{table*}

In 2003, while studying $B^+\to K^+\ppjp$, 
Belle~\cite{Choi:2003ue} discovered
an unexpected enhancement in the \ppjp\
invariant mass spectrum near 3872\mevcc.
This sighting of the $X$ in $B$-decays
was later confirmed by \babar~\cite{Aubert:2008gu}.
The $X\to\ppjp$ decay was also observed 
inclusively in prompt production 
from $\bar{p}p$ collisions at the Tevatron
by both CDF~\cite{Acosta:2003zx,Abulencia:2006ma,Aaltonen:2009vj}
and \DZero~\cite{Abazov:2004kp}. 
CDF~\cite{Abulencia:2006ma} studied the
angular distributions and correlations of the \ppjp\ final state,
finding that the dipion was favored to
originate as a $\rho^0$, and that
only $J^{PC}$ assignments of $1^{++}$ and $2^{-+}$
explained their measurements adequately.
Belle~\cite{Abe:2005ix} reported 
evidence
for the $\gamma\jpsi$ decay,
which \babar~\cite{Aubert:2006aj,Aubert:2008rn} confirmed 
at $4\sigma$ significance.
The radiative decay 
verifies the positive $C$-parity assignment of CDF.
It also bolsters the $1^{++}$ assignment
because a $2^{-+}$ state would have to undergo a high-order
multipole transition which would be 
more strongly suppressed than the observed rates allow.

\begin{table}[t]
\caption{$X(3872)$ mass and width measurements by decay mode and experiment.
The $\chi^2$/d.o.f.~values given in parentheses refer
to weighted averages of the masses only. The lines marked
$(B^\pm)$ and $(B^0)$ represent mass values quoted by \babar\
in charged and neutral $B$-decays, respectively
}
\label{tab:Spec_XMass}
\setlength{\tabcolsep}{0.20pc}
\begin{center}
\begin{tabular}{cccc}
\hline\hline
\rule[10pt]{-1mm}{0mm}
 Mode & Mass &  Width &Ref.\\
      & (MeV)& (MeV) &  ($\chi^2$/d.o.f.)\\[0.7mm]
\hline
\rule[10pt]{-1mm}{0mm}
\ppjp & 3871.46$\pm$0.37$\pm$0.07 & 1.4$\pm$0.7 &Belle~\cite{Adachi:2008te} \\[0.7mm]
$(B^\pm)$ & 3871.4$\pm$0.6$\pm$0.1 & 1.1$\pm$1.5$\pm$0.2 &\babar~\cite{Aubert:2008gu}\\[0.7mm]
$(B^0)$ & 3868.7$\pm$1.5$\pm$0.4 & - &\babar~\cite{Aubert:2008gu}\\[0.7mm]
 & 3871.8$\pm$3.1$\pm$3.0 & - &\DZero~\cite{Acosta:2003zx}   \\[0.7mm]
 & 3871.61$\pm$0.16$\pm$0.19 & - &CDF~\cite{Aaltonen:2009vj}   \\[0.7mm]
    & 3871.52$\pm$0.20 &1.3$\pm$0.6 &Avg$^3$ (2.1/4)\\[0.7mm]
\hline
\rule[10pt]{-1mm}{0mm}
\DstnDn & 3875.1$^{+0.7}_{-0.5}$$\pm$0.5 & 3.0$^{+1.9}_{-1.4}$$\pm$$0.9$&\babar~\cite{Aubert:2007rva} \\[0.7mm]
 & 3872.9$^{+0.6\,+0.4}_{-0.4\,-0.5}$ & 3.9$^{+2.8\,+0.2}_{-1.4\,-1.1}$ &Belle~\cite{Aushev:2008su} \\[0.7mm]
                   & 3874.0$\pm$1.2 & 3.5$^{+1.6}_{-1.0}$&Avg$^3$ (4.7/1)\\[0.7mm]
\hline\hline
\end{tabular}
\end{center}
\end{table}

\begin{table}[t]
\caption{Mass measurements relevant to the $X(3872)$. 
We define $\delta m_0\equiv m(\Dstn)-m(\Dze)$
and $\Delta m_{thr}\equiv m[X(3872)]- [ m(\Dstn)+m(\Dze) ]$ }
\label{tab:Spec_DMass}
\setlength{\tabcolsep}{0.10pc}
\begin{center}
\begin{tabular}{ccc}
\hline\hline
\rule[10pt]{-1mm}{0mm}
Quantity & Mass~(MeV)  & Ref. ($\chi^2$/d.o.f.)\\[0.7mm]
\hline
\rule[10pt]{-1mm}{0mm}
$m(D^0)$ & 1864.6$\pm$0.3$\pm$1.0 & ACCMOR~\cite{Barlag:1990fv} \\[0.7mm]
         & 1864.847$\pm$0.150$\pm$0.095 & CLEO~\cite{Cawlfield:2007dw} \\[0.7mm]
         & 1865.3$\pm$0.33$\pm$0.23 & KEDR~\cite{Anashin:2009ir} \\[0.7mm]
 & 1864.91$\pm$0.16 & Avg$^3$ (1.2/2) \\[0.7mm]
\hline
\rule[10pt]{-1mm}{0mm}
$\delta m_0$ & 142.12$\pm$0.07 & PDG08~\cite{Amsler:2008zzb}\\[0.7mm]
\hline
\rule[10pt]{-1mm}{0mm}
$2m(D^0)+ \delta m_0$ & 3871.94$\pm$0.33 & -  \\[0.7mm]
$m[X(3872)]$ & 3871.52$\pm$0.20 & \Tab{tab:Spec_XMass} (\ppjp) \\[0.7mm]
$\Delta m_{thr}$ & $-$0.42$\pm$0.39 & - \\[0.7mm]
&$\in[-0.92, 0.08]$ & @90\%~CL \\[0.7mm]
\hline\hline
\end{tabular}
\end{center}
\end{table}

\begin{figure}[t]
    \includegraphics[width=\figwid]{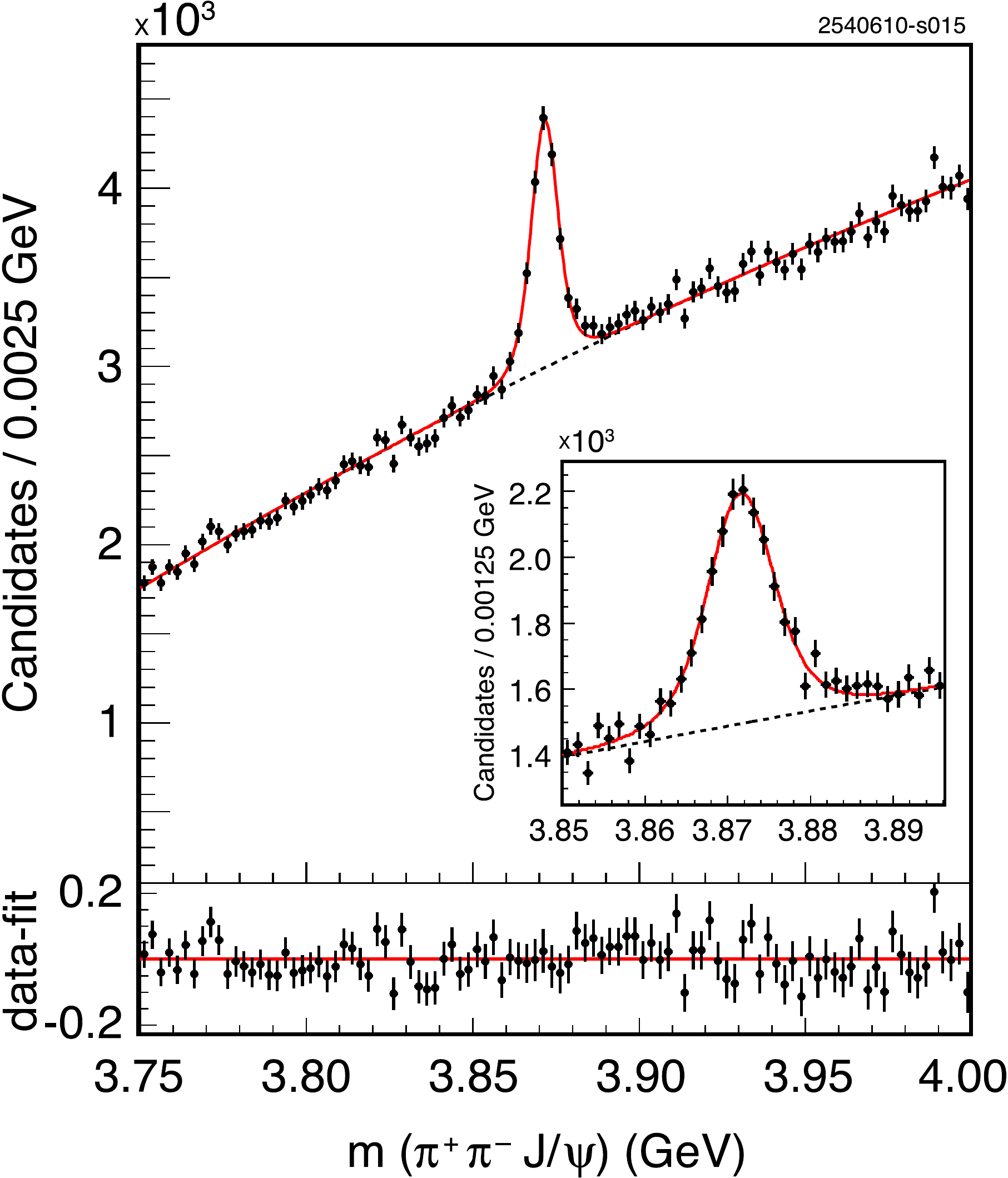}
     \caption{From CDF~\cite{Aaltonen:2009vj}, the \ppjp\ invariant 
              mass distribution for $X(3872)$ candidates, 
              showing the data {\it (points)}, the projection of the unbinned 
              maximum-likelihood fit {\it (solid line)} and its smooth 
              background component {\it (dashed line)}, and the inset
              which enlarges the peak region with finer binning. 
              The lower panel shows residuals of the data with respect to the fit. 
              \AfigPermAPS{Aaltonen:2009vj}{2009} }
    \label{fig:Spec_X3872Mass_cdf}
\end{figure}

From the beginning, the proximity of the $X$ mass
to \DstnDn\ threshold was conspicuous,
and eventually decays to \DstnDn\ were observed by
\babar~\cite{Aubert:2007rva} and Belle~\cite{Aushev:2008su}.
Interest in the relationship of $X$ to \DstnDn\ fueled improvements in
measurements of its mass, as shown in
\Tab{tab:Spec_XMass}, and of the $\Dze$ mass,
as shown in \Tab{tab:Spec_DMass}.
The $X$ mass measurements based upon
the \ppjp\ decay are 
consistent with one another. 
The world-average $X$ mass,
restricted to measurements using \ppjp\ decays,
is dominated by the CDF~\cite{Aaltonen:2009vj}
inclusive result,
illustrated in \Fig{fig:Spec_X3872Mass_cdf}.
The CDF systematic uncertainty on the mass was obtained from 
studies of $\psip\to\ppjp$ decays,
which have a similar topology and a well-known \psip\ mass to match.
The measured mass discrepancy was extrapolated from the \psip\ mass to the $X$
mass to obtain error estimates.
The world-average \Dze\ mass precision is dominated by a 
CLEO~\cite{Cawlfield:2007dw}
measurement that uses the decay chain $\Dze$$\to$$\phi K_S^0$,
$\phi\to K^+K^-$, $K_S^0\to\pi^+\pi^-$,
and is limited by statistics. 
Despite all these advances, the \DstnDn\ mass threshold
test remains ambiguous, with 
$m[X(3872)]- [ m(\Dstn)+m(\Dze) ]=-0.42\pm0.39$\mevcc.
This limits the hypothetical \DstnDn\ binding energy 
to be $<$0.92~MeV at 90\%~CL and does not foreclose
the possibility that the $X(3872)$ is {\it above} \DstnDn\ threshold.
Further clarity here would require much more precise mass
measurements for both the $X$ and the $D^0$.

Both Belle and \babar\ have reported $X(3872)$ signals
in the \DstnDn\ final state with branching fractions
about ten times higher than for \ppjp.
Both used $\Dstn\to\Dze\pi^0$ and $\Dze\gamma$ decays, 
both selected and kinematically constrained
a \Dstn\ candidate in each event,
and both performed unbinned 
maximum likelihood fits to the 
\DstnDn\ mass. (Belle's fit is two-dimensional, the 
second dimension being a $B$-meson-consistency
kinematic variable; \babar\ cuts on $B$-meson consistency.) 
Both results appear in \Tab{tab:Spec_XMass}.
(An earlier Belle publication~\cite{Gokhroo:2006bt}
used a dataset smaller by one-third than in \cite{Aushev:2008su}, 
made no \Dstn-mass constraint,
and measured a mass value of $3875.2\pm0.7^{+0.3}_{-1.6}\pm0.8$\mevcc.)
Belle~\cite{Aushev:2008su} 
fit to a conventional Breit-Wigner signal shape convolved
with a Gaussian resolution function.
\babar~\cite{Aubert:2007rva} fit the data to an ensemble of MC samples,
each generated with different plausible $X$ masses and widths
and assuming a purely $S$-wave decay of a spin-1 resonance.
The \babar\ $X$ mass from \DstnDn\ decays is more than 3\mevcc\ larger than
the world average from \ppjp, which engendered speculation
that the \DstnDn\ enhancement might be a different state
than that observed in \ppjp, but the smaller value observed
by Belle in \DstnDn\ seems to make that possibility unlikely.
The two $X$ mass measurements using \DstnDn\ decays are inconsistent 
by 2.2$\sigma$, and are 1.8$\sigma$ and 4.7$\sigma$ higher than
the \ppjp-based mass. 
However, 
important subtleties pointed out by Braaten and 
co-authors~\cite{Braaten:2007dw,Stapleton:2009ey} 
appear to explain at least qualitatively why
masses extracted in this manner are larger than in \ppjp.

\begin{figure}[b]
    \includegraphics[width=\figwid]{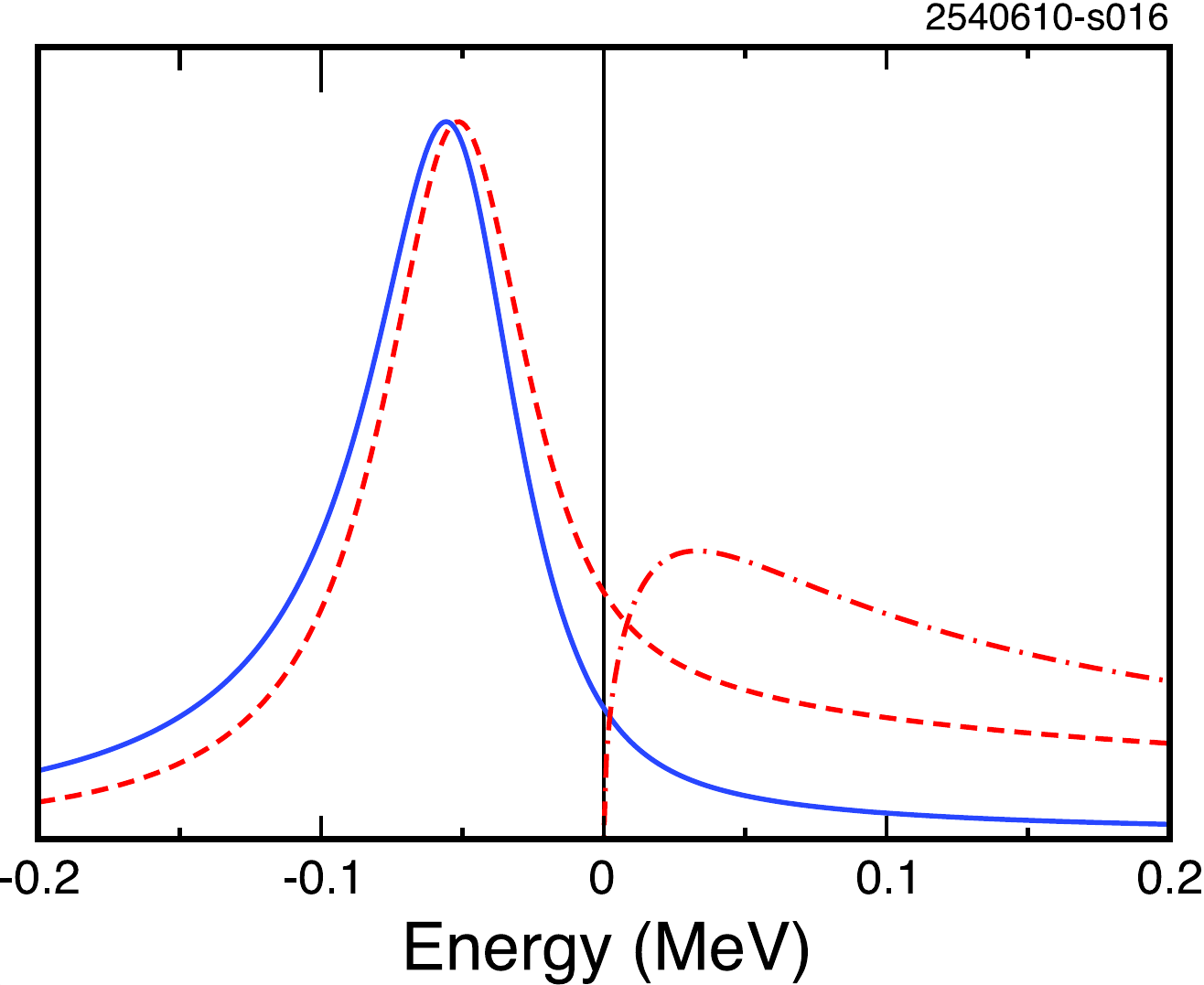}
     \caption{From Braaten and Stapleton~\cite{Stapleton:2009ey},
              the $X(3872)$ lineshapes extracted from a fit
              to the Belle~\cite{Aushev:2008su} \DstnDn\ events,
              unfolding the effects of experimental
              resolution, for \ppjp\ {\it (solid curve)}, $\DzDz\pi^0$
              {\it (dashed)}, and, when always 
              constraining
              one $\Dze\pi^0$ pairing per event to a 
              \Dstn\ mass, \DstnDn {\it (dot-dashed)}.
              The {\it horizontal axis} is the invariant mass
              of the decay products relative to \DstnDn\ threshold,
              and the {\it solid} and {\it dashed curves} are normalized
              so as to have the same peak height.
              \AfigPermAPS{Stapleton:2009ey}{2010}
              }
    \label{fig:Spec_X3872Mass_braaten8}
\end{figure}

Measuring the $X$ mass with the \DstnDn\ decay
is considerably more challenging than with \ppjp\ 
for several reasons~\cite{Braaten:2007dw,Stapleton:2009ey}. 
If conceived as a bound or virtual \DstnDn\ state~\cite{Goldberger:1964},
the $X$ lineshape in this decay mode is determined 
by the binding energy, the \Dstn\ natural width, 
and the natural width of the $X$ itself, which is at 
least as large as the \Dstn\ width~\cite{Braaten:2007dw}. 
Because the binding energy of the $X$ is less than 1\mev, 
whether or not its mass {\it peak} is below \DstnDn\ threshold,
substantial fractions of the {\it lineshape} will lie both
above and below that threshold.   The portion of
the $X$ lineshape below \DstnDn\ threshold, by definition, 
cannot decay to \DstnDn. However, $\DzDz\pi^0$ and 
$\DzDz\gamma$ final states are possible from decays of a
bound, effectively off-shell, \Dstn, as there is adequate
phase space available above $\DzDz\pi^0$ threshold. Due to imperfect
experimental resolution, these final states are 
indistinguishable from \DstnDn\ even though the \Dstn\
decay products have masses below that of \Dstn. Furthermore, the analysis
procedure which mass-constrains a \Dstn\ candidate in each event distorts the 
purported $X$ mass distribution for below-threshold decays.
Conversely, that portion of the $X$ lineshape above \DstnDn\
threshold can, of course, decay to \DstnDn, but the \DstnDn\ mass 
distribution should, by definition, be exactly zero below threshold. 
Therefore the kinematic constraint on the reconstructed
$\Dze\piz$ to the \Dstn-mass, as carried out by
Belle and \babar,
results in a broad \DstnDn\ mass peak above threshold that
should not be misconstrued as the true $X$ lineshape:
neither the mass nor width results from \DstnDn\ reflect
the true mass or width of the $X$.
Rather, the lineshapes for \ppjp, \DstnDn, and $\DzDz\pi^0$ (and
$\DzDz\gamma$) final states are related but slightly different from one
another, as shown in \Fig{fig:Spec_X3872Mass_braaten8}.
More data and more sophisticated analyses are required
to fully exploit what $\DzDz\pi^0$ and $\DzDz\gamma$ decays
can reveal about the nature of the $X$.

\begin{table}
\caption{For $X(3872)$, 
measured branching fractions and products
thereof, in units of $10^{-6}$:\newline
${\cal B}_{B^+}\equiv{\cal B}(B^+\to K^+ X)$,\newline
${\cal B}_{\gamma1}\equiv{\cal B}(B^+\to K^+ X)\times{\cal B}(X\to \gamma\jpsi)$,\newline
${\cal B}_{\gamma2}\equiv{\cal B}(B^+\to K^+ X)\times{\cal B}(X\to \gamma\psip)$,\newline
${\cal B}_0\equiv{\cal B}(B^0\to K X)\times{\cal B}(X\to f)$,\newline
${\cal B}_+\equiv{\cal B}(B^+\to K X)\times{\cal B}(X\to f)$, and\newline
${\cal B}_{0+}\equiv({\cal B}_0 + {\cal B}_+)/2$ for final state $f$.\newline
Branching fraction ratios are defined as:\newline
$R_{0+}\equiv{\cal B}(B^0\to K^0X)/{\cal B}(B^+\to K^+ X)$,\newline
$r_{DD\pi}\equiv{\cal B}_{0+}(\DzDz\pi^0)/{\cal B}_{0+}(\ppjp)$,\newline
$r_\omega\equiv {\cal B}(X\to\omega\jpsi)/{\cal B}(X\to\pi^+\pi^-\jpsi)$,\newline
$r_{\gamma 1}\equiv {\cal B}(X\to\gamma\jpsi)/{\cal B}(X\to\pi^+\pi^-\jpsi)$,
and\newline
$r_{\gamma 2}\equiv {\cal B}(X\to\gamma\psip)/{\cal B}(X\to\pi^+\pi^-\jpsi)$ 
}
\label{tab:Spec_RatBchgBneu}
\setlength{\tabcolsep}{0.20pc}
\begin{center}
\begin{tabular}{cccc}
\hline\hline
\rule[10pt]{-1mm}{0mm}
What & Mode &  Value  & Ref. ($\chi^2/$d.o.f.)\\
\hline
\rule[10pt]{-1mm}{0mm}
${\cal B}_{B^+}$ & & $<$320 & \babar~\cite{Aubert:2005vi}\\
\hline
\rule[10pt]{-1mm}{0mm}
${\cal B}_{\gamma 1}$ & $\gamma\jpsi$  & 1.8$\pm$0.6$\pm$0.1 & Belle~\cite{Abe:2005ix} \\
                      &                & 2.8$\pm$0.8$\pm$0.1 & \babar~\cite{Aubert:2008rn} \\
                      &                & 2.2$\pm$0.5         & Avg$^3$ (1.0/1)\\
\hline
\rule[10pt]{-1mm}{0mm}
${\cal B}_{\gamma 2}$ & $\gamma\psip$  & $<$3.4              &Belle~\cite{Bhardwaj:2010qwg} \\
                      &                & 0.8$\pm$2.0\footnote{
               Belle only quotes an upper limit
               for this preliminary result. From the
               information presented in \cite{Bhardwaj:2010qwg}, we have extracted
               an approximate central value and error for this table}         
                                                             &Belle~\cite{Bhardwaj:2010qwg} \\
                      &                & 9.5$\pm$2.7$\pm$0.6 &\babar~\cite{Aubert:2008rn} \\
                      &                & 3.8$\pm$4.1         &Avg$^3$ (6.4/1)\\
\hline
\rule[10pt]{-1mm}{0mm}
${\cal B}_{\gamma 2}/{\cal B}_{\gamma 1}$ & $\gamma\psip/\gamma\jpsi$  &
4.3$\pm$1.6 & Values above \\
\hline
\rule[10pt]{-1mm}{0mm}
${\cal B}_0$ & \DstnDn & 167$\pm$36$\pm$47  & \babar~\cite{Aubert:2007rva} \\
${\cal B}_+$ &         & 222$\pm$105$\pm$42 & \babar~\cite{Aubert:2007rva} \\
${\cal B}_0$ &         &  97$\pm$46$\pm$13  & Belle~\cite{Aushev:2008su} \\
${\cal B}_+$ &         &  77$\pm$16$\pm$10  & Belle~\cite{Aushev:2008su} \\
${\cal B}_{0+}$ &      &  90$\pm$19         & Avg$^3$ (3.6/3) \\
\hline
\rule[10pt]{-1mm}{0mm}
${\cal B}_{0+}/{\cal B}_{B^+}$ & \DstnDn & $>$28\% & Above \\
\hline
\rule[10pt]{-1mm}{0mm}
${\cal B}_0$ & \ppjp   & 3.50$\pm$1.90$\pm$0.40 & \babar~\cite{Aubert:2008gu} \\
${\cal B}_+$ &         & 8.40$\pm$1.50$\pm$0.70 & \babar~\cite{Aubert:2008gu} \\
${\cal B}_0$ &         & 6.65$\pm$1.63$\pm$0.55 & Belle~\cite{Adachi:2008te} \\
${\cal B}_+$ &         & 8.10$\pm$0.92$\pm$0.66 & Belle~\cite{Adachi:2008te} \\
${\cal B}_{0+}$ &      & 7.18$\pm$0.97          & Avg$^3$ (4.9/3) \\
\hline
\rule[10pt]{-1mm}{0mm}
${\cal B}_{0+}/{\cal B}_{B^+}$ & \ppjp & $>$2.2\% & Above \\
\hline
\rule[10pt]{-1mm}{0mm}
$R_{0+}$ & \ppjp        & 0.82$\pm$0.22$\pm$0.05 & Belle~\cite{Adachi:2008te}   \\
         & \ppjp        & 0.41$\pm$0.24$\pm$0.05 & \babar~\cite{Aubert:2008gu}  \\
         & \DstnDn      & 1.26$\pm$0.65$\pm$0.06 & Belle~\cite{Aushev:2008su} \\
         & \DstnDn      & 1.33$\pm$0.69$\pm$0.43 & \babar~\cite{Aubert:2007rva} \\
         & Both         & 0.70$\pm$0.16  & Avg$^3$ (3.0/3) \\
\hline
\rule[10pt]{-1mm}{0mm}
$r_{D^*D}$          &\DstnDn & 12.5$\pm$3.1           & Ratio of avgs \\
\hline
\rule[10pt]{-1mm}{0mm}
${\cal B}_0$ & $\omega\jpsi$ & 3.5$\pm$1.9$\pm$0.7 & Belle~\cite{Abe:2005ix} \\
${\cal B}_+$ &               & 8.5$\pm$1.5$\pm$1.7 & Belle~\cite{Abe:2005ix} \\
${\cal B}_0$ &               & 6$\pm$3$\pm$1 & Belle~\cite{delAmoSanchez:2010jr} \\
${\cal B}_+$ &               & 6$\pm$2$\pm$1 & Belle~\cite{delAmoSanchez:2010jr} \\
${\cal B}_{0+}$ &            & 5.8$\pm$1.2     & Avg$^3$ (2.7/3) \\
\hline
\rule[10pt]{-1mm}{0mm}
$r_\omega$ & $\omega\jpsi$   & 1.0$\pm$0.4$\pm$0.3 & Belle~\cite{Abe:2005ix} \\
           &                         & 0.8$\pm$0.3         & \babar~\cite{delAmoSanchez:2010jr} \\
           &                         & 0.85$\pm$0.26       & Avg$^3$ (0.1/1) \\
\hline
\rule[10pt]{-1mm}{0mm}
$r_{\gamma 1}$ & $\gamma\jpsi$ & 0.31$\pm$0.08 & Values above\\
$r_{\gamma 2}$ & $\gamma\psip$ & 0.53$\pm$0.57 & Values above\\
\hline
\rule[10pt]{-1mm}{0mm}
${\cal B}_0$ & $\pi^+\pi^0\jpsi$   & $-$5.7$\pm$4.9 & \babar~\cite{Aubert:2004zr} \\
${\cal B}_+$ &         & 2.0$\pm$3.8 & \babar~\cite{Aubert:2004zr} \\
${\cal B}_{0+}$ &      & $-$0.9$\pm$3.7          & Avg$^3$ (1.5/1) \\
\hline\hline
\end{tabular}
\end{center}
\end{table}

Branching-fraction-related measurements for $X(3872)$ appear
in \Tab{tab:Spec_RatBchgBneu}.
Note that
$X\to\DstnDn$ decays are an order of magnitude
more prevalent than $X\to\ppjp$,
and that experimental
information concerning the radiative decay $X\to\gamma\psip$
has recently become murky;
Belle's preliminary upper limit~\cite{Bhardwaj:2010qwg} is inconsistent with 
the \babar~\cite{Aubert:2008rn} measurement. 
Belle~\cite{Abe:2005ix} found 4.3$\sigma$ evidence
for $X\to\pi^+\pi^-\pi^0\jpsi$ in $B$-decays,
with the 3$\pi$ invariant mass clustered
near the kinematic endpoint, which is almost
one full $\Gamma_\omega$ below the $\omega$ mass peak. 
This suggests the decay $X\to\omega\jpsi$,
$\omega\to\pi^+\pi^-\pi^0$ on the $\omega$ low-side tail.
Despite this apparent phase-space suppression, the rate for 
$X\to\omega\jpsi$ was found to be comparable to that of \ppjp.  
In 2010 \babar~\cite{delAmoSanchez:2010jr} reported
corroborating evidence ($4.0\sigma$) of an $X(3872)$ peak
in $B^{0,+}\to (J/\psi\pi^+\pi^-\pi^0) K^{0,+}$ decays
using their full data sample, also finding
a decay rate comparable to that of \ppjp,
as shown in \Tab{tab:Spec_RatBchgBneu}.
Their analysis was able to identify the three-pion decay
as coming from an $\omega$-meson decay by weighting the
entries based on the pion opening angles in the $\omega$
rest-frame: phase-space weighting results in no net signal.
In a comparison of the observed $m(3\pi)$ 
mass distribution to that of MC 
simulations, \babar\ also found that the 
inclusion of one unit of orbital 
angular momentum in the $J/\psi\omega$ 
system, with its consequent negative
parity, substantially improves the 
description of the data. Hence the 
$X(3872)$ quantum number assignment~\cite{Abulencia:2006ma} 
of $J^{PC}=2^{-+}$ is 
preferred somewhat over the $1^{++}$ hypothesis
in the \babar\ analysis, leading
\babar\ to conclude that the $X(3872)$ can be interpreted
as an $\eta_{c2}(1D)$ charmonium state~\cite{Godfrey:1985xj,Eichten:2007qx}.
However, the $1^{++}$ assignment cannot be ruled out 
as unlikely by this analysis, just less likely than $2^{-+}$.
In addition, it has been shown~\cite{Kalashnikova:2010hv} that a
$2^{-+}$ assignment is not consistent with other properties of the 
$X(3872)$.

\babar~\cite{Aubert:2004zr}
searched for a charged partner state in the
decay $X^+\to\rho^+\jpsi$, finding the results
in \Tab{tab:Spec_RatBchgBneu}. The
average from charged and neutral $B$-decays
should be compared with the isospin-symmetry
prediction, which is double the rate
for $\rho^0\jpsi$. These rates disagree
by more than 4$\sigma$, making it most 
likely that the $X$ is an isosinglet.
The \babar~\cite{Aubert:2005vi} upper limit on
${\cal B}_{B^+}\equiv{\cal B}(B^+\to K^+ X)$
permits an inferred lower limit on
${\cal B}(X\to \ppjp)$, which, when combined
with the relative rates of \DstnDn\ and \ppjp,
yields $2.2\%<{\cal B}(X\to \ppjp)<10.5\%$
and  $28\%<{\cal B}(X\to\DstnDn)<94.2\%$.
Belle~\cite{Adachi:2008te} has studied
the question of whether or not the $X$, 
like conventional charmonia, tends
to be produced more strongly in $B^0\to K^{*0}X$
relative to nonresonant (NR)
$B^0\to (K^+\pi^-)_{\rm NR}X$. Using $X\to\ppjp$ decays,
they limit the $K^{*0}/(K^+\pi^-)_{\rm NR}$
ratio to be $<0.5$ at 90\%~CL, contrasted
with ratios closer to 3 for other charmonium states.

The possibility that the $X$ enhancement in \ppjp\
is composed of two different narrow states, $X_L$
and $X_H$, was 
addressed by CDF~\cite{Aaltonen:2009vj}.
By analyzing the observed lineshape,
$X_L$ and $X_H$ were found to have masses closer than 3.2~(4.3)\mevcc\
for a relative production fraction of unity (20\%).
Any mass difference between the $X$ appearing
in charged and neutral $B$-decays has also been limited
to $<$2.2\mevcc,
as detailed in \Tab{tab:Spec_DeltaMX}.

\begin{table}[b]
\caption{For $X(3872)$, 
$\Delta m_{0+}$ (in \mevcc), the difference
between the $X(3872)$ mass obtained
from neutral and charged $B$ decays;
and $\Delta m_{LH}$ (in \mevcc), the difference
in mass of two $X$ states produced
with equal strength in $p\bar{p}$ collisions
}
\label{tab:Spec_DeltaMX}
\setlength{\tabcolsep}{0.33pc}
\begin{center}
\begin{tabular}{cccc}
\hline\hline
\rule[10pt]{-1mm}{0mm}
Quantity & Mode &  Value  & Ref. ($\chi^2/$d.o.f.)\\
\hline
\rule[10pt]{-1mm}{0mm}
$\Delta m_{0+}$ & \ppjp & 2.7$\pm$1.6$\pm$0.4 & \babar~\cite{Aubert:2008gu} \\[0.7mm]
                &       & 0.18$\pm$0.89$\pm$0.26 & Belle~\cite{Adachi:2008te} \\[0.7mm]
                &       & 0.79$\pm$1.08 & Avg$^3$ (1.8/1) \\[0.7mm]
                &       & $<2.2$~at~90\%~CL & Above \\[0.7mm]
\hline
\rule[10pt]{-1mm}{0mm}
$\Delta m_{LH}$ & \ppjp & $<3.2$~at~90\%~CL & CDF~\cite{Aaltonen:2009vj} \\[0.7mm]
\hline\hline
\end{tabular}
\end{center}
\end{table}

Taking the totality of experimental information
on the $X(3872)$ at face value, the $X$ is a 
narrow resonant structure with the most probable 
quantum numbers $J^{PC}=1^{++}$ and $I=0$, and has mass within
1\mevcc\ of \DstnDn\ threshold. It may have comparable
decay rates to $\gamma\psip$ and 
(often-slightly-below-threshold) \DstnDn, but has
an order of magnitude smaller rate to both 
$\omega\jpsi$ and $\rho^0\jpsi$.
Decays to $\gamma\jpsi$ occur at roughly a quarter of 
the $\gamma\psip$ rate. If there
are two components of the observed
enhancements, they must be closer in mass
than a few\mevcc. It is produced and observed
in Tevatron $p\bar{p}$ collisions with a rate similar
to conventional charmonia, and at the $B$-factories
in $B\to K X$ decays. Unlike conventional
charmonia, $B\to K^* X$ is suppressed
with respect to $B\to K\pi X$.

The summarized properties of $X(3872)$ 
do not comfortably fit those of any plausible
charmonium state. Prominent decays to
\DstnDn\ and proximity to \DstnDn\ mass
threshold naturally lead to models~\cite{Goldberger:1964} which
posit the $X$ to be either a weakly bound
molecule of a \Dstn\ and a $\bar{\Dze}$
slightly below threshold or a virtual one
slightly unbound. Accommodating the large
radiative decay rates and substantial
\ppjp\ rate in such models creates
challenges because the
\Dstn\ and \Dze\ would be spatially separated
by large distances, suppressing the probability 
of overlap for annihilation.
This has led to the hypothesis
of mixing with a charmonium
state having the same quantum numbers.
Models of a tightly bound diquark-diantiquark
system $cu\bar{c}\bar{u}$ feature two neutral 
($cd\bar{c}\bar{d}$, $cd\bar{c}\bar{s}$) 
and one charged ($cu\bar{c}\bar{d}$) partner state,
which are limited by the corresponding mass difference
and null search measurements. 
Better understanding of the $X(3872)$
demands more experimental constraints and theoretical insight.

\subsubsection{Unconventional vector states}
\label{sec:SpecExp_UnconVector}

\subthreesection{$Y(4260)$, $Y(4360)$, $Y(4660)$, and $Y(4008)$}

\begin{figure}[t]
\includegraphics*[width=\figwid]{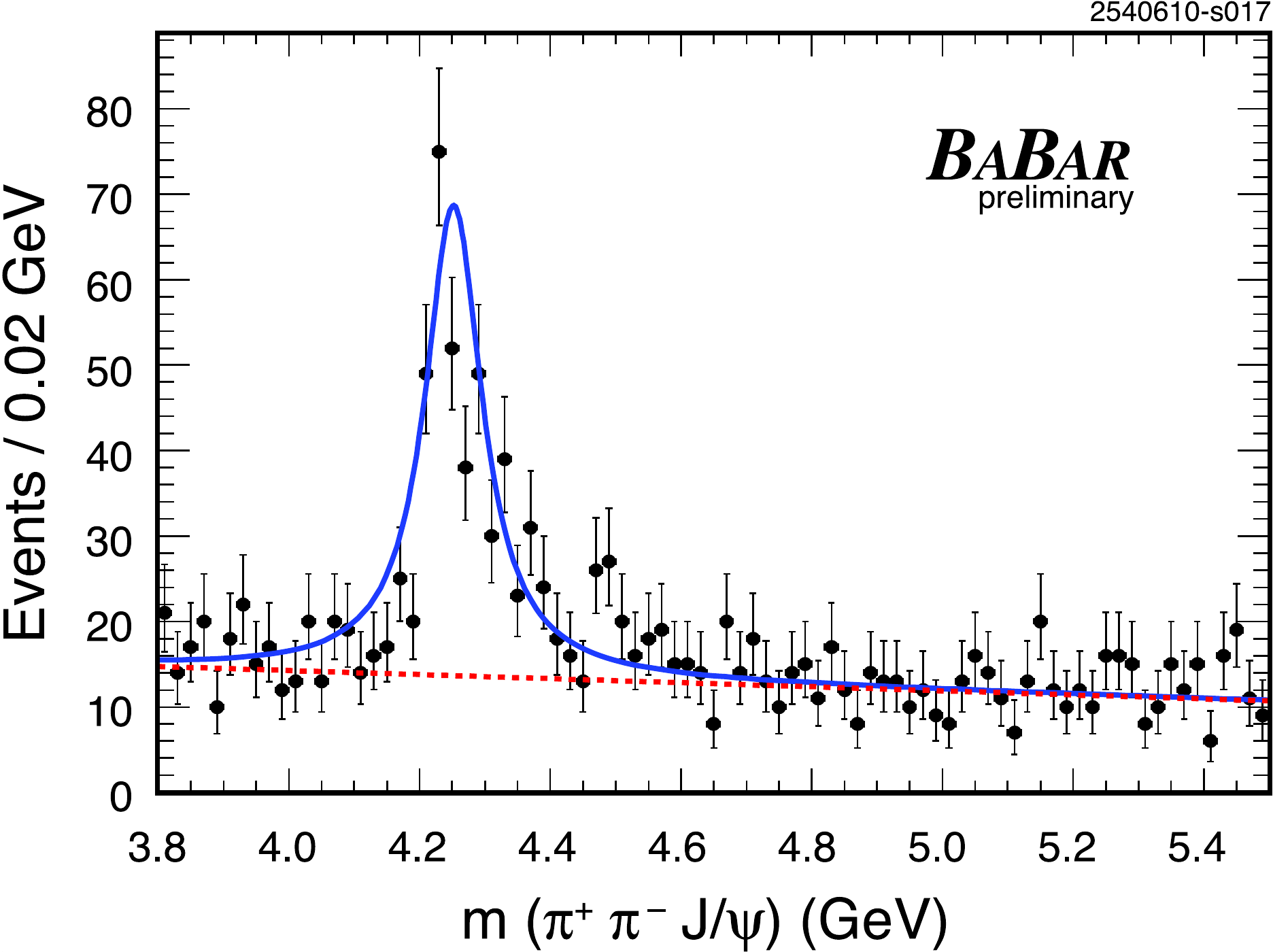}
\caption{From \babar~\cite{Aubert:2008ic},
         the invariant mass of $\pi^+\pi^-\jpsi$ 
         candidates produced in initial-state radiation,
         $e^+e^-\to\gamma_{ISR}\,\pi^+\pi^-\jpsi$.
         {\it Points with error bars} represent data,
         and the {\it curves} show the fit {\it (solid)} to
         a signal plus a linear background {\it (dashed)}
 }
\label{fig:Spec_Y4260Mass_babar}
\end{figure}

\begin{figure}[t]
\includegraphics*[width=\figwid]{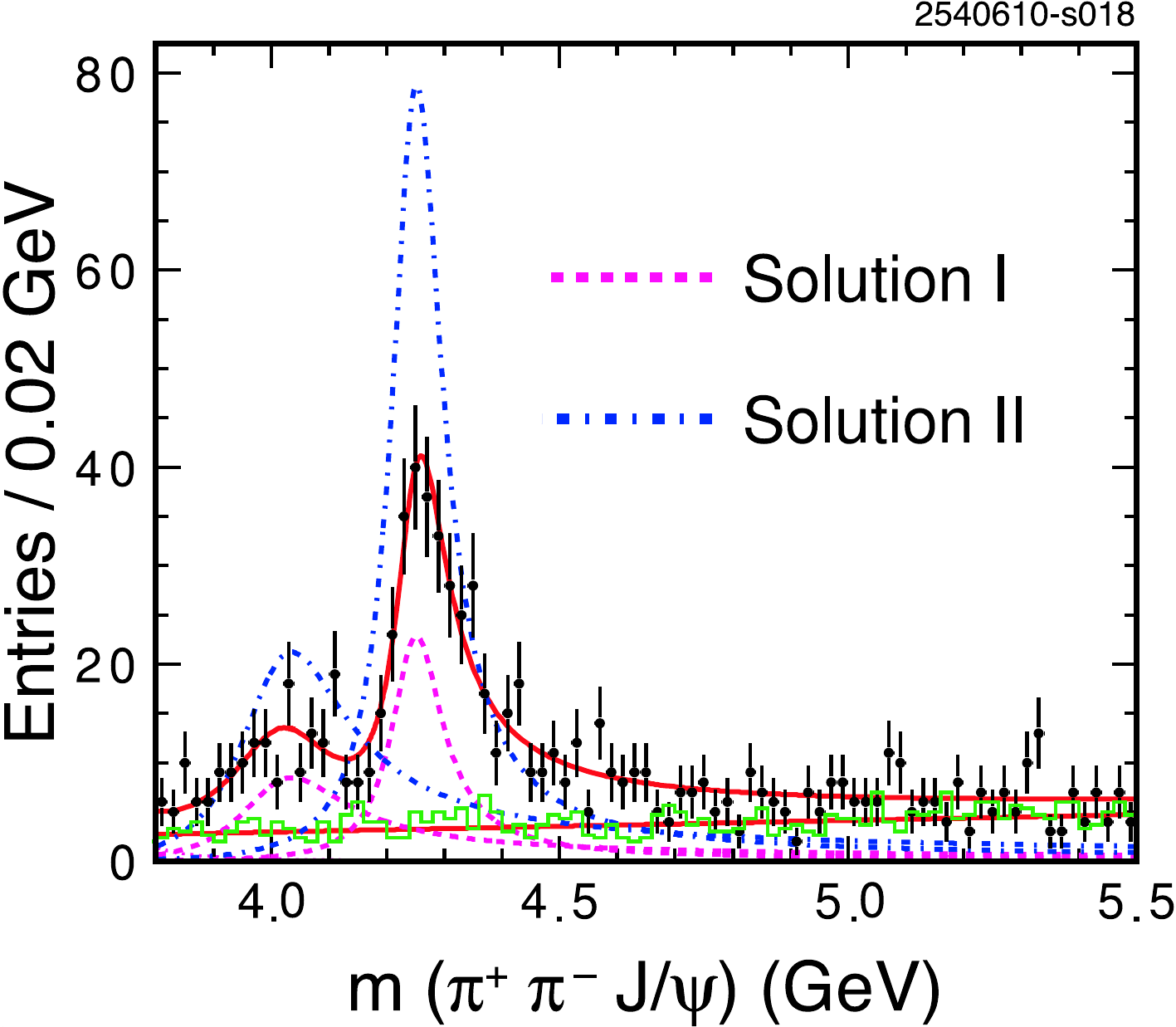}
\caption{From Belle~\cite{Belle:2007sj}, the invariant mass of $\pi^+\pi^-\jpsi$ 
         candidates produced in initial-state radiation,
         $e^+e^-\to\gamma_{ISR}\,\pi^+\pi^-\jpsi$,
         with \jpsi-sidebands already subtracted,
         unlike \Fig{fig:Spec_Y4260Mass_babar}.
         {\it Points with error bars} represent data,
         the {\it solid curve} shows the best fits to the data
         to two resonances including interference 
         with a floating phase, and the {\it dashed} and {\it dashed-dot curves} show
         the two pairs of individual resonance contributions for
         the two equally probable best-fit phases.
         \AfigPermAPS{Belle:2007sj}{2007} }
\label{fig:Spec_Y4260Mass_belle}
\end{figure}

\begin{table}[t]
\caption{Measured properties of the decay $Y(4260)\to\ppjp$,
including its mass $m$, width $\Gamma$, and branching fraction \Brat.
The Belle~\cite{Belle:2007sj} single-resonance fit result is quoted to
allow for comparison to the other two
}
\label{tab:Spec_Y4260}
\setlength{\tabcolsep}{1.3pc}
\begin{center}
\begin{tabular}{ccc}
\hline\hline
\rule[11pt]{-1mm}{0mm}
Quantity & Value & Ref. ($\chi^2$/d.o.f.)\\[0.7mm]
\hline
\rule[11pt]{-1mm}{0mm}
$m$  & 4259$\pm$8$^{+2}_{-6}$ & \babar~\cite{Aubert:2008ic} \\[0.7mm]
 (MeV)    & 4263$\pm$6             & Belle~\cite{Belle:2007sj}  \\[0.7mm]
     & 4284$^{+17}_{-16}$$\pm$4 & CLEO~\cite{He:2006kg}   \\[0.7mm]
     & 4263$\pm$5             & Avg$^3$ (1.8/2)\\[0.7mm]
\hline
\rule[11pt]{-1mm}{0mm}
$\Gamma$  & 88$\pm$23$^{+6}_{-4}$ & \babar~\cite{Aubert:2008ic} \\[0.7mm]
 (MeV)    &126$\pm$18             & Belle~\cite{Belle:2007sj}  \\[0.7mm]
          & 73$^{+39}_{-25}$$\pm$5& CLEO~\cite{He:2006kg}   \\[0.7mm]
          & 108$\pm$15            & Avg$^3$ (2.4/2)\\[0.7mm]
\hline
\rule[11pt]{-1mm}{0mm}
${\cal B}\times\Gamma_{ee}$  &5.5$\pm$1.0$^{+0.8}_{-0.7}$ & \babar~\cite{Aubert:2008ic} \\[0.7mm]
             (eV)            &9.7$\pm$1.1                 & Belle~\cite{Belle:2007sj}  \\[0.7mm]
                             &8.9$^{+3.9}_{-3.1}\pm1.8$   & CLEO~\cite{He:2006kg}   \\[0.7mm]
                             &8.0$\pm$1.4                 & Avg$^3$ (6.1/2)\\[0.7mm]
\hline\hline
\end{tabular}
\end{center}
\end{table}

\begin{table}[t]
\caption{Measured properties of the two enhancements
found in the $\pppsip$ mass distribution: $Y(4360)$ and $Y(4660)$.
Liu {\it et al.}~\cite{Liu:2008hja} performed a binned
maximum likelihood fit to the combined Belle and \babar\
cross section distributions (\Fig{fig:Spec_Y4360Fit_liu})
}
\label{tab:Spec_Y4360}
\setlength{\tabcolsep}{1.30pc}
\begin{center}
\begin{tabular}{ccc}
\hline\hline
\rule[10pt]{-1mm}{0mm}
Quantity & Value & Ref. ($\chi^2$/d.o.f.)\\[0.7mm]
\hline
\rule[11pt]{-1mm}{0mm}
$m$  & 4324$\pm$24 & \babar~\cite{Aubert:2006ge} \\[0.7mm]
(MeV)& 4361$\pm$9$\pm$9 & Belle~\cite{:2007ea}  \\[0.7mm]
     & 4353$\pm$15      & Avg$^3$ (1.8/1)\\[0.7mm]
     & 4355$^{+~9}_{-10}\pm$9 & Liu~\cite{Liu:2008hja}\\[0.7mm]
\hline
\rule[11pt]{-1mm}{0mm}
$\Gamma$  & 172$\pm$33 & \babar~\cite{Aubert:2006ge} \\[0.7mm]
   (MeV)  &  74$\pm$15$\pm$10 & Belle~\cite{:2007ea}  \\[0.7mm]
          &  96$\pm$42        & Avg$^3$ (6.8/1)\\[0.7mm]
          & 103$^{+17}_{-15}\pm$11 & Liu~\cite{Liu:2008hja}\\[0.7mm]
\hline
\rule[11pt]{-1mm}{0mm}
$m$  & 4664$\pm$11$\pm$5 & Belle~\cite{:2007ea}  \\[0.7mm]
(MeV)& 4661$^{+9}_{-8}\pm$6 & Liu~\cite{Liu:2008hja}\\[0.7mm]
\hline
\rule[11pt]{-1mm}{0mm}
$\Gamma$  &  48$\pm$15$\pm$3 & Belle~\cite{:2007ea}  \\[0.7mm]
   (MeV)  &  42$^{+17}_{-12}\pm$6 & Liu~\cite{Liu:2008hja}\\[0.7mm]
\hline\hline
\end{tabular}
\end{center}
\end{table}

\begin{figure}[tb]
\includegraphics*[width=\figwid]{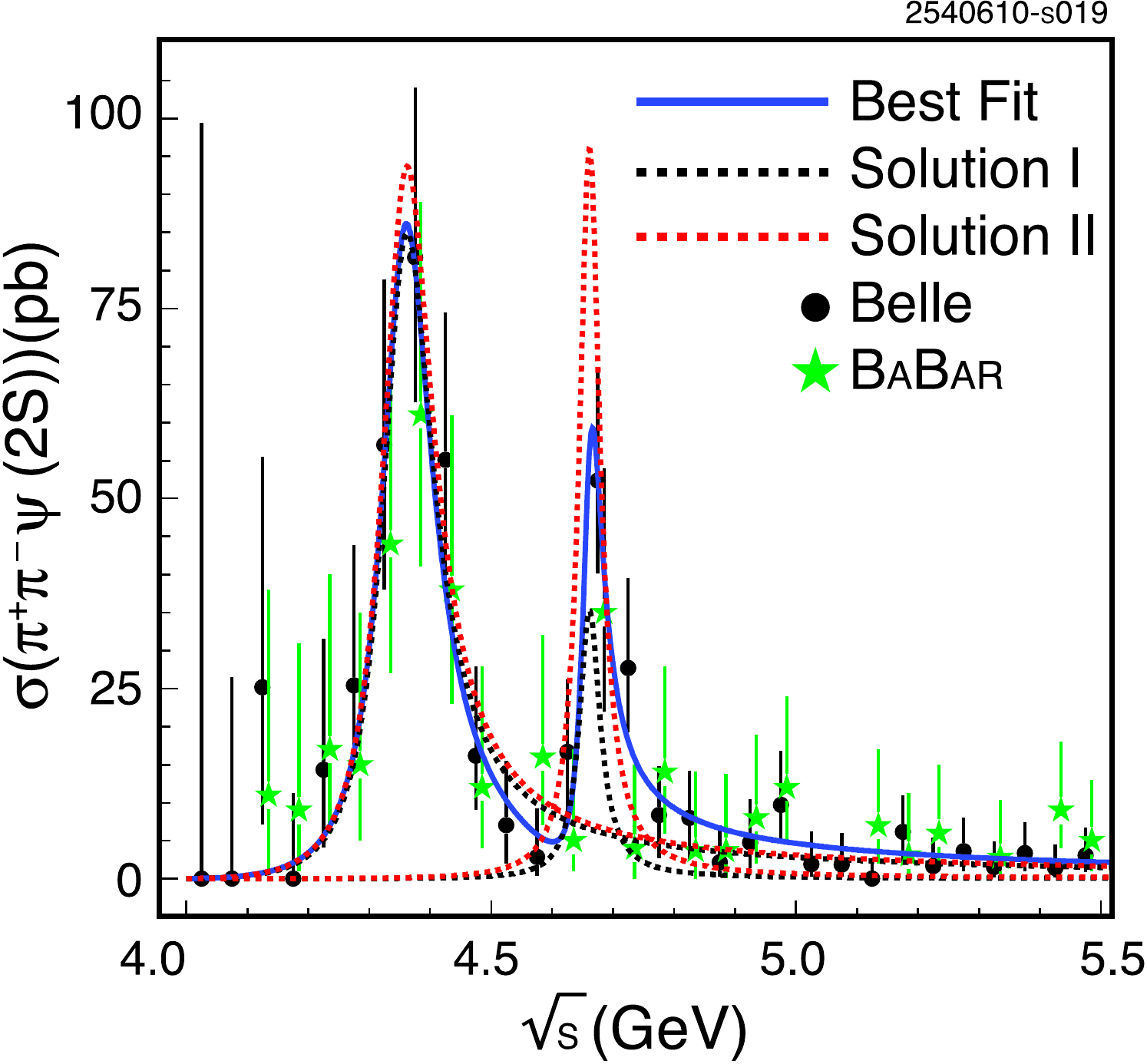}
\caption{From \cite{Liu:2008hja}, the $\pi^+\pi^-\psip$ 
         cross section as a function of $\sqrt{s}$,
         showing the result of a binned maximum-likelihood fit
         of combined Belle and \babar\ data,
         The {\it solid circles}
         and {\it stars} show the Belle and \babar\ data, respectively. 
         The {\it solid curve} shows the best fits to the data
         to two resonances including interference 
         with a floating phase, and the {\it dashed curves} show
         the two pairs of individual resonance contributions for
         the two equally probable best-fit phases.
         \AfigPermAPS{Liu:2008hja}{2008} }
\label{fig:Spec_Y4360Fit_liu}
\end{figure}

\begin{table}[tb]
\caption{Upper limits at 90\%~CL on the ratios $\sigma(\epem \to
  Y \to T) / \sigma(\epem \to Y \to \ppjp)$ at $\ecm =
  4.26\gev$ (CLEO~\cite{CroninHennessy:2008yi}) and 
  $\brat(Y \to T) / \brat(Y \to\ppjp)$ (for $Y(4260)$), and
  $\brat(Y \to T) / \brat(Y \to\pppsip)$ (for $Y(4360)$ and $Y(4660)$),
  from \babar~\cite{Aubert:2006mi,Aubert:2009xs,:2010vb} and Belle~\cite{Abe:2006fj}, 
  where $T$ is an open charm final state
}
\label{tab:Spec_Y4260_opencharm}
\setlength{\tabcolsep}{0.80pc}
\begin{center}
\begin{tabular}{lccc}
\hline\hline
\rule[10pt]{-1mm}{0mm}
  \ \ \ \ $T$  & $Y(4260)$ & $Y(4360)$ & $Y(4660)$ \\
\hline
\rule[10pt]{-1mm}{0mm}
\DDbar          & 4.0~\cite{CroninHennessy:2008yi},~7.6~\cite{Aubert:2006mi}   &   \\[0.7mm]
\DDst        & 45~\cite{CroninHennessy:2008yi},~34~\cite{Aubert:2009xs}    &   \\[0.7mm]
\DstDst      & 11~\cite{CroninHennessy:2008yi},~40~\cite{Aubert:2009xs}    &   \\[0.7mm]
$\DDst\pi$   & 15~\cite{CroninHennessy:2008yi},~9~\cite{Abe:2006fj} & 8~\cite{Abe:2006fj} & 10~\cite{Abe:2006fj} \\[0.7mm]
$\DstDst\pi$ & 8.2~\cite{CroninHennessy:2008yi}  & & \\[0.7mm]
\DsDs        & 1.3~\cite{CroninHennessy:2008yi},~0.7~\cite{:2010vb}  & &  \\ [0.7mm]
\DsDsts      & 0.8~\cite{CroninHennessy:2008yi},~44~\cite{:2010vb}  &&  \\ [0.7mm]
\DstsDsts    & 9.5~\cite{CroninHennessy:2008yi}  &&  \\[0.7mm]
\hline
\hline
\end{tabular}
\end{center}
\end{table}

The first observation of an unexpected vector charmonium-like
state was made by \babar~\cite{Aubert:2005rm} in ISR production 
of $Y(4260)\to\ppjp$, which was later updated~\cite{Aubert:2008ic} 
with twice the data, as shown in \Fig{fig:Spec_Y4260Mass_babar}.
CLEO~\cite{He:2006kg} and Belle~\cite{Belle:2007sj} confirmed the 
\babar\ result, but Belle also found a smaller, broader structure at 
4008\mevcc, as seen in 
\Fig{fig:Spec_Y4260Mass_belle}. Aside 
from the lower mass state, for which the updated \babar~\cite{Aubert:2008ic} 
analysis placed an upper limit, the three sets of measurements were 
quite consistent in mass and width, as shown in \Tab{tab:Spec_Y4260},
but only roughly so in strength.
\babar~\cite{Aubert:2006ge} found one more apparent enhancement,
$Y(4360)$, in \pppsip, which Belle~\cite{:2007ea} measured
with somewhat larger mass and smaller width,
as seen in \Tab{tab:Spec_Y4360}. 
Belle also found a second structure near 
4660\mevcc, as seen in \Fig{fig:Spec_Y4360Fit_liu}.
(A combined fit~\cite{Liu:2008hja} to Belle and \babar\ 
$\pppsip$ data found them to be consistent with one other.)
Because dipion transitions between vector quarkonia are commonplace
for charmonium and bottomonium, it was natural, then, that
the first inclination was to ascribe the $Y$'s to excited
vector charmonia. A number of additional features of these
states contradicted this hypothesis, however.
Only one, $Y(4660)$, is remotely near a predicted 
$1^{--}$ $c\bar{c}$ state ($3\,^3D_1$). The $Y(4260)$ and
$Y(4360)$ did not show up in inclusive hadronic cross section ($R$) measurements
as seen in \Fig{fig:Spec_RBES}, as would be expected
of such states. (There is no fine-grained $R$-scan data near $Y(4660)$.)

A comparison of the measured $\ppjp$ and total hadronic cross sections 
in the $\sqrt{s}\simeq 4260$~MeV region
yields a lower bound for $\Gamma(Y\to\ppjp)>508\kev$ at 90\%~CL, 
an order of magnitude higher than expected for conventional 
vector charmonium states~\cite{Mo:2006ss}. 
Charmonium would also feature dominant open charm decays,  
exceeding those of dipion transitions
by a factor expected to be $\gtrsim$100,
since this is the case for $\psi(3770)$ and $\psi(4160)$.
As summarized in \Tab{tab:Spec_Y4260_opencharm}, 
no such evidence has been found, significantly
narrowing any window for either charmonia
or, in some cases, quark-gluon hybrid interpretations.
CLEO~\cite{Coan:2006rv} studied direct production of
$Y(4260)$ in $e^+e^-$ collisions; verified the production cross section;
and identified the only non-\ppjp\ decay mode seen so far, 
$\pi^0\pi^0\jpsi$, 
occuring at roughly half of the \ppjp\ rate.

\begin{figure}[b]
\includegraphics*[width=\figwid]{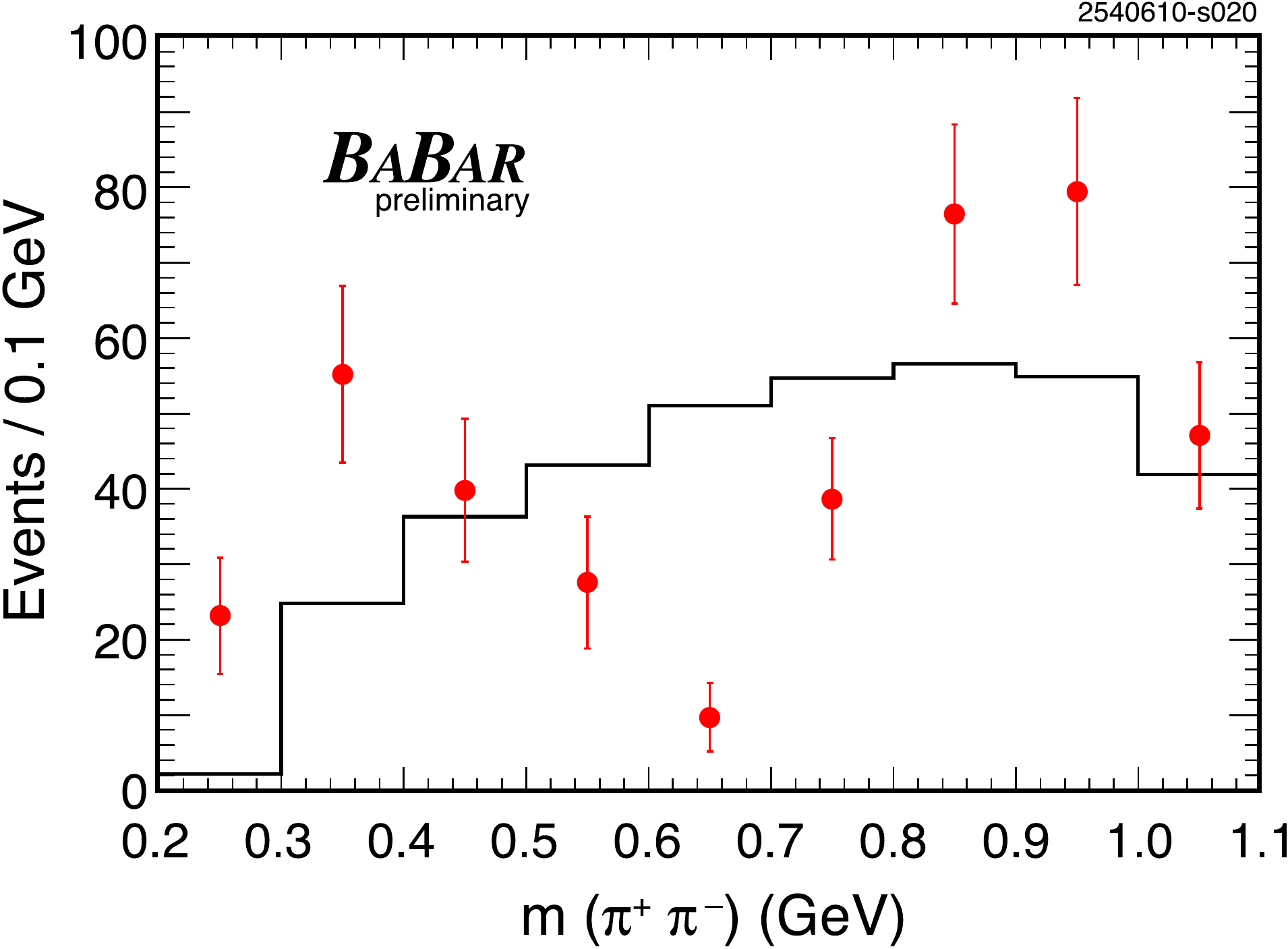}
\caption{From \babar~\cite{Aubert:2008ic}, the dipion invariant
         mass distribution of ISR-produced $Y(4260)\to\ppjp$ decays,
         where {\it points} represent data and the {\it line histogram} 
         is phase-space MC simulation
 }
\label{fig:Spec_Y4260_dipion_babar}
\end{figure}

\begin{figure}[t]
\includegraphics*[width=\figwid]{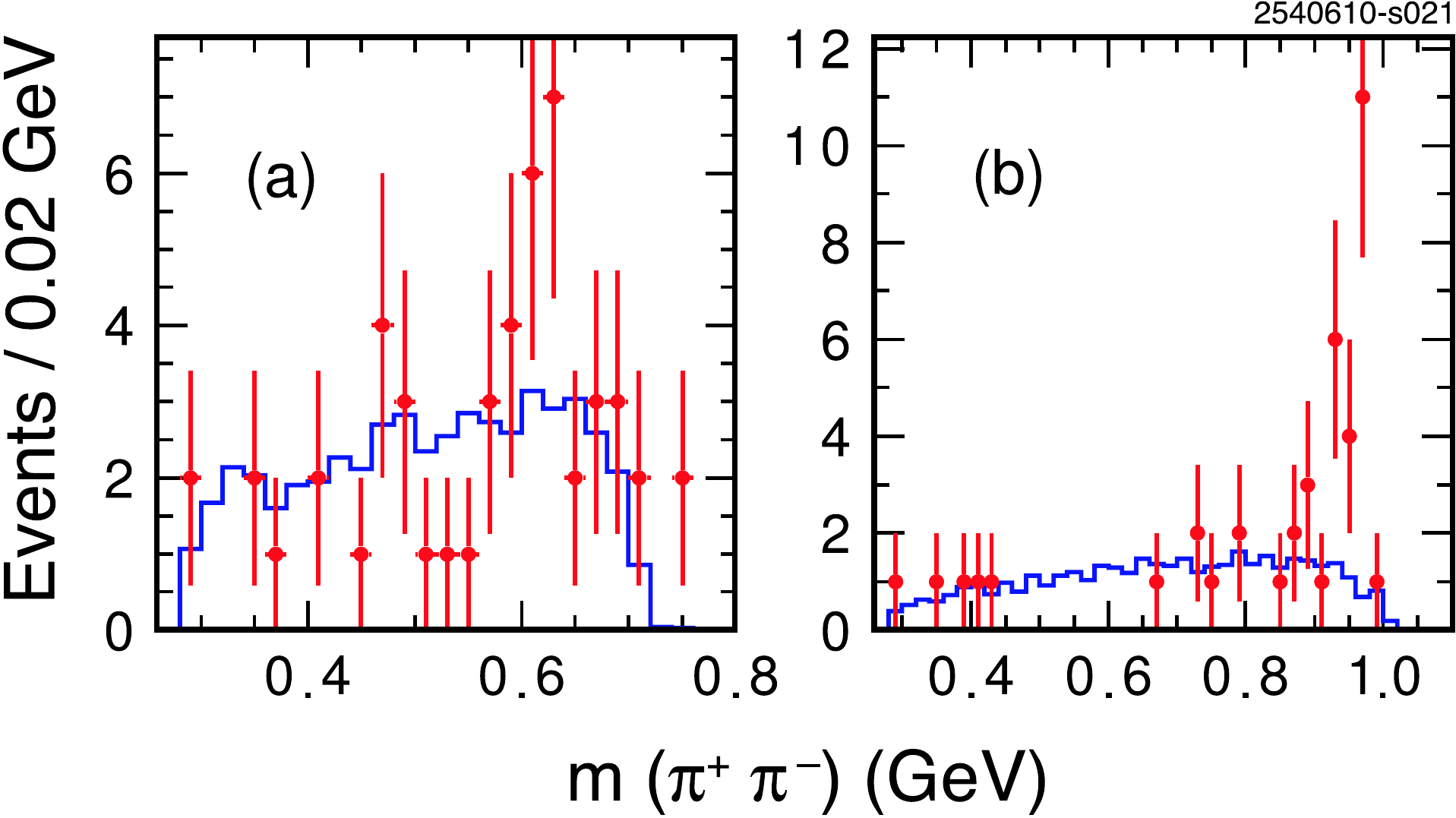}
\caption{From Belle~\cite{:2007ea}, the dipion
         invariant mass distribution of ISR-produced 
         (a)~$Y(4360)\to\pppsip$ and (b)~$Y(4660)\to\pppsip$,
         where {\it points} represent data 
         and the {\it line histograms} show phase-space
         MC simulations. \AfigPermAPS{:2007ea}{2007}
         }
\label{fig:Spec_Y43604660_dipion_belle}
\end{figure}

Any explanation for these vector states will 
have to describe their masses, widths, and manifest
reluctance to materialize in open charm or unflavored
light meson final states. The dipion invariant mass
spectra exhibit curious structures, as seen for $Y(4260)$ in 
\Fig{fig:Spec_Y4260_dipion_babar}~\cite{Aubert:2008ic},
$Y(4360)$ in \Fig{fig:Spec_Y43604660_dipion_belle}(a)~\cite{:2007ea},
and $Y(4660)$ in 
\Fig{fig:Spec_Y43604660_dipion_belle}(b)~\cite{:2007ea}.
The first shows a distinctly non-phase-space double-hump
structure which is qualitatively confirmed by Belle~\cite{Belle:2007sj},
the second exhibits a plurality of events at higher masses,
and the third indicates a quite dominant $f_0(980)$ component.

\subthreesection{$X(4630)$}

The \eell\ cross section was measured by Belle~\cite{Pakhlova:2008vn} using 
ISR and partial reconstruction (\Fig{fig:Spec_ExclXSec_BelleBaBar}(i)). 
A clear peak is evident 
near the threshold, and corresponds to 
\beqa
\Brat(X(4630)\to e^+e^-)\times{\cal B}(X(4630)\to\lala)=\hspace{0.3in}\non\\
(0.68\pm0.33)\times 10^{-6}\, .~~~~~\hspace{0.5in}
\eeqa
The
nature of this enhancement remains unclear. Although both mass and
width of the $X(4630)$ (see \Tab{tab:Spec_ExpSumUnc})
are consistent within errors with those of the
$Y(4660)$,
this could be coincidence and does not exclude
other interpretations.

\subthreesection{$Y_b(10888)$}

\begin{figure}[t]
\includegraphics*[width=\figwid]{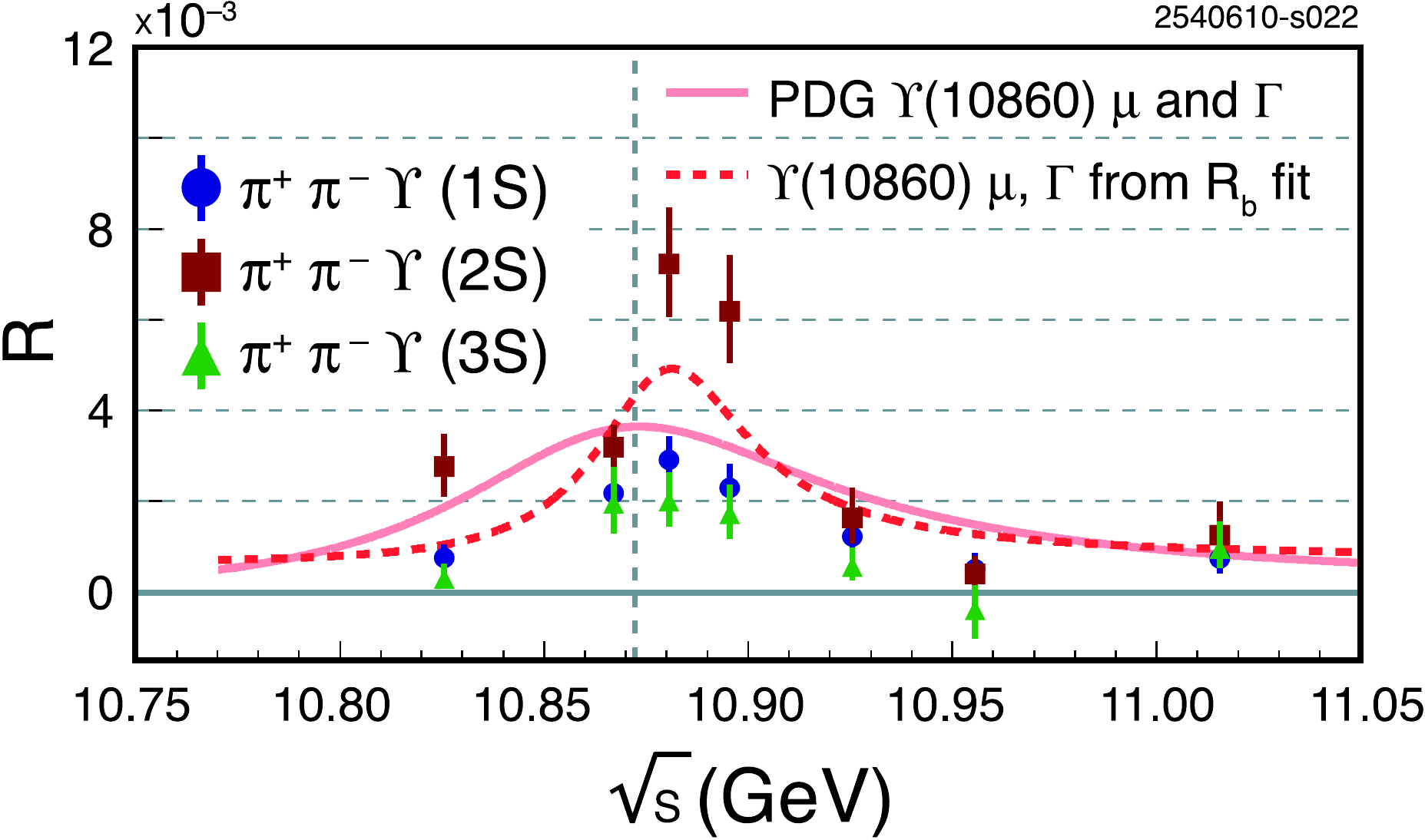}
\caption{From Belle~\cite{Chen:2008pu}, results of an
         energy scan near the $\Upsilon(5S)$. The {\it points
         with error bars} indicate the three sets of
         cross sections for the  $\pi^+\pi^-\Upsilon(nS)$
         ($n$=1, 2, 3), normalized to the point cross section.
         The {\it solid} and {\it dashed curves} show the $\Upsilon(5S)$
         lineshape from PDG08~\cite{Amsler:2008zzb} 
         and Belle~\cite{Chen:2008pu}, respectively
 }
\label{fig:Spec_YbMassFit_Belle}
\end{figure}

A recent Belle scan above the $\Upsilon(4S)$ was motivated by 
an earlier observation~\cite{Abe:2007tk} of anomalously large 
$\pi^+\pi^-\Upsilon(nS)$ ($n=1,2,3$) cross sections near
the $\Upsilon(5S)$ peak energy.
These new data allowed independent
determinations of the \UnS{5}\ lineshape and that
of the $\pi^+\pi^-\UnS{n}$ enhancement~\cite{Chen:2008pu}. 
A simultaneous fit to all three
measured $\pi^+\pi^-\Upsilon(nS)$ cross sections to a single Breit-Wigner
function represents the data well; this lineshape has 
somewhat\footnote{When Belle~\cite{Abe:2007tk} adds
the cross sections from $\dipi\UnS{n}$ events at each energy
scan point to the loosely selected hadronic events
for a fit to a single resonance, the quality of the
fit degrades by $2.0\sigma$ (where systematic uncertainties
are included) relative to the hadronic events alone. 
That is, the $\dipi\UnS{n}$ and hadronic events are consistent
with a single enhancement within two standard deviations.}
higher mass  
and narrower width (see \Tab{tab:Spec_ExpSumUnc}) 
than does the $\Upsilon(5S)$ resonance measured with
loosely selected hadronic events in the same experiment.
This suggests that the enhancement  
$\pi^+\pi^-\Upsilon(nS)$
could be a $1^{--}$ $Y_b$ state distinct from
$\Upsilon(5S)$ and perhaps of a similar origin as $Y(4260)$. 
The relevant cross sections and lineshapes are shown in
\Fig{fig:Spec_YbMassFit_Belle}. See also the discussion
in \Sec{sec:Dec_HT_Ups5Spipi}.

\subsubsection{Other positive $C$-parity states}
\label{sec:SpecExp_3940}

Of the multitude of new charmonium-like states, 
a puzzling cluster of them
from different production mechanisms and/or decay
chains gather near a mass of 3940\mevcc\  
($Z(3930)$, $Y(3940)$, $X(3940)$, $X(3915)$)
and have positive $C$-parity. 
Four others ($Y(4140)$,
$X(4160)$, $Y(4274)$, and $X(4350)$) also
have $C=+$ and have
related signatures. Definitive determination
of whether some of these are distinct from others
and whether any can be attributed to expected charmonia
requires independent confirmations, more precise
mass and branching fraction measurements, and
unambiguous quantum number assignments. 

\begin{figure}[t]
\begin{center}
   \includegraphics[width=\figwid]{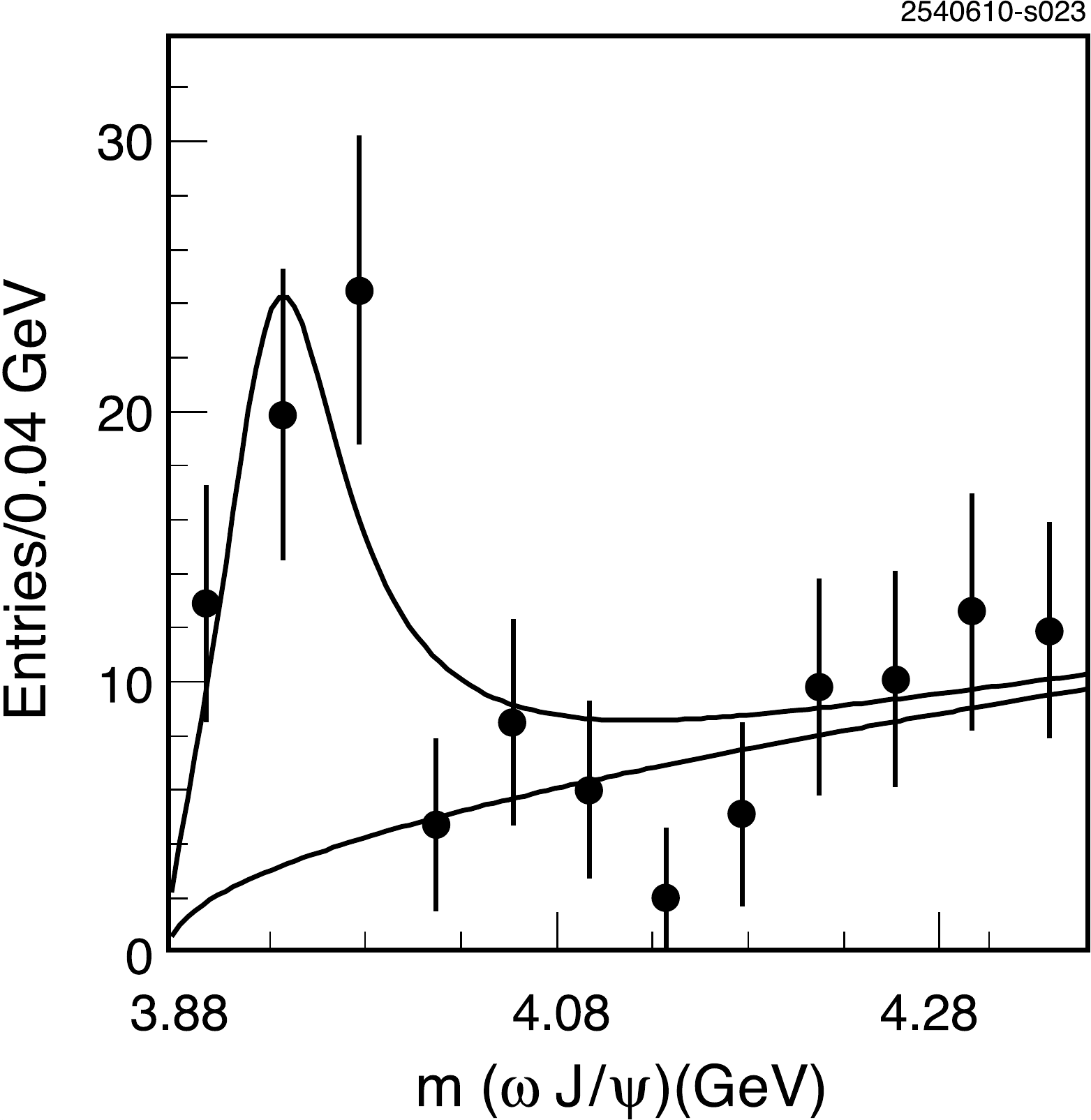}
   \caption{From Belle~\cite{Abe:2004zs}, the $\omega\jpsi$ mass
            distribution in $B^+\to K^+ \omega\jpsi$ decays.
            The {\it upper curve} is the total fit function, the {\it lower one}
            is the contribution of the phase-space-like threshold
            function. \AfigPermAPS{Abe:2004zs}{2005} }
   \label{fig:Spec_X3915_omegajpsimass_Belle}
\end{center}
\end{figure}

\begin{figure}[t]
\begin{center}
   \includegraphics[width=\figwid]{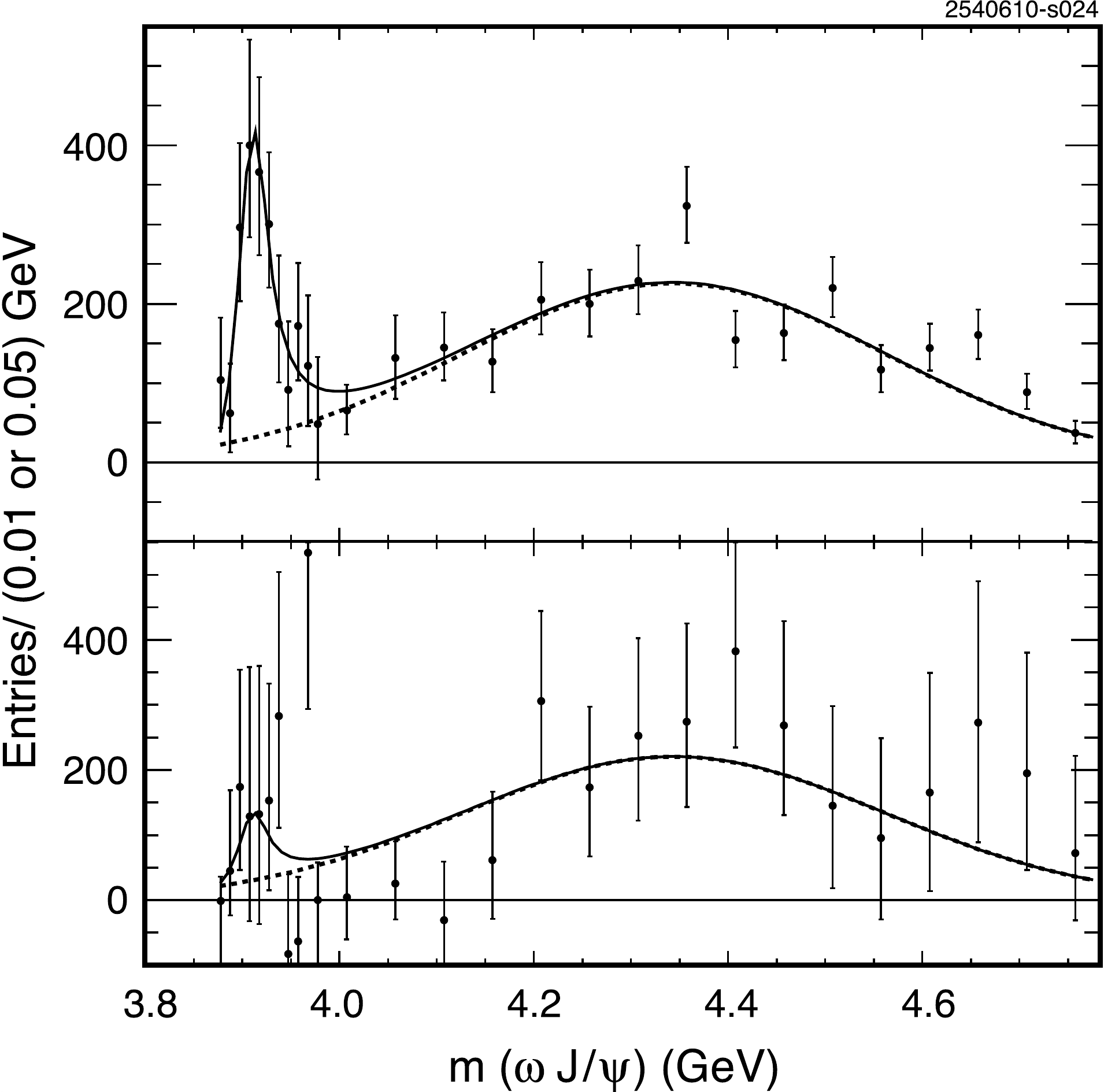}
   \caption{From \babar~\cite{Aubert:2007vj}, 
            the $\omega\jpsi$ mass distribution in $B^+\to K^+\omega\jpsi$
            {\it (upper)} and $B^0\to K^0\omega\jpsi$ 
            {\it (lower)} decays. The {\it solid (dashed) curve} represents the total
            fit (background) function. \AfigPermAPS{Aubert:2007vj}{2008} }
   \label{fig:Spec_X3915_omegajpsimass_Babar}
\end{center}
\end{figure}

\begin{figure}[t]
\begin{center}
   \includegraphics[width=\figwid]{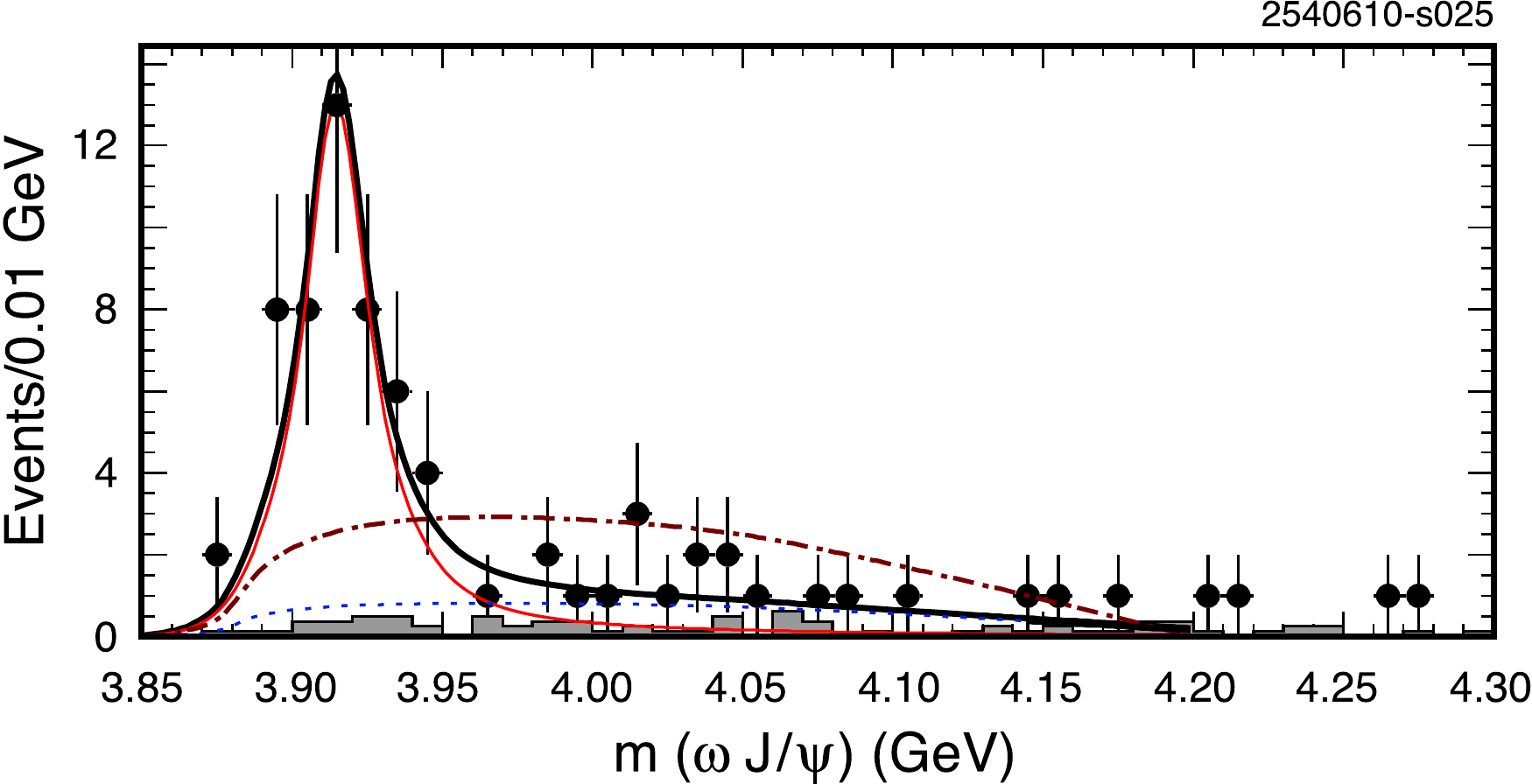}
   \caption{From Belle~\cite{Uehara:2009tx}, the $\omega\jpsi$
            invariant mass distribution for the $\gamma\gamma\to\omega\jpsi$ 
            data {\it (points)} and background 
            from scaled non-$\omega\jpsi$ sidebands {\it (shaded)}.
            The {\it bold solid, thinner solid,} and {\it dashed curves} are the
            total, resonance, and background contributions, respectively.
            The {\it dot-dashed curve} is
            the best fit with no resonance included.
            \AfigPermAPS{Uehara:2009tx}{2009} }
\label{fig:Spec_X3915_2gamma_Belle}
\end{center}
\end{figure}

\subthreesection{$Z(3930)$, $Y(3940)$, and $X(3915)$}

Some of this fog is clearing as more measurements appear.
As described previously, the state
decaying to \DDbar, previously known as $Z(3930)$, has
been identified as the $\chi_{c2}(2P)$. The $Y(3940)\to\omega\jpsi$
enhancement initially found by Belle~\cite{Abe:2004zs}  
in $B^+\to K^+ Y(3940)$ decays is shown in 
\Fig{fig:Spec_X3915_omegajpsimass_Belle}. 
It was confirmed by \babar~\cite{Aubert:2007vj}
with more statistics, albeit with somewhat smaller mass,
as shown in \Fig{fig:Spec_X3915_omegajpsimass_Babar}. 
But Belle~\cite{Uehara:2009tx} also found
a statistically compelling (7.7$\sigma$) resonant structure $X(3915)$
in $\gamma\gamma$ fusion decaying to $\omega\jpsi$,
as seen in \Fig{fig:Spec_X3915_2gamma_Belle}.
As the higher-mass Belle
$B\to K\,Y(3940)$ $(\to\omega\jpsi)$ sighting shares
the same production and decay signature as that of \babar's
$Y(3940)$, which has mass and width consistent 
with the $X(3915)$, the simplest interpretation is that
the $Y(3940)$ and $X(3915)$ are the same state,
and that the latter name should prevail as the mass is
closer to 3915\mevcc. It is this reasoning which motivates 
grouping them together in \Tabs{tab:Spec_ExpSumUnc}
and \ref{tab:Spec_X3915_mass}.
The $X(3915)$ clearly has $C=+$, but $J^P$ remains to be determined.

\begin{table}[t]
\caption{Measured properties of the $X(3915)\to\omega\jpsi$
(subsuming what has previously been called $Y(3940)$).
Here 
${\cal B}_B\times{\cal B}_X\equiv{\cal B}(B^+\to K^+X)\times{\cal B}(X\to\omega\jpsi)$ 
and
$\Gamma_{\gamma\gamma}\times {\cal
  B}_X\equiv\Gamma(X\to\gamma\gamma)\times{\cal B}(X\to\omega\jpsi)$
}
\label{tab:Spec_X3915_mass}
\setlength{\tabcolsep}{0.25pc}
\begin{center}
\begin{tabular}{cccc}
\hline\hline
\rule[10pt]{-1mm}{0mm}
Quantity & Value & Decay & Ref. ($\chi^2$/d.o.f.)\\
\hline
\rule[10pt]{-1mm}{0mm}
$m$  & 3942$\pm$11$\pm$13             & $B^+\to K^+ X$ & Belle~\cite{Abe:2004zs}\\[0.7mm]
(MeV)& 3914.6$^{+3.8}_{-3.4}$$\pm2.0$ & $B^+\to K^+ X$ & \babar~\cite{Aubert:2007vj} \\[0.7mm]
     & 3915$\pm$3$\pm$2 & $\gamma\gamma\to X$      & Belle~\cite{Uehara:2009tx} \\[0.7mm]
     & 3915.6$\pm$3.1                 & Both       & Avg$^3$ (2.7/2)\\[0.7mm]
\hline
\rule[10pt]{-1mm}{0mm}
$\Gamma$ & 87$\pm$22$\pm$26        & $B^+\to K^+ X$    & Belle~\cite{Abe:2004zs}\\[0.7mm]
  (MeV)  & 34$^{+12}_{-\ 8}$$\pm5$ & $B^+\to K^+ X$    & \babar~\cite{Aubert:2007vj} \\[0.7mm]
         & 17$\pm$10$\pm$3 & $\gamma\gamma\to X$   & Belle~\cite{Uehara:2009tx} \\[0.7mm]
         & 28$\pm$10               & Both          & Avg$^3$ (4.6/2)\\[0.7mm]
\hline
\rule[10pt]{-1mm}{0mm}
${\cal B}_B\times{\cal B}_X $ 
  & 7.1$\pm$1.3$\pm$3.1         &  $B^+\to K^+ X$ & Belle~\cite{Abe:2004zs}\\[0.7mm]
  $(10^{-5})$ & 4.9$^{+1.0}_{-0.9}$$\pm$0.5 &  $B^+\to K^+ X$ & \babar~\cite{Aubert:2007vj} \\[0.7mm]
  & 5.1$\pm$1.0 & $B^+\to K^+ X$ & Avg$^3$ (0.4/1) \\[0.7mm]
\hline
\rule[10pt]{-1mm}{0mm}
$\Gamma_{\gamma\gamma}\times {\cal B}_X$    
  & 61$\pm$17$\pm$8 & $\gamma\gamma\to X$ ($0^+$) & Belle~\cite{Uehara:2009tx} \\[0.7mm]
 (keV)  & 18$\pm$5$\pm$2  & $\gamma\gamma\to X$ ($2^+$) & Belle~\cite{Uehara:2009tx} \\[0.7mm]
\hline\hline
\end{tabular}
\end{center}
\end{table}

What more can be gleaned from the existing measurements?
The production rate and total width determinations 
summarized in \Tab{tab:Spec_X3915_mass}
are useful in testing whether the $X(3915)$ behaves
like other charmonia. Existing measurements for
product branching fractions in $B^+\to K^+ \psi$ decays,
where $\psi$= $\etac$, $\jpsi$, $\chi_{cJ}$, $\hsubc$,
$\etacp$, and $\psi(2S)$, indicate that such states
appear with branching fractions of $10^{-3}$ or below.
If the $X(3915)$ behaved at all similarly, this would, 
in turn, imply that $\Gamma(X(3915)\to\omega\jpsi)\gtrsim1\mevcc$,
whereas $\psip$ and $\psi(3770)$ have 
hadronic transition 
widths
at least ten times smaller.
Similarly, the $\gamma\gamma$-fusion results
in \Tab{tab:Spec_X3915_mass} imply a much larger
$\Gamma(X(3915)\to\gamma\gamma)$ than is typical
of the few-keV two-photon widths of $\chi_{c0,2}$,
unless $\omega\jpsi$ completely dominates its width
(which would also be surprising for a $c\bar{c}$ state).
In agreement with this pattern, $X(3915)$ does not appear to have prominent
decays to $\gamma\jpsi$~\cite{Aubert:2006aj}, \DDbar~\cite{Abe:2003zv}, or \DDst~\cite{Aushev:2008su}.
Hence any conventional $c\bar{c}$ explanation for $X(3915)$ would
likely have trouble accommodating
these quite {\it un}charmonium-like features.
More data and more analysis,
especially of angular distributions, are necessary to
firmly establish these conclusions.

\subthreesection{$X(3940)$}

\begin{figure}[b]
\begin{center}
   \includegraphics[width=\figwid]{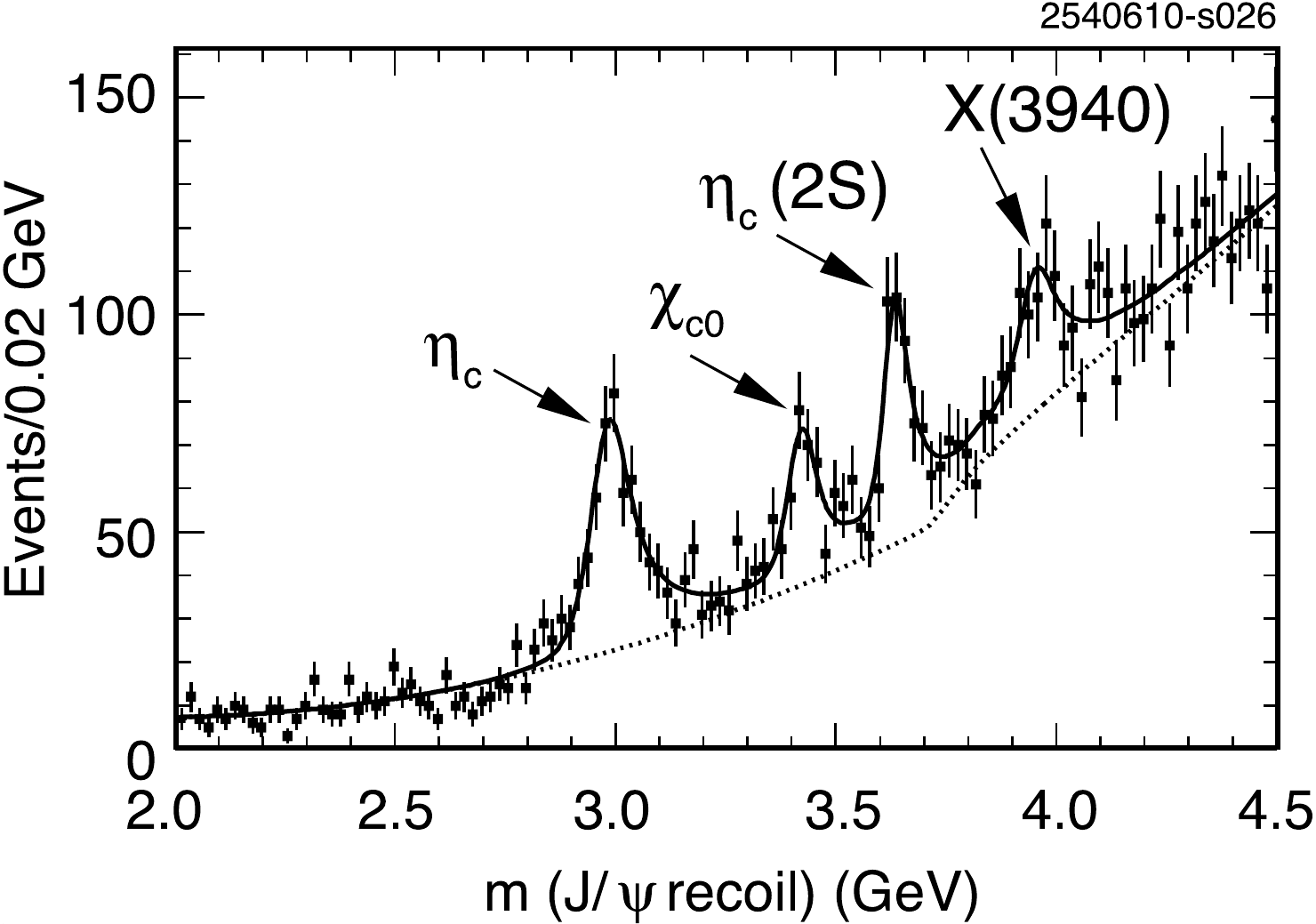}
   \caption{From Belle~\cite{Abe:2007jn}, the distribution of mass recoiling against
            the $\jpsi$ in $e^+e^- \to\jpsi (...)$ events
            {\it (points with error bars)}. Results of the fit are shown 
            by the {\it solid curve}, the {\it dashed curve} corresponds to the expected background
            distribution. \AfigPermAPS{Abe:2007jn}{2007} }
\label{fig:Spec_X3940_dcmass_Belle}
\end{center}
\end{figure}

\begin{figure}[t]
\begin{center}
   \includegraphics[width=\figwid]{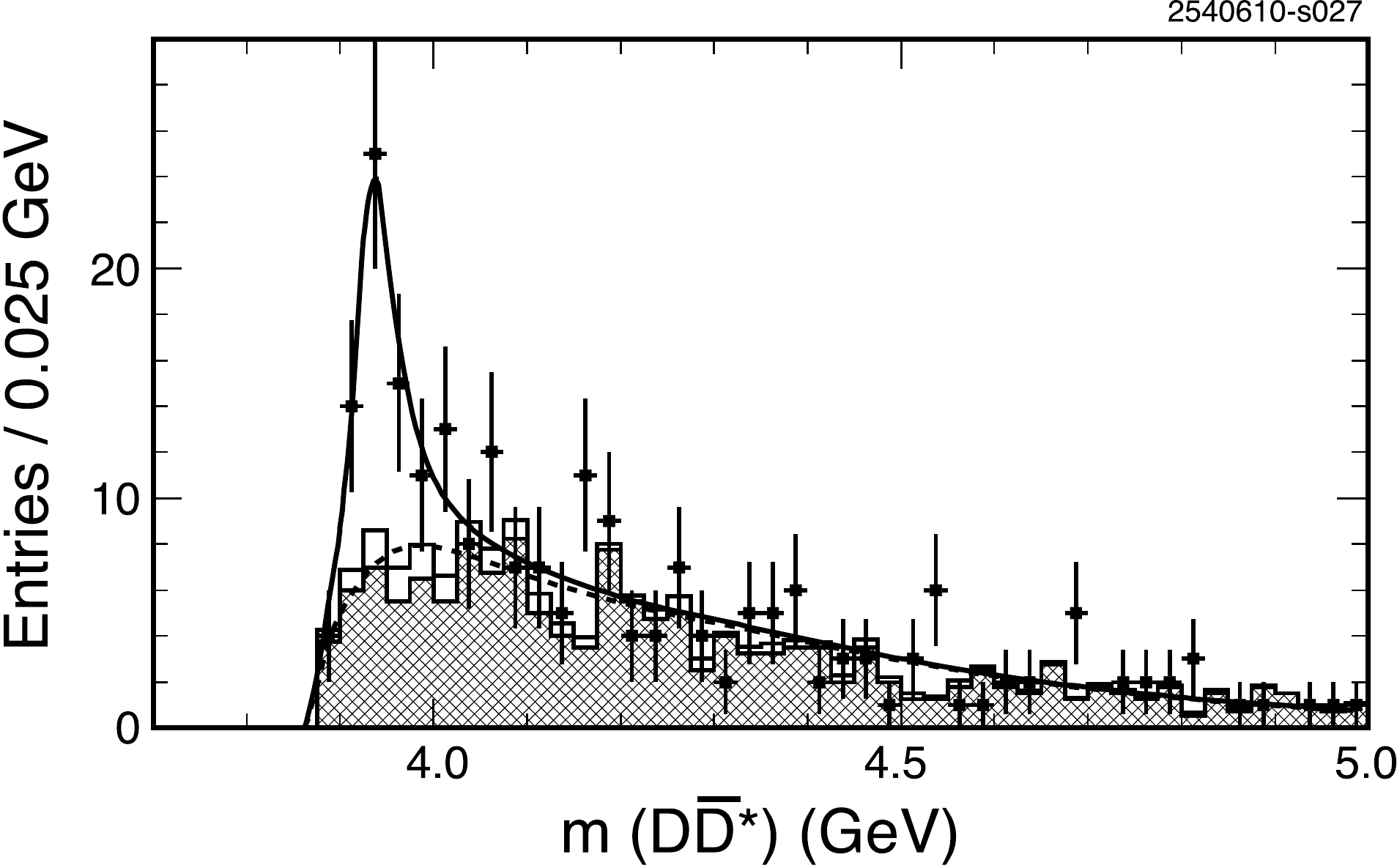}
   \caption{From Belle~\cite{Abe:2007sya}, 
            the distribution of \DDst\ mass recoiling against the
            \jpsi\ {\it (points with error bars)}. Results of the full
            (background only) fit are shown by the {\it solid (dashed) 
            curve}, and the {\it hatched histogram} is from the scaled 
            $D^{(*)}$-sidebands. \AfigPermAPS{Abe:2007sya}{2008} }
\label{fig:Spec_X3940_DDst_Belle}
\end{center}
\end{figure}

The situation for masses near 3940\mevcc\ gets even
messier when Belle's analyses of resonances in
$e^+e^-\to \jpsi (...)$~\cite{Abe:2007jn} and  
$e^+e^-\to \jpsi \DDst$~\cite{Abe:2007sya}
are considered. 
The former, as shown in \Fig{fig:Spec_X3940_dcmass_Belle}, 
which examined 
mass recoiling against a \jpsi\ in inclusive production,
was confirmed by the latter partial reconstruction
analysis with the mass and width shown in \Tab{tab:Spec_ExpSumUnc}.
The latter analysis reconstructed the \jpsi\ and a $D$ or $D^*$,
and then used the missing-mass spectrum to isolate events consistent
with the desired topology. 
The event
was then kinematically constrained to have the requisite missing
mass ($D^*$ or $D$), improving resolution on the missing momentum.
That four-momentum was then combined with that of the reconstructed
$D^{(*)}$ to form the \DDst\ invariant mass, 
as illustrated in \Fig{fig:Spec_X3940_DDst_Belle},
yielding a signal for $X(3940)$ with a significance of $6.0\sigma$.
From mass measurements alone, this state appears
distinct from the $X(3915)$ by 3.1$\sigma$.
Bolstering this notion,
explicit searches for a state with the appropriate
mass in $B$-decays (to \DDst~\cite{Aushev:2008su}) 
or double-charmonium production (decaying to 
$\omega\jpsi$~\cite{Abe:2007sya}) were negative,
yielding incompatible $\omega\jpsi/\DDst$
relative branching fractions for the two production 
mechanisms. 

\subthreesection{$X(4160)$}

Belle~\cite{Abe:2007sya} extended the $e^+e^-\to \jpsi \DDst$
analysis to also search for resonances decaying to \DstDst,
and found a broad enhancement $X(4160)$ just above \DstDst\ threshold
with mass and width shown in \Tab{tab:Spec_ExpSumUnc}. 
Double vector-charmonium production, which can
occur through two intermediate virtual photons instead of one, has not yet
been observed, so it seems unlikely that this state could
be $\psi(4160)$. And if
$e^+e^-\to\jpsi\, X(4160)$ is produced via annihilation, its $C$-parity
would necessarily be positive.

\newpage
\subthreesection{$Y(4140)$, $Y(4274)$, and $X(4350)$}

New measurements from CDF~\cite{Aaltonen:2009tz,collaboration:2010aa}
indicate at least one more $C=+$ state seen in 
$B$ decays and decaying to two vectors, one being a \jpsi, near
threshold. In inclusively selected $B^+\to K^+\, \phi\jpsi$ decays,
two enhancements in the $\phi\jpsi$ mass spectrum, 
with masses and widths as shown in 
\Tab{tab:Spec_ExpSumUnc}, are $Y(4140)$ and $Y(4274)$. 
The analysis requires both
that the final state be kinematically
consistent with $B$-decay, but also uses the CDF particle
identification system to require three charged kaons in the decay.
The signal significances are 5.0$\sigma$ for $Y(4140)$ and
$3.1\sigma$ for $Y(4274)$. Both remain unconfirmed.
However, Belle~\cite{Shen:2009vs} searched for production of $Y(4140)$
in two-photon fusion, $e^+e^-\to e^+e^- Y(4140)$, $Y(4140)\to\phi\jpsi$,
and found no evidence for it, obtaining a limit of
$\Gamma_{\gamma\gamma}\times{\cal B}(\phi\jpsi )<40$~eV for $J^P=0^+$
and $<5.9$~eV for $2^+$ at 90\%~CL. In that same analysis, 
Belle reported a 3.2$\sigma$
enhancement, $X(4350)\to\phi\jpsi$, 
with mass and width as in \Tab{tab:Spec_ExpSumUnc}
and a production rate measured to be
$\Gamma_{\gamma\gamma}\times{\cal B}(\phi\jpsi )=6.7^{+3.2}_{-2.4}\pm1.1$~eV
for $J^P=0^+$ and $1.5^{+0.7}_{-0.6}\pm0.3$~eV for $2^+$.

\subsubsection{Charged exotic mesons: the $Z$'s}
\label{sec:SpecExpChargedExotic}

The charmonium-like {\it charged} $Z$ states, seen by 
Belle 
in $Z^-\to\psip\pi^-$ and $\chi_{c1}\pi^-$
in $B\to Z^- K$ decays, are of special interest. 
If these states are mesons, they would necessarily
have a minimal quark substructure of $c\bar{c}u\bar{d}$ 
and therefore be manifestly exotic. In a manner similar 
to the first unearthing of $X(3872)$, $Z^-$ states were 
found in exclusively reconstructed $B$-decays in which a 
conventional charmonium state is a decay product of the 
$Z^-$. Here, a single charged pion accompanies either a 
\psip\ or \chicOne\ (compared to a $\pi^+\pi^-$ pair 
accompanying a \jpsi\ in the case of $X(3872)$).
The \psip\ is found via either $\psip\to\pi^+\pi^- J/\psi$ 
followed by $\jpsi\to\ell^+\ell^-$ or from direct dileptonic 
decay, $\psip\to\ell^+\ell^-$. The \chicOne\ is tagged by its 
decay $\chicOne\to\gamma\jpsi$, $\jpsi\to\ell^+\ell^-$. 
Statistics are gained in the \zBelle\ analysis by using 
both charged and neutral $B$ decays, combining each $Z^-$ 
candidate with either a neutral ($K_S^0\to\pi^+\pi^-$) or charged 
kaon candidate, if present. In all cases, background not from 
$B$-decays is small after the usual $B$-selection criteria
on expected $B$-candidate energy and mass have been applied, and 
is well-estimated using appropriately scaled sidebands in those 
variables. Each experiment finds consistency in various subsets 
of its own data (\eg from charged and neutral $B$ mesons 
and in different $\psip$ and $\jpsi$ decay modes) and thereby
justifies summing them for final results. Belle and \babar\ have 
comparable mass resolution and statistical power for studying these 
decays.

Belle found~\cite{Choi:2007wga} the first $Z^-$ state by 
observing a sharp peak near $\mpsitwospi=4430$\mevcc\ with 
statistical significance of $>6\sigma$.  The largest backgrounds 
are $B\to \psip K^*_i$, $K^*_i\to K\pi^-$, where $K_i^*\equiv\Ksone$ 
or \Kstwo. Hence \kpi\ mass regions around \Ksone\ and \Kstwo\ were 
excised in this Belle analysis. 
However, as {\it interference} between different partial 
waves in the \kpi\ system can produce fake ``reflection'' peaks in the
\mpsitwospi\ distribution, further attention is warranted. In the 
kinematically allowed \kpi\ mass range for this three-body $B$-decay, 
only $S$-, $P$- and $D$-partial waves in \kpi\ are significant. 
Belle found that no combination of interfering $L=0,$ 1, 2 partial 
waves can produce an artificial mass peak near 4430\mevcc\ without 
also producing additional, {\sl larger} structures nearby in 
\mpsitwospi. Such enhancements are absent in the Belle data, ruling out
such reflections as the origin of the apparent signal.

\begin{figure}[t]
\includegraphics*[width=\figwid]{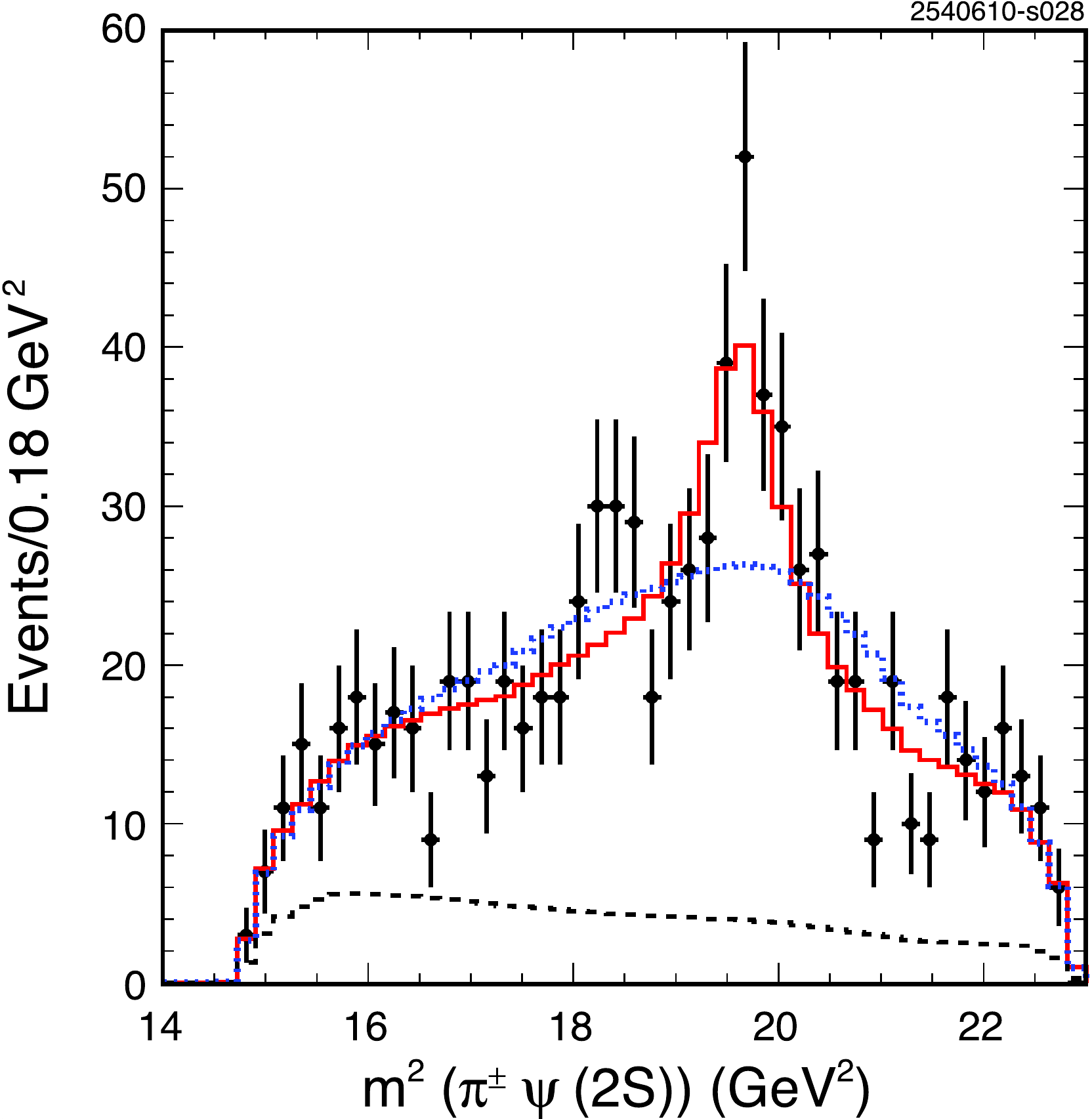}
\caption{From Belle~\cite{Mizuk:2009da}, for $B\to K\pi^-\psip$ candidates, 
         the data {\it points} show the $m^2(\psip\pi^-)$ projection of
         the Dalitz plot with the $K_i^\ast$ bands removed. The 
         {\it solid (dotted) histogram} shows the corresponding projections 
         of the fits with (without) a
         $\zBelle\to\psitwospi$ resonance term.
         The {\it dashed histogram} represents non-$B$-decay background
         estimated from energy-difference sidebands.
         \AfigPermAPS{Mizuk:2009da}{2009} }
\label{fig:Spec_Z4430_Belle_Second}
\end{figure}

In recognition of the role \kpi\ dynamics can play in the background 
shape and in response to the \babar\ non\-confirming analysis described 
below, Belle released a second analysis~\cite{Mizuk:2009da}  of 
\zBelle\ in their data. Here Belle modeled the $B\to \psitwospi K$
process as the sum of several two-body decays, with a $B\to \zBelle K$ 
signal component and, for background, $B\to \psip K_i^{\ast}$, where 
$K_i^{\ast}$ denotes all of the known $K^{\ast}\to$\kpi\ resonances 
that are kinematically accessible. Results of this second analysis 
are depicted in \Fig{fig:Spec_Z4430_Belle_Second}, which is an 
$m^2(\psip\pi^-)$ Dalitz plot projection with the prominent
$K^\ast$ bands removed. The data (points with error bars) are compared 
to the results of the fit with (solid histogram) or without (dashed
histogram) the \zBelle\ resonance; the former can be seen to be strongly 
favored over the latter, a $6.4\sigma$ effect. This Dalitz plot fit 
yields the mass and width shown in \Tab{tab:Spec_ExpSumUnc}
as well as the product branching fraction,
${\mathcal B}(B^0\to Z^- K)\times {\mathcal B}(Z^-\to\psip\pi^-) =
(3.2^{+1.8~+9.6}_{-0.9~-1.6})\times 10^{-5}$.
These values for the mass, width, and rate are consistent with the 
corresponding measurements reported in the initial Belle publication, 
but have larger uncertainties, which is indicative of the larger 
set of systematic variations in both signal and background properties 
that were considered. In the default fit, the \zBelle\ resonance was 
assumed to have zero spin. Variations of the fit included a $J=1$ 
assignment for the \zBelle, models with hypothetical $K^\ast\to$\kpi\ 
resonances with floating masses and widths, and radically different 
parametrizations of the \kpi\ $S$-wave amplitude.

\begin{figure}[t]
\begin{center}
\includegraphics*[width=\figwid]{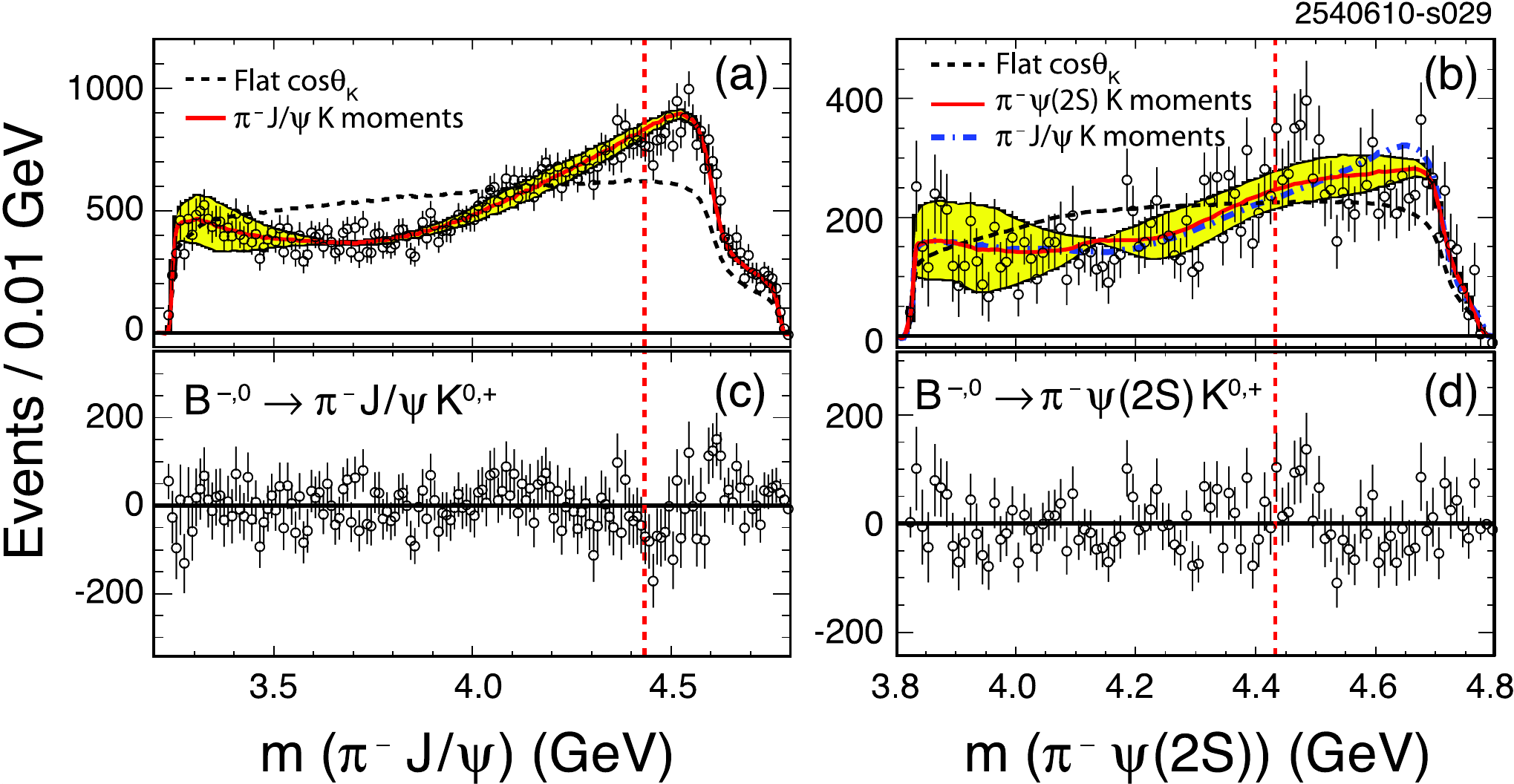}
\caption{From \babar~\cite{Aubert:2008nk}, the \psipi\ mass distributions for
         (a)~$B\to J/\psi\pi^- K$ and (b)~$B\to \psi^{\prime}\pi^- K$. 
         The {\it points} show the data 
         (integrated over all \kpi\/ regions) 
         after efficiency correction and background subtraction. The
         {\it dashed curves} show the \kpi\ reflection expected for a uniform decay angle 
         (\costhk) distribution, while the {\it solid curves} show the result when 
         accounting for the measured angular variation. The {\it shaded 
         bands} represent the effect of statistical uncertainty on the 
         normalized moments. In (b), the {\it dot-dashed curve} indicates the 
         result expected if the \kpi\ properties measured for $J/\psi\pi^-K$ 
         are used. The {\it dashed vertical lines} indicate the value of 
         $m(\psi\pi^-)=4433$\mevcc. In (c) and (d) appear
         the residuals {\it (data-solid curve)} for (a) and (b), respectively,
         after the {\it solid curves} are subtracted from the data.
         \AfigPermAPS{Aubert:2008nk}{2009} }
\label{fig:Spec_Z4430_Babar}
\end{center}
\end{figure}

The corresponding \babar\ search~\cite{Aubert:2008nk} added a
decay mode $\zBelle\to\jpsi\pi^-$ that was not considered in
either Belle study. Its inclusion increases statistics
for two purposes: first, potentially for more signal, since it 
is entirely reasonable to expect this mode to occur with at 
least as large a branching fraction as the discovery mode; and 
second, to study the \kpi\ resonance structure in the background, 
since the $J/\psi$ modes contain about six times more events than 
those with the $\psip$. The \babar\ analysis exploits this more
copious $B$-decay mode with exhaustive studies of \kpi\ partial 
wave dynamics, including fine-grained determination of angular
distributions and selection efficiencies over all regions of 
the Dalitz plot. The data were fit with floating $S$-, $P$-, 
and $D$-wave intensities. 
For both $\jpsi\pi^- K$ and $\psitwospi K$ samples, good fits
are obtained, as shown in the projections in \Fig{fig:Spec_Z4430_Babar}.
Hence both $\jpsi\pi^-$ and $\psip\pi^-$ mass distributions 
from \babar\ are well-represented by simulations with no extra 
resonant structure near $\mpsitwospi=4430\mevcc$. At the same time, 
if the \babar\ \mpsitwospi\ distribution with a \Ksone\ and \Kstwo\ 
veto (as done by Belle) is fit for the presence of a \zBelle\ at 
the same mass and width found by Belle, a signal with 2$\sigma$ 
statistical significance is found, indicating a statistical
consistency in the corresponding Belle and \babar\ mass 
distributions. This latter finding was verified with a direct 
bin-by-bin comparison between the Belle and \babar\ \mpsitwospi\ 
distributions after the \Ksone\ and \Kstwo\ veto: 
the two samples were found to be
statistically equivalent. That is, while no statistically 
significant signal for $Z(4430)^-$ in the \babar\ data has been 
found, neither does the \babar\ data refute the positive Belle 
observation of \zBelle.

\begin{figure}[tbp]
\includegraphics*[width=\figwid]{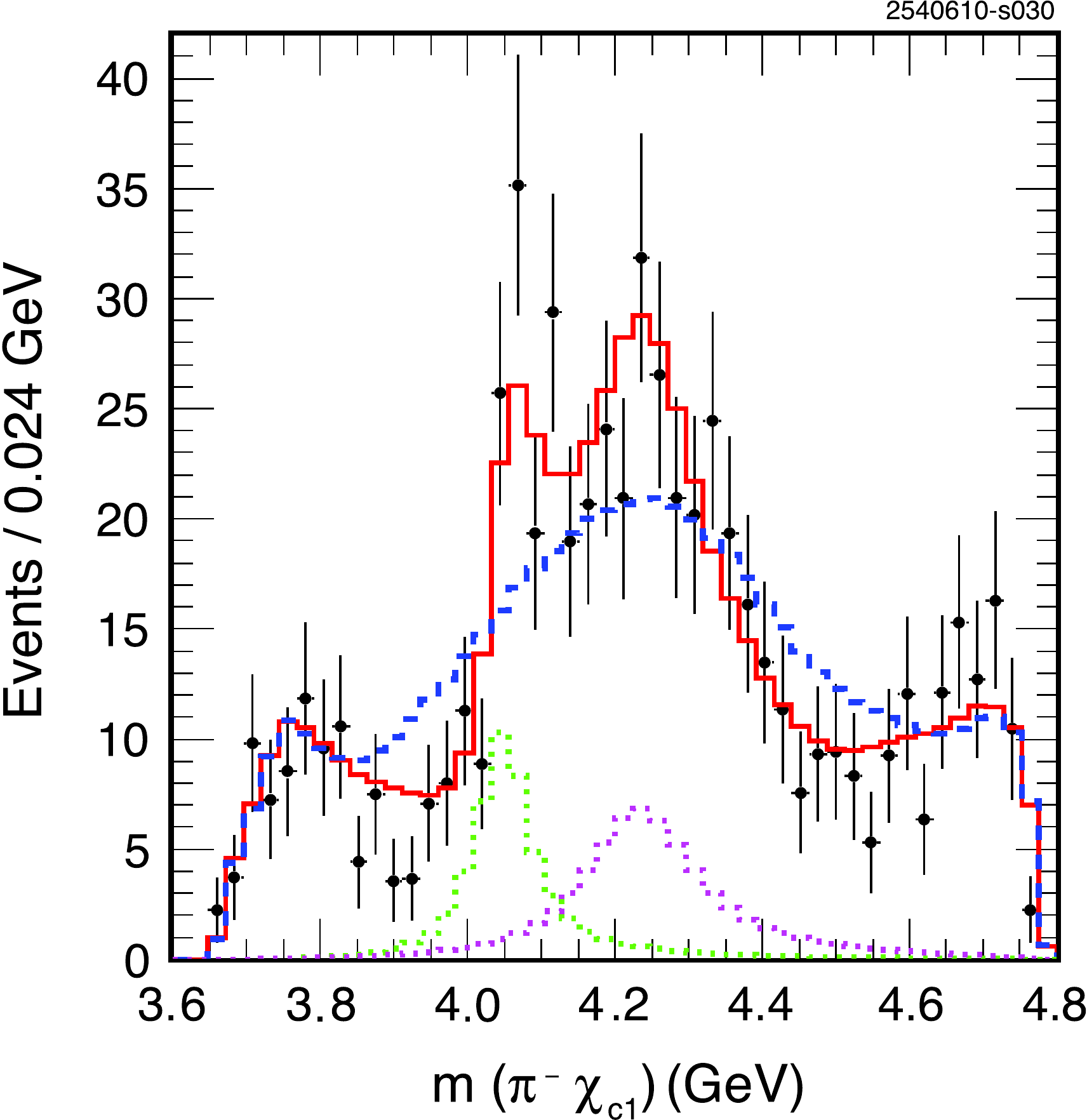}
\caption{From Belle~\cite{Mizuk:2008me}, 
         for $B\to K\pi^-\chi_{c1}$ candidates, the data {\it points} show the 
         $m(\pi^-\chi_{c1})$ projection of the Dalitz plot with the $K^\ast$ 
         bands removed. The {\it solid (dashed) histogram} shows the corresponding 
         projection of the fit with (without) the two $Z_i^-\to\chi_{c1}\pi^-$
         resonance terms. The {\it dotted histograms} show the fitted contributions
         of the two resonances. \AfigPermAPS{Mizuk:2008me}{2008} }
\label{fig:Spec_Z1Z2_Belle_NoKstar}
\end{figure}

Belle has also found~\cite{Mizuk:2008me} signals, dubbed \zOne\ and \zTwo,
in the $\chi_{c1}\pi^-$ channel, again using $B\to Z^-K$ decays.
Here the kinematically allowed mass range for the \kpi\
system extends beyond the $K^{\ast}_3(1780)$ $F$-wave resonance. Thus
$S$-, $P$-, $D$- and $F$-wave terms for the \kpi\ system are 
all included in the model. A Dalitz fit with a single resonance in the 
$Z^-\to\chi_{c1}\pi^-$ channel is favored over a fit with only $K_i^\ast$
resonances and no $Z^-$ by $>$10$\sigma$. Moreover, a fit with
two resonances in the $\chi_{c1}\pi^-$ channel is favored over the fit
with only one $Z^-$ resonance by $5.7\sigma$. 
Fitted mass values appear in \Tab{tab:Spec_ExpSumUnc}.
The product branching fractions
have central values similar to that for the \zBelle\ but with large errors.
\Figure{fig:Spec_Z1Z2_Belle_NoKstar} shows the $m(\chi_{c1}\pi^-)$
projection of the Dalitz plot with the $K^\ast$ bands excluded and the
results of the fit with no $Z^-\to\chi_{c1}\pi^-$ resonances and with
two $Z^-\to\chi_{c1}\pi^-$ resonances.

Thus, although two experiments have explored these
new states, we are left in the less than satisfying
situation of three claims of definitive observations
and one nonobservation which does not exclude
the positive measurement. If any or all of the three 
charged $Z^-$ states reported by Belle are in fact meson 
resonances, they would be ``smoking guns'' for exotics. 
It is therefore crucial that these states be confirmed 
or refuted with independent measurements. In particular, 
\babar\ should search for 
$Z_1(4050)^-$ and $Z_2(4250)^-$ 
in the $\chi_{c1}\pi^-$ channel and Belle should search 
for \zBelle\ in the $\jpsi\pi^-$ channel. 
That the purported \zBelle\ might decay copiously 
to $\psip\pi^-$ but barely or not at all to $\jpsi\pi^-$
is a theoretical puzzle worth addressing.
The \DZero\ and CDF experiments at the Tevatron could also
search for inclusive production of \zBelle.

\subsection{Characteristics of quarkonium systems} 

\begin{figure}[b]
   \begin{center}
      \includegraphics*[width=\figwid]{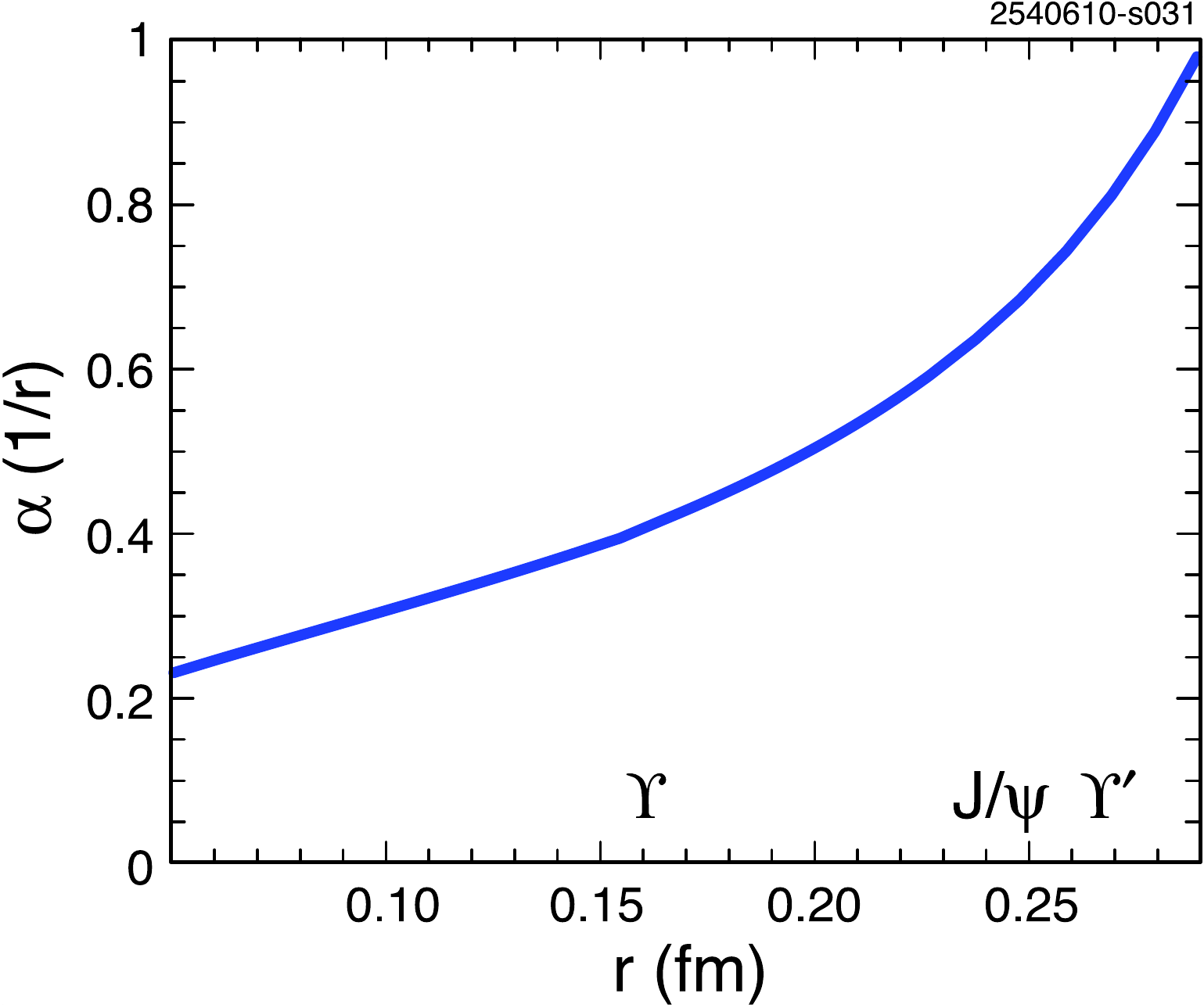}
      \caption{The strong coupling constant, $\als$, at one loop,
               as a function of quarkonium radius $r$,
               with labels indication approximate
               values of  $\hqm v$ 
               for \UnS{1}, \jpsi, and \UnS{2}}
      \label{fig:SpecTh_alpha-r}
   \end{center}
\end{figure}

Heavy quarkonia are systems composed of two heavy quarks,
each having mass $\hqm$ much larger than the 
QCD  confinement scale $\lamQ$. 
Because the system is nonrelativistic, quarkonium is characterized by 
the heavy-quark bound-state velocity, $v \ll 1$,
( $v^2 \sim 0.3$ for $c\bar{c}$, $v^2 \sim 0.1$ for $b\bar{b}$,  
$v^2 \sim 0.01$ for $t\bar{t}$)
and by a hierarchy of energy scales: the mass $\hqm$ (hard scale, H),
the relative momentum $p \sim \hqm v $ (soft scale, S),
and the binding energy $E \sim \hqm v^2$ (ultrasoft scale, US).
For energy scales close to $\lamQ$, perturbation theory breaks down  
and one has to rely on nonperturbative 
methods. Regardless, the nonrelativistic hierarchy of scales,
\beq
\hqm \gg p \sim 1/r \sim \hqm v  \gg E \sim \hqm v^2\,,
\eeq
where $r$ is the typical distance between the heavy quark 
and the heavy antiquark,
also persists below the scale $\lamQ$. 
Since $\hqm \gg \lamQ$, $\als(\hqm) \ll 1$, and phenomena occuring
at the scale $\hqm$ may be always treated perturbatively.
The coupling may also be small if $\hqm v \gg \lamQ$ and 
$\hqm v^2 \gg \lamQ$, 
in which case  $\als(\hqm v) \ll 1$ and $\als(\hqm v^2)\ll 1$, 
respectively. 
This is likely to happen only for the lowest charmonium and bottomonium
states (see \Fig{fig:SpecTh_alpha-r}). Direct
information on the radius of the quarkonia systems is not
available, and thus the
attribution of  some of the lowest bottomonia and charmonia states 
to the perturbative or the nonperturbative soft regime is at the
moment still ambiguous.
For $t\bar{t}$ threshold states  even the ultrasoft
scale may be considered perturbative.

\begin{figure}[t]
   \begin{center}
      \includegraphics*[width=\figwid]{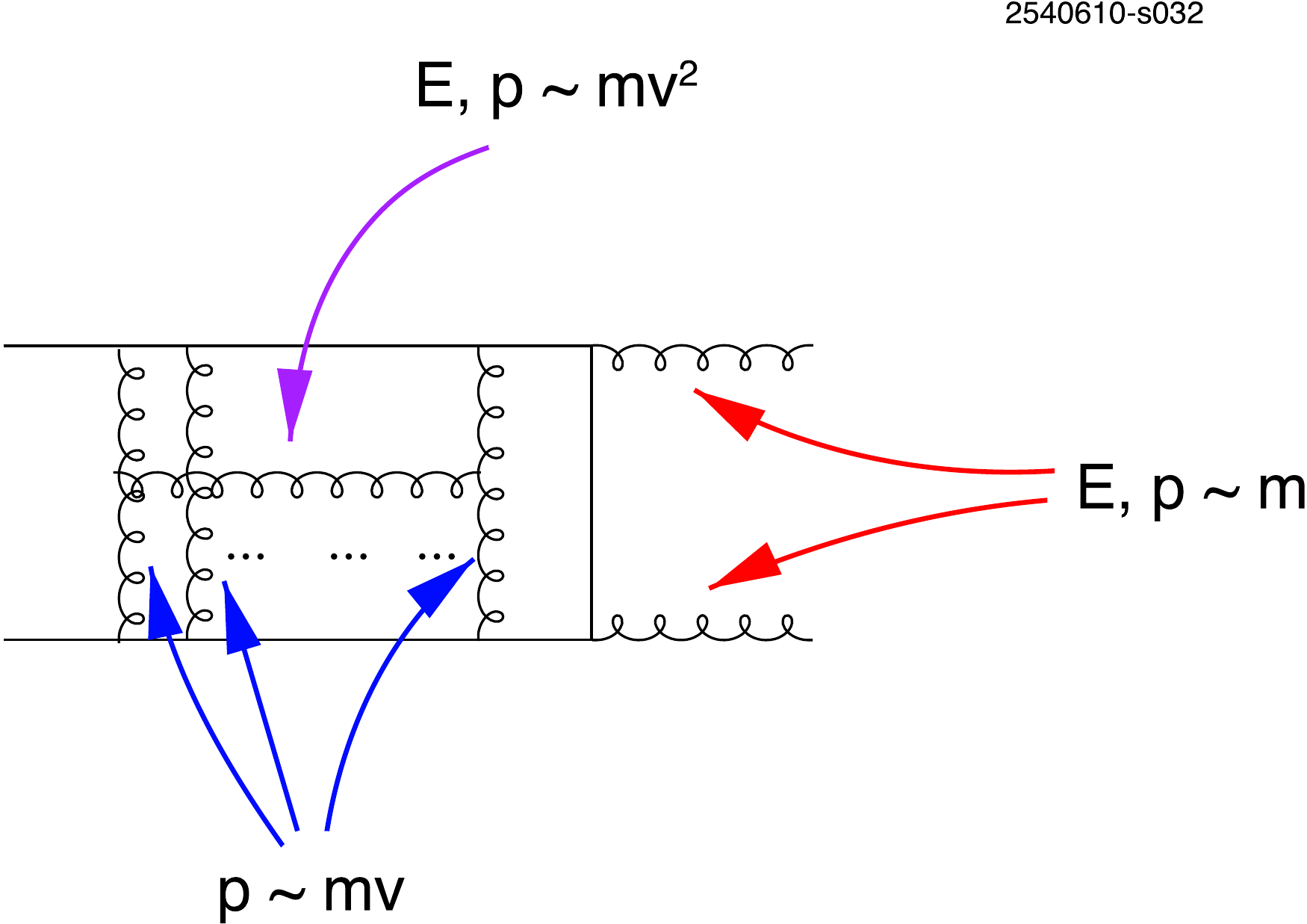}
      \caption{Typical scales appearing in a quarkonium annihilation diagram}
      \label{fig:SpecTh_gentangled}
   \end{center}
\end{figure}

This hierarchy of nonrelativistic scales 
separates quarkonia~\cite{Brambilla:2004wf}
from heavy-light mesons, the latter of which are characterized
by just two scales: $\hqm$ and 
$\lamQ$~\cite{Neubert:1993mb,Manohar:2000dt}.
This makes the theoretical description of 
quarkonium physics more complicated.
All quarkonium scales get entangled in a 
typical amplitude involving a quarkonium
observable, as illustrated in \Fig{fig:SpecTh_gentangled}. 
In particular, quarkonium annihilation 
and production take place at the scale $\hqm$, 
quarkonium binding takes place at the scale
$\hqm v$ (which is the typical momentum 
exchanged inside the bound state), while 
very low-energy gluons and light quarks 
(also called ultrasoft degrees of freedom)  
are sufficiently long-lived that a bound state has time 
to form and therefore are sensitive to the 
scale $\hqm v^2$. Ultrasoft gluons are responsible 
for phenomena like the Lamb shift in QCD.
The existence of several scales 
complicates the calculations.
In perturbative calculations of loop diagrams 
the different scales get entangled, challenging 
our abilities to perform higher-order 
calculations. In lattice QCD, the existence 
of several scales for quarkonium sets 
requirements on the lattice spacing 
($a < 1/\hqm$) and overall size ($L a > 1/(\hqm v^2)$) 
that are challenging to our present computational power.

\begin{figure}[t]
   \begin{center}
      \includegraphics*[width=\figwid]{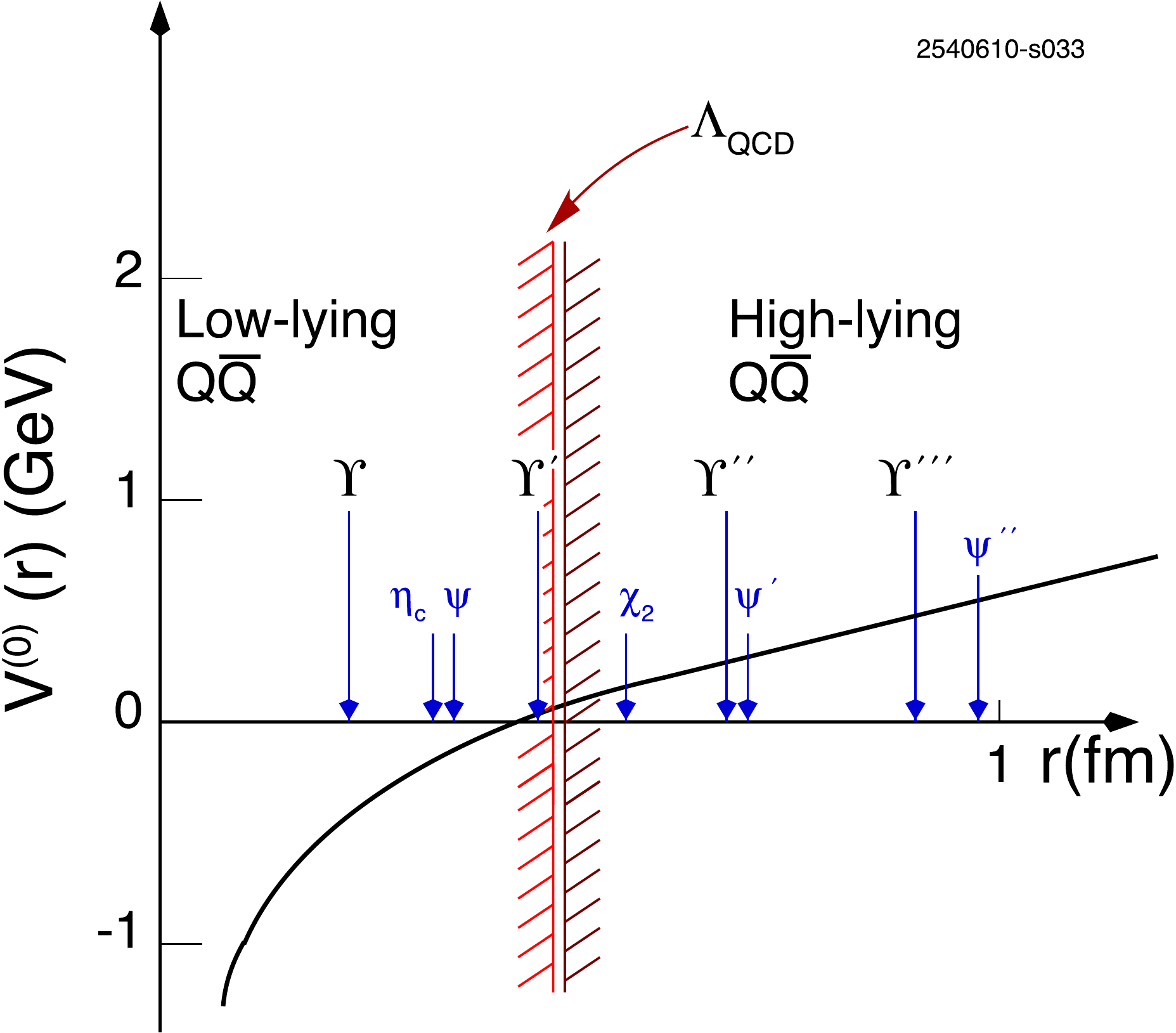}
      \caption{Static $Q\overline{Q}$  potential as a function of 
               quarkonium radius $r$}
      \label{fig:SpecTh_potheavy}
   \end{center}
\end{figure}

However, it is precisely the rich structure of 
separated energy scales that makes heavy 
quarkonium particularly well-suited to the study of the confined
region of QCD, its interplay 
with perturbative QCD, and of the behavior of the perturbation series in
QCD: heavy quarkonium is an ideal probe of confinement and deconfinement. 
Quarkonia systems with different radii have varying sensitivies to the 
Coulombic and confining interactions, as depicted in 
\Fig{fig:SpecTh_potheavy}.
Hence different quarkonia will dissociate in a medium at different
temperatures, providing, \eg a thermometer for the plasma,
as discussed in \Sec{sec:media_sec3}.

\subsection{Nonrelativistic effective field theories}
\label{sec:SpecTh_eft}

The modern approach to heavy quarkonium is provided by 
Nonrelativistic Effective Field Theories (NR~EFTs)~\cite{Brambilla:2004jw}. 
The idea is to
take advantage of the existence of a hierarchy of scales   
to substitute simpler but equivalent NR~EFTs for QCD.
A hierarchy of EFTs may be constructed by systematically integrating out 
modes associated with high-energy scales 
not relevant for the quarkonium system.
Such integration is performed as part of a matching procedure that 
enforces the equivalence between QCD and the EFT at a given 
order of the expansion in $v$.
The EFT realizes factorization between 
the high-energy contributions carried by the matching coefficients and 
the low-energy contributions carried by the dynamical degrees of freedom
at the Lagrangian level.
The Poincar\'e symmetry remains intact at the level of the NR~EFT
in a nonlinear realization that imposes exact relations among the 
EFT matching coefficients~\cite{Brambilla:2003nt,Brambilla:2001xk}.

\subsubsection{Physics at the scale $\hqm$: NRQCD}
\label{sec:SpecTh_nrqcd}

Quarkonium annihilation and production occur at the scale $\hqm$. 
The suitable EFT is Nonrelativistic
QCD~\cite{Caswell:1985ui,Bodwin:1994jh}, 
which follows from QCD by integrating out the scale $\hqm$. As a consequence, 
the effective Lagrangian is organized as an 
expansion in $1/\hqm$  and $\als(\hqm)$: 
\beq
{\cal L}_{\rm NRQCD}  = \sum_n \frac{c_n(\als(\hqm),\mu)}{\hqm^{n} } 
\times  O_n(\mu,\hqm v,\hqm v^2,...),
\eeq
where $O_n$ are the operators of NRQCD that 
are dynamical at the low-energy scales $\hqm v$
and $\hqm v^2$, $\mu$ is the NRQCD factorization 
scale, and $c_n$ are the Wilson
coefficients of the EFT that encode the 
contributions from the scale $\hqm$ and are
nonanalytic in $\hqm$.
Only the upper (lower) components  of the Dirac fields matter
for quarks (antiquarks) at energies lower than $\hqm$. 
The low-energy operators $O_n$  are constructed out of two or four 
heavy-quark/antiquark fields plus gluons. 
The operators bilinear in the fermion 
(or antifermion) fields are the same ones that can be obtained from a 
Foldy-Wouthuysen transformation of the QCD Lagrangian. 
Four-fermion operators have to be added. Matrix elements of $O_n$ 
depend on the scales $\mu$, $\hqm v$, $\hqm v^2$ and $\lamQ$.
Thus operators are counted in powers of $v$. 
The imaginary part of the coefficients of the four-fermion operators
contains the information on heavy quarkonium annihilation.
The NRQCD heavy quarkonium Fock state is given by a series of terms, 
where the leading term is a $Q\overline{Q}$ 
in a color-singlet state, and the first correction, suppressed in $v$,
comes from a $Q\overline{Q}$ in an octet state plus a gluon.
Higher-order terms are subleading in increasing powers of $v$.
NRQCD is suitable for spectroscopy studies on the lattice.  
The latest results on charmonium, bottomonium, and 
$B_c$ spectroscopy are reported in \Sec{sec:SpecTh_nrqcdlc}. For the latest results on 
NRQCD inclusive decay amplitudes, see \Sec{sec:Dec_RadLepTheory}.

\subsubsection{Physics at the scales $\hqm v$, $\hqm v^2$: pNRQCD}

Quarkonium formation occurs at the scale $\hqm v$. 
The suitable EFT is pNRQCD \cite{Pineda:1997bj,Brambilla:1999xf},  which 
follows from NRQCD by integrating out the scale $\hqm v \sim r^{-1}$.
The soft scale $\hqm v$ may or may not be larger than the confinement 
scale $\lamQ$ depending on the radius of the quarkonium system. 
When $\hqm v^2 \sim \lamQ$, we speak about weakly-coupled pNRQCD because 
the soft scale is perturbative and the matching of NRQCD to pNRQCD 
may be performed in perturbation theory. 
When $\hqm v  \sim \lamQ$, we speak about  
strongly-coupled pNRQCD because the soft scale 
is nonperturbative and the matching 
of NRQCD to pNRQCD is not possible in
perturbation theory. Below we will review 
recent results and applications
of the two EFTs.

\subsubsection{$\hqm v \gg \lamQ$: weakly-coupled pNRQCD}

The effective Lagrangian is organized as an expansion in $1/\hqm$  and $\als(\hqm)$, 
inherited from NRQCD, and an expansion in $r$  \cite{Brambilla:1999xf}: 
\beqa
&& {\cal L}_{\rm pNRQCD}  = \int d^3r\,  
\sum_n \sum_k \frac{c_n(\als(\hqm),\mu)}{\hqm^{n}}  
\non\\
&& \quad \times V_{n,k}(r,\mu^\prime, \mu) \; r^{k}  
\times O_k(\mu^\prime,\hqm v^2,...) ,
\label{eqn:SpecTh_pnrqcd}
\eeqa
where $O_k$ are the operators of pNRQCD that 
are dominant at the low-energy scale
$\hqm v^2$, $\mu^\prime$ is the pNRQCD factorization scale and $V_{n,k}$
are the Wilson coefficients of the EFT that encode the contributions 
from the scale $r$ and are nonanalytic in $r$.
The degrees of freedom that make up the operators $O_k$
are $Q\overline{Q}$ states, color-singlet $S$, color-octet $O_a T^a$,
and (ultrasoft) gluons. The operators are defined in a multipole expansion.
In the equations of motion of pNRQCD, we may identify 
$V_{n,0}= V_n$ with the $1/\hqm^n$ potentials that enter 
the Schr\"odinger equation and 
$V_{n,k\neq 0}$ with the couplings of the 
ultrasoft degrees of freedom that
provide corrections to the Schr\"odinger equation.
Since the degrees of freedom that enter the Schr\"odinger description 
are, in this case, both $Q\overline{Q}$  color singlet 
and $Q\overline{Q}$ color octets,
both singlet and octet potentials exist.

The bulk of the interaction is 
contained in potential-like terms, but non-potential interactions, 
associated with the propagation of low-energy degrees of freedom
are, in general, present as well and start 
to contribute at NLO in the multipole expansion.
They are typically related to nonperturbative effects. 
Matrix elements of $O_{n,k}$ depend on the 
scales $\mu^\prime$, $\hqm v^2$ and $\lamQ$.

If the quarkonium system is small, the soft scale is perturbative and the 
potentials can be {\it entirely} calculated in perturbation theory 
\cite{Brambilla:2004jw}. They are renormalizable, 
develop a scale dependence, and satisfy renormalization
group equations that eventually 
allow resummation of potentially large logarithms.

\subthreesection{Progress in the perturbative calculation of the potentials}
\label{sec:SpecTh_stpo} 

\subfoursection{$Q\overline{Q}$ singlet and octet potentials}

There has been much progress in the computation of the
perturbative corrections to the potential between a static quark and a
static antiquark in a color-singlet configuration. 
The subleading logarithmic term coming from ultrasoft
effects has been computed~\cite{Brambilla:2006wp}. The
resummation of those subleading ultrasoft logarithms has also been
performed~\cite{Brambilla:2009bi}, along with a comparison of the
static energy to lattice data~\cite{Brambilla:2009bi}. 
These calculations confirm that
perturbation theory can reproduce the lattice very accurately at short
distances (up to 0.25~fm) once the 
cancellation of the leading renormalon singularity
is implemented.

The three-loop correction to the static potential is now completely
known: the fermionic contributions to the three-loop 
coefficient~\cite{Smirnov:2008pn} first became available, and, more
recently, the remaining purely gluonic term has been 
obtained~\cite{Anzai:2009tm,Smirnov:2009fh}. The value found
for the $n_f$-independent three-loop coefficient is much lower than 
the previous, widely-used Pad\'e estimate and within the range
obtained by comparing to lattice data~\cite{Brambilla:2009bi}. 
The future implementation of the three-loop result may
improve the precision of some mass and strong-coupling determinations.
In particular, the 
recently obtained theoretical expression~\cite{Brambilla:2010pp} 
for the complete QCD static
energy at next-to-next-to-next-to-leading-logarithmic accuracy (NNNLL) has
been used 
to determine $r_0 \LMSb$ by comparison with available lattice
data, where $r_0$ is the lattice scale and $\LMSb$
is the QCD scale, obtaining 
\beq
r_0\LMSb=0.622^{+0.019}_{-0.015}
\eeq
for the zero-flavor case. 
This extraction was previously performed
at the NNLO level (including an estimate at NNNLO) in \cite{Sumino:2005cq}.
The same procedure can be used to obtain a precise evaluation of the
unquenched $r_0 \LMSb$ value after short-distance unquenched
lattice data for the $Q \overline{Q}$ exist.

At three-loop order a violation of the Casimir scaling
for the static QQbar potential is found, see
\cite{Anzai:2010td}.

The static octet potential is known up to two loops~\cite{Kniehl:2004rk}.
Relativistic corrections to the static singlet potential
have been calculated over the years and are 
summarized in \cite{Brambilla:2004jw}. 

\subfoursection{$QQQ$ and $QQq$ static potentials}

The three-quark static potential has been 
calculated in perturbation theory at 
next-to-leading order in the singlet, octet, and 
decuplet channels~\cite{Brambilla:2009cd,Brambilla:2005yk}.
Mixing between the octet representations has been found already at
tree level.
At next-to-next-to-leading order, the subset of diagrams 
producing three-body forces has been identified in Coulomb gauge and 
their contribution to the potential calculated. Combining it with the 
contribution of the two-body forces, which may be extracted from 
the quark-antiquark static potential, the complete 
next-to-next-to-leading order of
the three-quark static potential in the 
color-singlet channel has been obtained
in \cite{Brambilla:2009cd}. These results may be 
important for accurate calculations
of the lowest  $QQQ$ states as well as for comparison and study
of  the $QQQ$
lattice static energies in the domain of small interquark distance.

The same pNRQCD  description is also possible for 
$QQ$ states, which are relevant for doubly-charmed
baryons~\cite{Brambilla:2005yk,Fleming:2005pd}.  The $QQ$ antitriplet
static potential, relevant for the $QQq$ states, has been calculated at
next-to-next-to-leading order~\cite{Brambilla:2009cd}. 
First lattice calculations of 
the $QQq$ potential have also
become available~\cite{Yamamoto:2008fm,Suganuma:2008ej,Najjar:2009da}.

\subthreesection{Progress on the lowest spectra calculation}
\label{sec:SpecTh_lowestspectra}

For systems with small radii, precision calculations are possible.
In such cases, quarkonium may become a
benchmark for our understanding of QCD, in particular for the transition 
region between perturbative and nonperturbative QCD, and 
for the precise determination
of Standard Model parameters (\eg heavy quark masses, $\als$,
as described in \Sec{sec:SpecTh_param}).
When the soft scale is perturbative, the energy levels are given by the 
expectation value of the perturbative potentials, calculated to the 
appropriate order of the expansion in $\als$ plus nonperturbative 
terms that are not part of the potential, 
which start to contribute to the energy levels at order $\hqm \als^5$.
They enter energy levels and decay calculations in the form of local or 
nonlocal electric and magnetic condensates 
\cite{Brambilla:1999xj,Kniehl:1999ud,Voloshin:1978hc,Leutwyler:1980tn,Brambilla:2004jw}.  
A precise and systematic knowledge of such 
nonperturbative, purely glue-dependent objects is still lacking. 
It is therefore important that these condensates
be fixed, either by dedicated lattice determinations
or extracted from the data.
Within pNRQCD it is possible to relate the 
leading electric and magnetic nonlocal correlators  
to the gluelump masses and to some existing lattice (quenched) 
determinations~\cite{Brambilla:2004jw}.

However, since the nonperturbative contributions are suppressed in 
the power counting, it is possible to obtain precise determinations 
of the masses of the lowest quarkonium resonances within purely 
perturbative calculations, in the cases in which the perturbative 
series is convergent (\ie after the appropriate subtractions of 
renormalons have been performed), and large logarithms are
resummed~\cite{Pineda:2001ra,Pineda:2000gz,Hoang:2002yy,Manohar:1999xd}.

Once the potentials are known, energy levels can be computed.
The lowest heavy quarkonium states are suitable to obtain precise 
determinations of the $b$ and $c$ mass.
Such determinations are competitive with those coming from different systems 
and different approaches, as has been discussed at 
length~\cite{Brambilla:2004wf}. An update of recent result on mass extractions 
is given in \Sec{sec:SpecTh_param}.

Once the quark masses have been obtained, it is possible 
to obtain the energy levels of the lowest resonances.
However, which quarkonium state belongs to which regime is 
an open issue and no clear-cut method exists to decide this {\it a priori}, 
since we lack a direct method to determine the quarkonium 
radius~\cite{GarciaiTormo:2005bs,DomenechGarret:2008vk}.
Typically the lowest states $\Upsilon(1S), \eta_b$, 
$B_c$, and possibly $J/\psi$ and $\eta_c$
are assumed to be in the weakly-coupled regime.
The $S$-wave energy levels are known in perturbation theory at
next-to-next-to-next-to-leading order ($\hqm \als^5$)
\cite{Beneke:2005hg,Penin:2005eu,Penin:2002zv,Kniehl:2002br,Brambilla:1999xj,Kniehl:1999ud}.

A prediction of the $B_c$ mass has been obtained~\cite{Brambilla:2000db}. 
The NNLO calculation with finite charm mass 
effects~\cite{Brambilla:2001qk,Brambilla:2001fw} predicts a mass of 
6307(17)\mev, in agreement with the CDF measurement~\cite{Abulencia:2005usa}
and the lattice determination~\cite{Allison:2004be}.
A NLO calculation reproduces, in part, the $\chi_{bJ}(1P)$ fine splitting 
\cite{Brambilla:2004wu}. The same procedure seems to work at NNLO even 
for higher-mass bottomonium states 
(\ie measured masses match the predictions within
the respective theory errors, which are larger for 
higher-mass states)~\cite{Brambilla:2001qk}.
Including logs resummation at NLL, it is possible to obtain a 
prediction~\cite{Kniehl:2003ap}
\beq
\dmhf[\etab] = 41 \pm 11 ({\rm th}) ^{+9}_{-8} (\delta \als)\mevcc\,,
\label{eqn:SpecTh_etabNLL}
\eeq
in which the second error comes from the uncertainty in $\als$.
This value is consistent with another perturbative
prediction~\cite{Recksiegel:2003fm} but both
undershoot the average experimental value
from \Eq{eqn:SpecExp_HFSexp} by about $2\sigma$.
This discrepancy with experiment remains a challenge for theory. 
(There are further discussions of the \etab\ hyperfine splitting in 
\Secs{sec:SpecExp_etab},
\ref{sec:SpecTh_nrqcdlc},
\ref{sec:SpecTh_etab},
\ref{sec:SpecTh_alphasdec}, and
\ref{sec:SpecTh_mixing}, the
last of which offers the possibility of
new physics becoming manifest in $\eta_b(nS)$ mass splittings.)
Similar calculations yield a predicted $B_c$ 
hyperfine separation~\cite{Penin:2004xi}
\beq
m(B_c^*) - m(B_c) = 50 \pm 17 ^{+15}_{-12}\mevcc\,. 
\eeq

\subsubsection{$\hqm v \sim \lamQ$: strongly-coupled pNRQCD}

When $\hqm v \sim \lamQ$  the soft scale is nonperturbative  
and the matching from NRQCD to pNRQCD cannot be performed in
perturbation theory. 
Then the  potential matching coefficients
are obtained in the form of expectation values of gauge-invariant 
Wilson-loop operators. 
In this case, heavy-light meson pairs and heavy hybrids 
develop a mass gap of order $\lamQ$ with respect to the energy of the
$Q\overline{Q}$ pair, the second circumstance 
being apparent from lattice simulations.
Thus, away from threshold, 
the quarkonium singlet field $S$ is the only low-energy dynamical 
degree of freedom in the pNRQCD Lagrangian 
(neglecting ultrasoft corrections coming 
from pions and other Goldstone bosons).
This pNRQCD Lagrangian may be expressed 
as~\cite{Brambilla:2000gk,Pineda:2000sz,Brambilla:2004jw}:
\beq
\quad  {\cal L}_{\rm pNRQCD}=  { S}^\dagger
   \left(i\partial_0-\frac{{\bf p}^2}{2\hqm}-V_S(r)\right){S}\,. 
\label{eqn:SpecTh_sc}
\eeq
The singlet potential $V_S(r)$ can be expanded
in powers of the inverse of the quark mass;
static, $1/\hqm$ and $1/\hqm^2$ terms were calculated long 
ago~\cite{Brambilla:2000gk,Pineda:2000sz}.
They involve NRQCD matching coefficients (containing 
the contribution from the hard scale) and low-energy 
nonperturbative parts given in terms
of static Wilson loops and field-strength insertions in the static
Wilson loop
(containing the contribution from the soft scale).
In this regime of pNRQCD, we recover the quark potential singlet model. 
However, here the potentials are calculated in QCD by nonperturbative 
matching. Their evaluation requires calculations on the lattice 
or in QCD vacuum models.

Recent progress includes new, precise lattice calculations 
of these potentials 
obtained using the L\"uscher multi-level algorithm
(see \Sec{sec:SpecTh_latpnrqcd}
for more details).

Then, away from threshold, all the masses 
can be obtained by solving the Schr\"odinger equation
with such potentials. Some applications 
of these results to the spectrum calculation are 
ongoing~\cite{Laschka:2009um}.

A trivial example of application of this method is the mass of the \hsubc.
The lattice data show a vanishing long-range component 
of the spin-spin potential. Thus the potential appears to be entirely
dominated by its short-range, delta-like, part, 
suggesting that the $^1P_1$ state should be close to 
the center-of-gravity of the $^3P_J$ system.
Indeed, the measurements described in \Sec{sec:SpecExp_hc}
and summarized in \Tab{tab:Spec_hc}
are consistent with this expected value.

If we explicitly consider light quarks, each quarkonium 
state may develop a width due to decay through pion
emission. The neavy-light states develop a mass 
gap of order $\lamQ$ with respect to 
quarkonium which can be absorbed into the 
definition of the potentials or of the 
(local or nonlocal) condensates~\cite{Brambilla:2002nu}.

\subsection{Lattice QCD spectrum calculations}
\label{sec:SpecTh_latspectra} 

In quarkonia, the ultrasoft scale $mv^2$ is often of a similar
size as the scale $\lamQ$, where non-perturbative effects
become important. For all charmonium and
many bottomonium states the soft scale $mv\sim 1/r$ is not
much larger than $\lamQ$ either. The non-perturbative
contributions can be evaluated via computer simulations of
lattice-regularized QCD (Lattice QCD), where the lattice spacing
$a$ provides a hard ultraviolet cut-off on the available momenta
in a Euclidean space-time volume.

Light sea quarks are particularly expensive to simulate
numerically since
the computational effort increases as a large inverse power
of the corresponding pseudoscalar mass $m_{\rm PS}$.
The spatial lattice extent,
$La$, should be much larger than $m_{\rm PS}^{-1}$
to control finite-size effects, necessitating
large volumes. Only very recently
have the first simulations near the physical pion mass
$m_{\rm PS}\approx m_{\pi}$ become
possible~\cite{Davies:2008sw,Namekawa:2008ft,Taniguchi:2009hb,Gockeler:2009pe}.
In the first reference a variant of the staggered fermion
action was applied. These fermions are usually only defined for
$n_f$ mass-degenerate fermions, where $n_f$ is a multiple of four.
This restriction was circumvented by replacing the determinant of the
$n_f=8+4$ staggered action by a fourth root
that may then correspond to $n_f=2+1$. However, it remains
controversial whether this {\it ad hoc} prescription
recovers a unitary, local quantum field theory and thus
the correct continuum limit. Moreover, there are claims
that the additional so-called taste symmetry cannot be 
completely restored in the continuum limit~\cite{Creutz:2007yg,Creutz:2008nk},
some aspects of which have been refuted~\cite{Bernard:2006vv},
and others of which have yet to be clearly 
established~\cite{Golterman:2008gt}. 

The lattice regularization of QCD
is not unique, and many different discretized actions can be constructed
that should yield the same results after removing the cutoff
(continuum limit extrapolation: $a\rightarrow 0$). While the
Wilson action is subject to $O(a)$ discretization effects,
other actions are $O(a)$ improved, \eg chiral actions
fulfilling the Ginsparg-Wilson relation, staggered actions,
twisted-mass QCD or the nonperturbatively-improved
Sheikholeslami-Wilson (clover) action. Ideally, lattice
simulations are repeated at several values of the lattice spacing,
$a\ll\lamQ^{-1}$, and the results then
extrapolated to the continuum limit.
Lattice artifacts will be large if physics at scales
$q\not\ll a^{-1}$ becomes important,
spoiling the continuum limit extrapolation. In this respect,
the charm quark mass, $m_c$, and, even more so, the bottom quark
mass, $m_b$, pose challenges. By exploiting or ignoring
the multiscale nature of
quarkonium systems, different routes can be taken.

One way to proceed is to integrate out $\hqm$ as well as $\hqm v$
and to evaluate the resulting potential nonperturbatively
in lattice simulations
of pNRQCD~\cite{Pineda:1997bj,Brambilla:1999xf}.
Subsequently, the energy levels can be obtained by solving
a Schr\"odinger equation. This directly relates
lattice simulations to QCD vacuum models and to potential models.
However, at present this approach is
only semi-quantitative since no calculations of the matching
of lattice pNRQCD, where rotational symmetry is broken,
to QCD is available.

Another approach, which is more common,
is to integrate out only $\hqm$ and to simulate 
NRQCD~\cite{Caswell:1985ui,Bodwin:1994jh}
on the lattice~\cite{Lepage:1992tx}.
In this case, the lattice spacing, $a^{-1}$, plays the role of
the soft matching scale, $\mu$, used above. Unless this is
restricted to $\hqm\gtrsim a^{-1}\gtrsim \hqm v$, the NRQCD
matching coefficients will become uncontrollably large, and
the continuum limit cannot be taken. Therefore
$a$ must be sufficiently fine for discretization effects
to be smaller than the neglected higher-order terms in the
heavy-quark expansion. This restricts the applicability
of NRQCD methods to systems containing $b$ quarks.

Relativistic heavy quark actions can also be used
because the inequality $a^{-1}\gg \hqm$ usually still holds
so that spin-averaged
level-splittings can be obtained. To a lesser extent,
reliable results on the fine structure can be achieved as well.
The Fermilab effective field theory interpretation of the heavy quark
action~\cite{ElKhadra:1996mp,Kronfeld:2000ck} smoothly connects the region 
$\hqm\sim a^{-1}$ with the continuum limit $a^{-1}\gg \hqm$, allowing charm 
and bottom systems to be treated in the same set~up.

The following sections survey the present state-of-the-art. We start
with results obtained using a
relativistic heavy-quark action, continue with NRQCD,
and conclude with lattice pNRQCD results.

\subsubsection{Relativistic heavy quark actions}

One way of limiting the computational cost of small lattice spacings,
\ie of a large number of lattice points, is the use of anisotropic actions,
with a temporal lattice spacing, $a_{\tau}$, smaller than
the spatial lattice spacing, $a_{\sigma}=\xi a_{\tau}$, where
$\xi>1$. The spatial lattice
extent, $L_{\sigma}a_{\sigma}$,
still needs to be sufficiently large
to accommodate the quarkonium state, which has a size of order
$r\simeq (\hqm v)^{-1}$. In the presence of light sea quarks, one
would additionally wish to realize $L_{\sigma}a\gg m_{\rm PS}^{-1}$.
With sufficiently large $a_{\sigma}$, it is possible to
limit the number of lattice points, $\propto L_{\sigma}^3$.
It is then easy to realize
$a_{\tau}<m^{-1}$, where $m$ is the particle mass. 
Naively, such simulations are cheaper by
a factor $\xi^3$ relative
to the isotropic case. This method was successfully explored
in the quenched approximation~\cite{Liao:2002rj,Okamoto:2001jb},
and is reviewed in~\cite{Brambilla:2004wf}.

At tree level, it can be arranged that the lattice spacing
errors are $O[(\hqm a_{\tau})^{n}]$, where $n\in\{1,2\}$
depends on the heavy-quark action, but care is needed
to ensure it~\cite{Harada:2001ei}.
One-loop corrections may lead to $O[\als(\hqm a_{\sigma})^{n}]$ terms:
to the extent that $\als\xi^n$ is small,
the leading-order lattice effects can be regarded as
$O[(\hqm a_{\tau})^n]$. The anisotropy parameter $\xi$ must
be determined consistently for the quark and gluon contributions to
the QCD action. Within the quenched approximation,
where the feedback of quark fields onto the gluons is neglected,
this problem factorizes. The tuning
is much harder to achieve and numerically more costly in QCD with sea quarks.
Nevertheless such a program was pursued very successfully by
the Hadron Spectrum Collaboration~\cite{Lin:2008pr}.
It should be noted, however, that, in this case, the $O(a)$ improvement
is {\it not} nonperturbatively accurate.

Such anisotropic configurations have been employed to calculate
electromagnetic transition rates from
excited charmonium states, high-spin states,
and exotics~\cite{Dudek:2007wv,Dudek:2008sz,Dudek:2009kk}
with small volumes.
The lattice spacing and sea quark mass dependencies will
be investigated in the near future.
The same holds for an exploratory
study of the Regensburg group,
using the isotropic clover action for $n_f=2$
sea quarks (generated by the QCDSF Collaboration~\cite{Gockeler:2010yr})
and charm quarks, using a lattice
cutoff $a^{-1}\approx 1.7$\gev~\cite{Ehmann:2007hj}.
While the spin-averaged splittings are in qualitative agreement with
experiment, the fine structure is underestimated due to
unphysically-heavy sea quarks and the missing continuum limit
extrapolation. Another study with improved Wilson fermions was
performed by PACS-CS with $n_f=2+1$ sea quark flavours
at the physical $\pi$ mass~\cite{Namekawa:2008ft},
focusing only on very few $J^{PC}$ channels.
In this case, the $J/\psi$-$\etac$ splitting is 
underestimated by about 10\% relative to experiment, which can
probably be attributed to the finite lattice spacing. Note that
valence quark annihilation channels were omitted in
all these studies.

To this end, mixing effects
with noncharmed mesons and
flavor-singlet contributions to the spectrum
were evaluated~\cite{Ehmann:2009ki} and found to be
smaller than 10\mev\ in the pseudoscalar channel.
The latter effect was also investigated by the
MILC Collaboration~\cite{Levkova:2008qr},
with similar results. It should be noted that 
valence quark annihilation diagrams
were neglected in all other
unquenched studies. Finally, mixing effects
between radially-excited $c\bar{c}$ states and four-quark
molecules (or tetraquarks) were investigated at a sea pion mass,
$m_{\rm PS}<300\,$\mev, at $a^{-1}\approx 2.4$\gev~\cite{Bali:2009er}. 
Indications of attraction and
mixing were found in the $1^{++}$ channel, supporting
the interpretation of the $X(3872)$ as a \DstnDn\
bound state with a $\chi_{c1}(2P)$ admixture. Other
charmed tetraquark studies~\cite{Meng:2009qt,Chiu:2006hd}
were performed in
the quenched approximation and did not take into account
valence-annihilation diagrams.
A recent review on the tetraquark
topic is presented in~\cite{Prelovsek:2010ty}.

At present, all lattice studies of 
heavy quarkonium spectroscopy in which
a continuum limit is taken are based on $n_f\approx 2+1$ configurations
generated by the MILC Collaboration~\cite{Bernard:2001av} using
the so-called AsqTad improved staggered sea-quark action~\cite{Lepage:1998vj}.
Together with the Fermilab Lattice Collaboration,
charmonium and bottomonium spectra were investigated using the
clover action for the heavy quarks at four different
lattice spacings ranging from $(1.2\gev)^{-1}$ down to $(2.2\gev)^{-1}$,
and various light quark masses~\cite{Burch:2009az}.
Only $S$- and $P$-waves were studied. After extrapolating to
the physical limit, all spin-averaged splittings, with the exception
of the charmonium $2\overline{S}-1\overline{S}$ splitting, which is
muddled with threshold effects, are in agreement with experiment. Moreover,
the $1S$ and $1P$ fine structures are compatible with
the experimental values. The continuum limit extrapolation
was essential to achieve this.
The same combination of AsqTad sea-quark action and clover
charm-quark action was also used in a recent calculation
of singly- and doubly-charmed baryons~\cite{Liu:2009jc}.

\subsubsection{NRQCD lattice calculations}
\label{sec:SpecTh_nrqcdlc}

As discussed above, lattice NRQCD is a suitable method for
bottomonia studies, for which $am_b > 1$ and $\lamQ/(vm_b)\ll 1$.
However, the precision of the results for fine-structure
splittings is limited here by the fact that the QCD matching
coefficients are typically taken at tree level 
only in such lattice simulations.

The HPQCD and UKQCD Collaborations presented 
calculations~\cite{Gray:2005ur} of the
bottomonium spectrum using the $n_f=2+1$ MILC gauge
configurations~\cite{Bernard:2001av} described above,
expanding to order-$v^4$ in the heavy-quark velocity $v$.
Another lattice calculation for bottomonium~\cite{Meinel:2009rd}
that closely followed the methods of~\cite{Gray:2005ur}
was later released,
but using the $n_f=2+1$ configurations provided
by the RBC and UKQCD Collaborations~\cite{Allton:2008pn})
with the chiral domain-wall sea-quark action for a single lattice spacing 
$a^{-1}\approx 1.7$\gev, also expanding to order $v^4$. 
Both calculations~\cite{Gray:2005ur,Meinel:2009rd}
found agreement with experiment for
the spin-averaged splittings. The spin-dependent splitting
was seen to be
systematically underestimated in~\cite{Meinel:2009rd},
possibly due to the omission of relativistic, 
radiative, and discretization corrections (see \Sec{sec:SpecTh_etab}
for more on $\dmhf[\etab]$). The same method was then used
to calculate the spectrum of other states containing one or two
$b$-quarks, including 
baryons~\cite{Meinel:2009rd,Meinel:2009vv}. Again, the
underestimation of the experimental fine structure might be
due to either the coarse lattice spacing or the imprecise matching
between lattice NRQCD and QCD. 

Large scale simulations of the spectrum of
$\Upsilon$ states and $B_c$,
which include sea quark mass and continuum limit
extrapolations, were performed by the HPQCD
collaboration~\cite{Gregory:2009hq,Gregory:2009xz}. They
employed an NRQCD action for the $b$ quark and combined this
with the relativistic HISQ (highly improved staggered quark) action
for the charm quark. Their findings are very similar to those
of~\cite{Burch:2009az}; it will be very interesting to
perform a detailed comparison between results from
a relativistic $b$-quark action and NRQCD on the same ensemble
of gauge configurations.

The mass of the triply-heavy baryon $\Omega_{bbb}$ has been 
calculated~\cite{Meinel:2010pw} in lattice QCD 
with $2+1$ flavors of light sea quarks. The $b$-quark is implemented with 
improved lattice NRQCD. Gauge field ensembles from both the RBC/UKQCD and 
MILC collaborations with lattice spacings in the range from 0.08-0.12~fm 
are used. The final result for the mass, which includes 
an electrostatic correction, is 
\beqa
m(\Omega_{bbb}) = 14.371&\pm&0.004\,({\rm stat.})\pm0.011\,({\rm syst.})\non\\
&\pm&0.001\,({\rm exp.})\gevcc.~~~~
\eeqa
The hyperfine splitting 
between the physical $J={3\over2}$ state and a fictitious $J={1\over2}$ 
state is also presented~\cite{Meinel:2010pw}.

\subsubsection{pNRQCD lattice calculations}
\label{sec:SpecTh_latpnrqcd}

Another approach is offered by
pNRQCD \cite{Brambilla:2000gk,Pineda:2000sz}.
Unfortunately, lattice pNRQCD also suffers from the fact that
the matching coefficients between lattice NRQCD and QCD
are only known at tree level. pNRQCD bridges the gap between
QCD and potential models. Implemented on the lattice,
it amounts to calculating static potentials and spin- and
velocity-dependent corrections.
Recent and not so recent progresses in this area include the calculation
of string breaking in the static sector~\cite{Bali:2005fu},
which offers one entry point into the study of threshold states,
the calculation of potentials in baryonic static-static-light
systems~\cite{Yamamoto:2008jz,Najjar:2009da}, and the
calculation of interaction energies between two
static-light mesons~\cite{Detmold:2007wk}.
New very precise results on all leading
relativistic corrections to the static $Q\overline{Q}$ potential
have also been calculated~\cite{Koma:2006si,Koma:2006fw,Koma:2009ws}.

\subsection{Predictions for the \etab\ mass}
\label{sec:SpecTh_etab}

The calculation described above in \Sec{sec:SpecTh_lowestspectra}
with result in \Eq{eqn:SpecTh_etabNLL} gives a numerical result
for $\dmhf[\etab]$ ($41\pm14$\mevcc) that is 
typical of perturbative calculations
(\eg\ $44 \pm 11$\mevcc\ is given by~\cite{Recksiegel:2003fm}).
These values are somewhat smaller than those obtained
from lattice NRQCD as described in \Sec{sec:SpecTh_nrqcdlc} above
(\eg $61\pm14$\mevcc\ in~\cite{Gray:2005ur} and $52\pm1.5({\rm
stat.})$\mevcc\ in~\cite{Meinel:2009rd}).
However, it has been argued~\cite{Penin:2009wf} that additional 
short-range corrections of
$\delta^{\rm hard} \dmhf[\etab] \approx -20$\mev\ would
lower these unquenched lattice results to the level of
the perturbative predictions. 

Very recently a newer lattice prediction~\cite{Meinel:2010pv}
performed at order $v^6$ in the NRQCD velocity expansion
has been obtained at tree level
for the NRQCD matching coefficients and with domain-wall 
actions for sea quarks, including the spin splittings,
and based on the RBC/UKQCD gauge field ensembles.
This approach~\cite{Meinel:2010pv} addresses
the concerns in \cite{Penin:2009wf} (namely, that
radiative contributions in the calculation are missing
because the NRQCD matching coefficients are calculated at tree level)
by calculating appropriate {\it ratios} of spin splittings
and thereafter normalizing to a measured value.
With this method, and using
the experimental result for the $1P$ tensor splitting as input,
a $1S$ bottomonium hyperfine splitting of 
\beqa
\dmhf[\etab] = 60.3 &\pm& 5.5\,({\rm stat.})\non\\
&\pm& 5.0\,({\rm syst.})\non\\
&\pm& 2.1\,({\rm exp.})\mevcc\non\\
= 60.3&\pm&7.7\mevcc
\eeqa
is determined. This value is slightly smaller ($1.1\sigma$) than but 
consistent with the experimental measurements 
(\Eq{eqn:SpecExp_HFSexp} and \Tab{tab:Spec_etab}),
and somewhat larger ($1.2\sigma$) than but consistent with
earlier pQCD calculations~\cite{Kniehl:2003ap,Recksiegel:2003fm}
(see \Eq{eqn:SpecTh_etabNLL}).
However, this still leaves the weakly-coupled
pNRQCD (perturbative) and experimental values
$2.0\sigma$ apart, and it is not clear why;
the lattice NRQCD lies in between.
More study and better precision in the predictions
are required to gain confidence
in both perturbative and nonperturbative calculations.
(See further discussions of the
\etab\ hyperfine splitting in 
\Secs{sec:SpecExp_etab},
\ref{sec:SpecTh_lowestspectra},
\ref{sec:SpecTh_alphasdec}, and
\ref{sec:SpecTh_mixing}).

\subsection{Standard Model parameter extractions}
\label{sec:SpecTh_param}

Given the progress made in the effective field theories formulation, 
in the calculation of high order perturbative contributions 
and in the lattice simulations, 
quarkonium  appears to be  a very suitable system for precise 
determination of Standard Model parameters like the heavy quark masses 
and $\als$. Below we report about  recent determinations.
For a review and an introduction to the procedure and  methods used in this 
section, see \cite{Brambilla:2004wf}.

\subsubsection{$\als$ determinations}
\label{sec:SpecTh_alphas}

Below we review several extractions of the strong coupling
constant related to observables in heavy quarkonium physics.
All values for $\als$ are quoted in
the $\MSbar$ scheme with $n_f=5$, unless otherwise indicated.

\subthreesection{$\als$ from quarkonia masses on the lattice}
\label{sec:SpecTh_alphaslattice}

A precise determination of $\als$ from lattice simulations has
been presented by the HPQCD Collaboration~\cite{Davies:2008sw}. 
The mass difference $m[\UnS{2}]-m[\UnS{1}]$, the 
masses $m[\UnS{1}]$, $m[\etac]$, and light meson masses
were used to tune the bare parameters. Several short-distance
quantities, mainly related to Wilson loops, were computed to obtain
the original result~\cite{Davies:2008sw} 
\beq
\label{eqn:SpecTh_alphasHPQCD}
\als (m_{Z^0})= 0.1183\pm0.0008.
\eeq
An independent implementation 
of a similar lattice-based approach,
using the same experimental inputs as~\cite{Davies:2008sw},
yields~\cite{Maltman:2008bx}
\beq
\label{eqn:SpecTh_alphasMLMS}
\als (m_{Z^0})= 0.1192\pm0.0011\,.
\eeq
The HPQCD Collaboration updated their original result
of \Eq{eqn:SpecTh_alphasHPQCD}, superseding it with~\cite{McNeile:2010ji}
\beq
\label{eqn:SpecTh_alphasHPQCDUpdate}
\als (m_{Z^0})= 0.1184\pm0.0006\,.
\eeq
This value is not independent of the result in \Eq{eqn:SpecTh_alphasMLMS}.
The two results in \Eqs{eqn:SpecTh_alphasHPQCD} and 
(\ref{eqn:SpecTh_alphasMLMS}) are expected to be
nearly identical, as they use the same inputs
and different calculations of the same theoretical effects.

  Another new determination~\cite{Allison:2008xk,McNeile:2010ji} of $\als$
uses moments of heavy-quark correlators calculated on the lattice and
continuum perturbation theory. The same references also provide a
determination of the $c$- and $b$-quark masses, described in 
\Sec{sec:SpecTh_mass}. The result is~\cite{McNeile:2010ji} 
\beq
\label{eqn:SpecTh_alphasHPQCDCKSS}
\als (m_{Z^0}) = 0.1183\pm0.0007\,.
\eeq
These extractions of $\als$, based on lattice calculations
and quarkonia masses, are the most precise individual 
determinations among the many methods and measurements 
available~\cite{Bethke:2009jm}, and will thus tend to dominate any average
over different $\als$ determinations such as that in \cite{Bethke:2009jm}.

\newpage
\subthreesection{$\als$ from quarkonium radiative decays}
\label{sec:SpecTh_alphasdec}

The CLEO~\cite{Besson:2005jv} measurement of the inclusive radiative 
decay $\Upsilon(1S)\to \gamma X$ (see \Sec{sec:Dec_Gammaglueglue}), 
together with a theoretical description of the photon 
spectrum~\cite{GarciaiTormo:2005ch}, has made
it possible to obtain a precise determination of $\als$ from
\UnS{1} decays~\cite{Brambilla:2007cz}. 
A convenient observable is
the parton-level ratio 
\beq
R_\gamma\equiv \frac{ \Gamma(V\to\gamma gg) }{ \Gamma(V\to ggg) }
\label{eqn:SpecTh_Rgamma}
\eeq
for decays of heavy vector meson $V$.
The wave function of 
$V=\UnS{1}$ at the origin and the relativistic corrections cancel at
order $v^2$ in this ratio. Furthermore, one also needs to include color-octet
contributions in the decay rates, requiring an estimation of the
color-octet NRQCD matrix elements. The two color-octet matrix elements that
appear in the numerator of $R_{\gamma}$ at order $v^2$, $O_8(^1S_0)$ and
$O_8(^3P_0)$, also appear in the denominator, which includes
$O_8(^3S_1)$, thus decreasing the theoretical uncertainty associated
with the estimation of those matrix elements in the
$\als$ extraction. 
The theoretical expression for $R_{\gamma}$ at order $v^2$ is
\beq
\label{eqn:SpecTh_Rgammath}
R_\gamma = 
\frac{36}{5}\,\frac{e_b^2\,\alpha}{\als}\,\,\frac{1+C_{\gamma
i}\mathcal{R}_i}{1+C_{i}\mathcal{R}_i}\,.
\eeq
Here $e_b$ is the $b$-quark charge, $\alpha$ is the
fine structure constant, and $C_{\gamma i}\mathcal{R}_i$ and
$C_{i}\mathcal{R}_i$ represent the order $v^2$ corrections to the
numerator and denominator, respectively. 
The $v^2$ corrections~\cite{Brambilla:2007cz} 
account for radiative, relativistic, and octet effects. 
Experimental values of $R_\gamma$ 
for $V=\UnS{1S,2S,3}$ appear in \Tab{tab:DecGammaGG_model}.
The value of $R_{\gamma}$ for \UnS{1}\ 
obtained by CLEO~\cite{Besson:2005jv} using the Garcia-Soto (GS) QCD
calculation~\cite{GarciaiTormo:2005ch} for the photon 
spectrum\footnote{A theoretical description of the
                  photon spectrum is needed to extrapolate to the
		  experimentally inaccessible, 
                  low-energy part of the spectrum.}
is
\beq
\label{eqn:SpecTh_Rgammaexp}
R_\gamma[\UnS{1}]_{\rm exp} = 0.0245 \pm 0.0001 \pm 0.0013\,,
\eeq
where the first error is statistical and the second is systematic.
The value of $\als$ obtained
from \Eq{eqn:SpecTh_Rgammath} using \Eq{eqn:SpecTh_Rgammaexp}
as input is~\cite{Brambilla:2007cz}
\beqa
\label{eqn:SpecTh_alphasBGSV}
\als(m_{\Upsilon(1S)},n_f=4) & = & 0.184^{+0.015}_{-0.014}\,,\non\\
\als(m_{Z^0}) & = & 0.119^{+0.006}_{-0.005}\,. 
\eeqa
The recent lattice~\cite{Bodwin:2005gg} and 
continuum~\cite{GarciaiTormo:2004jw} estimates of the octet matrix
elements are used to obtain this result. The experimental systematic
uncertainty from \Eq{eqn:SpecTh_Rgammaexp} dominates the error in
$\als$ shown in \Eq{eqn:SpecTh_alphasBGSV}.

There are also CLEO measurements of $R_\gamma[\jpsi]$~\cite{Besson:2008pr} 
and $R_\gamma[\psip]$~\cite{Libby:2009qb} (see \Sec{sec:Dec_Gammaglueglue}
and \Tab{tab:DecGammaGG}). One could, in principle, extract
$\als$ in the same way
as for \UnS{1} above. However, the relativistic
and octet corrections are more severe than for the 
\UnS{1}, so terms of higher order than $v^2$ may not be small enough
to ignore. The effects due to the proximity of \psip\ to
open-charm threshold are difficult to estimate~\cite{GarciaiTormo:2007qs}.

\subthreesection{$\als$ from bottomonium hyperfine splitting}
\label{sec:SpecTh_alphasHFS}

The observation of \etab\ by \babar~\cite{:2008vj,:2009pz}
and CLEO~\cite{Bonvicini:2009hs} allows
$\als$ to be extracted from the singlet-triplet
hyperfine mass splitting, 
\beq
\dmhf[\etab]\equiv m[\UnS{1}] - m[\etab]\,.
\label{eqn:SpecTh_dmhf_etab}
\eeq 
The theoretical expression used for the
hyperfine splitting includes a perturbative component and a
non-perturbative one.
The perturbative component is given by the expression 
of~\cite{Kniehl:2003ap}, which includes order $\als$ corrections
to the leading order (LO) term, 
\beq
\dmhf[\etab]_{\rm LO}=\frac{C_F^4\,\als^4\,m_b}{3}\,,
\eeq
and resummation of the logarithmically-enhanced corrections up to
the subleading logarithms (which
are of the form $\alpha_s^n\ln^{n-1}\alpha_s$). The non-perturbative
part is parametrized in terms of the dimension-four gluon condensate,
which is fixed according to \cite{Colangelo:2000dp}. 

Using the average experimental value in \Tab{tab:Spec_etab}
and \Eq{eqn:SpecExp_HFSexp}
the resulting value of $\als$ is \cite{Colangelo:2009pb}
\beq
\label{eqn:SpecTh_alphasCSS}
\als (m_{Z^0})  =  0.125\pm0.001\pm0.001\pm0.001\,,
\eeq
where the first error is experimental, the second is associated with
the gluon condensate and the third accounts for the $b$-quark
mass uncertainty. 
This $\als$ value is slightly more than $3\sigma$ higher
than the updated HPQCD~\cite{McNeile:2010ji} 
lattice result in \Eq{eqn:SpecTh_alphasHPQCDUpdate}.

\subsubsection{Determinations of $m_b$ and $m_c$}
\label{sec:SpecTh_mass}

Below we review recent extractions of the heavy quark masses
related to observables in heavy quarkonium physics.

\subthreesection{Sum rules}
\label{sec:SpecTh_masssr}

\subfoursection{Moments}

The determination of the heavy quark masses from a sum-rule analysis
requires theoretical predictions for the $n^{\rm th}$ moments of the cross
section for heavy-quark production in \epem\ collisions. The
theoretical expression for the moments is related to derivatives of
the vacuum polarization function at $q^2=0$. 
Four-loop [$\mathcal{O}(\als^3)$] results for the vacuum
polarization function have appeared for the first 
moment~\cite{Chetyrkin:2006xg,Boughezal:2006px,Kniehl:2006bf},
second moment~\cite{Maier:2008he}, third 
moment~\cite{Maier:2009fz}, and approximate results for higher 
moments~\cite{Hoang:2008qy,Kiyo:2009gb}. All those four-loop results are 
used in the most recent low-momentum sum-rule determinations of the 
heavy-quark masses reported below.

\subfoursection{Low-n sum rules}

The most recent determination~\cite{Kuhn:2007vp,Chetyrkin:2009fv} 
of the $c$- and $b$-quark masses using low-momentum sum rules
incorporates four-loop
results for the derivatives of the vacuum polarization function along
with the most recent experimental data. 
The results are~\cite{Chetyrkin:2009fv}
\beqa
\label{eqn:SpecTh_mcCKMMMSS}
m_c^{\MSbar}(3\gev) & = & 0.986\pm 0.013\gev,\\
m_c^{\MSbar}(m_c^{\MSbar}) & = & 1.279\pm 0.013\gev,
\eeqa
and
\beqa
\label{eqn:SpecTh_mbCKMMMSS}
m_b^{\MSbar}(10\gev) & = & 3.610\pm 0.016\gev,\\
m_b^{\MSbar}(m_b^{\MSbar}) & = & 4.163\pm 0.016\gev.
\eeqa

For a critical discussion of the error attached to these determinations
and  a new (preliminary) mass determination using low-momentum sum rules
see \cite{hoangtalk}.

\subfoursection{Large-n sum rules}

A determination of the $b$-quark mass using 
nonrelativistic (large-$n$) sum rules,
including resummation of logarithms, has been
performed~\cite{Pineda:2006gx}. It incorporates next-to-next-to
leading order results along with the complete next-to-leading
logarithm resummation (and partial next-to-next-to-leading logarithm
resummation). Including logarithm resummation
improves the reliability of the theoretical computation. 
The value of the $\MSbar$ mass is \cite{Pineda:2006gx}
\beq
\label{eqn:SpecTh_mbPS}
m_b^{\MSbar}(m_b^{\MSbar})=4.19\pm0.06\gev.
\eeq
The $c$-quark mass has also been determined from a 
nonrelativistic sum-rules analysis~\cite{Signer:2008da}, with the result
\beq
\label{eqn:SpecTh_mcS}
m_c^{\MSbar}(m_c^{\MSbar})=1.25\pm0.04\gev.
\eeq

\subfoursection{Alternative approaches}

A determination of the $c$- and $b$-quark masses which uses
moments at $q^2\neq 0$ and includes the dimension-six gluon condensate
(also determined from the sum rules) has been 
reported~\cite{Narison:2010cg}:
\beqa
\label{eqn:SpecTh_mcmbN}
m_c^{\MSbar}(m_c^{\MSbar}) & = & 1.260\pm 0.018\gev,\\
m_b^{\MSbar}(m_b^{\MSbar}) & = & 4.220\pm 0.017\gev,
\eeqa
which employ an estimate of the four-loop contribution to the 
$q^2\neq 0$ moments.

\subthreesection{Quark masses from the lattice}
\label{sec:SpecTh_masslatt}

A determination
of the $b$-quark mass in full (unquenched) lattice QCD
using one-loop matching to
continuum QCD, finds~\cite{Gray:2005ur}
\beq
\label{eqn:SpecTh_mbHPUKQCD}
m_b^{\MSbar}(m_b^{\MSbar})=4.4\pm 0.3\gev\,.
\eeq
The $c$-quark mass was calculated by comparing lattice
determinations of moments of heavy-quark correlators to four-loop
continuum perturbation theory~\cite{Allison:2008xk,McNeile:2010ji}.
A $b$-quark mass calculation is also included in~\cite{McNeile:2010ji}. 
Due to the use of continuum, rather than lattice, perturbation
theory, a higher-order perturbative calculation can be used,
achieving very precise results~\cite{McNeile:2010ji}:
\beqa
\label{eqn:SpecTh_mcHPQCDCKSS}
m_c^{\MSbar}(3\gev) & = & 0.986\pm 0.006\gev,\\
m_c^{\MSbar}(m_c^{\MSbar}) & = & 1.273\pm 0.006\gev\,,
\eeqa 
and
\beqa
\label{eqn:SpecTh_mbHPQCDCKSS}
m_b^{\MSbar}(10\gev) & = & 3.617\pm 0.025\gev,\\
m_b^{\MSbar}(m_b^{\MSbar}) & = & 4.164\pm 0.023\gev\,.
\eeqa 

\subsubsection{$m_t$ determination}
\label{sec:SpecTh_topmass}

Determination of the top-quark mass $m_t$ at a
future $e^+e^-$ linear collider from a $t\bar{t}$ 
line-shape measurement (see \Sec{sec:Fut:ILC})
requires good theoretical knowledge of the
total $t\bar{t}$ production cross section in the threshold region. 
The threshold regime is characterized by $\als\sim v$. The
N$^k$LO result includes corrections of order $\als^nv^m$ with 
$n + m = k$.
The N$^2$LO result has been known
for some time now. Several contributions to the
N$^3$LO result have been calculated. Those include corrections to the
Green functions and wave function at the 
origin~\cite{Penin:2005eu,Beneke:2005hg,Beneke:2007gj,Beneke:2007pj,Beneke:2008cr};
matching coefficients of the effective theory 
currents~\cite{Marquard:2006qi,Marquard:2009bj};
electroweak effects in 
NRQCD~\cite{Eiras:2006xm,Kiyo:2008mh}; 
and corrections to the static potential (see \Sec{sec:SpecTh_stpo}).

Renormalization-group-improved expressions, which sum terms of the
type $\als\ln v$,
are also necessary to reduce the normalization uncertainties of the
cross section and improve the reliability of the calculation. Those
resummations, which were originally only done in the framework of
velocity NRQCD, have now been calculated within
pNRQCD~\cite{Pineda:2006ri}. The
terms at next-to-next-to-leading logarithmic accuracy are not
yet completely known~\cite{Hoang:2006ht}.
Consistent inclusion of all effects related to the instability
of the top quark is needed.
Some recent progress in this direction 
has been made~\cite{Hoang:2004tg,Hoang:2010gu,Beneke:2010mp}.
It is expected that a linear collider will provide an $m_t$
determination with uncertainties at the level of 100\mev\ 
(see \Sec{sec:Fut:ILC}). For comparison, the
Tevatron Electroweak Working Group has reported a best-current value of
$m_t=173.1\pm 1.3$\gev~\cite{TeV:2009ec}.

At the LHC, top quarks will be produced copiously.
It has been pointed out \cite{Fadin:1990wx}
that in the threshold region of
$t\bar{t}$ production at LHC, a significant amount of 
the (remnant of) color-singlet $t\bar{t}$ resonance states
will be produced, unlike at the Tevatron where the color-octet
$t\bar{t}$ states dominate. In fact, there appears the $1S$
peak in the $t\bar{t}$ invariant mass distribution
below $t\bar{t}$ threshold,
even after including the effects of initial-state radiation
and parton-distribution function,
and the position of this $1S$ peak is almost
the same as that in $e^+e^-$ collisions~\cite{Hagiwara:2008df,Kiyo:2008bv}.
Namely, theoretically there is a possibility of extracting the top quark
mass with high accuracy from this peak position, 
although experimentally it
is quite challenging to reconstruct the $t\bar{t}$
invariant mass with high accuracy.

Recently, a theoretical framework to compute the
fully-differential cross sections for top quark
production and its subsequent decays
at hadron colliders has been developed, incorporating the
bound-state effects which are important in the $t\bar{t}$
threshold region~\cite{Sumino:2010bv}.
A Monte Carlo event generator for LHC has been developed and
various kinematical distributions of the
decay products of top quarks have been computed.
In particular, it was found that a bound-state effect
deforms the $(bW^+)$-$(\bar{b}W^-)$ double-invariant-mass
distribution in a correlated manner, which
can be important in the top event reconstruction.

\subsection{Exotic states and states near or above threshold}

For states {\it away} from threshold, it has been shown that
appropriate EFTs to describe the quarkonium spectrum
can be constructed. In particular, in pNRQCD,
the relevant degrees of freedoms are clearly identified:
the leading order description coincides with the 
Schr\"odinger equation, the potentials
are the pNRQCD matching coefficients, and the energy 
levels are calculable in a well-defined procedure. 
{\it Close} to threshold, the situation 
changes drastically~\cite{Vairo:2006pc,Brambilla:2008zz}.
As described earlier in this section, 
the region close to and just above threshold is presently 
the most interesting, with a wealth of newly discovered states. 
Most new states do not fit potential-model expectations. 
This is to be expected, as we have seen that a potential
model description of quarkonium (strongly-coupled pNRQCD) 
emerges only for binding energies smaller than $\lamQ$.
Since the open heavy-flavor threshold is 
at the scale $\lamQ$ in HQET, a potential model 
description of states above that threshold cannot provide a 
reasonable approach. On the other hand, from a QCD point of view, 
a plethora of new states 
are expected. NRQCD is still a good EFT for states close to and 
just above threshold, at least 
when their binding energies remain much 
smaller than the heavy-flavor mass. 
The heavy quarks move slowly in these states, 
and the static limit should remain a good starting point.

Below we examine how things change close to threshold 
and  the new degrees of freedom that emerge.
First, \Sec{sec:SpecTh_gluon} considers the case in which 
there are only quarkonium and gluonic excitations. Away from threshold, the 
gluonic excitations have been integrated out to obtain strongly-coupled pNRQCD 
and the QCD nonperturbative potentials are the pNRQCD matching coefficients.
Close to threshold the gluonic excitations no longer develop a 
gap with respect to quarkonium and they have to be 
considered as dynamical degrees of freedom.
Next, \Sec{sec:SpecTh_lightq} considers the situation with 
dynamical ultrasoft light quarks and we discuss all the new degrees 
of freedom that may be generated. No QCD based theory description is yet 
possible in this situation, apart from systems like the $X(3872)$ that 
display universal characteristics and may be treated with EFTs methods.
Models for the description of states close to threshold just pick up 
some of the possible degrees of freedom and attribute to them  some 
phenomenological interaction.  These models will be described and 
their predictions contrasted.
Third, \Sec{sec:SpecTh_sum} will summarize the predictions of  sum rules,
a method that allows calculation of the masses of the states once an 
assumption on the operator content is made.
Lastly, all the new unconventional states
will be summarized along with possible interpretations
in \Sec{sec:SpecTh_sumnewstates}.

\subsubsection{Gluonic excitations}
\label{sec:SpecTh_gluon}

First, consider the case without light quarks. 
Here the degrees of freedom are heavy quarkonium, hybrids 
and glueballs. In the static limit, at and above the $\lamQ$ 
threshold, a tower of hybrid static 
energies (\ie of gluonic excitations) must be considered on top of 
the $Q\overline{Q}$ static singlet energy~\cite{Horn:1977rq,Hasenfratz:1980jv}.
The spectrum has been thoroughly studied on the lattice~\cite{Juge:2002br}.
At short distances, it is well described by the Coulomb potential
in the color-singlet or in the color-octet configurations.
At short distances, the spectrum of the hybrid static energies 
is described in the leading multipole expansion of pNRQCD by 
the octet potential plus a mass scale, which is called {\it gluelump mass}
\cite{Brambilla:1999xf,Bali:2003jq}.
At large distances the energies rise linearly in $r$.
The first hybrid excitation plays the role of the 
open heavy-flavor threshold, which does not exist in this case. 
If the Born-Oppenheimer approximation is viable, many 
states built on each of the hybrid potentials are expected. 
Some of these states may develop a width if decays to 
lower states with glueball emission (such as 
hybrid$\to$glueball~+~quarkonium) are allowed. 
The states built on the static potential (ground state) 
are the usual heavy quarkonium states. 

\begin{figure}[b]
   \begin{center}
      \includegraphics*[width=\figwid]{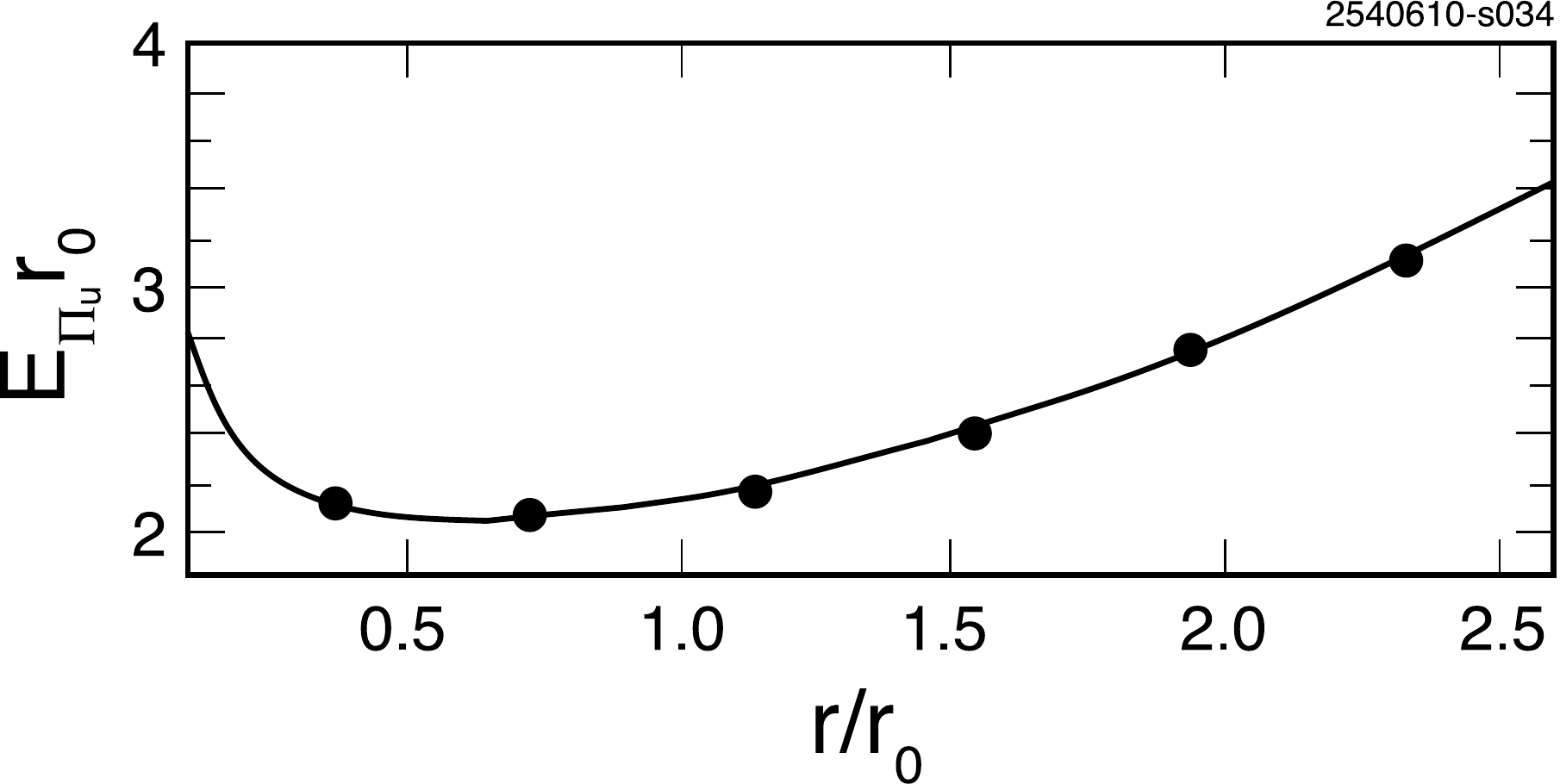}
      \caption{The hybrid static potential $\Pi_u$ at short and intermediate
               distances, $r_0 \approx 0.5$~fm. The {\it solid circles} are the
               lattice data\cite{Juge:2002br} and the {\it smooth curve}
               traces out \Eq{eqn:SpecTh_epiurzero} }
      \label{fig:SpecTh_figPiu}
   \end{center}
\end{figure}

Consider, for example, the $Y(4260)$, for which 
many interpretations have been proposed, including a
charmonium hybrid~\cite{Zhu:2005hp,Kou:2005gt,Close:2005iz}.
If the $Y(4260)$ is interpreted as a charmonium hybrid, 
one may rely on the heavy-quark expansion 
and on lattice calculations to study its properties. Decays into 
$D^{(*)} \bar{D}^{(*)}$ should be suppressed since 
they are forbidden at leading order in the heavy-quark 
expansion~\cite{Kou:2005gt} (see also \cite{Chiladze:1998ti}).
This is in agreement with the upper limit on $Y \to D \bar{D}$ 
reported by \babar\ (see \Tab{tab:Spec_Y4260_opencharm}).
The quantum numbers of the $Y(4260)$ are consistent with those of a
pseudoscalar $0^{-+}$ fluctuation $ |\phi\rangle$, belonging 
to the family of $\hqm v^2$ 
fluctuations around the gluonic excitation between a 
static quark and a static antiquark,
with quantum numbers $1^{+-}$, also known as $\Pi_u$,  
\beq
|Y\rangle  = |\Pi_u\rangle  \otimes  |\phi\rangle\,.
\eeq
It is suggestive that, according to lattice calculations~\cite{Juge:2002br}, 
$\Pi_u$ is the lowest gluonic excitation between a static quark and a static 
antiquark above the quark-antiquark color singlet. 
$|\phi\rangle$ is a solution to the Schr\"odinger equation 
with a potential that is the static energy of $\Pi_u$.
Fitting the static energy of $\Pi_u$ at short and 
intermediate distances, one finds
\beq
E_{\Pi_u}r_0 = \hbox{constant} + 0.11\, \frac{r_0}{r}  + 
0.24\, \left( \frac{r}{r_0}\right)^2\,,
\label{eqn:SpecTh_epiurzero}
\eeq
as illustrated in \Fig{fig:SpecTh_figPiu}.
Solving the corresponding Schr\"odinger equation, 
\beq
m_Y  = (2\times 1.48 + 0.87 + 0.53)\gev = 4.36\gev\,, 
\eeq
where 1.48\gev\ is the charm mass in the RS scheme~\cite{Pineda:2001zq} 
and 0.87\gev\ is the  $\Pi_u$
gluelump mass in the same scheme~\cite{Bali:2003jq}. 

\subsubsection{Spectrum with light quarks}
\label{sec:SpecTh_lightq}
Once light fermions have been incorporated into the spectrum,
new gauge-invariant states appear beside the heavy quarkonia,
hybrids, and glueballs. On
the one hand, we have the states with no heavy quark content. 
Due to chiral symmetry, there is a mass gap, of O($\Lambda_\chi$),
between the Goldstone bosons, which are massless in the
chiral limit, and the rest of the spectrum. 
The Goldstone bosons are considered as
ultrasoft degrees of freedom and $\Lambda_\chi \sim \Lambda_{\rm QCD}$, 
so that away from threshold the rest of the spectrum should
be integrated out. 
Besides these, there are also bound states made of one heavy quark and
light quarks, \ie the $\overline{Q}q–-Q\overline{q}$ system.
 The energy of this system
is, according to the HQET counting rules,
$m_{\overline{Q}q} + m_{Q\overline{q}} = 2m + 2 \overline{\Lambda}$.
Therefore, 
since the heavy-light binding 
energy $\overline{\Lambda} \sim \Lambda_{\rm QCD}$,
away from threshold these states have to be integrated out.
Close to threshold the situation is different.
In this case, there is
no mass gap between the heavy quarkonium and the creation of a 
$\overline{Q}q–-Q\overline{q}$  pair. Thus, 
for study of heavy quarkonium near threshold, these degrees of
freedom must be included in the spectrum, even if  
the mixing between the heavy quarkonium and the $\overline{Q}q–-Q\overline{q}$ 
is expected to be small, being suppressed in the large Nc counting.
Summarizing, light fermions contribute within this picture in three ways:
\begin{itemize}
\item {\it Hard light fermions,} 
which are encoded into the matching coefficients of the NRQCD Lagrangian
and obtained from the computation of perturbative Feynman diagrams at the scale
$\hqm$.
\item {\it Soft light fermions,} a term that denotes, in a generic way, all 
the fermions that are incorporated in the potentials. It is expected that 
their main effects can be simulated by a
variation of the value of the parameters in the potentials. They can be 
evaluated nonperturbatively via unquenched lattice calculation of the 
potentials.
\item {\it Ultrasoft light fermions,} which 
are the ones that will become pions and, since they are
also ultrasoft degrees of freedom, they 
should be incorporated in the effective Lagrangian
together with the heavy quarkonium. 
\end{itemize}

So the general picture is as follows:
The inclusion of light quarks
does not remove any states predicted
in the no-light-quarks scenario, but
the availability of decays via pion
emission does increase the width
of each such state in the spectrum.
Moreover, in addition to the regular quarkonium states, 
new states built using the light-quark quantum numbers may form. 
States made of two heavy and light quarks include 
those built on pairs of heavy-light mesons 
($D\bar{D}$, $B\bar{B}$, ...), like hadronic
molecular states~\cite{Tornqvist:1991ks,Swanson:2006st}; 
molecular states composed of   
the usual quarkonium states (built on the static potential); and 
light hadrons (hadro-quarkonium \cite{Dubynskiy:2008mq} ); 
pairs of heavy-light baryons \cite{Qiao:2005av}; 
tetraquark states~\cite{Jaffe:1976ig};
and likely many others.
It would be particularly interesting to have the 
spectrum of tetraquark potentials, 
or at least their ground states, from lattice QCD, since a tetraquark
interpretation of some of the newly discovered states 
has been advocated~\cite{Maiani:2004vq,Ebert:2005nc}
(see \Sec{sec:SpecTh_Tetraquarks}). If, again, 
the Born-Oppenheimer approximation is a reasonable 
approach, many states built on each of the tetraquark 
potentials may be expected, many developing (large) widths due 
to decays by emission of a pion (or other light hadron). 

How these different kinds of states 
``talk to each other'' is an important issue~\cite{Kalashnikova:2005ui}. 
Results on crosstalk of the static potential with a pair of 
heavy-light mesons on the lattice
have recently been reported (see \Sec{sec:SpecTh_latspectra}).
This explains why, from the QCD point of view,
so many states of a new nature appear
in this region of the spectrum. However,
a systematic QCD description of these states 
has not yet been developed.
For the time being, models are developed in order
to obtain more detailed information on these systems. 
Exceptional cases, \eg\ those for which the state is extremely close 
to a threshold (\eg $X(3872)$), allow for an 
effective field theory treatment~\cite{Braaten:2003he}
(see also \Sec{sec:SpecTh_molec}).

Results from some of the above-mentioned models follow:
There are differences among models 
involving four-quark fields, two heavy and two 
light. Given four-quark fields of the type, \eg
$c\bar{c}q\bar{q}$ (where $q$ represents a generic light quark), 
three quark-pair configurations are possible.
All of them have been exploited in the literature. However,
the resulting models are not equivalent
because different dynamics are attributed to different configurations.
Due to the absence of further theoretical input from QCD, many tetraquark 
studies rely on phenomenological models of the tetraquark interaction unless 
some special hierarchy of dynamical scales may be further exploited on the 
top of the nonrelativistic and perturbative expansions discussed so far.
In \cite{Hogaasen:2005jv,Buccella:2006fn}, it is assumed that 
\beq
X \sim (c\bar{c})^8_{S=1}\,   (q\bar{q})^8_{S=1}\,, 
\eeq
\ie that the dominant Fock-space component of the $X(3872)$ 
contains a $c\bar{c}$ pair and a $q\bar{q}$ pair
in a color-octet configuration with spin 1. 
Calculations have been based on a phenomenological interaction Hamiltonian. 
In \cite{Maiani:2004vq}, it is assumed that 
\beq
X \sim (cq)^{\bar{3}}_{S=1}\,  (\bar{c}\bar{q})^{3}_{S=0}\,
+ (cq)^{\bar{3}}_{S=0}\,   (\bar{c}\bar{q})^{3}_{S=1}\,.
\label{eqn:SpecTh_colortripletdiquark}
\eeq 
Here the clustering of quark pairs in tightly-bound color-triplet diquarks 
is not induced by a scale separation as it would happen in baryons with two 
heavy quarks~\cite{Brambilla:2005yk}, but is a dynamical assumption 
of the model. 
In~\cite{Tornqvist:1993ng,Tornqvist:2003na,Swanson:2003tb,Swanson:2004pp}, 
it is assumed that
\beqa
X &\sim& (c\bar{q})^1_{S=0}\,   (q\bar{c})^1_{S=1}\, 
+ (c\bar{q})^1_{S=1}\,   (q\bar{c})^1_{S=0}\non\\
&\sim& D^0\,\bar{D}^{*\,0} +
D^{*\,0}\, \bar{D}^0\,,
\eeqa
\ie that the dominant Fock-space component of the 
$X(3872)$ is a $D^0\, \bar{D}^{*\,0}$ and 
$D^{*\,0} \, {\bar D}^0$ molecule. Small short-range components 
of the type 
\beq
(c\bar{c})^1_{S=1}\, (q\bar{q})^1_{S=1}\,\simeq\,\jpsi \,\rho\,({\rm or~}\omega)
\eeq
are included as well. Predictions depend on the phenomenological
Hamiltonian, which typically contains short-range ($\sim 1/\lamQ$), 
potential-type interactions among the quarks and  
long-range ($\sim 1/m_\pi$) one-pion exchange.
In~\cite{Pakvasa:2003ea,Voloshin:2003nt,Voloshin:2004mh,Braaten:2003he}, 
it is assumed not only that the $X(3872)$ is a $D^0\bar{D}^{*0}$ and  
${\bar D}^0D^{*0}$ molecule but also that it is loosely~bound, 
\ie that the following hierarchy of scales is realized:
\beq
\lamQ \gg m_\pi\,\gg\, \frac{m_\pi^2}{m_{D^0}}\,\approx\, 
10\mev\,\gg\,\Ebind\,. 
\eeq
Indeed, the binding energy, \Ebind, which 
may be estimated from $m_X - (m_{D^{*0}}+m_{D^{0}})$,
is, as \Tab{tab:Spec_DeltaMX} shows, 
very close to zero, \ie much smaller than the natural scale 
$m_\pi^2/m_{D_0}$. Systems with a short-range interaction 
and a long scattering length have universal 
properties that may be exploited: 
in particular, production and decay 
amplitudes factorize into short-range 
and long-range component, where the latter 
depends only on a single parameter, 
the scattering length (see \Sec{sec:SpecTh_molec}). 

\subsubsection{Molecular states} 
\label{sec:SpecTh_molec}

\subthreesection{Loosely bound hadronic molecules}

The $X(3872)$ resonance (see \Sec{sec:SpecExpX3872}) 
cannot be easily explained as a standard charmonium 
excitation~\cite{Quigg:2004vf}. The close proximity 
of its mass to \DstnDn\ threshold suggested 
that it could be a good example of a hadronic molecule 
with $J^{PC}=1^{++}$ quantum numbers. 
A \DstnDn\ molecule would be 
characterized by an extremely small binding energy, 
as small as $\Ebind\approx 0.1$\mev. Indeed, the $B\to K X$ Belle 
production mechanism allows the formation of a 
nearly-at-rest \DstnDn\ system which could be very weakly bound.  
Several other bound states with similar 
properties have also been discovered.

The threshold proximity of many of the new states 
implies that, regardless of the binding mechanism, there 
should be a significant component of a hadronic molecule 
in the wave function of the state. The small
binding energy of this molecular component indicates 
that the \DstnDn\ scattering length, $a$, is
unnaturally large. This leads to some simplifications 
in the description of the properties of the molecule,
as its binding energy, \Ebind, and wave function are in fact 
largely determined by $a$, a phenomenon known as low-energy
{\it universality}. In the limit of a very shallow bound 
state, the scattering length is
\beq
a = \frac{1}{\sqrt{2\mred\Ebind\,}}\,,
\label{eqn:SpecTh_mola}
\eeq
where
\beq
\mred = \frac{m_{D^0}\, m_{D^{*0}}}{m_{D^0} + m_{D^{*0}}}
\eeq
is the reduced mass. 
Clearly, the scattering length $a \simeq 10$~fm is much 
larger than the natural length scale $R$ for a molecular 
state bound by pion exchange, $R\simeq 1/m_\pi = 1.5$~fm. 
There is a universal prediction for a wave function of the 
$S$-wave molecular state,
\beq
\psi(r)_{\rm mol} =  \frac{1}{\sqrt{2 \pi a}}\, \frac{e^{-\frac{r}{a}}}{r} \, ,
\label{eqn:SpecTh_molwf}
\eeq
where $r$ is the separation of the constituents
in the molecule's rest frame, 
which is correct up to terms of order $R/a$.
A proper investigation of near-threshold resonances 
is needed to address the appearance of this new length 
scale, the possibility of distinguishing between an 
``elementary'' particle ($q \bar{q}$, hybrid, 
or compact tetraquark) and a composite state 
(hadronic molecule), and how to estimate the admixture of the composite. 
It was suggested by 
Weinberg~\cite{Weinberg:1962hj,Weinberg:1963zz,Weinberg:1965zz} 
that the admixture fraction can be determined
model-independently for such near-threshold bound states. 
In this scheme, the admixture of a nonmolecular
component is parametrized in terms of a single parameter, 
$0\le \xi^2\le 1$, which measures the probability of finding 
the molecular component in the physical wave function of the 
state of interest. For $\xi < 1$, \Eq{eqn:SpecTh_molwf} becomes
\beq
\psi(r)_{\rm mol} =  \frac{\xi^2}{\sqrt{2 \pi a}}\, 
\frac{e^{-\frac{r}{a}}}{r} \,.
\label{eqn:SpecTh_molwf_mod}
\eeq
Accordingly, the expression for the scattering length, \Eq{eqn:SpecTh_mola},
is then
\beq
a = \frac{2\xi^2}{1+\xi^2}\,\,\frac{1}{\sqrt{2 \mred \Ebind}} \, .
\label{eqn:SpecTh_mola_mod}
\eeq
The expression acquires corrections of order $R$.
A simultaneous measurement of both binding
energy and scattering length can extract
the value of the parameter $\xi$, and
the nature of the state becomes an observable.  When this
formalism~\cite{Weinberg:1962hj,Weinberg:1963zz,Weinberg:1965zz} 
was applied to the deuteron, it was shown that, indeed, the
deuteron is a proton-neutron molecule.
The method just described can only be applied if
the particles forming the molecule are in a relative $S$-wave 
and if the state studied is sufficiently close to threshold,
\ie if $k \simeq \sqrt{2 \mred \Ebind}$ is the 
smallest momentum scale in the problem. The approach was 
generalized~\cite{Baru:2003qq,Baru:2004xg,Gamermann:2009uq} to include
inelastic channels, as well as an above-threshold resonance. It is stressed 
in~\cite{Baru:2003qq,Baru:2004xg} that the relevant quantity to be
studied is the effective coupling constant squared, or, 
equivalently, the residue at the bound-state pole that 
parametrizes the coupling strength of this state to
the relevant continuum channel, which can be shown to be
proportional to $\xi^2$ with a known factor of 
proportionality.  Thus this coupling constant, 
which is an observable, measures the
amount of molecular admixture in the sense defined 
in~\cite{Weinberg:1962hj,Weinberg:1963zz,Weinberg:1965zz}.  
A related approach uses pole-counting~\cite{Morgan:1992ge}, 
which studies the structure of the near-threshold singularities 
of the scattering amplitude. It
appears that the state is mostly elementary if there are two nearby poles in
the scattering amplitude, whereas composite particle corresponds to a single,
near-threshold pole. While these methods provide a diagnostic tool for
identifying near-threshold
molecular states, they do not provide information on the binding mechanism.
Some of these states might be interpreted as hadrocharmonia, discussed
in detail in \Sec{sec:SpecTh_hadrocharmonium}.

The analysis sketched above has been applied to various
states. Evidence supporting the identification of $Y(4660)$ as a
$\psip f_0(980)$~\cite{Guo:2008zg} bound system has been found. 
In addition, it has been proposed that
$X(3872)$~\cite{Voloshin:1976ap,De Rujula:1976qd,Tornqvist:1991ks,Swanson:2006st} 
is a \DstnDn\ bound system, and that $D_s(2317)$ and 
$D_s(2460)$~\cite{vanBeveren:2003kd,Kolomeitsev:2003ac,Browder:2003fk,Guo:2008gp} 
are bound states of $KD$ and $KD^*$, respectively. 
One may use unitarization schemes to investigate corresponding states 
{\it not} located near thresholds; \eg the $Y(4260)$ was 
suggested to be a $\jpsi f_0(980)$ bound
system~\cite{MartinezTorres:2009xb}, 
which would make it a close relative of $Y(4660)$. 

\subthreesection{Molecules and effective field theories}

 Effective field theory (EFT) techniques can be
used to study the dynamical properties of a threshold molecular state
independent of any particular model.
This is possible due to the multitude of scales present 
in QCD. The small binding energy
suggests that this state can play the role of the deuteron
in meson-antimeson interactions. Thus methods similar 
to those developed for the deuteron may be employed, with the added 
benefit of heavy-quark symmetry. A suitable effective 
Lagrangian describing the $X(3872)$
contains only heavy-meson degrees of freedom with 
interactions approximated by 
local four-boson terms constrained only by the 
symmetries of the theory. While the
predictive power of this approach is somewhat 
limited, several model-independent 
statements can be made. For instance, 
existence of \DstnDn\ molecule does not necessarily 
imply existence of $D^{*0}\bar{D}^{*0}$ 
or $D^{0}\bar D$ molecular states~\cite{AlFiky:2005jd}. 
First steps towards the development of a systematic 
EFT for the $X(3872)$ have been
taken~\cite{AlFiky:2005jd,Fleming:2007rp,Fleming:2008yn,Fleming:2009kp}.

Effective field theories can be used to study 
formation~\cite{Braaten:2006sy} and decays~\cite{Braaten:2005jj} 
of $X(3872)$ and other molecular states. 
In particular, it can be employed 
to study lineshapes~\cite{Braaten:2007xw,Braaten:2007dw}.
Those studies reveal that the spectral shape of the resonances
located near thresholds is the relevant observable. This is a
direct consequence of the importance of the effective coupling
constant for the nature of the state. In the case of the $Y(4660)$,
the spectrum shows a visible deviation from a symmetric
distribution~\cite{Guo:2008zg}, 
which, in the molecular picture, can be traced to the
increasing phase space available for the $\psip f_0(980)$ system. 
(Alternatively, the asymmetry might originate from interference of 
the resonance signal with that of the lower-lying 
$Y(4360)$~\cite{Cotugno:2009ys} --- see also
the discussion in \Sec{sec:SpecTh_Tetraquarks}.)
Only the mass of the $Y$ and the overall
normalization were left as free parameters in 
a fit to the experimental mass spectrum; the width can be calculated from the 
effective coupling constant under the assumption that the
$Y(4660)$ is indeed a $\psip f_0(980)$ bound system. 
This fit~\cite{Guo:2008zg} gives a
mass of $m_Y=(4665^{+3}_{-5})$\mev. From this fit, the
effective coupling constant
was found to be in the range 11-14\gev.
This, in turn, allowed a prediction of the width of 
$\Gamma_Y=(60\pm30)$\mev. 
The current quality of data allows for additional
decay channels, \eg $\lala$. 
To double check that the analysis is sensible, a
second fit to the experimental mass spectrum
is performed in which the effective coupling is allowed to float 
in addition to $m_Y$ and the overall normalization. 
This second fit calls for a coupling constant of 13\gev.
This result is interpreted in~\cite{Guo:2008zg} as 
strong evidence in favor of a
molecular interpretation for the $Y(4660)$.  Under this interpretation,
employing heavy-spin symmetry allows one to predict a
close relative, $Y_\eta$, 
to the $Y(4660)$, namely a bound state of $\etacp$ and
$f_0(980)$~\cite{Guo:2009id}. The mass difference between this state and
$Y(4660)$ is predicted to match that between $\psip$ and $\etacp$ up to
corrections of order $(\Lambda_{\rm QCD}/m_c)^2$, which gives
$m_{Y_\eta}= 4616^{+5}_{-6}$\mev. The width and spectral shape in the
$\etac\pi\pi$ channel are predicted to be equal to those of the
$Y(4660)$, respectively.
Further systematic studies are necessary to put these 
conjectures on firmer ground. This model-independent scheme has been
extended to states with one unstable constituent~\cite{Hanhart:2010wh}.  
Thus the method can now be applied to many more states in the spectrum.  
Measurements with higher statistics are needed to test these
predictions. For example,
an improved spectral shape measurement for $Y(4660)$ could determine
if the large predicted coupling to $\psip f_0(980)$ is present. 
Another test is a search for the decay 
$B^+\to K^+Y_\eta$, which has been estimated~\cite{Guo:2009id} to
have a branching fraction of $\sim 10^{-3}$. After
accounting for the possibility of final state interactions
in this picture, it has also been proposed~\cite{Guo:2010tk} 
that the $X(4630)$, which decays to $\lala$,
could be the same state as the $\psip f_0(980)$ molecule 
$Y(4660)$, which could be tested
with measurements of $B^+\to K^+\lala$ decays.

\subthreesection{$X(3872)$ as a \DstnDn\ molecule}

Analyses of published $X(3872)$ data in both the
\DstnDn\ and $\dipi\jpsi$ channels have shed light
on the nature of the $X$.  As for the
$Y(4660)$, the lineshape contains the important
information.  One approach~\cite{Hanhart:2007yq},
using then-existing data, concluded that the
$X(3872)$ is indeed generated by nonperturbative \DstnDn\
interactions, which are not sufficiently strong to form a
bound state, but only to produce a virtual state very close to 
\DstnDn\ threshold. A different approach~\cite{Braaten:2007xw} stressed that, 
if the $X(3872)$ is a bound state, there will be a resonant 
$D^0 \bar D^0 \pi^0$ peak {\it below} the \DstnDn\ threshold
attributable to the nonzero width of the $D^{*0}$. 
This latter approach, also using the initial lineshape measurements, 
identified the $X$ as a molecule, although a virtual state was not excluded.
Later, using the same formalism as~\cite{Hanhart:2007yq}
and additional data that had become available,
a fit to the measured lineshape found~\cite{Kalashnikova:2009gt}
that a significant admixture of a compact component inside the $X$
wavefunction is required. 
As discussed in \Sec{sec:SpecExpX3872}
and illustrated in \Fig{fig:Spec_X3872Mass_braaten8},
it was pointed out~\cite{Stapleton:2009ey} 
that the experimental lineshapes for these $D^0 \bar D^0 \pi^0$ events 
had been generated by constraining the measured particle momenta so that the
$D^0\piz$ (or $\bar D^0\piz$) candidates would have the \Dstn\ mass.
If the $X(3872)$ were a loosely bound molecule,
such an analysis procedure would cause the resonant 
invariant mass peak of $D^0 \bar D^0 \pi^0$ 
that is actually {\it below} \DstnDn\ threshold
to erroneously appear {\it above} the 
threshold and broadened. Taking this effect into account, 
the analysis~\cite{Stapleton:2009ey}
found the data to be consistent with the identification of the $X(3872)$
as a bound state with mass below \DstnDn\ threshold.
If future measurements have more
statistics in all relevant decay modes of $X(3872)$
and/or improve upon measured mass resolutions, while simultaneously 
avoiding the kinematic constraint to the \Dstn\ mass,
more definitive statements on the molecular nature
of the $X(3872)$ could be made.

Another avenue for studying the 
molecular nature of $X(3872)$ could become available at the LHC.
In particular, an EFT description~\cite{Canham:2009zq} of $X(3872)$ 
scattering off \Dze\ or \Dstn\ mesons has been developed,
and proposed for testing with either $B_c$ or $B\bar{B}$ decays
with final-state interactions.

\subthreesection{Questioning the $X(3872)$ molecular interpretation}

The production cross section of $X(3827)$ at the Tevatron
is a potential discriminant for its molecular interpretation.
Neither CDF nor \DZero\ have reported such a measurement
because it is not a trivial one. However, based on
published and unpublished-but-public CDF documents,
the product of the cross section of X(3872) and its branching fraction
into $\dipi\jpsi$ can be estimated.
The inclusive, prompt\footnote{The
   adjective {\it prompt} refers to X particles that are produced
   by QCD interactions and not by the weak decays of $b$-hadrons.}
production rate of $X\to\dipi\jpsi$
relative to $\psip\to\dipi\jpsi$, both with $\jpsi\to\dimu$,
for the same transverse momentum ($p_{\rm T}$) and
rapidity ($y$) restrictions,
{\it assuming equal selection efficiencies}, is estimated 
to be~\cite{Bignamini:2009sk}
\beqa
\frac{\sigma ( p\bar{p} \to X\, +{~\rm any}\, )_{\rm prompt}
\times\Brat(X\to\dipi\jpsi)}{
\sigma(p\bar{p}\to \psip\, +{~\rm any})_{\rm prompt}}\non\\
=(4.7\pm 0.8)\%\,.
\eeqa
This value is used in conjunction with a CDF~\cite{Aaltonen:2009dm}
measurement of the absolute cross section for
inclusive \psip\ production as a function of $p_{\rm T}$
for central rapidity to obtain~\cite{Bignamini:2009sk}
\beqa
\sigma(p\bar{p}\to X\, +{~\rm any})_{\rm prompt}
\times\Brat(X\to\dipi\jpsi)\non\\
=(3.1\pm0.7)~{\rm nb}\non\\
{\rm for~} p_{\rm T}>5\gev/c ~{\rm and~} |y|<0.6\,,
\label{eqn:SpecTh_Xcrosssection}
\eeqa
{\it assuming $X$ and \psip\ have the same rapidity distribution}.
Since the unknown branching fraction satisfies
$\Brat(X\to\dipi\jpsi)<1$,
\Eq{eqn:SpecTh_Xcrosssection} also provides a lower limit
on the prompt $X$ production cross section for
the transverse momentum and rapidity restrictions given.

The large magnitude of the prompt
production cross sections measured in $p\bar p$
collisions at the Tevatron came as a surprise to many. The original
Monte Carlo (MC) studies,
based in part on the generators HERWIG~\cite{Corcella:2000bw} and 
PYTHIA~\cite{Sjostrand:2000wi}, 
suggested that formation of loosely bound
\DstnDn\ molecules in this environment fall far short of the 
observed rates. Further theoretical work coupled with
MC studies~\cite{Bignamini:2009sk} reinforced this viewpoint, which was then
challenged by an independent examination~\cite{Artoisenet:2009wk} 
of the issues involved.
Both approaches allow formation of a molecule if its constituents,
after their initial production by the underlying generator,
have relative momentum up to a value \kmax. The two
approaches differ markedly in the values of \kmax\ that
are permitted. In~\cite{Bignamini:2009sk}, \kmax\ is chosen
to be comparable to the binding momentum $\kbind = \sqrt{2 \mred \Ebind}$,
whereas~\cite{Artoisenet:2009wk} 
argues that a value larger by an order of magnitude, and correspondingly
even larger prompt production cross section, is more appropriate
due to constituent rescattering effects.

Further justification for the choices in~\cite{Artoisenet:2009wk} 
was given in~\cite{Artoisenet:2010uu}, in which
deuteron production is taken as a case study 
to judge the efficacy of the arguments.
In addition, it allowed tuning of the underlying
generator for production of the molecule constituents in 
the required pairs. It is argued that the prescription 
in~\cite{Bignamini:2009sk} for the value of \kmax\ 
used in MC generation, as applied to deuterons,
is flawed for both fundamental and empirical 
reasons. Fundamentally, a loosely bound $S$-wave 
molecule does not satisfy the minimum uncertainty principle,
$\Delta r \Delta k \sim 1$.  Instead, it maximizes the 
uncertainty, satisfying $\Delta r \Delta k \gg 1$ with
$\Delta r \sim \kbind^{-1}$ and $\Delta k \gg \kbind$.
Empirically, the MC technique in~\cite{Bignamini:2009sk} underpredicts
the CLEO~\cite{Asner:2006pw} measurements of antideuterons in \UnS{1}\ decays
(see \Sec{sec:SpecExp_CLEOdeuteron}).

A rebuttal to~\cite{Artoisenet:2010uu} has been made~\cite{Burns:2010qq}. 
It argues that the deuteron and $X(3872)$ are not comparable.  First, the
discrepancy obtained between data and MC using a small \kmax\ (based 
mostly on the uncertainty principle, which limits the reasonable 
variations of $k$, and therefore \kmax, to be of order \kbind, not 
an order of magnitude larger) is modest for the deuteron 
(factors of 2-3) compared to the $X(3872)$ (factor of $\approx300$).
Moreover, MC studies with PYTHIA have found considerably better
agreement, within a factor of 2-3, between the estimated and observed 
cross sections of the deuteron. Discrepancies of this size in 
$e^+e^-$ collisions at LEP were considered reasonably close to 
the measurements and do not justify
rejecting the MC altogether~\cite{Sjostrand:2010private}. 
Finally, contends \cite{Burns:2010qq},
the deuteron is a system qualitatively different 
from a \DstnDn\ molecule because the $D^0$ is spinless
and cannot participate in spin interactions.
Conversely, spin interactions play an important role in the 
determination of the deuteron binding: 
the spin-singlet deuteron and its isospin partner, the dineutron, 
are not bound.

Much of the motivation for the $X(3872)$ molecular interpretation, 
which assumes $S$-wave binding, no longer applies 
if its quantum numbers are found to be $2^{-+}$,
as preferred by a recent \babar~\cite{delAmoSanchez:2010jr} analysis. 
A $\DstnDn$ $2^{-+}$ state would require a relative $P$-wave.
It is unlikely~\cite{Burns:2010qq}, that
$\pi$-exchange could bind such a state, given that, even in 
an $S$-wave configuration, it is
not clear that the attraction is sufficiently strong. Even if
such a state exists, there remains the further problem that unless
spin-dependent forces prevent the binding, one should expect
partner states with $0^{-+}$ and ($J^{PC}$-exotic) $1^{-+}$, 
for which there is no experimental evidence. 
A $P$-wave $2^{-+}$ molecule would also imply the existence of an
extremely narrow, more
deeply bound $S$-wave $1^{++}$ molecule.
Alternatively, forming a $2^{-+}$ $S$-wave bound state would require
a different molecule type such as $D_2D$ or $D_1D^*$, which would require
not only an immense binding energy of some 500\mev, but 
also loss of the appealing connection between the mass of
the $X(3872)$ and $\DstnDn$ threshold.

On the basis of a study by Cho and Wise~\cite{Cho:1994qp},
it is difficult to reconcile the observed $X(3872)$ 
prompt production cross section 
with the expectations for \polone\ standard charmonium. 
The integrated prompt cross section found using the Cho and Wise 
gluon fragmentation function in a \polone\ state is 
\beq
\sigma (p\bar p\to \polone+{\rm ~all})=0.6~{\rm nb}\,,
\eeq
some 50 and 120 times smaller than the estimated experimental 
cross section~\cite{Burns:2010qq}. 
As for the mass of the \polone\ charmonium, there are a number of 
studies available in the literature~\cite{Burns:2010qq}. A hadron-string 
calculation, with results that agree very well with previous determinations 
of charmonium and bottomonium levels,
is also proposed in~\cite{Burns:2010qq}. Most of these calculations 
indicate that the $X(3872)$ has the most 
difficulty matching quarkonium levels. In this model, 
while all charmonium and 
bottomonium levels agree with data within 
$\sim 10$\mev\ (excluding the \poltwo, which departs
from the experimental mass by 40\mev), the predicted \polone, the 
would-be $X(3872)$, falls short of the measured value by 80\mev.  
The mass mismatch and production cross section
jeopardize a $2^{-+}$ charmonium interpretation of the $X(3872)$. 
Clarification of the $X(3872)$
decay modes and relative branching fractions would help 
disentangle the possible explanations. The prominent radiative 
decay mode for a $D$-wave charmonium $X(3872)$ 
is expected to be $X(3872)\to h_c\gamma\to \jpsi \pi^0 \gamma$
while the $\etac\pi\pi$ channel should have the 
highest rate among hadronic modes.

The reader is also directed to 
\Secs{sec:SpecTh_Tetraquarks} and \ref{prod_section:pheno_with_QCD_corr}
for more information on the $X(3872)$.

\subsubsection{Tetraquark states}
\label{sec:SpecTh_Tetraquarks}

The nonstandard decay patterns of $X(3872)$ suggested other 
theoretical interpretations as well. Could the $X(3872)$ be a pointlike
hadron resulting from the binding of a diquark and an antidiquark?
This idea was discussed in Maiani~\etal~\cite{Maiani:2004vq}, 
following one interpretation of pentaquark baryons 
(antidiquark-antidiquark-quark states) 
proposed by Jaffe and Wilczek~\cite{Jaffe:2003sg} and the recent 
discussion of light scalar mesons in terms of tetraquarks 
by 't~Hooft~\etal~\cite{Maiani:2004uc,Hooft:2008we} (see 
also the review~\cite{Spanier:2008zz} and references therein). 
Some considerations by 't~Hooft on an 
open-string description of baryons~\cite{'tHooft:2004he} 
were also sources of inspiration.

\subthreesection{Diquarks} 

A spin-zero diquark operator in the attractive anti-triplet 
color (greek subscripts $\alpha$, $\beta$, $\gamma$) channel, 
antisymmetric in flavor (latin subscripts $i$, $j$, $k$) can be written as:
\beq
[qq]_{i\alpha}= \epsilon_{ijk}\,\,\epsilon_{\alpha \beta\gamma}\, \,\bar
q^{j\beta}_c\,\gamma_5 \,q^{k\gamma}
\label{eqn:SpecTh_defdq}
\eeq
where the subscript $c$ denotes charge conjugation.
The Fermi statistics of light quarks is respected in~\Eq{eqn:SpecTh_defdq}.
Spin-one diquarks (the so called `bad' ones) can be conceived 
but they are believed to have a smaller binding energy with respect 
to spin-zero candidates (see, \eg\cite{Alexandrou:2006cq}). 
A `bad' diquark (spin-one) operator can be written as:
\beq
[qq]^{ijr}_{\alpha}= \epsilon_{\alpha \beta\gamma}\, (\,\bar q^{i \beta }_c\, { \gamma^r} \, q^{j\gamma}\,+\,\bar q^{j \beta }_c\, {\gamma^r}\,  q^{i\gamma}\,) 
\label{eqn:SpecTh_defdq2}
\eeq
which is a ${\bf 6}$ in flavor space (and has three spin components) as
required by Fermi statistics ($r=1,2,3$).
Both represent positive parity states, $0^+$ and $1^+$, respectively.
Similarly one can construct $0^-$ and $1^-$ operators, $\bar q_c q$
and $\bar  q_c \gamma^r \gamma_5q$. The latter are  identically zero in
the `single mode configuration': quarks that are unexcited with respect
to one another. In fact, the most solid tetraquark candidates are scalar
mesons made up of `good' diquarks. 
The spin-zero light diquarks are very effective at reducing the number
of expected four-quark states. A $qq\bar q \bar q$  multiplet should contain 
81$={\bf 3}\otimes {\bf 3} \otimes {\bf \bar 3}\otimes {\bf \bar 3}$
particles for three quark flavors. But if the diquark
degrees of freedom are the relevant ones, the number of states is
reduced to 9$={\bf 3}\otimes {\bf \bar 3}$ for a diquark that behaves as an
antiquark and an antidiquark as a quark (see \Eq{eqn:SpecTh_defdq}). 
In the case of light scalar mesons this represents a way of encrypting 
the exoticity (we would have 9 light scalar mesons even if their
structure were $q\bar q$~\cite{Maiani:2004uc,Hooft:2008we}). Spin-one
diquarks of \Eq{eqn:SpecTh_defdq2} would 
enlarge the flavor structure as they are ${\bf 6}_f$ operators.

\subthreesection{Tetraquarks} 

The tetraquark model provides fertile ground for 
investigations of heavy-flavored states. 
One of the features of the diquark-antidiquark model 
proposed in~\cite{Maiani:2004vq}, 
which could also be considered a drawback, is the proliferation of
predicted states. Another is the paucity of insight from {\it selection rules} 
that could explain why many of these states are not observed (for a recent 
account, see, \eg\cite{Drenska:2009cd}). 
It is quite possible that those states are waiting to be discovered.

A tetraquark in the diquark-antidiquark incarnation is a state like
$[qq]_{i\alpha}[\bar q \bar q]^{j\alpha}$, if spin-zero 
diquarks are concerned, as is the case for the tetraquark interpretation
of light scalar mesons. To use a notation making flavor explicit one 
can write $[q_1q_2][\bar q_3 \bar q_4]$.
There is no real distinction between `good' and `bad' diquarks once one
of the quarks in the bound state 
is heavy: spin-spin interactions between quarks are $1/\hqm$ suppressed
and $\hqm\to \infty$ with respect to the light quark mass scale. 
In other words, one can expect that tetraquarks like $[cq][\bar c\bar
q^\prime]$ have the same chances to be formed by spin-one
or spin-zero diquarks. Moreover, the considerations of Fermi statistics
made above are no longer valid here. This enlarges the 
spectrum of predicted states. 

There is also the question discussed in \Sec{sec:SpecTh_molec} of whether
an $X(3872)$ tetraquark state can better match observed cross sections 
at the Tevatron. If only spin-zero diquarks were allowed, 
a $J^{PC}=1^{++}$ $X(3872)$ could not be described as a
tetraquark. For a $2^{-+}$ $X(3872)$, however, the tetraquark interpretation is
still viable. Known problems with the proliferation of 
states are shifted to lower mass scales~\cite{Burns:2010qq}.

The $[cq][\bar c\bar q^\prime]$ states should also appear in {\it charged}
combinations. Although there is unconfirmed evidence of 
a \zBelle\ state (see \Sec{sec:SpecExpChargedExotic}) decaying into
charmonium plus charged pion, there is yet no 
evidence of charged, almost degenerate partners, of the $X(3872)$.  
Another example of a possible tetraquark meson is the
$Y(4260)$, a $1^{--}$ resonance decaying into $\pi\pi\jpsi$,
as described in \Sec{sec:SpecExp_UnconVector}. 
The dipion mass distribution is consistent with there
being a substantial $f_0(980)$ component in the $Y(4260)$ decay, 
suggesting an exotic tetraquark structure 
$[cs][\bar c \bar s]$. The tetraquark model suggests a
$Y(4260)\to \DsDs$ decay mode~\cite{Maiani:2005pe},
a mode for which the experimental upper limit at 90\%~CL 
on branching fraction, $<1.3$ relative
to $\dipi\jpsi$ (see \Tab{tab:Spec_Y4260_opencharm}), 
is not a particularly stringent one.

As stated above, the main drawback of the tetraquark model is the
proliferation of predicted particles. For example, using a naive
constituent diquark model, in the hidden-strange and hidden-charm 
sector one can predict a quite complex pattern of 
states~\cite{Drenska:2009cd}. But since no such states near thresholds, 
such as $\phi\,\jpsi$ or $f_0(980)\,\jpsi$, are 
predicted, this model cannot account for the unconfirmed evidence
($3.8\sigma$ significance) for a new resonance, $Y(4140)$,
decaying to $\phi\,\jpsi$ near threshold, as reported by 
CDF~\cite{Aaltonen:2009tz} (see \Tab{tab:Spec_ExpSumUnc} and 
\Sec{sec:SpecExp_3940}). The naive constituent diquark 
model~\cite{Drenska:2009cd} does predict, however, a $0^{-+}$ state
decaying to $\phi\,\jpsi$ at about 4277\mev. 
The measured $\phi\,\jpsi$ mass spectrum reported by
CDF~\cite{Aaltonen:2009tz} does show an intriguing
enhancement near this mass, but the statistical significance
reported by CDF for this structure, $<3\sigma$, 
leaves the possibility that it is an artifact or a fluctuation. 
If this peak becomes more significant with more
data, it could bolster the constituent diquark approach.

Constituent quark models can only give rough estimates of 
the expected mass values. In contrast, the 
most striking prediction of the model are particles 
decaying to charmonia plus charged pions or $\rho$-mesons. 
One or more of the unconfirmed \zBellep, \zOne, and \zTwo\   
(see \Tab{tab:Spec_ExpSumUnc} and \Sec{sec:SpecExpChargedExotic}) 
could be examples of such particles. 
If any of these were confirmed by CDF (or by LHC 
experiments), there would be a much stronger argument in favor of the
tetraquark model than any mass spectrum determination of neutral candidates. 

\subthreesection{Baryonia} 

Assuming an {\it open-string} hadron picture of 
diquark-antidiquark tetraquarks, compelling evidence for their existence  
should be found in experimental searches for 
narrow structures coupled preferentially to a baryon and an antibaryon. 
It has been proposed~\cite{Cotugno:2009ys} that
there is compelling experimental evidence for a single vector baryonium
candidate, $Y_B$, that explains {\it two} unconfirmed 
states reported by Belle~\cite{Pakhlova:2008vn,:2007ea}. These states, 
the $X(4630)$, observed in the decay to 
$\lala$, and $Y(4660)$, which decays to $\dipi\psip$
(see \Tab{tab:Spec_ExpSumUnc} and \Sec{sec:SpecExp_UnconVector}),
have reported masses that differ by only two standard deviations and
compatible widths. They can also be fit well by a single
resonance~\cite{Cotugno:2009ys}
\beqa
m(Y_B)      &=& 4661\pm9\mev\,,\non\\
\Gamma(Y_B) &=& 63\pm23\mev\,. 
\eeqa
The mass of the $Y_B$ is significantly 
above the decay threshold and a straightforward four-quark 
interpretation explains its decay modes. 
The ratio of branching ratios found,
\beq
\frac{\mathcal{B}(Y_B\rightarrow\Lambda^{+}_{c}\Lambda^{-}_{c})}{\mathcal{B}(Y_B\rightarrow\psi(2S)\pi^{+}\pi^{-})} = 25\pm 7\,,
\label{eqn:SpecTh_br}
\eeq
highlights a strong affinity of $Y_B$ to the baryon-antibaryon 
decay mode. The phase space involved in the 
decays in \Eq{eqn:SpecTh_br} are rather similar
because the $\pi^+\pi^-$ pair results from 
an $f_0(980)$ decay (see \Fig{fig:Spec_Y43604660_dipion_belle} in 
\Sec{sec:SpecExp_UnconVector}).
As the $f_0(980)$ can be identified as a diquark-antidiquark particle
(see 't~Hooft~\etal~\cite{Maiani:2004uc,Hooft:2008we}),  
the $Y_B$ can also be interpreted~\cite{Cotugno:2009ys} as such an 
exotic state. A $[cd][\bar c \bar d]$
assignment for $Y_B$ naturally explains~\cite{Cotugno:2009ys} the 
ratio found in \Eq{eqn:SpecTh_br}; in the baryon-antibaryon mode,
a string of two heavy quarks with angular momentum excitation $\ell\neq 0$
would break: 
\beq
[qq]\gluon [\bar q\bar q] \to [qq]\gluon q+\bar q\gluon[\bar q \bar q]\,.
\eeq
While baryonic decays of tetraquarks should be the most favorable 
according to the string-color picture, but these are typically 
phase-space forbidden. 

The $Y(4360)$ (see \Sec{sec:SpecExp_UnconVector}) could be the radial 
ground state ($1P$) of the $Y_B(4660)$ ($2P$). These two states both 
decay into $\dipi\psi(2S)$ rather than $\dipi\jpsi$, a puzzling 
characteristic that awaits explanation~\cite{Cotugno:2009ys}. 
An alternative 
molecular interpretation of the $Y(4660)$ can be found in~\cite{Guo:2008zg}.

The $Y(2175)$, observed by \babar~\cite{Aubert:2006bu},
BES~\cite{:2007yt}, and Belle~\cite{Shen:2009zze},
is another interesting baryonium-like candidate, albeit in the 
light-quark sector. Assuming that
$Y(2175)$ is a four-quark meson~\cite{Drenska:2008gr}, 
it should preferentially decay into
$\Lambda\bar \Lambda$. Because the $\Lambda\bar \Lambda$
threshold is $\approx2231$\mev, this decay 
proceeds through the high-mass tail of the $Y(2175)$. The \babar\
data are consistent with this hypothesis~\cite{Drenska:2008gr}.

What about doubly charged particles?
A diquark-antidiquark open-charm 
$[cu][\bar d \bar s]$ composition, denoted here by $A^{++}$, could 
exist~\cite{Polosa:2010pc} and decay, \eg into $D^+ K^+$. 
It is very unlikely that a loosely
bound molecule of this kind could be produced. There has not been a
search for a doubly charged particle like $A^{++}$ close to 
the $D^+K^+$ mass. Theory is still not able to reliably predict
the $A^{++}$ rate within the tetraquark model.

\subthreesection{Counting quarks in heavy-ion collisions} 

Heavy-ion collisions also provide a means for definitively determining
the quark nature of, \eg the $X(3872)$. The nuclear modification 
ratios $R_{AA}$ and $R_{CP}$
(see \Sec{sec:media_sec5} and \Eqs{eqn:Med_rab}
and (\ref{eqn:Med_rcpdef}), respectively) of the $X(3872)$ and
its anisotropy coefficient, $v_2$ (see \Sec{sec:media_sec5}), 
which can be measured by ALICE,
could be useful tools in this task. In the recombination picture,
the $X(3872)$ is expected to be produced with rates similar to charm
mesons and baryons. Thus the soft part of the spectrum, 
where recombination is more effective, can be highly populated with
$X$'s. On the other hand, the fragmentation functions 
of a tetraquark-$X$ are different from those of a \DstnDn\ molecule
since the $D$ fragmentation functions are the standard ones. 
This effect could be studied in a manner similar to those described 
for light scalar mesons in~\cite{Maiani:2006ia}.

To summarize, we refer to \Tab{tab:SpecTh_tetraq}, 
which lists the most significant tetraquark candidates.
The $Y_B$ and \zBellep\ are the most likely.
Experimental study of the $Y(4260)\to\DsDs$ decay, 
responsible for most of the width in the tetraquark model, 
would be an important discriminant to assess its nature. 
More work on the tetraquark picture has been done recently by 
Ali~\etal~\cite{Ali:2009es,Ali:2009pi} 
and Ebert~\etal~\cite{Ebert:2008kb}.

\begin{table}[t]
   \begin{center}
      \caption{For significant tetraquark candidates,
               their spin-parity ($J^{PC}$), decay modes,
               and quark content}
      \label{tab:SpecTh_tetraq}
      \setlength{\tabcolsep}{0.45pc}
      \begin{tabular}{cccc}
\hline\hline
\rule[12pt]{-1mm}{0mm}
   $4q$ candidate & $J^{PC}$ & Decay Modes & $(\,[qq]\, [\bar q\bar q]\,)_{nJ}$  \\ [0.9mm]
\hline
\rule[12pt]{-1mm}{0mm}
    $Y_B(4660)$ & $1^{--}$ & $\Lambda_c^+ \Lambda_c^-,\, \pi\pi\psi(2S)$ & $(\,[cd]\,[\bar c\bar d]\,)_{2P}$\\[0.9mm]
    \zBellep & $1^{+-}$ & $\pi^+\psi(2S)$ & $(\,[cu]\,[\bar c \bar d]\,)_{2S}$ \\[0.9mm]
    $Y(4260)$ & $1^{--}$ & $\pi\pi\,\jpsi ,\, \DsDs~(?)$ & $(\,[cs]\,[\bar c\bar s]\,)_{1P}$ \\[0.9mm]
    $X(3872)$ & $1^{++}$ & $\rho^0\jpsi,\, \omega\jpsi,\, D\bar D\pi$ & $(\,[cu]\,[\bar c\bar u]\,)_{1S}$ \\[0.9mm]
    \hline\hline
      \end{tabular}
   \end{center}
\end{table}

\subsubsection{Hadrocharmonium}
\label{sec:SpecTh_hadrocharmonium}

The decay pattern for six of the new states
($Y(4260)$, $Y(4360)$, $Y(4660)$, \zOne, \zTwo, and \zBellep)
could be interpreted~\cite{Voloshin:2007dx} as an indication
of an intact charmonium state within a more complex hadronic
structure. These states have only been observed decaying
to a single preferred charmonium state accompanied by one or more 
light mesons (see \Tab{tab:Spec_ExpSumUnc}). Decays
into a different, apparently nonpreferred, charmonium resonance
with the same quantum numbers (\eg $\jpsi$ instead of $\psip$ for
$Y(4660)$) or open-charm hadrons have not been observed.
In some cases there are meaningful experimental upper bounds 
on such decays, as shown in \Tab{tab:Spec_Y4260_opencharm}. 
One explanation~\cite{Voloshin:2007dx} is that each such state
consists of its preferred charmonium embedded in a shell of
light-quark and gluon matter, \ie a compact charmonium is
bound inside a spatially large region of excited light matter. 
The observed decays can then be viewed as the de-excitation 
of the light hadronic matter into light mesons and liberation
of the compact charmonium. This
structure is referred to as hadrocharmonium~\cite{Voloshin:2007dx}, or
more generally, hadroquarkonium.

The picture of a hadroquarkonium mesonic resonance is quite
similar to the much-discussed nuclear-bound quarkonium. The
primary difference is that, instead of a nucleus, an excited mesonic
resonance provides the large spatial configuration of 
light-quark matter. For the quarkonium to remain
intact inside hadroquarkonium, the binding has to be relatively weak. 
In complete analogy with the treatment of charmonium binding in
nuclei (in terms of the QCD multipole 
expansion)~\cite{Peskin:1979va,Bhanot:1979vb,Kaidalov:1992hd,Sibirtsev:2005ex},
the interaction between a
compact, colorless quarkonium and the soft, light matter can be
described~\cite{Dubynskiy:2008mq} by an effective Hamiltonian 
proportional to the quarkonium chromopolarizability $\alpha$,
\beq
H_{eff} = -{1 \over 2} \, \alpha \, {\vec E}^a \cdot {\vec E}^a\,,
\label{eqn:SpecTh_e1e1}
\eeq
where ${\vec E}^a$ is the operator of the chromoelectric field. The
chromopolarizability can be
estimated from the transition $\psip \to \pi \pi \jpsi$, 
giving~\cite{Voloshin:2004un}
$\alpha^{(12)} \approx 2\gev^{-3}$. The average of the gluonic
operator over light hadron matter ($h$) with mass $m_h$ can be found using the
conformal anomaly relation in QCD:
\beq
\langle h \, | {1 \over 2} \, {\vec E}^a \cdot {\vec E}^a | \, h \rangle \ge {8 \pi^2 \over 9} \, m_h\,,
\eeq
which provides an estimate of the strength of the van~der~Waals-type
quarkonium-light hadrons interaction.
The likelihood of binding charmonium in light hadronic matter
depends on the relation between the mass $m_h$ and the spatial extent of
$h$~\cite{Dubynskiy:2008mq}. In a particular 
model~\cite{Erlich:2005qh,Karch:2006pv}
of mesonic resonances based on an AdS/QCD correspondence,
it can be proven~\cite{Dubynskiy:2008di}
that a heavy quarkonium does form a bound state inside a sufficiently
excited light-quark resonance. The decay of such a bound state
into open heavy-flavor hadrons is suppressed in the 
heavy-quark limit as $\exp (-\sqrt{\lamQ/\hqm})$,
where $\hqm$ is the heavy-quark mass. This is consistent with
the nonobservation of $Y$ and $Z$ decays into charm meson pairs.
However, it is not clear whether the charm quark is 
heavy enough for the heavy-quark limit to be applicable.

If the $Y$ and $Z$ resonances are hadrocharmonia, 
it is expected~\cite{Dubynskiy:2008mq} that:
\begin{itemize}
\item{Bound states of $\jpsi$ and $\psip$ with light nuclei
      and with baryonic resonances should exist, 
      \eg baryocharmonium decaying into $p \jpsi (+ {\rm pions})$.}
\item{Resonances containing $\chi_{cJ}$ that decay
      into $\chi_{cJ}+{\rm pion(s)}$ should also exist.
      The as-yet-unconfirmed \zOne\ and \zTwo\ states
      reported by Belle~\cite{Mizuk:2008me} are candidates 
      (see \Tab{tab:Spec_ExpSumUnc} and \Sec{sec:SpecExpChargedExotic}).}
\item{Decays of hadrocharmonia candidates to
      nonpreferred charmonium states, \eg $Y(4260) \to \dipi\psip$,
      or $Y(4360) \to \dipi \jpsi$, should be suppressed
      relative to preferred charmonia.}
\item{Resonances containing excited bottomonia such as $\Upsilon(3S)$,
      $\chi_b(2P)$, and/or $\Upsilon(1D)$ should exist in the mass range 
      $11\text{-}11.5$\gev.}
\end{itemize}

\subsubsection{QCD sum rules}
\label{sec:SpecTh_sum}

QCD sum rules 
(QCDSR)~\cite{Shifman:1978bx,Reinders:1984sr,Narison:2002pw,Colangelo:2000dp}
provide a method to perform QCD calculations 
of hadron masses, form factors and decay widths.
The method is based on identities between 
two- or three-point correlation functions, 
which connect hadronic observables with QCD fundamental 
parameters, such as quark masses, the strong coupling constant, 
and the  quantities which 
characterize the QCD vacuum, \ie the condensates. 
In these identities ({\it sum rules}), the phenomenological side 
(which contains information about hadrons)
is related to the QCD or OPE (operator product expansion) side, 
where the information about the quark content is introduced. 
Since the correlation functions are written in terms of well-defined 
quark currents, the method is effective in establishing the 
nature of the exotic states (molecule, tetraquark, hybrid, \etc).

In principle, QCDSR allows first-principle 
calculations. In practice, however, in order to extract the result, 
it is necessary to make expansions, truncations, and 
other approximations that may reduce the power of the formalism and introduce 
large errors. In addition, the convergence of the method often critically 
depends upon the decay channel.

\subthreesection{Formalism}

QCD sum rule calculations of hadron masses are based on the 
correlator of two hadronic currents:
\beq
\Pi(q)\equiv i\int d^4 x\, e^{iq\cdot x}
\,\lag{\,0}\,|\, T\, [\,j(x)\,j^\dagger(0)\,]\,|\,0\,\rag\, ,
\label{eqn:SpecTh_cor}
\eeq
where $j(x)$ is a current with the appropriate quantum numbers. 
The phenomenological and OPE correlation functions must then 
be identified. The Borel 
transformation~\cite{Shifman:1978bx,Reinders:1984sr,Narison:2002pw,Colangelo:2000dp}, 
which converts the Euclidean four-momentum squared, $Q^2$, into the
variable $M^2$ (where $M$ is the {\it Borel mass}), is applied
to improve the overlap between the two sides
of this identity. Working in Euclidean 
space is necessary to avoid singularities in the propagators
in \Eq{eqn:SpecTh_cor}. More precisely, it is necessary
to be in the {\sl deep Euclidean} region, \ie $Q^2 = -q^2 \gg \lamQ$. 
The Borel transform is well-defined in this region and a good
OPE convergence is obtained, dominated by the perturbative term.

After equating the two sides of the sum rule, assuming quark-hadron 
duality~\cite{Shifman:1978bx,Reinders:1984sr,Narison:2002pw,Colangelo:2000dp}
and making a Borel transform, the sum rule can be written as: 
\beq
\lambda \, e^{-m^2/M^2} = \int^{s_0}_{s_{\rm min}} \, 
ds\, \rho^{\rm OPE}(s) \,,
\label{eqn:SpecTh_sregra}
\eeq
where $m$ is the mass of the particle, $M$ is the Borel mass, 
$\rho$ is the spectral density obtained from the OPE side 
(taking the imaginary part of the correlation function), 
$s_0$ is the parameter which separates the pole (particle) 
from the continuum (tower of excitations with the same 
quantum numbers), and $s_{\rm min}$ is determined by kinematical 
considerations. The parameter $\lambda$ represents the coupling 
of the current to the hadron. Solving \Eq{eqn:SpecTh_sregra} 
for the mass, a function which 
is approximately independent of $M$ should be obtained. In practice, 
the result depends on the Borel mass and a value of $M$ must be chosen
within a domain called the {\it Borel window}. 
In order to determine the Borel window,
the OPE convergence and the pole contribution
are examined: the minimum value of $M$
is fixed by considering the convergence of the OPE, while the 
maximum value of $M$ is determined by requiring that the pole 
contribution be larger than the continuum 
contribution. As pointed out in~\cite{Matheus:2007ta}, 
it becomes more difficult for tetraquarks to simultaneously satisfy 
pole dominance and OPE convergence criteria. 
Reasonably wide Borel windows in which these two conditions 
are satisfied can exist only for heavy systems. Increasing the 
number of quark lines in a given system, the OPE convergence becomes 
gradually more problematic. For example, when a change is made from a meson  
(two quark lines) to a baryon (three quark lines), the 
perturbative term goes from a single loop to a double loop, 
suppressed by a factor of $\pi^2$ with respect to the 
single loop.  At the same time, higher-order
quark condensates become possible, while the nonperturbative 
corrections grow larger. As a consequence, larger Borel masses 
must be used to obtain convergence. However, at higher 
Borel masses the correlation function is dominated by the 
continuum contribution. With more quark lines, it becomes 
difficult to find a Borel window where both OPE convergence 
and pole dominance are satisfied. When a heavy quark is present, 
its mass provides a hard scale, which helps to make the 
OPE convergent at lower Borel masses. In summary: the calculations 
seem to indicate that it is more difficult to keep a 
larger number of quarks together with small spatial separation.

\subthreesection{QCDSR and $X(3872)$}

The $X(3872)$ (see \Sec{sec:SpecExpX3872} and 
\Tabs{tab:Spec_ExpSumUnc}, \ref{tab:Spec_XMass},
\ref{tab:Spec_DMass}, \ref{tab:Spec_RatBchgBneu}, and \ref{tab:Spec_DeltaMX})
has quantum numbers $J^{PC} = 1^{++}$ or $2^{-+}$.
It decays with equal strength into $\dipi\jpsi$ and $\dipi\piz\jpsi$,
indicating strong isospin-violation, which is incompatible with a  
$c\bar{c}$ state.  Its mass and the isospin-violation could be understood in 
several four-quark approaches. However, this state has a decay width 
of less than 2.2\mev, which is sometimes difficult to accommodate. 
In order to discuss four-quark configurations in more detail, a
distinction between a {\it tetraquark} and a {\it molecule} will be made: 
the former is simply a combination of four quarks 
with the correct quantum numbers, whereas the latter is a combination 
of two meson-like color-neutral objects. This separation can 
easily be made at the start of a calculation when the current is chosen. 
However, performing a Fierz transformation on the currents
will mix tetraquarks with molecules. Having this ambiguity in 
mind, this notation will be used to clearly refer to the
employed currents. The treatment given
below applies only for the $J^{PC} = 1^{++}$ assignment for $X(3872)$.

A current can be constructed for the $X$ based on diquarks in the
color-triplet configuration with symmetric spin distribution: $[cq]_{S=1}
[\bar{c}\bar{q}]_{S=0}+[cq]_{S=0}[\bar{c}\bar{q}]_{S=1}$
(see \Eq{eqn:SpecTh_colortripletdiquark}). 
Therefore the corresponding lowest-dimension interpolating operator 
for describing $X_q$ as a tetraquark state is given by
\beqa
j^{(q,{\rm di})}_\mu&=&{i\,\epsilon_{abc}\,
\epsilon_{dec}\over\sqrt{2}}\,[\,(\,q_a^T\,C
\,\gamma_5\,c_b\,)\,(\,\bar{q}_d\,\gamma_\mu\, C\,\bar{c}_e^T\,)
\non\\
&+&(\,q_a^T\,C\,\gamma_\mu\, c_b\,)\,
(\,\bar{q}_d\,\gamma_5C\,\bar{c}_e^T\,)\,]\,,
\label{eqn:SpecTh_cur-di}
\eeqa
where $q$ denotes a $u$ or $d$ quark and $c$ is the charm quark. 
We can also construct a 
current describing $X_q$ as a molecular \DstnDn\ state:
\beqa
j^{(q,{\rm mol})}_{\mu}(x) & = & {1 \over \sqrt{2}}
\bigg[
\left(\,\bar{q}_a(x)\, \gamma_{5}\, c_a(x)\,
\bar{c}_b(x)\, \gamma_{\mu}\,  q_b(x)\,\right) \nonumber \\
& - &
\left(\,\bar{q}_a(x)\, \gamma_{\mu}\, c_a(x)\,
\bar{c}_b(x)\, \gamma_{5}\,  q_b(x)\,\right)\,.
\bigg]
\label{eqn:SpecTh_cur-mol}
\eeqa
The currents in \Eqs{eqn:SpecTh_cur-di}~\cite{Matheus:2006xi} 
and (\ref{eqn:SpecTh_cur-mol})~\cite{Lee:2008uy} have both been used. 
In each case it was possible to find a Borel window where the pole
contribution is bigger than the continuum contribution and with a reasonable
OPE convergence. On the OPE side, the calculations were done to leading order 
in $\als$ including condensates up to dimension eight.  
The mass obtained in \cite{Matheus:2006xi} considering the 
allowed Borel window and the uncertainties in the parameters was 
$m_X=(3.92\pm0.13)$\gev, compatible with the measured value.
For the current in \Eq{eqn:SpecTh_cur-mol}, the OPE convergence 
and pole contribution yield a similar Borel window,
resulting in a predicted mass~\cite{Lee:2008uy}
of $m_X=(3.87\pm0.07)$\gev, also consistent with
the measured value and more precise than that obtained with
the tetraquark current.

In principle, we might expect a large partial decay width 
for the decay $X\to\rho^0\jpsi$. The initial state already 
contains the four necessary quarks and
no rules prohibit the decay. Therefore this decay
is allowed, similar to the case of the light scalars $\sigma$ and $\kappa$
studied in \cite{Brito:2004tv}, with widths of 
order 400\mev. The decay width is essentially determined by the 
$X\,\jpsi\, V$ ($V = \rho,~\omega$) coupling constant.
The decay width was computed using QCDSR~\cite{Navarra:2006nd}
with $g_{X\,J/\psi\, V}$ evaluated assuming  
that the $X(3872)$ is described by the 
tetraquark current, \Eq{eqn:SpecTh_cur-di}. 
The QCDSR calculation for the vertex $X(3872)J/\psi V$ is 
based on the evaluation of the three-point correlation 
function, which is a straightforward extension of \Eq{eqn:SpecTh_cor} 
to the case of three currents representing the three particles in the decay.
In \cite{Matheus:2009vq} the current representing the $X$ was given by:
\beq
j_{\alpha}^{X}=\cos{\theta}
j_\alpha^{(u,{\rm di})}+\sin{\theta}j_\alpha^{(d,{\rm di})}\,,
\label{eqn:SpecTh_XQCDSRmix}
\eeq
with $j_\alpha^{(q,{\rm di})}$ given in \Eq{eqn:SpecTh_cur-di}. 
This mixing between diquarks with different light flavors was 
first introduced~\cite{Maiani:2004vq} to explain the
decay properties of the $X(3872)$. Using $\theta\approx 20^\circ$
it is possible to reproduce the measured ratio
$r_\omega$ given in \Tab{tab:Spec_RatBchgBneu}.
With the same angle, the $X\,J/\psi\,\omega$ coupling constant 
was calculated with QCDSR~\cite{Navarra:2006nd}
and found to be $g_{X\psi \omega}=13.8\pm2.0$. This value is  
much bigger than the estimate of~\cite{Maiani:2004vq}
and leads to a large partial decay width of 
$\Gamma(X\to J/\psi ~(n\pi))=(50\pm15)$\mev. A similar value 
was also obtained~\cite{Matheus:2009vq} using a
molecular current similar to \Eq{eqn:SpecTh_cur-mol}. 
Therefore it is not possible to explain the small width 
of the $X(3872)$ from a QCDSR calculation if the $X$ is a pure
four-quark state. In~\cite{Matheus:2009vq}, the $X(3872)$ was treated 
as a mixture of a $c\bar{c}$ current with a molecular
current, similar to the mixing considered in~\cite{Sugiyama:2007sg}  
to study the light scalar mesons: 
\beq
J_{\mu}^q(x)= \sin\alpha\, j^{(q,{\rm mol})}_{\mu}(x)\, +\, \cos\alpha \,
j^{(q,2)}_{\mu}(x)\,,
\label{eqn:SpecTh_cur24}
\eeq
with $j^{(q,{\rm mol})}_{\mu}(x)$ given in \Eq{eqn:SpecTh_cur-mol} and
\beq
j^{(q,2)}_{\mu}(x) = {1 \over 6 \sqrt{2}}\, \diq\, 
[\,\bar{c}_a(x)\, \gamma_{\mu}\, 
\gamma_5\, c_a(x)\,]\,. 
\eeq
The introduction of  the quark condensate, $\langle\bar{q} q\rangle$,  
ensures that 
$j^{(q,2)}_{\mu}(x)$ and $j^{(q,{\rm mol})}_{\mu}(x)$ have the same dimension. 

It is not difficult to reproduce the experimental mass of the $X(3872)$
\cite{Matheus:2006xi,Matheus:2009vq,Lee:2008uy,Brito:2004tv}. 
This is also true for the current in \Eq{eqn:SpecTh_cur24} 
for a wide range of  mixing angles, $\alpha$, but, as observed 
in~\cite{Matheus:2009vq}, it is not possible to match
the measured value of $r_\omega$. 
In order to reproduce the ratio $r_\omega$ given 
in \Tab{tab:Spec_RatBchgBneu}, 
it is also necessary to consider a mixture of $D^+D^{*-}$ and 
$D^-D^{*+}$ components~\cite{Maiani:2004vq}. 
In this case the current is given by
\beq
j_{\mu}^X(x)= \cos\theta J_{\mu}^u(x)+\sin\theta J_{\mu}^d(x)\,,
\label{eqn:SpecTh_4mix}
\eeq
where $J_{\mu}^u(x)$ and $J_{\mu}^d(x)$ are given by \Eq{eqn:SpecTh_cur24}.
With this particular combination one obtains:
\beq
{\Gamma(X\to J/\psi\,\pi^+\pi^-\pi^0) \over \Gamma(X\to J/\psi\,\pi^+\pi^-)}
\simeq0.15\left({\cos\theta+\sin\theta \over \cos\theta-\sin\theta}\right)^2\,,
\label{eqn:SpecTh_ratioal}
\eeq
exactly the same relation~\cite{Navarra:2006nd,Maiani:2004vq} that
imposes $\theta\sim20^\circ$ in \Eq{eqn:SpecTh_XQCDSRmix} and compatible with 
the measured value of $r_\omega$ given in \Tab{tab:Spec_RatBchgBneu}.

It was shown~\cite{Matheus:2009vq} that, with \Eq{eqn:SpecTh_4mix} and 
a mixing angle  $\alpha=(9\pm4)^\circ$ in \Eq{eqn:SpecTh_cur24},
it is possible to describe the measured $X(3872)$ mass with
a decay width of $\Gamma(X\to J/\psi ~(n\pi))=(9.3\pm6.9)$\mev, which is  
compatible with the experimental upper limit.  The 
same mixing angle was used to evaluate the ratio
\beq           
r_{\gamma 1}\equiv
\frac{\Gamma(X\to\,\gamma\,\jpsi)}{\Gamma(X\to \dipi\jpsi)}\,,
\eeq
obtaining $r_{\gamma 1}=0.19\pm0.13$~\cite{Nielsen:2010ij}, in
excellent agreement with the measured value given in
\Tab{tab:Spec_RatBchgBneu}. Hence QCDSR calculations strongly 
suggest that the $X(3872)$ can be well described 
by a $c\bar{c}$ current with a small, but fundamental, 
admixture of molecular ($D \bar{D}^{*}$) or tetraquark 
$[cq][\bar{c}\bar{q}]$ currents. In connection with the 
discussion in \Sec{sec:SpecTh_molec}, a
possible \DsDsts\ molecular state, such an $X_s$ state, 
was also considered. The $X_s$ mass obtained, 3900\mev, 
was practically degenerate with the $X(3872)$. Therefore 
QCDSR indicate a larger binding energy for $X_s$ 
than for the $X(3872)$, leading to a smaller mass than
predicted in \cite{Drenska:2009cd}.

It is straightforward to extend the analysis done for the $X(3872)$ to the 
case of the bottom quark. Using the same interpolating field of
\Eq{eqn:SpecTh_cur-di} with the charm quark replaced by the bottom quark, the 
analysis done for $X(3872)$ was repeated~\cite{Matheus:2006xi} for an
analogous $X_b$. Here there is also a good Borel window. The prediction
for the mass of the state that couples to a tetraquark 
$[bq][\bar{b}\bar{q}]$ with $J^{PC}=1^{++}$ current is 
$m_{X_b}=(10.27\pm0.23)$\gev. The central value is close to the 
mass of $\UnS{3}$ and appreciably below the $B^*\bar{B}$ threshold 
at about 10.6\gev. For comparison, the molecular 
model predicts a mass for $X_b$ which is about 50-60\mev\ 
below this threshold~\cite{Swanson:2006st}, 
while a relativistic quark model without explicit 
$(b\bar{b})$ clustering predicts a value about 133\mev\ 
below threshold~\cite{Ebert:2005nc}.  

Summarizing, in QCDSR it is possible to 
satisfactorily explain all the $X(3872)$ properties with a
mixture of a $\approx97$\% $c\bar{c}$ component 
and a $\approx3$\% meson molecule component. This molecular 
component must be a mixture of 88\% $D^0 \, D^{*0}$ 
and 12\% $D^+ \, D^{*-}$~\cite{Matheus:2009vq}. 
These conclusions hold only for the quantum
number assignment $J^{PC} = 1^{++}$.

\subthreesection{QCDSR and the $Y$ states}

The states $Y(4260)$, $Y(4360)$ and $Y(4660)$ do not 
easily fit in the predictions of the standard quark model. 
The $Y(4260)$ has a $\dipi\jpsi$ decay width of $\simeq 100$\mev, 
and no isospin-violating decay such as
$Y  \, \to \, \jpsi \,  \pi^+\pi^-\pi^0 $ 
has been observed. With these features, 
the $Y$ is likely to be a meson molecule or a hybrid state. QCDSR calculations 
of the mass strongly suggest the 
$D^* \bar{D}_{0} - \bar{D}^* D_0$ and 
$D \bar{D}_1 - \bar{D} D_1 $ as 
favorite molecular combinations, where the symbols
$D$, $D^*$, $D_0$ and $D_1$ represent 
the lowest-lying pseudoscalar, vector, scalar, and axial-vector
charm messons, respectively. From now 
on we shall omit the combinations arising from 
symmetrization and use the short forms 
for these states, \eg $D_0 \bar{D}$. 

The vector $Y$ states can be described by molecular or tetraquark
currents, with or without an $s\bar{s}$ pair. QCD sum rule calculations 
have been performed~\cite{Albuquerque:2008up,Lee:2008gn} 
using these currents.
A possible interpolating operator representing  a $J^{PC}=1^{--}$ tetraquark   
state with the symmetric spin distribution
\beq
[cs]_{S=0}\,[\bar{c}\bar{s}]_{S=1}+[cs]_{S=1}\,[\bar{c}\bar{s}]_{S=0}
\eeq
is given by: 
\beqa
j_\mu&=&{\epsilon_{abc}\,\epsilon_{dec}\over\sqrt{2}}\,
[\,(s_a^T\,C\gamma_5\,c_b\,)\,
(\,\bar{s}_d\,\gamma_\mu\,\gamma_5\, C\,\bar{c}_e^T\,)
\non\\
&+&(\,s_a^T\,C\,\gamma_5\,\gamma_\mu\, c_b\,)\,
(\,\bar{s}_d\,\gamma_5\,C\,\bar{c}_e^T\,)\,]\,.
\label{eqn:SpecTh_fieldyt}
\eeqa
This current has good OPE convergence and 
pole dominance in a given Borel window. The result for the mass 
of the state described by the current in \Eq{eqn:SpecTh_fieldyt} is 
$m_Y = (4.65\pm0.10)$~\gev~\cite{Albuquerque:2008up}, 
in excellent agreement with the mass of the $Y(4660)$ meson,
lending credence to the conclusion~\cite{Albuquerque:2008up}
that the $Y(4660)$ meson can be described
with a diquark-antidiquark tetraquark current with a spin configuration 
given by scalar and vector diquarks. The quark content of the current
in \Eq{eqn:SpecTh_fieldyt}) 
is also consistent with the experimental dipion invariant 
mass spectra, which give some 
indication that the $Y(4660)$ has a well-defined dipion intermediate state 
consistent with $f_0(980)$. 

Replacing the strange quarks in \Eq{eqn:SpecTh_fieldyt} 
by a generic light quark $q$, the mass obtained for 
a $1^{--}$ state described with the symmetric spin 
distribution
\beq
[cq]_{S=0}\,[\bar{c}\bar{q}]_{S=1}+[cq]_{S=1}\,[\bar{c}\bar{q}]_{S=0}
\eeq 
is $m_Y=  (4.49\pm0.11)$\gev~\cite{Albuquerque:2008up},  
which is slightly larger than but consistent with the measured $Y(4360)$ mass.

The $Y$ mesons can also be described by molecular-type currents. 
A $D_{s0}(2317)\bar{D}_s^*(2110)$ molecule  with $J^{PC}=1^{--}$
could also decay into $\psip\pi^+\pi^-$ with a dipion mass 
spectrum consistent with $f_0(980)$. A current with $J^{PC}=1^{--}$ and a 
symmetric combination of scalar and vector mesons is
\beq
j_\mu={1\over\sqrt{2}}\,[\,(\,\bar{s}_a\,\gamma_\mu\, c_a\,)\,
(\,\bar{c}_b\,{s}_b\,)\,+\,(\,\bar{c}_a\,
\gamma_\mu\, s_a\,)\,(\,\bar{s}_b\,{c}_b\,)\,]\,. 
\label{eqn:SpecTh_mol}
\eeq
The mass obtained in~\cite{Albuquerque:2008up} for this current is
$m_{D_{s0}\bar{D}_s^*}=  (4.42\pm0.10)$\gev, 
which is in better agreement with $Y(4360)$ than $Y(4660)$.

To consider a molecular $D_{0}\bar{D}^*$ current with $J^{PC}=1^{--}$, 
the strange quarks in \Eq{eqn:SpecTh_mol} must be replaced with a 
generic light quark $q$. The mass obtained with such current 
is \cite{Albuquerque:2008up}
$m_{D_{0}\bar{D}^*}=  (4.27\pm0.10)$\gev, 
in excellent agreement with the $Y(4260)$ mass.
In order to associate this molecular state with $Y(4260)$, a better
understanding of the dipion invariant mass spectra
in $Y(4260)\to \dipi\jpsi$ is needed. 
From the measured spectra, it seems that 
the $Y(4260)$ is consistent with a nonstrange 
molecular state ${D_{0}\bar{D}^*}$.
Using a $D_0$ mass of  $m_{D_0}=2352\pm50$\mev, the ${D_{0}\bar{D}^*}$ 
threshold is  $\approx4360$\mev, 100\mev\ above the 
$4.27\pm0.10$\gev\ quoted above, indicating the possibility of a bound state.

A $J^{PC}=1^{--}$ molecular current can also be constructed with
pseudoscalar and  axial-vector mesons. A molecular $D\bar{D}_1$ 
current was used in~\cite{Lee:2008gn}. The mass obtained with this current is
$m_{D\bar{D}_1}=(4.19\pm0.22)$\gev. Thus, taking the mass uncertainty
into account, the molecular $D\bar{D}_1$ assignment for the $Y(4260)$ 
is also viable, in agreement with a meson-exchange model~\cite{Ding:2008gr}.
The $D\bar{D}_1$ threshold is $\approx$4285\mev, close to the $Y(4260)$ 
mass, indicating the possibility of a loosely bound molecular state.

Summarizing, the $Y$ states can be understood as 
charmonium hybrids, tetraquark states, and  a $D_0\bar{D}^*$ or $D\bar{D}_1$ 
molecular state for $Y(4260)$. 
Also possible are a tetraquark state with two axial $[cs]$ $P$-wave 
diquarks, or two scalar $[cs]$ $P$-wave diquarks for $Y(4360)$.

\subthreesection{\zBellep}

A current describing the \zBellep\ as a $D^*D_1$ molecule with $J^P=0^-$ 
is~\cite{Lee:2007gs}
\beqa
j={1\over\sqrt{2}}\,\left[\right. &\left(\right.& 
\bar{d}_a\,\gamma_\mu\, c_a\,\left.\right)\,
\left(\right.\,\bar{c}_b\,\gamma^\mu\,\gamma_5\,
u_b\,\left.\right)+\non\\
& \left(\right. & \bar{d}_a\,\gamma_\mu\,\gamma_5\, c_a\,\left.\right)\,
\left(\right. \,\bar{c}_b\,\gamma^\mu\, u_b\,\left.\right)\,\left.\right]\, .
\label{eqn:SpecTh_zmol}
\eeqa
This current corresponds to a symmetric
$D^{*+}\bar{D}_1^0+\bar{D}^{*0} D_1^+$ state with 
positive $G$-parity, consistent with the 
observed decay $\zBellep\to\pi^+\,\psip$.
The mass obtained in a QCDSR calculation using 
such a current is~\cite{Lee:2007gs} $m_{D^*D_1}=(4.40\pm0.10)$\gev,
in an excellent agreement with the measured mass. 

To check if the \zBellep\ could also be described as
a diquark-antidiquark state with $J^P=0^-$, the current~\cite{Bracco:2008jj}
\beqa
j_{0^-}={i\,\epsilon_{abc}\,\epsilon_{dec}\over\sqrt{2}}\,
[&(&u_a^T\,C\,\gamma_5c_b\,)\,
(\,\bar{d}_d\,C\,\bar{c}_e^T\,)\,-\non\\
&(&u_a^T\,C\, c_b\,)\,(\,\bar{d}_d\,\gamma_5\,C\,\bar{c}_e^T\,)\,]
\label{eqn:SpecTh_z0-}
\eeqa
was used to obtain the mass
$m_{Z_{(0^-)}} = (4.52\pm0.09)$\gev~\cite{Bracco:2008jj},
somewhat larger than, but consistent with, the experimental value. 
The result using a molecular-type current is in slightly 
better agreement with the experimental value. However, since 
there is no one-to-one correspondence between the structure of the current 
and the state, this result cannot be used to conclude that the \zBellep\  
is favored as a molecular state over a diquark-antidiquark state. 
To get a measure of the coupling between the state and the current, the 
parameter $\lambda$, defined in \Eq{eqn:SpecTh_sregra}, is evaluated as
$\lambda_{D^*D_1} \,  \simeq 1.5 \, \lambda_{Z_{(0^-)}}$. 
This suggests that a physical particle with
$J^P=0^-$ and quark content $c\bar{c}u\bar{d}$ has a stronger coupling
to the molecular $D^*D_1$-type current than with that of \Eq{eqn:SpecTh_z0-}.

A diquark-antidiquark interpolating 
operator  with $J^P=1^-$ and positive $G$-parity
was also considered~\cite{Bracco:2008jj}:
\beqa
j_\mu^{1^-}&=&{\epsilon_{abc}\,\epsilon_{dec}\over\sqrt{2}}\,
[\,(\,u_a^T\,C\,\gamma_5\,c_b\,)\,
(\,\bar{d}_d\,\gamma_\mu\,\gamma_5\, C\,\bar{c}_e^T\,)
\non\\
&+&(\,u_a^T\,C\,\gamma_5\,\gamma_\mu\, c_b\,)\,
(\,\bar{d}_d\,\gamma_5\,C\,\bar{c}_e^T\,)\,]\,.
\label{eqn:SpecTh_z1-}
\eeqa
In this case the Borel stability obtained is worse than 
for the $Z^+$ with $J^P=0^-$ \cite{Bracco:2008jj}. The mass obtained
is $m_{Z_{(1^-)}}=  (4.84\pm0.14)$\gev, 
much larger than both the measured value
and that obtained using the $J^P=0^-$ current.
Thus, it is possible to 
describe the \zBellep\ as a diquark-antidiquark or molecular state
with $J^P=0^-$, and $J^P=1^-$ configuration is disfavored.

It is straightforward to extend the $D^*D_1$-molecule analysis
to the bottom quark. Using the same interpolating field of
\Eq{eqn:SpecTh_zmol} but replacing the charm quark with bottom, 
an investigation of the hypothetical $Z_b$ is performed~\cite{Lee:2007gs}.
The OPE convergence is even better than for \zBellep.
The predicted mass is $m_{Z_{B^*B_1}}=  (10.74\pm0.12)$\gev,
in agreement with that of~\cite{Cheung:2007wf}.
For the analogous strange meson $Z_{s}^+$, considered as a 
pseudoscalar $D_s^*D_1$ molecule, the current 
is obtained by replacing the $d$ quark in \Eq{eqn:SpecTh_zmol} 
with an $s$ quark. The predicted mass is \cite{Lee:2007gs} 
$m_{D_s^*D_1}=  (4.70\pm0.06)$\gev,
larger than the $D_s^{*}D_1$ threshold of $\sim4.5$\gev, indicating
that this state is probably very broad and therefore might be
difficult to observe.

Summarizing, the \zBellep\ has been successfully described by both 
a $D^*D_1$ molecular current and a diquark-antidiquark current with 
$J^P = 0^-$. 

\begin{table}[tb]
   \begin{center}
      \caption{Summary of QCDSR results~\cite{Nielsen:2009uh}. The labels $1$, $3$, 
               and $\bar{3}$ refer to 
               singlet, triplet and anti-triplet color 
               configurations, respectively. The symbols 
               S, P, V, and A refer to scalar, pseudoscalar, 
               vector and axial-vector $q\bar{q}$, 
               $qq$ or $\bar{q}\bar{q}$ combinations, respectively }
     \label{tab:SpecTh_tab1}
     \setlength{\tabcolsep}{1.05pc}
     \begin{tabular}{ccc} 
\hline\hline
\rule[10pt]{-1mm}{0mm}
   State    & Configuration       & Mass~(GeV) \\ [0.9mm]
\hline
\rule[10pt]{-1mm}{0mm}
 $X(3872)$ & $[cq]_{\bar{3}} \, [\bar{c}\bar{q}]_{{3}}$ (S + A)  & $3.92 \pm 0.13$ \\[0.9mm]
           & $[c\bar{q}]_{1} \, [\bar{c}{q}]_{1}$ (P + V)  & $3.87 \pm 0.07$ \\[0.9mm]
           & $[c \bar{c}]_1 (A) \, + $~~~~~~~~~~~~~~ \\[0.9mm]
           & $[c\bar{q}]_{1} \, [\bar{c}{q}]_{1}$ (P + V)  & $3.77 \pm 0.18$ \\[0.9mm]
 $Y(4140)$ & $[c\bar{q}]_{1} \, [\bar{c}{q}]_{1}$ (V + V) &  $4.13 \pm 0.11$ \\[0.9mm]
               & $[c\bar{s}]_{1} \, [\bar{c}{s}]_{1}$ (V + V) &  $4.14 \pm 0.09$ \\[0.9mm]
     $Y(4260)$ &$[c\bar{q}]_{1} \, [\bar{c}{q}]_{1}$ (S + V)  &  $4.27 \pm 0.10$ \\[0.9mm]
               &$[c\bar{q}]_{1} \, [\bar{c}{q}]_{1}$ (P + A)  &  $4.19 \pm 0.22$ \\[0.9mm]
     $Y(4360)$ &$[cq]_{\bar{3}} \, [\bar{c}\bar{q}]_{{3}}$ (S + V) & $4.49 \pm 0.11$  \\ [0.9mm]
               &$[c\bar{s}]_{1} \, [\bar{c}{s}]_{1}$ (S + V)  & $4.42 \pm 0.10$  \\ [0.9mm]
     $Y(4660)$ &$[cs]_{\bar{3}} \, [\bar{c}\bar{s}]_{{3}}$  (S + V) & $4.65 \pm 0.10$ \\ [0.7mm]
     $Z(4430)$ &$[c\bar{d}]_{1} \, [\bar{c}{u}]_{1}$ (V + A)  & $4.40 \pm 0.10$ \\ [0.9mm]
               &$[cu]_{\bar{3}} \, [\bar{c}\bar{d}]_{{3}}$ (S + P) & $4.52 \pm 0.09$ \\ [0.9mm]
\hline\hline
     \end{tabular} 
  \end{center}
\end{table}

\Tab{tab:SpecTh_tab1} summarizes the QCDSR results 
for the masses and corresponding quark configurations of the 
new states~\cite{Nielsen:2009uh}. Masses obtained 
with QCDSR in the molecular approach can be found
in \cite{Zhang:2009vs} (they agree with those in
\Tab{tab:SpecTh_tab1} from~\cite{Nielsen:2009uh}
apart from small discrepancies, which should be addressed).  
The principal input parameters used to obtain the values in
\Tab{tab:SpecTh_tab1} are
\beqa
\label{eqn:SpecTh_qcdparam}
m_c(m_c)&=&(1.23\pm 0.05)\gev,\non\\
\lag\bar{q}q\rag&=&\,-(0.23\pm0.03)^3\gev^3,\non\\
\lag\bar{q}g\sigma.Gq\rag&=&m_0^2\lag\bar{q}q\rag,\non\\
m_0^2&=&0.8\,\GeV^2,\non\\
\lag g^2G^2\rag&=&0.88~\GeV^4\,.
\eeqa
Uncertainties in \Tab{tab:SpecTh_tab1} come from several sources:
quark masses and $\als$ are varied by their
errors around their central values in \Eq{eqn:SpecTh_qcdparam};
the condensates are taken from previous QCDSR analyses and 
their uncertainties propagated;
particle masses are extracted for a range of Borel masses
throughout the Borel window;
and in the case of mixing, the masses and widths are
computed for several values of the mixing angle centered
on its optimal value.

Some of the states discussed above can be understood as both tetraquark
and molecular structures. This freedom will be reduced once a 
comprehensive study of the decay width is performed. 
At this point, it appears that the only decay width calculated with QCDSR
is in~\cite{Matheus:2009vq}. 
The next challenge for the QCDSR community is to understand 
the existing data on decays. Explaining these decays will impose severe 
constraints in the present picture of the new states~\cite{Matheus:2009vq}.

\begin{table*}[tb]
\caption{As in Table~\ref{tab:Spec_ExpSumUnc}, 
new {\it unconventional} states in the $c\bar{c}$, $b\bar{c}$, and
$b\bar{b}$ regions, ordered by mass, with possible interpretations
(which do not apply solely to the decay modes listed alongside).
References are representative, not necessarily exhaustive.
The QCDSR notation is explained in the caption to \Tab{tab:SpecTh_tab1}} 
\setlength{\tabcolsep}{0.32pc}
\begin{center}
\begin{tabular}{lcccl|cc}
\hline\hline
\rule[10pt]{-1mm}{0mm}
 State & $m$~(MeV) & $\Gamma$~(MeV) & $J^{PC}$ & Modes & Interpretation & Reference(s) \\
\hline
\rule[10pt]{-1mm}{0mm}
$X(3872)$& 3871.52$\pm$0.20 & 1.3$\pm$0.6 & $1^{++}/2^{-+}$ 
  & $\pi^+\pi^-\jpsi$ & \DstnDn\ molecule (bound) &
    \cite{Braaten:2007dw,Stapleton:2009ey}\\ 
 & & & & & & \cite{Dong:2008gb,Dong:2009yp,Lee:2009hy}\\ 
 & & & & \DstnDn& \DstnDn\ unbound & \\
 & & & & $\gamma J/\psi$, $\gamma \psi(2S)$ &if $1^{++}$, $\chi_{c2}(2P)$ & \cite{Dudek:2009kk} \\
 & & & & $\omega J/\psi$ & if $2^{-+}$, $\eta_{c2}(1D)$ & \cite{delAmoSanchez:2010jr,Godfrey:1985xj,Eichten:2007qx} \\
 & & & & & charmonium + mesonic-molecule mixture & \cite{Nielsen:2009uh} \\
 & & & & & QCDSR:~$[cq]_{\bar{3}} \, [\bar{c}\bar{q}]_{{3}}$~(S+A) &\cite{Nielsen:2009uh} \\
 & & & & & QCDSR:~$[c\bar{q}]_{1} \, [\bar{c}{q}]_{1}$~(P+V) &\cite{Nielsen:2009uh} \\
 & & & & & QCDSR:~$[c \bar{c}]_1 {\rm (A)} \, +[c\bar{q}]_{1} \, [\bar{c}{q}]_{1}$~(P+V)  &\cite{Nielsen:2009uh} \\[0.97mm]

$X(3915)$ & $3915.6\pm3.1$ & 28$\pm$10 & $0,2^{?+}$ &
 $\omega \jpsi$ & $\Dstp\Dstm + \Dstn\bar{D}^{*0}$ &\cite{Branz:2009yt} \\[0.97mm]
$Z(3930)$ & $3927.2\pm2.6$ & 24.1$\pm$6.1 & $2^{++}$ &
     $D\bar{D}$ & $\chi_{c2}(2P)$ (\ie $2\,^3P_2$ $c\bar c$) & \\
 & & & & & $1\,^3F_2$ $c\bar c$ & \cite{Dudek:2009kk} \\[0.97mm]
$X(3940)$ & $3942^{+9}_{-8}$ & $37^{+27}_{-17}$ & $?^{?+}$ &
 $\DDst$ & & \\[0.97mm]
$G(3900)$ & $3943\pm21$ & 52$\pm$11 & $1^{--}$ &
$\DDbar$ & Coupled-channel effect &\cite{Eichten:1979ms} \\[0.97mm]
$Y(4008)$ & $4008^{+121}_{-\ 49}$ & 226$\pm$97 & $1^{--}$ &
$\pi^+\pi^-J/\psi$ & & \\[0.97mm]
$Z_1(4050)^+$ & $4051^{+24}_{-43}$ & $82^{+51}_{-55}$ & ?&
 $\pi^+\chi_{c1}(1P)$ & hadrocharmonium & \cite{Voloshin:2007dx,Dubynskiy:2008mq}\\[0.97mm]
$Y(4140)$ & $4143.0\pm3.1 $ & $11.7^{ +9.1}_{-6.2}$ & $?^{?+}$ &
     $\phi J/\psi$ & QCDSR:~$[c\bar{q}]_{1} \, [\bar{c}{q}]_{1}$~(V+V) &\cite{Nielsen:2009uh} \\
 & & & & & QCDSR:~$[c\bar{s}]_{1} \, [\bar{c}{s}]_{1}$~(V+V)
 &\cite{Nielsen:2009uh} \\
 & & & & & $\Dstsp\Dstsm$ &\cite{Branz:2009yt}\\[0.97mm]
$X(4160)$ & $4156^{+29}_{-25} $ & $139^{+113}_{-65}$ & $?^{?+}$ &
     $\DDst$ & & \\[0.97mm]
$Z_2(4250)^+$ & $4248^{+185}_{-\ 45}$ &
     177$^{+321}_{-\ 72}$ &?&
     $\pi^+\chi_{c1}(1P)$ & hadrocharmonium & \cite{Voloshin:2007dx,Dubynskiy:2008mq}\\[0.97mm]
$Y(4260)$ & $4263\pm5$ & 108$\pm$14 & $1^{--}$ & $\pi^+\pi^-\jpsi$ &
                             charmonium hybrid &\cite{Zhu:2005hp,Kou:2005gt,Close:2005iz} \\
 & & & & $\pi^0\pi^0\jpsi$ &  $J/\psi f_0(980)$ bound state & \cite{MartinezTorres:2009xb} \\
 & & & & & $D_0\bar{D}^*$ molecular state & \cite{Albuquerque:2008up} \\
 & & & & & $[cs][\bar{c}\bar{s}]$ tetraquark state & \cite{Maiani:2005pe,Nielsen:2009uh} \\
 & & & & & hadrocharmonium & \cite{Voloshin:2007dx,Dubynskiy:2008mq}\\
 & & & & & QCDSR:~$[c\bar{q}]_{1} \, [\bar{c}{q}]_{1}$~(S+V)&\cite{Nielsen:2009uh} \\
 & & & & & QCDSR:~$[c\bar{q}]_{1} \, [\bar{c}{q}]_{1}$~(P+A)&\cite{Nielsen:2009uh} \\[0.97mm]
$Y(4274)$ & $4274.4^{+8.4}_{-6.7}$ & $32^{+22}_{-15}$ & $?^{?+}$ &
     $B\to K (\phi J/\psi)$ & (see $Y(4140)$) &\\[0.97mm]
$X(4350)$ & $4350.6^{+4.6}_{-5.1}$ & $13.3^{+18.4}_{-10.0}$ & 0,2$^{++}$  &$\phi\jpsi$&& \\[0.97mm]
$Y(4360)$ & $4353\pm11$ & 96$\pm$42 & $1^{--}$ & $\pi^+\pi^- \psip$ & hadrocharmonium & \cite{Voloshin:2007dx,Dubynskiy:2008mq}\\
 & & & & & crypto-exotic hybrid & \cite{Dudek:2009kk}\\
 & & & & &$Y_B(4360)=[cd][\bar c \bar d](1P)$, baryonium & \cite{Cotugno:2009ys}\\
 & & & & &QCDSR:~$[cq]_{\bar{3}} \, [\bar{c}\bar{q}]_{{3}}$~(S+V)&\cite{Nielsen:2009uh} \\
 & & & & &QCDSR:~$[c\bar{s}]_{1} \, [\bar{c}{s}]_{1}$~(S+V)&\cite{Nielsen:2009uh} \\[0.97mm]
$Z(4430)^+$ & $4443^{+24}_{-18}$ & $107^{+113}_{-\ 71}$ & ?&
     $\pi^+\psi(2S)$ & $\Dstp\bar{D}_1^0$ molecular state & \cite{Lee:2007gs,Branz:2010sh} \\
  & & & & & $[cu][\bar{c}\bar{d}]$ tetraquark state & \cite{Bracco:2008jj}  \\
 & & & & & hadrocharmonium & \cite{Voloshin:2007dx,Dubynskiy:2008mq}\\
 & & & & & QCDSR:~$[c\bar{d}]_{1} \, [\bar{c}{u}]_{1}$~(V+A)&\cite{Nielsen:2009uh} \\
 & & & & & QCDSR:~$[cu]_{\bar{3}} \, [\bar{c}\bar{d}]_{{3}}$~(S+P)&\cite{Nielsen:2009uh} \\[0.97mm]
$X(4630)$ & $4634^{+\ 9}_{-11}$ & $92^{+41}_{-32}$ & $1^{--}$ &
     $\lala$ &  $Y_B(4660)=[cd][\bar c \bar d](2P)$, baryonium & \cite{Cotugno:2009ys} \\
 & & & & & $\psip f_0(980)$ molecule & \cite{Guo:2010tk} \\[0.97mm]
$Y(4660)$ & 4664$\pm$12 & 48$\pm$15 & $1^{--}$ &
     $\pi^+\pi^- \psi(2S)$ & $\psip f_0(980)$ molecule & \cite{Guo:2008zg} \\   
   & & & & & $[cs][\bar{c}\bar{s}]$ tetraquark state & \cite{Albuquerque:2008up}  \\   
 & & & & & hadrocharmonium & \cite{Voloshin:2007dx,Dubynskiy:2008mq}\\
 & & & & &$Y_B(4660)=[cd][\bar c \bar d](2P)$, baryonium & \cite{Cotugno:2009ys}\\
 & & & & &QCDSR:~$[cs]_{\bar{3}} \, [\bar{c}\bar{s}]_{{3}}$~(S+V)&\cite{Nielsen:2009uh} \\[0.97mm]
$Y_b(10888)$ & 10888.4$\pm$3.0 & 30.7$^{+8.9}_{-7.7}$ & $1^{--}$ &
      $\pi^+\pi^- \Upsilon(nS)$ & \UnS{5} & \cite{Hou:2006it} \\
 & & & & & $b$-flavored~$Y(4260)$&\cite{Chen:2008pu,Abe:2007tk} \\
\hline\hline
\end{tabular}
\end{center}
\label{tab:Spec_ExpHyp}
\end{table*}

\subsubsection{Theoretical explanations for new states}
\label{sec:SpecTh_sumnewstates}

  Table~\ref{tab:Spec_ExpHyp} lists the new states with
their properties and proposed theoretical explanations.
The theoretical hypothesis list is far from exhaustive.

\subsection{Beyond the Standard Model}
\label{sec:SpecTh_bsm} 

\subsubsection{Mixing of a light CP-odd Higgs and $\eta_b(nS)$ resonances}
\label{sec:SpecTh_mixing}

This section explores the possibility 
that the measured $\etab$ mass 
(\Tab{tab:Spec_etab} and \Eq{eqn:SpecExp_HFSexp})
is smaller than the
predictions (see 
\Secs{sec:SpecTh_lowestspectra},
\ref{sec:SpecTh_nrqcdlc}, 
\ref{sec:SpecTh_etab}, and
\ref{sec:SpecTh_alphasdec})
due to mixing with a CP-odd Higgs scalar $A$, and
predictions for the spectrum of the $\eta_b(nS)$-$A$ system
and the branching fractions into $\tau^+\tau^-$ as functions of
$m_A$ are made. Such mixing can cause
masses of the $\eta_b$-like eigenstates
of the full mass matrix to differ considerably from their values in
pure QCD \cite{Drees:1989du,Fullana:2007uq,Domingo:2008rr}. Thus the 
mass of the state interpreted as the $\eta_b(1S)$ can be smaller 
than expected if $m_A$ is slightly above 9.4\gev. 
The masses of the states interpreted 
as $\eta_b(2S)$ and $\eta_b(3S)$ can also be affected.
Furthermore, all $\eta_b(nS)$ states can acquire
non-negligible branching ratios into $\tau^+\,\tau^-$ due to their
mixing with $A$. 

A relatively light, CP-odd Higgs scalar 
can appear, \eg in nonminimal supersymmetric 
extensions of the Standard Model (SM) as the NMSSM
(Next-to-Minimal Supersymmetric Standard Model);
see \cite{Ellwanger:2009dp} and references therein. 
Its mass must satisfy constraints from LEP, where it
could have been produced in $\epem \to Z^* \to Z\,H$ and $H \to A\,A$
(where $H$ is a CP-even Higgs scalar). For $m_A > 10.5$\gev, where $A$
would decay dominantly into $b\bar{b}$, and $m_H < 110$\gev,
corresponding LEP constraints are quite strong~\cite{Schael:2006cr}. If 
$2m_\tau < m_A < 10.5$\gev, $A$ would decay dominantly into
$\tau^+\tau^-$ and values for $m_H$ down to $\sim 86$\gev\ are allowed 
\cite{Schael:2006cr} even if $H$ couples to the $Z$~boson with the
strength of a SM Higgs boson. 
A possible explanation of an excess
of $b\bar{b}$ events found at LEP~\cite{Dermisek:2005gg,Dermisek:2006wr}
provides additional motivation 
for a CP-odd Higgs scalar with a mass
below 10.5\gev. However, a recent (but preliminary) analysis from 
ALEPH has found no further evidence of such an excess.

\begin{figure}[b]
   \begin{center}
      \includegraphics*[width=\figwid]{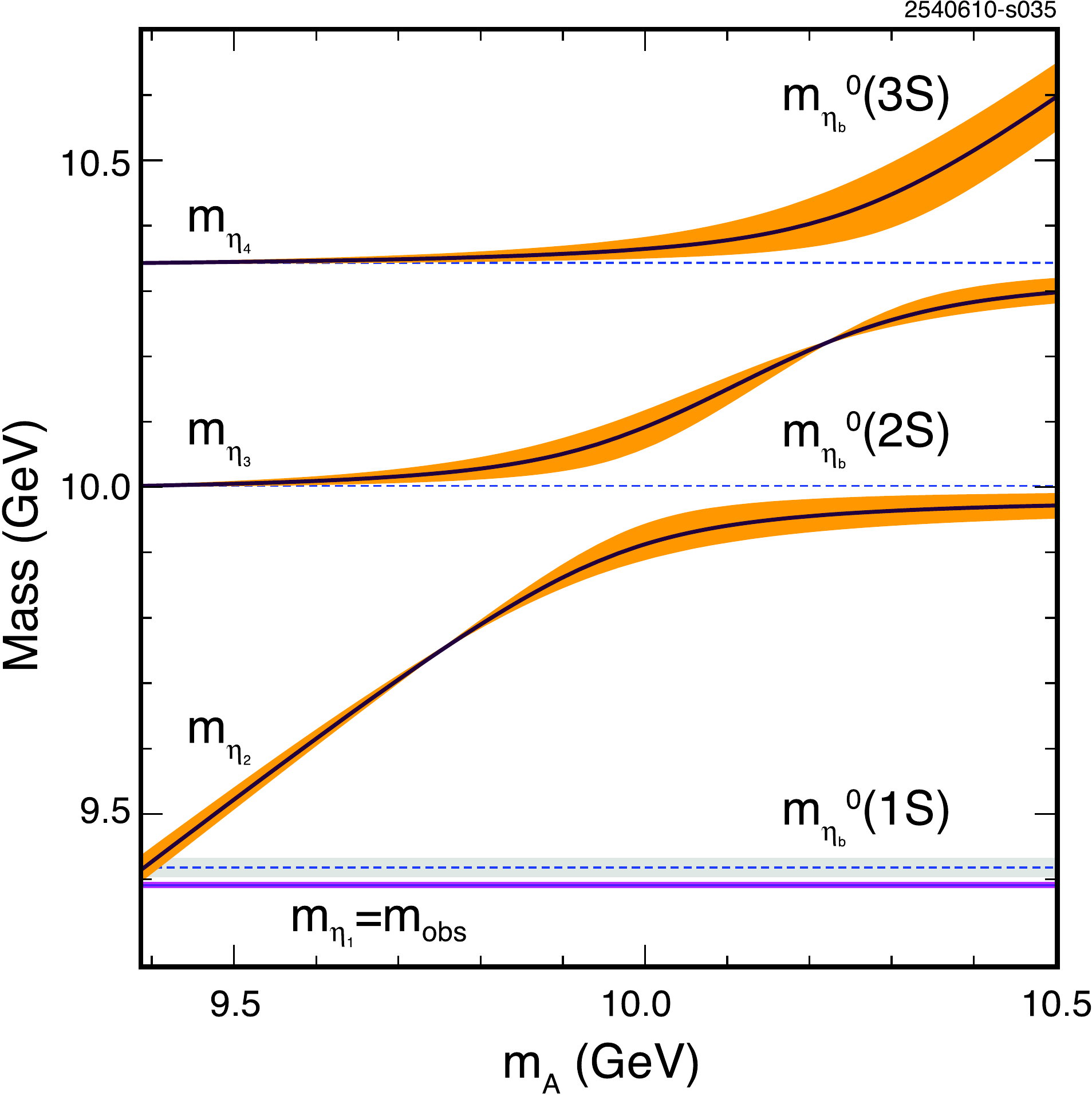}
      \caption{The masses of all eigenstates as function of $m_A$}
      \label{fig:SpecTh_heavy_masses}
   \end{center}
\end{figure}

The masses of all 4 physical states 
(denoted by $\eta_i$, $i=1\dots 4$) as functions of $m_A$ 
are shown together with the uncertainty bands in
\Fig{fig:SpecTh_heavy_masses} \cite{Domingo:2009tb}. 
By construction, $m_{\eta_1} \equiv m_{\rm obs}[\etab]$ is constrained
to the measured value. For clarity the assumed values for
$m_{\eta_b^0(nS)}$ are indicated as horizontal dashed lines. 
For $m_A$ not far above 9.4\gev, the effects of the mixing on
the states $\eta_b^0(2S)$ and $\eta_b^0(3S)$ are negligible, but for
larger $m_A$ the spectrum can differ considerably from the
standard one.

\begin{figure}[t]
   \begin{center}
      \includegraphics*[width=\figwid]{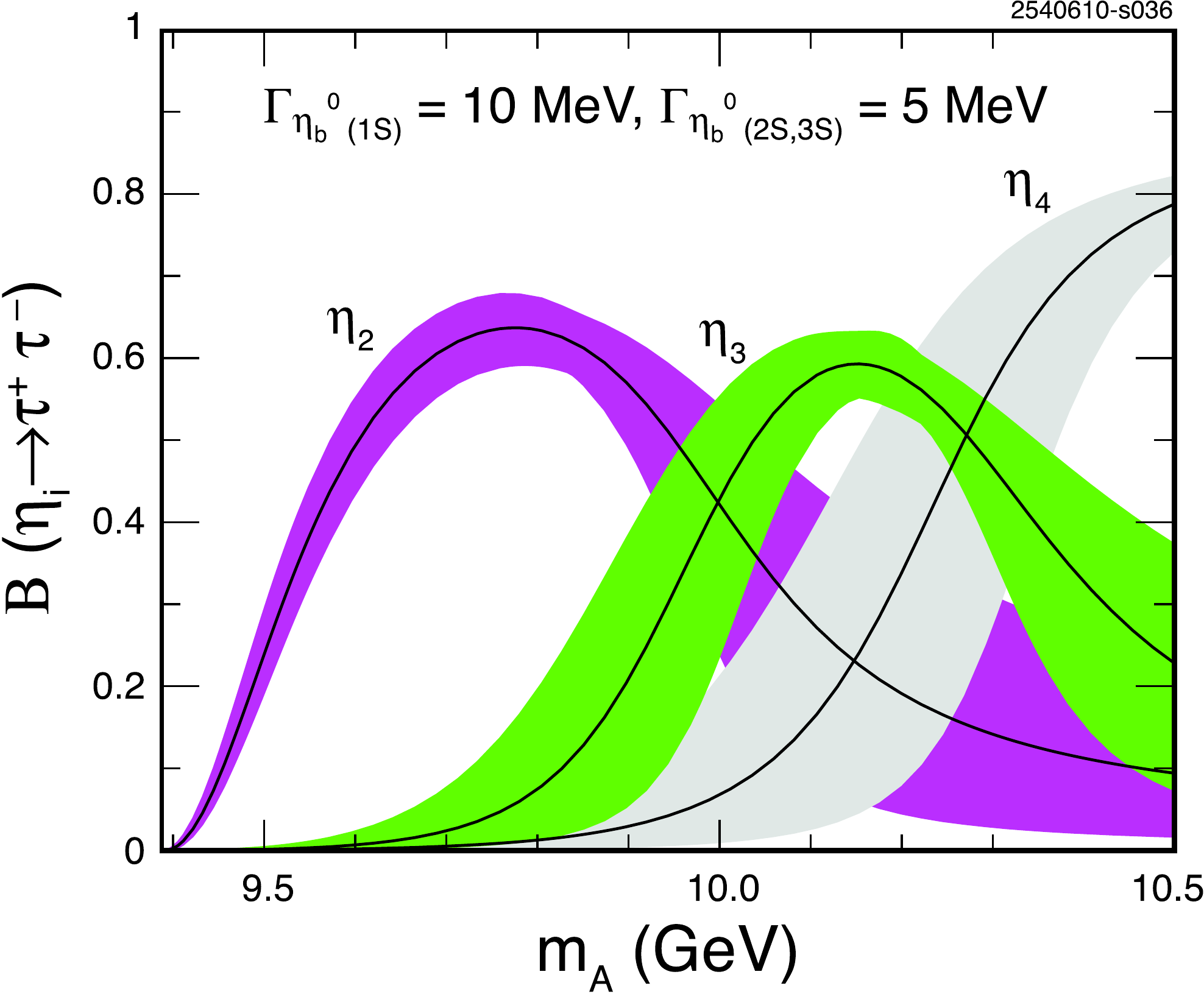}
      \caption{The branching ratios into $\tau^+\,\tau^-$ for the eigenstates
               $\eta_2$, $\eta_3$ and $\eta_4$ as functions of $m_A$}
      \label{fig:SpecTh_BrmA}
   \end{center}
\end{figure}

Now turning to the tauonic branching ratios of the $\eta_i$ 
states induced by their $A$-components, 
assuming $\Gamma_{\eta_b^0(1S)} \sim 5-20$\mev,
the predicted branching ratio of $\eta_b(1S)\to\tau^+\tau^-$ 
is compatible with the \babar\ upper
limit of 8\% at $90\%$ CL \cite{Aubert:2009cka}. 
For the heavier $\eta_i$ states,   
the corresponding branching fractions vary with $m_A$
as shown in \Fig{fig:SpecTh_BrmA}, where
$\Gamma_{\eta_b^0(1S)} \sim 10$\mev\ and 
$\Gamma_{\eta_b^0(2S)} \sim \Gamma_{\eta_b^0(3S)} \sim 5$\mev. 
With larger (smaller) total widths, these branching fractions would be 
smaller (larger).

The predicted masses and branching fractions
in \Figs{fig:SpecTh_heavy_masses} and \ref{fig:SpecTh_BrmA},
respectively, together with an accurate test of 
lepton-universality-breaking in
$\Upsilon$ decays~\cite{SanchisLozano:2002pm,Domingo:2009tb},
can play an important role both in the experimental search
for excited $\eta_b(nS)$ states and subsequent 
interpretation of the observed spectrum.
Such comparisons would test the hypothesis of 
$\eta_b$ mixing with a light CP-odd Higgs boson.

\subsubsection{Supersymmetric quarkonia}

As the top quark is too heavy to form QCD bound states 
(it decays via the electroweak transition $t \to b W$), 
the heaviest quarkonia that can be formed in the 
Standard Model are restricted to the 
bottom-quark sector. This, however, does not 
preclude the existence of the heavier quarkonium-like
structures in theories beyond the Standard Model. 

In particular, there exists some interest in detection of 
bound states of a top {\it squark}, $\widetilde t$, \ie 
{\it stoponium}. From a practical point of view, 
observation of such a state would allow precise
determination of squark masses, since stoponium decays via 
annihilation into SM particles do not have 
missing energy signatures with weakly-interacting 
lightest supersymmetric particles (LSPs). From 
a theoretical point of view such a scenario could be 
interesting because it could generate correct relic 
abundance for neutralino Dark Matter and baryon number 
asymmetry of the universe~\cite{Balazs:2004ae}.

As it turns out, some supersymmetric models allow for a 
relatively light top squark that can form bound states. One 
condition for such an occurrence involves forbidding two-body decay channels
\beq
\widetilde t_1 \to b \widetilde C_1, \qquad  \widetilde t_1 \to t 
\widetilde N_1,
\eeq
where $\widetilde t_1$ is the lighter mass eigenstate of a 
top squark, and $\widetilde C_1$ and  $\widetilde N_1$ 
are the lightest chargino and neutralino states, respectively. 
These decay channels can be kinematically forbidden if
$m_{\widetilde t_1}-m_{\widetilde C_1} < m_b$ and 
$m_{\widetilde t_1}-m_{\widetilde N_1} < m_t$, which can be 
arranged in SUSY models (although not in mSUGRA). Other decay 
channels are either flavor-violating or involve more
than two particles in the final state and are therefore 
suppressed enough to allow formation of stoponia. Since 
top squarks are scalars, the lowest-energy bound state 
$\eta_{\widetilde t}$ has spin zero and mass predicted in the 
range of 200-800\gev~\cite{Drees:1993uw,Hagiwara:1990sq,Martin:2008sv}.

The $\eta_{\widetilde t}$ spin-zero state can be most effectively 
produced at the LHC in the $gg \to \eta_{\widetilde t}$ 
channel. Possible decay channels of $\eta_{\widetilde t}$ 
include $\gamma\gamma$, $gg$, $ZZ$, $WW$, as well as 
hadronic and leptonic final states. It is the $\gamma\gamma$ 
decay channel that has recently received most attention. 

The production cross section for a stoponium $\eta_{\widetilde t}$ 
detected in the $\gamma\gamma$ channel can be written as~\cite{Martin:2008sv}
\beqa
\label{eqn:SpecTh_StopGG}
\sigma(pp \to \eta_{\widetilde t} \to \gamma\gamma) &=& \frac{\pi^2}{8 m^3_{\eta_{\widetilde t}}}
{\cal B}(\eta_{\widetilde t} \to gg) \Gamma (\eta_{\widetilde t} \to \gamma \gamma)
\nonumber \\
&\times & \int_\tau^1dx \frac{t}{x} g(x,Q^2) g (\tau/x, Q^2),~~~
\eeqa
where $g(x,Q^2)$ is the gluon parton distribution function, and 
${\cal B}(\eta_{\widetilde t} \to gg) = 
\Gamma(\eta_{\widetilde t} \to gg) /\Gamma_{tot}$ 
is the branching fraction for 
$\eta_{\widetilde t} \to gg$ decay. 
Note that \Eq{eqn:SpecTh_StopGG} contains $\eta_{\widetilde t}$ decay widths 
into $\gamma\gamma$ and $gg$ channels that depend on the 
value of the $\eta_{\widetilde t}$ wave function at the origin,
\beqa
 \Gamma (\eta_{\widetilde t} \to \gamma \gamma) &=& \frac{4}{3} \frac{8}{9}
 \frac{4 \pi \alpha^2}{m^2_{\eta_{\widetilde t}}}
 \left| \psi_{\eta_{\widetilde t}}(0) \right|^2 , 
 \nonumber \\
 \Gamma (\eta_{\widetilde t} \to gg) &=&  
\frac{9 \ \als}{8 \ \alpha} \Gamma (\eta_{\widetilde t} \to \gamma \gamma)\,. 
\eeqa
These decay widths have recently been evaluated taking into 
account one-loop QCD corrections~\cite{Martin:2009dj,Younkin:2009zn}. 
These next-to-leading order QCD corrections to 
$\Gamma (\eta_{\widetilde t} \to \gamma \gamma)$ 
significantly decrease~\cite{Martin:2009dj} the rate
while NLO corrections to the production cross section tend to 
increase it~\cite{Younkin:2009zn}, leaving the product in 
\Eq{eqn:SpecTh_StopGG} largely unchanged from the 
leading-order prediction.

It appears that, if possible, observation of stoponium would 
require significant statistics at the LHC. For example, for stoponium 
masses on the order of or less than 300\gev, an integrated luminosity
of 10~fb$^{-1}$~\cite{Martin:2008sv} is required. 
Larger mass values would 
require even greater statistics. Thus discovery of stoponium 
decays could only improve determinations of the squark mass, which 
would probably be already available using other methods. 
The possibility of learning about bound-state QCD dynamics in a 
regime where calculations are under better theoretical 
control makes this study worthwhile.

\subsubsection{Invisible decays of $\psi$ and $\Upsilon$}

Measurements of invisible, meaning undetected, decay rates of
$\psi$ and $\Upsilon$ resonances 
can discover or place strong constraints on dark
matter scenarios where candidate dark matter constituents
are lighter than the $b$-quark \cite{McElrath:2005bp}. 

According to the SM, invisible decays of the \UnS{1}\ 
can proceed via $b\bar{b}$ annihilation into
a $\nu\bar{\nu}$ pair with a branching fraction 
$\Brat(\UnS{1}\to \nu\bar{\nu})\ \simeq 10^{-5}$ \cite{Chang:1997tq}, 
well below current experimental sensitivity. However,
models containing low-mass dark matter components
might enhance such invisible decay modes up to observable
rates \cite{Fayet:1979qi,Fayet:1991ux,Fayet:2006sp,Yeghiyan:2009xc}. 
Interestingly, from the astroparticle and cosmological side, 
DAMA and CoGeNT experiments have reported the observation of events 
compatible with a Light Dark-Matter (LDM) 
candidate with a mass inside the interval 
[5,10]\gev~\cite{Fitzpatrick:2010em}. Such a mass range for 
LDM constituents is especially attractive in 
certain cosmological scenarios, \eg the Asymmetric
Dark Matter model where the cosmological dark matter density 
arises from the baryon asymmetry of the universe and is expected
to have an LDM mass of order [1-10]\gev\ (see 
\cite{Kaplan:2009ag,Cohen:2010kn} and references therein).

Searches for the $\UnS{1}\to \mathrm{invisible}$ mode have been carried out
by Argus~\cite{Albrecht:1986ht}, CLEO~\cite{Rubin:2006gc}, 
Belle~\cite{Tajima:2006nc}, and \babar~\cite{:2009qd}
using the cascade decay, $\UnS{2,3} \to \dipi\UnS{1}$, in which
the dipion tags the invisibly-decaying $\UnS{1}$.
The most recent experimental data from \babar~\cite{:2009qd} yield 
an upper limit at the 90\%~CL of
\beq
\Brat(\UnS{1}\to\mathrm{invisible})\, <\,3 
\times 10^{-4}~{\rm at~}90\%~{\rm CL}\,,
\eeq
which is only one order of magnitude above the SM expectation. 
This bound only applies to LDM candidates with masses less than
half the $\UnS{1}$ mass. The corresponding limit from 
BES~\cite{Ablikim:2007ek} is
\beq
\Brat(\jpsi \to \mathrm{invisible})\ <\ 7.2 \times 10^{-4}~{\rm at~}90\%~{\rm CL}\,.
\eeq

However, the invisible decay mode induced by scalar or pseudoscalar mediators
(like CP-odd or CP-even Higgs bosons) actually vanishes,
independent of the character of the dark matter candidate
(either scalar, Dirac or Majorana fermion), if
the decaying resonance has $J^{PC}=1^{--}$~\cite{Fayet:2009tv}. 
In order to get a non-vanishing decay rate, a new vector $U$-boson 
associated with gauging an extra $U(1)$ symmetry (\ie a new kind 
of interaction) would be required~\cite{Fayet:2009tv}.
Hence a different type of search has been performed by looking at the decay 
$\UnS{1} \to \gamma + \mathrm{invisible}$, which can proceed via a light 
scalar or pseudoscalar Higgs mediator decaying into a LDM pair. 
As discussed in \Sec{sec:SpecTh_mixing}, a light CP-odd Higgs boson 
$A_1$ mixing with $\eta_b$ resonances could naturally become the
mediator of the decay into undetected dark matter particles:
$\Upsilon \to \gamma\ A_1 (\to \mathrm{invisible})$.
The result for \UnS{3}\ decays from \babar~\cite{:2008st} is
\beqa
\Brat(\UnS{3} \to \gamma+\mathrm{invisible})\, <\, (0.7-30) \times 
10^{-6}\non\\
{\rm at~}90\%~{\rm CL},~{\rm for~} s^{1/2}_{\rm inv} < 7.8\gev\,,
\eeqa
where $s_{\rm inv}$ denotes the invariant-mass squared of the 
hypothetical LDM pair. The corresponding limit from CLEO~\cite{Insler:2010jw}
for \jpsi\ decays is
\beqa
\Brat(\jpsi\to \gamma+\mathrm{invisible})\, <\, (2.5-6.5) \times 
10^{-6}\non\\
{\rm at~}90\%~{\rm CL},~{\rm for~} s^{1/2}_{\rm inv} < 960\mev\,.
\eeqa
Experimental systematic effects make
improving limits from searches of this type at higher mass
difficult. The energy of the final state photon gets progressively 
smaller for larger invisible mass;
at low energy, the energy resolution and number of fake photon 
candidates are all typically less favorable than at higher
photon energy. Nevertheless, such improvements, if achieved,
would provide important constraints on theoretical
possibilities by reaching higher LDM masses.

\section[Decay]{Decay$^8$}

\addtocounter{footnote}{1}
\footnotetext{Contributing authors:
E.~Eichten$^\dag$, R.~E.~Mitchell$^\dag$, A.~Vairo$^\dag$, 
A.~Drutskoy, S.~Eidelman,
C.~Hanhart, B.~Heltsley,  G.~Rong, and C.-Z.~Yuan}
\label{sec:DecChapter}

\subsection{Radiative transitions}
\label{sec:Dec_radtrans}

  An electromagnetic transition between quarkonium states, which occurs via emission
of a photon, offers the distinctive experimental signature of
a monochromatic photon, a useful production mechanism for discovery and
study of the lower-lying state, and a unique window on the dynamics
of such systems. Below we first review the status and open questions
regarding the relevant theoretical frameworks and tools, and then describe
important measurements of charmonium and bottomonium electromagnetic transitions.

\subsubsection{Theoretical status}
\label{sec:Dec_EMT1}

The non\-relativistic nature of heavy quarkonium may be exploited 
to calculate electromagnetic transitions.
Non\-rela\-ti\-vi\-stic effective field theories provide a way 
to systematically implement the expansion in the 
relative heavy-quark velocity, $v$.
Particularly useful are 
non\-relativistic QCD (NRQCD) coupled to 
electromagnetism~\cite{Caswell:1985ui,Bodwin:1994jh},
which follows from QCD (and QED) by integrating out 
the heavy quark mass scale $m$,  and 
potential NRQCD coupled to 
electromagnetism~\cite{Pineda:1997bj,Brambilla:1999xf,Brambilla:2004jw,Brambilla:2005zw}, 
which follows from NRQCD (and NRQED) by integrating 
out the momentum transfer scale $mv$.

Electromagnetic transitions may be classified in terms of 
electric and magnetic transitions between eigenstates
of the leading-order pNRQCD Hamiltonian. The states are classified in terms of 
the radial quantum number, $n$, the orbital angular momentum, $l$, 
the total spin, $s$, and the total angular momentum, $J$. 
In the non\-relativistic limit,  
the spin dependence of the quarkonium wave function decouples from the spatial
dependence. The spatial part of the wave function, $\psi(x)$, can
be expressed in terms of a radial wave function, $u_{nl}(r)$, and 
the spherical harmonics, $Y_{lm}$, as  $\psi (x) = Y_{lm}(\theta, \phi) u_{nl}(r)/r$.
The spatial dependence of the electromagnetic transition amplitudes reduces to
expectation values of various functions of quark position and momentum
between the initial- and final-state wave functions~\cite{Brambilla:2004wf}. 

Magnetic transitions flip the quark spin.
Transitions that do not change the orbital angular momentum are 
called magnetic dipole, or M1, transitions.
In the non\-relati\-vi\-stic limit, the spin-flip transition 
decay rate between an initial state   
$i= n\sLj{2s+1}{l}{J}$ and a final state 
$f = n^{\prime}\sLj{2s^\prime+1}{l}{J^{\prime}}$ is:
\beqa
\Gamma (\, i  \stackrel{\mathrm{M1}}{\longrightarrow} \gamma + f\, ) =\hspace{2in}\non\\
\frac{16}{3} \alpha\, e_{Q}^{2}\,\frac{E_\gamma^{3}}{m_i^{2}}\,(\,2J^{\prime}+\,1)
\,{\rm S}^{\rm M}_{if}\,|\mathcal{M}_{if}|^{2}~,~~
\label{eqn:Dec_magnetic}
\eeqa
where $e_Q$ is the electrical charge of the 
heavy quark $Q$ ($e_b  = -1/3$, $e_c=2/3$), 
$\alpha$ the fine structure constant,  
$E_\gamma = (m_i^2-m_f^2)/(2m_i)$ is the photon energy, and 
$m_{i}$, $m_f$ are the masses of the 
initial- and final-state quarkonia, respectively.
The statistical factor ${\rm S}^{\rm M}_{if}={\rm S}^{\rm M}_{fi}$ reads
\beqa
{\rm S}^{\rm M}_{if} = 6\,(2s+1)(2s^{\prime}+1)\times\hspace{1in}\non\\
\left\{ \begin{array}{ccc}
                               J & 1 & J^{\prime}  \\
                               s^{\prime} & l & s
                                    \end{array}\right\}^{2} 
                        \left\{ \begin{array}{ccc}
                               1 & \frac{1}{2} & \frac{1}{2} \\
                               \frac{1}{2} & s^{\prime} & s 
                                    \end{array}\right\}^{2}~~.~~~
\eeqa
For $l=0$ transitions, $S^{\rm M}_{if} = 1$. 
For equal quark masses $m$, the overlap integral ${\cal M}_{if}$ is given by
\beqa
\label{eqn:Dec_mtxm1}
{\cal M}_{if} = (1+\kappa_Q)\times\hspace{2in}\non\\
 \int_0^\infty dr \,\, u_{nl}(r)\,
u_{n^{\prime}l}^{\prime}(r) \,\, j_0\left(\frac{E_\gamma \,r}{2}\right)~~,~~~
\eeqa
where $j_n$ are spherical Bessel functions and  
$\kappa_Q$ is the anomalous magnetic moment of a heavy quarkonium 
$Q\bar{Q}$. In pNRQCD, the quantity $1 + \kappa_Q$ is the 
Wilson coefficient of the operator 
$S^\dagger {\bf \sigma}\cdot e_Q{\bf B}^{\emrm}/(2m) S$, 
where ${\bf B}^{\emrm}$ is the magnetic field 
and $S$ is a $Q\bar{Q}$ color-singlet field.

Electric transitions do not change the quark spin.  
Transitions that change the orbital angular momentum 
by one~unit are called electric dipole, or E1, transitions. 
In the non\-rela\-ti\-vi\-stic limit, the spin-averaged 
electric transition rate between
an initial state $i=n\sLj{2s+1}{l}{J}$ and a final state
$f=n^{\prime}\sLj{2s^{\prime}+1}{l^{\prime}}{J^{\prime}}$ 
($l = l^\prime \pm 1$) is
\begin{equation}
\Gamma(\,\,i  \stackrel{\mathrm{E1}}{\longrightarrow} \gamma + f\,\,) =
\frac{4}{3}\,\alpha\, e_{Q}^{2}\,E_\gamma^{3}\,(2J^{\prime}+1)\,{\rm S}^{\rm E}_{if}  
\,|\mathcal{E}_{if}|^{2}~~,~ 
\label{eqn:Dec_electric}
\end{equation}
where the statistical factor ${\rm S}^{\rm E}_{if}={\rm S}^{\rm E}_{fi}$ is
\begin{equation}
{\rm S}^{\rm E}_{if} = \max{(l,l^{\prime})} \left\{
          \begin{array}{ccc}
            J & 1 & J^{\prime}  \\
            \l^{\prime} & s & l
            \end{array}\right\}^{2}~~.~
\end{equation}
The overlap integral ${\cal E}_{if}$ for equal quark masses $m$ is given by
\begin{eqnarray}
\label{eqn:Dec_Eifsize}
{\cal E}_{if} & = &
\frac{3}{E_\gamma}\int_0^\infty \!\!\!\!dr \,\, u_{nl}(r)\,
  u_{n^{\prime}\,l^{\prime}}(r)\,\,\times\nonumber\\
&& \left[\frac{E_\gamma\, r}{2} j_0\left(\frac{E_\gamma\, r}{2}\right) 
- j_1\left(\frac{E_\gamma\, r}{2}\right)\right]~~.~
\end{eqnarray}
Since the leading-order operator responsible for the electric transition 
does not undergo renormalization, the electric transition rate does 
not depend on a Wilson coefficient, analogous to the case of the
quarkonium magnetic moment appearing 
in the magnetic transitions.

If the photon energy is smaller than the typical inverse radius of the quarkonium, 
we may expand the overlap integrals in $E_\gamma r$, generating electric and magnetic 
multipole moments. At leading order in the multipole expansion, 
the magnetic overlap integral reduces to ${\cal M}_{if} = \delta_{nn'}$.
Transitions for which $n=n'$ are called {\it allowed} M1 transitions, 
transitions for which $n\neq n'$ are called {\it hindered} transitions. 
Hindered transitions happen only because of higher-order  
corrections and are suppressed by at least $v^2$ with respect to the allowed ones. 
At leading order in the multipole expansion
the electric overlap integral reduces to 
\beq
 {\cal E}_{if} = 
\int_0^\infty \!\!\!\!dr \,\, u_{nl}(r)\,r\,u_{n^{\prime}l^{\prime}}(r)~~.~~
\eeq
Note that E1 transitions are more copiously observed than allowed M1 transitions, because 
the rates of the electric transitions are enhanced by $1/v^2$ with respect to the magnetic ones.
Clearly, the multipole expansion is always allowed for transitions between states 
with the same principal quantum numbers ($E_\gamma  \sim mv^4 \hbox{ or } mv^3 \ll mv$) 
or with contiguous principal quantum numbers ($E_\gamma  \sim mv^2 \ll mv$).
For  transitions that involve widely separated states, the hierarchy $E_\gamma  \ll mv$ 
may not be realized. For example, in $\UnS{3}\to \gamma\etab$, we have  
$E_\gamma  \approx 921\mev$, which is smaller than the 
typical momentum transfer in the $\etab$, 
about $1.5\gev$~\cite{Kniehl:2003ap}, but may be comparable to or 
larger than the typical momentum 
transfer in the $\UnS{3}$. On the other hand, 
in $\psip\to\gamma\chi_{c1}$, we have 
$E_\gamma  \approx 171\mev$, which is smaller than 
the typical momentum transfer in both the $\psip$ 
and the $\chi_{c1}$. 

Beyond the non\-relativistic limit, 
\Eqs{eqn:Dec_magnetic} and (\ref{eqn:Dec_electric}) get 
corrections. These are radiative corrections counted in 
powers of $\als(m)$ and relativistic 
corrections counted in powers of $v$. 
These last ones include proper relativistic 
corrections of the type $(mv)^2/m^2$, 
recoil corrections and, for weakly-coupled  
quarkonia, also corrections of the type $\lamQ/(mv)$. 
Finally, we also have corrections 
of the type $E_\gamma/(mv)$ that 
involve the photon energy. In the charmoniun system, $v^2 \approx 0.3$,
and corrections may be as large as 30\%. Indeed, a negative correction
of about 30\% is required to bring the non\-relativistic prediction of
${\cal B}(\jpsi \to \gamma \etac)$, which is about 3\%,
close to the experimental value, which is about 2\%.
We will see that this is actually the case.
In the bottomonium system, $v^2 \approx 0.1$
and corrections may be as large as 10\%.

For a long time, corrections to the electromagnetic transitions 
have been studied almost entirely within pheno\-me\-nological 
models~\cite{Feinberg:1975hk,Sucher:1978wq,Eichten:1978tg,Kang:1978yw,
Sebastian:1979gq,Karl:1980wm,Grotch:1982bi,Moxhay:1983vu,McClary:1983xw,
Grotch:1984gf,Fayyazuddin:1993eb,Lahde:2002wj,Ebert:2002pp,Barnes:2005pb} 
(a sum rule analysis appears in \cite{Khodjamirian:1979fa}). 
We refer to reviews in \cite{Eichten:2007qx,Brambilla:2004wf}; 
a textbook presentation can be found in \cite{LeYaouanc:1988fx}. 
In contrast to models, the effective field theory approach 
allows a systematic and rigorous 
treatment of the higher-order corrections. 
The use of EFTs for electromagnetic transitions was initiated in 
\cite{Brambilla:2005zw}, in which 
a study of  magnetic transitions was performed.
The results of that analysis may be summarized in the following way.
\begin{itemize}
\item{The quarkonium anomalous magnetic moment $\kappa_Q$ does not get 
contributions from the scale $mv$: it is entirely determined by the quark anomalous 
magnetic moment. Since the quark magnetic moment appears at the scale $m$, it is 
accessible by perturbation theory: $\kappa_Q =  2\als(m)/(3\pi) + {\cal O}{(\als^2)}$.
As a consequence, $\kappa_Q$ is a small positive quantity, about 0.05 in the bottomonium case 
and about 0.08 in the charmonium one. This is confirmed by lattice 
calculations~\cite{Dudek:2006ej} 
and by the analysis of higher-order multipole amplitudes 
(see \Sec{sec:Dec_hm}).
}
\item{QCD does not allow for a scalar-type contribution to 
the magnetic transition rate. A scalar interaction is 
often postulated in phenomenological models.}
\end{itemize}
The above conclusions were shown to be valid at any order of 
perturbation theory as well as {\it non}\-perturbatively.
They apply to magnetic transitions from any quarkonium state.
For ground state magnetic transitions, 
we expect that perturbation theory may be used 
at the scale $m v$. Under this assumption, 
the following results were found at relative order $v^2$.
\begin{itemize}
\item{The magnetic transition rate between the vector and pseudoscalar quarkonium 
ground state, including the leading relativistic correction (parametrized by $\als$ at the 
typical momentum-transfer scale $m_i\als/2$) and the leading anomalous magnetic moment
(parametrized by $\als$ at the mass scale $m_i/2$), reads 
\begin{eqnarray}
&\Gamma(i \to \gamma + f) =
\dfrac{16}{3}\,\alpha\, e_{Q}^{2}\,\dfrac{E_\gamma^3}{m_i^2}\,\times&\nonumber\\
 & \left[\, 1 + \dfrac{4}{3}\,\dfrac{\als(m_i/2)}{\pi}
 - \dfrac{32}{27}\,\als^2(m_i\als/2)\,\right]~~,~  &
\label{eqn:Dec_magneticrel}
\end{eqnarray}
in which $i= 1^30_1$ and $f = 1^10_1$.
This expression is not affected by non\-perturbative contributions. 
Applied to the charmonium and 
bottomonium case it gives: 
$\mathcal{B}(\jpsi \to \gamma \etac) = (1.6 \pm 1.1) \%$ 
 (see \Sec{sec:Dec_1Sc} for the experimental situation) and 
$\mathcal{B}(\UnS{1} \to \gamma \etab) = (2.85 \pm 0.30) \times 10^{-4}$ 
 (see \Sec{sec:Dec_1Sb} for some experimental perspectives).
}
\item{A similar perturbative analysis, performed for hindered 
magnetic transitions, mischaracterizes the experimental 
data by an order of magnitude, 
pointing either to a breakdown of the 
perturbative approach for quarkonium states with principal 
quantum number $n >1$, or 
to large higher-order relativistic corrections.}
\end{itemize}

The above approach is well suited to studying the lineshapes 
of the \etac\ and $\etab$ in the 
photon spectra of $\jpsi \to \gamma \etac$ and 
$\UnS{1} \to \gamma \etab$, respectively.
In the region of $E_\gamma \ll m\als$, at leading order, 
the lineshape is given by~\cite{Brambilla:2010}
\begin{eqnarray}
\dfrac{d\Gamma}{dE_\gamma}\left(i \to \gamma + f\right) &=&
\dfrac{16}{3}\dfrac{\alpha\, e_Q^2}{\pi}\dfrac{E_\gamma^3}{m_i^2}\,\times\nonumber\\ 
&&\dfrac{\Gamma_f/2}{(m_i-m_f-E_\gamma)^2+\Gamma_f^2/4}\,,~~~~
\label{eqn:Dec_photonlineshape}
\end{eqnarray}
which has the characteristic asymmetric 
behavior around the peak seen in the data 
(compare with the discussion in \Sec{sec:Dec_1Sc}). 

No systematic analysis is yet available for relativistic corrections to 
electromagnetic transitions involving higher quarkonium states, 
\ie states for which $\lamQ$ is larger than the typical 
binding energy of the quarkonium.
These states are not described in terms of a Coulombic potential. 
Transitions of this kind include magnetic transitions between states with $n>1$ 
and all electric transitions, $n=2$ bottomonium states being on the boundary. 
Theoretical determinations rely on phenomenological models, 
which we know do not agree with QCD in the perturbative regime and 
miss some of the terms at relative order $v^2$~\cite{Brambilla:2005zw}. 
A systematic analysis is, in principle, possible in the same 
EFT framework developed for magnetic transitions. Relativistic corrections 
would turn out to be factorized in some high-energy coefficients, which 
may be calculated in perturbation theory, and in 
Wilson-loop amplitudes similar to those that encode the relativistic corrections 
of the heavy quarkonium potential~\cite{Pineda:2000sz}.
At large spatial distances, Wilson-loop amplitudes cannot be calculated in 
perturbation theory but are well-suited for lattice measurements.
Realizing the program of systematically factorizing relativistic corrections 
in Wilson-loop amplitudes and evaluating them on the lattice, would, for the first 
time, produce model-independent determinations of quarkonium electromagnetic 
transitions between states with $n>1$. These are the vast majority of transitions 
observed in nature. Finally, we note that, for near-threshold states such as
\psip, intermediate meson loops may provide important 
contributions~\cite{Li:2007xr}, which should be systematically accounted for.

Higher-order multipole transitions have been observed in 
experiments (see \Sec{sec:Dec_hm}), 
Again, a systematic treatment is possible in the EFT 
framework outlined above, but 
has not yet been realized.

\subsubsection{Study of $\psi(1S,2S)\to\gamma\etac$}
\label{sec:Dec_1Sc}

\begin{figure}[b]
  \begin{center}
    \includegraphics[width=\figwid]{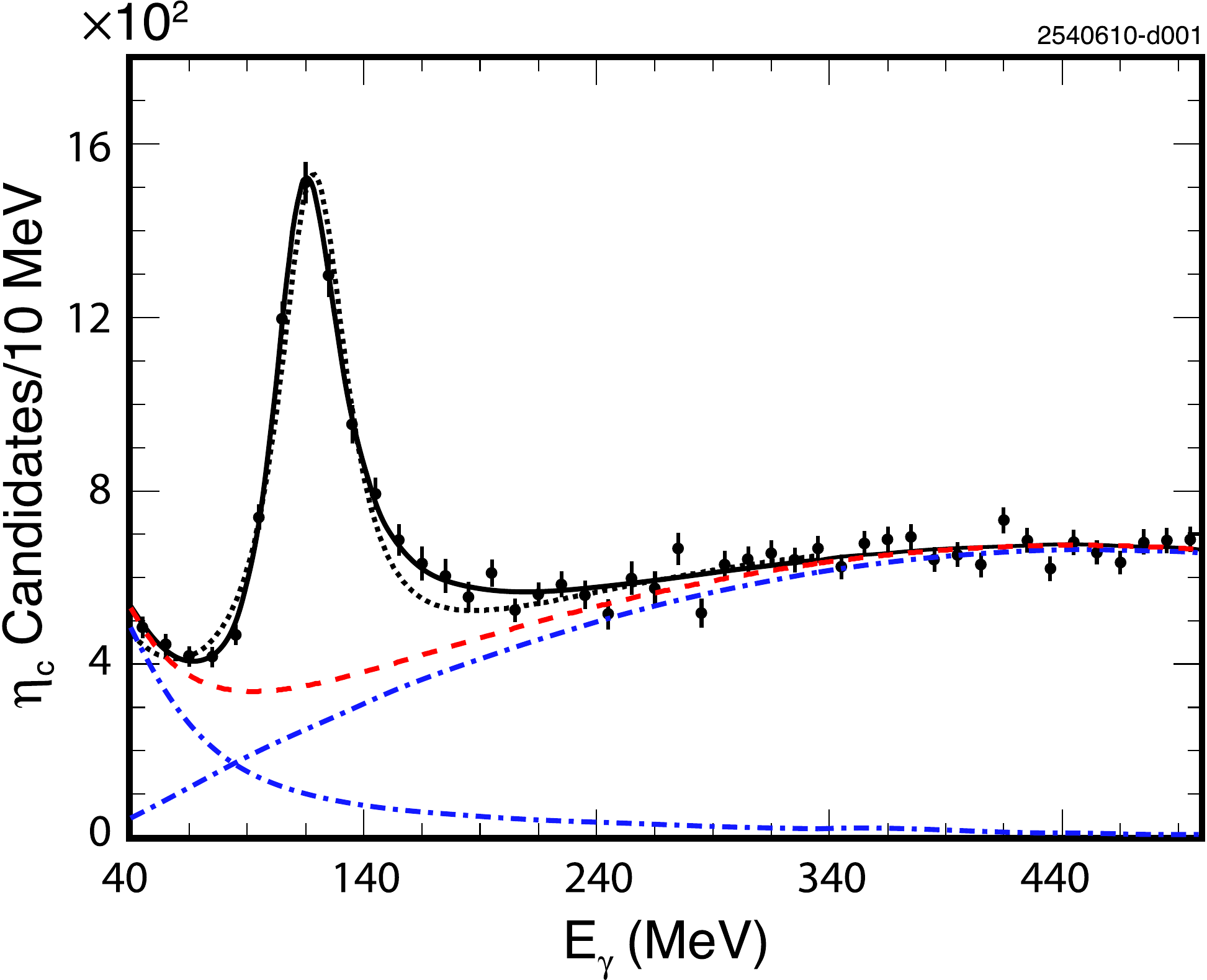}
      \caption{From CLEO~\cite{:2008fb}, 
               the photon energy spectrum from $\jpsi\to\gamma\etac$.
               The \etac\ is reconstructed in 12 different exclusive decay 
               modes.  The {\it dotted curve} represents a fit with 
               a relativistic Breit-Wigner function; the 
               {\it solid curve} uses a relativistic 
               Breit-Wigner function distorted by the energy dependence of 
               the phase-space term and the matrix element. The 
               {\it dash-dotted curves} show two components of 
               the background, which when summed become the {\it dashed curve}.
               \AfigPermAPS{:2008fb}{2009}}
      \label{fig:Dec_JPsi_GammaEtac}
   \end{center}
\end{figure}

Using a combination of inclusive and
exclusive techniques, CLEO~\cite{:2008fb} has recently measured
\begin{eqnarray}
\Brat(\jpsi\to\gamma\etac) &=& (1.98\pm0.09\pm0.30)\%\nonumber\\
\Brat(\psip\to\gamma\etac) &=& (0.432\pm0.016\pm0.060)\%\,.~~~~~~ 
\end{eqnarray}
The lineshape of the \etac\ in these M1
transitions was found to play a crucial role.  Because the width of
the \etac\ is relatively large, the energy dependence of the phase
space term and the matrix element distort the lineshape 
(see \Eq{eqn:Dec_photonlineshape}). Indeed, the photon
spectrum measured by CLEO shows a characteristic asymmetric behaviour 
(see \Fig{fig:Dec_JPsi_GammaEtac}).  The theoretical uncertainty in
this lineshape represents the largest systematic error in the
branching ratios.  Both M1 transitions are found to be larger than previous
measurements due to a combination of a larger \etac\ width and 
the first accounting for the pronounced asymmetry in the lineshape.  This process
has also been recently measured by KEDR~\cite{Anashin:2010nr}.

The distortion of the \etac\ lineshape also has implications for
the mass of the \etac.  As of 2006, there was a $3.3\sigma$
discrepancy between \etac\ mass measurements made from
$\psi(1S,2S)\to\gamma\etac$ (with a weighted average of 
$2977.3 \pm 1.3\mevcc$) and
from $\gamma\gamma$ or $p\overline{p}$ production (averaging
$2982.6\pm1.0\mevcc$).  The CLEO~\cite{:2008fb} 
analysis suggests that the solution to this problem may
lie in the lineshape of the \etac\ in the M1 radiative transitions.
When no distortion in the lineshape is used in the CLEO fit to
$\jpsi\to\gamma\etac$, the resulting \etac\ mass is consistent
with other measurements from M1 transitions ($2976.6\pm0.6\mevcc$,
statistical error only); however, when a distorted lineshape is taken
into account, the mass is consistent with those from $\gamma\gamma$ or
$p\overline{p}$ production ($2982.2\pm0.6\mevcc$, statistical error
only).  Recent measurements of the \etac\ mass in $\gamma\gamma$
production are consistent with this general
picture.  An \etac\ mass of $2986.1\pm1.0\pm2.5\mevcc$ is reported
in a Belle analysis of $\gamma\gamma\to h^{+}h^{-}h^{+}h^{-}$, where
$h = \pi,K$~\cite{:2007vb}, consistent with the higher \etac\ mass.
Also, \babar\ measures a mass of $2982.2\pm0.4\pm1.6\mevcc$ 
in $\gamma\gamma \to K_S K^{\pm} \pi^{\mp}$~\cite{Lees:2010de}.

\subsubsection{Observation of $\hsubc\to\gamma\etac$}
\label{sec:Dec_hcradtrans}

The decay chain $\psip\to\piz h_c(1P)$, $\hsubc\to\gamma\etac$
was first observed by CLEO~\cite{Rosner:2005ry,Dobbs:2008ec} using
24.5~million $\psip$ events, and later confirmed with higher
statistics by BESIII~\cite{Ablikim:2010rc} using 106~million \psip.
While the mass difference of the $h_c(1P)$ and $\chi_{cJ}(1P)$ states
is a measure of the hyperfine splitting in the $1P$ $c\overline{c}$
system, the product branching fraction can be used to glean
information about the size of the E1 transition
$h_c(1P)\to\gamma\etac$.
The product branching fraction
$\Brat (\psip\to\piz\hsubc)\times\Brat ( h_c(1P)\to\gamma\etac )$ 
was measured to be 
\beqa
&(4.19\pm0.32\pm0.45)\times10^{-4}~~&{\rm CLEO}~\text{\cite{Dobbs:2008ec}}\non\\
&(4.58\pm0.40\pm0.50)\times10^{-4}~~&{\rm BESIII}~\text{\cite{Ablikim:2010rc}}\, ,~~~
\label{eqn:Dec_hcgammaetac}
\eeqa
where both CLEO and BESIII used 
an inclusive technique (requiring reconstruction of just the $\piz$
and the transition photon and imposing appropriate
kinematic constraints), but
CLEO also utilized a fully exclusive technique
(in addition to the $\piz$ and $\gamma$,
reconstructing the \etac\ in multiple exclusive decay channels).
BESIII has also measured $\Brat(\psip\to\piz\hsubc)$
(see \Sec{sec:Dec_PsipToPi0Hc}), allowing extraction of
\beq
\Brat(\hsubc\to\gamma\etac) = ( 54.3\pm6.7\pm5.2 )\%\,.
\eeq

\begin{figure}[t]
  \begin{center}
    \includegraphics[width=\figwid]{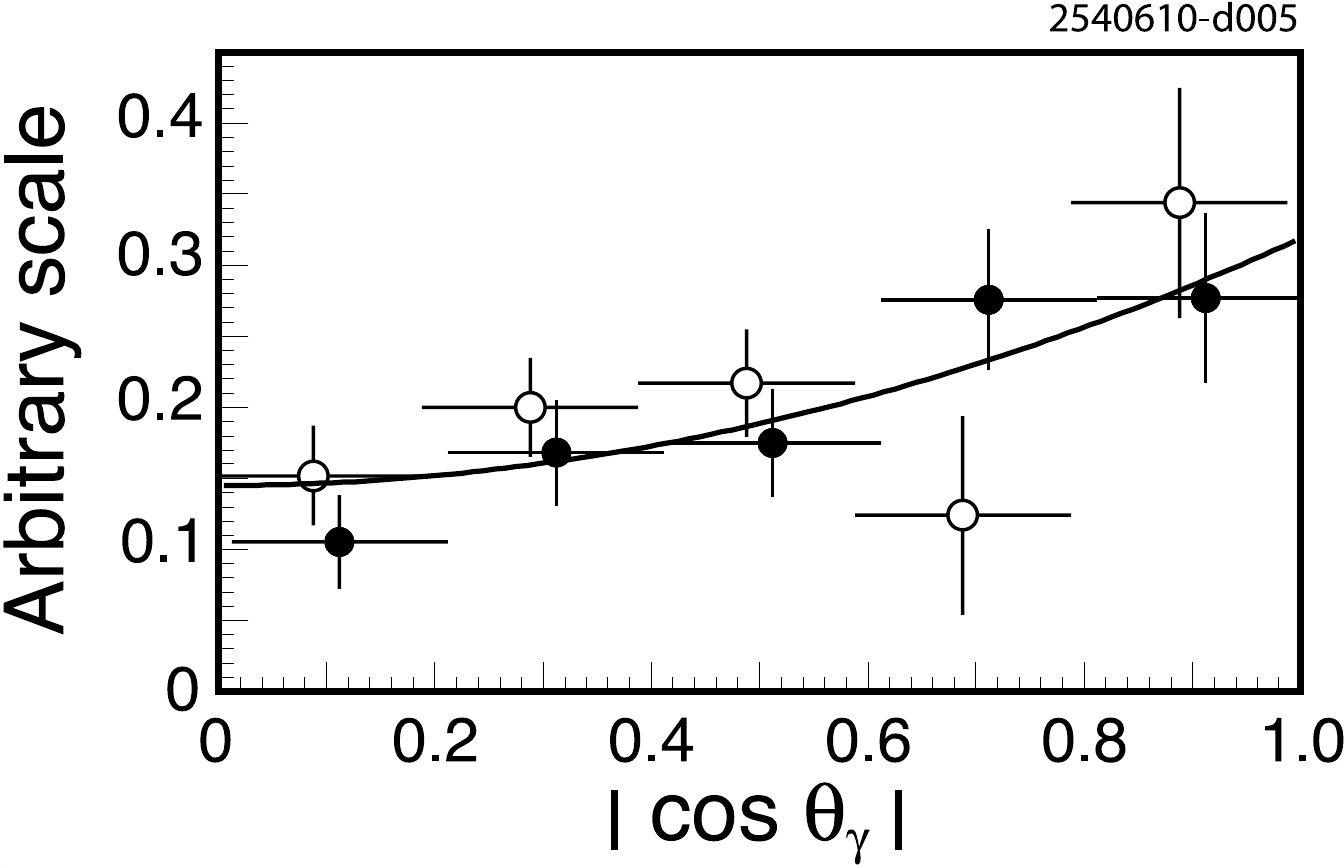}
      \caption{From CLEO~\cite{Dobbs:2008ec},
               the distribution of the photon polar
               angle in the $e^+e^-$ center-of-mass frame 
               from the transition sequence 
               $\psip\to\piz h_c(1P)$, $h_c(1P)\to\gamma\etac$.
               {\it Open circles} represent data from 
               an inclusive \etac\ decays and {\it solid circles} 
               data from exclusive \etac\ decays (see text).  
               The {\it solid curve} represents a fit of 
               both inclusive and exclusive data to $N(1+\alpha\cos^2\theta)$, 
               from which $\alpha$ was found to be $1.20\pm0.53$. 
               \AfigPermAPS{Dobbs:2008ec}{2008} }
       \label{fig:Dec_Hc_Angles}
  \end{center}
\end{figure}

As part of the same study, CLEO has measured the
angular distribution of the transition photon from
$h_c(1P)\to\gamma\etac$ (\Fig{fig:Dec_Hc_Angles}).
Fitting to a curve of the form $N(1+\alpha\cos^2\theta)$, and
combining the results from the inclusive and exclusive analyses, it
was found that $\alpha=1.20\pm0.53$, consistent with $\alpha=1$, the
expectation for E1 transitions.

\subsubsection{Nonobservation of $\psip\to\gamma\etacp$}
\label{sec:Dec_PsipToGEtac}

  After years of false alarms, the \etacp\ was finally
observed in $B$-decays and two-photon fusion 
(see \Sec{sec:SpecExp_etac2s}).
In an attempt
to both discover new decay modes and to observe it in a
radiative transition, CLEO~\cite{:2009vg} modeled
an \etacp\ analysis after its effort on the $\etac$,
wherein a systematic study of many exclusive hadronic decay modes
aided in measuring the lineshapes and branching fractions
in $\psi(1S,2S)\to\gamma\etac$ transitions~\cite{:2008fb}. Eleven
modes, which were chosen based in part upon success
in finding similar \etac\ decays, were sought
in the exclusive decay chain $\psip\to\gamma\etacp$,
$\etacp\to\,{\rm hadrons}$ in CLEO's
26~million \psip\ sample. One of the modes sought was the
dipion transition $\etacp\to\dipi\etac$, which used
proven hadronic decay modes of the $\etac$. No
\etacp\ signals were found, and
eleven product branching fraction upper limits were set.
None but one of these products can be used to directly
set a limit on the transition because none but one
have a measured \etacp\ branching fraction.
Using the \babar~\cite{Aubert:2008kp} branching fraction 
for $\etacp\to KK\pi$ allows CLEO to set an 
upper limit 
\beq
{\cal B}(\psip\to\gamma\etacp)<7.6\times10^{-4}\, .~~
\eeq
This value is greater than a phenomenological limit 
obtained~\cite{:2009vg} by assuming the matrix element is the same
as for $\jpsi\to\gamma\etac$ and correcting
the measured \jpsi\ branching fraction by the
ratio of total widths and phase-space factors,
$(3.9\pm 1.1)\times 10^{-4}$.

\subsubsection{$\psip\to\gamma\gamma\jpsi$ through $\chi_{cJ}$}
\label{sec:Dec_Bgchicj}

  One component of the CLEO~\cite{Mendez:2008kb,Adam:2005uh}
$\psip\to X \jpsi$ branching fraction analysis described in 
\Sec{sec:DecPsipBRs} addresses the $\gamma\gamma\jpsi$ final 
states that proceed through the doubly-radiative decays
$\psip\to\gamma\chi_{cJ}$, $\chi_{cJ}\to\gamma\jpsi$.
The resulting product branching fractions, measured
using CLEO's dataset of 27M \psip\ decays, are
shown in \Tab{tab:Dec_CLEO_BggJpsi}. These
inputs are used to determine $\Brat(\chi_{cJ}\to\gamma\jpsi)$ 
branching fractions, which are calculated by the Particle 
Data Group~\cite{Nakamura:2010pdg} from world averages of 
the quantities in \Tab{tab:Dec_CLEO_BggJpsi} and those
for $\psip\to\gamma\chi_{cJ}$.

\begin{table}[tb]
\caption{Results from CLEO~\cite{Mendez:2008kb,Adam:2005uh} 
for the $\psip\to\gamma\gamma\jpsi$ branching fractions 
through either $\chi_{cJ}$ intermediate states or a nonresonant (nr) 
channel }
\label{tab:Dec_CLEO_BggJpsi}
\setlength{\tabcolsep}{2.15pc}
\begin{center}
\begin{tabular}{lcc}
\hline \hline
\rule[10pt]{-1mm}{0mm}
Final state & \Brat~(\%)  \\[0.8mm]
\hline
\rule[10pt]{-1mm}{0mm}
$\gamma\,(\gamma\jpsi)_{\chi_{c0}}$ &$0.125\pm0.007\pm0.013$\\[0.8mm]
$\gamma\,(\gamma\jpsi)_{\chi_{c1}}$ &$3.56\pm0.03\pm0.12$\\[0.8mm]
$\gamma\,(\gamma\jpsi)_{\chi_{c2}}$ &$1.95\pm0.02\pm0.07$\\[0.8mm]
$\gamma\,(\gamma\jpsi)_{\rm nr}$    &$\le 0.1$\\[0.8mm]
\hline \hline
\end{tabular}
\end{center}
\end{table}

  A substantial difference between the original~\cite{Adam:2005uh} and
final~\cite{Mendez:2008kb} CLEO analyses,
aside from an eightfold increase in statistics, is the treatment of the 
transition through $\chi_{c0}$, which has the smallest rate of the three.
The primary systematic challenge for this mode is dealing with its 
small $\gamma\jpsi$ branching fraction relative to that of $\chi_{c1}$. 
In the energy spectrum of the lower energy photon in such decays,
there are peaks near 128, 172, and 258\mev, corresponding
to the intermediate $\chi_{c2}$,  $\chi_{c1}$, and $\chi_{c0}$
states, respectively (see Fig.~4 of ~\cite{Mendez:2008kb}).
However, due to nonzero photon energy-measurement resolution
and nonzero natural widths of the $\chi_{cJ}$, 
the three peaks overlap one another.
In particular, the high-side tail of
the lower-energy photon's spectrum for the transition through $\chi_{c1}$ has a 
significant contribution in the $\chi_{c0}$ energy region
(relative to the small $\chi_{c0}$ signal), a fact which 
introduces subtleties to the analysis. Because of the proximity of the 
photon lines to one another and the large disparity of rates, the 
measured product branching fraction for the $\chi_{c0}$ transition 
is sensitive to $\Gamma(\chi_{c1})$ and to the detailed lineshape 
in these decays, \ie some of the apparent $\chi_{c0}$ signal
is actually feedacross from $\chi_{c1}$. The second CLEO 
analysis~\cite{Mendez:2008kb} implemented the $E_\gamma^3$-weighting 
in its MC simulation of the transition,
as required from phase-space considerations for E1 decays given in 
\Eq{eqn:Dec_electric}. The inclusion of the $E_\gamma^3$ factor 
significantly increases the contribution
of $\chi_{c1}$ near the $\chi_{c0}$ peak (relative to not using this factor).
However, even if the MC simulation of these decays
allows for a $\chi_{c1}$ natural width 
of up to 1\mev (about $2.5\sigma$ higher than the nominal 
value~\cite{Amsler:2008zzb}) and uses the
expected E1-transition energy dependence of the lineshape, the region 
in the lower-energy photon spectrum in the valley 
between the $\chi_{c1}$ and $\chi_{c0}$ peaks
showed an excess of events in the CLEO data over the number expected. 
One hypothesis for filling that deficit suggested by CLEO is
the {\it nonresonant} (nr) decay $\psip\to\gamma\gamma\jpsi$.
CLEO estimated that a branching fraction of $\approx 0.1$\%
or smaller could be accommodated by the data. However,
CLEO did {\it not} claim observation of this nonresonant decay mode
due to a combination of limited statistics, uncertainties about $\chi_{cJ}$ widths,
and possible dynamical distortion of the lineshape,
all matters deserving further attention.
In fact, such a distortion was later observed in radiative transitions
from \psip\ to \etac\ (see \Sec{sec:Dec_1Sc}),
raising the importance of verifying the assumed photon lineshapes
with data for any future analysis.

A preliminary analysis of nonresonant 
$\psip\to\gamma\gamma\jpsi$ decays from BESIII 
has been released~\cite{besggjpsiICHEP,besggjpsiMESON},
which uses a dataset of 106M \psip\ decays, nearly
four times that of CLEO.
A statistically significant signal is claimed, and a 
branching fraction of $\approx0.1$\% is measured,
compatible with the CLEO upper bound. BESIII
also raised the issue of how interference between the 
resonant $\gamma\gamma\jpsi_{\chi_{cJ}}$ final states
might affect the lower-energy photon spectrum, 
and what the effect might be from
interference between the nonresonant and resonant channels.

\subsubsection{Higher-order multipole amplitudes}
\label{sec:Dec_hm}

\begin{table}[b]
\caption{From CLEO~\cite{:2009pn}, normalized magnetic dipole~(M2) amplitudes
         from an analysis of $\psip\to\gamma\chi_{cJ}$,
         $\chi_{cJ}\to\gamma\jpsi$ decays. For the $J=2$ values, 
         the electric octupole~(E3) moments were fixed to zero
}
\label{tab:Dec_tabmamp}
\setlength{\tabcolsep}{0.98pc}
\begin{center}
\begin{tabular}{lcc}
\hline\hline
\rule[10pt]{-1mm}{0mm}
Decay & Quantity & Value~($10^{-2}$)  \\
\hline
\rule[12pt]{-1mm}{0mm}
$\psip\to\gamma\chi_{c1}$ & $b^{J=1}_2$ & $2.76\pm0.73\pm0.23$\\[0.6mm]
$\psip\to\gamma\chi_{c2}$ & $b^{J=2}_2$ & $1.0\pm1.3\pm0.3$\\[0.6mm]
$\chi_{c1}\to\gamma \jpsi$  & $a^{J=1}_2$ & $-6.26\pm0.63\pm0.24$\\[0.6mm]
$\chi_{c2}\to\gamma \jpsi$  & $a^{J=2}_2$ & $-9.3\pm1.6\pm0.3$\\[0.6mm]
\hline\hline
\end{tabular}
\end{center}
\end{table}

\begin{figure}[t]
  \begin{center}
    \includegraphics[width=\figwid]{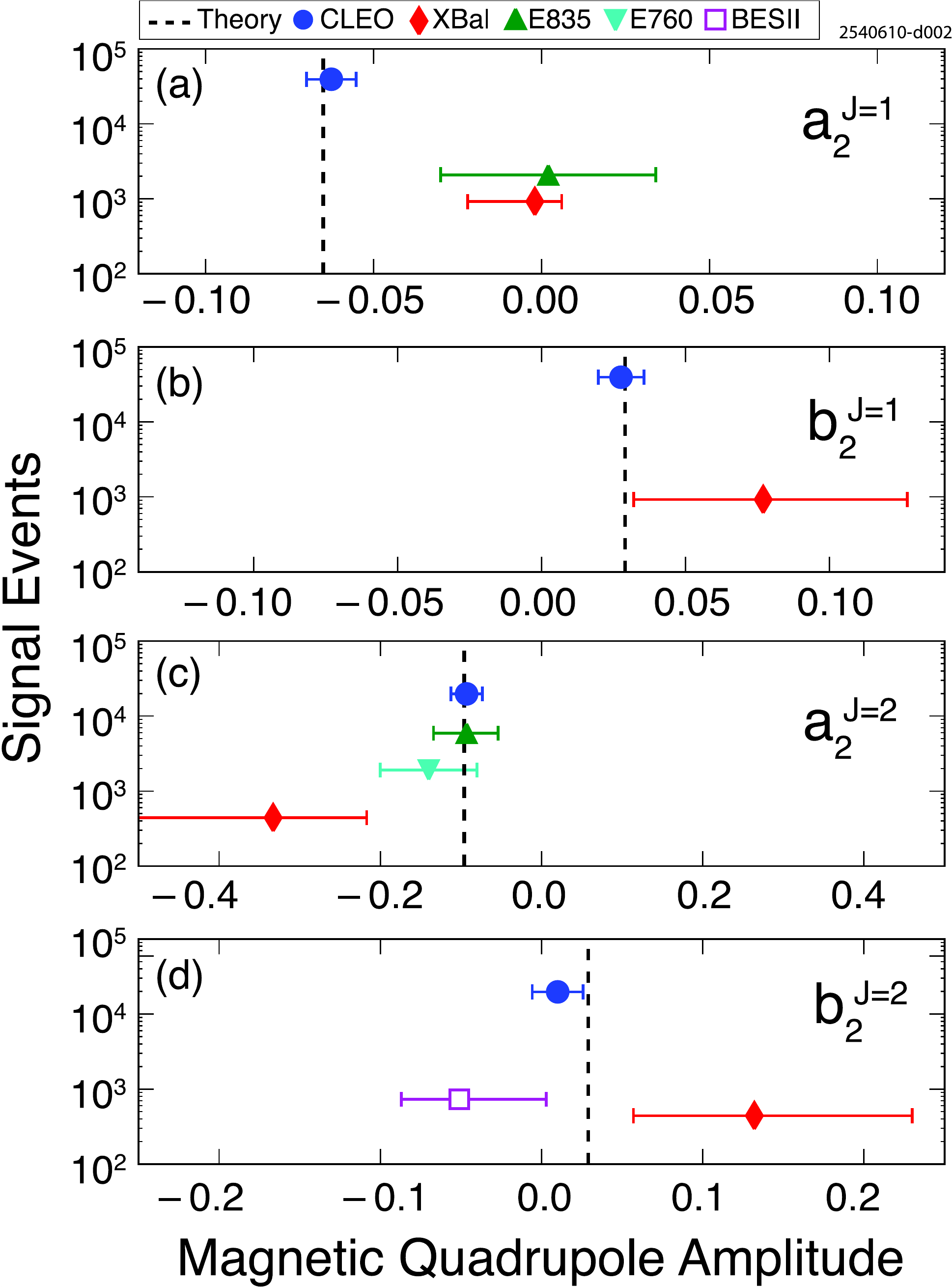}
      \caption{A compilation of measurements of normalized 
               magnetic dipole amplitudes from
               $\chi_{c1}\to\gamma \jpsi$ $(a^{J=1}_2)$,
               $\psip\to\gamma\chi_{c1}$ $(b^{J=1}_2)$,
               $\chi_{c2}\to\gamma \jpsi$ $(a^{J=2}_2)$, and 
               $\psip\to\gamma\chi_{c2}$ $(b^{J=2}_2)$.  
               The {\it solid circles} represent data  
               from CLEO~\cite{:2009pn}, which show consistency with
               predictions {\it (dashed vertical lines)}, unlike some
               earlier measurements.  
               The non\-relativistic theoretical expectations are 
               calculated with an anomalous magnetic moment of the 
               charm quark of zero and an assumed $1.5\gevcc$ 
               charm quark mass.  
               \AfigPermAPS{:2009pn}{2009}}
       \label{fig:Dec_Multipoles}
  \end{center}
\end{figure}

The radiative decays $\psip\to\gamma\chi_{c1,2}$ and
$\chi_{c1,2}\to\gamma \jpsi$ are dominated by electric dipole~(E1)
amplitudes.  However, they are expected to have a small additional
contribution from the higher-order magnetic quadrupole~(M2)
amplitudes.  Previous measurements of the relative sizes of the M2
amplitudes have disagreed with theoretical expectations.  CLEO~\cite{:2009pn} has
recently revisited this issue with a high-statistics analysis of the
decay chains $\psip\to\gamma\chi_{cJ};\chi_{cJ}\to\gamma \jpsi;
\jpsi\to l^+l^- (l=e,\mu)$ for $J=1,2$.  Starting with
$24\times10^6$ $\psip$ decays, CLEO observes approximately
$40\,000$ events for $J=1$ and approximately $20\,000$ events for
$J=2$, significantly larger event samples than previous measurements.
Using an unbinned maximum-likelihood fit to angular distributions,
CLEO finds the normalized M2 admixtures in \Tab{tab:Dec_tabmamp}. For
the quoted $J=2$ measurements, the electric octupole~(E3) moments were
fixed to zero. As shown in \Fig{fig:Dec_Multipoles},
these new measurements agree well with theoretical
expectations when the anomalous magnetic moment of the charm quark is
assumed to be zero and the mass of the charm quark is assumed to be
$1.5\gevcc$.

\subsubsection{Observation of $\psit\to\gamma\chi_{cJ}(1P)$}

The existence of the $\psit$ has long been established, and it
has generally been assumed to be the $1^3D_1$ charmonium state with
a small admixture of $2^3S_1$.  However, because it predominantly
decays to \DDbar, its behavior as a state of charmonium has
gone relatively unexplored in comparison to its lighter partners.  The
charmonium nature of the $\psit$ is especially interesting given
the unexpected discoveries of the $X$, $Y$, and $Z$~states, opening up
the possibility that the $\psit$ could include more exotic
admixtures.  The electromagnetic transitions,
$\psit\to\gamma\chi_{cJ}$, because they are
straightforward to calculate assuming the $\psit$ is the $^3D_1$
state of charmonium, provide a natural testing ground for the nature
of the $\psit$~\cite{Rosner:2001nm,Eichten:2004uh,Barnes:2005pb}.

\begin{figure}[b]
  \begin{center}
    \includegraphics[width=\figwid]{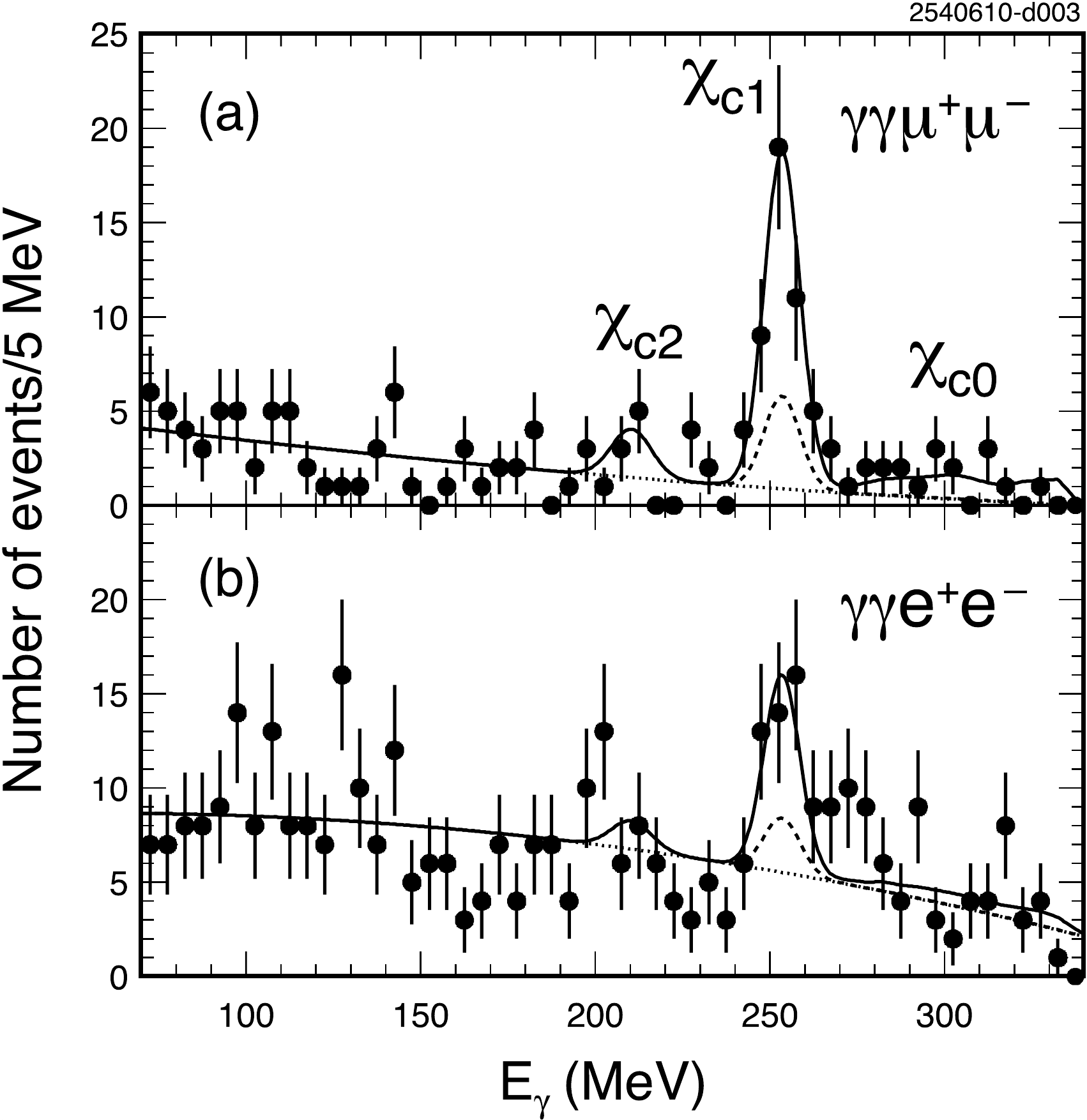}
      \caption{From CLEO~\cite{Coan:2005ps}, 
               the energy of the transition (lower energy) photon from 
               $\psit\to\gamma\chi_{cJ}$ found 
               when reconstructing $\chi_{cJ}\to\gamma \jpsi$ 
               and requiring the $\jpsi$ decay to (a)~$\mu^+\mu^-$  or 
               (b)~$e^+e^-$.  {\it Solid circles} represent data,
               the {\it dotted curve} shows the smooth fitted background,
               the {\it dashed curve} shows
               the sum of the smooth background fit
               and an estimated contribution from the tail 
               of the $\psip$ (events individually indistinguishable 
               from signal), and the {\it solid curve} is a result 
               of a fit of the data to all background and signal components.
               Background saturates the data at the $\chi_{c2}$ and 
               $\chi_{c0}$, but a significant $\chi_{c1}$ signal is obtained.
               \AfigPermAPS{Coan:2005ps}{2006} } 
      \label{fig:Dec_Psi3770_GammaChic1}
  \end{center}
\end{figure}

\begin{figure}[b]
  \begin{center}
    \includegraphics[width=\figwid]{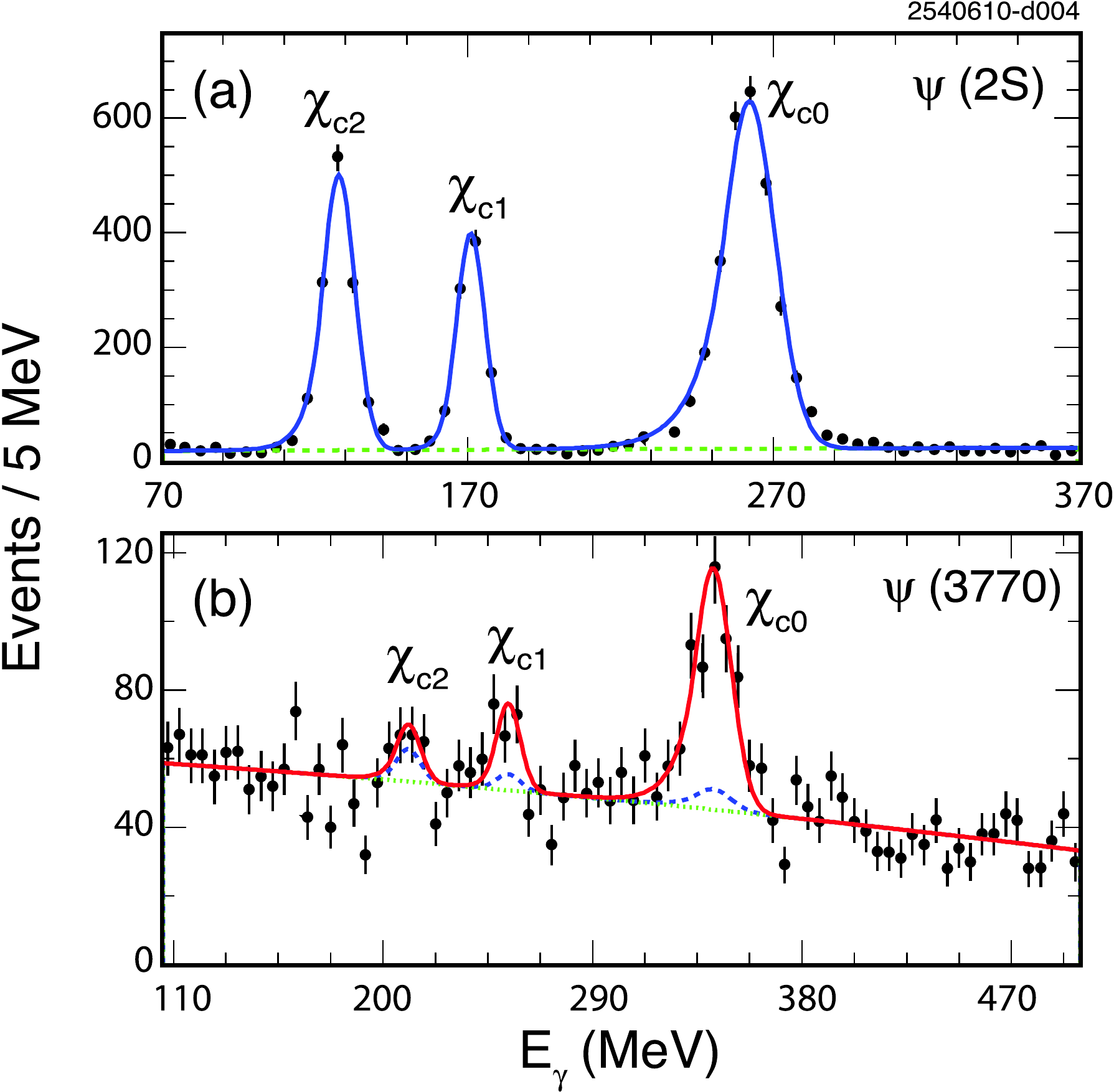}
      \caption{From CLEO~\cite{Briere:2006ff},
               the energy of the transition photon from 
               (a)~\psip, or (b)~$\psit$ decaying
               to $\gamma\chi_{cJ}$ when the 
               $\chi_{cJ}$ are reconstructed in exclusive 
               hadronic modes. {\it Short-dashed curves}
               represent fits to the smooth background.
               The {\it long-dashed curve} in (b) represents 
               estimated background from the tail of the 
               $\psip$, for which events are individually 
               indistinguishable from signal. A significant
               $\psit\to\gamma\chi_{c0}$ signal
               is obtained.
               \AfigPermAPS{Briere:2006ff}{2006}}
      \label{fig:Dec_Psi3770_GammaChic2}
  \end{center}
\end{figure}

CLEO has observed these transitions in two independent
analyses.  In the first~\cite{Coan:2005ps}, 
the $\chi_{cJ}$ were reconstructed exclusively in the decay chain
$\psit\to\gamma\chi_{cJ}$, 
$\chi_{cJ}\to\gamma \jpsi$, 
$\jpsi\to \ell^+\ell^-$,
with results depicted in \Fig{fig:Dec_Psi3770_GammaChic1}.  In the
second~\cite{Briere:2006ff}, the $\chi_{cJ}$ were reconstructed in
several exclusive hadronic modes and then normalized to
$\psip\to\gamma\chi_{cJ}$ using the same exclusive modes,
with results shown in \Fig{fig:Dec_Psi3770_GammaChic2}.  
Due to differing relative rates of the $\chi_{cJ}$ 
decay modes employed, the first method 
has more sensitivity to the transitions to $\chi_{c1,2}$
whereas the second is more suited to $\chi_{c0}$.  
Combining the results of the two analyses, the partial
widths of $\psit\to\gamma\chi_{cJ}$ were found to be
$172\pm30\kev$ for $J=0$, $70\pm17\kev$ for $J=1$, and 
$<21\kev$ at 90\%~CL for $J=2$.  These measurements
are consistent with relativistic calculations assuming the
$\psit$ is the $^3D_1$ state of charmonium.

\subsubsection{Observation of $\UnS{2S,3}\to\gamma\etab$}
\label{sec:Dec_1Sb}

The recent discovery of the $\etab$ state by \babar~\cite{:2008vj} 
has, through a measurement of the
$\etab$ mass, given us our first measurement of the $1S$~hyperfine
splitting in the bottomonium system.  This is obviously an important
accomplishment for spectroscopy as it provides a window into the
spin-spin interactions within the $b\bar{b}$ system.  However, in
addition to its contributions to spectroscopy, the observation of the
decays $\UnS{2S,3}\to\gamma\etab$ has resulted in our first
measurements of M1 radiative transition rates in the bottomonium
system.  A large range of theoretical predictions have been made for
these rates~\cite{Godfrey:2001eb}, especially for the hindered
transitions, to which the experimental measurement brings much-needed
constraints.

Using 109~million $\UnS{3}$ decays and 92~million $\UnS{2}$
decays, \babar~\cite{:2008vj,:2009pz} has measured
\begin{eqnarray}
\Brat(\UnS{3}\to\gamma\etab) &=& (4.8\pm0.5\pm0.6)\times10^{-4}\nonumber\\
\Brat(\UnS{2}\to\gamma\etab) &=&
(4.2^{+1.1}_{-1.0}\pm0.9)\times10^{-4}\, . ~~~~~
\end{eqnarray}
Both
measurements assume an $\etab$ width of $10\mevcc$.  The large
systematic errors in the branching fractions are due to the difficulty
in isolating the small $\etab$ signal from other nearby photon lines
($\chi_{bJ}(2P,1P)\to\gamma\UnS{1}$ and
$\UnS{3,2}\to\gamma\UnS{1}$) and from the large
background in the energy spectrum of inclusive photons.

In addition to the M1 transition rates, the energy dependence of the
matrix elements is also of interest.  In the case of charmonium (see
below), this energy dependence can introduce a non\-trivial distortion
of the \etac\ lineshape which can artificially pull the mass
measurement several\mev\ from its true value.  It is expected that the
same distortion mechanism will hold in the bottomonium system.  This
effect must then be understood if M1 transitions are to be used for
precision $\etab$ mass measurements.

Studying the $\etab$ lineshape in M1 transitions will require a
large reduction in background levels.  One possibility would be to
study exclusive $\etab$ decays.  Using exclusive $\etab$
decays could also allow a measurement of
$\mathcal{B}(\UnS{1}\to\gamma\etab)$, the allowed M1
transition, since background levels in the $\UnS{1}$ inclusive
photon energy spectrum are likely prohibitively large.

\subsection{Radiative and dileptonic decays}
\label{sec:Dec_radlepdec}

 Here we review theoretical status and experimental results
for radiative and dileptonic decays of heavy quarkonia.
The simplest parton-level decay of any heavy quarkonium vector
state occurs through annihilation into a virtual photon and thence into 
dilepton or quark-antiquark pairs. The latter can be difficult
to isolate from $ggg$ decay at the charmonium and bottomonium
mass scales, but fortunately is known to have a rate proportional to
$R\equiv\sigma(e^+e^-\to~{\rm hadrons})/\sigma(e^+e^-\to\mu^+\mu^-)$,
a quantity that is well-measured with off-resonance data
and which has a well-understood energy dependence. Conversely, dilepton
pairs have distinctive experimental signatures for which most
modern detectors are optimized, offering the prospect of high precision.
This high precision is quite useful in studies of both production and
decays of vector charmonium and bottomonium states. Dileptonic
widths also offer relative and absolute measures of wave function 
overlap at the origin. For all these reasons, decays to $\ell^+\ell^-$
are heavily studied and used to characterize the most basic
features of each vector state.

The simplest three-body decays of vector quarkonia are to
$\gamma\gamma\gamma$, $\gamma gg$, and $ggg$, and
their relative rates should reflect directly upon the value
of $\als$ at the relevant mass scale: naively, 
$R_\gamma\equiv {\cal B}(\gamma gg)/{\cal B}(ggg)
\simeq\alpha/\als(m)$. Although the $\gamma\gamma\gamma$ 
final state is experimentally straightforward to isolate, 
its rate is exceedingly small. 
Conversely, while $\gamma gg$ and
$ggg$ decays are abundant, distinguishing them from each
other and from transitions involving final state hadrons
is quite challenging. Indeed, $ggg$ decays cannot be
effectively
differentiated on an event-to-event basis
from quarkonium annihilation into
light quark-antiquark pairs nor from $\gamma gg$ final states
with soft photons. Experimental study of
$\gamma\gamma\gamma$, $\gamma gg$, and $ggg$
quarkonium decays has progressed substantially, but
has not yet entered the realm of precision.

The simplest decay or production mechanism 
of any scalar or tensor quarkonium state
is to and from a pair of photons. 
Two-photon decay
offers clean experimental signatures and 
a measure of the frequently small two-photon branching fraction,
whereas production via two-photon fusion
in $e^+e^-$ collisions offers the
prospect of determination of the diphotonic width.
As with vector states and three photons, the two-photon
coupling provides quite basic information about
scalar and tensor quarkonia.

For completeness, we mention that radiative decays
offer convenient production mechanisms for scalar and
tensor states such as light-quark or hybrid mesons, glueball candidates,
and non-Standard-Model Higgs or axion searches.
Treatment of such decays is beyond the scope of this review,
but some examples can be found in \Sec{sec:Dec_gammaP}.

\subsubsection{Theoretical status}
\label{sec:Dec_RadLepTheory}

Quarkonium annihilation happens at the heavy-quark mass scale $m$. 
Processes that happen at the scale $m$ are best described by NRQCD. 
The NRQCD factorization formula for the quarkonium annihilation width 
into light hadrons or photons or lepton pairs
reads~\cite{Bodwin:1992ye,Bodwin:1994jh}:
\begin{equation}
\Gamma_{H \hbox{-annih.}} = 
\sum_n \frac{2 \, {\rm Im} \, c_n }{ m^{(d_{O_n} - 4)}}\, \langle H | 
O^{\rm 4\text{-}fermion}_n |H \rangle\, , 
\end{equation}
where $O^{\rm 4\text{-}fermion}_n$ are four-fermion operators, 
$c_n$ are their Wilson coefficients,
$d_{O_n}$ their dimensions, and
$|H\rangle$ is the state that describes the quarkonium in NRQCD.
The Wilson coefficients $c_n$ are series in powers of the strong coupling constant
$\als$, evaluated at the heavy-quark mass, 
and the matrix elements  $\langle H | O^{\rm 4\text{-}fermion}_n |H \rangle$  are 
counted in powers of the heavy quark velocity $v$. The matrix elements live at the scale $mv$: 
they are non\-perturbative  if $mv \simg \lamQ$, 
while they may be evaluated in perturbation theory 
if $mv^2 \simg \lamQ$.

Substantial progress has been made in the evaluation of the factorization 
formula at order $v^7$~\cite{Brambilla:2006ph,Brambilla:2008zg}, 
in the lattice evaluation of the NRQCD matrix elements 
$\langle H | O^{\rm 4\text{-}fermion}_n |H \rangle$~\cite{Bodwin:2005gg},
and in the data of many hadronic 
and electromagnetic decays (see \cite{Brambilla:2004wf} 
and subsequent sections). 
As discussed in \cite{Brambilla:2004wf}, 
the data are clearly sensitive to NLO corrections in the Wilson coefficients $c_n$ 
(and presumably also to relativistic corrections). For an updated list of
ratios of $P$-wave charmonium decay widths, see \Tab{tab:Dec_tabdec}.

\begin{table}
\caption{Comparison of measured $\chi_{cJ}$ decay-width ratios (using 
PDG10~\cite{Nakamura:2010pdg})
with LO and NLO determinations~\cite{Brambilla:2004wf}, assuming
$m_c = 1.5\gev$ and $\als(2m_c) = 0.245$, but without 
corrections of relative order $v^2$. LH~$\equiv$~light hadrons }
\label{tab:Dec_tabdec}
\setlength{\tabcolsep}{0.65pc}
\begin{center}
\begin{tabular}{cccc}
\hline\hline
\rule[10pt]{-1mm}{0mm}
Ratio & PDG & LO & NLO  \\
\hline
\rule[20pt]{-1mm}{0mm}
$\displaystyle \frac{\Gamma(\chi_{c0}\to\gamma\gamma)}{
\Gamma(\chi_{c2}\to\gamma\gamma)}$
& 4.5 &   3.75  & 5.43   \\
\rule[20pt]{-1mm}{0mm}
$\displaystyle \frac{\Gamma(\chi_{c2}\to\lhad) - \Gamma(\chi_{c1}\to\lhad)}{
\Gamma(\chi_{c0}\to\gamma\gamma)}$
& 450 &  347  & 383   \\
\rule[20pt]{-1mm}{0mm}
$\displaystyle \frac{\Gamma(\chi_{c0}\to\lhad) -
\Gamma(\chi_{c1}\to\lhad)}{\Gamma(\chi_{c0}\to\gamma\gamma)}$ 
& 4200 &  1300  & 2781   \\
\rule[20pt]{-1mm}{0mm}
$\displaystyle \frac{\Gamma(\chi_{c0}\to\lhad ) - 
\Gamma(\chi_{c2}\to\lhad)}{
\Gamma(\chi_{c2}\to\lhad ) - \Gamma(\chi_{c1}\to\lhad)}$ 
& 8.4 &  2.75  & 6.63   \\
\rule[20pt]{-1mm}{0mm}
$\displaystyle \frac{\Gamma(\chi_{c0}\to\lhad) - \Gamma(\chi_{c1}\to\lhad)}
{\Gamma(\chi_{c2}\to {\lhad}) - \Gamma(\chi_{c1}\to \lhad)}$ 
& 9.4 &  3.75  & 7.63   \\
~\\
\hline\hline
\end{tabular}
\end{center}
\end{table}

In \cite{Brambilla:2007cz}, the high precision of data 
and matrix elements has been exploited to 
provide a new determination of $\als$ from
\beq 
\frac{\Gamma(\UnS{1}\to \gamma \; \lhad)}{\Gamma(\UnS{1}\to \lhad)}\, :
\eeq
\beq
\als (m_{\UnS{1}})= 0.184^{+0.015}_{-0.014}\, ,
\eeq
implying 
\beq
\als(m_{Z^0})=0.119^{+0.006}_{-0.005}\, .
\eeq

The NRQCD factorization formulas for 
electromagnetic and inclusive hadronic decay widths 
lose their predictive power as soon as we go to higher orders in $v$, due to 
the rapid increase in the number of non\-perturbative matrix 
elements~\cite{Brambilla:2008zg,Brambilla:2006ph,Brambilla:2004wf}. 
Quarkonia, with typical binding energies much smaller than $\lamQ$,  
conservatively include all quarkonia above the ground state.
Matrix elements of these states are inherently nonperturbative 
and may be evaluated 
on the lattice~\cite{Bodwin:2005gg}; few, however, are known. 
A way to reduce the number
of these unknown matrix elements 
is to go to the lower-energy EFT, pNRQCD, and to exploit the hierarchy 
$mv \gg mv^2$. In pNRQCD, NRQCD matrix elements factorize into two
parts: one, the quarkonium wave-function or its derivative at the origin,
and the second, gluon-field correlators 
that are universal, \ie\ independent of the quarkonium state.
The pNRQCD factorization has been exploited for P-wave and S-wave decays
in \cite{Brambilla:2002nu}.

  Quarkonium ground states have 
typical binding energy larger than or of the same order as $\lamQ$. 
Matrix elements of these states may be evaluated in perturbation theory
with the nonperturbative contributions being small corrections encoded in 
local or non\-local condensates.
Many higher-order corrections to spectra, 
masses, and wave functions have been 
calculated in this manner~\cite{Kniehl:1999ud}, 
all of them relevant to the quarkonium 
ground state annihilation into light hadrons and its electromagnetic decays.
For some recent reviews about applications, 
see \cite{Vairo:2006pc,Vairo:2006nq}.
In particular, $\UnS{1}$, $\etab$, $\jpsi$, and \etac\ 
electromagnetic decay widths at NNLL have been 
evaluated~\cite{Penin:2004ay,Pineda:2006ri}.
The ratios of electromagnetic decay widths were calculated for the 
ground state 
of charmonium and bottomonium at NNLL order~\cite{Penin:2004ay},
finding, \eg
\beq
\frac{\Gamma(\etab\to\gamma\gamma)}{\Gamma(\UnS{1}\to e^+e^-)} = 
0.502\pm 0.068 \pm 0.014\, . 
\label{eqn:Dec_etabgamgam}
\eeq
A partial NNLL-order analysis of the absolute widths of $\UnS{1} \to
e^+e^-$ and $\etab\to\gamma\gamma$ can be found in \cite{Pineda:2006ri}.

As the analysis of $\Gamma(\UnS{1}\to e^+e^-)$ of \cite{Pineda:2006ri} 
illustrates, for this fundamental quantity there 
may be problems of convergence 
of the perturbative series. 
Problems of convergence are common and severe for 
all the annihilation observables of ground state quarkonia and may be traced 
back to large logarithmic contributions, to be resummed by solving suitable 
renormalization group equations, and to large $\beta_0 \als$ contributions of either 
resummable or non\-resummable nature (these last ones are known as renormalons). 
Some large $\beta_0 \als$ contributions were 
successfully treated~\cite{Bodwin:2001pt} to provide a more reliable 
estimate for 
\beq
\frac{\Gamma(\etac\to \lhad)}{\Gamma(\etac \to \gamma\gamma)} = (3.26
\pm 0.6)\times 10^3\, ,
\label{eqn:Dec_etacLH}
\eeq
or  $(3.01 \pm 0.5)\times 10^3$ in a different resummation scheme. 
A similar analysis could be performed for the $\etab$, which
combined with a determination of $\Gamma(\etab\to\gamma\gamma)$
would then provide a theoretical determination of the $\etab$ width.
At the moment, without any resummation or renormalon subtraction performed,  
\beq
\frac{\Gamma(\etab\to \lhad)}{\Gamma(\etab \to \gamma\gamma)} \simeq
(1.8\text{--}2.3)\times 10^4\,.
\label{eqn:Dec_gametabratio}
\eeq
Recently a new resummation scheme has been suggested for electromagnetic
decay ratios of heavy quarkonium and applied to determine the $\etab$
decay width into two photons \cite{Kiyo:2010jm}:
\begin{equation}
\Gamma(\etab \to \gamma\gamma) = 0.54 \pm 0.15 ~{\rm keV}\,.
\label{eqn:Dec_GammaetabKiyo}
\end{equation}
Substituting \Eq{eqn:Dec_GammaetabKiyo} into \Eq{eqn:Dec_gametabratio} gives
$\Gamma(\eta_b(1S) \to \hbox{LH}) = 7$-16\mev.

\subsubsection{Measurement of $\psi, \Ups\to\gamma gg$}
\label{sec:Dec_Gammaglueglue}

In measurements of the $\gamma gg$ rate from \jpsi~\cite{Besson:2008pr},
\psip~\cite{Libby:2009qb}, and 
\UnS{1S,2S,3}~\cite{Besson:2005jv},
CLEO finds that the most effective experimental strategy
to search for $\gamma gg$ events is to focus solely upon
those with energetic photons (which are less prone to many backgrounds),
then to make the inevitable large subtractions of $ggg$, $q\bar{q}$,
and transition backgrounds on a statistical 
basis, and finally to extrapolate the radiative photon energy
spectrum to zero with the guidance of both theory and the measured high
energy spectrum. The most troublesome background remaining
is from events with energetic $\piz\to\gamma\gamma$ decays which result
in a high-energy photon in the final state. One of several 
methods used to estimate
this background uses the measured {\it charged} pion
spectra and the assumption of isospin invariance to simulate
the resulting photon spectrum with Monte Carlo techniques;
another measures the exponential shape of the photon-from-$\piz$
distribution at low photon energy, where $\gamma gg$ decays
are few, and extrapolates to the full energy range.
Backgrounds to $\gamma gg$ from transitions require the input of
the relevant branching fractions and their uncertainties.
The rate for $ggg$ decays is then estimated as that fraction
of decays that remains after all dileptonic, transition,
and $q\bar{q}$ branching fractions are subtracted, again
requiring input of many external measurements and
their respective uncertainties. Not surprisingly, 
the relative errors on the results of 10-30\%
are dominated by the systematic uncertainties
incurred from background subtraction methods,
photon-spectrum model-dependence, and
external branching fractions.

  The CLEO measurements of the observable 
$R_\gamma$ defined in \Eq{eqn:SpecTh_Rgamma} are shown in 
\Tabs{tab:DecGammaGG_model} and \ref{tab:DecGammaGG}. It should be noted that
the uncertainties in $R_\gamma$ for $V=\UnS{1S,2S,3}$
are partially correlated with one another because 
of shared model dependence and analysis systematics.
In addition, the {\sl shape} of the measured
direct photon spectrum from \jpsi\ is quite similar to
that of the $\Ups$'s, in contrast to the \psip, for
which the spectrum appears to be softer.
Absolute values for the branching
fractions have been calculated with input of
other world-average branching fractions in
PDG10~\cite{Nakamura:2010pdg},
and are reproduced in \Tab{tab:DecGammaGG}.
See \Sec{sec:SpecTh_alphasdec} for discussion
of extraction of $\als$ from $R_\gamma$ measurements.

\subsubsection{Observation of $\jpsi\to\gamma\gamma\gamma$}
\label{sec:Dec_jpsito3gamma}

Orthopositronium, the $^3S_1$ $e^+e^-$ bound state, decays to
$3\gamma$ almost exclusively and has long been used for precision
tests of QED~\cite{Karshenboim:2005iy}. The rate of its analog for
QCD, three-photon decay of vector charmonium,
in particular that of the $\jpsi$, acts as a probe of the strong
interaction~\cite{Voloshin:2007dx}. Due to similarities at the parton
level, relative rate measurements of the branching fractions for $\jpsi \to
3\gamma$, $\jpsi \to \gamma gg$, $\jpsi \to 3g$, and $\jpsi \to
l^+l^-$ provide crucial grounding for QCD
predictions~\cite{Voloshin:2007dx,Kwong:1987ak,Petrelli:1997ge}.
Previous searches for the quarkonium decay to $3\gamma$ have 
yielded only upper limits:  
${\cal B}(\omega \to 3\gamma) < 1.9 \times 10^{-4}$  
at 95\%~CL~\cite{:1997uc} and  
${\cal B}(\jpsi \to 3\gamma) < 5.5 \times 10^{-5}$ 
at 90\%~CL~\cite{Partridge:1979tn}.

\begin{figure}[t]
  \begin{center}
    \includegraphics[width=\figwid]{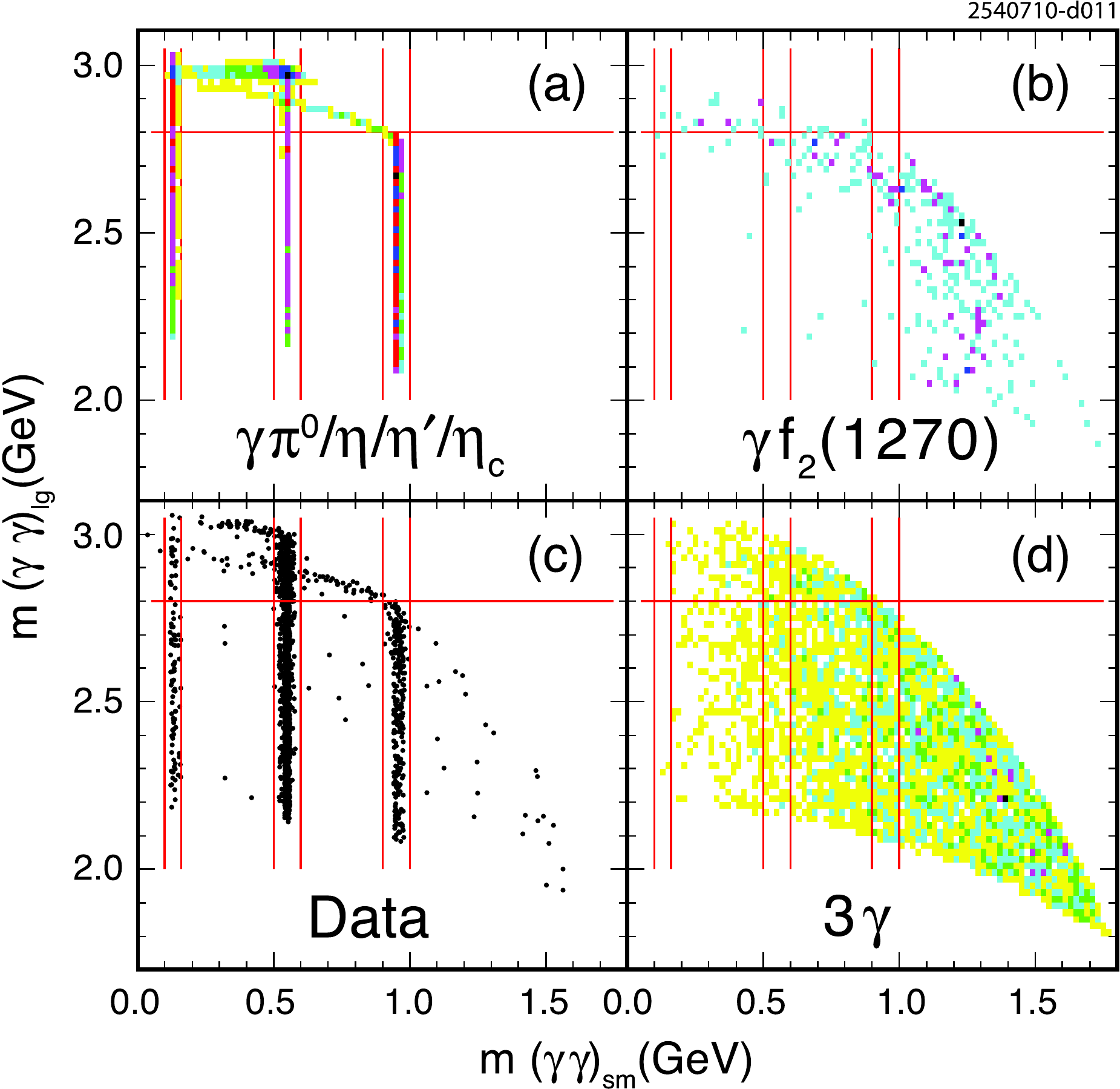}
      \caption{From CLEO~\cite{:2008ab},
               distributions of the largest vs.~the 
               smallest diphoton mass combination per event
               in $\psip\to\dipi\jpsi$, $\jpsi\to 3\gamma$ candidates.
               Parts~(a), (c), and (d) are from MC simulations
               of $\jpsi\to\gamma\eta/\etap/\etac\to 3\gamma$, 
               $\gamma f_2(1270)\to\gamma\dipiz\to 5\gamma$, 
               and the signal process, $\jpsi\to 3\gamma$, respectively.
               In these three MC plots {\it darker shading} of bins signifies
               relatively larger event density than {\it lighter shades}.
               Part~(b) shows the CLEO data, where
               each {\it solid circle} indicates a single event.
               {\it Solid lines} demarcate regions excluded from the
               $\jpsi\to 3\gamma$ event selection. 38 data events
               populate the signal region, of which $24.2^{+7.2}_{-6.0}$
               were estimated to be from $\jpsi\to 3\gamma$,
               with the remainder due to background,
               mostly various sources of $\jpsi\to\gamma\dipiz$.
               \AfigPermAPS{:2008ab}{2008} }
       \label{fig:Dec_3Gamma}
  \end{center}
\end{figure}

CLEO~\cite{:2008ab} performed a search for the all-photon
decays of the $\jpsi$ using tagged $\jpsi$ mesons from $9.6\times10^6$
$\psip \to \jpsi \dipi$ decays. After
excluding backgrounds from $\jpsi\to\gamma\eta/\etap/\etac\to 3\gamma$
with restrictions on photon-pair masses, and minimizing
$\jpsi\to\gamma\dipiz$ backgrounds in which two of the photons
are very soft by imposing stringent energy-momentum conservation
via a kinematic fit, CLEO
reported the first observation of the decay $\jpsi \to 3\gamma$ 
by finding 38 events, 12.8 of which were estimated to be 
background, mostly 
from various sources of $\jpsi\to\gamma\dipiz\to 5\gamma$ decays.
The branching fraction is measured to be 
\beq
{\cal B}(\jpsi \to 3\gamma) = (1.2 \pm 0.3 \pm 0.2) \times 10^{-5}
\eeq
with statistical significance for the signal 
of $6.3\sigma$. Diphoton mass plots
are shown in \Fig{fig:Dec_3Gamma} for data, signal MC,
and two possible sources of background.
The measured
three-photon branching fraction lies between the zeroth-order
predictions~\cite{Kwong:1987ak} for 
${\cal B}_{3\gamma}/{\cal B}_{\gamma gg}\approx(\alpha/\als)^2/3$ and 
${\cal B}_{3\gamma}/{\cal B}_{3g}\approx(\alpha/\als)^3$ and is
consistent with both, but is a factor of $\approx 2.5$ below that for
${\cal B}_{3\gamma}/{\cal B}_{ll}\approx\alpha/14$, all assuming
$\als(m_{\jpsi})\approx 0.3$. 
Upper limits on the branching fractions for
$\jpsi \to \gamma\gamma$, $4\gamma$, and $5\gamma$
were also determined, setting more stringent 
restrictions than previous experiments.

\begin{table}[tb]
\caption{
Measured values of $R_\gamma$, as defined in \Eq{eqn:SpecTh_Rgamma},
assuming the Garcia-Soto (GS)~\cite{GarciaiTormo:2005ch} 
or Field~\cite{Field:1983cy} models
of the direct photon spectrum from \Ups\ decays,
and their averages, from CLEO~\cite{Besson:2005jv}.
The uncertainty on each average includes
a component to account for model dependence
(which dominates the other uncertainties);
statistical errors are negligible
}
\label{tab:DecGammaGG_model}
\setlength{\tabcolsep}{0.87pc}
\begin{center}
\begin{tabular}{cccc}
\hline\hline
\rule[10pt]{-1mm}{0mm}
State & & $R_\gamma$~(\%) &\\[0.7mm]
& GS & Field & Average \\[0.7mm]
\hline
\rule[10pt]{-1mm}{0mm}
\UnS{1} & $2.46\pm0.13$ & $2.90\pm0.13$ & $2.70\pm 0.27$\\[0.6mm]
\UnS{2} & $3.06\pm0.22$ & $3.57\pm0.22$ & $3.18\pm 0.47$ \\[0.6mm]
\UnS{3} & $2.58\pm0.32$ & $3.04\pm0.32$ & $2.72\pm 0.49$ \\[0.6mm]
\hline \hline 
\end{tabular}
\end{center}
\end{table}

\begin{table}
\caption{
Measured values of $R_\gamma$
from CLEO, as defined in \Eq{eqn:SpecTh_Rgamma},
and the respective absolute branching fractions from
PDG10~\cite{Nakamura:2010pdg}
}
\label{tab:DecGammaGG}
\setlength{\tabcolsep}{0.34pc}
\begin{center}
\begin{tabular}{ccccc}
\hline\hline
\rule[10pt]{-1mm}{0mm}
State &  $R_\gamma$~(\%) & ${\cal B}(\gamma gg)$~(\%) &
 ${\cal B}(ggg)$~(\%)& Source\\[0.7mm]
\hline
\rule[10pt]{-1mm}{0mm}
\jpsi   & $13.7 \pm 1.7$  & $8.8 \pm 0.5$  & $64.1\pm 1.0$ & \cite{Besson:2008pr,Nakamura:2010pdg}\\[0.6mm]
\psip   &  $9.7 \pm 3.1$  & $1.02\pm 0.29$ & $10.6\pm 1.6$ & \cite{Libby:2009qb,Nakamura:2010pdg}\\[0.6mm]
\UnS{1} &  $2.70\pm 0.27$ & $2.21\pm 0.22$ & $81.7\pm 0.7$ & \cite{Besson:2005jv,Nakamura:2010pdg}\\[0.6mm]
\UnS{2} &  $3.18\pm 0.47$ & $1.87\pm 0.28$ & $58.8\pm 1.2$ & \cite{Besson:2005jv,Nakamura:2010pdg}\\[0.6mm]
\UnS{3} &  $2.72\pm 0.49$ & $0.97\pm 0.18$ & $35.7\pm 2.6$ & \cite{Besson:2005jv,Nakamura:2010pdg}\\[0.6mm]
\hline \hline 
\end{tabular}
\end{center}
\end{table}

\subsubsection{Nonobservation of $\psip, \UnS{1}\to\gamma\eta$}
\label{sec:Dec_gammaP}

\begin{table}[b]
\caption{Branching fractions (in units of $10^{-4}$) for charmonium
         decays to $\gamma\,(\piz/\eta/\etap)$
         from CLEO~\cite{:2009tia}
         and PDG08~\cite{Amsler:2008zzb}, the latter
         of which is dominated by BESII~\cite{Ablikim:2005je}.
         The rightmost column shows the difference
         between the two in units of standard error ($\sigma$).
         Upper limits are quoted at 90\%~CL. 
         Entries in the last two rows include the effects
         of estimated continuum background and ignore
         (include) maximal destructive interference
         between $\psit$ and continuum sources
}
\label{tab:Dec_gampseudo}
\setlength{\tabcolsep}{0.30pc}
\begin{center}
\begin{tabular}{lccc}
\hline
\hline
\rule[10pt]{-1mm}{0mm}
Mode & CLEO & PDG08 & \#$\sigma$ \\[0.7mm]
\hline
\rule[10pt]{-1mm}{0mm}
$\jpgpiz$ & $     0.363 \pm      0.036 \pm      0.013$ & $      0.33^{+0.06}_{-0.04}$ & $  0.4$ \\ [0.7mm]
$\ttgeta$ & $     11.01 \pm       0.29 \pm       0.22$ & $       9.8 \pm        1.0$ & $  1.2$ \\ [0.7mm]
$\ttgepr$ & $      52.4 \pm        1.2 \pm        1.1$ & $      47.1 \pm        2.7$ & $  1.7$ \\ [0.7mm]
$\ppgpiz$ & $ <0.07$ & $       <54$ & -  \\ [0.7mm]
$\ttgeta$ & $ <0.02$ & $      <0.9$ & -  \\ [0.7mm]
$\ttgepr$ & $      1.19 \pm       0.08 \pm       0.03$ & $      1.36 \pm       0.24$ & $ -0.7$ \\ [0.7mm]
$\ptgpiz$ & $ <3$ & - & -  \\ [0.7mm]
$\ttgeta$ & $ <0.2~(1.5)$ & - & -  \\[0.7mm] 
$\ttgepr$ & $ <0.2~(1.8)$ & - & -  \\ [0.7mm]
\hline
\hline
\end{tabular}
\end{center}
\end{table}

Both BESII~\cite{Ablikim:2005je} and CLEO~\cite{:2009tia}
report studies on the exclusive final states
$\gamma\,(\piz/\eta/\etap)$ from charmonium,
BESII from \jpsi~decays alone, and
CLEO from decays of \jpsi, \psip, and \psit.
Resulting branching fractions appear in \Tab{tab:Dec_gampseudo},
where it can be seen that the results from the
two experiments are consistent with each other,
and that precision has steadily improved. From
the perspective of charmonium physics, the most
striking feature of these numbers is the
nonobservation of $\psip\to\gamma\eta$.

  CLEO and BESII have consistent values of the ratio 
\beq
r_n\equiv \frac{{\cal B}(\psi(nS)\to\gamma\eta)}{
                {\cal B}(\psi(nS)\to\gamma\etap)}
\label{eqn:Dec_etaetaprime}
\eeq
for\jpsi\  of $r_1\approx0.2$ within a few percent; the
naive expectation would be that, for \psip,
$r_2\approx r_1$. Yet the CLEO
result implies that $r_2<1.8\%$, or $r_2/r_1<8\%$, 
both at 90\%~CL.
Specifically, the rate of $\gamma\eta$ relative to $\gamma\etap$
from \psip\ is {\it at least an order of magnitude smaller}
than that from \jpsi. If instead we characterize
the effect in terms of ``the 12\% rule'' 
(\Sec{sec:Dec_RhoPiPuzzle}), we note
that, relative to its dileptonic width,
$\psip\to\gamma\etap$ is suppressed
by a factor of five with respect to $\jpsi\to\gamma\etap$,
but $\psip\to\gamma\eta$ is suppressed
by at least {\it two} orders of magnitude
with respect to $\jpsi\to\gamma\eta$.

Do we see such transitions at expected rates in $\Upsilon$ decays?
The CLEO~\cite{Athar:2007hz} search for
$\UnS{1}\to\gamma\eta^{(\prime)}$ failed to find evidence for either
pseudoscalar meson in radiative decays, setting the limits
\begin{eqnarray}
&{\cal B}(\UnS{1}\to\gamma\eta)&<1.0\times 10^{-6} \nonumber\\
&{\cal B}(\UnS{1}\to\gamma\etap)&<1.9\times 10^{-6}\,.~~
\end{eqnarray}
These values rule out the predictions of Chao~\cite{Chao:1990im},
which are based on mixing of $\eta$, \etap, and \etab,
but not sensitive enough to probe those of Ma~\cite{Ma:2002ww},
which uses a QCD-factorization approach,
nor those of Li~\cite{Li:2007dq}, which posits
a substantial two-gluon component in $\eta$ and \etap\ 
within a perturbative QCD framework.

What dynamical effect is
present in \psip\ decays that is absent in \jpsi\ decays
that can explain this large of a suppression? Is it
related to $\rho\pi$ suppression 
in \psip\ decays relative to \jpsi\ (\Sec{sec:Dec_RhoPiPuzzle}),
another vector-pseudoscalar final state?
Is there a connection between the suppression of $\gamma\eta$ and
$\gamma\etap$ in \UnS{1}\ decay and that of $\gamma\eta$ in \psip\ decay?
These questions remain unanswered.

Events selected from
this charmonium analysis have been used to address physics questions
other than those directly associated with $c\bar{c}$ bound states:
\begin{itemize}
\item{The flavor content of $\eta$ and \etap\ mesons,
which are commonly thought to be mixtures
of the pure SU(3)-flavor octet and singlet states,
with a possible admixture of 
gluonium~\cite{Rosner:1982ey,Rosner:1985fz,Gilman:1987ax};
if $\jpsi\to\gamma\eta^{(\prime)}$ occurs through
$c\bar{c}\to\gamma gg$, which is expected to
fragment in a flavor-blind manner, the mixing angle can be extracted
from $r_1$ as defined in \Eq{eqn:Dec_etaetaprime}.
The measured value~\cite{Ablikim:2005je} from charmonium is
consistent with that obtained from other 
sources~\cite{Bai:1998ny,Ambrosino:2006gk,Escribano:2007cd}.}

\item{The high-statistics sample of $\jpsi\to\gamma\etap$ decays
was also used by CLEO to perform the first simultaneous
measurement of the largest five $\etap$ branching fractions~\cite{:2009tia},
attaining improved precisions, and to
improve the measurement precision of the $\etap$ mass~\cite{Libby:2008fg}.}

\item{Although CLEO~\cite{:2009tia} only set limits for
\psit\ decays to these final states, clean signals
for both $\gamma\eta$ and $\gamma\etap$ final states
were observed. However, these rates were seen
to be consistent with that expected from continuum
production as extrapolated from the bottomonium
energy region using the only other measurement
of $e^+e^-\to\gamma\eta^{(\prime)}$, by \babar~\cite{Aubert:2006cy}.}

\item{Rosner~\cite{Rosner:2009bp} explores the implications
of these measurements on production mechanisms 
for $\gamma P$ ($P$=pseudoscalar) final states, in particular
the contribution of the vector dominance model (through $\rho\pi$)
to $\jpsi\to\gamma\piz$.}
\end{itemize}

\subsubsection{Two-photon widths of charmonia}
\label{sec:Dec_twophoton}

Considerable progress has been made with
measurements of the two-photon width of the \etac\ and
$\chi_c(1P)$ states. Two different approaches have been used. The
first one uses the formation of a charmonium state in two-photon
collisions followed by the observation of its decay products.  In this
case the directly measured quantity is the product of the charmonium
two-photon width and the branching ratio of its decay to a specific
final state.  A two-photon width can then be computed from the product 
if the corresponding branching fraction has been measured. Such
measurements were performed at Belle for the following decays:
$\chi_{c0(2)} \to \dipi,~K^+K^-$~\cite{Nakazawa:2004gu}, 
$\etac\to p\bar{p}$~\cite{Kuo:2005nr}, 
$\chi_{c0(2)} \to K^0_SK^0_S$~\cite{Chen:2006gy}, $\chi_{c0(2)} \to
\dipiz$~\cite{Uehara:2008pf,:2009cka} and at CLEO for the
$\chi_{c2} \to \jpsi\gamma$ decay~\cite{Dobbs:2005yk}.  Belle also
applied this method to study the two-photon formation of the
$\etac$, $\chi_{c0}(1P)$ and $\chi_{c2}(1P)$ via various final
states with four charged particles ($2\pi^+2\pi^-$,
$\dipi K^+K^-$, $2K^+2K^-$) and quasi-two-body final states
($\rho\rho,~\phi\phi,\eta\eta,~\ldots$)~\cite{:2007vb,Uehara:2010mq}. 
In the latter study~\cite{:2007vb},
Belle also sought the two-photon production of the $\etacp$
and did not find a significant signal in any four-body final state

\begin{figure}[t]
  \begin{center}
    \includegraphics[width=\figwid]{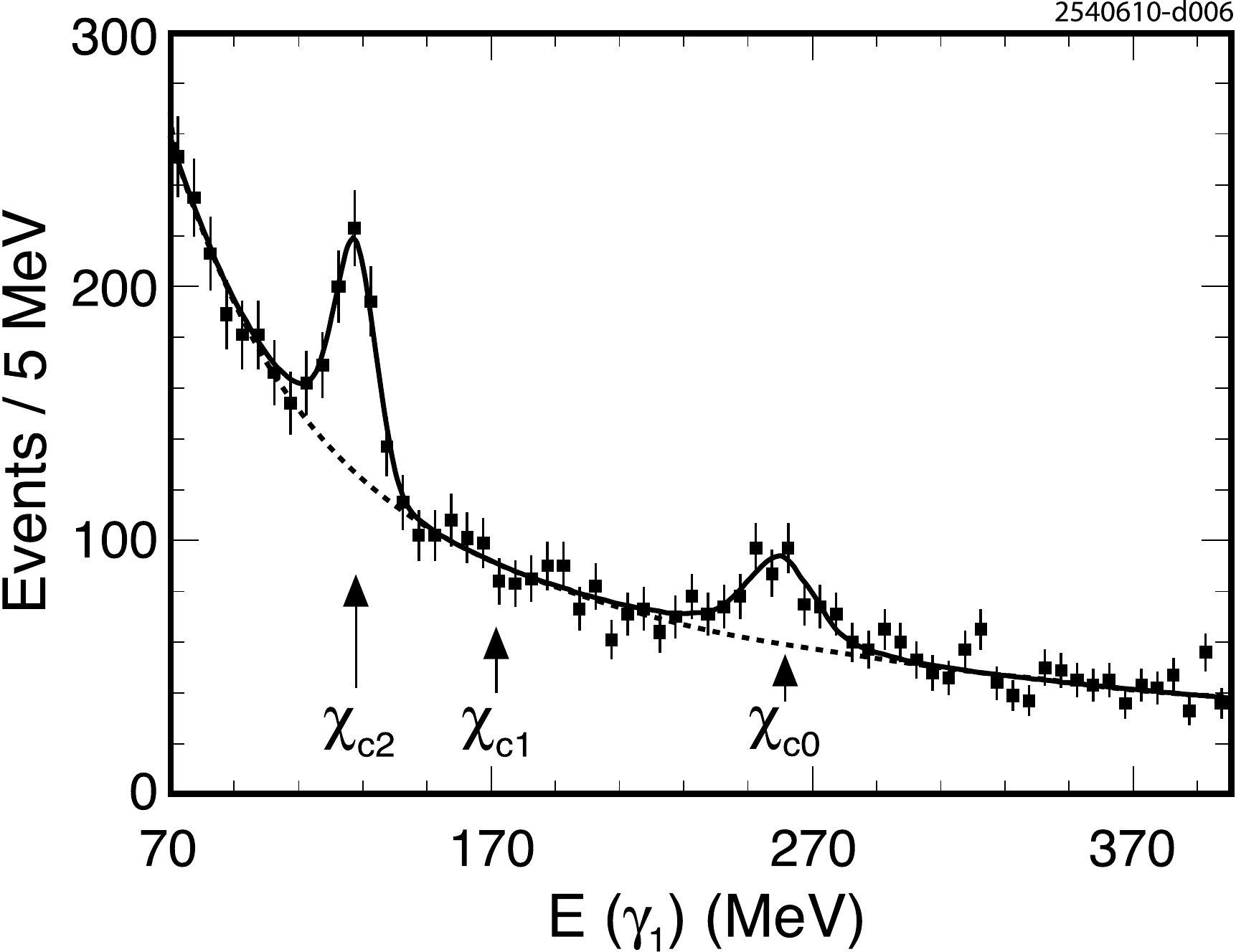}
      \caption{From CLEO~\cite{Ecklund:2008hg},
               the fitted spectrum for $E(\gamma_1)$ from the 
               reaction $\psip \to \gamma_1\chi_{cJ},~~\chi_{cJ} 
               \to \gamma_2\gamma_3$, where $\gamma_1$ is the 
               least energetic photon in the \psip\
               center-of-mass frame. The expected positions
               of $E(\gamma_1)$ from $\chi_{c0},~\chi_{c1},~\chi_{c2}$
               are marked with {\it arrows}. The {\it dashed curve} represents 
               a fit of the non\-peaking background to a polynomial. 
               \AfigPermAPS{Ecklund:2008hg}{2008} }
      \label{fig:Dec_Chic_2Gamma}
  \end{center}
\end{figure}

The second method is based on the charmonium decay into two photons. 
CLEO performed~\cite{Ecklund:2008hg} such measurements  using the reactions
$\psip \to \gamma_1\chi_{cJ},~~\chi_{cJ} \to \gamma_2\gamma_3$,
where $\gamma_1$ is the least-energetic final-state photon in the \psip\
center-of-mass frame. Clear signals were observed for the $\chi_{c0}$ 
and $\chi_{c2}$, as shown in \Fig{fig:Dec_Chic_2Gamma}. 
(Two-photon decay of spin-one 
states is forbidden by the Landau-Yang 
theorem~\cite{landau,Yang:1950rg}).
Using the measured signal yield and 
the previously-determined number of $\psip$ produced (24.6
million), the product of the 
branching fractions is determined as: 
\begin{eqnarray}
{\cal B}(\psip \to \gamma\chi_{c0})\times
{\cal B}(\chi_{c0} \to \gamma\gamma) = \hspace{0.8in}&\nonumber\\
(\, 2.17 \pm 0.32 \pm 0.10\, ) \times 10^{-5}\, ,&~~~~~~~\nonumber \\ 
{\cal B}(\psip \to \gamma\chi_{c2})\times 
{\cal B}(\chi_{c2} \to \gamma\gamma) = \hspace{0.8in}&\nonumber\\
(\, 2.68 \pm 0.28 \pm 0.15\, ) \times 10^{-5}\,.~~&
\end{eqnarray}

World-average values for the $\psip\to\gamma\chi_{cJ}$ branching 
fractions and total widths of the $\chi_{c0(2)}$ states 
can then used to calculate the two-photon widths for 
the corresponding charmonia. In \Tab{tab:Dec_CC_2Gamma} 
we list these two-photon widths, which result
from a constrained fit to all relevant 
experimental information from PDG08~\cite{Amsler:2008zzb}.
Using those values, the experimental ratio becomes
$\Gamma_{\gamma\gamma}(\chi_{c2})/\Gamma_{\gamma\gamma}(\chi_{c0})=0.22\pm 0.03$.
The LO and a NLO determination are shown in the first row of 
\Tab{tab:Dec_tabdec};
some relevant theoretical issues are discussed in 
\Sec{sec:Dec_RadLepTheory}.

\begin{table}
\caption{World-average values of the two-photon width for
         various charmonium states from PDG08~\cite{Amsler:2008zzb}
}
\label{tab:Dec_CC_2Gamma}
\setlength{\tabcolsep}{0.48pc}
\begin{center}
\begin{tabular}{lccc}
\hline\hline
\rule[10pt]{-1mm}{0mm}
State & $\chi_{c0}(1P)$ &  $\chi_{c2}(1P)$ & \etac\  \\
\hline
\rule[10pt]{-1mm}{0mm}
$\Gamma_{\gamma\gamma}$~(keV) & $2.36 \pm 0.25$ & $0.515 \pm 0.043$ 
& $7.2 \pm 0.7 \pm 2.0$  \\ 
\hline \hline 
\end{tabular}
\end{center}
\end{table}

As part of the $\jpsi\to 3\gamma$ analysis
described in \Sec{sec:Dec_jpsito3gamma},
CLEO~\cite{:2008ab}
failed to find evidence for the decay $\etac\to\gamma\gamma$.
The product branching fraction for the decay chain
$\jpsi\to\gamma\etac$, $\etac \to \gamma\gamma$
was measured to be $(1.2^{+2.7}_{-1.1}\pm0.3)\times 10^{-6}$
($<6\times 10^{-6}$ at 90\%~CL), which, using 
the CLEO~\cite{:2008fb}  
measurement of ${\cal B}(\jpsi\to\gamma\etac)=(1.98\pm0.31)\%$
(discussed in \Sec{sec:Dec_1Sc}),
can be rewritten as 
\begin{eqnarray}
{\cal B}(\etac\to\gamma\gamma)&=&(6.1^{+13.7}_{-5.6})
\times10^{-5}~{\rm (CLEO)}\nonumber\\
&<&2.4\times10^{-4}~{\rm at~90\%~CL}\, . 
\end{eqnarray}
This central value is
smaller than but 
consistent with the only other direct measurement
of $\etac\to\gamma\gamma$ measured in $B^\pm\to K^\pm\etac$ decays
by Belle~\cite{Abe:2006gn}, which obtains 
\beq
{\cal B}(\etac\to\gamma\gamma)=(2.4^{+1.2}_{-0.9})
\times 10^{-4}~{\rm (Belle)}\, ,
\eeq
for which the
Belle product branching fraction was unfolded by PDG08~\cite{Amsler:2008zzb}.

\subsubsection{Dileptonic widths in the $\psi$ family}

\begin{table}[b]
\caption{Measurements of 
         $\Gamma(\jpsi \to e^+e^-)\times{\cal B}(\jpsi \to \ell^+\ell^-)$ 
         and their relative accuracies $\delta$; the weighted
         average and its ($\chi^2/$d.o.f.) are shown assuming lepton
         universality. The two values
         from KEDR have some systematic uncertainties in common
         and are shown separately below the average
}
\label{tab:Dec_JPsi_ll}
\setlength{\tabcolsep}{0.31pc}
\begin{center}
\begin{tabular}{lccc}
\hline\hline
\rule[10pt]{-1mm}{0mm}
Experiment & $\ell^+\ell^-$ & $\Gamma(\jpsi \to e^+e^-)\times$ & $\delta$~(\%) \\
\rule[10pt]{-1mm}{0mm}
           &                & ${\cal B}(\jpsi \to \ell^+\ell^-)$~(keV) & \\
\hline
\rule[10pt]{-1mm}{0mm}
\babar~\cite{Aubert:2003sv}          & $\mu^+\mu^-$ & $0.3301 \pm 0.0077 \pm 0.0073$ & 3.2 \\
CLEO~\cite{Adams:2005mp}             & $\mu^+\mu^-$ & $0.3384 \pm 0.0058 \pm 0.0071$ & 2.7 \\
KEDR~\cite{Anashin:2009pc}           &$\ell^+\ell^-$& $0.3321 \pm 0.0041 \pm 0.0050$ & 1.9 \\
Avg (0.43/2)                         &$\ell^+\ell^-$& $0.3334 \pm 0.0047$ & 1.4 \\[0.8mm]
KEDR~\cite{Anashin:2009pc}           & $\mu^+\mu^-$ & $0.3318 \pm 0.0052 \pm 0.0063$ & 2.5 \\
KEDR~\cite{Anashin:2009pc}           & $e^+e^-$     & $0.3323 \pm 0.0064 \pm 0.0048$ & 2.4 \\
\hline\hline
\end{tabular}
\end{center}
\end{table}

\babar~\cite{Aubert:2003sv},
CLEO~\cite{Adams:2005mp,Li:2005uga,Adam:2005mr}, and
KEDR~\cite{Anashin:2009pc}
have all performed measurements of dileptonic decays 
from the narrow members of the $\psi$ resonance family 
with much-improved precisions.  
\babar~\cite{Aubert:2003sv} pioneered the use of
vector charmonia produced through initial-state radiation 
from $e^+e^-$ collisions collected for other purposes
at energies higher than
the state being studied, in particular for the \jpsi\
from $\sqrt{s}\approx 10.58\gev$.
CLEO followed \babar's lead,
studying \jpsi~\cite{Adams:2005mp} and 
\psip~\cite{Adam:2005mr} mesons from $\sqrt{s}=3770\mev$. 
KEDR~\cite{Anashin:2009pc} followed
the more straightforward route, which requires a dedicated
scan of the resonance.
In both kinds of analyses the directly measured
quantity is $\Gamma(\jpsi \to e^+e^-)\times{\cal B}(\jpsi \to\ell^+\ell^-)$.
The dileptonic branching fraction can then be divided
out from the result once to obtain \Gee\ and twice for \Gtot.
CLEO and \babar\ used only $\jpsi \to \mu^+\mu^-$ decays
because the $e^+e^-$ final state has larger backgrounds
from radiative Bhabha events,
whereas KEDR used dimuons and dielectrons,
obtaining the best precision to date.
In all three cases $e^+e^-\to\ell^+\ell^-$
and its interference with the resonant signal must be addressed.
The results are listed in \Tab{tab:Dec_JPsi_ll},
where it can be seen that the four measurements are consistent
with one another and have combined precision of 1.4\%.

CLEO~\cite{Li:2005uga} also provided an independent
measurement of \jpsi\ dileptonic branching fractions using 
$\psip \to \jpsi\dipi$, $\jpsi\to\ell^+\ell^-$
decay chains, and normalizing to all \jpsi\ decays produced
via the $\dipi$ transition by fitting the dipion
recoil mass distribution.
CLEO obtained 
\begin{eqnarray}
{\cal B}(\jpsi \to \mu^+\mu^-)&=&(5.960 \pm 0.065\pm 0.050)\%\nonumber\\ 
{\cal B}(\jpsi \to e^+e^-)&=&(5.945 \pm 0.067 \pm 0.042)\%\nonumber\\ 
\frac{{\cal B}(\jpsi \to e^+e^-)}{{\cal B}(\jpsi \to \mu^+\mu^-)}&=&
(99.7\pm1.2\pm0.6)\%\nonumber\\ 
{\cal B}(\jpsi \to \ell^+\ell^-)&=&(5.953 \pm 0.070)\%\, ,
\end{eqnarray}
consistent with and having uncertainties at least factor of two smaller than
previous determinations. With this measurement, the
assumption of lepton universality, and the weighted average
from \Tab{tab:Dec_JPsi_ll}, we obtain
$\Gee(\jpsi)=5.60\pm0.10\kev$~(1.8\%) and
 $\Gtot(\jpsi)=94.1\pm2.6\kev$~(2.7\%),
compared to relative uncertainties on these
quantities before any of these measurements
(PDG04~\cite{Eidelman:2004wy})
of 3.1\% and 3.5\%, respectively.

\begin{table}[b]
\caption{Measurements of $\Gamma_{ee}(\psip)$  and their relative
         accuracies $\delta$. The CLEO value quoted has been
         updated with branching fractions from \cite{Mendez:2008kb}
         and the weighted-average with ($\chi^2/$d.o.f.) 
         of the top three measurements is given
}
\label{tab:Dec_Psi2_ll}
\setlength{\tabcolsep}{0.73pc}
\begin{center}
\begin{tabular}{lcc}
\hline\hline
\rule[10pt]{-1mm}{0mm}
Experiment & $\Gamma(\psip \to e^+e^-)$~(keV) & $\delta$~(\%) \\[0.7mm]
\hline
\rule[10pt]{-1mm}{0mm}
CLEO~\cite{Adam:2005mr,Mendez:2008kb} & $2.407 \pm 0.083$ & 3.4 \\[0.7mm]
BES~\cite{Ablikim:2006zq}    & $2.330 \pm 0.036 \pm 0.110$ & 5.0 \\[0.7mm]
BES~\cite{Ablikim:2008zzb}   & $2.388 \pm 0.037 \pm 0.096$ & 4.3\\[0.7mm]
Avg~(0.29/2)                 & $2.383\pm0,056$ & 2.3 \\[0.7mm]
PDG04~\cite{Eidelman:2004wy} &$2.12 \pm 0.12$ & 5.7\\[0.7mm]
\hline \hline 
\end{tabular}
\end{center}
\end{table}

For \psip, CLEO~\cite{Adam:2005mr} used 
the decays $\psip\to X_i\jpsi$, $\jpsi\to\ell^+\ell^-$, where
$X_i=\dipi$, $\dipiz$, and $\eta$, directly
measuring the products ${\cal B}(\psip\to X_i\jpsi)\times\Gee[\psip]$.
CLEO initially used ${\cal B}(\psip\to X_i\jpsi)$ values from
\cite{Adam:2005uh} to extract $\Gee[\psip]=2.54\pm0.11\kev$~(4.3\%), but
if instead we use CLEO's updated values~\cite{Mendez:2008kb},
described below in \Sec{sec:DecPsipBRs},
we obtain $\Gee[\psip]=2.407\pm0.083\kev$~(3.4\%).
We then obtain the ratio $\Gee[\psip]/\Gee(\jpsi)=0.43\pm0.02$,
a quantity which might be more precisely predicted in
lattice QCD than either $\Gamma_{ee}$ alone~\cite{Gray:2005ur},
in which we have used our updated $\Gee[\psip]$ and 
our world-average \Gee[\jpsi] from above.
     
    BES~\cite{Ablikim:2006zq,Ablikim:2008zzb} 
studied the energy range $\sqrt{s}=$3.660-3.872\gev\ 
to determine the resonance parameters of $\psip$ and
$\psit$. From the fit of the cross sections
for $D^0\bar{D}^0,~D^+D^-$ and non-$D\bar{D}$ production
the branching fractions and partial widths for $\psip\to e^+e^-$
and $\psit \to e^+e^-$ decays were determined. 
The results of these measurements
of $\Gamma_{ee}[\psip]$ are shown in
\Tab{tab:Dec_Psi2_ll}, where the improvement
since 2004 can be observed.

Decays of the $\psip$ into $\tau$-lepton pairs are less probable
and therefore less studied. BES~\cite{Ablikim:2006iq} 
used a sample of 14~million produced $\psip$
to measure the corresponding branching
fraction. The result,
${\cal B}(\psip \to\tau^+\tau^-)=(3.08 \pm 0.21 \pm 0.38) \times 10^{-3}$,
has better relative precision (14\%) than previous
measurements~\cite{Brandelik:1977xz,Bai:2000pr}.  Using part of their
statistics KEDR measured $\Gamma(\psip \to e^+e^-)\times{\cal B}(\psip
\to \tau^+\tau^-)$ to be $9.0 \pm 2.6$ eV~\cite{Anashin:2007zz},
which, using the average from \Tab{tab:Dec_Psi2_ll},
implies a value ${\cal B}(\psip \to\tau^+\tau^-)=(3.8 \pm 1.1) \times
10^{-3}$, consistent with but considerably less precise than the result
from BES.

\begin{table}[b]
\caption{Measurements of $\Gamma_{ee}(\psit)$ and their 
         relative accuracies $\delta$. The BES and CLEO
         results listed do not include the potential effect
         of interference with higher-mass $\psi$-states
} 
\label{tab:Dec_Psi3770_ll} 
\setlength{\tabcolsep}{0.91pc}
\begin{center} 
\begin{tabular}{lcc} 
\hline\hline 
\rule[10pt]{-1mm}{0mm}
Experiment & $\Gamma(\psit \to e^+e^-)$~(keV) & $\delta$~(\%) \\ [0.7mm]
\hline 
\rule[10pt]{-1mm}{0mm}
PDG04~\cite{Eidelman:2004wy} & $0.26 \pm 0.04$ & 15.4 \\[0.7mm]
CLEO~\cite{Besson:2005hm} & $0.203 \pm 0.003^{+0.041}_{-0.027}$ & $^{+20.0}_{-13.3}$ \\ [0.7mm]
BES~\cite{Ablikim:2006zq} & $0.251 \pm 0.026 \pm 0.011$ & 11.2 \\ [0.7mm]
BES~\cite{Ablikim:2006md} & $0.277 \pm 0.011 \pm 0.013$ & 6.1 \\[0.7mm]
PDG10~\cite{Nakamura:2010pdg} & $0.259 \pm 0.016$ & 6.2 \\[0.7mm]
\hline\hline 
\end{tabular} 
\end{center} 
\end{table}

\begin{table}[t]
\caption{$\Gamma_{ee}$ of the higher $\psi$ states
         from PDG04~\cite{Eidelman:2004wy} and the 
         BES~\cite{Ablikim:2007gd} global fit
}
\label{tab:Dec_Psi_ll}
\setlength{\tabcolsep}{1.34pc}
\begin{center}
\begin{tabular}{lcc}
\hline\hline
\rule[10pt]{-1mm}{0mm}
Resonance & $\Gamma_{ee}$~(keV)               & $\Gamma_{ee}$~(keV) \\[0.7mm]
\rule[10pt]{-1mm}{0mm}
          & from PDG04 & from BES \\[0.7mm]
\hline
\rule[10pt]{-1mm}{0mm}
$\psit$ & $0.26 \pm 0.04$ & $0.22 \pm 0.05$ \\ [0.7mm]
$\psi(4040)$ & $0.75 \pm 0.15$ & $0.83 \pm 0.20$ \\ [0.7mm]
$\psi(4160)$ & $0.77 \pm 0.23$ & $0.48 \pm 0.22$ \\ [0.7mm]
$\psi(4415)$ & $0.47 \pm 0.10$ & $0.35 \pm 0.12$ \\[0.7mm]
\hline\hline 
\end{tabular}
\end{center}
\end{table}

CLEO~\cite{Besson:2005hm} measured the hadronic cross section 
at a single energy point near the peak of the $\psit$,
$\sqrt{s}=3773\mev$, taking interference between the
final states of resonance decays and non\-resonant $e^+e^-$
annihilation into account. From the observed cross section,
which is significantly smaller than some of the previous
measurements~\cite{Rapidis:1977cv,Abrams:1979cx}, 
$\Gamma_{ee}[\psit]$ is also obtained. 
In a scan over the \psit\ energy region,
68 energy points in the range 3.650-3.872\gev,
BES~\cite{Ablikim:2006md} measured
$R\equiv\sigma(e^+e^-\to\ {\rm hadrons})/\sigma(e^+e^-\to\mu^+\mu^-)$,
determining the parameters of the $\psit$
resonance, including the leptonic width.  The
results of the described measurements of \Gee[\psit] are
shown in \Tab{tab:Dec_Psi3770_ll}, where
it can be seen that world-average
uncertainty improved by more than a factor of two
between 2004 and 2010.

Finally, BES~\cite{Bai:2001ct} performed a global fit of 
$R$ in the energy range 3.7-5.0\gev,
covering the four resonances, $\psit$, $\psi(4040)$, $\psi(4160)$,
and $\psi(4415)$~\cite{Ablikim:2007gd}. 
Interference between the four $\psi$ states was accounted for
(which was not the case for the $\Gee[\psit]$ measurements 
in \Tab{tab:Dec_Psi3770_ll})
and an energy-dependent width based on all accessible two-body decay
channels was used. The results are shown in \Tab{tab:Dec_Psi_ll}. It can
be seen that the new results have larger uncertainties than previous ones,
which ignored interference with higher-mass $\psi$ states.

\subsubsection{Dileptonic widths in the $\Ups$ family}

CLEO~\cite{Adams:2004xa,Rosner:2005eu,Besson:2006gj}  
has made a systematic study of dileptonic decays  
of the narrow states in the $\Ups$ family. 
To determine dimuonic branching fractions, the quantities 
$\Btmumu \equiv \Gmumu/\Ghad$ are measured for each \UnS{n},
where \Gmumu~(\Ghad) is the rate for  $\Ups$ 
decay to $\mu^+\mu^-$~(hadrons). 
Assuming lepton universality, 
\beq
\Bmumu = \frac{\Gmumu}{\Gtot} = \frac{\Btmumu}{1+3\Btmumu}\, . 
\eeq
The results of this analysis, based on much larger data samples 
than available to previous experiments, are summarized in 
\Tab{tab:Dec_Ups_MM}.  
While the result for the $\UnS{1}$ is in good agreement with the
world average, the CLEO results for the $\UnS{2}$ and
$\UnS{3}$ are about $3\sigma$ larger than previous
world averages. However, the CLEO values are confirmed by 
their proximity to $\tau^+\tau^-$ branching fraction
measurements, which are discussed  below.

\begin{table}[b]
\caption{Dimuonic branching fractions of the narrow $\Ups$ states
         from PDG04~\cite{Eidelman:2004wy} and CLEO~\cite{Adams:2004xa}
}
\label{tab:Dec_Ups_MM}
\setlength{\tabcolsep}{1.09pc}
\begin{center}
\begin{tabular}{lcc}
\hline\hline
\rule[10pt]{-1mm}{0mm}
Resonance & 
\Bmumu~(\%) & \Bmumu~(\%) \\[0.7mm]
\rule[10pt]{-1mm}{0mm}
& from PDG04 & from CLEO \\[0.7mm]
\hline
\rule[10pt]{-1mm}{0mm}
$\UnS{1}$ & $2.48 \pm 0.06$ & $2.49 \pm 0.02 \pm 0.07$ \\[0.7mm]
$\UnS{2}$ & $1.31 \pm 0.21$ & $2.03 \pm 0.03 \pm 0.08$ \\[0.7mm]
$\UnS{3}$ & $1.81 \pm 0.17$ & $2.39 \pm 0.07 \pm 0.10$ \\[0.7mm]
\hline\hline
\end{tabular}
\end{center}
\end{table}

\begin{table}[t]
\caption{From CLEO~\cite{Rosner:2005eu}, measured values of 
         $\Gee\Ghad/\Gtot$
         for the narrow $\Ups$ states 
}
\label{tab:Dec_Ups_EE}
\setlength{\tabcolsep}{2.30pc}
\begin{center}
\begin{tabular}{lc}
\hline\hline
\rule[10pt]{-1mm}{0mm}
Resonance & $\Gee\Ghad/\Gtot$~(keV) \\[0.7mm]
\hline
\rule[10pt]{-1mm}{0mm}
$\UnS{1}$ & $1.252 \pm 0.004 \pm 0.019$ \\[0.7mm]
$\UnS{2}$ & $0.581 \pm 0.004 \pm 0.009$ \\[0.7mm]
$\UnS{3}$ & $0.413 \pm 0.004 \pm 0.006$ \\[0.7mm]
\hline\hline
\end{tabular}
\end{center}
\end{table}

In order to measure a quantity very close to \gee,
CLEO~\cite{Rosner:2005eu} performed
dedicated scans to measure the integral of the  \Ups\ production cross 
section over incident $e^+e^-$ energies to determine  
\begin{eqnarray}
\Gtee&\equiv&\frac{\Gee\Ghad}{\Gtot}\nonumber\\
 \Gtee &=& 
 \dfrac{m^2_\Ups}{6\pi^2}\int\sigma(e^+e^-\to\Ups\to{\rm hadrons})\,dE\,.~
\label{eqn:DecGtee}
\end{eqnarray}
The resulting values, listed in \Tab{tab:Dec_Ups_EE}, are consistent with, 
but more precise than, the PDG world averages. 

\begin{table}[t]
\caption{From CLEO~\cite{Besson:2006gj}, 
         measured $\tau$-pair branching fractions of the narrow $\Ups$
         states and ratios to corresponding dimuonic rates
}
\label{tab:Dec_Ups_TT}
\setlength{\tabcolsep}{0.75pc}
\begin{center}
\begin{tabular}{lcc}
\hline\hline
\rule[10pt]{-1mm}{0mm}
Resonance  & \Btt/\Bmumu & \Btt~(\%)\\ [0.7mm]
\hline
\rule[10pt]{-1mm}{0mm}
 $\UnS{1}$ & $1.02 \pm 0.02 \pm 0.05$  & $2.54 \pm 0.04 \pm 0.12$\\[0.7mm]
 $\UnS{2}$ & $1.04 \pm 0.04 \pm 0.05$  & $2.11 \pm 0.07 \pm 0.13$\\[0.7mm]
 $\UnS{3}$ & $1.05 \pm 0.08 \pm 0.05$  & $2.52 \pm 0.19 \pm 0.15$\\[0.7mm]
\hline\hline
\end{tabular}
\end{center}
\end{table}

CLEO~\cite{Besson:2006gj} also addressed the third dileptonic width,
\Gtt~(making the
first observation of the decay $\UnS{3} \to \tau^+\tau^-$) 
and measured precise values of \Btt/\Bmumu\ for
$\UnS{n},~n=1,2,3$, as shown in \Tab{tab:Dec_Ups_TT}.  
Using the CLEO values of \Bmumu~\cite{Adams:2004xa}
(discussed above)
allowed reporting of the absolute \Btt\ values as well.
The results obtained are consistent with the 
expectations from the Standard Model, 
and \Btt\ values for \UnS{1S,2}\ have much-improved precision
over previous measurements.

\begin{table}[t]
\caption{Values of \Gtot\ for the \UnS{1S,2S,3}\
from PDG04~\cite{Eidelman:2004wy}, 
CLEO~\cite{Adams:2004xa}, and 
PDG08~\cite{Amsler:2008zzb}
}
\label{tab:Dec_Ups_Tot}
\setlength{\tabcolsep}{0.90pc}
\begin{center}
\begin{tabular}{lccc}
\hline\hline
\rule[10pt]{-1mm}{0mm}
\UnS{n}  &       & \Gtot~(keV) & \\[0.7mm] 
         & PDG04 & CLEO & PDG08 \\ [0.7mm]
\hline
\rule[10pt]{-1mm}{0mm}
 \UnS{1} & $53.0\pm 1.5$ & $52.8\pm 1.8$ & $54.02\pm 1.25$ \\[0.7mm]
 \UnS{2} & $43  \pm 6  $ & $29.0\pm 1.6$ & $31.98\pm 2.63$ \\[0.7mm]
 \UnS{3} & $26.3\pm3.4 $ & $20.3\pm 2.1$ & $20.32\pm 1.85$ \\[0.7mm]
\hline\hline
\end{tabular}
\end{center}
\end{table}

The total width of the resonances can be expressed as  
\beq
\Gtot=\dfrac{\Gtee}{\Bmumu~( 1 - 3\Bmumu )}\, ,
\eeq
where \Gtee\ is measured as in \Eq{eqn:DecGtee},
and which, when combined with the more precise \Bmumu\ 
also described above, yields improved measurements of the total widths
\Gtot. 
The larger \Bmumu\ for $\UnS{2S,3}$ as
determined by CLEO (validated by consistent values of 
${\cal B}_{\tau\tau}$) leads to  
smaller and more precise $\Gtot(2,3S)$, as seen
in \Tab{tab:Dec_Ups_Tot}.

Improved measurements of the $\UnS{4}$ parameters were
reported by \babar~\cite{Aubert:2004pwa}. Three 
scans of the energy range $\sqrt{s}=10.518-10.604$\gev\ were
performed with 11, 7, and 5 energy points, respectively, with 
integrated luminosity of typically 0.01~fb$^{-1}$ per point. This information was 
complemented by a large data sample of 76~fb$^{-1}$ 
collected at the peak of the $\UnS{4}$, from which 
the cross section at the peak was determined. 
The nominal $\sqrt{s}$ values of the scans were corrected using 
an energy calibration based on the dedicated run at the 
$\UnS{3}$. A fit of the energy dependence allows a 
determination of the $\UnS{4}$ parameters, among 
them mass, total, and electronic width,
obtaining 
\begin{eqnarray}
m(\UnS{4})&=&10579.3\pm0.4\pm1.2\mevcc\nonumber\\
\Gtot(\UnS{4})&=&20.7\pm1.6\pm2.5\mevcc\nonumber\\
\Gee(\UnS{4})&=&321 \pm 17 \pm 29\evcc\, ,
\end{eqnarray}
all of which dominate the PDG08~\cite{Amsler:2008zzb} world averages.

\subsection{Hadronic transitions}
\label{sec:Dec_hadtrans}

\subsubsection{Theoretical status}
\label{sec:Dec_HadTranTheory}
 
The general form for a hadronic transition is
\begin{equation}
\Phi_i\to\Phi_f+h
\label{eqn:Dec_HT-1}
\end{equation}
where $\Phi_i$, $\Phi_f$
and $h$ stand for the initial state, final state quarkonia, and the
emitted light hadron(s).  In the $c\bar{c}$ and $b\bar{b}$ systems, the 
mass $m_{\Phi_i}-m_{\Phi_f}$ 
varies from a few hundred\mev\ to slightly over a\gev,  
so the kinematically allowed final light hadron(s) $h$ 
are dominated by single-particle 
($\piz$, $\eta$, $\omega$, ...) or two-particle 
($2\pi$ or $2 K$) states.  The low momenta of the 
light hadrons in these transitions allow the 
application of chiral lagrangian methods.  To date, 
over twenty hadronic transitions have been observed experimentally.  

Hadronic transitions are important decay modes for low-lying 
heavy quarkonium states.  In fact, the first observed hadronic 
transition, $\psip\to \pi\pi\jpsi$, has a branching fraction
recently measured by CLEO~\cite{Mendez:2008kb} to be
$(52.7\pm1.3)\%$ (see \Sec{sec:DecPsipBRs}).
Calculating such transitions requires nonperturbative 
QCD. The standard approach is a 
QCD Multipole Expansion (QCDME) for gluon
emission, which is modeled after the 
multipole expansion used for 
electromagnetic transitions (see \Sec{sec:Dec_EMT1}).

Many contributed to the early development of the QCDME 
approach~\cite{Gottfried:1977gp,Bhanot:1979af,Voloshin:1978hc},  
but Yan~\cite{Yan:1980uh} was the first to present
a gauge-invariant formulation within QCD.  For a heavy $Q\bar{Q}$ 
bound state, a {\it dressed} ({\it constituent}) quark 
is defined as
\begin{equation}
\tilde{\psi}({\mathbf x},t)\equiv U^{-1}({\mathbf x},t)\,\psi(x)\, , 
\end{equation}
where $\psi(x)$ is the usual quark field and
$U$ is defined as a path-ordered exponential 
along a straight line from
${\mathbf X}\equiv {(\mathbf x_1+\mathbf x_2)}/2$ 
(the center-of-mass coordinate of $Q$ and $\bar{Q}$) to $\mathbf x$, 
\begin{equation} 
U({\mathbf x},t)=P\exp\bigg[ig_s\int^{\mathbf x}_{\mathbf X}
{\mathbf A}({\mathbf x}^\prime,t)\cdot d{\mathbf x}^\prime\bigg]~.~~
\end{equation}
Gluon-field color indices have been suppressed. The {\it dressed}
gluon field 
is defined by
\begin{eqnarray}
\tilde{A}_\mu({\mathbf x},t)\equiv &
U^{-1}({\mathbf x},t)\, A_\mu(x)\,U({\mathbf x},t)\nonumber\\
&-\dfrac{i}{g_s}\,U^{-1}({\mathbf x},t)\,\partial_\mu U({\mathbf x},t)\, .
\end{eqnarray}
Now we can make the QCD multipole expansion, in powers of
$~ \mathbf{(x - X)}\cdot\mathbf\nabla~$ operating on the gluon field in exact
analogy with QED:
\begin{eqnarray}
\tilde{A}_0({\mathbf x},t)&=&A_0({\mathbf X},t)-({\mathbf x}-{\mathbf X})\cdot
{\mathbf E}({\mathbf X},t)+\cdots \nonumber \\
\tilde{{\mathbf A}}({\mathbf X},t)&=&-\frac{1}{2}({\mathbf x}-{\mathbf X})
\times {\mathbf B}({\mathbf X},t)+\cdots \;
\end{eqnarray}
where ${\mathbf E}$ and ${\mathbf B}$ are color-electric and
color-magnetic fields, respectively. The resulting
Hamiltonian for a heavy $Q\bar Q$ system is then~\cite{Yan:1980uh}
\begin{equation}
H^{\rm eff}_{\rm QCD}=H^{(0)}_{\rm QCD}+H^{(1)}_{\rm QCD}+ H^{(2)}_{\rm QCD}\, ,
\end{equation}
with $H^{(0)}_{\rm QCD}$  taken as the zeroth-order 
Hamiltonian, even though it does not represent free fields 
but instead the sum of the kinetic and potential 
energies of the heavy quarks. We also define
\begin{equation}
H^{(1)}_{\rm QCD}\equiv Q_a\, A^a_0(\mathbf X,t)\, ,
\end{equation}
in which $Q_a$ is the color charge of the $Q\bar{Q}$ system 
(zero for color-singlets), and
\begin{equation}
H^{(2)}_{\rm QCD}\equiv -{\mathbf d}_a\cdot
{\mathbf E}^a(\bf X,t)-{\mathbf m}_a\cdot
{\bf B}^a(\mathbf X,t)+\cdots\hspace{0.5cm}\, ,
\label{eqn:Dec_HT-H2}
\end{equation}
which is treated perturbatively. The quantities
\beq
{\mathbf d}^i_a= g_E \int d^3x\,\, \tilde{\psi}^{\dagger}\, 
{\mathbf (x-X)}^i \, t_a\, \tilde{\psi}
\eeq
and 
\beq
{\mathbf m}^i_a = g_M/2  \int d^3x\,\, \tilde{\psi}^{\dagger} \,
\epsilon^{ijk}\,{\mathbf (x- X)}_j\, \gamma_k\, t_a\,\tilde{\psi}
\eeq
are the color-electric dipole moment (E1) and the
color-magnetic dipole moment (M1) of the $Q\bar{Q}$ system, 
respectively. Higher-order terms (not shown) 
give rise to higher-order electric (E2, E3, ...) 
and magnetic moments (M2, ...).
Because $H^{(2)}_{\rm QCD}$ in \Eq{eqn:Dec_HT-H2} 
couples color-singlet to octet $Q\bar Q$ states, 
the transitions between eigenstates $|i\rangle$ and 
$|f\rangle$ of $H^{(0)}_{\rm QCD}$ are at least 
second-order in $H^{(2)}_{\rm QCD}$.  The leading-order 
term is given by:
\begin{eqnarray}
\lefteqn{\big\langle  f\;h\big|\,H_2\,\frac{1}{E_i-H^{(0)}_{\rm QCD}+
i\partial_0-H_{\rm QCD}^{(1)}}\,\big|i\big \rangle = } \nonumber \\
 && \sum_{KL} \big\langle  f\;h \big|H_2\big|KL\big\rangle\, 
\frac{1}{E_i-E_{KL}}\, \big\langle KL \big|H_2\big|i\big \rangle\, ,
\label{eqn:Dec_HT-2g}
\end{eqnarray}
where the sum $KL$ is over a complete set of color-octet 
$Q\bar Q$ states $|KL\rangle$ with associated energy $E_{KL}$. 
Finally, a connection is made to the physical hadronic transitions 
in \Eq{eqn:Dec_HT-1} by assuming factorization of the 
heavy-quark interactions and the production of light 
hadrons. For example, the leading order E1-E1 transition amplitude is:
\begin{eqnarray}
\lefteqn{{\cal M}(\Phi_i\to\Phi_f+h)= }\nonumber \\
& & \frac{1}{24} \sum_{KL} \frac{\big
\langle  f\big|d_{m}^{ia}\big|KL\big\rangle 
\langle KL\big|d_{ma}^{j}\big|i\big \rangle }{E_i-E_{KL}}\,
        \big\langle h \big| {\mathbf  E}^{ai}  {\mathbf E}_a^j
        \big|0\big\rangle\, .~~
\label{eqn:Dec_QCS}
\end{eqnarray}
The allowed light hadronic final state $h$ is determined by the quantum
numbers of the gluonic operator. The leading order term E1-E1 in 
\Eq{eqn:Dec_HT-2g} has $CP=++$ and $L=0,2$ and hence couples to $2\pi$ 
and $2K$ in $I=0$ states. Higher-order terms (in powers of $v$) couple 
as follows:  E1-M1 in $O(v)$ with ($CP=-\hspace{0.3mm}-$) couples to
$\omega$;  E1-M1, E1-E2 in $O(v)$ and M1-M1, E1-M2 in $O(v^2)$ with 
($CP=+\hspace{0.3mm}-$) couples to both $\piz$ (isospin-breaking) and $\eta$ 
(SU(3)-breaking); and M1-M1, E1-E3, E2-E2 ($CP=++$) are 
higher-order corrections to the E1-E1 terms.

Applying this formulation to observed hadronic transitions requires 
additional phenomenological assumptions. 
Following Kuang and Yan~\cite{Yan:1980uh,Kuang:1981se},  
the heavy $Q\bar{Q}$ bound states spectrum of $H^{(0)}_{\rm QCD}$ is calculated
by solving the state equation with a given potential model.  The intermediate octet
$Q\bar Q$ states are modeled by the Buchmueller-Tye quark-confining 
string (QCS) model~\cite{Buchmuller:1980su}. Then chiral symmetry 
relations can be employed to parametrize the light hadronic matrix element.  
The remaining unknown coefficients in the light hadron matrix 
elements are set by experiment or calculated using a duality argument between
the physical light hadron final state and associated two-gluon final
state. A detailed discussion of all these assumptions can be found in the 
QWG review~\cite{Brambilla:2004wf}. 

For the most common transitions, $h = \pi_1 + \pi_2$, the effective 
chiral lagrangian form is~\cite{Brown:1975dz}
\begin{eqnarray}
\frac{g_E^2}{6}\big\langle \pi_1\pi_2 \big|\, {\mathbf  E}^{a}_i  \,
   {\mathbf E}_{aj}\, \big|0\big\rangle = 
   \frac{1}{\sqrt{(2\omega_1)(2\omega_2)}}
   \Bigg[\, C_1\, \delta_{ij}\, q_1^{\mu}q_{2\mu} + \nonumber  \\
      C_2\,(\,q_{1k}\, q_{2l} + q_{1l}\, q_{2k} - 
      \frac{2}{3}\,\delta_{ij}\, q_1^{\mu}\,q_{2\mu}\,)\,\Bigg]\, .~~
\label{eqn:Dec_HT-C12}
\end{eqnarray}
If the  polarization of the heavy $Q\bar Q$ initial and 
final states is measured, more information can be extracted 
from these transitions and a
more general form of \Eq{eqn:Dec_HT-C12} is 
appropriate~\cite{Dubynskiy:2007ba}.

Important {\it single} light-hadron transitions include 
those involving the $\eta, \piz$, or $ \omega$ mesons.  
The general form of the light-hadronic factor for
the $\eta$ transition, which is dominantly E1-M2, 
is~\cite{Kuang:2006me}
\begin{equation}
 \frac{g_e\,g_M}{6} \big\langle \eta\big| {\mathbf E}^a_i\, \partial_i 
 {\mathbf B}^a_j\big| 0 \big\rangle = i (2\pi)^\frac{3}{2}\, C_3\, q_j\, .
\label{eqn:Dec_HT-C3}
\end{equation}
The $\piz$ and $\eta$ transitions are related by the 
structure of chiral symmetry-breaking~\cite{Ioffe:1980mx}.
Many more details for these and other transitions within the 
context of the Kuang-Yan model can be found in the review 
of Kuang~\cite{Kuang:2006me}. 

A summary of all experimentally observed hadronic transitions and 
their corresponding theoretical expectations within the Kuang-Yan (KY)
model  is presented in \Tab{tab:Dec_KYExp}.  The experimental 
partial widths are determined from the measured branching fractions 
and total width of the initial 
state.  If the total width is not well-measured, the 
theoretically-expected width is used, as indicated. 
The theory expectations are adjusted using the current 
experimental inputs to rescale the model parameters 
$|C_1|$ and $|C_2|$ in \Eq{eqn:Dec_HT-C12} and $|C_3|$ in
\Eq{eqn:Dec_HT-C3}.

\begin{table}[htdp]
\caption{Partial widths for observed hadronic transitions. 
         Experimental results are from PDG08~\cite{Amsler:2008zzb} 
         unless otherwise noted. Partial widths determined from 
         known branching fractions and total widths. 
         Quoted values assume total widths of
         $\Gamma_{\rm tot}(\chi_{b2}(2P)) = 138 \pm 19$\kev~\cite{Cawlfield:2005ra}, 
         $\Gamma_{\rm tot}(\chi_{b1}(2P)) =  96 \pm 16$\kev~\cite{Cawlfield:2005ra},
         $\Gamma_{\rm tot}(\UoneDT) = 28.5$\kev~\cite{Kwong:1988ae,He:2009bf}
         and
         $\Gamma_{\rm tot}(\UnS{5}) = 43 \pm 4$\mev~\cite{Aubert:2008hx}.
         Only the charged dipion transitions are shown here, 
         but the corresponding measured $\dipiz$ rates,
         where they exist,
         are consistent with a parent state of $I=0$.
         Theoretical results are given using the 
         Kuang and Yan (KY)
         model~\cite{Kuang:1981se,Kuang:1989ub,Kuang:2006me}.  
         Current experimental inputs were used to rescale the parameters 
         in the theory partial rates. 
         ($|C_1| = 10.2\pm0.2 \times 10^{-3}$, $C_2/C_1 = 1.75\pm 0.14$,
         $C_3/C_1 = 0.78 \pm 0.02$ for the Cornell case)
}
\label{tab:Dec_KYExp}
\setlength{\tabcolsep}{0.30pc}
\begin{center}
\begin{tabular}{lcc}
\hline\hline
\rule[10pt]{-1mm}{0mm}
Transition & $\Gamma_{\rm partial}$~(keV) & $\Gamma_{\rm partial}$~(keV) \\
\rule[10pt]{-1mm}{0mm}
 & (Experiment) & (KY Model) \\
\hline 
\rule[10pt]{-1mm}{0mm}
& & \\[-3mm]
 $\psip$  & &  \\
\rule[10pt]{-1mm}{0mm}
    $~~\to \jpsi + \pi^+ \pi^- $  & $102.3 \pm 3.4$& input~($|C_1|$) \\
    $~~\to \jpsi + \eta $  & $10.0 \pm 0.4$ & input~($C_3/C_1$)  \\
    $~~\to \jpsi + \piz $ & $0.411 \pm 0.030${~\cite{Mendez:2008kb}} & 0.64~\cite{Ioffe:1980mx} \\
    $~~\to  h_c(1P) + \piz $ & $0.26 \pm 0.05${~\cite{Ablikim:2010rc}} & 0.12-0.40 ~\cite{Kuang:2002hz} \\[1mm]
\rule[10pt]{-1mm}{0mm}
 $\psit$ & & \\
\rule[10pt]{-1mm}{0mm}
    $ ~~\to \jpsi + \pi^+ \pi^-$ & $52.7 \pm 7.9$ & input~($C_2/C_1$) \\
    $~~\to \jpsi + \eta $ & $24 \pm 11 $ & \\[1mm]
 $\psi(3S)$ & & \\
    $~~\to \jpsi + \pi^+ \pi^- $ & $< 320$ (90\%~CL)& \\[1mm]  
 & & \\[-3mm]
\rule[10pt]{-1mm}{0mm}
 $\UnS{2}$ & & \\
\rule[10pt]{-1mm}{0mm}
    $~~\to \UnS{1} + \pi^+ \pi^- $ & $5.79 \pm 0.49$ & $8.7$~\cite{Ke:2007ih}\\ 
    $~~\to \UnS{1} + \eta $ & $(6.7 \pm 2.4) \times
    10^{-3} $ & $0.025$~\cite{Kuang:2006me}\\[1mm]  
    
\rule[10pt]{-1mm}{0mm}
 $\UoneDT$ & & \\
\rule[10pt]{-1mm}{0mm}
    $~~\to \UnS{1} + \pi^+ \pi^- $ & $0.188 \pm 0.046${~\cite{Sanchez:2010kz}}  & $0.07$~\cite{Moxhay:1987ch}  \\ [1mm]
    
\rule[10pt]{-1mm}{0mm}
$\chi_{b1}(2P)$ & & \\
\rule[10pt]{-1mm}{0mm}
    $~~\to  \chi_{b1}(1P) + \pi^+ \pi^- $ & $0.83 \pm 0.33${~\cite{Cawlfield:2005ra}} & $0.54$~\cite{Kuang:1988bz}  \\
    $~~\to \UnS{1} + \omega $ & $1.56 \pm 0.46$ & \\
\rule[10pt]{-1mm}{0mm}
$\chi_{b2}(2P)$ & & \\
\rule[10pt]{-1mm}{0mm}
    $~~\to \chi_{b2}(1P) + \pi^+ \pi^- $ & $0.83 \pm 0.31${~\cite{Cawlfield:2005ra}}  & $0.54$~\cite{Kuang:1988bz}  \\
    $~~\to \UnS{1} + \omega $     & $1.52 \pm 0.49$ & \\[1mm]
    
\rule[10pt]{-1mm}{0mm}
 $\UnS{3}$ & & \\
\rule[10pt]{-1mm}{0mm}
   $~~\to \UnS{1} + \pi^+ \pi^-$  & $0.894\pm 0.084$ & $1.85$~\cite{Ke:2007ih}\\ 
   $~~\to \UnS{1} + \eta $        & $<3.7\times10^{-3}$ & $ 0.012 $~\cite{Kuang:2006me}\\
   $~~\to \UnS{2} + \pi^+ \pi^-$  & $0.498 \pm 0.065$ & $0.86$~\cite{Ke:2007ih}\\[1mm]
 
\rule[10pt]{-1mm}{0mm}
 $\UnS{4}$ & & \\
\rule[10pt]{-1mm}{0mm}
   $~~\to \UnS{1} + \pi^+ \pi^- $ & $1.64 \pm 0.25$ & $4.1$~\cite{Ke:2007ih} \\ 
   $~~\to \UnS{1} + \eta$         & $4.02 \pm 0.54$ & \\
   $~~\to \UnS{2} + \pi^+ \pi^- $ & $1.76 \pm 0.34$ & $1.4$~\cite{Ke:2007ih} \\[1mm]
   
\rule[10pt]{-1mm}{0mm}
 $\UnS{5}$ & & \\
\rule[10pt]{-1mm}{0mm}
   $~~\to \UnS{1} + \pi^+ \pi^-$  & $228 \pm 33$ &  \\
   $~~\to \UnS{1} +  K^+ K^- $    & $26.2 \pm 8.1$ &  \\
   $~~\to \UnS{2} + \pi^+ \pi^-$  & $335 \pm 64$ & \\
   $~~\to \UnS{3} + \pi^+ \pi^-$  & $206 \pm 80$ & \\[1mm]
\hline\hline 
\end{tabular}
\end{center}
\end{table}
  
The multipole expansion works well for transitions of heavy $Q\bar Q$ states
below threshold~\cite{Eichten:2007qx}.  Within the specific KY model a fairly
good description of the rates for the two-pion transitions is observed.  The
partial width $\Gamma(\UnS{3} \to \UnS{1} \dipi)$ was predicted
to be suppressed due to cancellations between the various QCS intermediate
states~\cite{Kuang:1981se}, allowing nonleading terms, $O(v^2)$, to
contribute significantly.  The non-$S$-wave behavior of the $m_{\dipi}$
dependence in $\UnS{3}$ decays, also observed in the $\UnS{4} \to
  \UnS{2} \dipi$ transitions, may well reflect this influence of
higher-order terms.  Other possibilities are discussed in
\Sec{sec:Dec_HT_Ups5Spipi}.  
For single light-hadron transitions some puzzles
remain. For example, the ratio 
\beq
\frac{\Gamma(\UnS{2}\to\eta\UnS{1})}{\Gamma(\psip\to\eta \jpsi(1S))}
\eeq
is much smaller than expected from theory (see \Sec{sec:Dec_HT_U2eta}).

The situation is more complicated for above-threshold, 
strong open-flavor decays.  The issues are manifest 
for $\UnS{5}$ two-pion transitions
to $\UnS{n}~(n=1,2,3)$.  First, states above threshold do not have sizes
that are small compared to the QCD scale 
(\eg $\sqrt{\langle r^2\rangle_{\UnS{5}}} = 1.2 {\rm~fm}$),
making the whole QCDME approach less
reliable. Second, even within the KY model, the QCS intermediate 
states are no longer far away from the initial-state mass.  
Thus the energy denominator, $E_i - E_{KL}$ in \Eq{eqn:Dec_QCS}, 
can be small, leading to large enhancements in
the transition rates that are sensitive to the exact position of the
intermediate states~\cite{Ke:2007ih}. This is the reason for the large theory
widths seen in \Tab{tab:Dec_KYExp}. Third, a number of 
new states 
(see \Secs{sec:SpecExp_Unanticipated})
that do not fit into the
conventional $Q\bar Q$ spectra have been observed, implying
additional degrees of freedom appearing in the QCD spectrum beyond 
naive-quark-model counting.  Hence the physical quarkonium 
states have open-flavor meson-pair contributions and possible 
hybrid ($Q\bar Qg$) or tetraquark contributions.  
The effect of such terms on hadronic transitions is not yet
understood~\cite{Simonov:2008qy}.  A possibly-related puzzle is the
strikingly-large ratio 
\beq
R_\eta[\UnS{4}] \equiv \frac{\Gamma(\UnS{4} \to \UnS{1}\, \eta)}{
\Gamma(\UnS{4} \to \UnS{1}\, \dipi)}\approx 2.5~.~~
\eeq
This ratio is over a hundred times larger than one would expect  
within the KY model, which is particularly surprising, since  the  
similarly-defined ratio, 
$R_\eta[\UnS{2}]\approx 10^{-3}$,  is actually less than half of the KY  
model expectations (see \Tab{tab:Dec_KYExp}) 
and the experimental upper bound
on $R_\eta[\UnS{3}]$ is already slightly
below KY-model expectations.
Much theoretical work remains 
in order to understand the hadronic transitions of the 
heavy $Q\bar Q$ systems above threshold.

Many of the new $XYZ$ states
(see \Secs{sec:SpecExp_Unanticipated})
are candidates for
so-called hadronic molecules. If this were the case, and they indeed owe
their existence to non\-perturbative interactions among heavy mesons, the
QCDME needs to be extended by heavy-meson loops. These loops 
provide non\-multipole, long-ranged contributions, so as to allow 
for the inclusion of their influence.
However, if hadron loops play a significant, sometimes even non\-perturbative,
role above $\bar DD$ threshold, one should expect them to be at least of some
importance below the lowest inelastic threshold. Correspondingly, one should
expect to find some systematic deviations between quark-model
predictions and data.  
By including intermediate heavy-meson effects within the framework of
QCDME~\cite{Moxhay:1988ri,Zhou:1990ik}, improved agreement 
with the experimental data on
dipion transitions in the $\psi$ and $\Ups$ systems was obtained.

Alternatively, a non\-relativistic effective field
theory (NREFT) was
introduced~\cite{Guo:2009wr} 
that allows one to study the effect of heavy-meson loops on
charmonium transitions with controlled uncertainty. In this work, it was argued
that the presence of meson loops resolves the long-standing discrepancy 
between, on the one hand, the
values of the light-quark mass-differences extracted from the masses of the
Goldstone bosons, and on the other,
the ratio of selected charmonium transitions, namely
$\psip\to \jpsi \piz/ \psip\to \jpsi \eta$.
NREFT uses the velocity of the
heavy mesons in the intermediate state,
$v\sim\sqrt{|m-2m_D|/m_D}$,  as expansion parameter.
Thus, for transitions of states below \DDbar\ threshold, the
analytic continuation of the standard expression is to be used.
For low-lying charmonium transitions, $v$ is found to be of order 0.5.
A typical transition via a $D$-meson loop may then be counted as
\beq
v^3/(v^2)^2 \ \times \ \mbox{vertex factors}~.~~
\eeq
For the transition between two $S$-wave charmonia, which decay
into $D^{(*)}\bar D^{(*)}$ via a $P$--wave vertex, the vertex
factors scale as $v^2$. Thus the loop contributions appear to
scale as order $v$, and,
for values of the velocity small relative to those that can be
captured by QCDME, are 
typically suppressed. However, in certain cases enhancements
may occur. For example, for $\psip\to \jpsi \piz$ and
$\psip\to \jpsi \eta$, flavor symmetry is broken, and therefore
the transition matrix element needs to scale as $\delta$, the energy
scale that quantifies the degree of flavor-symmetry violation in
the loop and which originates from the mass differences of charged and
neutral $D$-mesons. However, if an energy scale is pulled out of
the integral, this needs to be balanced by removing the energy scale $v^2$
from the power counting. Thus the estimate for the
loops contribution scales as $\delta/v$ compared to the piece of order $\delta$
that emerges from QCDME. Hence, for certain transitions, meson loops
are expected to significantly influence the rates.

The contributions of heavy-meson loops to charmonium decays
follow a very special pattern. They are expected to be more important
for charmonia pairs close to the lowest two-meson threshold.  In addition,
loops appear to be suppressed for transitions between $S$ and $P$ wave
charmonia~\cite{Guo:2010zk}.  These conjectures
can be tested experimentally by systematic, high-precision 
measurements of as many transitions as possible.
Once the role of meson loops is established 
for transitions between conventional heavy quarkonia below
or very close to open-flavor threshold,
an extension from perturbative to 
a non\-perturbative treatment might lead to
a combined analysis of excited charmonia and
at least a few of the $XYZ$ states.

\subsubsection{Branching fractions for $\psip\to X\jpsi$}
\label{sec:DecPsipBRs}

Precision measurements of the hadronic transitions $\psip\to X\jpsi$
are important for a number of reasons beyond
the obvious one of providing an accurate normalization
and accounting for \psip\ decays.  First, they
allow experimental comparisons with increasingly sophisticated
theoretical calculations (see \Sec{sec:Dec_HadTranTheory}).  
Second, they can be used in comparisons
with the analogous $\Ups$ transitions in the bottomonium system.
Finally, the transitions provide a convenient way to access the
$\jpsi$, where the transition pions in $\psip\to\dipi\jpsi$,
for example, can be used in tagging a clean and well-normalized 
\jpsi\ sample.  
Using its full sample of 27~million $\psip$ decays, 
CLEO~\cite{Mendez:2008kb} measured these rates 
with substantially improved precision over
previous measurements.  
The analysis measures $\mathcal{B}(\psip \to
\dipi\jpsi)$ {\it inclusively}, selecting events
based upon the mass recoiling against the 
transition dipion and placing no restriction on the \jpsi\ decay.  
Its uncertainty is dominated by a 2\% systematic error on
the produced number of \psip.
Other transitions to \jpsi, the {\it exclusive} 
modes through \dipiz, $\eta$, \piz, and
$\psip\to\gamma\chi_{cJ}\to\gamma\gamma\jpsi$, 
as well as the {\it inclusive} rate for $\psip\to\jpsi\,+\,{\rm any}$,
are all measured {\it relative} to the
\dipi\ transition and use $\jpsi\to\ell^+\ell^-$,
thereby reducing systematic error from the 
number of \psip, which cancels in the ratios.
These relative rates are measured with precision from 2--6\%.
The absolute branching fractions determined from this
analysis are shown in \Tab{tab:Dec_CLEO_BXJpsi}
which are higher values than most previous measurements.
The ratio of $\dipiz$ to $\dipi$ transitions
is consistent with one-half,
the expectation from isospin invariance, which was not
the case in some earlier analyses.
See \Sec{sec:Dec_Bgchicj} for a discussion of
the $\gamma\gamma\jpsi$ portion of this analysis,
which addresses the product branching fractions
$\Brat(\psip\to\gamma\chi_{cJ})\times\Brat(\chi_{cJ}\to\gamma\jpsi)$
and whether any nonresonant such final states are present.

\begin{table}[t]
\caption{Results from the branching fraction
analyses for \psip\ and \psit\ decays to $X\jpsi$ 
from CLEO~\cite{Mendez:2008kb,Adam:2005mr}}
\label{tab:Dec_CLEO_BXJpsi}
\setlength{\tabcolsep}{0.18pc}
\begin{center}
\begin{tabular}{lcc}
\hline \hline
\rule[10pt]{-1mm}{0mm}
Quantity & \psip~(\%) & \psit~(\%) \\[0.8mm]
\hline
\rule[10pt]{-1mm}{0mm}
$\Brat(\dipi\jpsi)$     &$35.04\pm0.07~\pm0.77~$
 &  $0.189 \pm 0.020\pm0.020$\\[0.8mm]
$\Brat(\dipiz\jpsi)$    &$17.69\pm0.08~\pm0.53~$
 &  $0.080\pm0.025\pm0.016$\\[0.8mm]
$\Brat(\eta\jpsi)$      &$~3.43\pm0.04~\pm0.09~$
 &  $0.087\pm0.033\pm0.022$\\[0.8mm]
$\Brat(\piz\jpsi)$      &$0.133\pm0.008\pm0.003$
 &  $<0.028$ at 90\%~CL\\[0.8mm]
$\Brat(\jpsi+{\rm any})$&$62.54\pm0.16~\pm1.55~$ & - \\[1.0mm]
$\dfrac{\Brat(\dipiz\jpsi)}{\Brat(\dipi\jpsi)}$&$50.47\pm1.04$
 & $42\pm17$\\[3.5mm]
\hline \hline
\end{tabular}
\end{center}
\end{table}

\subsubsection{Observation of $\psip\to\piz\hsubc$}
\label{sec:Dec_PsipToPi0Hc}

The hadronic transition $\psip\to\piz h_c(1P)$
was first observed by CLEO
for $\hsubc\to\gamma\etac$~\cite{Rosner:2005ry,Dobbs:2008ec}
and later seen for $h_{c}(1P)\to 2(\dipi)\piz$~\cite{:2009dp}
(see \Sec{sec:Dec_hchaddec}).
However, those analyses could only measure
product branching fractions.
BESIII~\cite{Ablikim:2010rc} has used its much
larger 106~million \psip\ sample to inclusively
measure $\Brat(\psip\to\piz\hsubc)$ by
observing a significant enhancement at the $\hsubc$ 
in the mass recoiling against a reconstructed \piz,
with none of the $\hsubc$ decay products reconstructed. The result
is $\Brat(\psip\to\piz\hsubc)=(8.4\pm1.3\pm1.0)\times10^{-4}$,
in agreement with the range predicted in~\cite{Kuang:2002hz}
(see \Tab{tab:Dec_KYExp}).

\subsubsection{Nonobservation of $\etacp\to\dipi\etac$}

The historically uncooperative nature of the \etacp\ 
has continued into the modern era, despite its discovery
(\Sec{sec:SpecExp_etac2s}).
CLEO~\cite{:2009vg} failed not only to see \etacp\ in
radiative transitions (\Sec{sec:Dec_PsipToGEtac}),
but also failed to observe the
expected dipion transition to \etac, setting the
upper limit 
\begin{eqnarray}
&{\cal B}(\psip\to\gamma\etacp)\times{\cal
B}(\etacp\to\dipi\etac)&\nonumber\\
&<1.7\times 10^{-4}{\rm ~at~90\%~CL}\, .&~
\end{eqnarray}

\subsubsection{Observation of $\psit\to X \jpsi$}

Hadronic transitions of the $\psit$ to $\jpsi$ are sensitive to
the relative sizes of the $2^3S_1$ and $1^3D_1$ admixture contained in
the $\psit$, and thus, at least in principle, contain information
about the nature of the $\psit$.  BESII~\cite{Bai:2003hv} first
reported evidence for 
the transition $\psit \to \dipi\jpsi$ 
at $\sim3\sigma$ significance using
approximately $2\times10^5$ $\psit$ decays.
CLEO~\cite{Adam:2005mr}, with roughly $2\times10^6$ $\psit$ decays,
later observed the $\dipi$ transition at $11.6\sigma$,
and also found
evidence for the $\dipiz\jpsi$~($3.4\sigma$) and $\eta\jpsi$~($3.5\sigma$)
transitions. 
The CLEO results appear in \Tab{tab:Dec_CLEO_BXJpsi}.
With more data, these hadronic transitions could
be used to shed light on the $c\overline{c}$ purity of the $\psip$
and $\psit$ states~\cite{Voloshin:2005sd}.

\subsubsection{Observation of $\UnS{2}\to\eta\,\UnS{1}$}
\label{sec:Dec_HT_U2eta}

\begin{figure}[t]
  \begin{center}
    \includegraphics[width=\figwid]{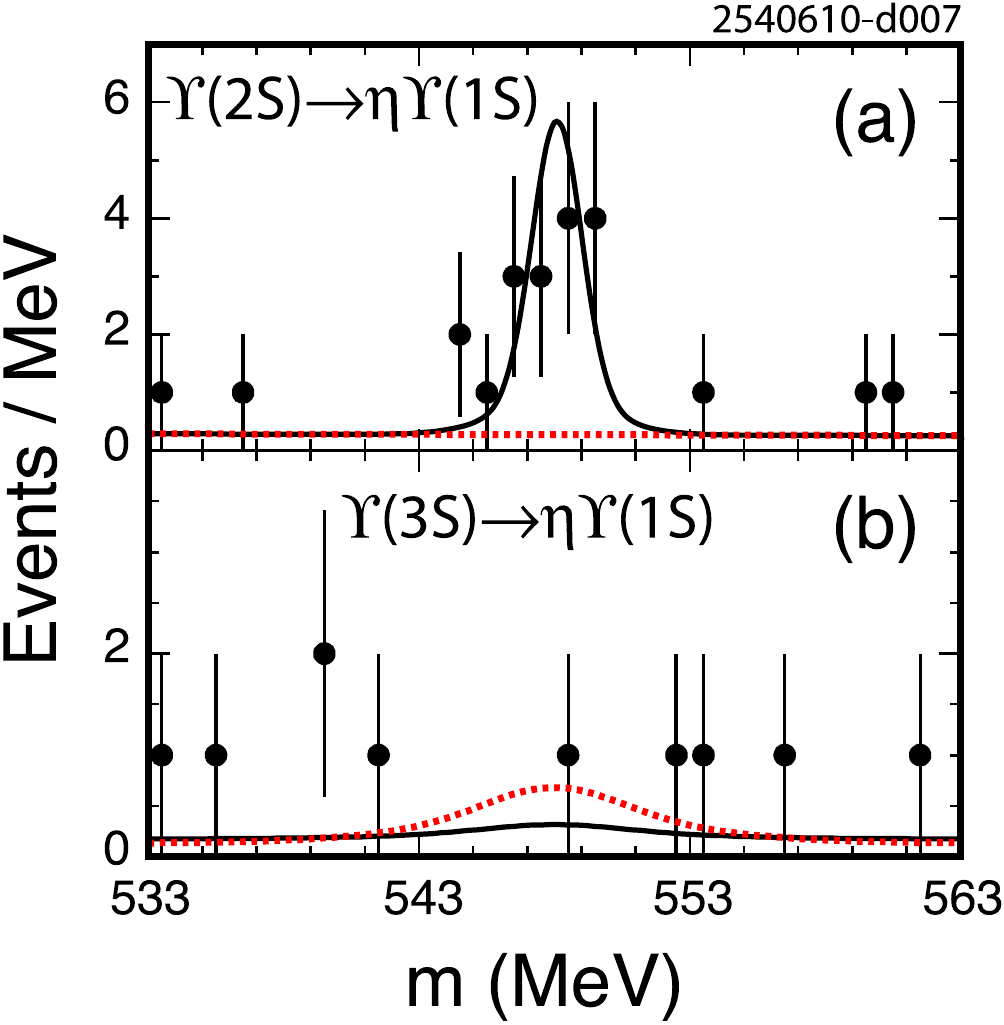}
      \caption{From CLEO~\cite{He:2008xk}, the 
               invariant mass of the $\eta$ candidate 
               for (a)~$\UnS{2}\to\eta\,\UnS{1}$ 
               and (b)~$\UnS{3}\to\eta\,\UnS{1}$ for  
               $\eta$-mesons exclusively reconstructed in the
               three decay modes $\eta\to\gamma\gamma$, $\dipi\piz$, 
               and $3\piz$.  In each case, the {\it solid curve} 
               corresponds to the best fit to a flat background and
               signal.
               The {\it dotted curve} represents (a)~the best-fit
               background, and (b)~the 90\%~CL upper 
               limit for a signal.
               \AfigPermAPS{He:2008xk}{2008} }
       \label{fig:Dec_Ups_EtaUps}
  \end{center}
\end{figure}

The observation of $\UnS{2}\to\eta\,\UnS{1}$ using 9~million
$\UnS{2}$ decays collected with the CLEO~III detector represents
the first observation of a quarkonium transition involving the spin
flip of a bottom quark.  The transition rate is
sensitive to the chromomagnetic moment of the $b$ quark.  Using three
decay modes, $\eta\to\gamma\gamma$, $\dipi\piz$, and $3\piz$,
CLEO~\cite{He:2008xk} observes $13.9^{+4.5}_{-3.8}$ events of the form
$\UnS{2}\to\eta\,\UnS{1}$ (\Fig{fig:Dec_Ups_EtaUps}).
Using the difference in log-likelihood for fits with and without
signal, it was determined that this corresponds to a $5.3~\sigma$
observation.  Correcting for acceptance and incorporating systematic
errors, the observed number of events translates to a branching
fraction of
\begin{eqnarray}
\Brat(\UnS{2}\to\eta\,\UnS{1})&=&(2.1^{+0.7}_{-0.6}\pm0.3)\times10^{-4}\nonumber\\
\Brat(\UnS{3}\to\eta\,\UnS{1})&<&1.8\times10^{-4}{\rm ~at~90\%~CL}\, .~~~~~~
\end{eqnarray}
The $2S$ rate is a factor of four smaller than one would expect scaling
from the analogous charmonium transition rate, $\psip\to\eta\jpsi$.  
Similarly, the $3S$ limit is
already two times smaller than what one would expect from the same
scaling.  The interpretation of this pattern is still unclear.

\subsubsection{Observation of $\chi_{b1,2}(2P)\to\omega\UnS{1}$}
\label{sec:Dec_HT_chib12omegaUps}

 In 2003, CLEO~\cite{Severini:2003qw} reported the 
observation of the decay chain $\UnS{3}\to\gamma\chi{bJ}(2P)$,
$\chi{bJ}(2P)\to\omega\UnS{1}$, for $J=1,2$, with $\omega\to\dipi\piz$ and
$\UnS{1}\to\ell^+\ell^-$, for which there is marginal phase space
available, and measured
\begin{eqnarray}
\Brat(\chi_{b1}\to\omega\UnS{1})=(1.63^{+0.35\,+0.16}_{-0.31\,-0.15})\%&\nonumber\\
\Brat(\chi_{b2}\to\omega\UnS{1})=(1.10^{+0.32\,+0.11}_{-0.28\,-0.10})\%&\,.~
\end{eqnarray}
The relative rates are comparable, in agreement with Voloshin's
prediction~\cite{Voloshin:2003zz} 
on the basis of $S$-wave phase space factors and the
$E1*E1*E1$  gluon configurations expected by the multipole expansion. 

\subsubsection{Dipion transitions from \UnS{2S,3}}
\label{sec:Dec_HT_Ups2S3Spipi}

  The double-hump dipion invariant mass distribution
in \upsDec{3}{1} transitions has been thought to be
at least puzzling, and frequently thought to be 
anomalous, indicative of either new physics or
an intermediate scalar dipion resonance. However, a
CLEO~\cite{CroninHennessy:2007zz} analysis 
has offered an alternative to these characterizations;
simply, that dipion-quarkonium dynamics 
expected to occur within QCD are responsible.
Brown and Cahn~\cite{Brown:1975dz}
derived the general matrix element for dipion 
transitions from heavy vector quarkonia; it
is constrained by PCAC and 
simplified by treating it as a multipole 
expansion~\cite{Gottfried:1977gp,Yan:1980uh,Voloshin:1980zf}.
This general matrix element has three terms:
one proportional to $(m_{\pi\pi}^2-2m_\pi^2)$,
one proportional to the product $E_1E_2$ of the 
two pion energies in the parent rest frame,
and a third which characterizes the transition
requiring a $b$-quark spin-flip. Although
the third chromomagnetic term is thought to be
highly suppressed, and therefore ignorable, 
even the second term had generally
been neglected prior to the CLEO analysis.
Sensitivity to the second term, which 
is also proportional to $\cos\theta_X$, $\theta_X$ being
the dipion helicity angle, is greatest at
low $m_{\pi\pi}$, where experiments using only
charged pions can only reconstruct very few events:
soft charged particles curl up in the
detector magnetic field before reaching
any tracking chambers. The notable
aspect of the CLEO analysis is not only that
it fits for all three terms with complex form factors
\clg{A}, \clg{B}, and \clg{C}, respectively, but
also that it performs a simultaneous fit to dipion
transitions through both charged {\it and neutral} 
pion pairs. The latter subset of the data
enhances sensitivity to the $\cos\theta_X$ dependence because
even when neutral pions are slow, frequently both 
\piz-decay photons can 
be reconstructed in the calorimeter.

\begin{figure}[b]
  \begin{center}
    \includegraphics[width=\figwid]{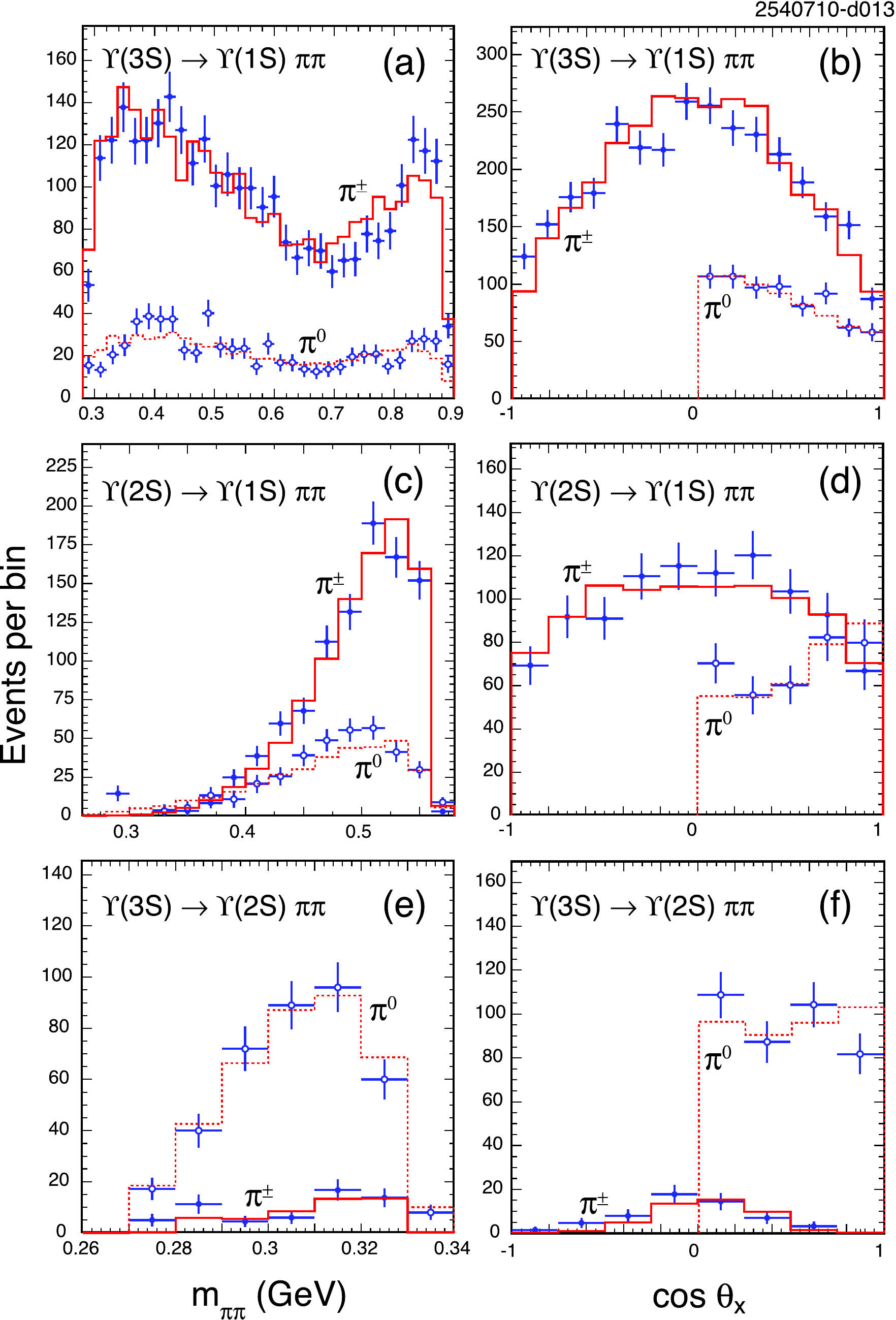}
      \caption{From CLEO~\cite{CroninHennessy:2007zz},
               (a), (c), (e) the dipion mass distributions
               and (b), (d), (f) dipion helicity angle $\cos\theta_X$
               distributions for the
               transitions (a), (b) \upsDec{3}{1},
               (c), (d) \upsDec{2}{1}, and (e), (f) \upsDec{3}{2},
               in which {\it closed (open) circles} represent the 
               \dipi~(\dipiz) data, and {\it solid (dotted) line histogram}
               the MC simulation for \dipi~(\dipiz) transitions
               generated with the best-fit parameters in 
               \Tab{tab:Dec_CLEOMpipiUps}.
               \AfigPermAPS{CroninHennessy:2007zz}{2007} }
       \label{fig:Dec_CLEOdipi}
  \end{center}
\end{figure}

Results are shown in
\Fig{fig:Dec_CLEOdipi} and \Tab{tab:Dec_CLEOMpipiUps},
from which the following conclusions are drawn:
\begin{itemize}
\item{The CLEO data in all three \Ups\ dipion transitions
can be represented well by a two-term form of
the matrix element, provided the form factors
are allowed to be different for all three decays.}
\item{Sensitivity of the CLEO data to
the chromomagnetic term, \ie to a nonzero value
of \clg{C}/\clg{A}, is small because the functional
dependencies of the \clg{B}\ and \clg{C}\ terms
are quite similar; a much larger dataset
of \upsDec{3}{1} is required to probe this component.
None of the three \Ups\ dipion transitions
require a nonzero \clg{C}/\clg{A}; \ie there
is no evidence yet that spin-flips play a significant
role in these decays. In the case of \upsDec{3}{1},
where a slight sensitivity to this term exists,
an upper limit of $|\clg{C}/\clg{A}|<1.09$ at 90\%~CL is given.}
\item{The dynamics in \upsDec{2}{1}
are reproduced by the fits, but only if nonzero
values of \clg{B}/\clg{A}\ are allowed. The data
in this transition
are described well without any complex component in \clg{B}/\clg{A}.}
\item{The two-peak structure in \upsDec{3}{1}
is reproduced without any new physics or
any intermediate dipion resonances, \clg{B}/\clg{A}
is found to have a significant {\it complex} component,
and the dipion-helicity form factor is nearly
three times larger than that of the dipion-mass-dependent term.}
\end{itemize}
Dubynskiy and Voloshin~\cite{Dubynskiy:2007ba} comment
further, on both the formalism and prospects for
learning more from such decays. They cast doubt
upon the possibility that \clg{B}\ can have
have a complex component when $\clg{C}\equiv 0$, 
on general principles; conclude
that \clg{B}\ and \clg{C}\ terms are degenerate
unless \clg{C}\ is set to zero;
and suggest resolving the \clg{B}-\clg{C}\ terms' 
degeneracy by including initial- and final-state
polarization information in fits to experimental data.

\begin{table}[t]
\caption{Results from fits to the CLEO~\cite{CroninHennessy:2007zz}
         $\UnS{m}\to\pi\pi\UnS{n}$, $(m,\, n)=(3,\,2),\,(3,\,1),\,(2,\,1)$ 
         transitions data for the complex form factors (\clg{A}, \clg{B},
         \clg{C}) of
         the three terms,
         and the associated phase angle, $\delta_{\clg{BA}}$.
         The phase angles are quoted in 
         degrees, and have a two-fold
         ambiguity of reflection in the real axis.
         The first three fits constrain the chromomagnetic
         coefficient to be zero ($\clg{C}\equiv 0$),
         but \clg{C} floats in the fourth fit. The operators
         \clg{R}\ and \clg{I}\ denote real and imaginary parts, respectively}
\label{tab:Dec_CLEOMpipiUps}
\setlength{\tabcolsep}{0.85pc}
\begin{center}
\begin{tabular}{llc}
\hline \hline
\rule[10pt]{-1mm}{0mm}
Transition & Quantity & Value \\
\hline
\rule[10pt]{-1mm}{0mm}
\upsDec{3}{1}
   & \ReBA         &  $-2.52  \pm  0.04$ \\ 
   & \ImBA         &  $\pm 1.19\pm 0.06$ \\
$\qquad(\clg{C}\equiv0)$   & $|\BoAf|$       &  $2.79  \pm   0.05$ \\
   & $\delta_{\clg{BA}}~(^\circ)$ &  $155\,(205)\pm 2$  \\[2mm]
\upsDec{2}{1}
   & \ReBA   & $-0.75\pm 0.15$ \\
   & \ImBA   & $0.00 \pm 0.11$ \\
$\qquad(\clg{C}\equiv0)$   & $|\BoAf|$ & $0.75 \pm 0.15$ \\
   & $\delta_{\clg{BA}}~(^\circ)$ & $ 180\pm 9$ \\[2mm]
\upsDec{3}{2}
   & \ReBA & $-0.40\pm 0.32$ \\ 
$\qquad(\clg{C}\equiv0)$   & \ImBA & $0.00 \pm 1.1$ \\[2mm]
\upsDec{3}{1}
  &  $|\BoAf|$ & $2.89 \pm 0.25$ \\ 
  & $|\CoAf|$ & $0.45 \pm 0.40$ \\
\qquad(\clg{C} floats)&&$<1.09~(90\%~{\rm CL})$ \\[1mm]
\hline \hline
\end{tabular}
\end{center}
\end{table}

\begin{table}[b]
\caption{Branching fractions for bottomonium dipion
transitions $\UnS{n}\to\pi\pi\UnS{m}$ for $n=2,3$ and $m<n$,
as compiled by the Particle Data Group
as indicated and as measured by CLEO and \babar;
the PDG10~\cite{Nakamura:2010pdg}
{\it (PDG08~\cite{Amsler:2008zzb})}
numbers include {\it (do not include)} 
the CLEO~\cite{Bhari:2008rb} and \babar~\cite{Aubert:2008bv} results}
\label{tab:Dec_CLEOBRpipiUps}
\setlength{\tabcolsep}{0.35pc}
\begin{center}
\begin{tabular}{lcc}
\hline \hline
\rule[10pt]{-1mm}{0mm}
Transition & \clg{B}~(\%) & Source\\[0.7mm]
\hline
\rule[10pt]{-1mm}{0mm}
$\UnS{2}\to\dipi\,\UnS{1} $ & $18.8\pm0.6$          & PDG08~\cite{Amsler:2008zzb} \\[0.7mm]
                          & $18.02\pm0.02\pm0.61$ & CLEO~\cite{Bhari:2008rb} \\[0.7mm]
                          & $17.22\pm0.17\pm0.75$ & \babar~\cite{Aubert:2008bv} \\[0.7mm]
                          & $18.1\pm0.4$          & PDG10~\cite{Nakamura:2010pdg} \\[2mm]
$\UnS{2}\to\dipiz\,\UnS{1}$ & $9.0\pm0.8$           & PDG08~\cite{Amsler:2008zzb} \\[0.7mm]
                          & $8.43\pm0.16\pm0.42$  & CLEO~\cite{Bhari:2008rb} \\[0.7mm]
                          & $8.6\pm0.4$           & PDG10~\cite{Nakamura:2010pdg} \\[2mm]
$\UnS{3}\to\dipi\,\UnS{1} $ & $4.48\pm0.21$         & PDG08~\cite{Amsler:2008zzb} \\[0.7mm]
                          & $4.46\pm0.01\pm0.13$  & CLEO~\cite{Bhari:2008rb} \\[0.7mm]
                          & $4.17\pm0.06\pm0.19$  & \babar~\cite{Aubert:2008bv} \\[0.7mm]
                          & $4.40\pm0.10$         & PDG10~\cite{Nakamura:2010pdg} \\[2mm]
$\UnS{3}\to\dipiz\,\UnS{1}$ & $2.06\pm0.28$         & PDG08~\cite{Amsler:2008zzb} \\[0.7mm]
                          & $2.24\pm0.09\pm0.11$  & CLEO~\cite{Bhari:2008rb} \\[0.7mm]
                          & $2.20\pm0.13$         & PDG10~\cite{Nakamura:2010pdg} \\[2mm]
$\UnS{3}\to\dipi\,\UnS{2} $ & $2.8\pm0.6$           & PDG08~\cite{Amsler:2008zzb} \\[0.7mm]
                          & $2.40\pm0.10\pm0.26$  & \babar~\cite{Aubert:2008bv} \\[0.7mm]
                          & $2.45\pm0.23$         & PDG10~\cite{Nakamura:2010pdg} \\[2mm]
$\UnS{3}\to\dipiz\,\UnS{2}$ & $2.00\pm0.32$         & PDG08~\cite{Amsler:2008zzb} \\[0.7mm]
                          & $1.82\pm0.09\pm0.12$  & CLEO~\cite{Bhari:2008rb} \\[0.7mm]
                          & $1.85\pm0.14$         & PDG10~\cite{Nakamura:2010pdg} \\[0.7mm]
\hline \hline
\end{tabular}
\end{center}
\end{table}

  CLEO~\cite{Bhari:2008rb} used the matrix elements determined above
to obtain correct efficiencies in the first new measurement of
these transition branching ratios in a decade.
\babar~\cite{Aubert:2008bv} followed suit soon after,
using radiative returns from $e^+e^-$ collisions
at 10.58~GeV and the nearby continuum to the narrow
\Ups\ states as a source. Both groups used
$\Ups\to\ell^+\ell^-$ decays; \babar\ used
only charged, fully-reconstructed dipions, whereas
CLEO used both charged and neutral, except for
\upsDec{3}{2}, where the low efficiency
for charged track reconstruction made the
charged mode too difficult to normalize precisely.
CLEO gained some statistical power by performing
both exclusive ($\Ups\to\ell^+\ell^-$) and
inclusive ($\Ups\to{\rm any}$) versions 
for each transition; in the latter case
the signal was obtained by a fit to the
invariant mass recoiling against the dipion
for a smooth background and signal term peaking
at the appropriate \Ups\ mass. \babar\
extracted signals by fitting distributions
of $\Delta m\equiv m(\dipi\ell^+\ell^-)-m(\ell^+\ell^-)$ 
for a smooth background and signal at
the mass difference appropriate for the
desired parent \Ups,
after cutting on $m(\ell^+\ell^-)$ around
the mass appropriate for the desired daughter \Ups.
The combination of CLEO and \babar\ branching
fractions, as shown in \Tab{tab:Dec_CLEOBRpipiUps},
which are consistent with each other and previous measurements,
made significant improvements
to the branching fraction precisions and
dominate the PDG10~\cite{Nakamura:2010pdg} world averages shown.

\subsubsection{Observation of $\UoneDJ\to\dipi\UnS{1}$}
\label{sec:Dec_HT_Ups1Dpipi}

The $\UoneDJ\to\dipi\UnS{1}$ measurements from 
\babar~\cite{Sanchez:2010kz} (see \Sec{sec:SpecExp_Ups1D})
represent the only available data on
hadronic transitions of the $\UoneDJ$ states. 
Partial rates 
are expected to be independent of $J$~\cite{Kuang:1989ub}.
Using {\it predicted}~\cite{Kwong:1988ae} branching fractions 
for $\UnS{3}\to\gamma\chi_{bJ^\prime}(2P)$ and 
$\chi_{bJ^\prime}(2P)\to\gamma\UoneDJ$, 
\babar\ quotes
\begin{eqnarray}
{\cal B}(\UoneDO\to\dipi\UnS{1}) =\hspace{3cm}&&\nonumber\\
( 0.42^{+0.27}_{-0.23} \pm 0.10)\%~(<0.82\%)\, ,~~~~~\nonumber\\
{\cal B}(\UoneDT\to\dipi\UnS{1}) =\hspace{3cm}&&\nonumber\\
 ( 0.66^{+0.15}_{-0.14} \pm 0.06 )\%\, ,~{\rm and}~~~~~~~~~~~~\nonumber\\
{\cal B}(\UoneDH\to\dipi\UnS{1}) =\hspace{3cm}&&\nonumber\\
 ( 0.29^{+0.22}_{-0.18} \pm 0.06 )\%~(<0.62\%)\, ,~~
\end{eqnarray}
where upper limits are given at 90\%~CL and include systematic uncertainties.
This hadronic 
\UoneDT\ transition provides an important benchmark for 
comparing various theoretical predictions for partial 
rates~\cite{Kwong:1988ae,Moxhay:1987ch,Kuang:1989ub}. 
Furthermore, comparing 
with the observed  $\psit \to \dipi \jpsi$ transition 
may give insight into threshold
effects in the $\psit$ state.

\subsubsection{Dipion and $\eta$ transitions from \UnS{4}}
\label{sec:Dec_HT_Ups4Spipi}

\begin{figure}[b]
  \begin{center}
    \includegraphics[width=\figwid]{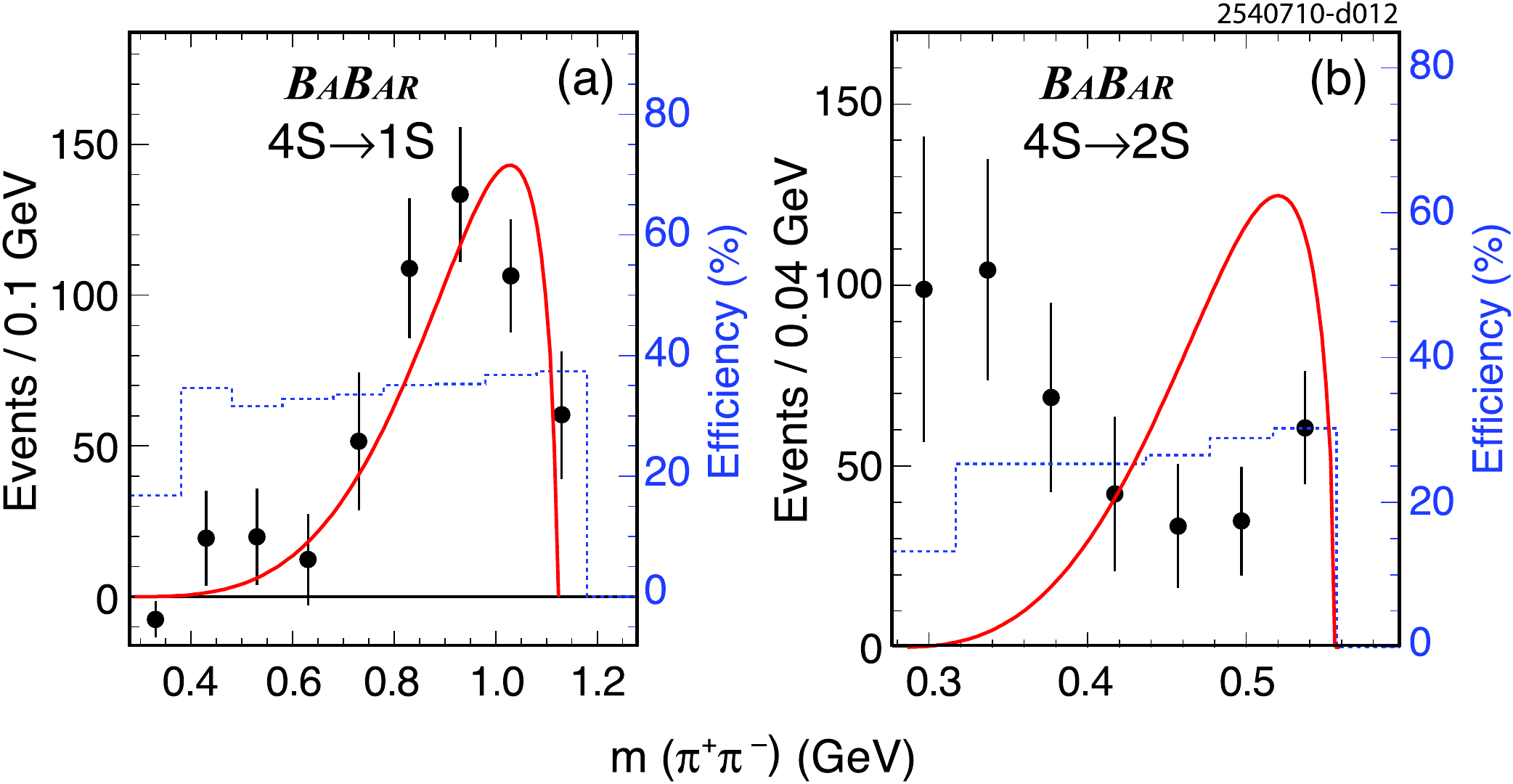}
      \caption{From \babar~\cite{Aubert:2006bm},
               efficiency-corrected dipion mass distributions
               for (a) $\UnS{4}\to\dipi\UnS{1}$,
               and (b) $\UnS{4}\to\dipi\UnS{2}$,
               in which {\it solid circles} denote
               data, {\it dotted hitograms} denote
               reconstruction and event selection efficiencies, which
               follow the scales on the {\it right-side axes},
               and the {\it solid curves} represent 
               theoretical predictions~\cite{Aubert:2006bm}.
               Experimental resolution on $m_{\dipi}$ is less than 5\mevcc.
               \AfigPermAPS{Aubert:2006bm}{2006} }
       \label{fig:Dec_BABAR4Sdipi}
  \end{center}
\end{figure}

Even above open-bottom threshold, the $b\bar{b}$ 
vector bound state undergoes hadronic transitions,
as first measured by \babar~\cite{Aubert:2006bm},
which was later updated~\cite{Aubert:2008bv}.
The latter analysis is done in a manner
similar to that described in \Sec{sec:Dec_HT_Ups2S3Spipi}, 
using mass-difference windows around $\Delta m=m[\UnS{4}]-m[\UnS{j}]$,
with $j=1,\,2$.
Belle~\cite{:2009zy} reported a similar analysis, but only
on $\UnS{4}\to\dipi\UnS{1}$. The two resulting branching fractions 
for the latter transition are consistent with one other;
PDG10 reports (averaged for $\UnS{4}\to\dipi\UnS{1}$) branching fractions
\begin{eqnarray}
\Brat(\UnS{4}\to\dipi\UnS{1})=\hspace{3cm}\nonumber\\
(0.810\pm0.06)\times 10^{-4}&\nonumber\\
\Brat(\UnS{4}\to\dipi\UnS{2})=\hspace{3cm}\nonumber\\
(0.86\pm0.11\pm0.07)\times 10^{-4}&\nonumber\\
\frac{\Brat(\UnS{4}\to\dipi\UnS{2})}{\Brat(\UnS{4}\to\dipi\UnS{1})}=
1.16\pm0.16\pm0.14\, .&~~
\end{eqnarray}
Of particular note is that while the dipion
mass spectrum for $\UnS{4}\to\dipi\UnS{1}$ has
a typical spectrum with a single peak, that of
$\UnS{4}\to\dipi\UnS{2}$ appears to have a 
double-peak structure like $\UnS{3}\to\pi\pi\UnS{1}$,
as seen in \Fig{fig:Dec_BABAR4Sdipi}.

In the same analysis, \babar~\cite{Aubert:2006bm} sought
$\eta$ transitions in the \Ups\ system as well, via $\eta\to\dipi\piz$.
While no evidence is found for such transitions from
\UnS{2}\ or \UnS{3}\ (consistent with the CLEO
results discussed
in \Sec{sec:Dec_HT_U2eta}), a quite significant signal
was observed for $\eta\UnS{1}$ final states
in the \UnS{4}\ data sample, which, if attributed
fully to resonant production, corresponds to
\begin{eqnarray}
\Brat(\UnS{4}\to\eta\UnS{1})=(1.96\pm0.06\pm0.09)\times 10^{-4}\nonumber\\
\frac{\Brat(\UnS{4}\to\eta\UnS{1})}{\Brat(\UnS{4}\to\dipi\UnS{1})}=
2.41\pm0.40\pm0.12\, .~~
\end{eqnarray}
The \babar\ continuum sample taken just below \UnS{4} 
is only a small fraction of that taken on-\UnS{4}, so
even though no $\eta\UnS{1}$ are observed in that sample,
the possibility that the observed events are attributable
to $e^+e^-\to \eta\UnS{1}$ can only be excluded at
the $2.7\sigma$ level.

\subsubsection{Dipion transitions near \UnS{5}}
\label{sec:Dec_HT_Ups5Spipi}

As described in \Sec{sec:SpecExp_UnconVector},
Belle~\cite{Abe:2007tk} reported first observation
of apparent dipion transitions from the \UnS{5}
to the lower mass narrow \Ups\ states. The caveat
``apparent'' is applied because a later analysis,
also by Belle~\cite{Chen:2008pu}, and also described
in \Sec{sec:SpecExp_UnconVector}, suggests
that these transitions may originate not
from the enhancement in the hadronic cross section
from $e^+e^-$ collisions near $\sqrt{s} \approx 10.87\gev$ known
as ``\UnS{5}'', but from a {\it separate
new state} $Y_b$ close by in mass, much as $Y(4260)$
was observed via dipion transitions to lower-mass
charmonium.

 If interpreted as \UnS{5}\ transitions,
the measured dipion decay widths are more than two orders of
magnitude larger than those in the similar 
$\UnS{4}\to \UnS{1}\,\dipi$~\cite{Aubert:2008bv,:2009zy}
(see \Sec{sec:Dec_HT_Ups4Spipi}), and
$\UnS{2S,3}\to \UnS{1}\,\dipi$
(see \Sec{sec:Dec_HT_Ups2S3Spipi}) transitions.
The reason for such large differences is not clear.
Two theoretical ideas have been proposed. The first
approach~\cite{Simonov:2008ci} assumes that the bottomonium
transitions with two-pion emission come through intermediate virtual
$B\bar{B} \pi (\pi)$ formation, followed by the recombination 
of the $B$ pair into the $\Ups$. The difference
between the $\UnS{5}$ and $\UnS{4}$ decays results from a
large imaginary part in the $\UnS{5}$ decay amplitude appearing
due to the positive difference between the $\UnS{5}$ mass and the
sum of the masses of the $B\bar{B} \pi \pi$ system. A similar approach
was also used~\cite{Meng:2007tk,Meng:2008dd}, in which specific decay
parameters were predicted with good accuracy. An alternative
theoretical idea was proposed~\cite{Hou:2006it}, in which the large
$\UnS{5}\to \UnS{1}\,\dipi$ decay width was
explained as a possible indication of the production of a $b\bar{b}g$
hybrid state $Y_b$ with a mass close to the mass of the $\UnS{5}$
resonance.

\subsubsection{Observation of $\chi_{bJ}(2P)\to\pi\pi\chi_{bJ}(1P)$}

The decay $\chi_{bJ}(2P)\to\pi\pi\chi_{bJ}(1P)$ ($J=1,2$) was observed
by CLEO~\cite{Cawlfield:2005ra} using 6~million decays of the
$\UnS{3}$.  This is the only observed
hadronic transition from the $\chi_{bJ}(nP)$ states
aside from the surprisingly large 
$\chi_{bJ}(2P)\to\omega\UnS{1}$; as such, it provides an important
benchmark.  The following decay chain was used in the search:
$\UnS{3}\to\gamma\chi_{bJ^\prime}(2P)$;
$\chi_{bJ^\prime}(2P)\to\pi\pi\chi_{bJ}(1P)$;
$\chi_{bJ}(1P)\to\gamma\UnS{1}$; $\UnS{1}\to \ell^+\ell^-$.
Both charged and neutral pion pairs were sought, with
results combined assuming isospin invariance.  The data
were not adequate to distinguish between allowed
values of $J^\prime$ and $J$; hence it was assumed  that $J^\prime=J$.  The
$J=0$ transition was inaccessible due to the smallness of the
branching fractions in the decay chain.  Assuming the $J=2$ and
$J=1$ transitions have the same partial width, it was found that
\begin{eqnarray}
\Gamma(\chi_b(2P)\to\pi\pi\,\chi_b(1P) = 0.83 \pm 0.22 \pm 0.21\kev.~~~
\end{eqnarray}
This rate is consistent with theoretical
expectations~\cite{Kuang:1981se}.

\subsection{Hadronic decays}
\label{sec:Dec_haddec}

In general, the nonrelativistic quark model does a remarkable job also for
decays of the lowest (below $\bar DD$ threshold) charmonia -- see,
\eg \cite{Barnes:2005pb}.  However, there are some striking
discrepancies where additional experimental as well as theoretical
work is necessary, as will be outlined below -- see, \eg
\Sec{sec:Dec_RhoPiPuzzle}.  Improved data, \eg sensitive to
lineshapes due to improved resolution and statistics, as well as data on
additional transitions, should shed important light on the structure of
the light charmonia.

As described at the end of \Sec{sec:Dec_HadTranTheory}, for some
transitions heavy meson loops might play a significant role. The
natural question that arises asks what their influence on decays could be.
Here, unfortunately, the NREFT described above is not applicable anymore,
for the momenta of the final state particles as well as the intermediate
heavy meson velocities get too large for a controlled expansion.
However, those studies can be performed within phenomenological 
models~\cite{Li:2007xr,Liu:2009dr,Zhang:2009kr,Chen:2010re}. 

\subsubsection{The 12\% rule and $\rho\pi$ puzzle}
\label{sec:Dec_RhoPiPuzzle}

From perturbative QCD (pQCD), it is expected that both 
$\jpsi$ and $\psip$ decay into any exclusive
light-hadron final state with a 
width proportional to the square of the wave function 
at the origin~\cite{Appelquist:1974zd,De Rujula:1974nx}. 
This yields the pQCD ``12\% rule'',
\begin{eqnarray}
 Q_h &\equiv&\frac{{\cal B}({\psip\to h})}{
  {\cal B}({\jpsi\to h})} \nonumber\\
&=&\frac{{\cal B}({\psip\to e^+e^-})}{
   {\cal B}({\jpsi\to e^+e^-})} \approx 12\%\,.~~
\end{eqnarray}
A large violation of this rule was first observed in 
decays to $\rho\pi$ and $K^{*+}K^-+c.c.$ by 
Mark~II~\cite{Franklin:1983ve}, and became known as {\it the $\rho\pi$ puzzle}. 
Since then, many two-body decay modes of the $\psip$
 (and some multibody ones) 
have been measured by 
BES~\cite{Bai:2003vf,Bai:2003ng,Ablikim:2004kv,Ablikim:2004sf,Ablikim:2004ky,Ablikim:2006aw,Ablikim:2006aha} 
and CLEO~\cite{Adam:2004pr}; some decays obey the rule 
while others violate it to varying degrees.

The $\rho\pi$ mode is essential for this study -- the recent measurements, together with the old information, show us a new picture of the charmonium decay dynamics~\cite{Yuan:2005es}.
With a weighted average of
$\mathcal{B}(\jpsi\to\dipi\piz)=(2.00\pm 0.09)\%$ from the
existing measurements, and an estimation of
\beq
\frac{\mathcal{B}(\jpsi\to\rho\pi)}{
      \mathcal{B}(\jpsi\to\dipi\piz)}=1.17\times(1\pm 10\%)
\eeq
using the information given in \cite{Chen:1991sf}, one gets
$\mathcal{B}(\jpsi\to\rho\pi)=(2.34\pm 0.26)\%$. This is
substantially larger than the world average listed by
PDG08~\cite{Amsler:2008zzb}, which is $(1.69\pm 0.15)\%$, from a simple
average of many measurements.
The branching fraction of $\psip\to \dipi\piz$ is measured to
be $(18.1\pm 1.8\pm 1.9)\times 10^{-5}$ and $(18.8^{+1.6}_{-1.5}\pm
1.9)\times 10^{-5}$ at BESII~\cite{Ablikim:2005jy} and
CLEO~\cite{Adam:2004pr}, respectively. To extract the 
$\rho\pi$ component, however, the experiments make 
different choices, which in turn lead 
to different answers. BESII uses a partial wave
analysis (PWA), while CLEO counts the number of events by applying a
$\rho$ mass cut. The branching fraction from BESII is $(5.1\pm 0.7\pm
1.1)\times 10^{-5}$, while that from CLEO is $(2.4^{+0.8}_{-0.7}\pm
0.2)\times 10^{-5}$. If we take a weighted
average and inflate the resulting uncertainty
with a PDG-like scale factor accounting for the disagreement, we obtain
$\mathcal{B}(\psip\to \rho\pi)=(3.1\pm 1.2)\times 10^{-5}$.
With the results from above, one gets 
\beq 
Q_{\rho\pi} = \frac{\mathcal{B}(\psip\to \rho\pi)} 
{\mathcal{B}(\jpsi\to \rho\pi)} = (0.13\pm 0.05)\%\,.
\eeq
The suppression compared to the 12\% rule is about two orders of magnitude.

There are enough measurements of $\psip$ and $\jpsi$ decays for an
extensive study of the ``12\%
rule''~\cite{Bai:2003vf,Bai:2003ng,Ablikim:2004kv,Ablikim:2004sf,Ablikim:2004ky,Ablikim:2006aw,Ablikim:2006aha,Adam:2004pr,Amsler:2008zzb,Ablikim:2005ir,Briere:2005rc,Pedlar:2005px},
among which the Vector-Pseudoscalar (VP) modes, like the $\rho\pi$,
have been measured with the highest priority. The ratios of the
branching fractions are generally suppressed relative to the 12\% rule for the 
non-isospin-violating VP and
Vector-Tensor (VT) modes (\ie excluding modes like $\omega\piz$
and $\rho^0\eta$), while
Pseudoscalar-Pseudoscalar (PP) modes are enhanced. The multihadron
modes and the baryon-antibaryon modes cannot be simply
characterized, as some are enhanced, some are suppressed, 
and some match the expectation of the rule. Theoretical
models, developed for interpreting specific modes, have not yet
provided a solution for all of the measured channels. For a recent
review, see \cite{Brambilla:2004wf,Mo:2006cy}.

\subsubsection{Observation of $h_{c}(1P)\to 2(\dipi)\piz$}
\label{sec:Dec_hchaddec}

\begin{figure}[b]
  \begin{center}
    \includegraphics[width=\figwid]{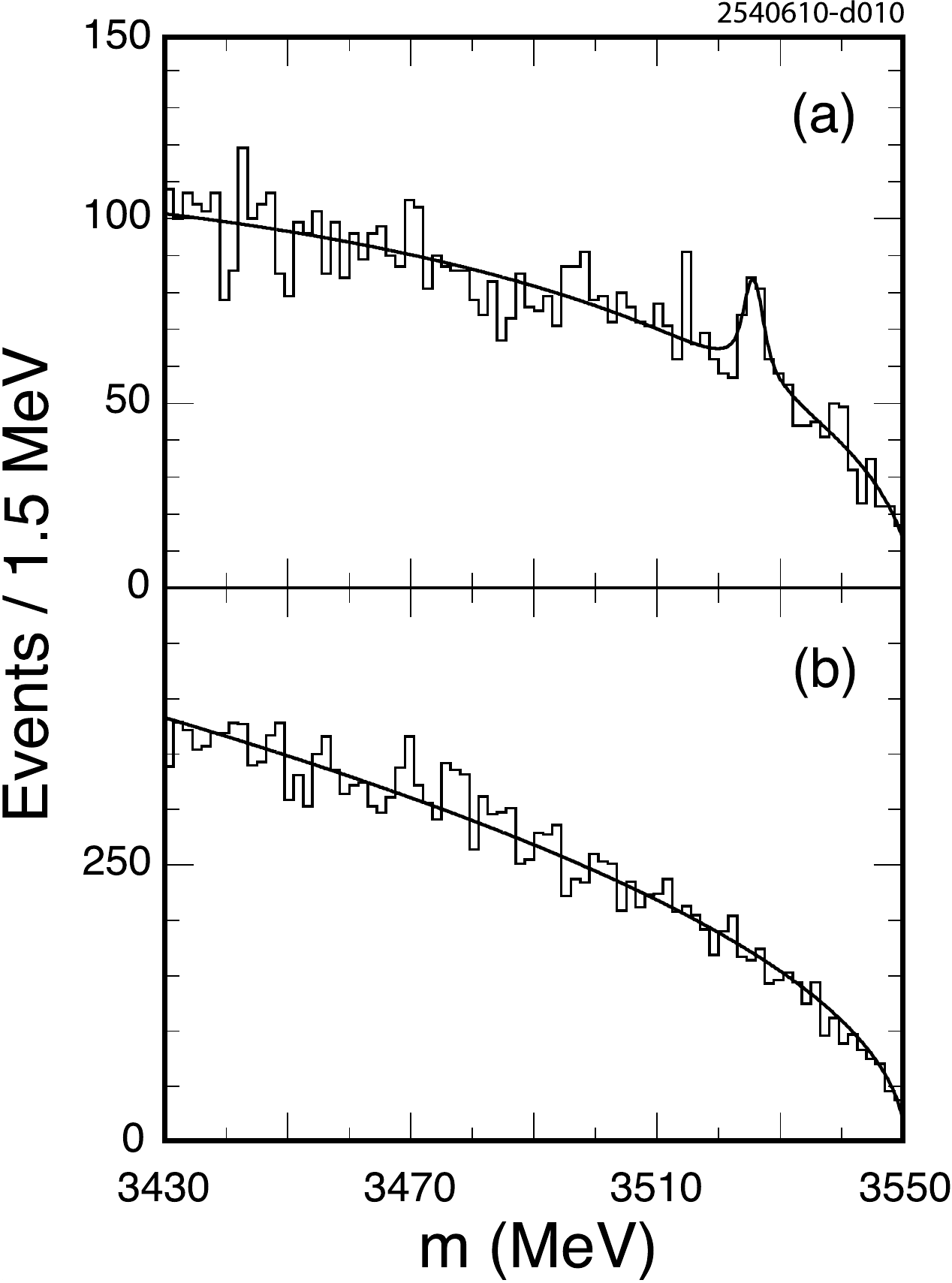}
      \caption{From CLEO~\cite{{:2009dp}}, the $2(\dipi)\piz$ mass 
               distribution showing 
               evidence for the $\psip\to\piz h_c(1P)$ followed by 
               $h_c(1P)\to 2(\dipi)\piz$.  
               {\it Solid line histograms} show (a)~data 
               and (b)~Monte Carlo simulation of backgrounds.
               {\it Solid curves} show the results of fits 
               to the respective histograms for a smooth background
               and $h_c(1P)$ signal.
               \AfigPermAPS{:2009dp}{2009} }
       \label{fig:Dec_Hc_Hadronic}
  \end{center}
\end{figure}

CLEO reported the first evidence for an exclusive
hadronic decay mode of the $h_c(1P)$~\cite{:2009dp}, previously only
seen through its radiative transition to the $\etac$.  Using
$25.7\times10^6$ $\psip$ decays, CLEO performed a search for
hadronic decays of the $h_c(1P)$ in the channels $\psip\to\piz
h_c(1P); h_c(1P)\to n(\dipi)\piz$, where $n=1,2,3$.  Upper
limits were set for the $3\pi$ and $7\pi$ decay modes, but evidence
for a signal was found in the $5\pi$ channel with a significance of
$4.4~\!\sigma$.  The $5\pi$ mass distribution from data is shown in
\Fig{fig:Dec_Hc_Hadronic}(a); the corresponding spectrum from
background Monte Carlo is shown in \Fig{fig:Dec_Hc_Hadronic}(b).
The measured mass is consistent with previous measurements.  The
measured product branching fraction is
\begin{eqnarray}
&\Brat(\psip\to\piz h_c(1P))\times\Brat(h_c(1P)\to2(\dipi)\piz)=&\nonumber\\
& (1.88^{+0.48+0.47}_{-0.45-0.30})\times10^{-5}\, .&~
\end{eqnarray}
This value is approximately 5\% of that given in
\Eq{eqn:Dec_hcgammaetac},
indicating the total hadronic decay width of the $h_c(1P)$ is likely
of the same order as its radiative transition width to the
$\etac$.

\subsubsection{$\chi_{cJ}(1P)$ hadronic decays}
\label{sec:Dec_chicjhaddec}

Precision in study of hadronic $\chi_{cJ}(1P)$
decays continues to improve with larger datasets,
and is beginning to approach that achieved for
\jpsi\ and \psip.
Decays of the $P$-wave states provide
information complementary to that from the
$S$-wave states, which probe short-range processes.  At CLEO and
BESII the $P$-wave states are accessed through the predominant E1
radiative transitions $\psip\to\gamma\chi_{cJ}(1P)$.  Recent
measurements have also come from Belle through the process
$\gamma\gamma\to\chi_{c0,2}(1P)$.

A large number of different hadronic decay modes of the
$\chi_{cJ}(1P)$ states have recently been measured, many 
for the first time.  These include two-meson decays at
BESII~\cite{Ablikim:2005ds}, CLEO~\cite{Asner:2008nw}, and
Belle~\cite{Nakazawa:2004gu,Chen:2006gy,Uehara:2008pf,:2009cka,Uehara:2010mq}; 
two-baryon decays at CLEO~\cite{Naik:2008dk}; 
three-meson decays at CLEO~\cite{Athar:2006gh} and BESII~\cite{Ablikim:2006vm}; 
four-meson decays at 
BESII~\cite{Ablikim:2004cg,Ablikim:2005kp,Ablikim:2004ks,Ablikim:2006vp},
CLEO~\cite{:2008vna}, and BELLE~\cite{:2007vb}; 
as well as others 
like $4\pi p\overline{p}$ and $p\overline{n}\piz$ at
BESII~\cite{Ablikim:2006nm,Ablikim:2006aha}.

More work is required on the theoretical front to understand the wide
variety of hadronic $\chi_{cJ}(1P)$ decays now experimentally
observed. There are indications that the pattern of decays may
require the introduction of a color octet mechanism that includes
contributions from the subprocess $c\overline{c}g\to
q\overline{q}$~\cite{Bodwin:1992ye}.  Hadronic decays of the
$\chi_{cJ}(1P)$ may also provide insight into the relative
contributions of singly and doubly-OZI violating
processes~\cite{Zhao:2007ze}.

\subsubsection{Non-\DDbar\ \psit\ hadronic decays} 
\label{sec:Decay_nonDDbar}

\begin{figure}[b]
  \begin{center}
    \includegraphics[width=\figwid]{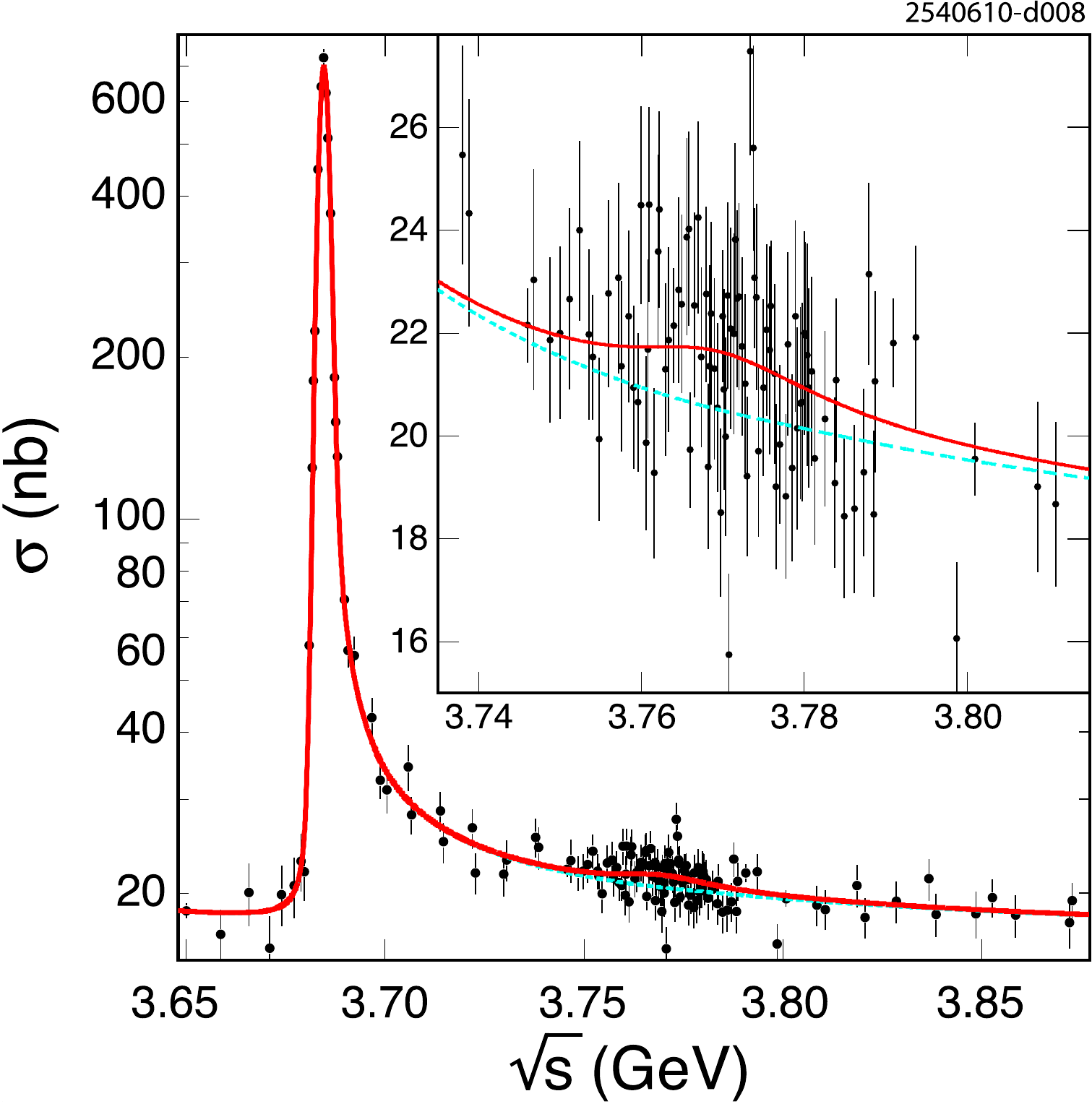}
      \caption{From BES~\cite{Ablikim:2008zzb}, 
               the hadronic cross section
               versus $\sqrt{s}$, extracted from counting 
               inclusively selected hadronic events with 
               a charged kaon of
               energy 1.15-2.00\gev,
               a range which excludes \DDbar\ events. 
               {\it Solid circles} represent data,
               the {\it dashed curve} represents the 
               contributions from \jpsi, \psip, and continuum
               hadron production, and
               the {\it solid curve} the best fit to the data
               of expected background plus a floating 
               $\psit\to\nonDDbar$ component (see text). 
               \AfigPermPLB{Ablikim:2008zzb}{2008} }
       \label{fig:Dec_Psi3770_Hadronic}
  \end{center}
\end{figure}

Experimental measurements of the branching fractions for
$\psit\to\nonDDbar$ provide important
information about the nature of the $\psit$ resonance.  
While the $\psit$ is thought to be dominantly the $1^3D_1$ state
of charmonium, its dilepton width indicates it should also contain
a substantial $n^3S_1$ component.  The size of this component is related to the
\nonDDbar-width of the \psit~\cite{Eichten:1978tg,Eichten:2004uh}.
Several measurements of the cross sections for $\psit\to\nonDDbar$
final states have recently been made,
but there is a disagreement between those from 
BES~\cite{Ablikim:2006zq,Ablikim:2006aj,Ablikim:2007zz,Ablikim:2008zzb} 
and that of CLEO~\cite{Besson:2005hm}. 

\begin{table}
\caption{Measurements of branching fractions for
         $\psit\to\DDbar$ and
         \nonDDbar\ at BES, CLEO, and as averaged
         by PDG10. Rates for
         \DzDz\ and \DpDm\ are constrained
         to sum to that of \DDbar, and
         the rates for \DDbar\ and \nonDDbar\ 
         are constrained to sum to unity.
         Hence within each set of measurements,
         uncertainties are highly correlated
}
\label{tab:Dec_BF_3770}
\setlength{\tabcolsep}{1.20pc}
\begin{center}
\begin{tabular}{lccc}
\hline\hline
\rule[10pt]{-1mm}{0mm}
 Final state &  $\cal B$~(\%) & Source \\[0.7mm]
\hline
\rule[10pt]{-1mm}{0mm}
\DzDz         & $49.9 \pm 1.3 \pm 3.8$ & BES~\cite{Ablikim:2006zq}\\[0.7mm]
\DpDm         & $35.7 \pm 1.1 \pm 3.4$ & BES~\cite{Ablikim:2006zq}\\[0.7mm]
\DDbar        & $85.5 \pm 1.7 \pm 5.8$ & BES~\cite{Ablikim:2006zq}\\[0.7mm]
\nonDDbar     & $14.5 \pm 1.7 \pm 5.8$ & BES~\cite{Ablikim:2006zq}\\[0.7mm]
\rule[15pt]{-1mm}{0mm}
\DzDz         & $46.7 \pm 4.7 \pm 2.3$ & BES~\cite{Ablikim:2006aj}\\[0.7mm]
\DpDm         & $36.9 \pm 3.7 \pm 2.8$ & BES~\cite{Ablikim:2006aj}\\[0.7mm]
\DDbar        & $83.6 \pm 7.3 \pm 4.2$ & BES~\cite{Ablikim:2006aj}\\[0.7mm]
\nonDDbar     & $16.4 \pm 7.3 \pm 4.2$ & BES~\cite{Ablikim:2006aj}\\[0.7mm]
\rule[15pt]{-1mm}{0mm}
\DDbar        & $86.6 \pm 5.0 \pm 3.6$ & BES~\cite{Ablikim:2008zzb}\\[0.7mm]
\nonDDbar     & $13.4 \pm 5.0 \pm 3.6$ & BES~\cite{Ablikim:2008zzb}\\[0.7mm]
\rule[15pt]{-1mm}{0mm}
\DDbar        & $103.3 \pm 1.4 ^{+4.8}_{-6.6}$ & CLEO~\cite{Besson:2005hm}\\[0.7mm]
\nonDDbar     & $-3.3 \pm 1.4 _{-4.8}^{+6.6}$  & CLEO~\cite{Besson:2005hm}\\[0.7mm]
\rule[15pt]{-1mm}{0mm}
\DDbar        & $93_{-9}^{+8}$        &PDG10~\cite{Nakamura:2010pdg}\\[0.7mm]
\nonDDbar     & $7^{+9}_{-8}$          &PDG10~\cite{Nakamura:2010pdg}\\[0.7mm]
\hline \hline
\end{tabular}
\end{center}
\end{table}

The most recent and precise result from BES~\cite{Ablikim:2008zzb} is
$\mathcal{B}(\psit\to\nonDDbar) = (15.1\pm5.6\pm1.8)\%$.
This was obtained by
counting inclusively selected hadronic events with 
a charged kaon of energy 1.15-2.00\gev, 
a range which excludes \DDbar\ events,
in the range of center-of-mass energies 3.650-3.872\gev,
and correcting the observed number for efficiency and integrated luminosity
at each energy point.
These cross sections were then fit for a 
smooth background and the expected lineshape
of \psit\ with floating normalization. 
\Figure{fig:Dec_Psi3770_Hadronic} shows these measured
cross sections versus the center-of-mass energy together
with the best fit, which indicates evidence for \nonDDbar\ decays.
Combining this result with another one, also by BES~\cite{Ablikim:2007zz},
which measured cross sections at just two points 
(3.773\gev\ and 3.650\gev) gives
a $4.8\sigma$ signal significance for 
$\psit\to\nonDDbar$ decays.  
Other measurements from BES -- which include variations 
on the above-described techniques -- are listed in \Tab{tab:Dec_BF_3770}.

CLEO~\cite{Besson:2005hm} measured the cross section 
for $\psit\to\nonDDbar$ by comparing the cross sections for 
$\psit\to\,{\rm hadrons}$ to that measured for 
$\psit\to\DDbar$~\cite{Dobbs:2007zt}.  The former
quantity is obtained by subtracting the hadronic cross section measured
on the ``continuum''
at $\sqrt{s}=3671\mev$, as extrapolated to $\sqrt{s}=3773\mev$ and
corrected for interference with \psip\ decays, from the
hadronic cross section measured near the peak of
the \psit, $\sqrt{s}=3773\mev$.
The net \nonDDbar\ cross section obtained by CLEO is smaller
than those of BES and is consistent with zero.  
These \nonDDbar\ cross sections are summarized 
in \Tab{tab:Dec_XS_3770},
and resulting branching fractions in \Tab{tab:Dec_BF_3770}.
As the CLEO and BES results are in conflict, the
PDG10 average in \Tab{tab:Dec_BF_3770} averages
between the two and inflates the combined uncertainty
so as to be consistent with both.

\begin{table}
\caption{Measurements of \nonDDbar\ cross sections for \psit\
         decays and the experimentally observed cross section for \psit\
         production at 3.773\gev
}
\label{tab:Dec_XS_3770}
\setlength{\tabcolsep}{0.64pc}
\begin{center}
\begin{tabular}{lcc} 
\hline\hline
\rule[10pt]{-1mm}{0mm}
Experiment  &  $\sigma^{\rm obs}_\nonDDbar$~(nb)
            &  $\sigma^{\rm obs}_{\psit}$~(nb)  \\ [1mm]
\hline
\rule[10pt]{-1mm}{0mm}
CLEO~\cite{Besson:2005hm}    &  $-0.01\pm 0.08^{+0.41}_{-0.30}$ & $6.38\pm 0.08^{+0.41}_{-0.30}$ \\ [0.7mm]
BESII~\cite{Ablikim:2006zq}  &  $1.14\pm 0.08\pm 0.59$ & $7.18\pm 0.20 \pm 0.63$ \\[0.7mm]
BESII~\cite{Ablikim:2006aj}  &  $1.04\pm 0.23\pm 0.13$ & $6.94\pm 0.48 \pm 0.28$ \\[0.7mm]
BESII~\cite{Ablikim:2008zzb} &  $0.95\pm 0.35\pm 0.29$ & $7.07\pm 0.36 \pm 0.45$ \\[0.7mm]
BESII~\cite{Ablikim:2007zz}  &  $1.08\pm 0.40\pm 0.15$ &               ---       \\[0.7mm]
MARKII~\cite{Abrams:1979cx}  & --                      & $9.1\pm 1.4$\\[0.7mm]
\hline \hline
\end{tabular}
\end{center}
\end{table}

To search for light hadron decays of \psit, both 
BES~\cite{Ablikim:2007ss,:2007wg,:2007ht,:2008vi,Ablikim:2008vk,Ablikim:2009zz,Ablikim:2010zz}
and CLEO~\cite{Huang:2005fx,CroninHennessy:2006su,Adams:2005ks}
extensively studied various exclusive light hadron decay modes
for $\psit \to {\rm LH}$ (LH$~\equiv~$light~hadron), but 
for only one channel was found to have a significant signal:
CLEO~\cite{Adams:2005ks} measured the branching fraction
$\mathcal{B}(\psit\to \phi\eta)=(3.1\pm 0.6\pm 0.3)\times10^{-4}$. 
This branching fraction is obtained by subtracting the
extrapolated continuum cross
section for $e^+e^- \to \phi\eta$ measured at $\sqrt{s}=3671\mev$
from that measured at 3.773\gev. CLEO's measurement 
explicitly ignored the possible interference among
amplitudes for this final state from \psit, continuum, and \psip.

Although CLEO did not claim observations for other light hadron decay
modes, some evidence for such decays can be found
in the CLEO cross sections.
\Tab{tab:Dec_3770_LH} lists some cross sections for
$e^+e^-\to {\rm LH}$ measured at 3.773 and 3.671\gev\ by
CLEO~\cite{Adams:2005ks}. Curiously, the final states
\dipi\piz, $\rho\pi$, and $\omega\eta$ have {\it smaller}
cross sections at $\sqrt{s}=3773\mev$ than at 3671\mev,
suggesting that an interference effect has come into play
to reduce the observed cross section. If this is the
cause, this would imply \psit\ branching fractions for
these modes of order $10^{-4}\text{--}10^{-3}$.
It is also noteworthy that $K^{*0}\bar {K^0}$ is
produced more copiously 
than $K^{*+}{K^-}$ (by a factor of at least 20) at {\it both}
energies. Whether these two phenomena are related remains 
an open question.

\begin{table}[b]
\caption{Measurements from CLEO~\cite{Adams:2005ks} of cross sections for 
         $e^+e^- \to \dipi\piz$  and 
         $e^+e^- \to {\rm VP}$ channels on the continuum
         just below the \psip\ at $\sqrt{s}=3.671$\gev\ and
         on the \psit\ resonance at $\sqrt{s}=3.773$\gev
}
\label{tab:Dec_3770_LH}
\setlength{\tabcolsep}{0.92pc}
\begin{center}
\begin{tabular}{lcc} 
\hline\hline
\rule[10pt]{-1mm}{0mm}
Final state        & $\sigma$~(pb)              &$\sigma$~(pb) \\
                   & $\sqrt{s}=3.671$\gev       & $\sqrt{s}=3.773$\gev\\[0.7mm]
\hline
\rule[10pt]{-1mm}{0mm}
$\dipi\piz$  &        $13.1^{+1.9}_{-1.7} \pm 2.1$   & $7.4 \pm 0.4 \pm 2.1$ \\[0.7mm]
$\rho\pi$          &        $ 8.0^{+1.7}_{-1.4} \pm 0.9$   & $4.4 \pm 0.3 \pm 0.5$ \\[0.7mm]
~~~$\rho^0\piz$   &        $ 3.1^{+1.0}_{-0.8} \pm 0.4$   & $1.3 \pm 0.2 \pm 0.2$ \\[0.7mm]
~~~$\rho^+\pi^-$   &        $ 4.8^{+1.5}_{-1.2} \pm 0.5$   & $3.2 \pm 0.3 \pm 0.2$ \\[0.7mm]
$\omega\eta$       &        $ 2.3^{+1.8}_{-1.0} \pm 0.5$   & $0.4 \pm 0.2 \pm 0.1$ \\[0.7mm]
$\phi\eta$         &        $ 2.1^{+1.9}_{-1.2} \pm 0.2$   & $4.5 \pm 0.5 \pm 0.5$ \\[0.7mm]
$K^{*0}\bar {K^0}$ &        $ 23.5^{+4.6}_{-3.9}\pm 3.1$   & $23.5\pm 1.1 \pm 3.1$ \\[0.7mm]
$K^{*+}{K^-}$      &        $  1.0^{+1.1}_{-0.7}\pm 0.5$   & $<0.6$ \\[0.7mm]
\hline\hline
\end{tabular}
\end{center}
\end{table}

BES~\cite{Ablikim:2008zzc} has observed a
\psit-lineshape anomaly in measurement 
of inclusive cross sections measured in the range
$\sqrt{s}=3.70-3.87\gev$ (see also \Sec{sec:SpecExp_ConVecCharm}). 
It was suggested by BES and by Dubynskiy and
Voloshin~\cite{Dubynskiy:2008vs}
that a second structure near 3765\mev\ could be responsible.
If such a structure exists, and it decays into
some of the low multiplicity LH states discussed
above, yet another amplitude comes into play
which could interfere and cause observed cross sections
to be smaller near the \psit\ than on the continuum.
A very recent preliminary analysis by KEDR~\cite{TodyshevICHEP:2010} 
of its \epem\ scan data near \psit\ 
finds inconsistency with this lineshape anomaly.

The conflict about the fraction of
\nonDDbar\ decays from \psit\ is 
not restricted to being between
the BESII and CLEO experiments;
an inclusive-exclusive rift also
remains to be bridged.
Inclusive-hadronic measurements
alone have difficulty supporting
a $\psit\to\nonDDbar$ branching fraction
of more than several percent without some
confirmation in exclusive mode measurements,
which currently show a large number of
modes, including the leading candidate
low-multiplicity transitions and decays, to be quite small.
The large datasets expected at BESIII
offer an opportunity for such a 
multifaceted approach to $\psit\to\nonDDbar$.

\subsubsection{Observation of $\UnS{1}\to\,$antideuteron\,$+X$}
\label{sec:SpecExp_CLEOdeuteron}

The appearance of deuterons in fragmentation has been
addressed theoretically in the framework of a coalescence
model~\cite{Gutbrod:1988gt,Sato:1981ez}, via the binding
of a nearby neutron and proton. Experimental constraints
on the process from measured deuteron production 
are limited. ARGUS~\cite{Albrecht:1989ag} found evidence
in \UnS{1S,2}\ decays for such production but with very
low statistics. These results have been accommodated
in a string-model calculation~\cite{Gustafson:1993mm}.
Further experimental information is essential.

CLEO~\cite{Asner:2006pw} addressed this issue with
a measurement of antideuteron production in samples
of 22, 3.7, and 0.45~million \UnS{1S,2S,4}\ decays, respectively.
Only antideuterons were sought due to the presence of a large 
deuteron background. This background arises from
nuclear interactions with matter (such as gas,
vacuum chambers, and beam collimators) initiated by either particles
created in the \epem\ annihilation or errant $e^\pm$ from the 
colliding beams. These interactions result in
the appearance of 
neutrons, protons, and deuterons
in the detector. Antideuterons are identified primarily by 
their distinctive energy loss ($dE/dx$) as a function of momentum,
as measured in tracking chambers, but residual pion and proton
backgrounds are additionally suppressed by requiring deuteron-appropriate
response in the RICH detectors. This identification is
relatively background-free over the momentum range of 0.45-1.45\gevc,
and is found to correspond to a branching fraction
\beq
\Brat(\UnS{1}\to\antid X)=(2.86\pm0.19\pm0.21)\times10^{-5}\,.
\eeq

How often
is the baryon-number conservation for the \antid\  
accomplished with a deuteron? Deuteron background becomes
tolerable in \antid-tagged events, and three \deut\antid\ 
candidate events are found in the CLEO \UnS{1}\ data sample, 
meaning that baryon-number compensation
occurs with a deuteron about 1\% of the time. By counting 
events with zero, one, or two protons accompanying an identified
\antid, CLEO finds that compensation by each of
$pn$, $np$, $nn$, or $pp$ occurs at roughly the same rate, about 
a quarter of the time.

  CLEO also measures antideuteron fractions in 
the data samples of \UnS{2}\ and \UnS{4}.
The \UnS{2}\ result can be used to calculate the
rate of $\chi_{bJ}\to\antid X$ by subtracting a
scaled \UnS{1}\ \antid-fraction 
to account for $\UnS{2}\to X\UnS{1}$ transitions,
and again, with a different scaling factor,
to account for $\UnS{2}\to ggg,~\gamma gg$ decays,
assuming they have the same
\antid-fraction as \UnS{1}; the balance
are attributed to appearance through
$\UnS{2}\to\gamma\chi_{bJ}$, $\chi_{bJ}\to\antid X$. 
No significant excess is observed,
either here nor in \UnS{4}\ decays,
so CLEO reports the upper limits 
$\Brat(\chi_{bJ}\to\antid X) < 1.1\times 10^{-4}$,
averaged over $J=0,\,1,\,2$, and
$\Brat(\UnS{4}\to\antid X)<1.3\times 10^{-5}$,
both at 90\%~CL.

Artoisenet and Braaten~\cite{Artoisenet:2010uu} find this
CLEO result useful in tuning parameters of event generators
in a study of production of loosely bound hadronic molecules
(see \Sec{sec:SpecTh_molec} for a discussion of the relevance
of this measurement to the nature of the $X(3872)$).
Brodsky~\cite{Brodsky:2009bg} comments on how measurements
such as these should be extended in order to probe the
hidden-color structure of the antideuteron wave function.

\subsubsection{Observation of $\Ups,\,\chi_{bJ}\to$~open charm}

Very little is known about the heavy-flavor (\ie open charm) content
of bottomonium hadronic decays, which can be used as a tool to 
probe of the post-$b\bar{b}$-annihilation fragmentation processes.
\UnS{n}\ hadronic decays are dominated by
those materializing through three gluons ($ggg$), $\chi_{b0,2}$ through $gg$,
and $\chi_{b1}$ through $q\bar{q}g$~\cite{Barbieri:1975am};
each of these processes is expected to have its own characteristic open-charm
content.

CLEO~\cite{Briere:2008cv} first selected events 
from its \UnS{2S,3}\ data samples to have
a \Dze\ (or $\bar{\Dze}$) meson exclusively reconstructed,
and with momentum $>2.5$\gevc; the momentum cut is
required to suppress backgrounds. The single-photon energy spectra
for these events were then fit for
the presence of narrow peaks corresponding to the radiative transitions
$\UnS{m}\to\gamma\chi_{bJ}(nP)$.
Significant signals for \Dze\ production only from $\chi_{b1}(1P,2P)$
are observed with branching fractions
({\it not} correcting for the \Dze\ momentum cut) of about 10\%, 
with upper limits at 90\%~CL for the others ranging from 2-10\%.
CLEO then combines these measured fractions with some assumptions and theoretical
input to simultaneously extract $\rho_8$,
an NRQCD non-perturbative parameter, and branching
fractions for \Dze\ mesons produced with {\it any} momenta;
the assumptions and $\rho_8$ directly affect the spectrum
of \Dze\ mesons.
With these assumptions, CLEO obtains
$\rho_8\approx0.09$ and 
\beq
\Brat(\chi_{b1}(1P,2P)\to\Dze X)\approx 25\%\, ,
\eeq
both in agreement with 
predictions~\cite{Barbieri:1979gg,Bodwin:2007zf}.
Upper limits at 90\%~CL are set for the remaining four $\chi_{bJ}$ that are
also consistent with the predictions, with the exception of
$\chi_{b2}(2P)$, which has an upper limit that is half the predicted value of 12\%.
The largest branching fractions occur for $\chi_{b1}(1P,2P)$,
as expected, which are the states expected to decay via $q\bar{q}g$.

\babar~\cite{:2009wm} searched its \UnS{1}\ data sample
for the presence of \Dstp\ (and \Dstm) mesons.
\babar\ measures 
\beq
\Brat(\UnS{1}\to\Dstp X)=(2.52\pm0.13\pm0.15)\%\, ,
\eeq
the first such observation of charm in \UnS{1} decays,
and extracts the resulting \Dstp\ momentum spectrum. \babar\
also finds this rate to be considerably in excess (\ie roughly double) 
of that expected from $b\bar{b}$ annihilation into a single photon.
The excess is seen to be in agreement with a prediction~\cite{Kang:2007uv}
based on splitting a virtual gluon, but appears to be 
too small to accommodate an octet-state contribution~\cite{Zhang:2008pr}.

\subsubsection{Observation of $\chi_{bJ}(1P,2P)\to$~hadrons}

\begin{figure}[b]
  \begin{center}
    \includegraphics[width=\figwid]{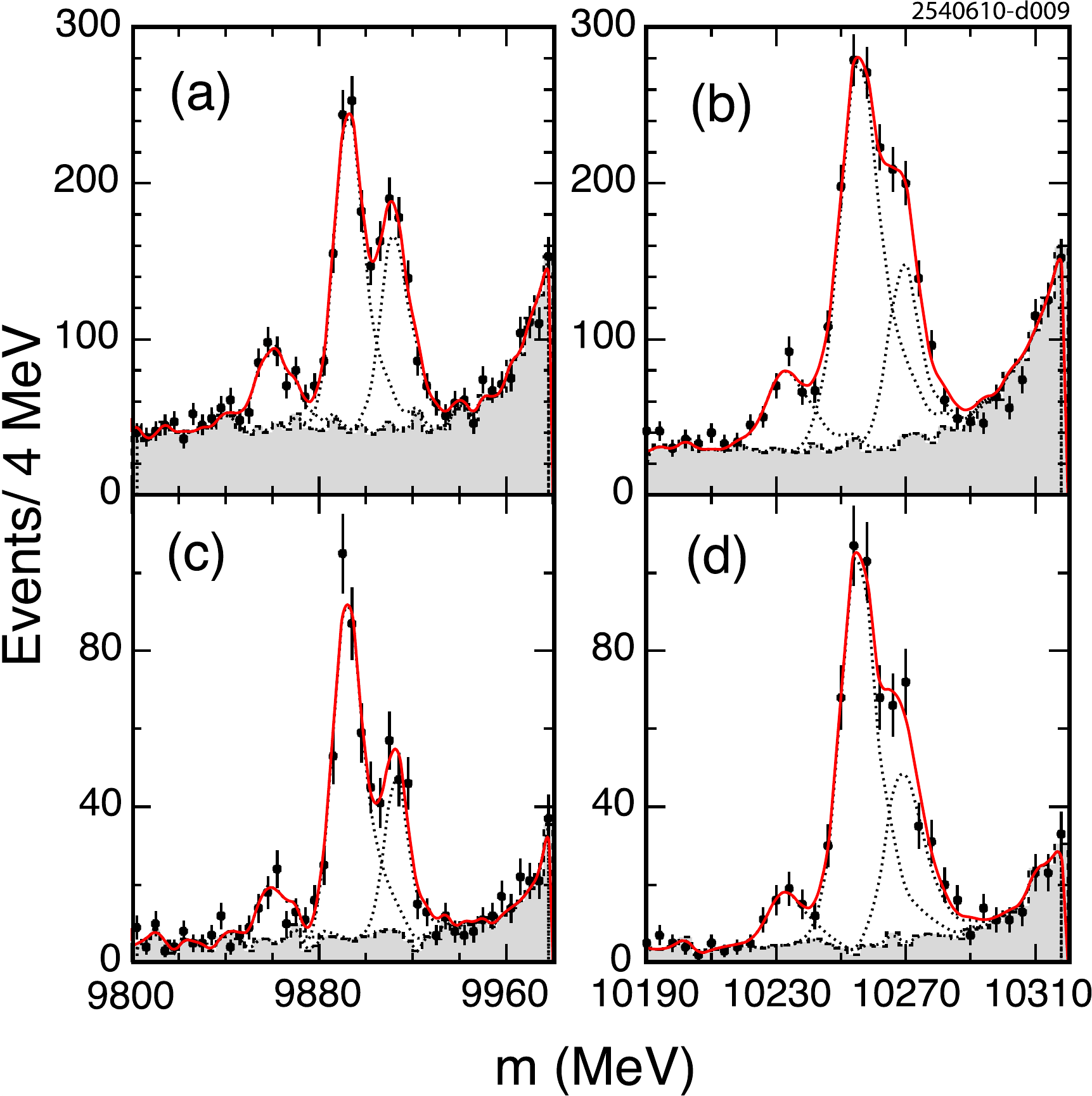}
      \caption{From CLEO~\cite{:2008sx}, 
               the invariant mass of the sum of exclusive 
               decay modes of the $\chi_{bJ}(1P)$ from 
               (a),~(c)~$\UnS{2}$ and (b),~(d)~$\UnS{3}$
               radiative decays. (a) and (b) show the sum 
               of 659 exclusive decay modes, while (c) and (d) 
               show the sum of the 14  
               decay modes with significant branching fractions,
               the values of which were measured. {\it Solid circles}
               represent data, {\it shaded histograms} the 
               backgrounds determined from \UnS{1}\ data,
               {\it dotted curves} the contributions
               of the individual $\chi_{bJ}$ signals as determined
               from fits to the data, and {\it solid curves}
               the sum of background and fitted signals. 
	       \AfigPermAPS{:2008sx}{2008} }
      \label{fig:Dec_Chib_Hadronic}
  \end{center}
\end{figure}

Using its full sample of 9~million $\UnS{2}$ and 6~million
$\UnS{3}$ decays collected with the CLEO~III detector, CLEO 
made the first measurements of branching fractions of
exclusive decays of the $\chi_{bJ}(1P,2P)$ states~\cite{:2008sx},
which were accessed through the allowed E1
radiative transitions $\UnS{3} \to \gamma\chi_{bJ}(2P)$ and
$\UnS{2} \to \gamma\chi_{bJ}(1P)$.  A comprehensive set of
659~decay exclusive modes was included in the search, where each decay mode was
a combination of $\pi^{\pm}$, $\piz$, $\eta$, $K^{\pm}$, $K_S$, and
$p^{\pm}$.  Of these 659, 14 were found that have significances of
greater than $5\sigma$ for both $\chi_{bJ}(1P)$ and $\chi_{bJ}(2P)$
decays (combined for $J = 0, 1, 2$).
\Figure{fig:Dec_Chib_Hadronic} shows the invariant mass of the
exclusive decay modes for all 659 and the selected 14 decay modes of
the $\chi_{bJ}(1P)$ and $\chi_{bJ}(2P)$.  Branching fractions were
measured for the 14 decay modes from both the $\chi_{bJ}(1P)$ and
$\chi_{bJ}(2P)$ for $J = 0, 1, 2$.  The branching fractions ranged
from approximately $(1\text{--}20)~\times10^{-4}$.  The largest branching
fractions measured were to $6\pi2\piz$ and $8\pi2\piz$.  This survey
of branching fractions, besides being useful for testing models of
bottomonium decays, also gives some indication of which exclusive
channels might be most fruitful in searches for new states in the
bottomonium region.

\section[Production]{Production$^9$}

\addtocounter{footnote}{1}
\footnotetext{
P.~Artoisenet, 
A.~Bertolin, 
G.~T.~Bodwin$^\dag$, 
C.-H.~Chang, 
K.-T.~Chao, 
J.-P.~Lansberg, 
F.~Maltoni,
A.~Meyer$^\dag$, 
V.~Papadimitriou$^\dag$, and
J.-W.~Qiu
}
\label{sec:ProdChapter}

\subsection{Introduction to theoretical concepts}
\label{prod_subsec:fac}

In this subsection, we provide an introduction to some of the
theoretical ideas that will appear in subsequent subsections, many of
which are based on various factorization formulas for quarkonium
production and decay. This subsection also serves to establish notation
and nomenclature that is used in subsequent subsections.

\subsubsection{Momentum scales and factorization}

A heavy quarkonium has at least three intrinsic momentum scales: the
heavy-quark mass $m_Q$; the momentum of the heavy quark or antiquark in
the quarkonium rest frame, which is of order $m_Qv$; and the binding
energy of the heavy quark-antiquark ($Q\bar Q$) pair, which is of order
$m_Qv^2$. Here $v$ is the typical velocity of the heavy quark or
antiquark in the quarkonium rest frame. ($v^2\approx 0.3$ for the $J/\psi$
and $v^2\approx 0.1$ for the $\Upsilon$.)

If a heavy quarkonium is produced in a hard-scattering process, then, in
addition to the intrinsic scales of the quarkonium, the hard-scattering
scale $p$ enters into the description of the production process. The
hard-scattering scale $p$ is usually set by a large momentum transfer in
the production process. In quarkonium production in hadron-hadron
collisions (hadroproduction) and hadron-lepton collisions, $p$ is
usually taken to be of order $p_T$, the transverse momentum of the
quarkonium, while in quarkonium production in $e^+e^-$ collisions, $p$
is usually taken to be of order $p^*$, the quarkonium momentum in the
$e^+e^-$ center-of-mass frame.

One might expect intuitively that the production process could be
understood in terms of two distinct steps: the production of the $Q\bar
Q$ pair, which would occur at the scale $p$, and the subsequent
evolution of the $Q\bar Q$ pair into the quarkonium, which would involve
the smaller dynamical scales $m_Qv$ and $m_Qv^2$. The first step would
be calculable in an expansion in powers of $\alpha_s(p)$, while the
second step would typically involve inherently nonperturbative physics.
The term ``short distance'' is often used to refer to the momentum scale
$p$ (distance scale $1/p$), while the term ``long distance'' is often
used to refer to typical hadronic momentum scales, such as $m_Qv$,
$m_Qv^2$, or $\Lambda_{\rm QCD}$. The term ``short distance'' is also
sometimes used to refer to the scale $m_Q$ in the context
of NRQCD.

In order to establish that this intuitive picture of quarkonium
production is actually a property of QCD, one must demonstrate that the
short-distance, perturbative effects at the scale $p$ can be separated
from the long-distance, nonperturbative dynamics. Such a separation is
known as ``factorization.''  In proving a ``factorization theorem,'' one 
must show that an amplitude or cross section can be expressed as a sum of
products of infrared-safe, short-distance coefficients with well defined
operator matrix elements. Such short-distance coefficients are
perturbatively calculable. The operator matrix elements would contain
all of the long-distance, nonperturbative physics. They might be
determined phenomenologically or, possibly, through lattice simulations.
If it can be further demonstrated that the long-distance matrix
elements are universal, {\it i.e.}, process independent, then
factorization formulas yield much greater predictive power.

The nonperturbative evolution of the $Q\bar Q$ pair into a quarkonium
has been discussed extensively in terms of models and in terms of the
language of effective theories of QCD
\cite{Bodwin:1994jh,Brambilla:2004jw,Brambilla:2004wf}. Different
treatments of this evolution have led to various theoretical models for
inclusive quarkonium production. Most notable among these are the color-singlet
model (CSM), the color-evaporation model (CEM), the nonrelativistic QCD
(NRQCD) factorization approach, and the fragmentation-function approach.

\subsubsection{The color-singlet model}
\label{prod_sec:CSM}

The CSM was first proposed shortly after the discovery of the $J/\psi$
\cite{Einhorn:1975ua,Ellis:1976fj,Carlson:1976cd,Chang:1979nn,Berger:1980ni,Baier:1981uk,Baier:1981zz,Baier:1983va}.
In this model, it is assumed that the $Q\bar Q$ pair that evolves into
the quarkonium is in a color-singlet state and that it has the same spin
and angular-momentum quantum numbers as the quarkonium. In the CSM, the
production rate for each quarkonium state is related to the absolute
values of the color-singlet $Q\bar Q$ wave function and its derivatives,
evaluated at zero $Q\bar Q$ separation. These quantities can be
extracted by comparing theoretical expressions for quarkonium decay
rates in the CSM with experimental measurements. Once this extraction
has been carried out, the CSM has no free parameters. The CSM was
successful in predicting quarkonium production rates at relatively low
energy \cite{Schuler:1994hy}. Recently, it has been found that, at high
energies, very large corrections to the CSM appear at next-to-leading
order (NLO) and next-to-next-to-leading order (NNLO) in $\alpha_s$
\cite{Artoisenet:2007xi,Campbell:2007ws,Artoisenet:2008fc}. (See
\Sec{prod_section:higherorderchannels}). Consequently, the possibility
that the CSM might embody an important production mechanism at high
energies has re-emerged. However, given the very large corrections at
NLO and NNLO, it is not clear that the perturbative expansion in
$\alpha_s$ is convergent. Furthermore, in the production and decay of
$P$-wave and higher-orbital-angular-momentum quarkonium states, the CSM
is known to be inconsistent because it leads to uncanceled infrared
divergences. (See Ref.~\cite{Brambilla:2004wf} and references therein.)
As we will describe below, the NRQCD factorization approach encompasses
the color-singlet model, but goes beyond it.

\subsubsection{The color-evaporation model}  
\label{prod_sec:CEM}

The CEM
\cite{Fritzsch:1977ay,Halzen:1977rs,Gluck:1977zm,Barger:1979js,Amundson:1995em,Amundson:1996qr}
is motivated by the principle of quark-hadron duality. In the CEM, it
is assumed that every produced $Q\bar Q$ pair evolves into a quarkonium
if it has an invariant mass that is less than the threshold for
producing a pair of open-flavor heavy mesons. It is further assumed that
the nonperturbative probability for the $Q\bar Q$ pair to evolve into a
quarkonium state $H$ is given by a constant $F_H$ that is
energy-momentum and process independent. Once $F_H$ has been fixed by
comparison with the measured total cross section for the production of
the quarkonium $H$, the CEM can predict, with no additional free
parameters, the momentum distribution of the quarkonium production rate. The
CEM predictions provide rough descriptions of the CDF data for $J/\psi$,
$\psi(2S)$, and $\chi_{cJ}$ production at $\sqrt{s}=1.8$~TeV
\cite{Amundson:1996qr}. In Ref.~\cite{Bodwin:2005hm}, the CEM
predictions are fit to the CDF data for $J/\psi$, $\psi(2S)$, and
$\chi_{cJ}$ production at $\sqrt{s}=1.8$~TeV \cite{Abe:1997yz}. The
quality of these fits is generally poor, with $\chi^2/\hbox{d.o.f.}$ for
the $J/\psi$ fits of about $7$--$8$ without initial-state $k_T$ smearing and
$2$--$4.5$ with initial-state $k_T$ smearing. In contrast, the NRQCD
factorization approach, which we are about to describe, yields fits to
the CDF $J/\psi$ data with $\chi^2/\hbox{d.o.f.}$ of about $1$.

\subsubsection{The NRQCD factorization approach}
\label{prod_sec:NRQCD-fact}%

The NRQCD factorization approach \cite{Bodwin:1994jh} to
heavy-quarkonium production is by far the most sound theoretically
and most successful phenomenologically. NRQCD is an effective theory of
QCD and reproduces full QCD dynamics accurately at momentum scales of
order $m_Q v$ and smaller. Dynamics involving momentum scales of order
$m_Q$ or larger are taken into account through the short-distance
coefficients of the operators that appear in the NRQCD action. The NRQCD
factorization approach expresses the probability for a $Q\bar Q$ pair to
evolve into a quarkonium in terms of matrix elements of NRQCD operators.
These matrix elements can be characterized in terms of their scaling
with the heavy-quark velocity $v$ \cite{Bodwin:1994jh}. In the NRQCD
factorization approach, the inclusive cross section for the direct
production of a quarkonium state $H$ is written as a sum of products of
these NRQCD matrix elements with the corresponding $Q\bar Q$
production cross sections:
\begin{equation}
\sigma(H)=\sum_n \sigma_n(\Lambda) \langle {\cal O}_n^H(\Lambda)\rangle 
\,.
\label{prod_eq:nrqcd_fac}
\end{equation}
Here $\Lambda$ is the ultraviolet cutoff of the effective theory, the
$\sigma_n$ are expansions in powers of $v$ of the cross sections to
produce a $Q\bar Q$ pair in the color, spin, and orbital-angular
momentum state $n$. The $\sigma_n$ are convolutions of parton-level
cross sections at the scale $p$ with parton distribution functions. (The
former are short-distance quantities, while the latter are long-distance
quantities that depend on the nonperturbative dynamics of the initial
hadrons.) The matrix elements $\langle {\cal O}_n^H(\Lambda)\rangle$ are
vacuum-expectation values of four-fermion operators in NRQCD.  We
emphasize that Eq.~(\ref{prod_eq:nrqcd_fac}) represents  both processes in
which the $Q\bar Q$ pair is produced in a color-singlet state and
processes in which the $Q\bar Q$ pair is produced in a color-octet
state. It is conjectured that the NRQCD factorization expression in
Eq.~(\ref{prod_eq:nrqcd_fac}) holds when the momentum transfer $p$ in the
hard-scattering production process is of order $m_Q$ or larger.

Unlike the CSM and the CEM expressions for the production cross section,
the NRQCD factorization formula for heavy-quarkonium production depends
on an infinite number of unknown matrix elements.  However, the sum in
Eq.~(\ref{prod_eq:nrqcd_fac}) can be organized as an expansion in powers of
$v$. Hence, the NRQCD factorization formula is a double expansion in
powers of $v$ and powers of $\alpha_s$. In phenomenological
applications, the sum in Eq.~(\ref{prod_eq:nrqcd_fac}) is truncated at a
fixed order in $v$, and only a few matrix elements typically enter into
the phenomenology. The predictive power of the NRQCD factorization
approach is based on the validity of such a truncation, as well as on
perturbative calculability of the $Q\bar Q$ cross sections and the
universality of the long-distance matrix elements.

If one retains in Eq.~(\ref{prod_eq:nrqcd_fac}) only the color-singlet
contributions of leading order in $v$ for each quarkonium state, then
one obtains the CSM. As we have mentioned, such a
truncation leads to inconsistencies because the omission of color-octet
contributions results in uncanceled infrared divergences in the production 
rates of $P$-wave and higher-orbital-angular-momentum quarkonium states.

The CEM implies that certain relationships must hold between the NRQCD
long-distance matrix elements \cite{Bodwin:2005hm}. These 
relationships are generally inconsistent with the scaling of the matrix 
elements with $v$ that is predicted by NRQCD. The shortcomings of 
the CEM in describing the Fermilab Tevatron data can be traced, at least 
in part, to these inconsistencies \cite{Bodwin:2005hm}.

As we will explain in more detail below, in the case of inclusive
quarkonium production, a compelling proof of NRQCD factorization is
still lacking. A further difficulty with the NRQCD factorization formula
in Eq.~(\ref{prod_eq:nrqcd_fac}) is that a straightforward perturbative
expansion of the short-distance coefficients may not yield an optimal
organization of the expression for the cross section. The difficulty
with such a straightforward expansion is that it ignores the fact that
different orders in $\alpha_s$ in the perturbative expansion may have
different dependences on $m_Q/p$. Consequently, at large $p/m_Q$, higher
orders in the perturbation expansion may be more important than lower
orders. Therefore, it may be useful to organize the production cross
section in powers of $p/m_Q$ before expanding the short-distance
coefficients in powers of $\alpha_s$ \cite{Nayak:2005rw,Nayak:2005rt,kang-qiu-sterman}.

Although the application of NRQCD factorization to heavy-quarkonium
production processes has had many successes, there remain a number of
discrepancies between its predictions and experimental measurements. The
most important of these successes and discrepancies are discussed in the 
remainder of \Sec{sec:ProdChapter}.

\subsubsection{The fragmentation-function approach}
\label{prod_sec:fragmentation}%

In the fragmentation-function approach to factorization for inclusive
quarkonium production \cite{Nayak:2005rw,Nayak:2005rt,kang-qiu-sterman},
one writes the production cross section in terms of convolutions of
parton production cross sections with light-cone fragmentation functions.
This procedure provides a convenient way to organize the contributions
to the cross section in terms of powers of $m_Q/p$. As we will explain
below, it might also represent the first step in proving NRQCD
factorization \cite{Nayak:2005rw,Nayak:2005rt,kang-qiu-sterman}. In the
second step, one would establish that the light-cone fragmentation
functions could be expanded in terms of NRQCD matrix elements.

We now describe the fragmentation-function approach for the specific
case of single inclusive heavy-quarkonium production at transverse
momentum $p_T \gg m_Q$. The contribution to the cross section at the
leading power in $m_Q/p_T$ is given by the production of a single parton
({\it e.g.}, a gluon), at a distance scale of order $1/p_T$, which
subsequently fragments into a heavy quarkonium
\cite{Braaten:1996pv}.  The contribution to the cross section at the
first subleading power in $m_Q/p_T$ is given by the production of a
$Q\bar Q$ pair in a vector- or axial-vector state, at a distance scale of
order $1/p_T$, which then fragments into a heavy quarkonium
\cite{kang-qiu-sterman}.  It was shown in the perturbative-QCD
factorization approach \cite{Nayak:2005rw,kang-qiu-sterman} that the
production cross section can be factorized as
\begin{eqnarray}
&& d\sigma_{A+B\to H+X}(p_T) =
\nonumber\\
&& \hskip 0.2in
 \sum_i d\hat{\sigma}_{A+B\to i+X}(p_T/z,\mu)
\otimes D_{i\to H}(z,m_Q,\mu)
\nonumber\\
&& 
~+ \sum_{[Q\bar{Q}(\kappa)]}
d\hat{\sigma}_{A+B\to [Q\bar{Q}(\kappa)]+X}(P_{[Q\bar{Q}(\kappa)]}=p_T/z,\mu)
\nonumber\\
&& \hskip 0.8in
  \otimes\, D_{[Q\bar{Q}(\kappa)]\to H}(z,m_Q,\mu) 
\nonumber\\
&& \hskip 0.3in
+~ {\cal O}(m_Q^4/p_T^4)\,,
\label{prod_eq:pqcd_fac}
\end{eqnarray}
where the first term in Eq.~(\ref{prod_eq:pqcd_fac}) gives the contribution
of leading power in $m_Q/p$, and the second term gives the first
contribution of subleading power in $m_Q/p$. $A$ and $B$ are the initial
particles in the hard-scattering process and  $\otimes$ represents a
convolution in the momentum fraction $z$. In the first term in
Eq.~(\ref{prod_eq:pqcd_fac}), the cross section for the inclusive
production of a single particle $i$, $d\hat{\sigma}_{A+B\to i+X}$,
contains all of the information about the incoming state and
includes convolutions with parton distributions in the cases in which
A or B is a hadron. The cross section $d\hat{\sigma}_{A+B\to i+X}$ is
evaluated at the factorization scale $\mu\sim p_T$. The quantity
$D_{i\to H}$ is the fragmentation function for an off-shell parton of
flavor $i$ to fragment into a quarkonium state $H$
\cite{Collins:1981uw}. The argument $m_Q$ indicates explicitly the
dependence of $D_{i\to H}$ on the heavy-quark mass. Similarly, in the
second term in Eq.~(\ref{prod_eq:pqcd_fac}), $d\hat{\sigma}_{A+B\to
[Q\bar{Q}(\kappa)]+X}$ is the inclusive cross section to produce an
on-shell $Q\bar Q$ pair with spin and color quantum numbers $\kappa$.
The cross section $d\hat{\sigma}_{A+B\to [Q\bar{Q}(\kappa)]+X}$ is also
evaluated at the factorization scale $\mu\sim p_T$, but it is suppressed
by a factor $m_Q^2/p_T^2$ relative to $d\hat{\sigma}_{A+B\to i+X}$. The
quantity $D_{[Q\bar{Q}(\kappa)]\to H}$ is the fragmentation function for
an off-shell $Q\bar Q$ pair with quantum numbers $\kappa$ to fragment
into a quarkonium state $H$ \cite{kang-qiu-sterman}. The predictive
power of the factorization formula in Eq.~(\ref{prod_eq:pqcd_fac}) relies on
the perturbative calculability of the single-particle inclusive and
$Q\bar Q$ inclusive cross sections and the universality of the
fragmentation functions.

The dependences of the single-parton and the $Q\bar Q$-pair
fragmentation functions on the factorization scale $\mu$ are given by
their respective evolution equations. These evolution equations can be
used to express the fragmentation functions at the scale $\mu\sim p$ in
terms of the fragmentation functions at the scale $\mu_0\sim 2m_Q$,
thereby resumming the logarithms of $\mu/m_Q$ that are contained in the
fragmentation functions.

\subsubsection{Relationship of the fragmentation-function approach to 
the NRQCD factorization approach}

If the NRQCD factorization formula in Eq.~(\ref{prod_eq:nrqcd_fac}) is valid
for the leading and first subleading power of $m_Q^2/p_T^2$, then it
implies that the fragmentation functions in Eq.~(\ref{prod_eq:pqcd_fac}) can
be expanded in terms of NRQCD matrix elements
\cite{Nayak:2005rw,Nayak:2005rt,kang-qiu-sterman}:
\begin{eqnarray}
D_{i\to H}(z,m_Q,\mu_0)
= \sum_n d_{i\to n}(z,m_Q,\mu_0) \langle {\cal O}_n^H\rangle 
\,\hskip0.15in  \nonumber\\
D_{[Q\bar{Q}(\kappa)]\to H}(z,m_Q,\mu_0)
=\hskip1.4in\nonumber\\
 \sum_n  d_{[Q\bar{Q}(\kappa)]\to n}(z,m_Q,\mu_0) 
    \langle {\cal O}_n^H\rangle\,.\hskip 0.4in 
\label{prod_eq:frag_fac}
\end{eqnarray}
Here, the short-distance coefficients $d_{i\to n}(z,m_Q,\mu_0)$ and
$d_{[Q\bar{Q}(\kappa)]\to n}(z,m_Q,\mu_0)$ describe, respectively, the
perturbative evolution at the scale $\mu_0$ of an off-shell parton of
flavor $i$ and a $Q\bar Q$ pair with quantum numbers $\kappa$ into a
$Q\bar Q$ pair in the nonrelativistic state $n$. Viewed in this way, the
factorization formula in Eq.~(\ref{prod_eq:pqcd_fac}) is simply a
reorganization of the sum over $n$ in Eq.~(\ref{prod_eq:nrqcd_fac}). (The
contributions denoted by ${\cal O}(m_Q^4/p_T^4)$ in
Eq.~(\ref{prod_eq:pqcd_fac}) are the difference between the NRQCD expression
in Eq.~(\ref{prod_eq:nrqcd_fac}) and the first two terms in
Eq.~(\ref{prod_eq:pqcd_fac}) expanded as a series in $\alpha_s$
\cite{Berger:2001wr}.) Although 
Eqs.~(\ref{prod_eq:nrqcd_fac}) and (\ref{prod_eq:pqcd_fac}) are equivalent if the
NRQCD factorization formalism is valid for heavy-quarkonium production,
the formula in Eq.~(\ref{prod_eq:pqcd_fac}) provides a systematic
reorganization of the cross section in term of powers of $m_Q/p_T$ and
a systematic method for resumming potentially large logarithms of
$p_T/m_Q$. That reorganization and resummation may make the
$\alpha_s$ expansion more convergent.

\subsubsection{Difficulties in establishing NRQCD factorization}
\label{prod_sec:NRQCD-difficulties}%

The fragmentation functions $D_{i\to H}$ and $D_{[Q\bar{Q}(\kappa)]\to
H}$ include certain contributions whose compatibility 
with NRQCD factorization is not obvious. These contributions arise from 
processes which, when viewed in the quarkonium rest frame,
involve the emission of a gluon with momentum of order $m_Q$ from the
fragmenting parton. That relatively hard gluon can exchange soft
gluons with the color-octet $Q\bar Q$ pair that evolves into the
quarkonium. Such soft interactions can produce logarithmic infrared
divergences, which must be absorbed into the NRQCD matrix elements in
Eq.~(\ref{prod_eq:frag_fac}) in order to obtain short-distance coefficients
$d_{i\to n}$ and $d_{[Q\bar{Q}(\kappa)]\to n}$ that are infrared safe.
The interactions of soft gluons with the gluon that has momentum of
order $m_Q$ can be represented by interactions of the soft gluons
with a lightlike eikonal line (gauge-field link) \cite{Nayak:2005rt}.
Similar lightlike eikonal lines are required in order to render the
color-octet NRQCD long-distance matrix elements gauge invariant
\cite{Nayak:2005rw,Nayak:2005rt,kang-qiu-sterman}. If it can be shown
that the color-octet NRQCD matrix elements are independent of the
directions of these eikonal lines, then it follows that the
infrared-divergent soft interactions with gluons that have momenta of order
$m_Q$ can be absorbed into universal ({\it i.e.}, process independent)
NRQCD long-distance matrix elements. It has been shown that this is the
case through two-loop order and, therefore, that the NRQCD factorization
in Eq.~(\ref{prod_eq:frag_fac}) is valid through two-loop order
\cite{Nayak:2005rw,Nayak:2005rt,kang-qiu-sterman}. However, the NRQCD
factorization in Eq.~(\ref{prod_eq:frag_fac}) has not been verified at higher
orders and, therefore, it is not known if the NRQCD factorization
formula in Eq.~(\ref{prod_eq:nrqcd_fac}) is valid
\cite{Nayak:2005rw,Nayak:2005rt,kang-qiu-sterman}. Note that, because
the potential violations of NRQCD factorization at higher loop orders
involve gluons with arbitrarily soft momenta, such violations are not
suppressed by powers of $\alpha_s$ and, consequently, they could
completely invalidate the NRQCD factorization formula in
Eq.~(\ref{prod_eq:nrqcd_fac}).

It is clear that the NRQCD factorization formula cannot apply directly to
reactions in which an additional heavy $Q$ or $\bar Q$ is produced
nearly co-moving with the $Q\bar Q$ pair that evolves into the heavy
quarkonium. That is because the NRQCD factorization formula is designed
to take into account only a heavy quark and a heavy antiquark at small
relative velocity \cite{Nayak:2007mb,Nayak:2007zb}. If an additional
heavy $Q$ or $\bar Q$ is nearly co-moving with a $Q\bar Q$ pair, a
color-octet $Q\bar Q$ pair could evolve into a color-singlet $Q\bar Q$
pair by exchanging soft gluons with the additional $Q$  or $\bar Q$.
Such nonperturbative color-transfer processes could be taken into
account by generalizing the existing NRQCD factorization formalism to
include long-distance matrix elements that involve additional heavy
quarks or antiquarks. Such color-transfer processes might be
identified experimentally by looking for an excess of heavy-flavored
mesons near the direction of the quarkonium.

\subsubsection{$k_T$ factorization}

The $k_T$-factorization approach is an alternative to standard
collinear factorization that has been applied to analyses of inclusive
hard-scattering processes. In the case of quarkonium production, the
$k_T$-factorization approach has usually been applied within the CSM
\cite{Hagler:2000dd,Hagler:2000eu,Yuan:2000qe,Baranov:2002cf,Baranov:2007ay,Baranov:2007dw}.
In $k_T$-factorization formulas, the parton distributions for the
initial-state hadrons depend on the parton transverse momentum, as well
as on the parton longitudinal momentum fraction. The leading-order
$k_T$-factorization expressions for hard-scattering rates contain some
contributions that appear in the standard collinear-factorization
formulas in higher orders in $\alpha_s$ and $\Lambda_{\rm QCD}/p$. In
some kinematic situations, these higher-order corrections might be
important numerically, and the $k_T$-factorization predictions in
leading order might, in principle, be more accurate than the
collinear-factorization predictions in leading order. (An example of
such a kinematic situation is the high-energy limit $s\gg \hat s$, where
$s$ ($\hat s$) denotes the square of the total four-momentum of the
colliding hadrons (partons).) On the other hand, the $k_T$-dependent
parton distributions are less constrained by phenomenology than are the
standard parton distributions, and the uncertainties in the
$k_T$-dependent parton distributions are not yet well quantified in
comparison 
with the uncertainties in the standard parton
distributions. Consequently, in 
practice, the $k_T$-factorization predictions may be more uncertain than
the 
corresponding collinear-factorization predictions.

\subsubsection{Factorization in exclusive quarkonium production}
\label{prod_sec:fact-exclusive}%

NRQCD factorization has been proven for the amplitudes for two
exclusive quarkonium production processes
\cite{Bodwin:2008nf,Bodwin:2010fi}: exclusive production of a quarkonium
and a light meson in $B$-meson decays and exclusive production of
two-quarkonium states in $e^+e^-$ annihilation. The proofs begin by
factoring nonperturbative processes that involve virtualities of order
$\Lambda_{\rm QCD}$ or smaller from hard processes that involve
virtualities of order $p$. (Here, $p$ is the $e^+e^-$ center-of-mass
energy in $e^+e^-$ annihilation, and $p$ is the $B$-meson mass in
$B$-meson decays.) At this stage, the quarkonia enter through
gauge-invariant quarkonium distribution amplitudes. It is then argued
that each quarkonium distribution amplitude can be written as a sum of
products of perturbatively calculable short-distance coefficients
with NRQCD long-distance matrix elements. The difficulties that
occur in establishing this step for inclusive quarkonium production do
not appear in the case of exclusive quarkonium production because only
color-singlet $Q\bar Q$ pairs evolve into quarkonia in exclusive
production. The proofs of factorization for exclusive quarkonium
production reveal that the violations of factorization are generally
suppressed by a factor $m_Qv/p$ for each final-state quarkonium
and, therefore, vanish in calculations of quarkonium production
at order $v^0$ in the velocity expansion.

The factorization proofs in Refs.~\cite{Bodwin:2008nf,Bodwin:2010fi}
also establish factorized forms in which the light-cone distribution
amplitudes, rather than NRQCD long-distance matrix elements, account for
the nonperturbative properties of the quarkonia. In contrast with an
NRQCD long-distance matrix element, which is a single number, a
light-cone distribution amplitude is a function of the heavy-quark
longitudinal momentum fraction. Hence, the light-cone distribution
amplitudes are incompletely determined by phenomenology, and predictions
that are based on light-cone factorization formulas
\cite{Ma:2004qf,Bondar:2004sv,Braguta:2005kr,Braguta:2006nf,Braguta:2007ge}
must rely on constrained models for the light-cone distribution
amplitudes. Generally, the quantitative effects of the model assumptions
on the light-cone factorization predictions are not yet known.

\subsubsection{Factorization in quarkonium decays}
\label{prod_sec:fact-decay}%

There are NRQCD factorization formulas for exclusive quarkonium decay
amplitudes and for inclusive quarkonium decay rates
\cite{Bodwin:1994jh}. As in the NRQCD factorization formula for
inclusive quarkonium production in Eq.~(\ref{prod_eq:nrqcd_fac}), these decay
formulas consist of sums of products of NRQCD matrix elements with
short-distance coefficients. In the cases of decays, the short-distance
coefficients are evaluated at a scale $\mu$ of order $m_Q$ and are
thought to be calculable as power series in $\alpha_s(\mu)$. It is
generally believed that the NRQCD factorization formula for quarkonium
decays can be proven by making use of standard methods for establishing
perturbative factorization. The color-singlet NRQCD long-distance
production matrix elements are proportional, up to corrections of
relative order $v^4$, to the color-singlet NRQCD long-distance decay
matrix elements. However, there is no known relationship between the
color-octet production and decay matrix elements.

\subsubsection{Future opportunities}
\label{prod_sec:factorization-future}%

One of the crucial theoretical issues in quarkonium physics is the
validity of the NRQCD factorization formula for inclusive quarkonium
production. It is very important either to establish that the NRQCD
factorization formula is valid to all orders in perturbation theory or
to demonstrate that it breaks down at some fixed order in perturbation
theory.

The NRQCD factorization formula is known to break down when an
additional heavy quark or antiquark is produced in close proximity to a
$Q\bar Q$ pair that evolves into a quarkonium. It would help in
assessing the numerical importance of such processes if experimental
measurements could determine the rate at which heavy-flavored
mesons are produced nearby in phase space to a heavy quarkonium. If such
processes prove to be important numerically, then it would be useful
to extend the NRQCD factorization formalism to include them.

\subsection{Production at the Tevatron, RHIC and the
LHC}\label{prod_sec:hadroproduction}%

The first measurements by the CDF collaboration of
the {\it direct} production\footnote{``Prompt
production'' excludes quarkonium production from weak decays of more
massive states, such as the $B$ meson. ``Direct production'' further
excludes quarkonium production from feeddown, via the electromagnetic
and strong
interactions, from more massive states, such as higher-mass quarkonium states.}
 of the $J/\psi$ and the $\psi(2S)$ at
$\sqrt{s}=1.8$~TeV \cite{Abe:1997jz,Abe:1997yz} revealed a striking
discrepancy with the existing theoretical calculations: The observed
rates were more than an order of magnitude greater than the calculated rates
at leading order (LO) in $\alpha_s$ in the CSM.
(See \Sec{prod_sec:CSM} for a discussion of the CSM.) This discrepancy
has triggered many theoretical studies of quarkonium hadroproduction,
especially in the framework of NRQCD factorization. (See
\Sec{prod_sec:NRQCD-fact} for a discussion of NRQCD factorization.) In
the NRQCD factorization approach, mechanisms beyond those in the CSM
arise, in which the production of charmonium states proceeds through the
creation of a $c\bar c$ pair in a color-octet state. For the
specific case of the production of the $J/\psi$ or the $\psi(2S)$
(henceforth denoted collectively as ``$\psi$''), these color-octet
transitions take place at higher orders in $v$. Depending on the
convergence of the expansions in $\alpha_s$ and $v$ and the validity of
the NRQCD factorization formula, the NRQCD factorization approach may
provide systematically improvable approximations to the inclusive
quarkonium production rates. For some recent reviews, see
Refs.~\cite{Brambilla:2004wf,Lansberg:2006dh,Lansberg:2008gk,Kramer:2001hh}.
For some perspectives on quarkonium production at the LHC, see
Ref.~\cite{Lansberg:2008zm}.

Despite recent theoretical advances, which we shall detail below, we are
still lacking a clear picture of the mechanisms at work in quarkonium
hadroproduction. These mechanisms would have to explain, in a consistent
way, both the cross section measurements and the polarization
measurements for charmonium production at the Tevatron
\cite{Abe:1997yz,Abe:1997jz,Abachi:1996jq,Affolder:2000nn,Acosta:2004yw,Abulencia:2007us,Aaltonen:2009dm}
and at RHIC
\cite{Adare:2006kf,Adler:2003qs,Atomssa:2008dn,Abelev:2009qaa,daSilva:2009yy,Adare:2009js}.
For example, the observed $p_T$ spectra in prompt $\psi$ production
seem to suggest that a dominant contribution at large $p_T$
arises from a color-octet process in which a gluon fragments into a
$Q\bar Q$ pair, which then evolves nonrelativistically into a
quarkonium. Because of the approximate heavy-quark spin symmetry of
NRQCD, the dominance of such a process would lead to a substantial
transverse component for the polarization of $\psi$'s produced  at large
$p_T$ \cite{Cho:1994ih,Leibovich:1996pa,Braaten:1999qk}. This prediction
is clearly challenged by the experimental measurements
\cite{Abulencia:2007us}.

A possible interpretation of such a failure of NRQCD factorization is
that the charmonium system is too light for relativistic effects to be
small and that, in phenomenological analyses, the velocity expansion of
NRQCD \cite{Bodwin:1994jh} may have been truncated at too low an order.
However, such an explanation would seem to be at odds with other
successful predictions of the NRQCD approach to charmonium physics. If
the convergence of the velocity expansion of NRQCD is indeed an issue,
then one would expect better agreement between theory and the available
experimental data on hadroproduction in the case of the bottomonium
states and, in particular, in the case of the $\Upsilon$. Better
convergence of the velocity expansion might explain, for example, why a
computation that retains only color-singlet contributions
\cite{Artoisenet:2008fc} (that is, only contributions of leading order
in $v$) seems to be in better agreement with the data for $\Upsilon$
production
\cite{Affolder:1999wm,Acosta:2001gv,Abazov:2005yc,Abazov:2008za} than
with the data for $\psi$ production \cite{Aaltonen:2009dm}. We will
discuss this comparison between theory and experiment in greater detail
later.

In efforts to identify the mechanisms that are at work in inclusive
$\psi$ or $\Upsilon$ production, it is important to have control of the
higher-order perturbative corrections to the short-distance coefficients
that appear in the NRQCD factorization formula. Several works have been
dedicated to the study of the corrections of higher-order in $\alpha_s$
and their phenomenological implications for the differential production
rates. We summarize these results in
\Secs{prod_section:higherorderchannels}  and
\ref{prod_section:pheno_with_QCD_corr}. An important observable that has been
reanalyzed in the context of higher-order perturbative corrections is
the polarization of the quarkonium. We review the analyses of the
quarkonium polarization in \Sec{prod_section:polarization}. In addition
to the rates and polarizations in inclusive quarkonium production, other
observables have been shown to yield valuable information about the
production mechanisms. We discuss them in
\Sec{prod_section:newobservables}. Finally, we summarize the future
opportunities for theory and experiment in inclusive quarkonium
hadroproduction in \Sec{prod_sec:hadroproduction-fo}.

\subsubsection{Channels at higher-order in $\alpha_s$}
\label{prod_section:higherorderchannels}

At the LHC, the Tevatron, and RHIC, quarkonium production proceeds
predominantly via gluon-fusion processes. The production cross sections
differential in $p_T$ for $S$-wave quarkonium states have been
calculated only recently at NLO in $\alpha_s$
(Refs.~\cite{Artoisenet:2007xi,Campbell:2007ws,Gong:2008sn,Gong:2008hk,Gong:2008ft}).
These NLO calculations also provide predictions for the quarkonium
polarization differential in $p_T$. In the case of the production of
spin-triplet $P$-wave quarkonium states, the NLO corrections to the
production rate differential in $p_T$ have been calculated even more
recently \cite{Ma:2010vd}.

One common outcome of these calculations is that the total cross
sections are not much affected by corrections of higher order in
$\alpha_s$ \cite{Campbell:2007ws,Brodsky:2009cf}. That is, the
perturbation series for the total cross sections seem to exhibit normal
convergence. However in the case of $\psi$, $\Upsilon$, or $\chi_{cJ}$
production via color-singlet channels, very large corrections appear in
the cross sections differential in $p_T^2$ at large $p_T$. This
behavior, which has also been seen in photoproduction
\cite{Kramer:1995nb}, is well understood. QCD corrections to the
color-singlet parton cross section open new production channels, whose
contributions fall more slowly with $p_T$ than do the LO contributions.
Hence, the contributions from the new channels increase substantially the
cross section in the large-$p_T$ region.

\begin{figure}[t]
\centering
{\includegraphics[width=\figwid]{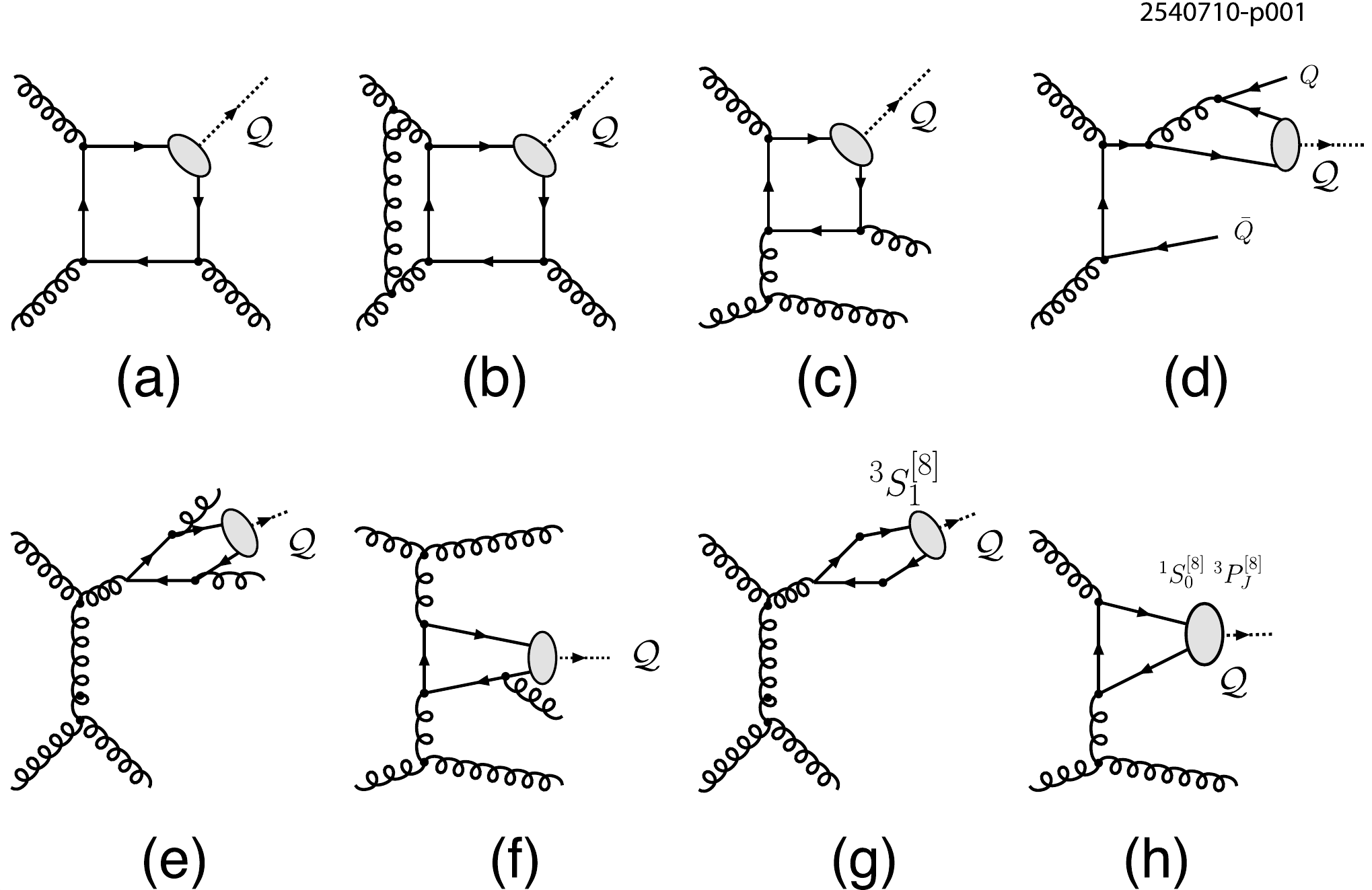}}
\caption{Representative diagrams that contribute to the hadroproduction
of ${}^3S_1$ quarkonium states via color-singlet channels at orders
$\alpha_s^3$ (a), $\alpha_s^4$ (b,c,d), and $\alpha_s^5$ (e,f), and via
color-octet channels at order $\alpha_s^3$ (g,h). The quark and
antiquark that are attached to the ellipses are taken to be on shell,
and their relative velocity is set to zero. \figPermsM{(a)--(f),        
                  (g)--(h)}{Artoisenet:2008fc,Lansberg:2008gk}          
                {2008, 2009}{\APS\ and \SPV} }
\label{prod_diagrams}
\end{figure}

We now discuss this phenomenon briefly. If only the LO
(order-$\alpha_s^3$) contribution to the production of a color-singlet
${}^3S_1$ $Q\bar Q$ state is taken into account, then the partonic cross
section differential in $p_T^2$ scales as $p_T^{-8}$ 
(Refs.~\cite{Chang:1979nn,Berger:1980ni,Baier:1981uk,Baier:1983va,Kartvelishvili:1978id}).
This behavior comes  from the contributions that are associated with
``box'' graphs, such as the one in Fig.~\ref{prod_diagrams}(a). At NLO in
$\alpha_s$ (order $\alpha_s^4$),  several contributions with distinct
kinematic properties arise. The loop corrections, illustrated in
Fig.~\ref{prod_diagrams}(b), are not expected to have substantially
different $p_T$ scaling than the Born contributions. However, the
$t$-channel gluon-exchange diagrams, such as the one depicted in
Fig.~\ref{prod_diagrams}(c),  yield contributions that scale as
$p_T^{-6}$ and are, therefore, kinematically enhanced in comparison with
the Born contribution. At sufficiently large $p_T$, this contribution
has a kinematic enhancement that compensates for its $\alpha_s$
suppression, and it is expected to dominate over the Born contribution.
At NLO, there is also a contribution that arises from the process in
which a second $Q\bar Q$ pair is produced, in addition to the $Q\bar Q$
pair that evolves into a quarkonium. In this ``associated production''
process, both $Q\bar Q$ pairs have the same flavor. In the limit $p_T
\gg m_Q$, the associated-production mechanism reduces to heavy-quark
fragmentation. Therefore, in this limit, the corresponding partonic
cross section differential in $p_T^2$ scales as $p_T^{-4}$. The
associated-production contribution eventually provides the bulk of the
color-singlet parton cross section at sufficiently large $p_T$. In the
$p_T$ region that is covered by the current experiments, though, this
contribution is relatively small, owing to phase-space suppression
\cite{Artoisenet:2007xi}.

Similar arguments can be used to understand the impact of QCD
corrections on the rates differential in $p_T^2$ for the production of a
$Q\bar Q$ pair in a color-octet state. At LO in $\alpha_s$, the
production of a color-octet $^3S_1$ $Q\bar Q$ pair proceeds, at large
$p_T$, predominantly through gluon fragmentation
[Fig.~\ref{prod_diagrams}(g)]. Thus, the rate differential in $p_T^2$
scales as $p_T^{-4}$. This is the smallest power of $1/p_T$ that is
possible in partonic cross sections. Hence, in this case, the NLO
correction cannot contain a kinematically enhanced channel and does not
affect substantially the shape of the differential rate
\cite{Gong:2008ft}. The situation is different for the production of a
$C$-even color-octet $Q\bar Q$ pair because, in this case, there is no
fragmentation process at LO in $\alpha_s$. [See
Fig.~\ref{prod_diagrams}(h).] At LO in $\alpha_s$, the rates for these
channels, differential in $p_T^2$, scale as $p_T^{-6}$. The
fragmentation channels appear at NLO in $\alpha_s$. Consequently the NLO
correction to the differential rate is expected to yield a substantial
enhancement at large $p_T$. This feature has been checked explicitly in
Ref.~\cite{Gong:2008ft} in the specific case of the production of a
color-octet ${}^1S_0$ $Q\bar Q$ state.

In view of the strong impact of the correction at NLO in $\alpha_s$ on
the color-singlet differential rate at large transverse momentum, it is
natural to examine the QCD corrections that appear at even higher orders
in the $\alpha_s$ expansion. At NNLO (order $\alpha_s^5$), new, important
channels with specific kinematic properties continue to appear. Some of
these channels have actually been studied for some time in specific
kinematic limits in which one can take advantage of large separations
between different perturbative energy scales. The color-singlet
gluon-fragmentation channel [Fig.~\ref{prod_diagrams}(e)] has been
investigated in the framework of the fragmentation approximation
\cite{Braaten:1995cj}, which is relevant in the limit $p_T \gg m_Q$. The
processes in which the $Q\bar Q$ pair is produced by the exchange of two
gluons in the $t$ channel [Fig.~\ref{prod_diagrams}(f)] were
investigated in the $k_T$-factorization approach
\cite{Yuan:2000qe,Baranov:2002cf,Baranov:2007ay,Baranov:2007dw}. This
approach is relevant in the high-energy limit $s\gg \hat s$, where $s$
($\hat s$) denotes the square of the total four-momentum of the
colliding hadrons (partons). In that kinematic limit, other enhanced
processes, which are initiated by a symmetric two-gluon color-octet
state and an additional gluon, have been investigated more recently in
Ref.~\cite{Khoze:2004eu}. These processes correspond to higher-order 
contributions in the framework of the (standard) collinear approximation
of perturbative QCD. The advantage in considering either of these
kinematic limits is that the perturbation expansion can be reorganized
in such a way as to simplify the evaluation of the dominant
contribution. Furthermore, the convergence of the perturbation expansion
is improved because large logarithms of the ratio of the disparate
energy scales are resummed. Away from the asymptotic regime, the
corrections to each of these approaches may be important. The impacts of
these corrections in the kinematic region that is covered at the current
hadron colliders is not known accurately. There is an alternative
method that has been proposed for estimating the NNLO corrections to the
color-singlet differential rate that is known as the NNLO$^\star$ method
\cite{Artoisenet:2008fc}. The NNLO$^\star$ method does not attempt to
separate the various energy scales. Instead, it considers only NNLO
corrections involving real gluon emission and imposes an infrared cutoff
to control soft and collinear divergences. The NNLO$^\star$ estimates
suffer from large uncertainties, which arise primarily from the
sensitivities of these estimates to the infrared cutoff and to the
choice of renormalization scale.

A specific higher-order process that has been investigated is the
so-called ``$s$-channel $Q\bar Q$-cut'' process
\cite{Lansberg:2005pc,Haberzettl:2007kj}. In this process, an on-shell
$Q\bar Q$ pair is produced. That pair then rescatters into a quarkonium
state. The contribution of LO in $\alpha_s$ to the amplitude for this
process is given by the imaginary part of a specific set of one-loop
diagrams \cite{Artoisenet:2009mk}, and the square of this amplitude
contributes to the cross section at order $\alpha_s^5$.

In addition to these new results for the QCD corrections to
inclusive quarkonium production, there have also been new results
for the QED and relativistic corrections to inclusive quarkonium production. 
The QED correction to the inclusive $J/\psi$ production rate in hadron-hadron
collisions has been computed \cite{He:2009zzb,He:2009cq} and turns out to
be small in the region of $p_T$ that is covered by the
current experiments. Relativistic corrections to the color-singlet rate
for inclusive $J/\psi$ hadroproduction have been shown to be negligible
over the entire $p_T$ range that is currently accessible \cite{Fan:2009zq}.

\begin{figure}[b]\centering
\includegraphics[width=\figwid]{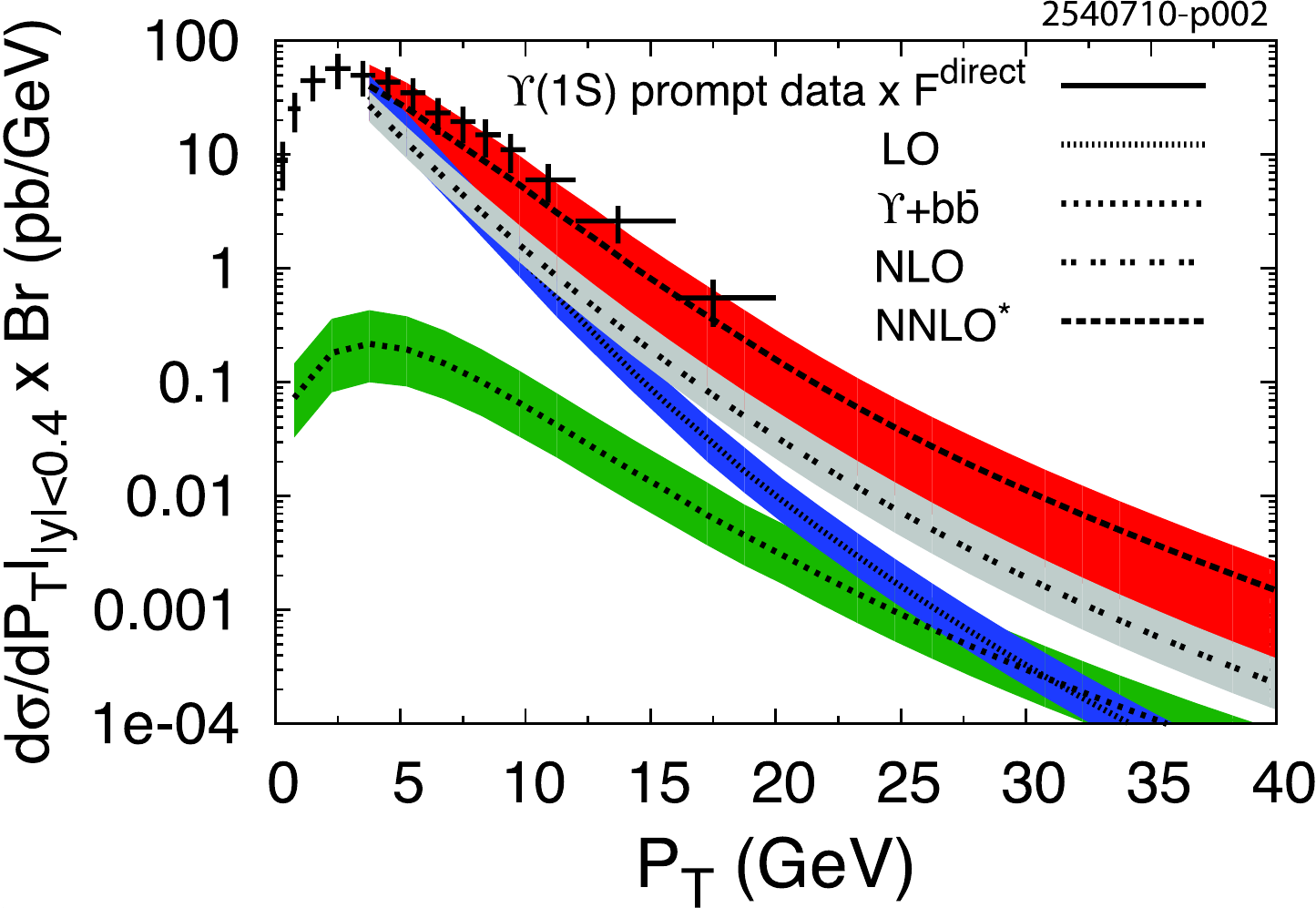}

\caption{Comparison between the CSM predictions at NLO and NNLO$^\star$
accuracy for the $\Upsilon$ cross section as a function of the
$\Upsilon$ transverse momentum at the Tevatron at $\sqrt{s}=1.80$~TeV
(Ref.~\cite{Artoisenet:2008fc}) and the CDF data for $\Upsilon(1S)$
production \cite{Acosta:2001gv}. The crosses are the CDF data for prompt
$\Upsilon(1S)$ production, multiplied by $F^{\rm direct}$, the fraction
of direct $\Upsilon(1S)$'s in prompt $\Upsilon(1S)$ events, as measured
by the CDF collaboration using an older event sample
\cite{Affolder:1999wm}. The {\it lines} show the central values of the
theoretical predictions, and the {\it bands} depict the theoretical
uncertainties. The theoretical uncertainty bands for the LO, NLO, and
$\Upsilon +b\bar b$ contributions were obtained by combining the
uncertainty from $m_b$ with the uncertainties that are obtained by
varying the renormalization scale $\mu_f$ and the factorization scale
$\mu_r$ between $2m_T$ and $m_T/2$, where $m_T=\sqrt{4m_b^2+p_T^2}$. The
error band for the NNLO$^\star$ contribution is obtained by varying the
infrared cutoff $s^{\rm min}_{ij}$ between $2 m_b^2$ and $m_b^2/2$ and
by varying $\mu_f$ and $\mu_r$ between $2m_T$ and $m_T/2$. 
\figPermXAPS{Artoisenet:2008fc}{2008} }
\label{prod_fig:dsdpt_NNLO}
\end{figure}

\subsubsection{Phenomenology, including QCD corrections}
\label{prod_section:pheno_with_QCD_corr}

\subthreesection{$\psi$, $\Upsilon$, and $\chi_c$ production at the Tevatron}

In the case of $\Upsilon$ production, the contributions of the NLO
corrections to the color-singlet channels reduce the discrepancy between
the color-singlet contribution to the inclusive cross section and the
data collected by the CDF collaboration, as is illustrated in
Fig.~\ref{prod_fig:dsdpt_NNLO}.\footnote{We note that no
phenomenological analysis of $\chi_{bJ}$ production is available yet at
NLO accuracy. Such an analysis would be necessary in order to
predict the prompt $\Upsilon$ cross section to NLO accuracy. As a
makeshift, one could multiply the available data by the measurement
of the fraction of direct $\Upsilon$'s in the total rate (integrated
over $p_T$) \cite{Affolder:1999wm}. Note, however, that the NLO
calculation of the $\chi_{cJ}$ production rate \cite{Ma:2010vd}
indicates that the fraction of direct $J/\psi$'s may depend rather
strongly on $p_T$.}  However, the predicted NLO rate drops too
rapidly at large $p_T$, indicating that another production mechanism is
at work in that phase-space region. A recent study
\cite{Artoisenet:2008fc} has shown that contributions from channels that
open at NNLO (order $\alpha_s^5$) may fill the remaining gap between the
color-singlet contribution at NLO and the data. The estimate of the NNLO
contribution from this study, called the ``NNLO$^\star$ contribution'',
is shown in the (red) band labeled NNLO$^\star$ in Fig.
\ref{prod_fig:dsdpt_NNLO}. Owing to the large theoretical uncertainties,
this improved prediction for the color-singlet contribution does not
imply any severe constraint on other possible contributions, such as a
color-octet contribution. However, in contrast with previous LO
analyses, in which color-octet contributions were {\it required} in
order to describe the data, the NNLO$^\star$ estimate of the
color-singlet contribution shows that color-octet contributions are now
merely {\it allowed} by the rather large theoretical uncertainties in
the NNLO$^\star$ estimate.
 
\begin{figure}[t]\centering
\includegraphics[width=\figwid]{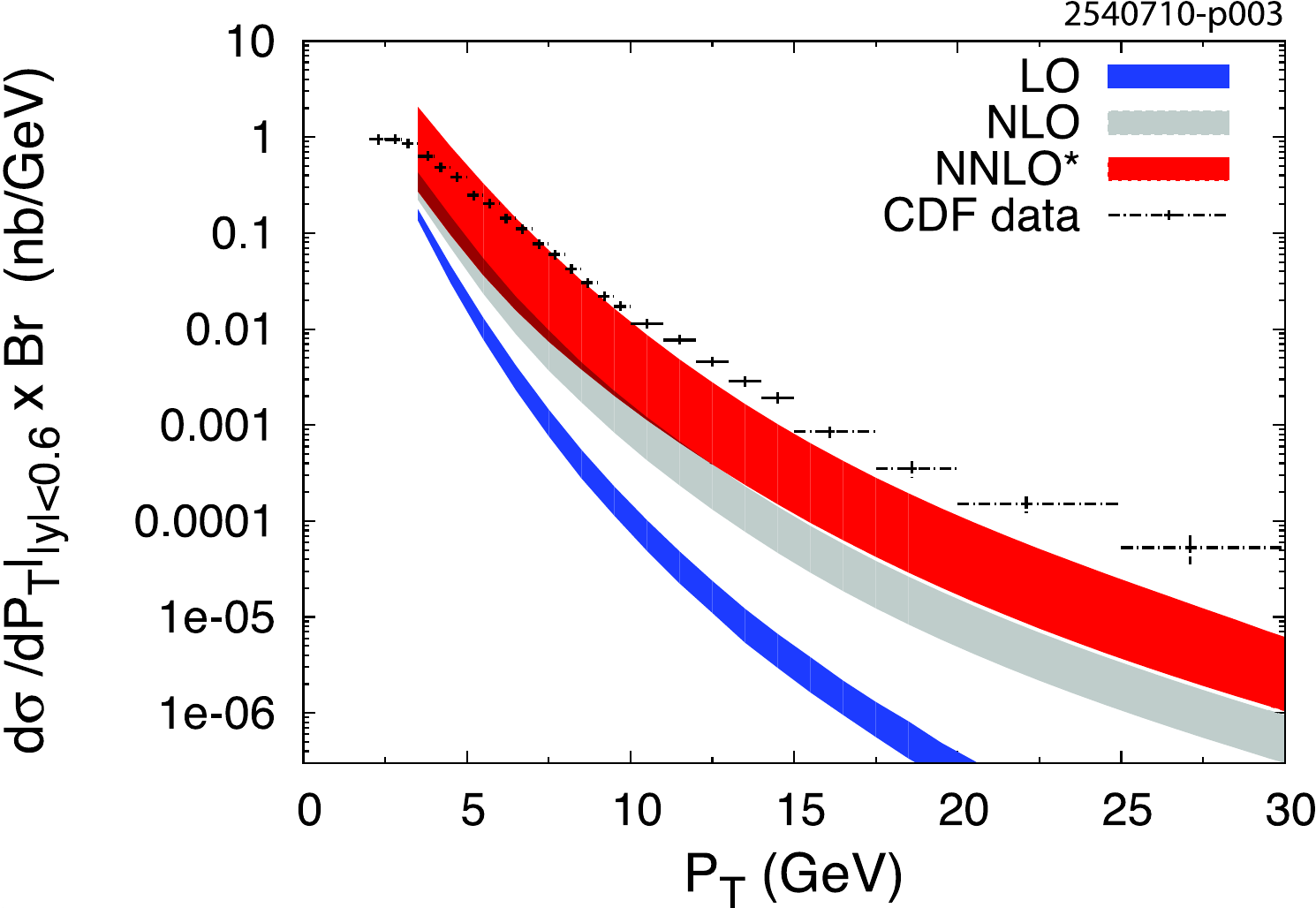}
\caption{Comparison between the CSM predictions for the $\psi(2S)$ cross
sections at LO, NLO, and NNLO$^\star$ accuracy as a function of the
$\psi(2S)$ $p_T$ at the Tevatron at $\sqrt{s}=1.96$~TeV
(Ref.~\cite{Lansberg:2008gk}) and the CDF prompt $\psi(2S)$ data
\cite{Aaltonen:2009dm}. The theoretical uncertainty bands for the LO and
NLO contributions were obtained by combining the uncertainty from $m_c$
with the uncertainties that are obtained by varying the renormalization
scale $\mu_f$ and the factorization scale $\mu_r$ between $2m_T$ and
$m_T/2$, where $m_T=\sqrt{4m_c^2+p_T^2}$. The theoretical uncertainty
band for the NNLO$^\star$ contribution was obtained by varying the
infrared cut-off $s^{\rm min}_{ij}$ between $4 m_c^2$ and $m_c^2$ and
by varying $\mu_f$ and $\mu_r$ between $2m_T$ and $m_T/2$. 
\figPermXSPV{Lansberg:2008gk}{2009} }
\label{prod_fig:psi2Sdsdpt_NNLO}
\end{figure}

The impact of the QCD corrections on the color-singlet contribution has
also been studied in the case of $\psi$ hadroproduction
\cite{Lansberg:2008gk}.
The comparison with the data is simpler in the case of the 
$\psi(2S)$ than in the case of the $J/\psi$, owing to the absence of 
significant feeddown from excited charmonium states to the $\psi(2S)$.
In a recent paper, the CDF collaboration has reported a new measurement of the 
inclusive $\psi(2S)$ cross section \cite{Aaltonen:2009dm}. The rates for 
the prompt production of the $\psi(2S)$ and for the production of the 
$\psi(2S)$ in $B$-meson decays were also extracted in that analysis. 
The reconstructed differential rate for the prompt component is compared to 
the prediction for the color-singlet rate at LO, NLO and NNLO$^\star$ 
accuracy in Fig.~\ref{prod_fig:psi2Sdsdpt_NNLO}.
At medium values of  $p_T$, the upper limit of the NNLO$^\star$ rate
is compatible with the CDF results. 
At larger values of $p_T$, a gap appears between the color-singlet 
rate and the data \cite{Lansberg:2008gk}.
The $J/\psi$ differential production rate has the same qualitative
features as the $\psi(2S)$ differential production rate
\cite{Artoisenet:2008zza}. It is worth emphasizing that the current
discrepancy between the  color-singlet rate and the Tevatron data has
been dramatically reduced by the inclusion of higher-order QCD
corrections.

NLO QCD corrections to the color-octet production channels ${}^3S_1$ and
${}^1S_0$ have been analyzed for $J/\psi$ production \cite{Gong:2008ft}
and for $\Upsilon$ production \cite{Gong:2010bk}. In both cases, these
corrections proved to be small when the $p_T$ of the produced quarkonium
is less than $20$~GeV. In Ref.~\cite{Gong:2008ft}, values of the NRQCD
long-distance matrix elements $\langle O^{J/\psi} \big( ^3S_1^{[8]}\big)
\rangle$ and $\langle O^{J/\psi} \big( ^1S_0^{[8]}\big) \rangle$ were
obtained by fitting the theoretical prediction to the prompt production
rate that was measured by the CDF collaboration \cite{Acosta:2004yw}.
The values of the NRQCD matrix elements that were extracted in this
analysis are compatible with the values that were extracted in LO
analyses. In the analysis of Ref.~\cite{Gong:2008ft}, feeddown
contributions were ignored and the $P$-wave color-octet long-distance
matrix elements were set to zero. A satisfactory fit could not be
obtained for the experimental data points that have $p_T<6$~GeV, and so
these points were not included in the fit. In this regard, it should be
kept in mind that resummation of large logarithms may be needed at small
$p_T$ and that NRQCD factorization may break down at small $p_T$.

The $s$-channel $c\bar c$-cut ($s$CC) contributions to $\psi$
hadroproduction have been investigated in
Ref.~\cite{Lansberg:2005pc,Haberzettl:2007kj} in the framework of a
phenomenological model. A first analysis of this model
\cite{Haberzettl:2007kj}, which was based on the data of
Refs.~\cite{Abe:1997yz,Acosta:2004yw} and incorporated
constraints for the small- and large-$p_T$ regions, supported rates that
are significantly larger than those of the CSM prediction. This
analysis, which did not include resummation of initial-gluon
contributions, yielded a good fit to the $p_T$ dependence of the RHIC data.
However, it has been shown recently, by evaluating the leading-order
contribution of the $s$CC amplitude in the framework of NRQCD, that the
$s$CC contributions can account only for a negligible fraction of the
$J/\psi$ production rate that is measured by the CDF collaboration
\cite{Artoisenet:2009mk}.

Some specific NNLO contributions in the CSM were considered in
Ref.~\cite{Khoze:2004eu}. These contributions are referred to in the
literature as the ``gluon-tower model.'' They can be viewed as LO BFKL
contributions, and, thus, they are expected to be enhanced in comparison
to other NNLO contributions by a factor $\log s/\hat{s}$. At large $s$,
this logarithm may compensate for the $\alpha_s^2$ suppression of these
contributions. At $\sqrt{s}=1.96$~TeV and $|y| < 0.6$, the gluon-tower
model predicts the $J/\psi$ cross section, integrated over $p_T$, to be
$\sigma(|y| < 0.6)=2.7~\mu \hbox{b}$ (Ref.~\cite{Khoze:2004eu}), in near
agreement with the CDF measurement $\sigma(|y| < 0.6) =
4.1^{+0.6}_{-0.5}~\mu \hbox{b}$ (Ref.~\cite{Shears:2003ic}).\footnote{We
note that the LO CSM result for the $p_T$-integrated direct $J/\psi$
cross section $d\sigma/dy$, evaluated at $y=0$
(Ref.~\cite{Lansberg:2010kh}), is
compatible with the CDF \cite{Acosta:2004yw} and PHENIX
\cite{Adare:2006kf,daSilva:2009yy} measurements.} However, the
theoretical prediction is somewhat sensitive to an effective gluon mass
that is introduced as an infrared cutoff in the model. The comparison
does not take into account feeddown from $P$-wave states in the measured
cross section, and the model cannot, at present, predict the $p_T$
dependence of the cross section.

At LO in $\alpha_s$, NRQCD factorization predicts that the ratio of
production cross sections,
$R_{\chi_c}=\sigma_{\chi_{c2}}/\sigma_{\chi_{c1}}$, is dominated by the
color-octet contribution at large $p_T$ and approaches the value
$R_{\chi_c}=5/3$ as $p_T$ increases. This LO prediction is in sharp
disagreement with the CDF measurement \cite{Abulencia:2007bra}, which
finds that $R_{\chi_c}\approx 0.75$ at large $p_T$. Recently, NLO
corrections to $\chi_{cJ}$ production have been calculated in
Ref.~\cite{Ma:2010vd}. It is found that the NLO corrections are large at
large $p_T$. They make the contributions of the color-singlet ${}^3P_J$
channels negative and comparable to the color-octet contribution for
large $p_T$. They also cause the ${}^3P_1$ color-singlet contribution to
fall at a slower rate than the ${}^3P_2$ color-singlet contribution as
$p_T$ increases. Taking into account the large NLO correction, the
authors of Ref.~\cite{Ma:2010vd} were able to fit the measured $p_T$
distribution of $R_{\chi_c}$, using a plausible value for the ratio of
the relevant NRQCD long-distance matrix elements. Hence, there may now
be a resolution of this outstanding conflict between theory and
experiment. One interesting prediction of the fit to $R_{\chi_c}$ in
Ref.~\cite{Ma:2010vd} is that the feeddown from the $\chi_{cJ}$
states to the $J/\psi$ state may be quite large---perhaps 30\% of the
prompt $J/\psi$ rate at $p_T=20$~GeV. Such a large proportion of feeddown
events in the prompt $J/\psi$ rate could have an important effect on the
prompt $J/\psi$ polarization.

\subthreesection{New NLO hadroproduction results}

As this article was nearing completion, two papers
\cite{Ma:2010yw,Butenschoen:2010rq} appeared that give complete
calculations of the corrections of NLO in $\als$ for the color-octet
production channels through relative order $v^4$, that is, for the
color-octet ${}^3S_1$, ${}^1S_0$, and ${}^3P_J$ channels  We now give a
brief account of these results.

The calculations in Refs.~\cite{Ma:2010yw,Butenschoen:2010rq} are in
numerical agreement for the short-distance coefficients for $J/\psi$
production at the Tevatron. The calculations confirm that the
corrections to the color-octet ${}^3S_1$ and ${}^1S_0$ channels are
small, but also show that there is a large, negative $K$ factor 
in the color-octet ${}^3P_J$ channel.

In Ref.~\cite{Ma:2010yw}, the NLO calculation was fit to the CDF data
for the prompt production of the $J/\psi$ and the $\psi(2S)$
\cite{Acosta:2004yw,Aaltonen:2009dm}, and values were obtained for two
linear combinations of NRQCD long-distance matrix elements. In the case
of the $J/\psi$, these fits took into account feeddown from the
$\psi(2S)$ state and the $\chi_{cJ}$ states, where the latter was
obtained from the NLO calculation of $\chi_c$ production
\cite{Ma:2010vd} that was described above. Satisfactory fits could not
be obtained to the experimental data points for the $J/\psi$ and the
$\psi(2S)$ with $p_T<7$ GeV, and so these points were excluded from
the fits. The fitted values of the linear combinations of matrix
elements were used to predict the cross section for $J/\psi$ production
at CMS, and good agreement with the CMS data \cite{prod_CMS-J/psi} was
obtained. This analysis suggests the possibility that that the cross
section is dominated by the color-octet ${}^1S_0$ contribution, rather
than by the color-octet ${}^3S_1$ contribution, in contrast with
conclusions that had been drawn on the basis of LO fits to the Tevatron
data.

In Ref.~\cite{Butenschoen:2010rq}, values of the NRQCD long-distance
matrix elements were extracted by using the NLO calculation of $J/\psi$
hadroproduction of Ref.~\cite{Butenschoen:2010rq} and an NLO calculation
of $J/\psi$ photoproduction from Ref.~\cite{Butenschoen:2009zy} to make
a combined fit to the CDF Run~II data for prompt $J/\psi$ production
\cite{Acosta:2004yw} and to the HERA~I and HERA~II H1 data for prompt
$J/\psi$ photoproduction \cite{Adloff:2002ex,Aaron:2010gz}. In this fit,
only CDF data with $p_T>3$~GeV were used, as the flattening of the cross
section at smaller values of $p_T$ cannot be described by fixed-order
perturbation theory. Feeddown of the $\psi(2S)$ and $\chi_{cJ}$ states to
the $J/\psi$ was not taken into account in the fits. This is the first
multiprocess fit of NRQCD long-distance matrix elements for quarkonium
production. 
The values of the NRQCD long-distance matrix elements that were obtained
in this fit do not differ greatly from those that were obtained in LO
fits. They were used to predict the cross sections for prompt $J/\psi$
production at PHENIX \cite{Adare:2009js} and CMS \cite{prod_CMS-J/psi},
and good agreement with the data was achieved in both cases.

The values of the linear combinations of $J/\psi$ NRQCD long-distance
matrix elements that were obtained in Ref.~\cite{Ma:2010yw} are not
consistent with the values of the NRQCD long-distance matrix elements
that were obtained in Ref.~\cite{Butenschoen:2010rq}. Since the
calculations of Refs.~\cite{Ma:2010yw,Butenschoen:2010rq} are in
agreement on the short-distance cross sections, any discrepancies in the
extracted NRQCD long-distance matrix elements must be due to differences
in the fitting procedures. Clearly, it is necessary to understand the
significance of the various choices that have been made in the fitting
procedures before any definite conclusions can be drawn about the sizes
of the NRQCD long-distance matrix elements.

\subthreesection{$J/\psi$ production at RHIC}

Recently, the STAR collaboration at RHIC has reported an analysis of
prompt $J/\psi$ production for values of $p_T$ up to 12~GeV
(Ref.~\cite{Abelev:2009qaa}). In Ref.~\cite{Abelev:2009qaa}, the measured
production rate as a function of $p_T$ is compared with predictions
based on NRQCD factorization at LO \cite{Nayak:2003jp} and the CSM up to
NNLO$^\star$ accuracy \cite{Artoisenet:2008fc}. The calculations
do not include feeddown from the $\psi(2S)$ and the $\chi_{c}$ states. The
data clearly favor the  NRQCD factorization prediction over the CSM
prediction. However, no definite conclusions can be drawn because the
effects of feeddown have not been taken into account.

\begin{figure}[b!]\centering
\includegraphics[width=\figwid]{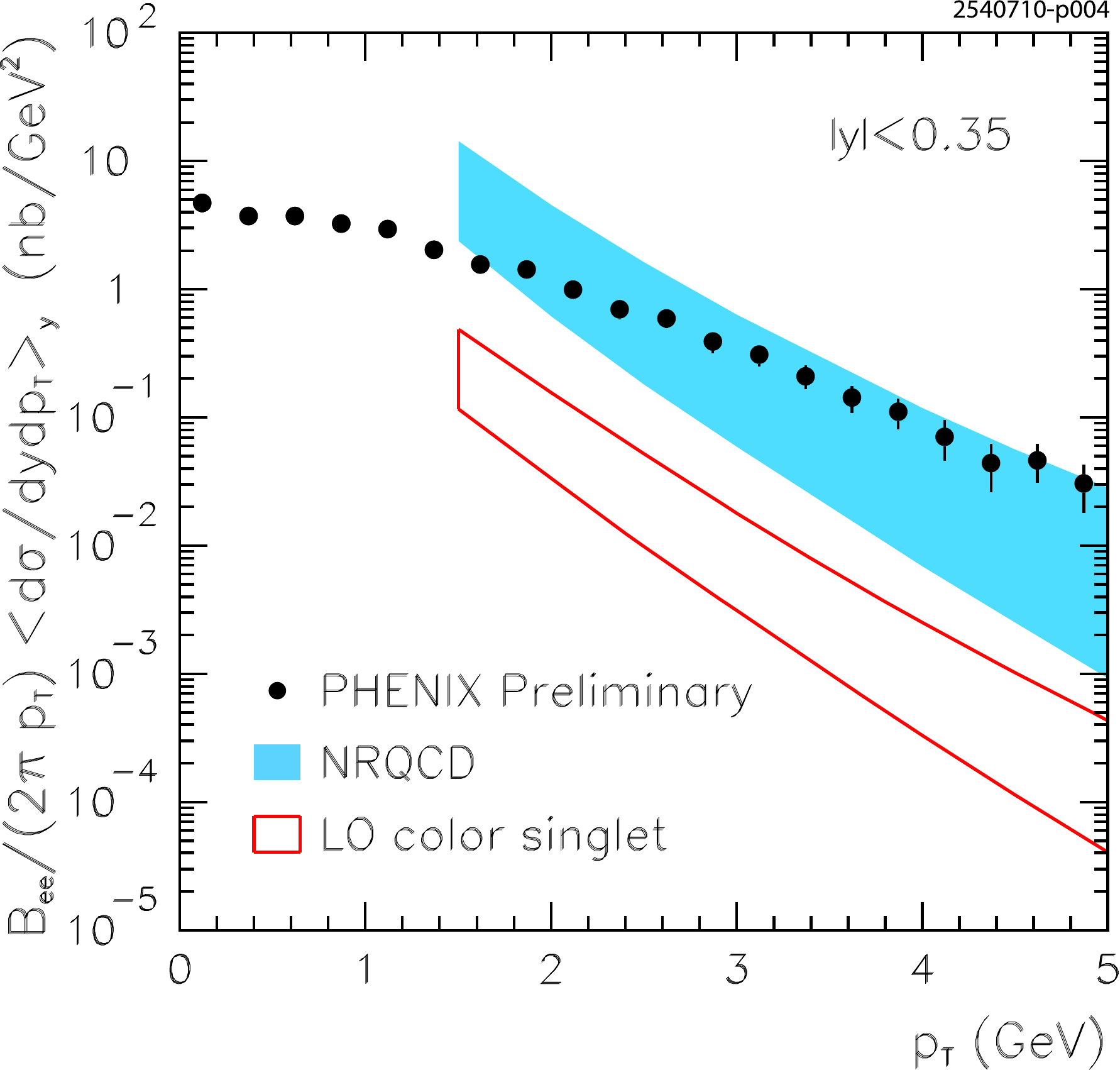}
\caption{Comparison of the LO NRQCD and the LO CSM predictions for the
$J/\psi$ cross section as a function of the $J/\psi$ transverse momentum
\cite{Chung:2009xr} with the data from the PHENIX collaboration
\cite{daSilva:2009yy,Adare:2009js}. The theoretical uncertainty bands
were obtained by combining the uncertainties from $m_c$ and the
NRQCD long-distance matrix elements with the uncertainties that are
obtained by varying the renormalization scale $\mu_r$ and the
factorization scale $\mu_f$ between $2m_T$ and $m_T/2$. Here
$m_T=\sqrt{4m_c^2+p_T^2}$. \figPermXAPS{Chung:2009xr}{2010} }
\label{prod_fig:rhic_p_T} 
\end{figure}

A calculation of prompt $J/\psi$ production at RHIC, including feeddown
from the $\psi(2S)$ and $\chi_{c}$ states, has been carried out in
Ref.~\cite{Chung:2009xr} in the CSM and the NRQCD factorization
formalism at LO. In Fig.~\ref{prod_fig:rhic_p_T}, we show a comparison
between the predictions of Ref.~\cite{Chung:2009xr} for the prompt
$J/\psi$ cross section as a function of $p_T$ and data from the PHENIX
collaboration \cite{daSilva:2009yy,Adare:2009js}. Again, the NRQCD
predictions are favored over the CSM predictions. However, in this case,
the small values of $p_T$ involved may call into question the validity of
perturbation theory, and the omission of higher-order corrections to the
CSM, which are known to be large, also undermines the comparison.

\begin{figure}[b]\centering
\includegraphics[width=\figwid]{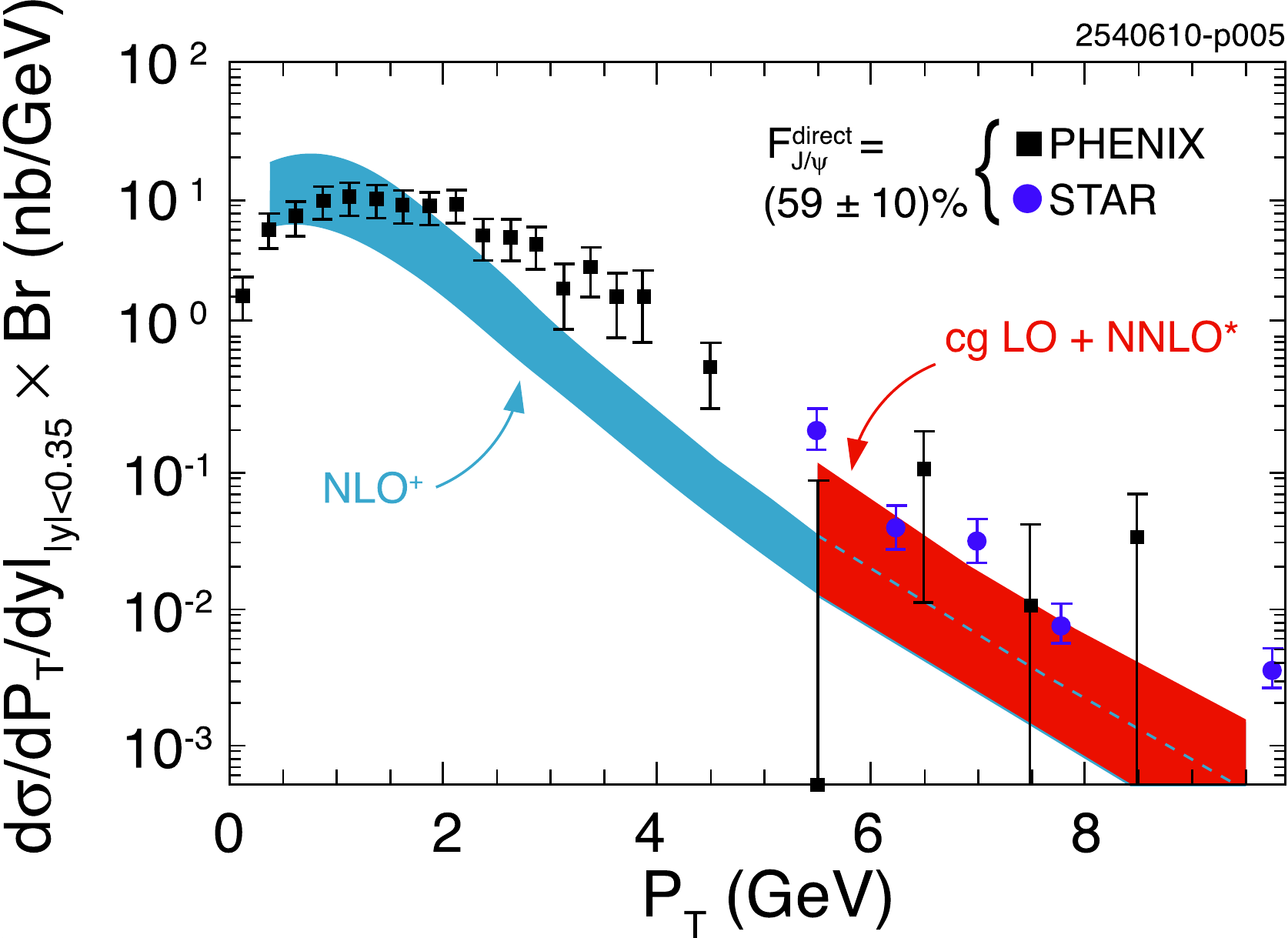}
\caption{Comparison between CSM prediction for the $J/\psi$ cross
section at NLO accuracy as a function of the $J/\psi$ transverse
momentum in $pp$ collisions at RHIC at $\sqrt{s_{NN}} = 200$~GeV and
$|y|<0.35$ (Ref.~\cite{Lansberg:2010vq}) and the PHENIX
\cite{Adare:2006kf} and STAR \cite{Abelev:2009qaa} prompt $J/\psi$ data.
The NLO$^+$ contribution contains the contributions from $gg$ and $gq$
fusion at NLO accuracy, where $q$ is a light quark, plus the
contribution from $cg$ fusion at LO accuracy. The theoretical
uncertainty band for the NLO$^+$ contribution was obtained by
combining the uncertainties that are obtained by varying $m_c$ in the
range $1.4~{\rm GeV}< m_c < 1.6~{\rm GeV}$ and by varying the factorization
scale $\mu_f$ and the renormalization scale $\mu_r$ through the values
$((0.75,0.75);(1,1);(1,2);(2,1);(2,2))\times m_T$. Here
$m_T=\sqrt{4m_c^2+p_T^2}$. The theoretical uncertainty band for the
NNLO$^\star$ contribution was obtained by combining the uncertainties
that are obtained by varying $m_c$ in the range $1.4~{\rm GeV}< m_c
<1.6~{\rm GeV}$, by varying $\mu_f$ and $\mu_r$ in the range  $0.5 m_T
<\mu_f=\mu_r< 2 m_T$, and by varying the infrared cutoff $s_{ij}^{\rm
min}$ in the range $2.25~{\rm GeV}^2 <s_{ij}^{\rm min}< 9.00~{\rm
GeV}^2$. From Ref.~\cite{Lansberg:2010vq} }
\label{prod_fig:rhic_p_T_NLO} 
\end{figure}

Higher-order corrections to the color-singlet contribution to $J/\psi$
production at RHIC have been considered in Ref.~\cite{Lansberg:2010vq}
and were found to be large. A comparison between the predictions of
Ref.~\cite{Lansberg:2010vq} for the cross section differential in $p_T$
and the PHENIX and STAR prompt $J/\psi$ data is shown in
Fig.~\ref{prod_fig:rhic_p_T_NLO}. The color-singlet contributions
through NLO agree with the PHENIX prompt $J/\psi$ data for $p_T$ in the
range $1$--$2$~GeV, but fall substantially below the PHENIX and STAR
prompt $J/\psi$ data for larger values of $p_T$. The NNLO$^\star$
color-singlet contribution can be computed reliably only for
$p_T>5$~GeV. The upper limit of the theoretical uncertainty band for the
NNLO$^\star$ contribution is compatible with the PHENIX and STAR data,
although the theoretical uncertainties are very large.

As we have mentioned above, the NLO analysis of
Ref.~\cite{Butenschoen:2010rq}, which includes the ${}^3S_1$, ${}^1S_0$,
and ${}^3P_J$ color-octet channels, uses matrix elements that are
extracted from a combined fit to CDF data \cite{Acosta:2004yw} and H1
data \cite{Adloff:2002ex,Aaron:2010gz} to predict the cross sections for
$J/\psi$ production at HERA. This prediction agrees well with 
PHENIX data \cite{Adare:2009js}.

\subthreesection{Exclusive production of charmonia}

The {\it exclusive} production of charmonium states (plus beam
particles) has also been observed at hadron-hadron colliders. In most
current theoretical models, the exclusive production of states with
charge parity $-1$, such as the $J/\psi$ or the $\psi(2S)$, is dominated
by the process of photon-Pomeron fusion (photoproduction), while the
exclusive production of states with charge parity $+1$, such as the
$\chi_{c0}$, is dominated by the process of Pomeron-Pomeron fusion. In
perturbative model calculations, the Pomeron is represented as an
exchange of two or more gluons in a color-singlet state. Exclusive
quarkonium production could provide an important tool with which to
probe these mechanisms.

The CDF Collaboration has measured exclusive $J/\psi$, $\psi(2S)$, and
$\chi_{c0}$ production in $p\bar p$ collisions at $\sqrt{s}=1.96$~TeV
(Ref.~\cite{Aaltonen:2009kg}). The CDF measurements are in agreement
with theoretical predictions that are based on models for the Pomeron
\cite{Khoze:2000jm,Yuan:2001nu,Klein:2003vd,Khoze:2004yb,Goncalves:2005yr,Bzdak:2005rp,Bzdak:2007cz,Schafer:2007mm,Motyka:2008ac}.
The PHENIX collaboration has measured exclusive $J/\psi$ production in
Au$+$Au collisions at $\sqrtsNN=200$~GeV
(Ref.~\cite{ConesadelValle:2009up}) and also finds agreement with
theoretical predictions
\cite{Klein:1999qj,Baltz:2002pp,Nystrand:2004vn,Strikman:2005ze,Goncalves:2005sn,Goncalves:2007qu,Ivanov:2007ms,AyalaFilho:2008zr}.
See \Sec{sec:media_subsec57} for more on photoproduction in nuclear collisions.

\subthreesection{$X(3872)$ production at the Tevatron and the LHC}

Since the discovery of the $X(3872)$ by the Belle collaboration in 2003
(Ref.~\cite{Choi:2003ue}), this state has attracted a large
interest in the particle-physics community. The $X(3872)$ is the exotic
state for which the largest experimental data set is available. The
production of the $X(3872)$ at the Tevatron has been analyzed by both the
CDF collaboration \cite{Acosta:2003zx,Abulencia:2006ma,Aaltonen:2009vj}
and the \DZero\ collaboration \cite{Abazov:2004kp} in the $J/\psi\,
\pi^+ \pi^-$ decay channel. The CDF collaboration has shown that most of
the $X(3872)$'s at the Tevatron are produced promptly, rather than
through $b$-hadron decays \cite{Bauer:2004bc}.

As is discussed in \Secs{sec:SpecExpX3872}, \ref{sec:SpecTh_molec},
and \ref{sec:SpecTh_Tetraquarks} of this article, the exact
nature of the $X(3872)$ is still subject to debate. Nevertheless, it is
plausible that the prompt production rate of the $X(3872)$ can be
predicted correctly in the factorization framework of NRQCD. The reason
for this is that, in all the viable hypotheses as to the nature of the
$X(3872)$, the particle content of the state includes a charm-quark pair
with a relative momentum $q \ll m_c$. The small size of $q$ suggests
that one can make use of the NRQCD expansion of the production rate in
powers of $q/m_c$. It follows that the expression for the cross section
is given by the NRQCD factorization formula in
Eq.~(\ref{prod_eq:nrqcd_fac}). It was argued in
Refs.~\cite{Braaten:2004jg,Artoisenet:2009wk} that it is reasonable to
truncate the NRQCD series so that it includes only contributions to the
$X(3872)$ production rate from $c \bar c$ pairs that are created in an
$S$-wave configuration. Moreover, in the case of hadroproduction, it was
argued that a truncation that retains only the color-octet ${}^3S_1$
channel would provide a reliable prediction at large transverse
momentum. It follows that the corresponding NRQCD long-distance matrix
element can be extracted from the measured production rate at the
Tevatron in the $J/\psi\, \pi^+ \pi^-$ decay channel. In
Ref.~\cite{Artoisenet:2009wk}, the aforementioned simplifying
assumptions are used in the NRQCD factorization framework to predict
the prompt production rate for $X(3872) \rightarrow J/\psi\, \pi^+
\pi^-$ as a function of $p_T$ for various LHC experiments. In the same
work, the production of the $X(3872)$ from $b$-hadron decays is
discussed. The data samples at the LHC are predicted to be large,
suggesting that the $X(3872)$ can be studied very effectively at the
LHC. Measurements of the prompt production rate at the LHC as a function
of $p_T$ would provide a key test of the NRQCD factorization approach to
$X(3872)$ hadroproduction.

\subsubsection{Quarkonium polarization: a key observable}
\label{prod_section:polarization}

Measurements of quarkonium polarization observables may yield
information about quarkonium production mechanisms that is not available
from the study of unpolarized cross sections alone.

The three polarization states of a $J=1$ quarkonium can be specified in
terms of a particular coordinate system in the rest frame of the
quarkonium. This coordinate system is often called the
``spin-quantization frame.'' In a hadron collider, the $J/\psi$,
$\psi(2S)$ and $\Upsilon$ resonances are reconstructed through their
electromagnetic decays into a lepton pair. The information about the
polarization of the quarkonium state is encoded in the angular
distribution of the leptons. This angular distribution is usually
described in the quarkonium rest frame with respect to a particular
spin-quantization frame. In that case, the angular distribution of the
quarkonium can be expressed in terms of three real parameters that are
related to the spin-polarization amplitudes of the $J=1$ quarkonium
state.

In hadron-hadron collisions, polarization analyses are often restricted
to the measurement of the distribution as a function of the polar angle
with respect to the chosen spin-quantization axis. This distribution is
parametrized as
\begin{equation}
1+\alpha \cos^2 \theta_{\ell\ell}. \label{prod_polar_asymmetry}
\end{equation}
The parameter $\alpha$ in Eq.~(\ref{prod_polar_asymmetry}) is directly
related to the fraction of the cross section that is longitudinal (or
transverse) with respect to the chosen spin-quantization axis: 
$\alpha=1$ corresponds to 100\% transverse polarization;
$\alpha=-1$ corresponds to 100\% longitudinal polarization.

In experimental analyses, knowledge of the angular distribution of
dileptons from quarkonium decay is important because, typically, detector
acceptances fall as dileptons are emitted more along the direction of
the quarkonium momentum---especially at small $p_T$. This effect is
included in the corrections to the experimental acceptance. However, it
induces systematic experimental uncertainties.

In theoretical calculations, polarization parameters, such as the polar
asymmetry $\alpha$, can be expressed in terms of ratios of polarized
quarkonium cross sections. In some cases, these ratios are less
sensitive than the production cross sections to the theoretical
uncertainties from quantities such as the factorization scale, the
renormalization scale, the heavy-quark mass, and the NRQCD long-distance 
matrix elements. For example, in the cases of the production of the
$J/\psi$ or the $\psi(2S)$ in the NRQCD factorization formalism, the
polarization parameter $\alpha$ depends, to good approximation, on
ratios of color-octet long-distance matrix elements, but not on their
magnitudes.

One should keep in mind that a measurement of the polar asymmetry
parameter $\alpha$ alone does not give complete information about the
polarization state of the produced quarkonium. The importance of
measuring all of the parameters of the dilepton angular distribution for
a variety of choices of the spin-quantization frame has been emphasized
in Refs.~\cite{Faccioli:2008dx,Faccioli:2010ji}. The significance of the
information that is obtained in measuring $\alpha$ alone depends very
much on the orientation of the spin-quantization axis. So far, most of
the theoretical studies of polarization in quarkonium production have
been carried out for the case in which the spin-quantization axis is
taken to be along the direction of the quarkonium momentum in the
laboratory frame
\cite{Artoisenet:2007xi,Artoisenet:2008fc,Lansberg:2008gk,Cho:1994ih,Leibovich:1996pa,Braaten:1999qk,Gong:2008sn,Gong:2008ft,Yuan:2000qe,Baranov:2002cf,Baranov:2007ay,Baranov:2007dw,Khoze:2004eu,Lansberg:2005pc,Haberzettl:2007kj,Li:2008ym,Lansberg:2009db}.
That choice of spin-quantization axis \cite{Cho:1994ih} is often
referred to as the ``helicity frame.'' In Ref.~\cite{Braaten:2008xg}, it
is shown that one can make more sophisticated choices of the
spin-quantization axis, which involve not only the kinematics of the
quarkonium state, but also the kinematics of other produced particles.
These alternative choices of spin-quantization axis can increase the
significance of the measurement of $\alpha$. However, their optimization
requires knowledge of the dominant quarkonium production mechanism.

Experimental measurements of quarkonium polarization have been made for
a variety of spin-quantization frames. Measurements by the
CDF \cite{Affolder:2000nn,Abulencia:2007us,Acosta:2001gv},
\DZero\ \cite{Abazov:2008za}, and
PHENIX \cite{Atomssa:2008dn,daSilva:2009yy,Adare:2009js} collaborations
were carried out in the helicity frame, while some measurements at
fixed-target experiments \cite{Brown:2000bz,Chang:2003rz} were carried
out in the Collins-Soper frame \cite{Collins:1977iv}. Recently, the
Hera-B collaboration has analyzed quarkonium polarizations
\cite{Abt:2009nu} not only in the helicity and Collins-Soper frames, but
also in the Gottfried-Jackson frame \cite{Gottfried:1964nx}, in which
the spin-quantization axis is along the direction of the incident beam.
In Ref.~\cite{Faccioli:2008dx}, a global analysis was made of
polarization measurements that were carried out in the Collins-Soper and
helicity frames. That analysis shows that the results that were obtained
in these two spin-quantization frames are plausibly compatible when the
experimental rapidity ranges are taken into account. However, it is
clear that additional analyses in different spin-quantization frames
would be very informative.

\begin{figure}[b]
\centering
\includegraphics[width=\figwid]{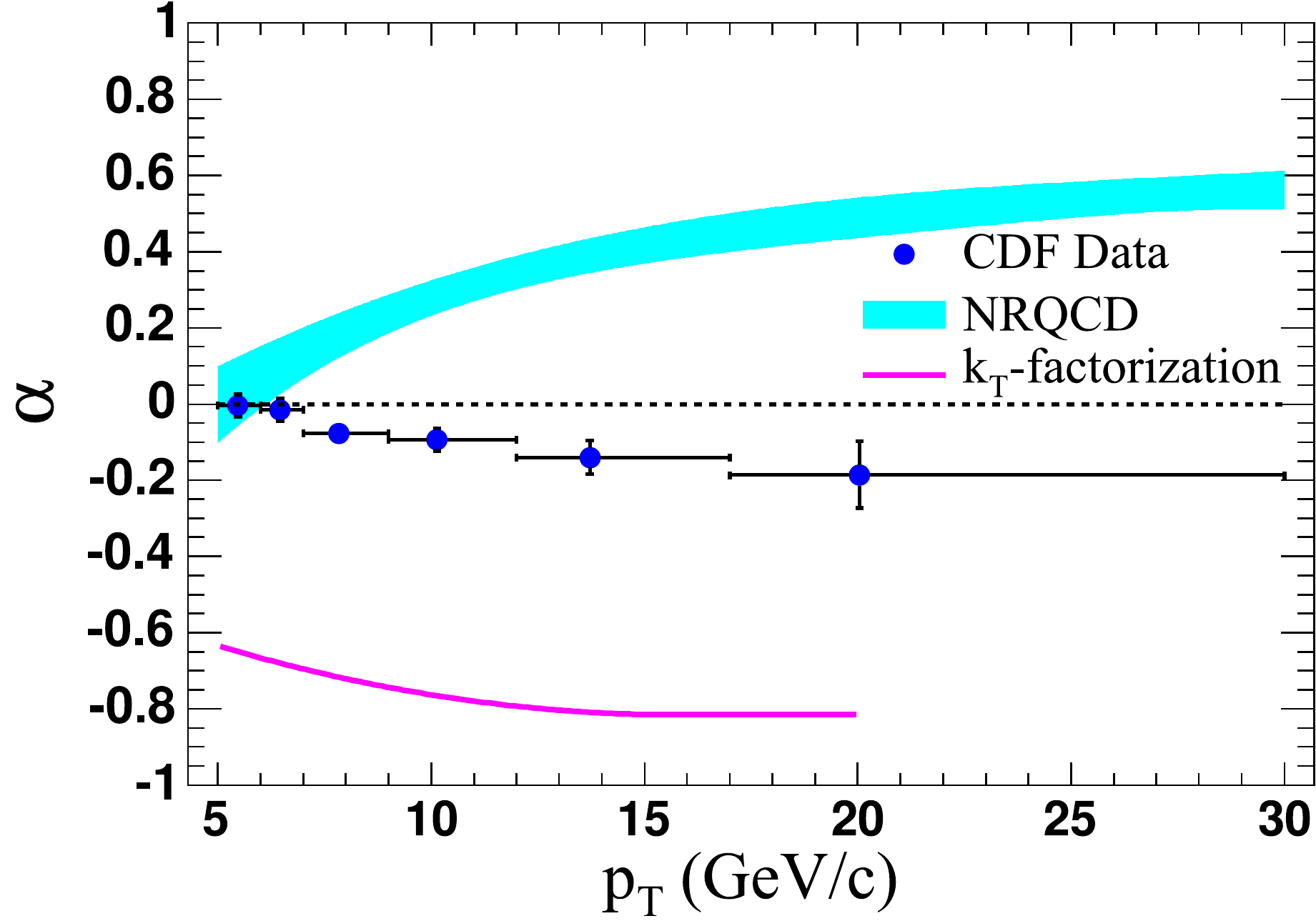}
\caption{The polarization parameter $\alpha$ for prompt $J/\psi$
production in $p\bar p$ collisions at $\sqrt{s}=1.96$ TeV as a function
of $p_T$. The {\it points} are the CDF data \cite{Abulencia:2007us}, the
{\it band} is the prediction from LO NRQCD factorization
\cite{Braaten:1999qk}, and the {\it line} is the prediction from $k_T$
factorization~\cite{Baranov:2002cf}. The theoretical
uncertainty in the LO NRQCD factorization prediction was obtained by
combining the uncertainties from the parton distributions (estimated by
comparing the MRST98LO (Ref.~\cite{Martin:1998sq}) and the CTEQ5L
(Ref.~\cite{Lai:1999wy}) distributions), the uncertainties from the
color-octet NRQCD long-distance matrix elements, the uncertainties that
are obtained by varying $m_c$ in the range $1.45~{\rm GeV}< m_c
<1.55~{\rm GeV}$, and the uncertainties that are obtained by varying the
factorization and renormalization scales in the range  $0.5 m_T
<\mu_f=\mu_r< 2 m_T$. Here $m_T=\sqrt{4m_c^2+p_T^2}$.
\figPermXAPS{Abulencia:2007us}{2007} }
\label{prod_polarizationJpsi}
\end{figure}

According to the CDF Run~II measurement of the $\psi$ polarization in
the helicity frame \cite{Abulencia:2007us}, the prompt $\psi$ yield
becomes increasingly longitudinal as $p_T$ increases. The disagreement
of this result with a previous CDF polarization measurement that was
based on Run~I data~\cite{Affolder:2000nn} has not been resolved. In
Fig.~\ref{prod_polarizationJpsi} the CDF measurement of the polarization
parameter $\alpha$ for the prompt $J/\psi$ production at the Tevatron in
Run~II is compared with the NRQCD factorization prediction at LO in
$\alpha_s$ (Ref.~\cite{Braaten:1999qk}). This prediction ignores
possible violations of the heavy-quark spin symmetry, which appear at
relative order $v^3$. The effects of feeddown from the $\psi(2S)$ and
the $\chi_{cJ}$ states are taken into account in the NRQCD factorization
prediction. However, it should be kept in mind that the
corrections at NLO in $\alpha_s$ to the $\chi_c$ production rate are
large \cite{Ma:2010vd} and are not taken into account in the NRQCD
prediction in Fig.~\ref{prod_polarizationJpsi}. The solid line in
Fig.~\ref{prod_polarizationJpsi} is the prediction from the $k_T$
factorization approach~\cite{Baranov:2002cf}, which includes only
color-singlet contributions.

At LO accuracy in $\als$, the NRQCD factorization prediction for the
$J/\psi$ polarization clearly disagrees with the observation of a very
small polar asymmetry in the helicity frame. One obvious issue is the
effect of corrections of higher order in $\als$ on the NRQCD
factorization prediction.

\begin{figure}[b] \centering
\includegraphics[width=\figwid]{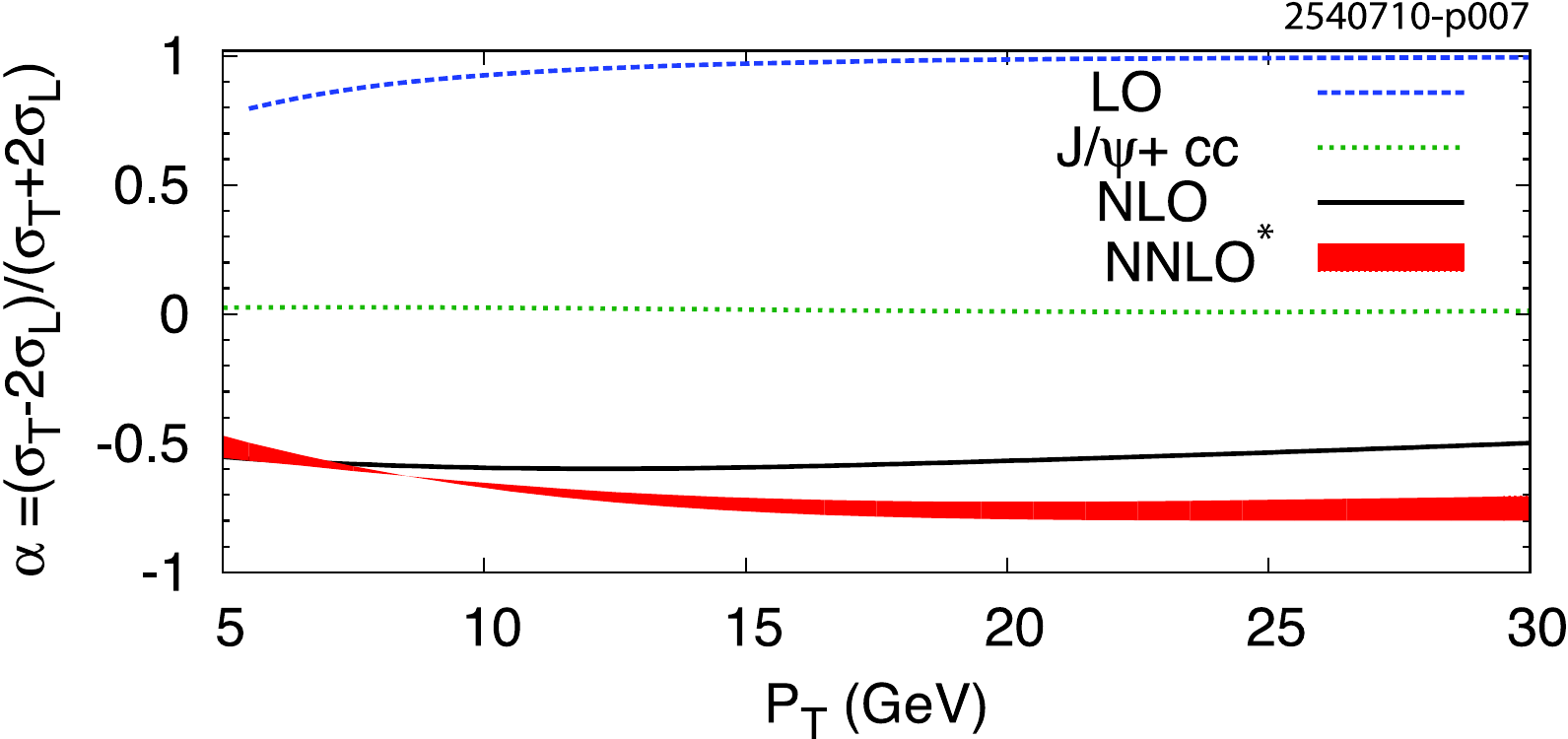}
\caption{Predictions for the polarization parameter $\alpha$ for
direct $J/\psi$ production in the color-singlet channel in $p \bar p$
collisions at the Tevatron at $\sqrt{s}=1.96$ TeV at LO, NLO and
NNLO$^\star$ accuracy \cite{Lansberg:2008gk}. Most of the uncertainties
in $\alpha$ for the LO, $J/\psi +c\bar c$, and NLO cases cancel. The
theoretical uncertainty band for the NNLO$^\star$ case was obtained by
varying the infrared cutoff $s^{\rm min}_{ij}$ between $2m_c^2$ and
$m_c^2/2$ }
\label{prod_polarizationCS}
\end{figure}

Corrections of higher order in $\als$ to $J/\psi$ production via the
color-singlet channel dramatically affect the polarization in that
channel. While the prediction at LO in $\als$ for the helicity of the
$J/\psi$ in the color-singlet channel is mainly transverse at medium and
large $p_T$, calculations at NLO or NNLO$^\star$ accuracy for the
color-singlet channel reveal a polarization that is increasingly
longitudinal as $p_T$ increases, as can be seen in
Fig.~\ref{prod_polarizationCS}. A similar trend for the polarization as
a function of $p_T$ is found in some other analyses of the color-singlet
channel, such those in the $k_T$ factorization approach
\cite{Yuan:2000qe,Baranov:2002cf,Baranov:2007ay,Baranov:2007dw} (see
Fig.~\ref{prod_polarizationJpsi}), the gluon-tower approach
\cite{Khoze:2004eu}, and the $s$-channel-$Q\bar Q$-cut 
approach \cite{Lansberg:2005pc,Haberzettl:2007kj}.

In the case of the color-octet $^3S_1$ channel, the NLO correction to
the helicity of the $J/\psi$ is very small \cite{Gong:2008ft}. This NLO
correction would not change substantially the comparison between the
NRQCD factorization prediction and the experimental data that is shown
in Fig.~\ref{prod_polarizationJpsi}. As we have explained in
\Sec{prod_section:pheno_with_QCD_corr}, the NLO analysis of
Ref.~\cite{Ma:2010yw} suggests the possibility that the $J/\psi$
direct-production cross section is dominated by the color-octet
${}^1S_0$ contribution, rather than by the color-octet ${}^3S_1$
contribution, even at the largest values of $p_T$ that are accessed in
the Tevatron measurements. However, the NLO analysis in
Ref.~\cite{Butenschoen:2010rq} concludes that the color-octet ${}^3S_1$
contribution at NLO is not very different from that at LO. A complete
NLO analysis of the direct $J/\psi$ polarization, including the
contribution of the color-octet ${}^3P_J$ channel is still lacking and
is an important theoretical goal. Further progress in determining the
relevant production mechanisms would be aided significantly by
high-statistics measurements of the polarizations of the $J/\psi$, the
$\chi_{cJ}$, and the $\psi(2S)$ in direct production.

\begin{figure}[tb]\centering
\includegraphics[width=\figwid]{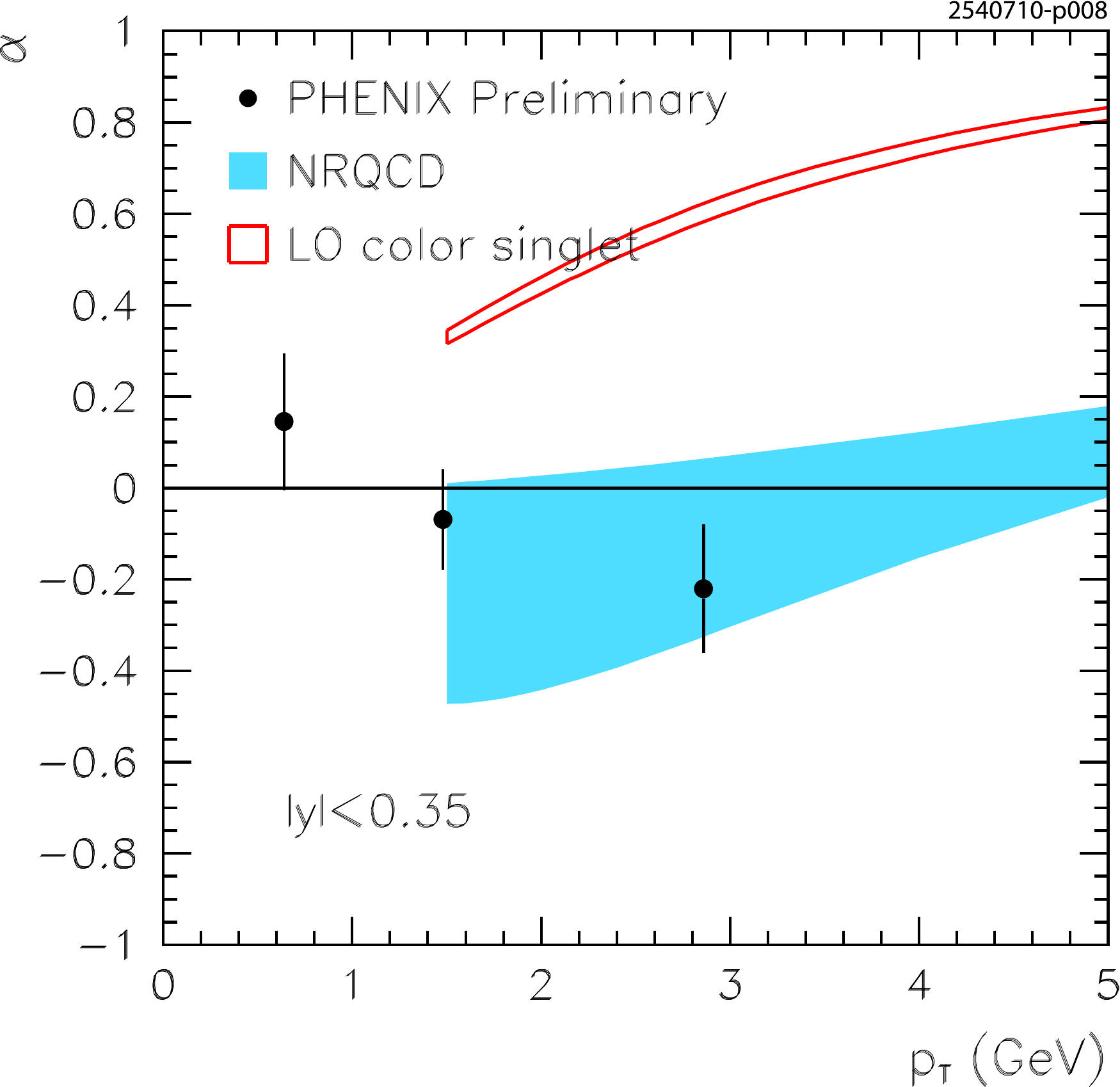}
\caption{Comparison of the LO NRQCD and LO CSM predictions for the prompt
$J/\psi$ polarization in $pp$ collisions at RHIC at $\sqrt{S_{nn}} =
200$~GeV and $|y|<0.35$ (Ref.~\cite{Chung:2009xr}) with the data from
the PHENIX collaboration \cite{daSilva:2009yy,Adare:2009js}. The
theoretical uncertainty bands were obtained by combining the
uncertainties from $m_c$ and the NRQCD long-distance matrix elements
with the uncertainties that are obtained by varying the renormalization
scale $\mu_f$ and the factorization scale $\mu_r$ between $2m_T$ and
$m_T/2$. Here $m_T=\sqrt{4m_c^2+p_T^2}$. 
 \figPermXAPS{Chung:2009xr}{2010} }
\label{prod_fig:rhic_pol} 
\end{figure}

\begin{figure}[t]\centering
\includegraphics[width=\figwid]{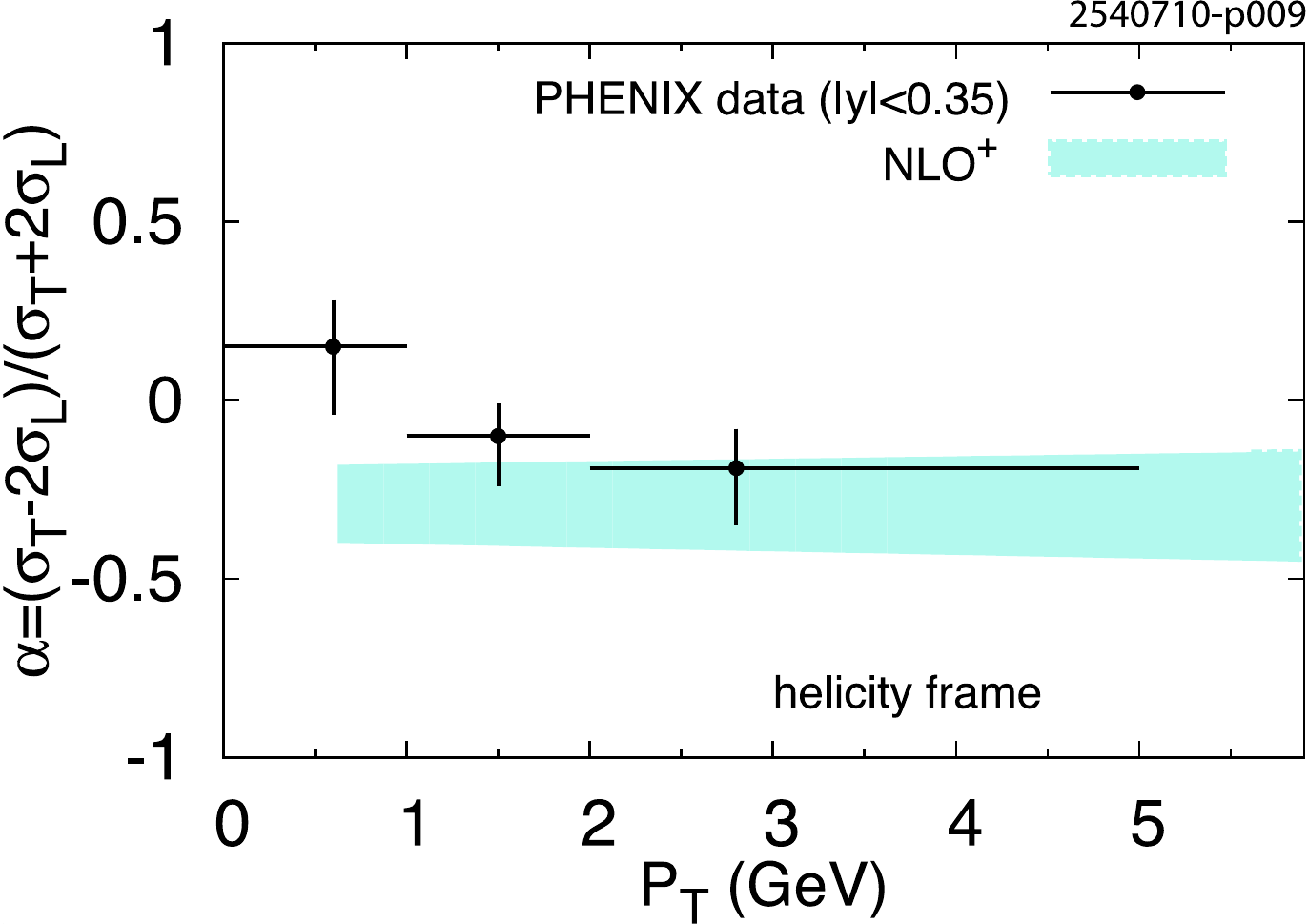}
\caption{Comparison of the NLO CSM calculation of $J/\psi$ polarization
in $pp$ collisions at RHIC at $\sqrt{S_{NN}} = 200$~GeV and $|y|<0.35$
(Ref.~\cite{Lansberg:2010vq}) with the prompt $J/\psi$ polarization data
from the PHENIX collaboration \cite{Atomssa:2008dn,Adare:2009js}. The
theoretical uncertainty band was obtained by combining the uncertainties
that are obtained by varying $m_c$ in the range $1.4~{\rm GeV}< m_c <
1.6~{\rm GeV}$, by varying the factorization scale $\mu_f$ and the
renormalization scale $\mu_r$ over the values
$((0.75,0.75);(1,1);(1,2);(2,1);(2,2))\times m_T$. Here
$m_T=\sqrt{4m_c^2+p_T^2}$. From Ref.~\cite{Lansberg:2010vq} }
\label{prod_fig:rhic_pol_NLO}
\end{figure}

The polarization of the $J/\psi$ has also been measured in hadronic
collisions at $\sqrt{s}=200$~GeV. Data from the PHENIX collaboration for
prompt $J/\psi$ polarization as a function of $p_T$
(Refs.~\cite{daSilva:2009yy,Adare:2009js}) exist in the range
$0<p_T<3$~GeV and indicate a polarization that is compatible with zero,
with a trend toward longitudinal polarization as $p_T$ increases.
Comparisons of the data with LO calculations in the CSM and the NRQCD
factorization formalism \cite{Chung:2009xr} are given in
Refs.~\cite{Adare:2009js,Chung:2009xr}. As can be seen from
Fig.~\ref{prod_fig:rhic_pol}, the data favor the NRQCD factorization
prediction and are in agreement with it. However, the small values of
$p_T$ involved may call into question the validity of the NRQCD
factorization formula. The NLO color-singlet contribution to the
$J/\psi$ polarization at RHIC has been computed in
Ref.~\cite{Lansberg:2010vq}. A comparison of this prediction with the
PHENIX prompt $J/\psi$ data differential in $p_T$
\cite{Atomssa:2008dn,Adare:2009js} is shown in
Fig.~\ref{prod_fig:rhic_pol_NLO}. As can be seen, the CSM contributions
to the polarization through NLO are in agreement with the PHENIX prompt
$J/\psi$ data.

\begin{figure}[t]
\centering
\includegraphics[width=\figwid]{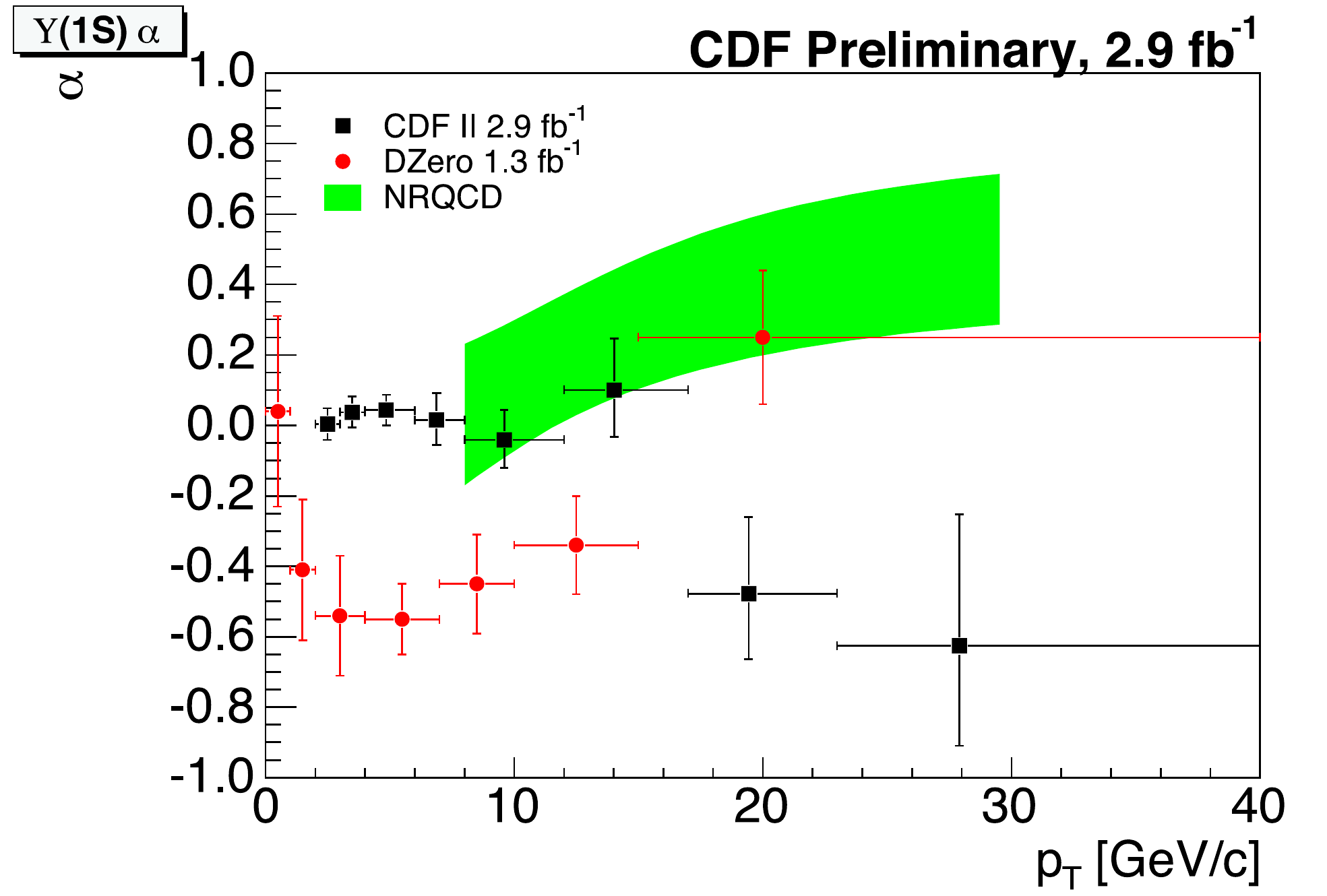}
\caption{The polarization parameter $\alpha$ in the helicity frame for
prompt $\Upsilon(1S)$ production in $p \bar p$ collisions at
$\sqrt{s}=1.96$~TeV. The NRQCD factorization prediction at LO in
$\alpha_s$ (Ref.~\cite{Braaten:2000gw}) is compared with the data of the
CDF collaboration \cite{CDF_public_note_9966} and the \DZero\
collaboration \cite{Abazov:2008za}. The CDF measurement was made over the
rapidity interval $|y|<0.6$, while the \DZero\ measurement was made
over the rapidity interval $|y|<1.8$. The NRQCD factorization prediction
in Ref.~\cite{Braaten:2000gw} was integrated over the range $|y|<0.4$
(Ref.~\cite{prod_J_Lee_private}) and includes feeddown from the
$\Upsilon(2S)$, $\Upsilon(3S)$, $\chi_b(1P)$, and $\chi_b(2P)$ states. 
The theoretical uncertainty band was obtained by combining the
uncertainties from the NRQCD long-distance color-singlet and color-octet
matrix elements, $m_b$, the parton distributions, and the quarkonium
branching fractions with uncertainties that are obtained by varying the
renormalization and factorization scales from $\mu_T/2$ to $2\mu_T$.
Here $\mu_T=\sqrt{m_b^2+p_T^2}$. Figure provided by Hee Sok Chung, using
Ref.~\cite{prod_cdf_upsilon_alpha}, which is based on the analysis of
Ref.~\cite{CDF_public_note_9966} }
\label{prod_fig:comp_polarization_D0_CDF}
\end{figure}

\begin{figure}[tb]
\centering
\includegraphics[width=\figwid]{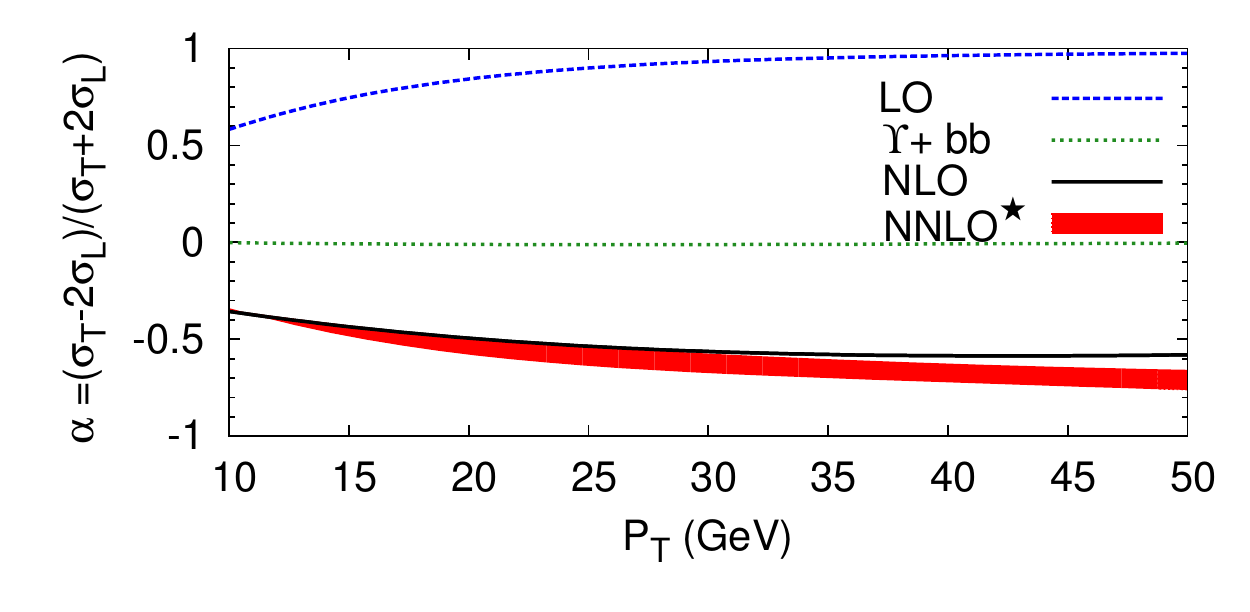}
\caption{Polarization parameter $\alpha$ for direct $\Upsilon$
production in the color-singlet channel in $p \bar p$ collisions at
$\sqrt{s}=1.96$ TeV at LO, NLO and NNLO$^\star$ accuracy. Most of the
uncertainties in $\alpha$ for the LO, $\Upsilon +b\bar b$, and NLO cases
cancel. The theoretical uncertainty band for the NNLO$^\star$ case was
obtained by varying the infrared cutoff $s^{\rm min}_{ij}$ between $2
m_b^2$ and $m_b^2/2$ and by varying $\mu_f$ and $\mu_r$ between $m_T/2$
and $2m_T$. Here $m_T=\sqrt{4m_c^2+p_T^2}$. The variations with respect
to $\mu_f$ and $\mu_r$ are negligible in comparison with the variation
with respect to $s^{\rm min}_{ij}$. \figPermXAPS{Artoisenet:2008fc}{2008} 
}
\label{prod_fig:alpha_Upsilon}
\end{figure}

The polarization of the $\Upsilon(1S)$ in prompt production has been
measured by both the CDF collaboration (Run~I) \cite{Acosta:2001gv} and
by the \DZero\ collaboration (Run~II) \cite{Abazov:2008za}. The
\DZero\ measurement has substantially larger experimental
uncertainties than the CDF measurement. The CDF collaboration has
recently reported a new preliminary measurement of the polarization of
prompt $\Upsilon$'s that is based on a larger data set from Run~II
\cite{CDF_public_note_9966}. This Run~II measurement is consistent with
the CDF Run~I measurement. The results of the CDF Run~II and \DZero\
measurements are shown in Fig.~\ref{prod_fig:comp_polarization_D0_CDF},
along with the NRQCD factorization prediction at LO in $\alpha_s$
(Ref.~\cite{Braaten:2000gw}). The origin of the large discrepancy
between the CDF and \DZero\ data is unclear. However, we note that
the CDF measurement was made over the rapidity interval $|y|<0.6$, while
the \DZero\ measurement was made over the rapidity interval
$|y|<1.8$. The NRQCD factorization prediction in
Ref.~\cite{Braaten:2000gw} was integrated over the range $|y|<0.4$
(Ref.~\cite{prod_J_Lee_private}). The LO NRQCD factorization prediction
is marginally compatible with the CDF data at medium $p_T$ and
incompatible with the CDF data at large $p_T$, while the LO NRQCD
factorization prediction is incompatible with the \DZero\ data at
medium $p_T$ and compatible with the \DZero\ data at large $p_T$.
The effects of higher-order QCD corrections on the polarization of
direct $\Upsilon$'s produced via the color-singlet channel are shown in
Fig.~\ref{prod_fig:alpha_Upsilon}. The higher-order corrections in this
case have the same qualitative features as in the case of $\psi$: The
higher-order corrections change the polarization from nearly 100\%
transverse to substantially longitudinal. A complete computation of the
prompt $\Upsilon$ polarization at NLO in the NRQCD factorization,
including feeddown from $\chi_{bJ}$ states, is not yet available.

\subsubsection{New observables in hadroproduction}
\label{prod_section:newobservables}

It may be useful, in order to progress in the understanding of the
mechanisms that are responsible for heavy-quarkonium production, to
identify, to compute, and to measure new observables.

As we have seen from previous discussions, the $\psi$ and $\Upsilon$
production rates in hadron-hadron collisions, differential in $p_T$, are
complicated to calculate because the dominant color-singlet channels at
large $p_T$ arise beyond LO in $\alpha_s$. In the case of the
$\Upsilon$, higher-order corrections bring the color-singlet
contribution close to the Tevatron data. However, in the case of $\psi$,
there is a gap, which increases with increasing $p_T$, between the
higher-order color-singlet contributions and the data. In both cases, the
uncertainties are very large and, in the $\Upsilon$ case, they are too
large to make a definite statement about the relative proportions of
color-singlet and color-octet contributions. However, it is worth noting
that the uncertainties in the higher-order color-singlet contributions
affect the normalization of the differential rates much more than the
shape. This suggests the use of observables that do not depend on the
total rate. One interesting proposal is to compare ratios of
differential cross sections at different values of $\sqrt{s}$, the
center-of-mass energy of the colliding hadrons~\cite{Bodwin:bw}. For a
given color channel, this ratio can be predicted quite accurately. If
the ratios that are associated with the color-singlet and 
color-octet channels are sufficiently  well separated, then the relative
sizes of the color-singlet and color-octet contributions could be
extracted by comparing the measured production rate at the Tevatron with
the measured production rate at the LHC.

One interesting observable is the hadronic activity near the
quarkonium direction \cite{Kraan:2008hb} or, more generally, the
$J/\psi$-hadron azimuthal correlation. The UA1 collaboration compared
their charged-track distributions with Monte Carlo simulations for a
$J/\psi$ produced in the decay of a $b$ hadron and a $J/\psi$ produced
in feeddown from a $\chi_{cJ}$ state
\cite{Albajar:1987ke,Albajar:1990hf}. At the time of the UA1 study,
$\chi_c$ feeddown was still expected to be the dominant source of prompt
$J/\psi$'s. In either the NRQCD factorization formalism or the CSM at
higher orders in $\alpha_s$, it is expected that the production process for
prompt $J/\psi$'s is more complex than the $\chi_{cJ}$-feeddown process
alone. Recently, the STAR collaboration reported the first measurement of
the $J/\psi$-hadron azimuthal correlation at RHIC
\cite{Abelev:2009qaa,Tang:2010zza}. The STAR collaboration compared its
measurement with up-to-date LO PYTHIA predictions\footnote{Note that one
expects such predictions to be affected by channels that appear beyond
LO in $\alpha_s$.} and found no significant hadronic yield in the
direction of the $J/\psi$ beyond that which is expected from the LO
PYTHIA predictions. Observation of the hadronic activity around the
quarkonium might help to disentangle color-octet contributions from
color-singlet contributions to the inclusive production process. One
expects additional hadronic activity around the quarkonium in
color-octet production. However, in practice, it may be difficult to
identify this hadronic activity \cite{Kraan:2008hb} because there
are competing effects, such as the suppression of collinear gluon
radiation from massive particles (dead-cone effect) and the suppression
of soft gluon radiation from color-singlet objects, both of which would be
difficult to compute reliably.

It would be useful to identify additional observables that can be
computed reliably in order to test the many production models that are
available \cite{Brambilla:2004wf,Lansberg:2006dh}. One such observable
could be the rate of production of heavy-flavor mesons in association
with a quarkonium. Final states that could be studied include $\psi + c
\bar c$ and $\Upsilon + b \bar b$. Associated production could be
investigated first in $pp$ collisions and subsequently in $pA$ and $AA$
collisions. The study of associated production in hadron collisions is
motivated by measurements that were carried out at the $B$~factories
that show that, in a surprisingly large fraction of $J/\psi$ events, a
second $c \bar c$ pair is produced. (See
\Sec{prod_sec:inclusive-e+e-}.) It is not yet known whether such a
large fraction of $J/\psi+c\bar c$ events occurs in hadroproduction.
Analyses at the Tevatron and at RHIC are already possible. The LO
prediction for associated production at the Tevatron at
$\sqrt{s}=1.96$~TeV has been computed in Ref.~\cite{Artoisenet:2007xi}
and shows that the integrated cross sections are significant:
\begin{eqnarray}
\sigma(\jpsi +c \bar c) \times \Brat(\jpsi\to \mu^+\mu^-) &\approx& 
1\hbox{~nb};\nonumber\\
\sigma(\Ups +b \bar b)\times \Brat(\Ups\to \mu^+\mu^-) &\approx&  
1\hbox{~pb}.
\end{eqnarray}
In order to illustrate the measurement potential at RHIC, the author of
Ref.~\cite{Lansberg:2008gk} computed the differential cross sections as
a function of $p_T$ and found them to be on the order of 1~pb/GeV at
$p_T=5$~GeV for the STAR kinematics.
Measurements of such processes would provide tests of the NRQCD
factorization formalism. They would also provide information about the
color-transfer mechanism \cite{Nayak:2007mb,Nayak:2007zb} (see
\Sec{prod_sec:NRQCD-difficulties}), which involves soft-gluon exchanges
between comoving heavy particles and is known to violate standard NRQCD
factorization. In this case, it would be useful to compare heavy-flavor 
activity near the quarkonium direction and away from the quarkonium 
direction.

A new observable that could be measured in existing and future
experiments is the rate of production of a photon in association with a
$J/\psi$ or an $\Upsilon$. The QCD corrections to the rates for these
processes have been computed recently at NLO \cite{Li:2008ym} and
NNLO$^\star$ \cite{Lansberg:2009db} accuracy. As is argued in
Ref.~\cite{Lansberg:2009db},  a measurement of such processes would
provide information on the quarkonium production mechanisms that is
complementary to that which is provided by measurements of inclusive
quarkonium production.

In order to facilitate phenomenological studies, an automated tree-level
amplitude generator MadOnia \cite{Artoisenet:2007qm} has been developed
for processes involving quarkonium production or decay. It is now
embedded in the online version of MadGraph/MadEvent \cite{Alwall:2007st}
and, thus, is publicly available. A  number of studies
\cite{Artoisenet:2007xi,Artoisenet:2008fc,Brodsky:2009cf,Lansberg:2009db,Artoisenet:2008tc,Klasen:2008mh}
have already taken advantage of the flexibility of MadOnia and of the
possibility to interface it with showering and hadronization programs.

\subsubsection{Future opportunities}
\label{prod_sec:hadroproduction-fo}%

The results of the past few years, on both the theoretical and
experimental fronts, have yielded important clues as to the mechanisms
that are at work in inclusive quarkonium hadroproduction. In general,
however, theoretical uncertainties remain too large to draw any definite
conclusions about the production mechanisms.

Regarding these uncertainties, one of the key issues is that, in
color-singlet production channels, the mechanisms that are dominant at
large $p_T$ appear only at higher orders in $\alpha_s$. (See
\Secs{prod_section:higherorderchannels} and
\ref{prod_section:pheno_with_QCD_corr}.) In addition to the new complete
calculations of NLO contributions, estimates have been made of the NNLO
contributions to $\psi$ and $\Upsilon$ production in the color-singlet
channels in several different frameworks (the fragmentation
approximation, the $k_T$ factorization approach, the gluon-tower model,
and the NNLO$^\star$ approach). A more accurate treatment of
higher-order corrections to the color-singlet contributions at the
Tevatron and the LHC is urgently needed. Here, the re-organization of
the perturbation series that is provided by the fragmentation-function
approach (\Sec{prod_sec:fragmentation}) may be an important tool.

Furthermore, the current theoretical predictions suffer from
uncertainties that are related to the long-distance dynamics that is
involved in quarkonium production. For example, the prediction of the
$\psi$ or $\Upsilon$ polarization relies on the approximate heavy-quark
spin symmetry of NRQCD. This approximate symmetry is based on the
application of the velocity-scaling rules of NRQCD to evaluate
the order of suppression of the spin-flip contribution. In the case of
inclusive quarkonium decays, calculations of the NRQCD long-distance 
matrix elements on the lattice \cite{Bodwin:2005gg} have constrained the
size of the spin-flip contribution. A similar constraint in case of
inclusive quarkonium production would obviously be very valuable.

More generally, lattice determinations of the NRQCD long-distance
production matrix elements would provide very useful constraints on the
theoretical predictions and would also serve to check the
phenomenological determinations of the long-distance matrix elements. An
outstanding theoretical challenge is the development of methods for
carrying out such lattice calculations, which are, at present, stymied
by fundamental issues regarding the correct lattice formulation of
single-particle inclusive rates in Euclidean space.

Further light could be shed on the NRQCD velocity expansion and its
implications for low-energy dynamics by comparing charmonium
and bottomonium production. The heavy-quark velocity $v$ is
much smaller in bottomonium systems than in charmonium systems. Hence,
the velocity expansion is expected to converge more rapidly for
bottomonium systems than for charmonium systems. In particular,
spin-flip effects and color-octet contributions are expected to be
smaller in bottomonium systems than in charmonium systems. The NRQCD
factorization formula for inclusive quarkonium production, if it is
correct, becomes more accurate as $p_T$ increases and probably holds
only for values of $p_T$ that are greater than the heavy-quark mass.
Therefore, the high-$p_T$ reach of the LHC may be crucial in studying
bottomonium production.

There are many unresolved theoretical issues at present that bear on
the reliability of predictions for prompt $J/\psi$ and $\Upsilon$
production. These issues may affect predictions for both the direct
production of the $J/\psi$ and the $\Upsilon$ and the production of the
higher-mass quarkonium states that feed down into the $J/\psi$ and the
$\Upsilon$. Therefore, it would be of considerable help in
disentangling the theoretical issues in $J/\psi$ and $\Upsilon$
production if experimental measurements could separate direct production
of the $J/\psi$ and the $\Upsilon$ from production via feeddown from
higher-mass charmonium and bottomonium states. Ideally, the direct
production cross sections and polarizations would both be measured
differentially in $p_T$. Measurement of the direct $J/\psi$ cross
section and polarization might be particularly important at large $p_T$,
given the large proportion of $\chi_{cJ}$ feeddown events in the prompt
$J/\psi$ rate at large $p_T$ that is predicted in Ref.~\cite{Ma:2010vd}.

Although it would be ideal to have measurements of direct quarkonium
production rates and polarizations, it is, of course, very important to
resolve the existing discrepancy between the CDF and \DZero\
measurements of the prompt $\Upsilon$ polarization. The CDF measurement
of the $\Upsilon(1S)$ polarization is for the rapidity range $|y|<0.6$,
while the \DZero\ measurement is for the rapidity range $|y|<1.8$.
It would be very useful for the two experiments to provide polarization
measurements that cover the same rapidity range.

It might also be useful to formulate new measurements and observables
that would provide information that is complementary to that which is
provided by the differential rates and polarization observables. The
large rates for $J/\psi$ and $\Upsilon$ production that are expected at
the LHC open the door to new analyses. As we have mentioned in
\Sec{prod_section:newobservables}, the possibilities include
studies of quarkonium production at different values of $\sqrt{s}$,
studies of hadronic energy near and away from the quarkonium direction,
and studies of the production of heavy-flavor mesons in association with
a quarkonium. It is important in all of these studies to identify
observables that are accessible under realistic experimental conditions
and that can be calculated accurately enough to allow meaningful
comparisons with experimental measurements. In this endeavor,
communication between the experimental and theoretical experts in these
areas will be crucial.

\subsection{$ep$ collisions}
\label{prod_sec:ep}%

Inelastic production of charmonia in $ep$ collisions at HERA proceeds
via photon-gluon fusion: A photon emitted from the incoming electron or
positron interacts with a gluon from the proton to produce a $c \bar c$
pair that evolves into a color-neutral charmonium state by the radiation
of soft and/or hard gluons.

The elasticity observable $z$ is defined as the fraction of energy of
the incoming photon, in the proton rest frame, that is carried by the
final-state charmonium. The kinematic region of inelastic charmonium
production is $0.05 \lesssim z < 0.9$. In the so-called ``photoproduction
regime,'' at low photon virtuality $Q^2$, the incoming electron is
scattered through a small angle, and the incoming photon is quasi real.
The invariant mass of the $\gamma p$ system $W_{\gamma p}$ depends on the
energy of the incoming photon. In photoproduction, photons can interact
directly with the charm quark (direct processes), or via their hadronic
component (resolved processes). Resolved processes are relevant at low
elasticities ($z \lesssim 0.3$). HERA has been a unique laboratory for the
observation of photoproduction in the photon-proton center-of-mass range
$20 < W_{\gamma p} < 320$~GeV.

In $ep$ scattering at HERA, toward high values of elasticity ($z > 0.95
$), another production mechanism that is distinctly different from
boson-gluon fusion becomes dominant.  In this mechanism, which applies
both to exclusive and diffractive charmonium production, the incoming
photon fluctuates into a $c\bar{c}$ QCD dipole state which, subsequently
interacts with the proton by the exchange of two or more gluons in a
colorless state. This colorless interaction transfers momentum that
allows the $c\bar{c}$ pair to form a bound quarkonium state. Experiments
distinguish between two categories of diffractive processes. In elastic
processes, the proton stays intact, {\it i.e.}, $\gamma p \rightarrow
J/\psi p$ ($z \approx 1$). In proton-dissociative processes, the proton
breaks up into a low-mass final state, {\it i.e.}, $\gamma p \rightarrow
J/\psi Y$. In proton-dissociative processes, $m_Y$, mass of the state
$Y$, is less than about $2$~GeV or $z$ lies in the $0.95 \lesssim z
\lesssim 1$. Many measurements of the diffractive production of the
$\rho^0$, $\omega$, $\phi$, $J/\psi$, $\psi(2S)$ and $\Upsilon$ states
have been performed at HERA
\cite{Aaron:2009xp,Kananov:2008zz,Chekanov:2005cqa,Breitweg:2000mu,Aktas:2005xu,Chekanov:2004mw,Chekanov:2002xi,Aktas:2003zi,Chekanov:2009tu,Adloff:2002re,Adloff:2000vm,Chekanov:2009zz}.\footnote{Here,
only the most recent measurements are cited.} These measurements were
crucial in reaching a new understanding of the partonic structure of
hard diffraction, the distributions of partons in the proton, the
validity of evolution equations in the low-$x$ limit, and the
interplay between soft and hard QCD scales. For a detailed report on
diffractive quarkonium production at HERA, we refer the reader to
Ref.~\cite{Ivanov:2004ax}. Possible future opportunities involving
measurements of diffractive quarkonium production 
are discussed in \Sec{sec:Fut_photoprod}.

The ZEUS and H1 collaborations have published several measurements of
inelastic $J/\psi$ and $\psi(2S)$ production that are based on data from
HERA Run~I \cite{Adloff:2002ex,Chekanov:2002at}. A new measurement,
making use of the full Run~II data sample, was published recently by the
H1 collaboration \cite{Aaron:2010gz}. The ZEUS collaboration has
published a new measurement of the $J/\psi$ decay angular distributions
in inelastic photoproduction, making use of the collaboration's full data
sample \cite{Chekanov:2009br}.

The data samples of $J/\psi$ events that result from the experimental
selection cuts are dominated by inelastic production processes in which
the $J/\psi$'s do not originate from the decay of a heavier resonance.
Sub-dominant diffractive backgrounds, as well as feeddown contributions
from the $\psi(2S)$, the $\chi_{cJ}$, and $b$-flavored hadrons are
estimated to contribute between 15\% and 25\% of the total $J/\psi$
events, depending on the kinematic region. These backgrounds are usually
neglected in theoretical predictions of direct $J/\psi$ production rates.
 
The measurements of $J/\psi$ cross sections and polarization parameters
reported by the ZEUS and H1 collaborations have been 
compared extensively to NRQCD factorization predictions at LO in $\alpha_s$. 
In these studies, a truncation of the NRQCD velocity expansion is used, 
in which the independent long-distance matrix elements
are $\langle \mathcal{O}_1^{J/\psi} \left(^3S_1 \right)   \rangle$,
$\langle \mathcal{O}_8^{J/\psi} \left(^3S_8 \right)   \rangle$,
$\langle \mathcal{O}_8^{J/\psi} \left(^1S_0 \right)   \rangle$ and
$\langle \mathcal{O}_8^{J/\psi} \left(^3P_0 \right)   \rangle$.
In the CSM, all of these matrix elements, except for the first one, are,
in effect, set to zero. Usually the values of the long-distance
color-octet matrix elements are extracted from the Tevatron data, in
which case the comparisons of the resulting predictions for $J/\psi$
production at HERA with the data offer the opportunity to assess the
universality of the long-distance matrix elements. The comparisons
between the predictions at LO in $\alpha_s$ and the data are summarized in
Ref.~\cite{Brambilla:2004wf}. 

In calculations in the NRQCD factorization formalism, large
uncertainties arise from the sensitivity to the input parameters: the
mass of the charm quark, the factorization and renormalization scales,
and the values of the color-octet matrix elements, which are obtained
from fits to the Tevatron data. One also expects sizable uncertainties
owing to the omission of corrections of higher-order in both $\alpha_s$
and $v$. As we shall see, because of these theoretical uncertainties,
the relative sizes of the color-singlet and color-octet contributions in
charmonium photoproduction are still unclear.

In fixed-order, tree-level predictions of the  color-octet contribution
to photoproduction, a large peak appears in the $z$ distribution near the
kinematic endpoint $z = 1$. This feature, which was first interpreted as
a failure of the universality of the long-distance matrix elements, has
since been attributed to the breakdown of the NRQCD velocity expansion
and the perturbation expansion in $\alpha_s$ near $z=1$.  In this
region, in order to obtain a reliable theoretical prediction, one must
resum large perturbative corrections to all orders in $\alpha_s$ and
large nonperturbative corrections to all orders in $v$. It is known that
the resummation of the color-octet contribution leads to a significant
broadening of the peak at large $z$~\cite{Fleming:2006cd}. However, the
effects of the nonperturbative resummation of the velocity expansion
cannot be determined precisely without further information about the
so-called ``shape function.'' In principle, that information could be
extracted from data on charmonium production in $e^+e^-$ collisions.

$J/\psi$ photoproduction at HERA has also been studied in the
$k_T$-factorization scheme \cite{Baranov:2002cf,Kniehl:2006sk}. In this
framework, it has been argued that the color-singlet contribution alone
can explain the HERA data. A prediction for the color-singlet yield at
LO in $\alpha_s$ in the $k_T$-factorization approach reproduces the
measured shapes of the $p_T$ and $z$ distributions reasonably well.
However, it should be kept in mind that these predictions rely on
unintegrated parton distributions, which are not well known at present.
This uncertainty might be reduced as more accurate unintegrated parton
distributions sets become available \cite{Martin:2009ii}.

The pioneering calculation of the correction of NLO in $\alpha_s$ to the 
color-singlet contribution to the 
direct $J/\psi$ cross section was presented in 
Ref.~\cite{Kramer:1994zi}. The NLO correction 
affects not only the normalization of the photoproduction rate, but also 
the shape of the $p_T$ distribution. Both of these effects bring the 
color-singlet contribution into better agreement with the data. The large 
impact of the NLO correction at large transverse momentum can be 
understood in terms of the kinematic enhancement of NLO color-singlet 
production processes relative to the LO color-singlet production 
processes. This effect is similar to the one that appears in 
hadroproduction in the color-singlet  channel. (See 
\Sec{prod_section:higherorderchannels}.)

The NLO calculation of Ref.~\cite{Kramer:1994zi} suggests that
production in the color-singlet channel might be the main mechanism at
work in $J/\psi$ photoproduction. However, in more recent work
\cite{Artoisenet:2009xh,Chang:2009uj,Butenschon:2009zz,Butenschoen:2009zy,Butenschoen:2010iy},
which confirms the calculation of Ref.~\cite{Kramer:1994zi}, it has been
emphasized that the factorization and renormalization scales in
Ref.~\cite{Kramer:1994zi} have been set a value ($m_c/\sqrt{2}$) that
is generally considered to be too low in the region of large $p_T$. As
we shall see, a more physical choice of scale, such as
$\sqrt{4m_c^2+p_T^2}$, leads to predictions for cross sections
differential in $p_T^2$ or in $z$ that lie considerably below the
H1 and ZEUS measurements. Hence, the size of the NLO color-singlet
contribution does not exclude the possibility that other contributions,
such as those from the color-octet channel, are at least as large as the
color-singlet contribution. In the region of low transverse
momentum, which is the dominant region for $J/\psi$ production at HERA,
the sensitivity to the factorization and renormalization scales is very
large and complicates the identification of the dominant production
mechanism at HERA.

\subsubsection{Phenomenology of the cross section, including NLO 
corrections}

\begin{figure}
\includegraphics[width=\figwid]{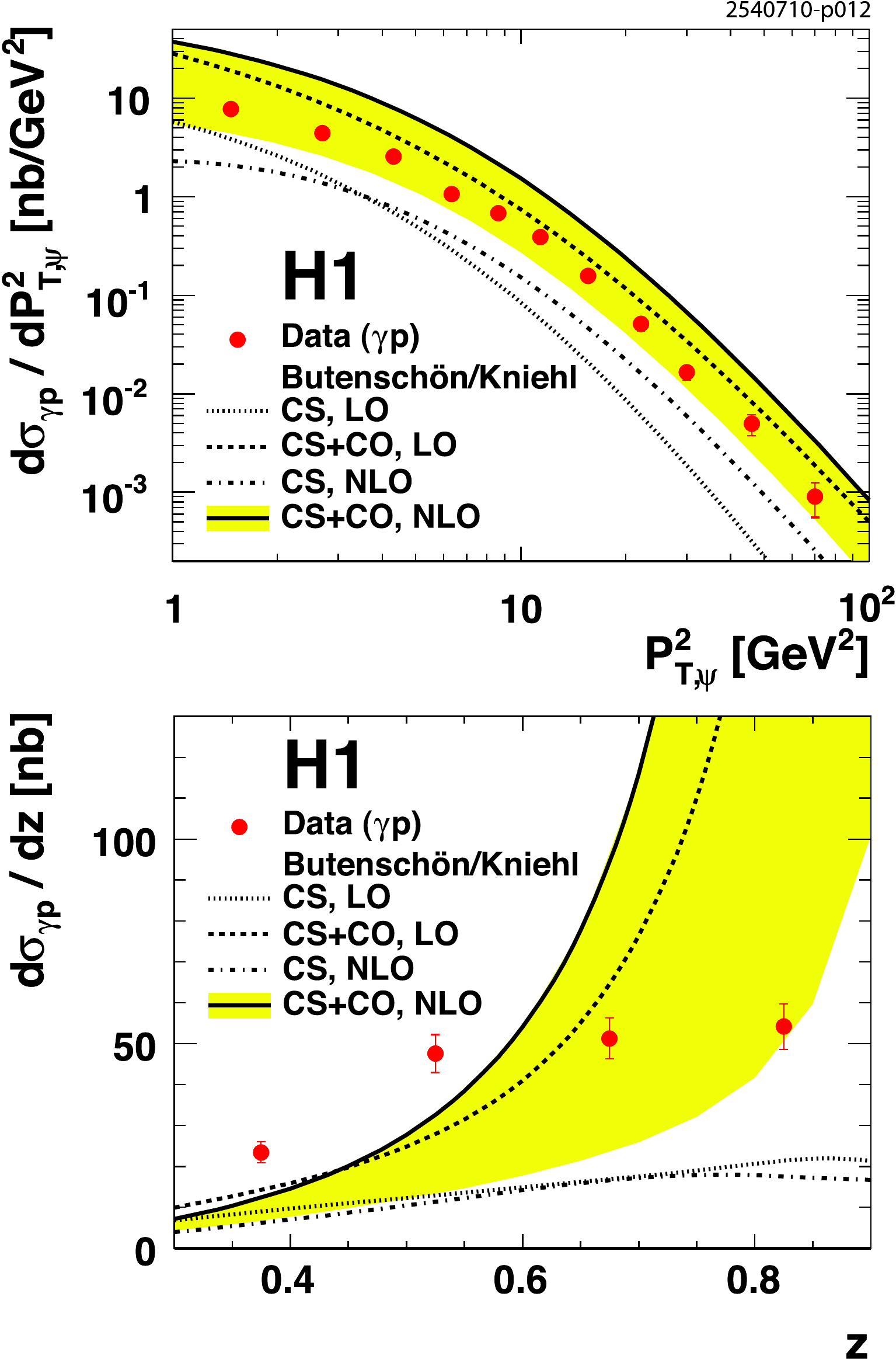}
\caption{Cross sections differential in $p_T^2$ and in $z$ for $J/\psi$
photoproduction at HERA. The measurement by the H1 collaboration
\cite{Aaron:2010gz} is compared to the CSM and NRQCD predictions
at LO and NLO in $\alpha_s$ from
Refs.~\cite{Butenschon:2009zz,Butenschoen:2009zy}. ``CS'' and ``CO''
denote the color-singlet and color-octet contributions, respectively.
The {\it dashed line} depicts the central values of the complete NRQCD
factorization prediction (including color-singlet and color-octet
contributions) at NLO in $\alpha_s$, and the {\it band} shows the uncertainty
in that prediction that arises from the uncertainties in the color-octet
long-distance NRQCD matrix elements. From
Ref.~\cite{Aaron:2010gz} }
\label{prod_photoproduction_diff_crossX}
\end{figure}

In Fig.~\ref{prod_photoproduction_diff_crossX} we show a comparison
between data from the H1 collaboration \cite{Aaron:2010gz} for
the $J/\psi$ photoproduction cross sections  differential in $p_T^2$ and
in $z$  and calculations in the NRQCD factorization formalism from
Refs.~\cite{Butenschon:2009zz,Butenschoen:2009zy,Butenschoen:2010iy}.\footnote{A
more detailed comparison of the H1 data with theory predictions can be
found in Ref.~\cite{Aaron:2010gz}.} The dashed line depicts the
central values of the complete NRQCD factorization prediction (including
color-singlet and color-octet contributions) at NLO in $\alpha_s$, and
the band shows the uncertainty in that prediction that arises from the
uncertainties in the color-octet long-distance NRQCD matrix elements.
Note that the contributions from resolved photoproduction, which are
important in the low-$z$ region, and the contributions from diffractive
production, which are important near $z=1$, are not included in the
NRQCD factorization prediction. The uncertainties that are shown arise
from the uncertainties in the NRQCD color-octet long-distance matrix
elements. These matrix elements were obtained through a fit to the
Tevatron hadroproduction data that used the NRQCD prediction at LO in
$\alpha_s$, augmented by an approximate calculation of some higher-order
corrections from multiple-gluon radiation \cite{Kniehl:1998qy}.

The NRQCD factorization prediction at NLO accuracy in $\alpha_s$ is
in better agreement with the H1 data than the color-singlet contribution
alone. However, it should be kept in mind that the mass and scale
uncertainties of the color-singlet contribution have not been displayed
here. As we have already mentioned, these uncertainties are large, even
at NLO in $\alpha_s$.

The NRQCD factorization prediction for the cross section differential in
$z$ shows a rise near $z=1$ that is characteristic of the color-octet
contributions. As we have mentioned, resummations of the series in
$\alpha_s$ and in $v$ are needed in order to obtain a reliable
theoretical prediction in this region. In the low-$z$ region, the NRQCD
factorization prediction undershoots the data. In this region, the
corrections to resolved photoproduction through NLO in $\alpha_s$ may be
needed in order to bring the theory into agreement with the data.

\subsubsection{Polarization}

In addition to the studies of $J/\psi$ differential cross sections that
we have mentioned, there have also been recent analyses of polarization
in $J/\psi$ photoproduction. The polarization observables may provide
additional information about the production mechanisms.

Experimentally, the $J/\psi$ polarization is extracted from the angular
distribution of the leptons that originate in $J/\psi$ decays. In
the $J/\psi$ rest frame, the distribution takes the general form
\begin{eqnarray}
\frac{d \Gamma(J/\psi \rightarrow l^+l^-)}{d\Omega}   & \propto &   1+\lambda \cos^2  \nonumber  \theta +  \mu \sin 2\theta \cos\phi \nonumber \\
& & + \frac{\nu}{2} \sin^2\theta \cos 2\phi,
\end{eqnarray}
where $\theta$ and $\phi$ are the polar and azimuthal angles of the
$l^+$ three-momentum with respect to a particular spin-quantization
frame. (See \Sec{prod_section:polarization}.) The ZEUS
collaboration has published a new measurement of the parameters
$\lambda$ and $\nu$ in the target spin-quantization frame
\cite{Beneke:1998re} that is based on an integrated luminosity of
$468$~pb$^{-1}$ \cite{Chekanov:2009br}. The H1 collaboration has
published new measurements of the parameters $\lambda$ and $\nu$ in both
the helicity and the Collins-Soper spin-quantization frames that are
based on an integrated luminosity of $165$~pb$^{-1}$
\cite{Aaron:2010gz}. The H1 collaboration uses a more restricted
range in the energy fraction $z$, namely, $0.3< z<0.9$, in order to
suppress possible contributions from diffractive or feeddown processes.
In both experiments, the polarization parameters in each bin are
extracted by comparing the data with Monte Carlo distributions for
different values of the polarization parameters, using a $\chi^2$
criterion to assess the probability of each distribution.

\begin{figure}
\includegraphics[width=\figwid]{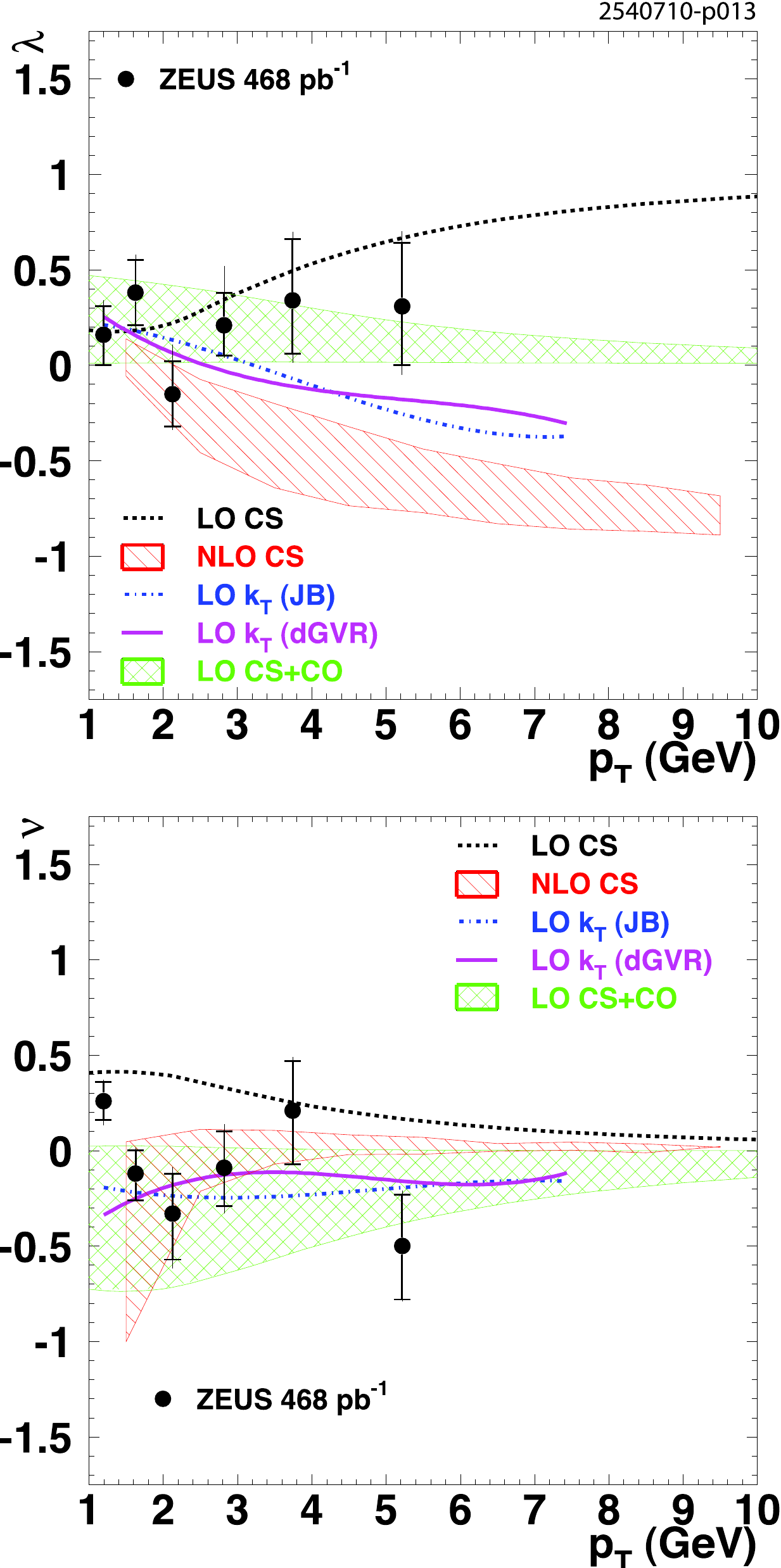}
\caption{Polarization parameters $\lambda$ and $\nu$ in the target frame
as a function of $p_T$ for $J/\psi$ photoproduction at HERA. The
measurement by the ZEUS collaboration \cite{Chekanov:2009br} of the
polarization parameters in the target frame is compared with the
color-singlet contribution at LO in $\alpha_s$ (labeled LO CS)
\cite{Beneke:1998re} and at NLO in $\alpha_s$ (labeled NLO CS)
\cite{Artoisenet:2009xh}, with predictions in the
$k_T$-factorization approach for two different sets of unintegrated
parton-distributions functions (labeled LO $k_T$ (JB) and LO $k_T$
(dGRV)) \cite{Baranov:2008zzc}, and with the complete NRQCD
factorization predictions (including color-singlet and color-octet
contributions) at LO in $\alpha_s$ (labeled LO CS+CO)
\cite{Beneke:1998re}. The theoretical uncertainty bands labeled NLO CS
were obtained by varying the factorization scale $\mu_f$ and the
renormalization scale $\mu_r$ in the range defined by $0.5\mu_0<\mu_f$,
$\mu_r<2\mu_0$, and $0.5<\mu_r/\mu_f<2$, where $\mu_0=4m_c$. The
theoretical uncertainty bands labeled LO CS+CO were obtained by
considering uncertainties in the values of the color-octet NRQCD
long-distance matrix elements. \figPermXSPV{Chekanov:2009br}{2009} }
\label{prod_photoproduction_pol_target}
\end{figure}

\begin{figure}
\includegraphics[width=\figwid]{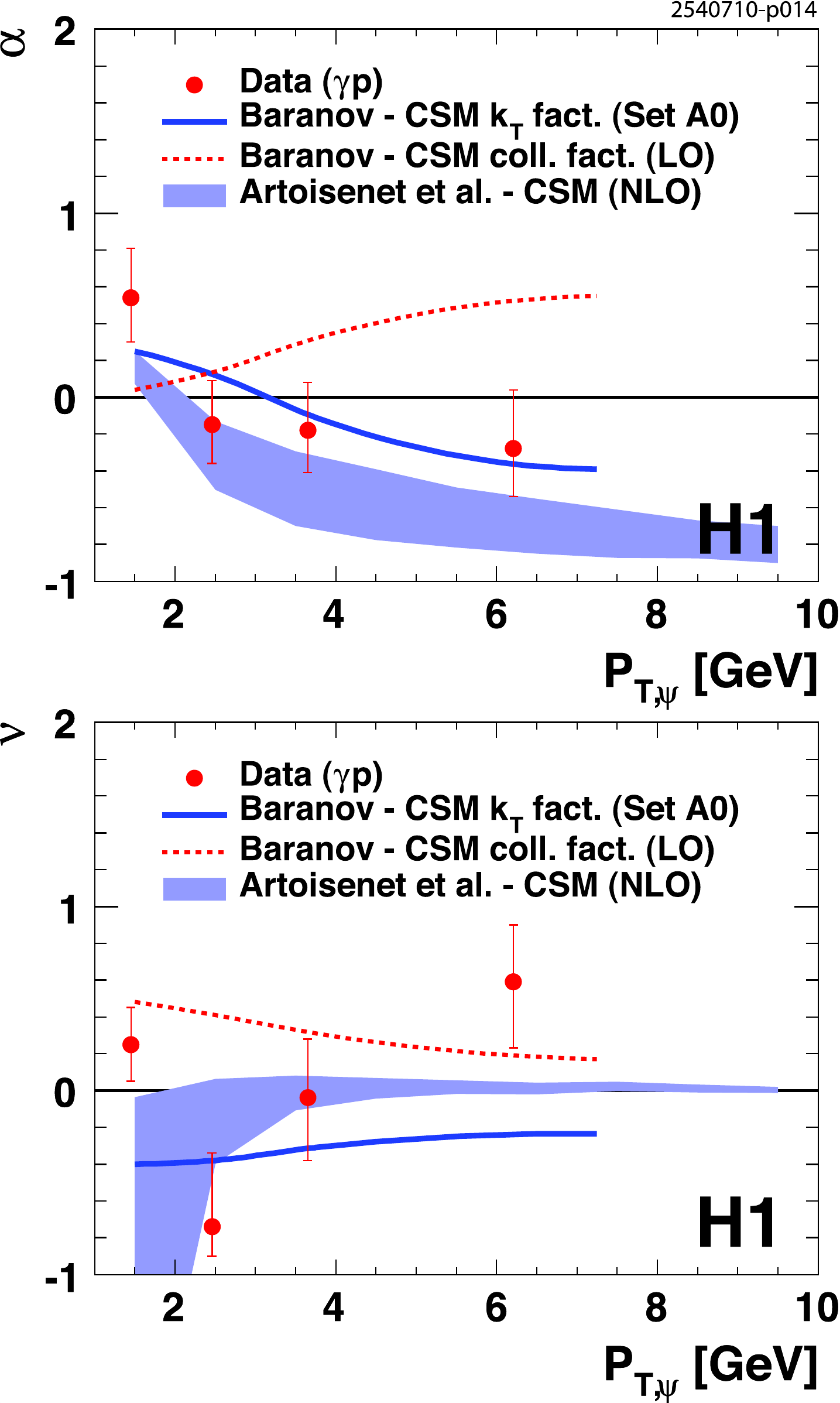}
\caption{Polarization parameters $\alpha=\lambda$ and $\nu$ in the helicity
frame as a function of $p_T$ for $J/\psi$ photoproduction at HERA. The
measurement by the H1 collaboration \cite{Aaron:2010gz} of the
polarization parameters in the helicity frame is compared with the
predictions in the $k_T$-factorization approach {\it (solid line)}
\cite{Baranov:2008zzc} and with the predictions in color-singlet model
(CSM) at LO in $\alpha_s$ {\it (dashed line)} \cite{Baranov:2008zzc} and at
NLO $\alpha_s$ {\it (filled band)}
\cite{Artoisenet:2009xh,prod_artoisenet_private}. The theoretical
uncertainty bands were obtained by varying the factorization scale
$\mu_f$ and the renormalization scale $\mu_r$ in the range defined by
$0.5\mu_0<\mu_f$, $\mu_r<2\mu_0$, and $0.5<\mu_r/\mu_f<2$, where
$\mu_0=4m_c$. From Ref.~\cite{Aaron:2010gz} }
\label{prod_photoproduction_pol_helicity}
\end{figure}

We show comparisons of several theoretical predictions with the ZEUS
data in Fig.~\ref{prod_photoproduction_pol_target} and with the H1 data
in Fig.~\ref{prod_photoproduction_pol_helicity}.

The curves in Fig.~\ref{prod_photoproduction_pol_target} that are labeled
``LO CS$+$CO'' are the complete NRQCD factorization predictions
(including color-singlet and color-octet contributions) at LO in
$\alpha_s$ \cite{Beneke:1998re}. These agree reasonably well with the
ZEUS data in the target frame, except for the value of $\nu$ in the
lowest-$p_T$ bin. However, at such a low value of $p_T$, the NRQCD
factorization formula is not expected to be valid.

The curves that are labeled ``LO $k_T$'' and ``CSM $k_T$'' are
predictions in the $k_T$-factorization scheme \cite{Baranov:2008zzc}.
The set of unintegrated parton distribution functions that is used is
indicated in parentheses. The $k_T$-factorization predictions bracket
the ZEUS data in the target frame for both $\lambda$ and $\nu$ and are
in reasonable agreement with the H1 data for $\lambda$ and $\nu$ in the
helicity frame, if one takes the difference between the two
$k_T$-factorization predictions to be a measure of the theoretical
uncertainty.

The bands that are labeled ``NLO CS'' and ``CSM NLO'' correspond to
predictions in the color-singlet model at NLO in $\alpha_s$
(Ref.~\cite{Artoisenet:2009xh}). (Similar results for the polarization at
NLO in $\alpha_s$ in the color-singlet model were obtained in
Ref.~\cite{Chang:2009uj}.) The uncertainties that are shown in these
bands arise from the sensitivity of the polarization to the
factorization and renormalization scales and are much larger than the
uncertainties that arise from the uncertainty in the value of $m_c$. A
comparison of the LO and NLO predictions of the CSM shows the large
impact of the NLO correction on the polarization parameters.  At NLO,
the parameter $\lambda$ is predicted to decrease with increasing $p_T$,
in both the target and the helicity frames. (This trend is also observed
in the LO prediction of the $k_T$ factorization formalism, which
effectively accounts for some topologies that occur at higher orders in
$\alpha_s$ in the collinear factorization scheme.) The NLO color-singlet
predictions for $\lambda$ are compatible with the H1 data in the
helicity frame, but differ significantly from the ZEUS data in the
target frame. In contrast, the NLO color-singlet prediction for the
parameter $\nu$ is in reasonable agreement with the ZEUS data in the
target frame, as well as with the H1 data in the helicity frame.

A more complete presentation of the comparison between theory and
experiment for the polarization parameters $\lambda$ and $\nu$,
including the distributions of these parameters as functions of $z$, can
be found in Refs.~\cite{Aaron:2010gz,Chekanov:2009br}.

\subsubsection{Future opportunities}
\label{prod_sec:ep-fo}

In spite of the recent advances that we have described, it is still
unclear which mechanisms are at work in $J/\psi$ photoproduction at
HERA. The new computations of NLO corrections to the differential
cross-sections and the polarization parameters in the
collinear-factorization scheme show that there is room for a color-octet
contribution. Indeed, the NLO NRQCD factorization prediction for the
cross section fits the data reasonably, with central values that are
closer to the data points than are the central values of the NLO
color-singlet contribution. Recent analyses of the color-singlet
contribution in the $k_T$ factorization scheme also show reasonable
agreement with the data.

In both $k_T$ factorization and in collinear factorization, theoretical
uncertainties remain substantial. Improvement of the situation for $k_T$
factorization will require better knowledge of the $k_T$-dependent
parton distributions. In collinear factorization, the large sensitivity
of the NLO color-singlet rates to the renormalization scale signals that
QCD corrections beyond NLO might be relevant, especially for the
description of the polarization parameters. Here, an analysis in the
fragmentation-function approach (see \Sec{prod_sec:fragmentation}) might
help to bring the perturbation series under better control. Theoretical
uncertainties in the NRQCD factorization prediction for the $J/\psi$
polarization could be reduced by computing the color-octet contributions
to the polarization parameters at NLO accuracy in $\alpha_s$. The large
uncertainties in the color-octet long-distance matrix elements dominate
the uncertainties in the NRQCD factorization prediction for the cross
section. Since these matrix elements are obtained by fitting NRQCD
factorization predictions to the Tevatron data, improvements in the
theoretical uncertainties in the Tevatron (or LHC) predictions are
necessary in order to reduce the uncertainties in the matrix elements.
Finally, theoretical uncertainties in the region near the kinematic
endpoint $z=1$ might be reduced through a systematic study of
resummations of the perturbative and velocity expansions in both $ep$
and $e^+e^-$ quarkonium production.

\subsection{Fixed-target production}
 
\subsubsection{Phenomenology of fixed-target production}

The NRQCD factorization approach has also been tested against the
charmonium production data that have been obtained from fixed-target
experiments and $pp$ experiments at low energies. Owing to the limited
statistics, quarkonium observables measured in these experiments have, in
general, been restricted to total cross sections for $J/\psi$ and $\psi(2S)$
production. Also, in the case of $J/\psi$ production, feeddown
contributions have not been subtracted.

In the most recent analysis of the fixed-target and low-energy $pp$
quarkonium production data \cite{Maltoni:2006yp}, experimental
results for the inclusive production rates of the $J/\psi$ and the
$\psi(2S)$ have been examined, along with the experimental results for
the ratios of these cross sections. A total of 29 experimental results
have been analyzed in order to extract the color-octet contribution to
the observed rates. By comparing the values of the color-octet matrix
elements that are extracted from the fixed-target data with the values
that are extracted from the Tevatron data, one can test the universality
of the NRQCD matrix elements.

The analysis of Ref.~\cite{Maltoni:2006yp} made use of the NRQCD
short-distance coefficients through order $\alpha_s^3$ for the $P$-wave
channels and for all of the color-octet channels that are of leading
order in $v$ (Ref.~\cite{Petrelli:1997ge}). The LO short-distance
coefficients for the color-singlet ${}^3S_1$ channel were employed, as
the computation of the QCD correction to these short-distance
coefficients in $pp$ collisions had not yet been completed at the
time of the analysis of Ref.~\cite{Maltoni:2006yp}.

In extracting the color-octet contribution from the production rates
that are measured in fixed-target experiments, the authors of
Ref.~\cite{Maltoni:2006yp} treated the NRQCD long-distance matrix
elements as follows: Heavy-quark spin symmetry was employed; the
color-singlet matrix elements were taken from the potential-model
calculation of Ref.~\cite{Eichten:1995ch}; the color-octet matrix
element for $P$-wave charmonium states was set equal to a value that was
extracted in Ref.~\cite{Nason:1999ta} from the CDF data
\cite{Abe:1997jz}. Regarding the color-octet matrix elements for the
$J/\psi$ and $\psi(2S)$, it was assumed that $\langle \mathcal{O}_8^{H}
\left( ^1S_0 \right) \rangle= \langle \mathcal{O}_8^{H} \left( ^3P_0
\right) \rangle /m_c^2$ and that the ratios of these matrix elements to
$\langle \mathcal{O}_8^{H} \left({}^3S_1 \right) \rangle$ are given by
the values that were extracted from the Tevatron data. The color-octet
contributions to the direct production of the $J/\psi$ and the
$\psi(2S)$ were multiplied by rescaling parameters $\lambda_{J/\psi}$
and $\lambda_{\psi'}$, respectively, which were varied in order to fit
the fixed-target data. In addition to these two parameters, the
factorization and renormalization scales were varied in the fit, while
the mass of the charm quark was held fixed at $1.5$~GeV.

\begin{figure}[b]
\includegraphics[width=\figwid]{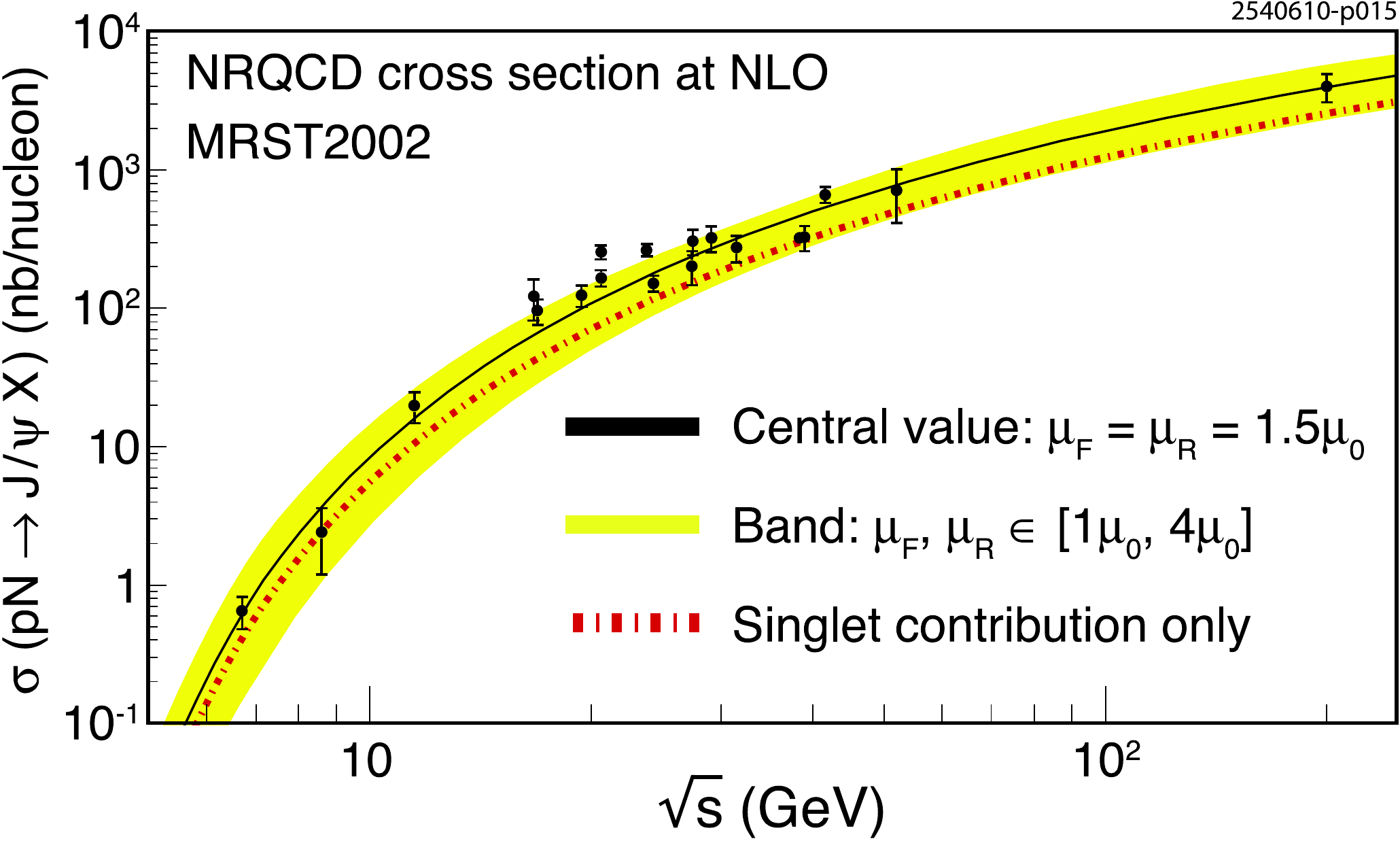}
\caption{Fit of the NRQCD factorization cross section for the production
of the $J/\psi$ as a function of the center-of-mass energy $\sqrt{s}$ to
data from fixed-target experiments and $pp$ experiments at low energies.
The {\it curves} are the result of the fitting procedure that is described in
the text. The color-singlet component is shown as a {\it dotted-dashed line}.
The theoretical uncertainty {\it band} was obtained by varying the
factorization scale $\mu_F$ and the renormalization scale $\mu_R$ in the
ranges $\mu_0<\mu_F<4\mu_0$ and $\mu_0<\mu_R<4\mu_0$, where $\mu_0=2m_c$.
\figPermXPLB{Maltoni:2006yp}{2006} }
\label{prod_Jpsi_cross_section}
\end{figure}

The $\chi^2$ of the fit favors a nonzero value for the color-octet
contribution to direct $J/\psi$ and $\psi(2S)$ production. This can be
seen in Fig.~\ref{prod_Jpsi_cross_section} for the case of the $J/\psi$: The
color-singlet contribution alone (dotted-dashed line) systematically
undershoots the data, while the fitted complete NRQCD contribution
(solid line) is in good agreement with the data. However, the fit also
indicates that the values of the color-octet matrix elements that are
needed to explain the fixed-target data are only about $10 \%$ of the
values that were extracted from fits to the Tevatron data. A similar
result was obtained in the case of the color-octet matrix elements for
the $\psi(2S)$.  In both cases, the stability of the fitting procedure
was checked by changing the selection of measurements used in the fits
and by changing the set of parton distribution functions that was used
in computing the cross sections. In any of these scenarios, the values
of the color-octet matrix elements that were extracted from the
fixed-target data are smaller than the values that were extracted from
the Tevatron data.

The analysis of fixed-target data that we have described could be
updated by making use of the recent results for the NLO corrections to the
hadroproduction of $S$-wave states \cite{Campbell:2007ws,Gong:2008ft}.
Furthermore, a more accurate value for the $J/\psi$ color-singlet matrix
element has been derived recently in Ref.~\cite{Bodwin:2007fz}.
However, it is expected, even with these improvements, that 
the qualitative conclusions of Ref.~\cite{Maltoni:2006yp} would 
still hold.

It is interesting to consider this fixed-target result in light of two
recent developments: (1) the possible impact of higher-order corrections
to the color-singlet yield on the phenomenology of $J/\psi$ production at
the Tevatron (\Sec{prod_section:pheno_with_QCD_corr}) and (2) the latest
$J/\psi$ polarization measurement by the CDF
collaboration~\cite{Abulencia:2007us} (\Sec{prod_section:polarization}).
Both of these developments suggest that the values of the color-octet
NRQCD matrix elements may be smaller than had previously been supposed.
However, in assessing the significance of the fixed-target result, it
should be remembered that the fixed-target data, integrated over $p_T$,
are dominated by data from the lowest values of $p_T$. At low $p_T$, one
would not expect NRQCD factorization to hold.

\subsubsection{Future opportunities}

Fixed-target experiments provide another means by which to test the
various theories of inclusive quarkonium production. In order to test
definitively theoretical hypotheses, such as NRQCD factorization, that
are based on hard-scattering factorization, it is necessary to make
measurements at values of $p_T$ that are much greater than the
heavy-quark mass. If fixed-target experiments with such a high reach in
$p_T$ could be devised, then it would be very useful to measure both the
cross section and the quarkonium polarization as functions of $p_T$.

\subsection{Exclusive production in {$e^+e^-$} collisions}
\label{prod_sec:exclusive}%
\subsubsection{Theory vs.~experiment}

As recently as three years ago, large discrepancies existed between
experimental measurements and theoretical predictions for the process
$e^+e^-\to J/\psi+\eta_c$ at the $B$-factory energy of
$\sqrt{s}=10.58$~GeV. In its initial measurement of this process, the
Belle collaboration obtained $\sigma(e^+e^-\to J/\psi+\eta_c)\times
\Brat_{\ge 4}=33_{-6}^{+7}\pm 9~\hbox{fb}$ (Ref.~\cite{Abe:2002rb}), where
$\sigma(e^+e^-\to J/\psi+\eta_c)$ is the cross section and $\Brat_{\ge 4}$
is the branching fraction of the $\eta_c$ into four or more charged
tracks. The first theoretical predictions for $\sigma(e^+e^-\to
J/\psi+\eta_c)$ were based on calculations in the NRQCD factorization
approach \cite{Bodwin:1994jh}. These initial calculations were carried
out at LO in $\alpha_s$ and $v$. They gave the following predictions for
$\sigma(e^+e^-\to J/\psi+\eta_c)$: $5.5~\hbox{fb}$
(Ref.~\cite{Liu:2002wq}), $3.78\pm 1.26~\hbox{fb}$
(Ref.~\cite{Braaten:2002fi}), and $2.3~\hbox{fb}$
(Ref.~\cite{Hagiwara:2003cw}).\footnote{In Ref.~\cite{Braaten:2002fi} a
cross section of $2.31\pm 1.09~\hbox{fb}$ was reported initially. Later,
a sign error in the QED interference term was corrected and the value
cited above was obtained.} The differences between the calculations of
Refs.~\cite{Liu:2002wq,Braaten:2002fi,Hagiwara:2003cw} arise from QED
effects, which are included only in Ref.~\cite{Braaten:2002fi}; from
contributions from an intermediate $Z$ boson, which are included only in
Ref.~\cite{Hagiwara:2003cw}; and from different choices of $m_c$, the
NRQCD long-distance matrix elements, and $\alpha_s$. The sensitivities
of the calculations to the values of these parameters are important
sources of theoretical uncertainties, which we will discuss later.

The Belle collaboration has, more recently, measured the quantity
$\sigma(e^+e^-\to J/\psi+\eta_c)\times \Brat_{>2}=25.6\pm 2.8\pm
3.4~\hbox{fb}$ (Ref.~\cite{Abe:2004ww}), where $\Brat_{>2}$ is the branching
fraction of the charmonium state that is recoiling against the $J/\psi$
(in this case the $\eta_c$) into more than two charged tracks. This
cross section times branching fraction has also been measured by the
\babar\ collaboration, which obtains $\sigma(e^+e^-\to
J/\psi+\eta_c)\times \Brat_{> 2}=17.6\pm 2.8\pm 2.1~\hbox{fb}$
(Ref.~\cite{Aubert:2005tj}). These more recent experimental results
reduced the discrepancy between theory and experiment, but did not
eliminate it. In assessing the size of the discrepancy, it is important
to recognize that the experimental results are cross sections times
branching fractions. Hence, they are lower bounds on the cross sections,
which are the quantities that appear in the theoretical predictions.

Table~\ref{prod_tab:jpsiplusH} contains a summary of experimental
measurements and NRQCD predictions for the process $e^+e^-\to J/\psi +
H$, where $H$ is $\eta_c$, $\chi_{c0}$, or $\eta_c(2S)$. As can be seen
from Table~\ref{prod_tab:jpsiplusH}, significant discrepancies exist between
LO NRQCD predictions and experiment, not only for exclusive production
of $J/\psi + \eta_{c}$, but also for exclusive production of $J/\psi
+ \chi_{c0}$ and $J/\psi + \eta_c(2S)$.
\begin{table*}
   \caption{Experimental measurements and NRQCD predictions for $e^+e^-\to
            \jpsi + H$, where $H$ is $\eta_c$, $\chi_{c0}$, or $\eta_c(2S)$. Cross
            sections are in units of fb. The quantity $\Brat_{>2}$ is the branching
            fraction of the charmonium state that is recoiling against the $\jpsi$
            into more than two charged tracks }
\label{prod_tab:jpsiplusH}
   \setlength{\tabcolsep}{2.1pc}
   \begin{center}
      \begin{tabular}{lccc}
      \hline\hline
      \rule[10pt]{-1mm}{0mm}
Quantity & $\eta_c(1S)$ & $\chi_{c0}(1P)$ & $\eta_c(2S)$\\[0.6mm]
\hline
      \rule[10pt]{-1mm}{0mm}
$\sigma\times \Brat_{>2}$ (Belle~\cite{Abe:2004ww}) 
        & $25.6\pm 2.8\pm 3.4$ 
        & $6.4\pm 1.7\pm 1.0$ & $16.5\pm 3.0\pm 2.4$\\[0.7mm]
$\sigma\times \Brat_{>2}$ (\babar~\cite{Aubert:2005tj}) 
        & $17.6\pm 2.8^{+1.5}_{-2.1}$ 
        & $10.3\pm 2.5^{+1.4}_{-1.8}$ 
        & $16.4\pm 3.7^{+2.4}_{-3.0}$ \\[0.7mm]
\hline
      \rule[10pt]{-1mm}{0mm}
$\sigma$ (Liu, He, Chao~\cite{Liu:2002wq}) & 5.5 & 6.9 & 3.7 \\[0.7mm]
$\sigma$ (Braaten, Lee~\cite{Braaten:2002fi}) & $3.78\pm 1.26$
                            & $2.40\pm 1.02$ & $1.57\pm 0.52$\\[0.7mm]
$\sigma$ (Hagiwara, Kou, Qiao~\cite{Hagiwara:2003cw}) & 2.3& &\\[0.7mm]
$\sigma$ (Bodwin {\it et al.}~\cite{Bodwin:2006ke}) & $17.5\pm 5.7$ & & \\[0.7mm]
$\sigma$ (He, Fan, Chao~\cite{He:2007te}) & $20.4$ & & \\[0.7mm]
$\sigma$ (Bodwin, Lee, Yu~\cite{Bodwin:2007ga}) & $17.6^{+8.1}_{-6.7}$ & &\\[0.7mm]
\hline\hline
      \end{tabular}
   \end{center}
\end{table*}
An important step toward resolving the discrepancy between theory and
experiment for $\sigma(e^+e^-\to J/\psi+\eta_c)$ was the calculation in
Ref.~\cite{Zhang:2005cha} of the corrections of NLO in $\alpha_s$. These
corrections yield a $K$~factor of about $1.96$. This result has been
confirmed in Ref.~\cite{Gong:2007db}. While this $K$~factor is
substantial, it does not, by itself, eliminate the discrepancy between
theory and experiment.

In the NRQCD factorization formalism there are, in addition to
corrections of higher order in $\alpha_s$, corrections of higher order
in $v$, {\it i.e.}, relativistic corrections. In $\sigma(e^+e^-\to
J/\psi+\eta_c)$, relativistic corrections can arise in two ways. First,
they can appear directly as corrections to the process $e^+e^-\to J/\psi
+\eta_c$ itself. Second, they can arise indirectly through the NRQCD
long-distance matrix elements that appear in the expression for
$\sigma(e^+e^-\to J/\psi +\eta_c)$. For example, the matrix element of
leading order in $v$ that appears in $J/\psi$ production can be
determined phenomenologically from the experimental value for the width
for $J/\psi \to e^+e^-$ and the theoretical expression for that width.
There are relativistic corrections to the theoretical expression for the
width, which affect the value of the long-distance matrix element that
one obtains.

The first relativistic correction appears at relative order $v^2$, where
$v^2\approx 0.3$ for charmonium. It has been known for some time that
this correction is potentially large: In Ref.~\cite{Braaten:2002fi}, it
was found that the order-$v^2$ correction is $1.95 \langle
v^2\rangle_{J/\psi}+ 2.37 \langle v^2\rangle_{\eta_c}$. Here, $\langle
v^2\rangle_H$ is the ratio of an order-$v^2$ NRQCD long-distance
matrix element to the LO matrix element in the quarkonium
state $H$. The authors of Ref.~\cite{Braaten:2002fi} estimated the
matrix elements of order $v^2$ by making use of the Gremm-Kapustin
relation \cite{Gremm:1997dq}, which follows from the NRQCD equations of
motion. On the basis of these estimates, they found that the $K$~factor 
for the relativistic corrections is about
$2.0_{-1.1}^{+10.9}$. The very large uncertainties in this $K$ factor
reflect the large uncertainties in the Gremm-Kapustin-relation estimates
of the matrix elements of order $v^2$.

In Ref.~\cite{Bodwin:2006dn}, significant progress was made in reducing the
uncertainties in the order-$v^2$ NRQCD matrix elements. The approach in
this work was to make use of a static-potential model to calculate the
quarkonium wave functions and, from those wave functions, to compute the
dimensionally regulated NRQCD matrix elements. If the static potential
is known accurately, for example, from lattice calculations, then the
corrections to the static potential model are of relative order $v^2$
(Ref.~\cite{Brambilla:1999xf,Pineda:2000sz}). Hence, one can regard the
potential-model calculation as a first-principles calculation with
controlled uncertainties.

Making use of the results of Ref.~\cite{Bodwin:2006dn}, the authors of
Ref.~\cite{Bodwin:2006ke} computed the relativistic corrections to
$\sigma(e^+e^-\to  J/\psi +\eta_c)$. Taking into account the corrections
of NLO in $\alpha_s$ from Ref.~\cite{Zhang:2005cha}, they obtained
$\sigma(e^+e^-\to J/\psi +\eta_c)=17.5\pm 5.7$~fb, where the quoted
uncertainty reflects only the uncertainties in $m_c$ and the
order-$v^2$ 
NRQCD long-distance matrix elements. This result includes the effects of
a resummation of a class of relativistic corrections that arise from the
quarkonium wave function. One might worry that the large relativistic
corrections that appear in this calculation, which result in a $K$
factor of about $2.6$, are an indication that the $v$ expansion of NRQCD
is out of control. However, this large correction is the result of
several corrections of a more modest size: a direct correction of about
$40\%$ and two indirect corrections (for the $J/\psi$ and the $\eta_c$)
of about $37\%$ each. Furthermore, higher-order terms in the resummation
of wave-function corrections change the direct correction by only about
$13\%$, suggesting that the $v$ expansion is indeed converging well.

The authors of Ref.~\cite{He:2007te} took a different approach to
calculating relativistic corrections, determining the
NRQCD long-distance matrix elements of LO in $v$ and of relative order
$v^2$ by using $\Gamma(J/\psi\to e^+e^-)$,
$\Gamma(\eta_c\to\gamma\gamma)$, and $\Gamma(J/\psi\to \textrm{light
hadrons})$ as inputs. Their result, $\sigma(e^+e^-\to J/\psi
+\eta_c)=20.04$~fb, is in agreement with the result of
Ref.~\cite{Bodwin:2006ke}. However, the values of the NRQCD matrix
elements that are given in Ref.~\cite{He:2007te} differ significantly
from those that were used in Ref.~\cite{Bodwin:2006ke}. This difference
probably arises mainly because of the very large relativistic
corrections to $\Gamma(J/\psi\to \textrm{light hadrons})$, which may not
be under good control.

The results of Refs.~\cite{Bodwin:2006ke,He:2007te} greatly
reduced the difference between the experimental and theoretical central
values for $\sigma(e^+e^-\to  J/\psi +\eta_c)$. However, in order to
assess the significance of these results, it is essential to have a
reliable estimate of the theoretical uncertainties. Such an estimate was
provided in Ref.~\cite{Bodwin:2007ga}. In this work, which was based on
the method of Ref.~\cite{Bodwin:2006ke}, uncertainties from various
input parameters, such as $m_c$ and the electromagnetic widths of the
$J/\psi$ and the $\eta_c$, were taken into account, as well as
uncertainties from the truncations of the $\alpha_s$ and $v$ expansions.
Correlations between uncertainties in various components of the
calculation were also taken into account. In addition, various
refinements were included in the calculation, such as the use of the
vector-meson-dominance method to reduce uncertainties in the
QED contribution and the inclusion of the
effects of interference between relativistic corrections and corrections
of NLO in $\alpha_s$. The conclusion of Ref.~\cite{Bodwin:2007ga} is
that $\sigma(e^+e^-\to  J/\psi +\eta_c)=17.6^{+8.1}_{-6.7}$~fb. This result
is in agreement, within uncertainties, with the \babar\ result, even if
one allows for the fact that the branching fraction $\Brat_{>2}$ could be as
small as $0.5$--$0.6$.

An alternative approach to theoretical calculations of exclusive
quarkonium production in $e^+e^-$ annihilation is the light-cone method
\cite{Ma:2004qf,Bondar:2004sv,Braguta:2005kr,Braguta:2006nf,Braguta:2007ge}.
Generally, the light-cone-method predictions for exclusive quarkonium
production cross sections are in agreement with the experimental
results. The light-cone approach to quarkonium production can be derived
from QCD.\footnote{Light-cone factorization formulas are derived in
Ref.~\cite{Bodwin:2008nf} in the course of proving NRQCD factorization
formulas.} In principle, the light-cone approach is as valid as the
NRQCD factorization approach. In practice, it is, at present, necessary
to model the light-cone wave functions of the quarkonia, possibly making
use of constraints from QCD sum rules
\cite{Braguta:2006wr,Braguta:2008qe}. Consequently, the existing
light-cone calculations are not first-principles calculations, and it is
not known how to estimate their uncertainties reliably. The light-cone
approach automatically includes relativistic corrections that arise from
the quarkonium wave function. As has been pointed out in
Ref.~\cite{Bodwin:2006dm}, the light-cone calculations contain
contributions from regions in which the quarkonia wave-function momenta
are of order $m_c$ or greater. In NRQCD, such contributions are
contained in corrections to the short-distance coefficients of higher
order in $\alpha_s$. Therefore, in order to avoid double counting, one
should refrain from combining light-cone results with NRQCD corrections
of higher order in $\alpha_s$. Finally, we note that, in
Ref.~\cite{Bodwin:2006dm}, it was suggested that resummations of
logarithms of $\sqrt{s}/m_c$ have not been carried out correctly in some
light-cone calculations.

\subsubsection{Future opportunities}

Clearly, it would be desirable to reduce both the theoretical and
experimental uncertainties in the rates for the exclusive production of
quarkonia in $e^+e^-$ annihilation and to extend theory and experiment
to processes involving additional quarkonium states.

On the experimental side, the central values of the Belle and \babar\
measurements of $\sigma(e^+e^-\to  J/\psi +\eta_c)\times \Brat_{>2}$ differ
by about twice the uncertainty of either measurement. Although those
uncertainties are small in comparison with the theoretical
uncertainties, the rather large difference in central values suggests
that further experimental work would be useful. Furthermore, it would be
very useful, for comparisons with theory, to eliminate the uncertainty
in the cross section that arises from the unmeasured branching fraction
$\Brat_{>2}$.

On the theoretical side, the largest uncertainty in
$\sigma(e^+e^-\to J/\psi +\eta_c)$ arises from the uncertainty in $m_c$.
One could take advantage of the recent reductions in the uncertainty in
$m_c$ \cite{Allison:2008xk,Chetyrkin:2009fv} to reduce the theoretical
uncertainty in $\sigma(e^+e^-\to  J/\psi +\eta_c)$ from this source. The
next largest source of theoretical uncertainty arises from the omission
of the correction to the $J/\psi$ electromagnetic width at NNLO in
$\alpha_s$ (Ref.~\cite{Beneke:1997jm,Czarnecki:2001zc}). Unfortunately,
the large scale dependence of this correction is a serious impediment to
progress on this issue. The uncalculated correction to $\sigma(e^+e^-\to
J/\psi +\eta_c)$ of relative order $\alpha_s v^2$ is potentially large,
as is the uncalculated correction of relative order $\alpha_s^4$. While
the calculation of the former correction may be feasible, the
calculation of the latter correction is probably beyond the current state
of the art. However, one might be able to identify large contributions
that could be resummed to all orders in $\alpha_s$.

Finally, it would be desirable to extend the theoretical calculations to
include $P$-wave and higher $S$-wave states. In the NRQCD approach, a
serious obstacle to such calculations is the fact that the relativistic
corrections become much larger for excited states, possibly spoiling the
convergence of the NRQCD velocity expansion.

\subsection{Inclusive production in {$e^+e^-$} collisions}
\label{prod_sec:inclusive-e+e-}%

\subsubsection{Experiments and LO theoretical expectations}

In 2001, the prompt $J/\psi$ inclusive production cross section
$\sigma(e^+e^-\rightarrow J/\psi+ X)$  was measured to be $\sigma_{\rm
tot}=2.52\pm 0.21 \pm 0.21 $~pb by the \babar\ collaboration
\cite{Aubert:2001pd}.  A smaller value, $\sigma_{\rm tot}=1.47\pm 0.10
\pm 0.13 $~pb was found by the Belle collaboration \cite{Abe:2001za}.
The color-singlet contribution to the prompt $J/\psi$ inclusive
production cross section at LO in $\alpha_s$, including contributions
from the processes $e^+e^-\to J/\psi+c\bar c$, $e^+e^-\to J/\psi+gg$,
and $e^+e^-\to J/\psi+q\bar q+gg$ ($q=u$, $d$, $s$), was estimated to be
only about $0.3$--$0.5$~pb
(Ref.~\cite{Yuan:1996ep,Cho:1996cg,Baek:1998yf,Schuler:1998az,Kiselev:1994pu,Liu:2003zr}),
which is much smaller than the measured value. This would suggest that
the color-octet contribution might play an important role in inclusive
$J/\psi$ production
\cite{Yuan:1996ep,Cho:1996cg,Baek:1998yf,Schuler:1998az,Kiselev:1994pu,Liu:2003zr}.
However, it was found by the Belle collaboration \cite{Abe:2002rb} that
the associated production cross section
\begin{eqnarray}
\sigma ( e^+e^-\to \jpsi + c \bar c ) = \left( 0.87^{+0.21}_{-0.19}\pm
0.17 \right)~{\rm pb}\, ,
\label{prod_bellecc}
\end{eqnarray}
is larger, by at least a factor of 5, than the LO NRQCD factorization
prediction, which includes both the color-singlet contribution
\cite{Cho:1996cg,Baek:1998yf,Schuler:1998az,Kiselev:1994pu,Liu:2003zr,Liu:2003jj} and the color-octet contribution \cite{Liu:2003jj}.
The ratio of the $J/\psi +c\bar c$ cross section to the $J/\psi$ inclusive
cross section was found by the Belle collaboration \cite{Abe:2002rb} to
be
\begin{eqnarray}
R_{c \bar c}=\frac{\sigma ( e^+e^- \to \jpsi + c \bar c )} 
{\sigma ( e^+e^-\to\jpsi +X ) } = 0.59 ^{+0.15}_{-0.13}\pm 0.12\, ,~~~~~~ 
\label{prod_R-cc}
\end{eqnarray} 
which
is also much larger than LO NRQCD factorization prediction. If one
includes only the color-singlet contribution at LO in $\alpha_s$, then 
the ratio is predicted to be $R_{c \bar c}=0.1$--$ 0.3$
(Ref.~\cite{Yuan:1996ep,Cho:1996cg,Baek:1998yf,Schuler:1998az,Kiselev:1994pu,Liu:2003zr,Hagiwara:2004pf}),
depending on the values of input parameters, such as $\alpha_s$, $m_c$,
and, especially, the color-singlet matrix elements. A large color-octet
contribution could enhance substantially the $J/\psi$ inclusive cross
section (the denominator) but could enhance only slightly the
$J/\psi+c\bar c$ cross section (the numerator)\cite{Liu:2003jj} and,
therefore, would have the effect of decreasing the prediction for $R_{c
\bar c}$. Thus, the LO theoretical results and the experimental results 
in Eqs.~(\ref{prod_bellecc}) and (\ref{prod_R-cc}) presented a serious challenge to
the NRQCD factorization picture.\footnote{The light-cone perturbative-QCD
approach \cite{Berezhnoy:2003hz} gives a prediction that $R_{c \bar
c}=0.1$--$0.3$, while the CEM gives a prediction that $R_{c \bar c}=0.06$
(Ref.~\cite{Kang:2004zj}), both of which are far below the measured
value of $R_{c \bar c}$.}

Several theoretical studies were made with the aim of resolving this puzzle in
$J/\psi$ production. The authors of Ref.~\cite{Fleming:2003gt}
used soft-collinear effective theory (SCET) to resum the color-octet
contribution to the $J/\psi$ inclusive cross section, the authors of
Ref.~\cite{Lin:2004eu} used SCET to analyze the color-singlet
contribution to $e^+e^- \to J/\psi+gg$, and the authors of
Ref.~\cite{Leibovich:2007vr} resummed the LO and NLO logarithms in the
color-singlet contribution to the $J/\psi$ inclusive cross section. 
These resummation calculations, while potentially useful, did not 
resolve the puzzle. 

Very recently, the Belle collaboration reported new measurements
\cite{Pakhlov:2009nj}:
\begin{eqnarray}
\sigma ( \jpsi+ X )
&=&(1.17\pm 0.02\pm 0.07)~{\rm pb}\, ,~~~~~~\label{prod_X}\\
\sigma ( \jpsi+c\bar c )
&=&(0.74\pm 0.08^{+0.09}_{-0.08})~{\rm pb}\, ,~~~~~~\label{prod_CCBar}\\
\sigma ( \jpsi+X_{{\rm non}\  c\bar c} ) 
&=&(0.43\pm 0.09\pm 0.09)~{\rm pb}\, .~~~~~~
\label{prod_eq:nonCCBar}
\end{eqnarray}
the value of the inclusive $J/\psi$ cross section in
Eq.~(\ref{prod_X}) is significantly smaller than the values that were obtained
previously by the \babar\ collaboration \cite{Aubert:2001pd} and by the
Belle collaboration \cite{Abe:2001za}, but it is still much larger than the
LO NRQCD prediction. The cross section $\sigma(e^+e^-\rightarrow
J/\psi+c\bar c)$ in Eq.~(\ref{prod_CCBar}) is also much larger than the LO
color-singlet and color-octet predictions.

\subsubsection{$e^+ e^- \to J/\psi+c \bar c$ at NLO }

\begin{figure}
\includegraphics[width=\figwid]{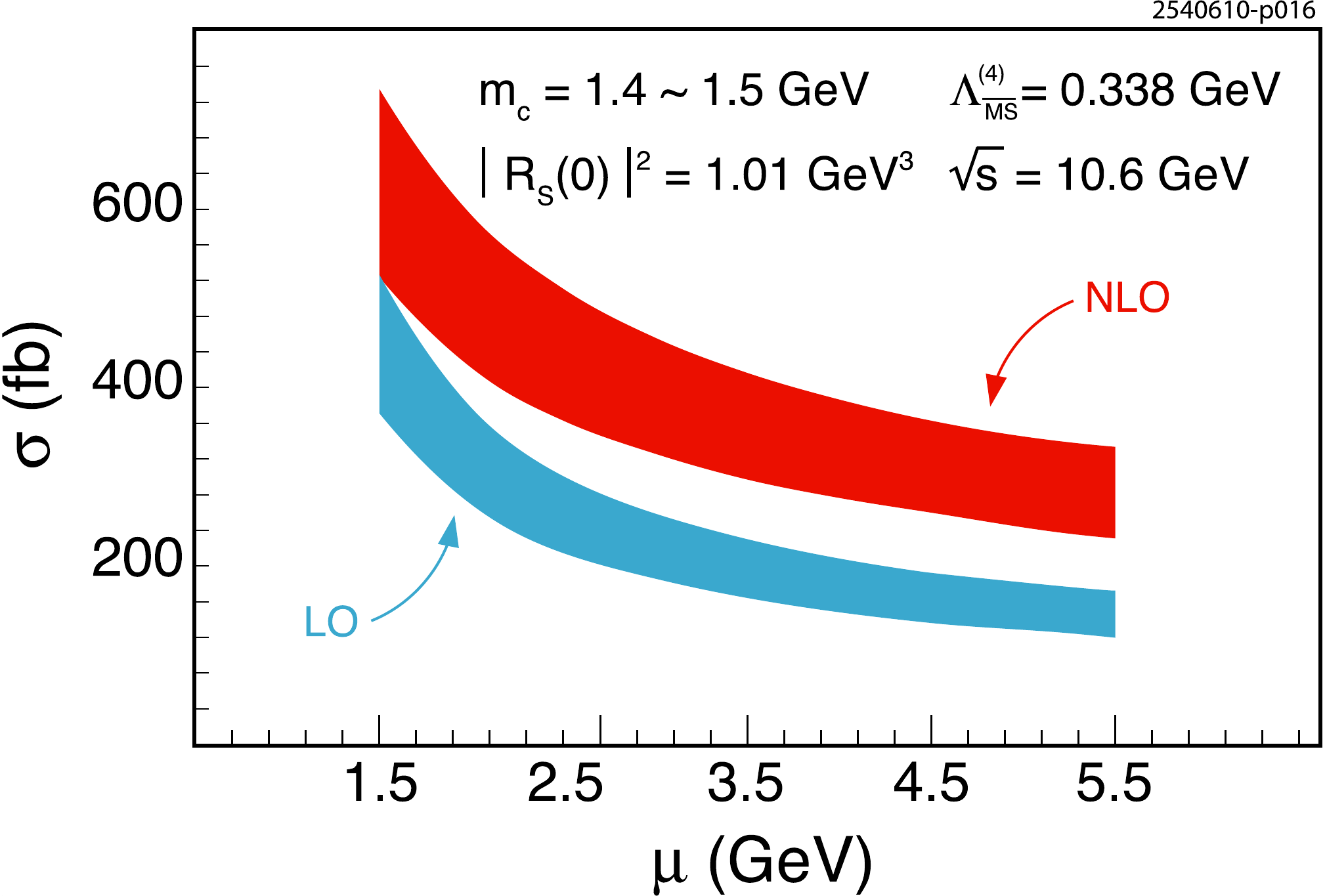}
\caption{Color-singlet contributions to the direct production cross
sections for $e^+ e^-\rightarrow J/\psi+c \bar c$ as functions of the
renormalization scale $\mu_R$. The following values of the input
parameters were taken: $|R_S(0)|^2=1.01~{\rm GeV}^3$,
$\Lambda_{\overline{\rm MS}}^{(4)}=0.338~{\rm GeV}$, $\sqrt{s}=10.6~{\rm
GeV}$. Results at NLO in $\alpha_s$ are represented by the 
{\it upper band}, and
results at LO in $\alpha_s$ are represented by the {\it lower band}. In each
case, the {\it upper border} corresponds to the input value $m_c=1.4~{\rm GeV}$
and the {\it lower border} corresponds to the input value $m_c=1.5~{\rm GeV}$.
\figPermXAPS{Zhang:2006ay}{2007} }
\label{prod_dependenceOFmu} 
\end{figure}

An important step toward resolving the puzzle of the $J/\psi$ production
cross section was taken in Ref.~\cite{Zhang:2006ay}, where it was found
that the correction of NLO in $\alpha_s$ to $e^+ e^- \to J/\psi+c \bar
c$ gives a large enhancement. In this work, a value for the square of
the $J/\psi$ wave function at the origin was obtained by comparing the
observed $J/\psi$ leptonic decay width ($5.55\pm 0.14\pm 0.02$~keV) with
the theoretical expression for that width, including corrections of NLO
in $\alpha_s$. (The square of the wave function at the origin is
proportional to the color-singlet NRQCD long-distance matrix element of
leading order in $v$.) The value $|R_S(0)|^2=1.01~\mbox{GeV}^3$ that was
obtained in this work is a factor of $1.25$ larger than the value
$|R_S(0)|^2=0.810~\mbox{GeV}^3$ that was obtained in potential-model
calculations and used in Ref.~\cite{Liu:2003jj} to calculate the $e^+
e^- \to J/\psi+c \bar c$ cross section.  Taking $m_c=1.5~$GeV, the
renormalization scale $\mu_R=2m_c$, and $\Lambda^{(4)}_{\overline {\rm
MS}}=0.338$~GeV, the authors of Ref.~\cite{Zhang:2006ay} found the
direct production cross section at NLO in $\alpha_s$ to be
\begin{equation}
\label{prod_jsetac} 
\sigma^{\rm NLO}(e^+e^-\rightarrow J/\psi+c \bar c+X)=0.33 \ \rm{pb},
\end{equation}
which is a factor of $1.8$ larger than the LO result ($0.18$ ~pb) that is
obtained with the same set of input parameters. Results for the NLO
cross section for other values of the input parameters can be found in
Table~\ref{prod_table:rc}. From Table~\ref{prod_table:rc} and
Fig.~\ref{prod_dependenceOFmu}, it can be seen that the
renormalization-scale dependence of the NLO cross section for 
$e^+e^-\rightarrow J/\psi+c \bar c$ is quite strong. This strong $\mu_R$
dependence is related to the large size of the NLO contribution relative to
the LO contribution. The cross section is also sensitive to the value of
the charm-quark mass. It is larger for smaller values of $m_c$.

\begin{table}[tb]
\caption{Color-singlet contributions to cross sections for $J/\psi$
production in $e^+e^-$ annihilation at the $B$-factory energy,
$\sqrt{s}=10.58$~GeV.  The table shows cross sections for the production
of $J/\psi+c\bar c$ (Ref.~\cite{Zhang:2006ay}) and $J/\psi +gg$
(Ref.~\cite{Ma:2008gq}), as well as the ratio $R_{c\bar c}$, which is
computed from the expression
in Eq.~(\ref{prod_R-cc}) by summing, in the denominator,
over only the $J/\psi+c\bar c$ and $J/\psi+gg$ 
cross sections. 
The cross
sections, in units of pb, are shown for different values of $m_c$ (1.4
and 1.5~GeV) and $\mu_R$ ($2m_c$ and $\sqrt{s}/2$), along with the
corresponding value of $\alpha_s(\mu_R)$. The prompt $J/\psi+c\bar c$
cross sections include feeddown from the $\psi(2S)$ and the $\chi_{cJ}$
states, while the prompt $J/\psi+gg$ cross sections include feeddown
from the $\psi(2S)$ state }
\label{prod_table:rc}
\setlength{\tabcolsep}{0.92pc}
\begin {center}
\begin{tabular}{lcccc}
 \hline\hline
\rule[10pt]{-1mm}{0mm}
$\mu_R$~(GeV)&$2.8$& $3.0$& $5.3$&$5.3$ \\[0.8mm]
$\alpha_s(\mu_R)$&0.267 & 0.259& 0.211& 0.211\\[0.8mm]
$m_c$~(GeV)&$1.4$ & $1.5$ & $1.4$ & $1.5$\\[0.8mm]
\hline
\rule[10pt]{-1mm}{0mm}
$\sigma^{\rm LO}(gg)$&0.42 & 0.32& 0.26& 0.22\\[0.8mm]
$\sigma^{\rm NLO}(gg)$&0.50 & 0.40& 0.39& 0.32\\[0.8mm]
$\sigma^{\rm NLO}_{\rm prompt}(gg)$&0.67 & 0.54& 0.53& 0.44\\[0.8mm]
\hline
\rule[10pt]{-1mm}{0mm}
$\sigma^{\rm LO}(c\bar c)$&0.27 & 0.18& 0.17&0.12 \\[0.8mm]
$\sigma^{\rm NLO}(c\bar c)$&0.47 & 0.33& 0.34&0.24 \\[0.8mm]
$\sigma^{\rm NLO}_{\rm prompt}(c\bar c)$&0.71 & 0.51& 0.53&0.39 \\[0.8mm]
\hline
\rule[10pt]{-1mm}{0mm}
$R^{\rm LO}_{c \bar c}$& 0.39& 0.36&0.40 & 0.35\\[0.8mm]
$R^{\rm NLO}_{c \bar c}$& 0.51& 0.49&0.50 & 0.47\\[0.8mm]
\hline\hline
\end{tabular}
\end {center}
\vspace{-0.5cm}
\end{table}

In Ref.~\cite{Zhang:2006ay} the QED contribution at order $\alpha_s
\alpha^3$ and the contribution from $e^+e^- \to 2\gamma^* \to \ J/\psi+
c {\bar c}$ were found to increase the $J/\psi +c\bar c$ cross section
by only a small amount. The feeddown contributions $e^+e^-
\to\psi(2S)+ c {\bar c} \to J/\psi+ c {\bar c}+X$ and $e^+e^-
\to\chi_{cJ}+ c {\bar c} \to J/\psi+ c {\bar c}+X$ were also estimated in
Ref.~\cite{Zhang:2006ay}. The primary feeddown contribution comes from
the $\psi(2S)$ and produces an enhancement factor of 1.355 for the prompt
$J/\psi c\bar c$ cross section.

Taking into account all of the aforementioned contributions, we
obtain the following estimate for the prompt cross section:
\begin{equation}
\label{prod_jsetacProm} \sigma^{\rm NLO}_{\rm prompt}(e^+ + e^-\rightarrow
J/\psi+c \bar c+X )=0.51\ \rm{pb},
\end{equation}
where the input values $m_c=1.5$~GeV and $\mu_R=2m_c$ have been used.
As is shown in Table~\ref{prod_table:rc}, despite the uncertainties in the
input parameters, the NLO correction to the color-singlet contribution to
$e^+ e^-\rightarrow J/\psi+c \bar c$ substantially increases the cross
section and largely reduces the discrepancy between experiment and
theory. Furthermore, the NLO relativistic correction to this process is
found to be negligible \cite{He:2007te}, in contrast with the NLO
relativistic correction to the process $e^+ + e^-\rightarrow
J/\psi+\eta_c$ (\Sec{prod_sec:exclusive}).

Recently, the authors of Ref.~\cite{Gong:2009ng} confirmed the results
of Ref.~\cite{Zhang:2006ay} and presented a more detailed analysis of
the $J/\psi$ angular distributions and polarization parameters, using
slightly different input parameters than those in
Ref.~\cite{Zhang:2006ay}.

\subsubsection{$e^+ e^- \to J/\psi+gg$ at NLO }
\label{prod_sec:psi-gg-nlo}

\begin{figure}[b]
\includegraphics[width=\figwid]{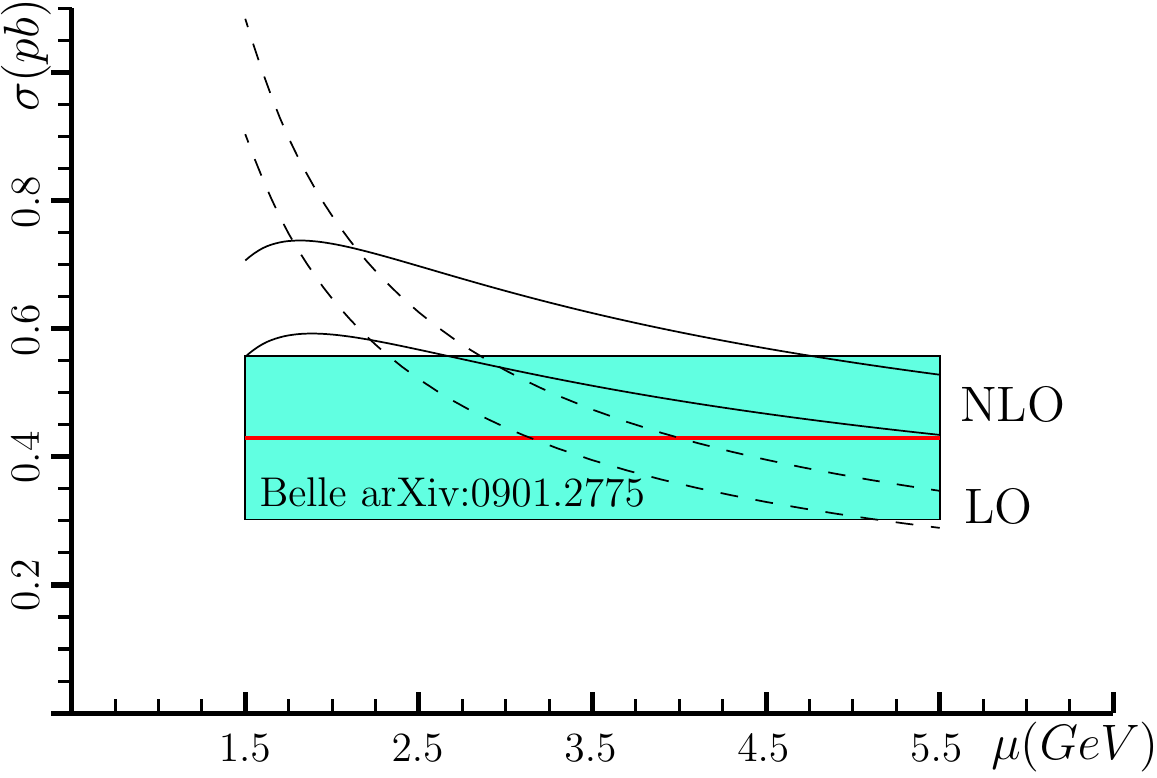}
\caption{Prompt cross sections for the process $e^+ e^- \to J/\psi +gg$.
The {\it curves} are the predictions of Ref.~\cite{Ma:2008gq}, plotted as
functions of the renormalization scale $\mu_R$ at LO {\it (dashed 
curves)} and
NLO {\it (solid curves)} in $\alpha_s$. In each case, the {\it upper 
curves}
correspond to $m_c=1.4$~GeV, and the {\it lower curves} correspond to
$m_c=1.5$~GeV. The experimental datum {\it (shaded band)} is for the process
$e^{+}e^{-}\to J/\psi+X_{{\rm non-}c\bar{c}}$
(Ref.~\cite{Pakhlov:2009nj}). \figPermXAPS{Ma:2008gq}{2009} }
\label{prod_fig:psiggdepmu}
\end{figure}

In NRQCD factorization, the production cross section for the $J/\psi$ in
association with light hadrons, $\sigma(e^+e^-\to J/\psi +X_{{\rm
non-}c\bar c})$, includes the color-singlet contribution  $\sigma(J/\psi
+gg)$, and the color-octet contributions
$\sigma(J/\psi({}^3P_J^{[8]},{}^1S_0^{[8]})+g)$. Contributions from
other Fock states are suppressed by powers of $\alpha_s$ or $v$. The
corrections of NLO in $\alpha_s$ to $\sigma( J/\psi +gg)$ were
calculated in Refs.~\cite{Ma:2008gq,Gong:2009kp} and found to enhance
the LO cross section by about 20--30\%. The prompt production cross
section $\sigma(e^+e^-\to J/\psi + gg)$ at NLO in $\alpha_s$, including
the feeddown contribution from the $\psi(2S)$, can be found in
Table~\ref{prod_table:rc}. From Fig.~\ref{prod_fig:psiggdepmu}, it can
be seen that the renormalization-scale dependence at NLO is moderate and
much improved in comparison with the renormalization-scale dependence at
LO. It can also be seen that the NLO result is consistent with the
latest Belle measurement of $\sigma(e^+e^-\rightarrow J/\psi+X_{{\rm
non-}c\bar c})$ (Ref.~\cite{Pakhlov:2009nj}), given the experimental
uncertainties. Resummation of the leading logarithms near the kinematic
endpoint of the $J/\psi$ momentum distribution is found to change the
endpoint momentum distribution, but to have only a small effect on the
total $J/\psi+gg$ cross section \cite{Ma:2008gq}.

Results for the color-singlet contributions to $\sigma(e^+ e^-\to 
J/\psi+c\bar{c})$ (Ref.~\cite{Zhang:2006ay}), $\sigma( J/\psi +gg)$
(Ref.~\cite{Ma:2008gq}), and the corresponding ratio $R_{c \bar c}$ are
summarized in Table~\ref{prod_table:rc}. In Table~\ref{prod_table:rc}, the
color-octet contribution $\sigma(J/\psi(^3P_J^{[8]},^1S_0^{[8]})+g)$ is
ignored. It can be seen that the NLO results significantly reduce the
discrepancies between theory and experiment.

\begin{figure}[b]
\includegraphics[width=\figwid]{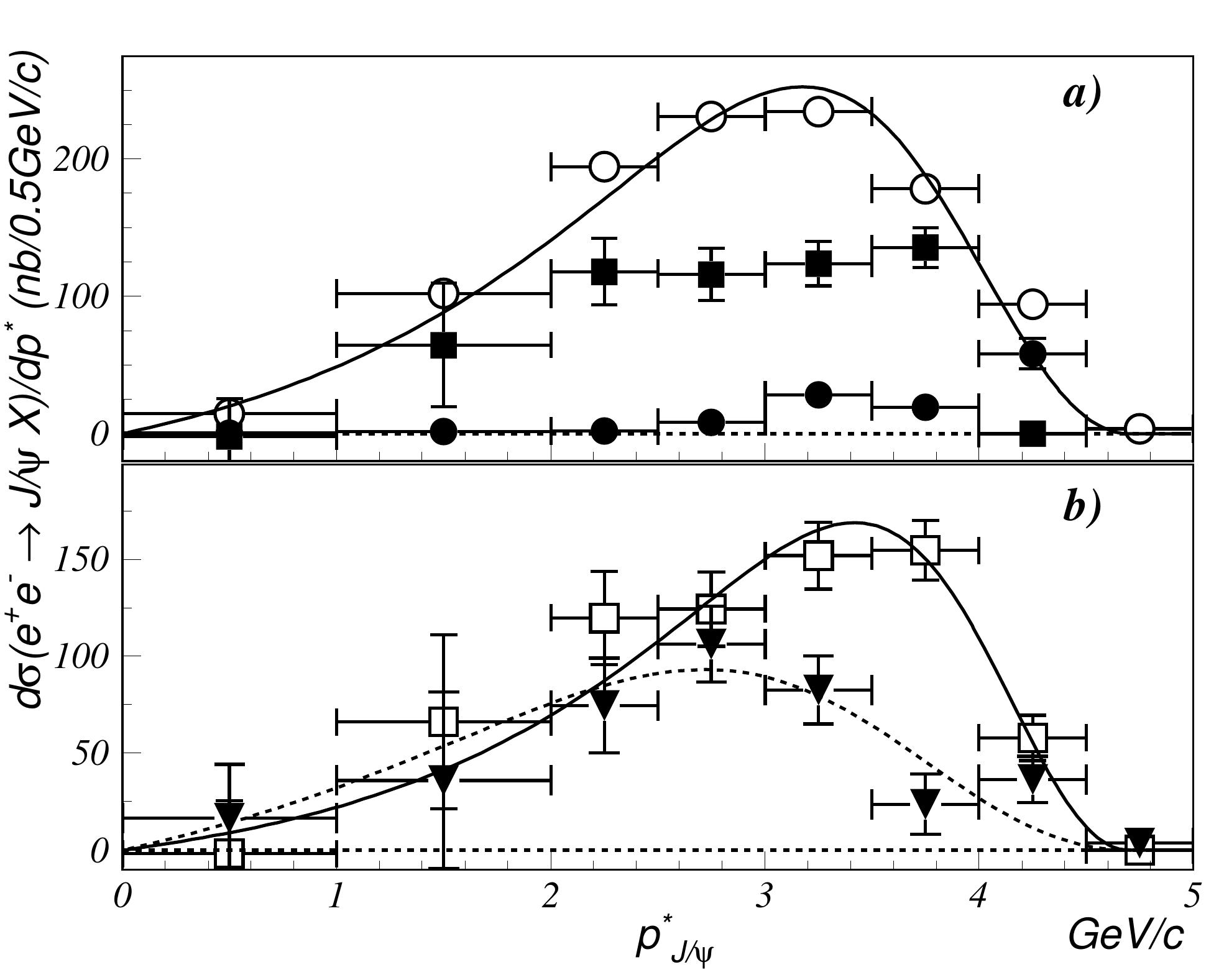}
\caption{$J/\psi$ momentum distributions: (a) 
the Belle measurements \cite{Pakhlov:2009nj} for the inclusive distribution
{\it (open circles)}, the distribution from $e^+e^-\to J/\psi +H_c +X$ 
{\it (filled squares)}, where $H_c$ is a charmed hadron, and the distribution from
double-charmonium production {\it (filled circles)}; (b) 
the Belle measurements \cite{Pakhlov:2009nj} for the distribution for
the sum of all $e^+e^-\to J/\psi+c\bar c$ processes {\it (open squares)} and
the distribution from the $e^+e^-\to J/\psi+X_{{\rm non-}c\bar c}$
processes {\it (filled triangles)}. The {\it curves} are the results of using the
Peterson function \cite{Peterson:1982ak} to fit the inclusive
distribution {\it [solid curve in (a)]}, the $e^+e^-\to J/\psi + c\bar c$
distribution {\it [solid curve in (b)]}, and the $e^+e^-\to J/\psi+ X_{{\rm
non-}c\bar c}$ distribution {\it [dashed curve in (b)]}. 
\figPermXAPS{Pakhlov:2009nj}{2009} }
\label{prod_fig:momentum}
\end{figure}

In Fig.~\ref{prod_fig:momentum}, the measured distributions in the $J/\psi$
momentum $p^*$ are shown for the processes $e^+e^-\to J/\psi +c\bar c$
and $e^+e^-\to J/\psi + X_{{\rm non-}c\bar c}$
(Ref.~\cite{Pakhlov:2009nj}). The calculated $J/\psi$ momentum
distributions for the process $e^+e^-\to J/\psi+gg$ at NLO in $\alpha_s$
(Refs.~\cite{Ma:2008gq,Gong:2009kp}) are roughly compatible with the
$e^+e^-\to J/\psi+X_{{\rm non-}c\bar c}$ data. We note that the NLO
$J/\psi$ momentum distribution is much softer than the LO $J/\psi$
momentum distribution (see Fig.~6 of Ref.~\cite{Ma:2008gq}), resulting
in better agreement with data.

Finally, it was found in Ref.~\cite{He:2009uf} that the
relative-order-$v^2$ relativistic correction can enhance the
$e^{+}e^{-}\to J/\psi+gg$ cross section by 20--30\%, which is comparable
to the enhancement that arises from the corrections of NLO in $\alpha_s$
(Ref.~\cite{Ma:2008gq,Gong:2009kp}). This relativistic correction has
been confirmed in Ref.~\cite{Jia:2009np}. If one includes both the
correction of NLO in $\alpha_s$ and the relative-order-$v^2$
relativistic correction, then the color-singlet contribution to
$e^{+}e^{-}\to J/\psi+gg$ saturates the latest observed cross section
for $e^{+}e^{-}\to J/\psi+X_{{\rm non-}c\bar c}$ in
Eq.(\ref{prod_eq:nonCCBar}), even if a significantly smaller color-singlet
matrix element is chosen. This leaves little room for the color-octet
contribution $\sigma(J/\psi(^3P_J^{[8]},^1S_0^{[8]}) +g)$ and may imply
that the true values of the color-octet matrix elements are much smaller than
those that have been extracted in LO fits to the Tevatron data or those
that would be expected from a naive application of the NRQCD
velocity-scaling rules.

\subsubsection{Color-octet process $e^+ e^- \to J/\psi({}^3P_J^{[8]},{}^1S_0^{[8]})+g$ }

The color-octet contribution to the cross section at LO in $\alpha_s$ is
given at LO in $v$ by $\sigma(J/\psi(^3P_J^{[8]},^1S_0^{[8]})+g)$. This
contribution was calculated in Ref.~\cite{Braaten:1995ez}, and an
enhancement near the kinematic endpoint, $z=1$, was predicted. Here,
$z=E_{c\bar c}/E_{c\bar c}^{\rm max}$ is the energy of the $c\bar c$
pair divided by the maximum possible energy of the $c\bar c$ pair. As
can be seen from Fig.~\ref{prod_fig:momentum}, measurements of the
$J/\psi$ momentum distribution do not show any enhancement near the
kinematic endpoint. In Ref.~\cite{Fleming:2003gt}, resummations of the
NRQCD velocity expansion near the endpoint and resummations of
logarithms of $1-z$ were considered. These resummations smear out the
peak near $z=1$ and shift it to smaller values of $z$, making the theory
more compatible with the data. However, the resummation results rely
heavily on a nonperturbative shape function that is not well known, and
so it is not clear if they can reconcile the theoretical and
experimental results. The non-observation of an enhancement near $z=1$
might also point to a possibility that we mentioned in
\Sec{prod_sec:psi-gg-nlo}, namely, that the ${}^3P_J$ and ${}^1S_0$
color-octet matrix elements are much smaller than would be expected from
the LO fits to the Tevatron data or the NRQCD velocity-scaling rules.

Very recently, the corrections at NLO in $\alpha_s$ to the color-octet
contribution to inclusive $J/\psi$ production have been calculated
\cite{Zhang:2009ym}. In comparison with the LO result, the NLO
contributions are found to enhance the short-distance coefficients in the
color-octet contributions 
$\sigma(e^+ e^-\to c \bar c({}^1S_0^{(8)})+g)$
and $\sigma(e^+ e^-\to c \bar c({}^3P_J^{(8)})+g)$ (with $J=0$, $1$, $2$)
by a factor of about $1.9$. Moreover, the NLO corrections smear the peak
at the endpoint in the $J/\psi$ energy distribution, although the bulk
of the color-octet contribution still comes from the region of large
$J/\psi$ energy. One can obtain an upper bound on the sizes of the
color-octet matrix elements by setting the color-singlet contribution to
be zero in $\sigma(e^{+}e^{-}\to J/\psi+X_{{\rm non-}c\bar{c}})$.  The
result, at NLO in $\alpha_s$, is 
\begin{eqnarray}
&&\langle 0| {\cal
O}^{J/\psi}({}^1S_0^{(8)})|0\rangle + 4.0\langle0| {\cal
O}^{J/\psi} ({}^3P_0^{(8)})|0\rangle/m_c^2 \nonumber\\
&&\qquad <(2.0 \pm 0.6)\times
10^{-2}~{\rm GeV}^3.
\label{prod_co-bound}
\end{eqnarray}
This bound is smaller by about a factor of $2$ than the values that were
extracted in LO fits to the Tevatron data for the combination of matrix
elements in Eq.~(\ref{prod_co-bound}).

\subsubsection{$J/\psi$ production in $\gamma\gamma$ collisions}

Photon-photon ($\gamma\gamma$) collisions can be can be studied at
$e^+e^-$ colliders by observing processes in which the incoming $e^+$
and $e^-$ each emit a virtual $\gamma$ that is very close to its mass
shell. The inclusive cross section differential in $p_T$ for the
production of $J/\psi$ in $\gamma\gamma$ collisions at LEP has been
measured by the DELPHI collaboration
\cite{TodorovaNova:2001pt,Abdallah:2003du}.

The cross section differential in $p_T$ has been calculated in the CSM 
and the NRQCD factorization approach at LO in $\alpha_s$
(Refs.~\cite{Ma:1997bi,Japaridze:1998ss,Godbole:2001pj,Klasen:2001mi,Klasen:2001cu}).
The computations include three processes: the direct-$\gamma$ process
$\gamma\gamma\to (c\bar c)+g$, which is of order $\alpha^2\alpha_s$; the
single-resolved-$\gamma$ process $i\gamma\to (c\bar c)+i$, which is of
order $\alpha\alpha_s^2$; and the double-resolved-$\gamma$ process
$ij\to (c\bar c)+k$, which is of order $\alpha_s^3$. Here, $ij=gg$,
$gq$, $g\bar q$, or $q\bar q$, where $q$ is a light quark. Because the
leading contribution to the distribution of a parton in a $\gamma$ is of
order $\alpha/\alpha_s$, all of the processes that we have mentioned
contribute to the $\gamma\gamma$ production rate in order
$\alpha^2\alpha_s$. The corrections of NLO in $\alpha_s$ to the
direct-$\gamma$ process have been computed in
Ref.~\cite{Klasen:2004tz}.

\begin{figure}[b]
\begin{center}
\includegraphics[width=\figwid]{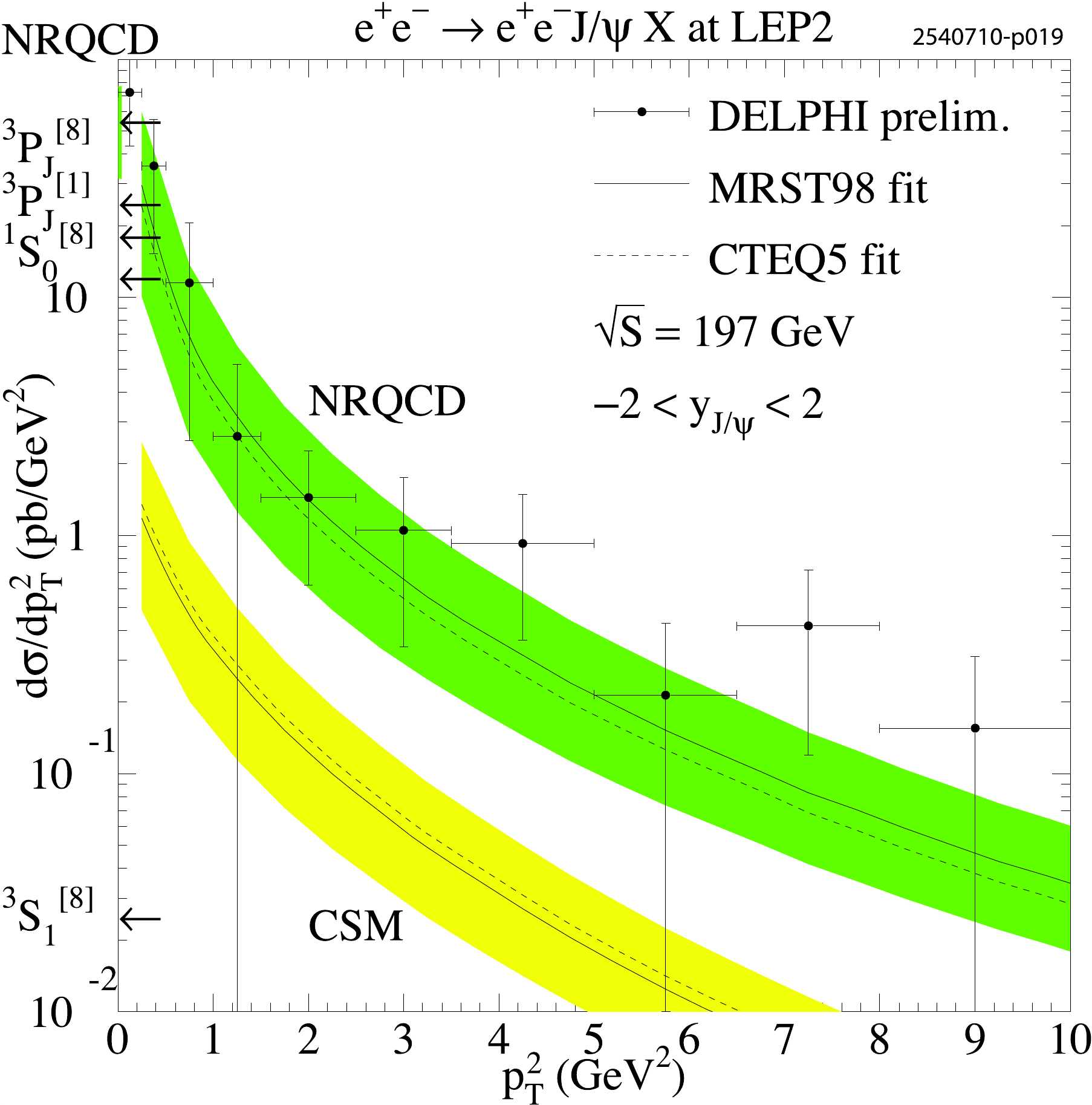}
\caption{Comparison of the inclusive cross section differential in
$p_T$ for $\gamma \gamma$ production of the $J/\psi$
(Refs.~\cite{TodorovaNova:2001pt,Abdallah:2003du}) with the predictions
at LO in $\alpha_s$ of the CSM and the NRQCD factorization approach. The
{\it upper two curves} are the NRQCD factorization predictions, and the 
{\it lower two curves} are the CSM predictions. The {\it solid and dashed 
curves} correspond to the MRST98LO (Ref.~\cite{Martin:1998sq}) and the CTEQ5L
(Ref.~\cite{Lai:1999wy}) parton distributions, respectively. The 
{\it arrows} indicate the NRQCD factorization predictions at $p_T=0$ for the
${}^3P_J^{[1]}$, ${}^1S_0^{[8]}$, ${}^3S_1^{[8]}$, and ${}^3P_J^{[8]}$
contributions. The theoretical uncertainty bands were obtained by
combining the uncertainties from $m_c=1.5\pm 0.1$~GeV, the decay
branching fractions of the $\psi(2S)$ and the $\chi_{cJ}$ states, the
parton distributions, and the NRQCD long-distance matrix elements with
the uncertainties that are obtained by varying the renormalization
factorization scales between $2m_T$ and $m_T/2$. Here
$m_T=\sqrt{4m_c^2+p_T^2}$. \figPermXAPS{Klasen:2001cu}{2002} }
\label{fig:gamma-gamma-psi}
\end{center}
\end{figure}

A comparison of the LO CSM and NRQCD factorization predictions
with the DELPHI data is shown in Fig.~\ref{fig:gamma-gamma-psi}. The data
clearly favor the NRQCD factorization prediction. However, it should be
kept in mind that there may be large NLO corrections to the
color-singlet contribution, as is the case in $pp$ and $ep$ charmonium
production. Therefore, no firm conclusions can be drawn until a complete
NLO calculation of the production cross section is available.

\subsubsection{Future opportunities}

The central values for the prompt $J/\psi$ inclusive production cross
section that were obtained by the \babar\ collaboration and the Belle
collaboration differ more than a factor of $2$. It would certainly be
desirable to clear up this discrepancy. Furthermore, the \babar\
collaboration has not presented results for $\sigma(e^+e^-\rightarrow
J/\psi+c\bar c)$ and $\sigma(e^+e^-\rightarrow J/\psi+X_{{\rm non}-c\bar
c})$. It is very important that the \babar\ collaboration check the Belle
results for these cross sections, which play a central role in efforts
to understand the mechanisms of inclusive quarkonium production.
Additionally, measurements of greater accuracy of the charmonium angular
distributions and polarization parameters would be useful in
understanding the mechanisms of quarkonium production. Measurements of
inclusive cross sections for the production of charmonium states other
than the $J/\psi$ might also yield important clues regarding the
production mechanisms.

In the theoretical prediction for the color-singlet contribution to
$\sigma(e^+e^-\to J/\psi +c\bar c)$, large uncertainties arise from the
uncertainties in the renormalization scale $\mu_R$ and $m_c$. (The
uncertainties in the theoretical prediction for the color-singlet
contribution to $\sigma(e^+e^-\to J/\psi +gg)$ seem to be under rather
good control.) The uncertainties in the color-singlet contribution to
$\sigma(e^+e^-\to J/\psi +c\bar c)$ might be reduced by understanding
the origins of the large corrections at NLO in $\alpha_s$ and by taking
advantage of recent progress in determining $m_c$
(Ref.~\cite{McNeile:2010ji}). It is important to examine further whether
the observed value of $\sigma(e^+e^-\rightarrow J/\psi+X_{{\rm
non}-c\bar c})$ is actually saturated by $\sigma(e^+e^-\to J/\psi +gg)$,
which would imply that the color-octet($^3P_J^{[8]},^1S_0^{[8]}$)
contributions are negligible. In this regard, one might obtain
additional information by examining the $J/\psi$ angular distribution
and polarization parameters, in addition to the total cross section
and momentum distribution. Finally, we mention, as we did in
\Sec{prod_sec:ep-fo}, that theoretical uncertainties in the region near
the kinematic endpoint $z=1$ might be reduced through a systematic study
of resummations of the perturbative and velocity expansions in both $ep$
and $e^+e^-$ quarkonium production.

Quarkonium production in $\gamma\gamma$ collisions provides yet another
opportunity to understand quarkonium production mechanisms. In order to
make a definitive comparison of the DELPHI data for the $J/\psi$ cross
section differential in $p_T$ with the NRQCD factorization prediction,
it is necessary to have a complete calculation of all of the direct and
resolved contributions at least through NLO in $\alpha_s$. Further
measurements of quarkonium production in $\gamma \gamma$ collisions
should be carried out at the next opportunity at an $e^+e^-$ or $ep$
collider. It may also be possible to measure quarkonium production in 
$\gamma\gamma$ and $\gamma p$ collisions at the LHC 
\cite{deFavereaudeJeneret:2009db}.

\subsection{$B_c$ production}
\label{sec:Prod_Bc}
\subsubsection{Experimental progress}

The first observation of the $B_c$ was reported by the CDF collaboration
in Ref.~\cite{Abe:1998wi}. Subsequently, this unique
double-heavy-flavored meson has been observed by both the CDF and 
\DZero\ collaborations at the Tevatron via two decay channels: $B_c\to
J/\psi+\bar{l}\nu_l$ and $B_c\to J/\psi+\pi^+$
(Refs.~\cite{Abulencia:2006zu,Abulencia:2005usa,Abazov:2008kv,Abazov:2008rba,Aaltonen:2007gv}).
Using an event sample corresponding to an integrated luminosity of
$360$~pb$^{-1}$ at $\sqrt s=1.96$~TeV, the CDF collaboration has
measured the $B_c$ lifetime in the decay $B_c^+ \to J/\psi e^+ \nu_e$
(Ref.~\cite{Abulencia:2006zu}) and obtained
\begin{equation}
\tau_{B_c}=0.463^{+0.073}_{-0.065} \pm 0.036 \hbox{~ps}\,.
\end{equation}
The $B_c$ lifetime has also been measured by the CDF collaboration in
the decay $B_c\to J/\psi+l^\pm +X$ (Ref.~\cite{prod_cdf_bc_semilept}).
The result of this measurement, which is based on an event sample
corresponding to an integrated luminosity of $1$~fb$^{-1}$ at $\sqrt
s=1.96$~TeV, is
\begin{equation}                                                         
\tau_{B_c}=0.475^{+0.053}_{-0.049} \pm 0.018 \hbox{~ps}\,.               
\end{equation}

The CDF collaboration has measured the $B_c$ mass in the decay $B_c^\pm
\to J/\psi \pi^\pm$ (Refs.~\cite{Abulencia:2005usa,Aaltonen:2007gv}),
obtaining in its most recent measurement \cite{Aaltonen:2007gv}, which
is based on an integrated luminosity of $2.4$~fb$^{-1}$ at
$\sqrt{s}=1.96$~TeV, the value
\begin{equation}
m_{B_c}=6275.6 \pm 2.9 \pm 2.5 \hbox{~MeV}\,.
\end{equation}
The \DZero\ collaboration, making use of an event sample based on an
integrated luminosity of $1.3$~fb$^{-1}$ at $\sqrt{s}=1.96$~TeV, has
also provided measurements of the  $B_c$ lifetime \cite{Abazov:2008kv},
\begin{equation}
\tau_{B_c}=0.448^{+0.038}_{-0.036} \pm 0.032 \hbox{~ps},
\end{equation}
and the $B_c$ mass \cite{Abazov:2008rba},
\begin{equation}
m_{B_c}=6300 \pm 14 \pm 5 \hbox{~MeV}\,.
\end{equation}
The results obtained by the two collaborations are consistent with each
other. (See \Sec{sec:SpecExp_Bc} of this article for a further
discussion of the $B_c$ mass and lifetime.)

Recently, the CDF collaboration has updated a previous measurement
\cite{prod_cdf_rbc_old} of the ratio
\begin{equation}
R_{B_c}=\frac{\sigma(B^+_c)\Brat(B_c\to
\jpsi+\mu^++\nu_\mu)}{\sigma(B^+)\Brat(B^+\to \jpsi+K^+)}.
\end{equation}
The new analysis \cite{prod_CDF3}, which used a data set that
corresponds an integrated luminosity of $1$~fb$^{-1}$, yielded the
results
\begin{equation}
R_{B_c}=0.295 \pm
0.040({\rm stat})^{+0.033}_{-0.026} ({\rm sys} )\pm0.036 (p_T),
\end{equation}
for $p_T > 4\hbox{~GeV}$, and
\begin{equation}
R_{B_c}=0.227 \pm
0.040 ({\rm stat})^{+0.024}_{-0.017}({\rm sys} )\pm0.014 (p_T),
\end{equation}
for $p_T > 6\hbox{~GeV}$,  where $p_T$ is the $B_c$ transverse momentum.
The measurements in Refs.~\cite{prod_cdf_rbc_old,prod_CDF3} provided the
first, indirect, experimental information on the production cross
section.

\subsubsection{Calculational schemes}

Experimental studies of $B_c$ production could help to further
theoretical progress in understanding the production mechanisms for
heavy-quark bound states. On the other hand, experimental observations
of the $B_c$ are very challenging and might benefit from theoretical
predictions of $B_c$ production rates, which could be of use in devising
efficient observational strategies, for example, in selecting decay
channels to study.

So far, two theoretical approaches have been used to obtain predictions
for $B_c$  hadroproduction. Both of them are based on the NRQCD
factorization approach.

\begin{figure}[b!]\centering
\includegraphics[width=\figwid]{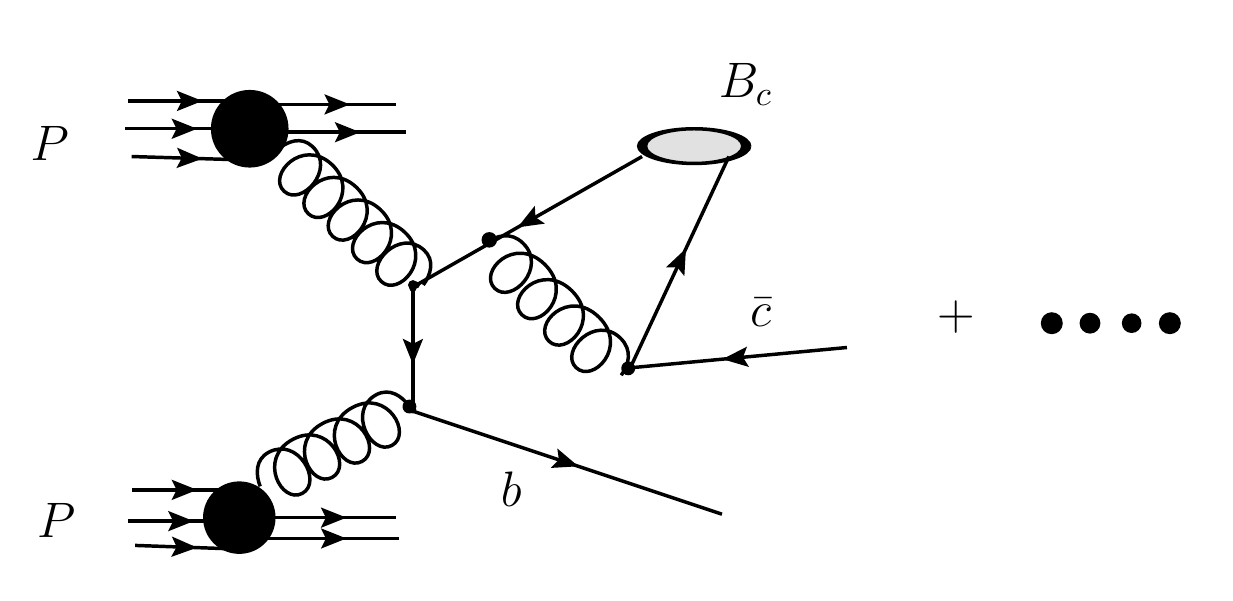}
\caption{Typical Feynman diagram for $B_c$ production via gluon-gluon
fusion in the fixed-order approach at order $\alpha_s^4$ }
\label{prod_fig:fixed-order} 
\end{figure}

The simplest approach conceptually is to calculate the contributions
from all of the hard subprocesses, through a fixed  order in $\alpha_s$,
that produce a $c\bar c$ pair and a $b\bar b$ pair. The $c$ quark
(antiquark) is required to be nearly co-moving with the $b$ antiquark
(quark) in order to produce a  $B_c$. We call this approach the
``fixed-order approach.'' A typical Feynman diagram that appears in this
approach is shown in Fig.~\ref{prod_fig:fixed-order}. If the
initial-state partons are light ({\it i.e.}, gluons, or $u$, $d$, or $s$
quarks or antiquarks), then the leading order for $B_c$ production in
the fixed-order approach is $\alpha_s^4$. LO computations in
this approach can be found in
Refs.~\cite{Beenakker:1988bq,Nason:1989zy,Chang:1992jb,Chang:1994aw,Kolodziej:1995nv,Berezhnoy:1996an,Baranov:1997sg,Berezhnoi:1997uz,Frixione:1997ma}.

\begin{figure}[b!]\centering
\includegraphics[width=\figwid]{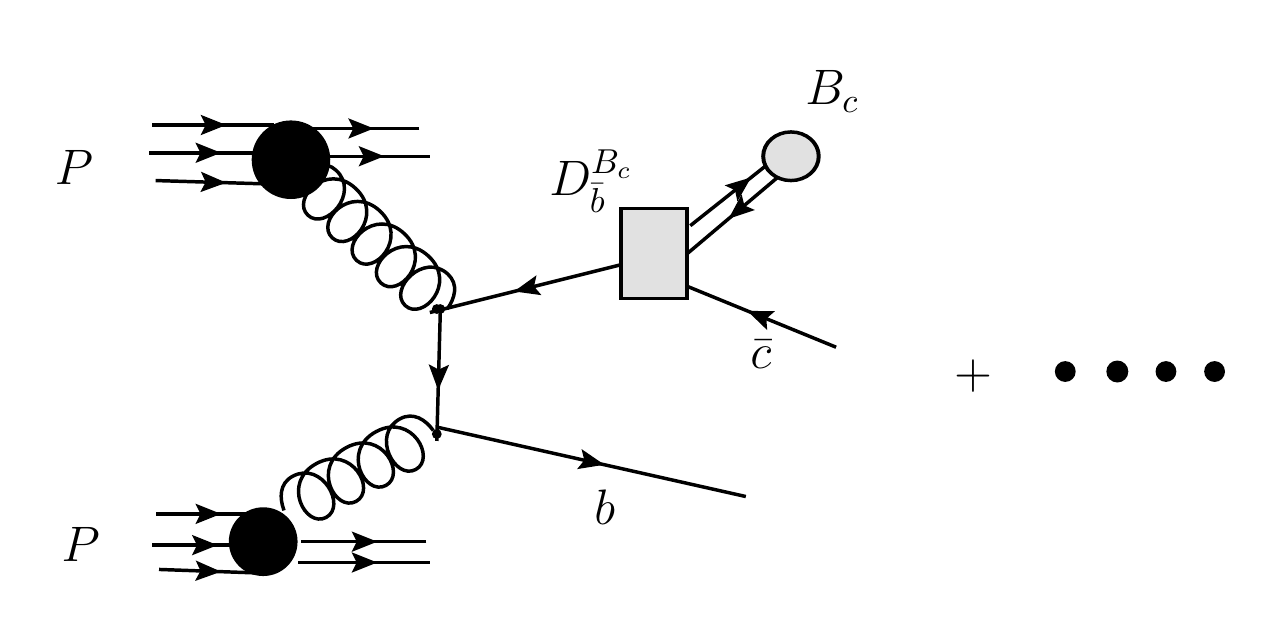} \caption{Typical
Feynman diagram for $B_c$ production via gluon-gluon fusion in the
fragmentation approach at order $\alpha_s^2$. $D_{\bar{b}}^{B_c}$ is the 
fragmentation function for a $b$ antiquark into a $B_c$ }
\label{prod_fig:frag} 
\end{figure}

An alternative approach is the fragmentation approximation, in which
the production process is factorized into convolutions of fragmentation
functions with a simple perturbative-QCD hard-scattering sub-process
\cite{Braaten:1993jn,Cheung:1999ir}. An example of such a factorized
contribution is one in which $pp \to b\bar b$, with the final-state $b$
or $\bar b$ fragmenting to $B_c + \bar c$. This process is illustrated
in Fig.~\ref{prod_fig:frag}. The fragmentation approximation drastically
simplifies the calculation and also provides a formalism with which to
resum large final-state logarithms of $p_T/m_b$. However, it has been
shown in Ref.~\cite{Chang:1996jt} that the fragmentation diagrams are not
dominant at the Tevatron unless the $B_c$ is produced at very large
(experimentally inaccessible) values of $p_T$. A similar behavior has
been found in Ref.~\cite{Artoisenet:2007xi} in the case of charmonium or
bottomonium production at the Tevatron and the LHC in association with a
heavy-quark pair. Therefore, we can conclude that there is no need to
resum logarithms of $p_T/m_b$ at the values of $p_T$ that are accessible
at the current hadron colliders. That is, a fixed-order calculation is
sufficient, and the fragmentation approach is not relevant.

Within either the fixed-order approach or the fragmentation approach,
various choices of factorization scheme are possible. In the simplest
scheme, which is known as the fixed-flavor-number (FFN) scheme, the
flavor content of the proton is held fixed and includes only the light
flavors. Specifically, it is assumed that only the light quarks are
``active'' flavors, which means that there are contributions involving
light-quark (and antiquark) parton distributions, but there are no
contributions involving $c$- and $b$-quark (and antiquark) parton
distributions.

\begin{figure}[t!]\centering
\includegraphics[width=\figwid]{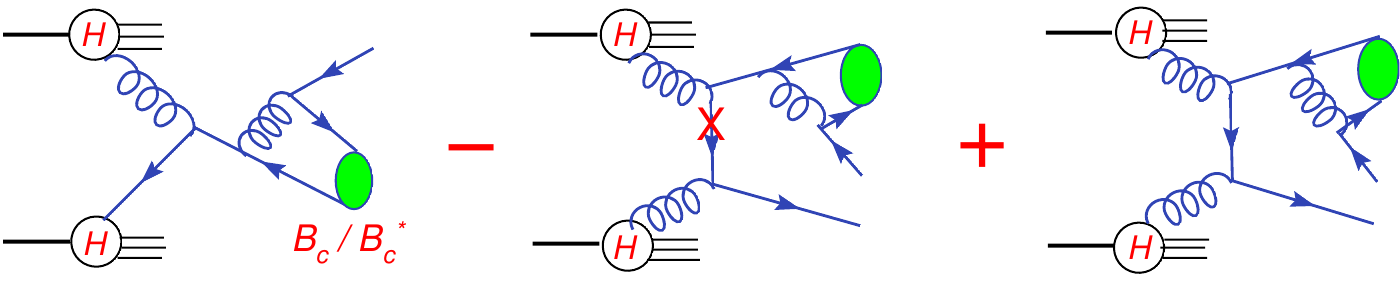} 
\caption{Typical Feynman diagrams for $B_c$ production via gluon-gluon
fusion in a GM-VFN scheme. The middle diagram represents a subtraction that
removes double-counting of the contribution in the right-hand diagram in
which the lower initial-state gluon produces a heavy quark-antiquark
pair that is collinear to that gluon. \figPermXAPS{Chang:2006eu}{2006} }
\label{prod_fig:VFN} 
\end{figure}

Alternatives to the FFN scheme have been proposed with the aim of
improving upon FFN calculations by resumming large initial-state
collinear logarithms of $Q/m_c$ and $Q/m_b$, where $Q$ is a kinematic
scale in the production process ({\it e.g.}, $p_T$). Such logarithms
arise, for example, in a process in which an initial-state gluon splits
into a quark-antiquark pair. The quark or antiquark then participates in
a hard scattering, giving rise to a $B_c$. (An example of such a process
is $g \to c \bar c$, followed by $g\bar c \to B_c \bar b$.) It has been
suggested that one can resum these initial-state collinear logarithms by
making use of heavy-flavor parton distributions. The basic idea is that
one can absorb logarithms of $Q/m_c$ or $Q/m_b$ into massless $c$-quark
(or antiquark) and $b$-quark (or antiquark) parton distributions, in
which they can be computed through DGLAP evolution.  In such methods,
there are contributions in which the $B_c$ is produced in a $2 \to 2$
scattering that involves a $c$ or $b$ initial-state parton. Because the
contributions involving massless $c$ and $b$ parton distributions are
good approximations to the physical process only at large $Q$, the
formalism must suppress these contributions when $Q$ is of the order of
the heavy-quark mass or less. A factorization scheme in which the number
of active heavy quarks varies with $Q$ in this way is called a
general-mass
variable-flavor-number (GM-VFN) scheme. Typical Feynman diagrams that enter
into the calculation of $B_c$ production in a GM-VFN scheme are shown in
Fig.~\ref{prod_fig:VFN}.

\begin{figure}
\centering
\includegraphics[width=\figwid]{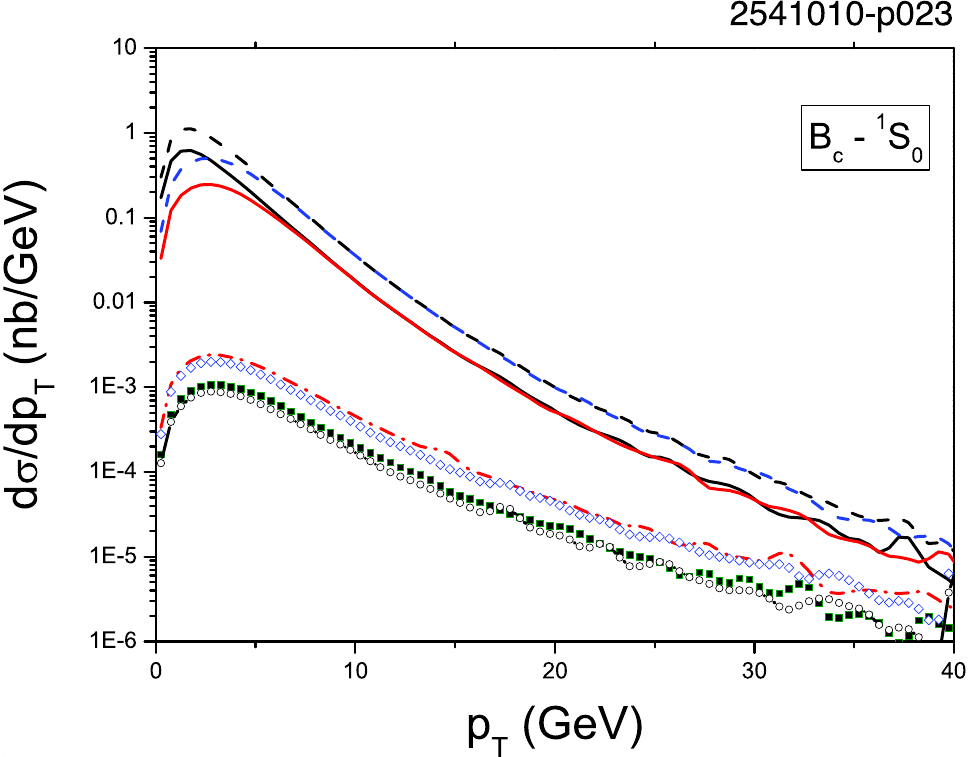}%
\caption{Predictions for $p_T$ distributions of the $B_c({}^1S_0)$ in
production at the Tevatron \cite{Chang:2005wd}. Four pairs of {\it 
curves} are
shown, corresponding, from top to bottom, to the following
contributions: $gg$-fusion with the cut $|y|<1.3$, $gg$-fusion with the
cut $|y|<0.6$, $q\bar q$-annihilation with the cut $|y|<1.3$, and $q\bar
q$-annihilation  with the cut $|y|<0.6$.  Here, $q$ is a light quark. 
In each pair of {\it curves}, the {\it upper curve} is the GM-VFN prediction, 
and the {\it lower curve} is the FFN prediction }
\label{prod_fig005}
\end{figure}

\begin{figure}
\centering
\includegraphics[width=\figwid]{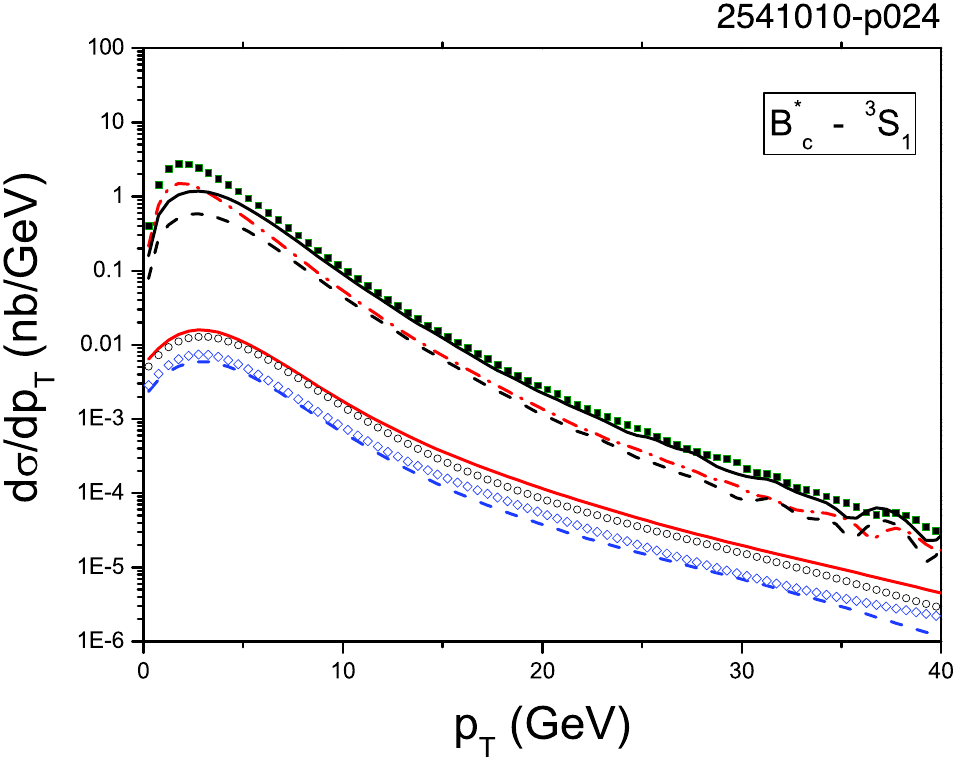}
\caption{Predictions for $p_T$ distributions of the $B^*_c({}^3S_1)$ in
production at the Tevatron \cite{Chang:2005wd}. Four pairs of curves are
shown, corresponding, from top to bottom, to the following
contributions: $gg$-fusion with the cut $|y|<1.3$, $gg$-fusion with the
cut $|y|<0.6$, $q\bar q$-annihilation with the cut $|y|<1.3$, and $q\bar
q$-annihilation  with the cut $|y|<0.6$. Here, $q$ is a light quark. In
each pair of {\it curves}, the {\it upper curve} is the GM-VFN prediction, and 
the {\it lower curve} is the FFN prediction }
\label{prod_fig006}
\end{figure}

The GM-VFN scheme was first applied to $B_c$ production in
Refs.~\cite{Chang:2006eu,Chang:2005wd}. 
(Earlier calculations  
\cite{Aivazis:1993pi,Aivazis:1993kh,Olness:1997yc,Amundson:2000vg} 
in the GM-VFN scheme did {\it not} address $B_c$ production.) 
Comparisons [845] of results from both FFN and GM-VFN schemes
for $B_c({}^1S_0)$ and $B_c({}^3S_1)$ production at the
Tevatron are shown in Figs.~\ref{prod_fig005} and \ref{prod_fig006},
respectively.  The FFN and GM-VFN approaches yield different results at
small $p_T$, possibly because, in the implementation of the GM-VFN
scheme in Ref.~\cite{Chang:2005wd}, the factorization scale
$\mu_F=\sqrt{p_T^2+m_{B_c}^2}$ was chosen, which implies that there are
contributions from initial $c$ and $b$ quarks down to $p_T=0$. Outside
the small-$p_T$ region, the FFN and GM-VFN approaches give very similar
results. This implies that the resummations that are contained in the
GM-VFN scheme have only a small effect and that there is no
compelling need to resum initial-state logarithms.

In conclusion, there is no evidence of a need to improve the FFN
order-$\alpha_s^4$ computation of $B_c$ production by resumming large
final-state or initial-state logarithms. On the other hand, a complete 
computation of the NLO corrections in the FFN approach would be very
welcome in order to allow one to improve the dependence of the
predictions on the renormalization scale and 
reduce the theoretical uncertainties.

\subsubsection{Phenomenology}

The study of $B_c$ production and decays will be possible at the LHC as
well as at the Tevatron. The LHCb experiment, in particular, has been
designed especially for the study of $b$ hadrons ({\it e.g.}, $B^\pm$,
$B^0$, $\bar{B^0}$, $B_s$, $B_c$, $\Lambda_b$). Thus, many new
results for the $B_c$ meson are expected. New data will be useful in
understanding the production mechanism itself and also in determining
the decay branching ratios.

Because the excitations of the $B_c$ carry both $b$ and $c$ flavor
quantum numbers, they must decay into the $B_c$ ground state with almost
100\% probability, either directly or via strong and/or electromagnetic
cascades. No experimental distinction among direct or indirect
production is possible at the moment. It is therefore important that
predictions for $B_c$ production include feeddown from the excitations
of the $B_c$, such as the $B_c^*(^3S_1)$, $B_{cJ,L=1}^*$,
$B_{cJ,L=2}^{**}$, etc.,  and the radially excited states. Predictions
in LO perturbative QCD for the production cross sections of
the $B^*_c$ and the low-lying $P$-wave excited states $B^*_{cJ,L=1}$ are
available in analytic and numerical form in
Refs.~\cite{Chang:2004bh,Chang:2005bf} and in numerical form in
Ref.~\cite{Artoisenet:2007qm}. These predictions show that the
contributions of the color-octet channels, $[(c\bar{b})_{\bf 8}, ^1S_0]$
and $[(c\bar{b})_{\bf 8}, ^3S_1]$,  to the production of low-lying
$P$-wave excited states, such as the $B^*_{cJ,L=1}$, are comparable to
the contributions of the leading color-singlet channels, provided that
the values of the color-octet matrix elements are consistent with NRQCD
velocity scaling.

\begin{figure}[b]
\includegraphics[width=\figwid]{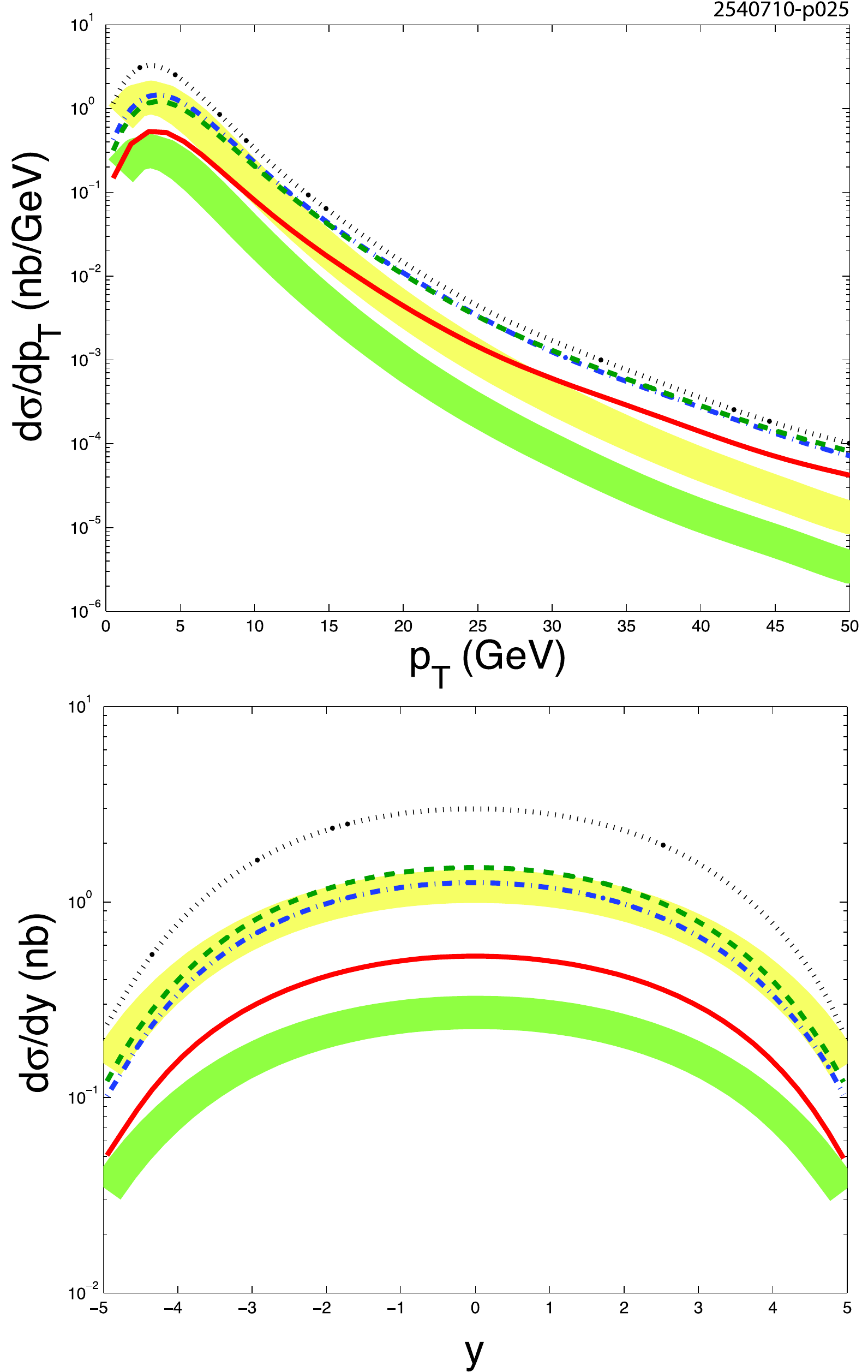}%
\caption{Distributions in $p_T$ and $y$ for production of $c\bar{b}$
mesons at the LHC. The {\it dashed}, {\it solid}, {\it dash-dotted}, and
{\it dotted lines} represent the color-singlet $^1P_1$, $^3P_0$,
$^3P_1$, and $^3P_2$ contributions, respectively. The {\it lower} and
{\it upper shaded bands} represent the color-octet $^1S_0$ and $^3S_1$
contributions, respectively. \figPermXAPS{Chang:2005bf}{2005} }
\label{prod_fig1} \vspace{-0mm}
\end{figure}

\begin{figure}[b]
\includegraphics[width=\figwid]{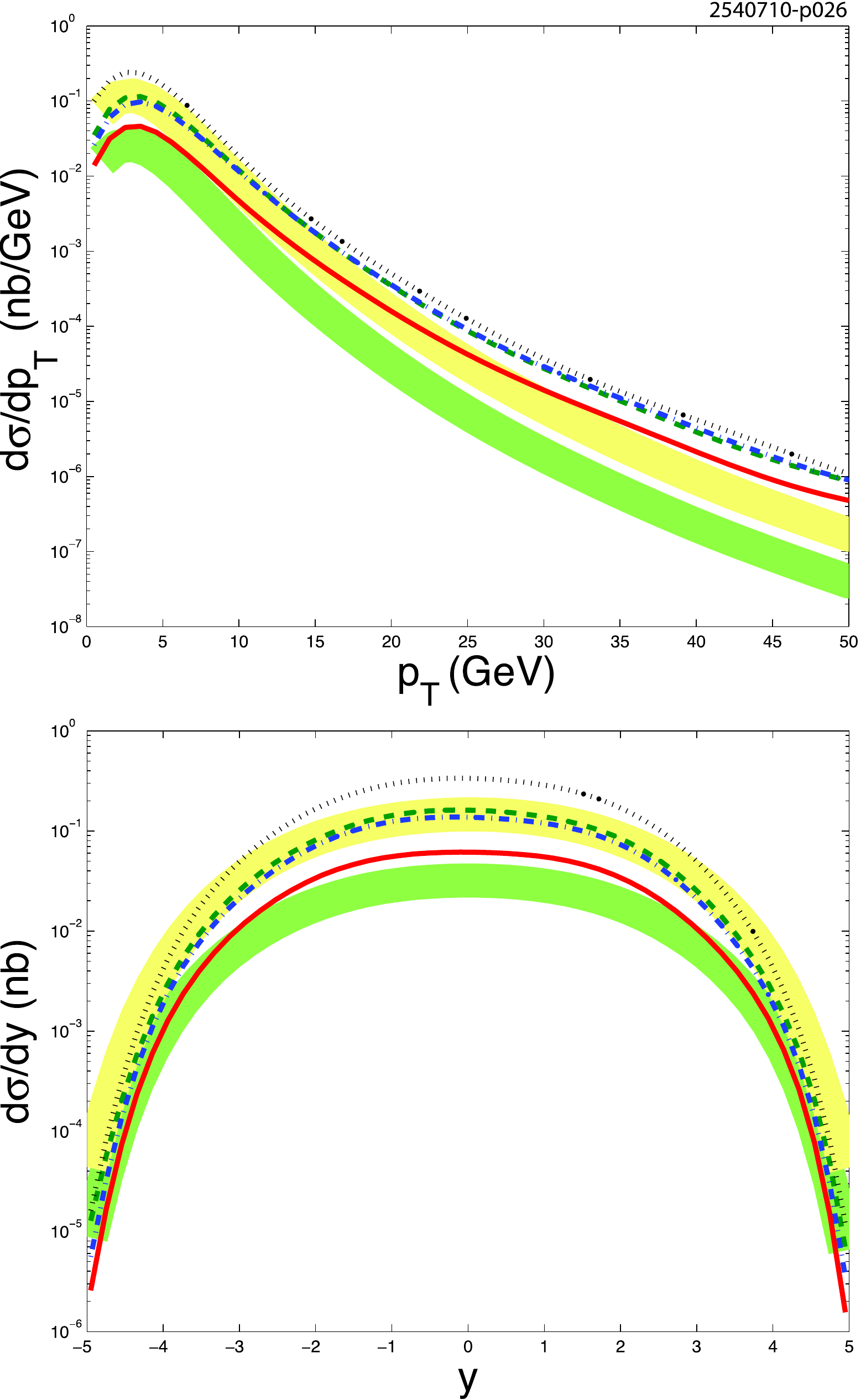}%
\caption{Distributions in $p_T$ and $y$ for production of $c\bar{b}$
mesons at the Tevatron. The {\it dashed}, {\it solid}, {\it
dash-dotted}, and {\it dotted lines} represent the color-singlet
$^1P_1$, $^3P_0$, $^3P_1$, and $^3P_2$ contributions, respectively. The
{\it lower} and {\it upper shaded bands} represent the color-octet
$^1S_0$ and $^3S_1$ contributions, respectively. 
\figPermXAPS{Chang:2005bf}{2005} }
\label{prod_fig2} \vspace{-0mm}
\end{figure}

Simulations of the production of $c\bar{b}$ and $\bar{c} b$ mesons are
now implemented in the programs BCVEGPY \cite{Chang:2006xka} and MadOnia
\cite{Artoisenet:2007qm}. Both programs are interfaced to PYTHIA
\cite{Sjostrand:2006za}.

In Figs.~\ref{prod_fig1} and \ref{prod_fig2}, we show typical
distributions in $y$ and $p_T$ for the production of excited
$(c\bar{b})$ mesons at the LHC and the Tevatron \cite{Chang:2005bf}.
The contributions from the various parton-level production channels are
shown separately.

\subsubsection{Future opportunities}

The observation and study of the $B_c$ mesons and their excitations are
new and exciting components of the quarkonium-physics plans for
both the Tevatron and the LHC. Much has yet to be learned about the
production and decay of these states.

A number of improvements in the theoretical predictions for $B_c$ mesons
are needed, such as computations of the NLO corrections to the
production cross sections. In addition, new mechanisms for $B_c$
production at high $p_T$ should be explored. These include production via
$Z^0$ or top-quark decays \cite{Chang:2007si}.

\section[In medium]{In medium$^{19}$}

\addtocounter{footnote}{1}
\footnotetext{Contributing authors:
A.~D.~Frawley$^\dag$, P.~Petreczky$^\dag$, R.~Vogt$^\dag$, 
R.~Arnaldi, N.~Brambilla, P.~Cortese,
S.~R.~Klein, C.~Louren\c{c}o, A.~Mocsy,
E.~Scomparin, and H.~K.~W\"ohri}
\label{sec:MedChapter}

\subsection{Quarkonia as a probe of hot and dense matter}
\label{sec:media_sec1}

It is expected that strongly-interacting matter shows qualitatively
new behavior at temperatures and/or densities which are
comparable to or larger than the typical hadronic scale.
It has been argued that under such extreme conditions
deconfinement of quarks and gluons should set in and the 
thermodynamics of strongly-interacting matter could then
be understood in terms of these elementary degrees of freedom.
This new form of matter is called
{\em quark-gluon plasma}~\cite{Shuryak:1980tp}, or QGP.
The existence of such a transition has indeed been demonstrated from first
principles using Monte Carlo simulations of lattice QCD.
The properties of this new state of matter have also been 
studied~\cite{Petreczky:2009at,Fodor:2009ax,DeTar:2009ef,Petreczky:2004xs}.

In addition to theoretical efforts, the deconfinement transition and
the properties of hot, strongly-interacting matter are also studied 
experimentally in heavy-ion collisions~\cite{Satz:2000bn,Muller:2006ee}. 
A significant part of the extensive experimental heavy-ion
program is dedicated to measuring quarkonium yields since Matsui and Satz
suggested that quarkonium suppression could be a signature of 
deconfinement~\cite{Matsui:1986dk}.
In fact, the observation of anomalous suppression was considered to be
a key signature of deconfinement at SPS energies~\cite{press_release}.

However, not all of the observed quarkonium suppression in 
nucleus-nucleus ($AB$) collisions relative to scaled proton-proton ($pp$)
collisions is due to quark-gluon plasma formation. 
In fact, quarkonium suppression was also observed in proton-nucleus ($pA$)
collisions, 
so that part of the nucleus-nucleus suppression is due to 
cold-nuclear-matter effects. Therefore it is necessary to disentangle hot-
and cold-medium effects. We first discuss cold-nuclear-matter effects
at different center-of-mass energies. Then we discuss what is known 
about the properties of heavy $Q \overline Q$ states in hot, deconfined media. 
Finally, we review recent experimental
results on quarkonium production from $pA$ collisions
at the SPS and from $pp$, d+Au, and $AA$ collisions 
at RHIC.

\subsection{Cold-nuclear-matter effects}
\label{sec:media_sec2}

The baseline for quarkonium production and suppression in heavy-ion collisions 
should be determined from studies of
cold-nuclear-matter (CNM) effects. The name cold matter 
arises because these effects are observed in hadron-nucleus interactions 
where no hot, dense matter effects are expected. There are several 
CNM effects. Modifications of the parton 
distribution functions in the nucleus, relative to the nucleon, 
(\ie {\it shadowing}) and
energy loss of the parton traversing the nucleus before the hard scattering 
are both assumed to be initial-state effects, intrinsic to the nuclear target. 
Another CNM effect is absorption (\ie destruction) 
of the quarkonium state as it 
passes through the nucleus. Since the latter occurs after the 
$Q\overline Q$ pair has been produced and while it is traversing the nuclear 
medium, this absorption is typically referred to as a final-state effect.
In order to disentangle the mechanisms affecting the produced $Q \overline Q$,
data from a variety of center-of-mass energies and different phase-space
windows need to be studied.
In addition, the inclusive \jpsi\ yield includes contributions 
from $\chi_c$ and $\psip$ decays to \jpsi\ at the 30-35\% level
\cite{Faccioli:2008ir}.  While there is some information
on the $A$ dependence of $\psip$ production, that 
on $\chi_c$ is largely unknown~\cite{Vogt:2001ky}.  

\begin{figure}[b]
   \begin{center}
      \includegraphics[width=\figwid]{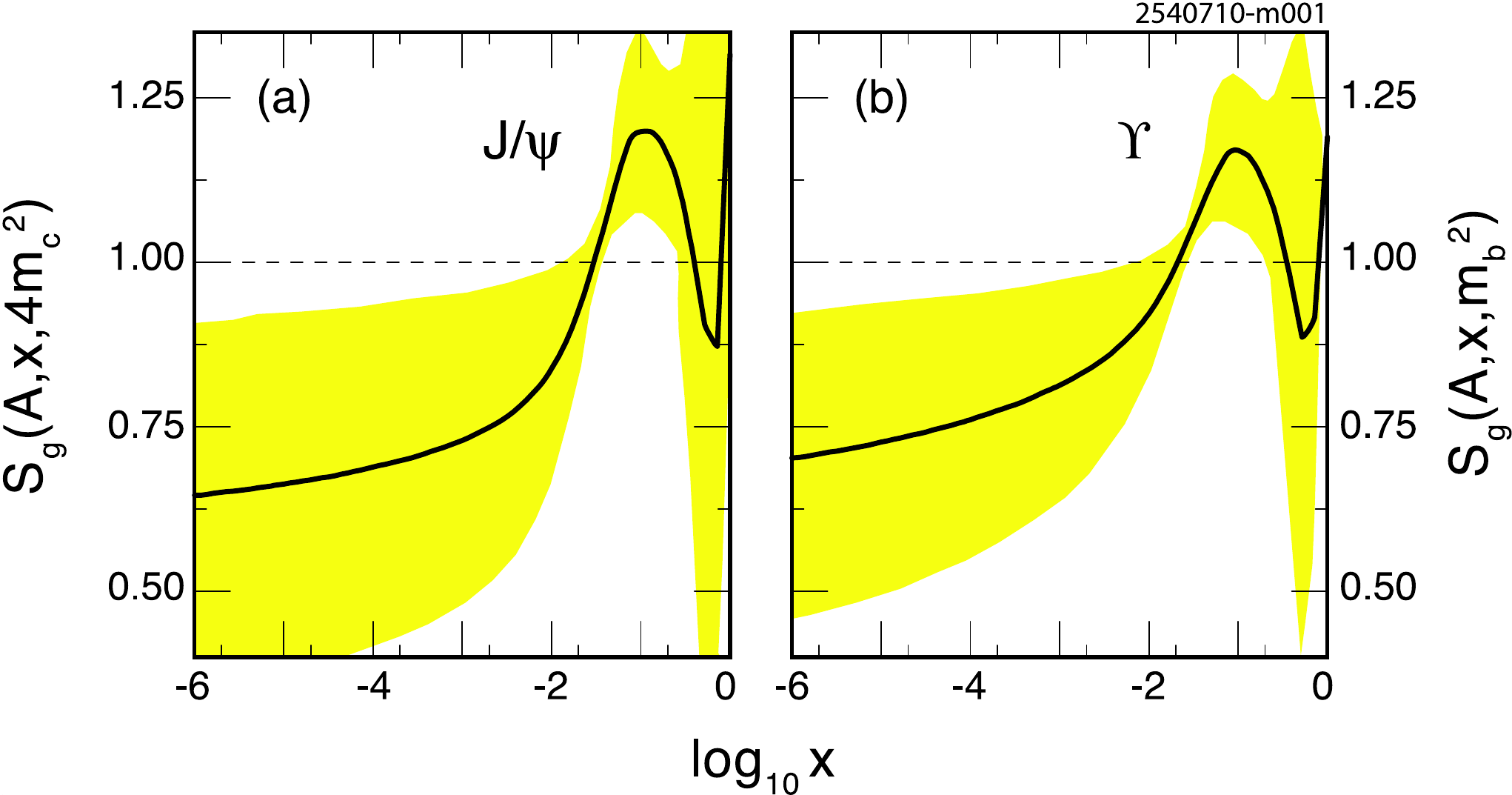}
      \caption{The EPS09 gluon-shadowing 
               parametrization~\cite{Eskola:2009uj} 
               at $Q = 2m_c$ and $m_b$.
               The central value {\it (solid curves)} and 
               the associated uncertainty
               {\it (shaded band)} are shown }
      \label{fig:media_fig1}
   \end{center}
\end{figure}

Even though the contributions to CNM effects may 
seem rather straightforward, there are a number of associated uncertainties. 
First, while nuclear modifications of the quark densities are relatively 
well-measured in nuclear deep-inelastic 
scattering (nDIS), the modifications of the 
gluon density are not directly measured. The nDIS measurements probe only the 
quark and antiquark distributions directly. The scaling violations in nDIS can 
be used to constrain the nuclear gluon density. Overall momentum conservation 
provides another constraint. However, more direct probes of the gluon 
density are needed. Current shadowing parametrizations are derived from 
global fits to the nuclear parton densities and
give wide variations in the nuclear gluon 
density, from almost no effect to very large shadowing at low-$x$, 
compensated 
by strong antishadowing around $x \sim 0.1$.  The range of the possible
shadowing effects is illustrated  in \Fig{fig:media_fig1} 
by the new EPS09~\cite{Eskola:2009uj} parametrization and 
its associated uncertainties, employing the scale values used to fix the 
\jpsi\ and \Ups\ cross sections below the open-heavy-flavor threshold
\cite{Frawley:2008kk}.

The color glass condensate (CGC) is expected to play an important role 
in quarkonium production at RHIC and
the LHC since the saturation scale $Q_{S,A}(x)$ is comparable to the 
charm quark mass~\cite{Kharzeev:2003sk}. 
In this picture, collinear factorization
of \jpsi\ production is assumed to break down and forward \jpsi\ 
production is suppressed.  Indeed, CGC suppression of
\jpsi\  formation may mask some QGP effects~\cite{Kharzeev:2008nw}.

The nuclear absorption survival probability depends on the quarkonium 
absorption cross section. There are more inherent uncertainties 
in absorption than in the shadowing parametrization, which is obtained 
from data on other processes and is independent of the final 
state. Typically an absorption cross section is fit to the $A$ dependence 
of \jpsi\ 
and/or $\psip$ production at a given energy. This is rather simplistic since 
it is unknown whether the object traversing the nucleus is a precursor 
color-octet state or a fully-formed color-singlet quarkonium state. 
If it is an octet state, it 
is assumed to immediately interact with a large, finite cross section 
since it is a colored object~\cite{Kharzeev:1995br}. 
In this case, it has often been assumed that all precursor 
quarkonium states will 
interact with the same cross section. If it is produced as a 
small color-singlet, the absorption cross section 
immediately after the production of the 
$Q \overline Q$ pair should be small and increasing with proper time until, 
at the formation time, it reaches its final-state size~\cite{Blaizot:1989de}.  
High-momentum
color-singlet quarkonium states will experience negligible nuclear 
absorption effects since they will be formed well outside the target.  See
\cite{Vogt:2001ky} for a discussion of the $A$ dependence of absorption
for all the quarkonium states.

\begin{figure}[tb]
   \begin{center}
      \includegraphics[width=\figwid]{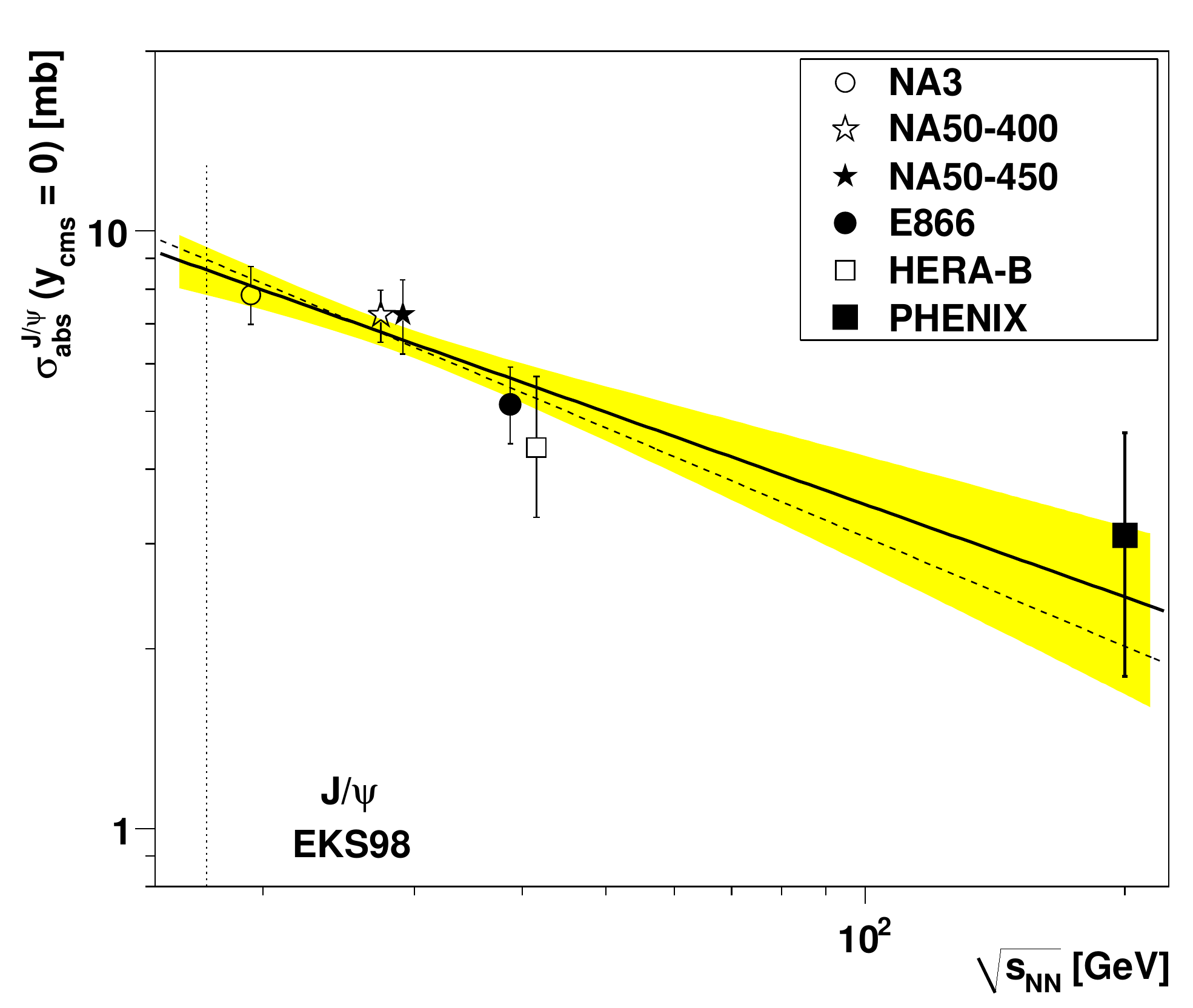}
      \caption{The extracted energy dependence of 
               $\sigma_{\rm abs}^{\jpsi}$ at
               midrapidity.  The {\it solid line} is a 
               power law approximation 
               to $\sigma_{\rm abs}^{\jpsi}(y=0,\sqrtsNN)$ using the 
               EKS98~\cite{Eskola:1998iy,Eskola:1998df}
               shadowing parametrization with the CTEQ61L parton densities
               \cite{Pumplin:2002vw,Stump:2003yu}.  
               The {\it band} 
               indicates the uncertainty in the extracted cross
               sections.  The {\it dashed curve} shows an 
               exponential fit for comparison.  The data at
               $y_{\rm cms} \sim 0$ from NA3~\cite{Badier:1983dg}, 
               NA50 at 400\,GeV\,\protect\cite{Alessandro:2006jt} and 
               450\,GeV\,\protect\cite{Alessandro:2003pc},
               E866~\cite{Leitch:1999ea}, 
               HERA-B~\cite{Abt:2008ya}, and
               PHENIX~\cite{daSilva:2009yy} are also shown.  
               The {\it vertical dotted line} indicates the energy 
               of the Pb+Pb and In+In collisions at the
               CERN SPS. 
               \AfigPermSPV{Lourenco:2008sk}{2009}}
      \label{fig:media_fig2}
   \end{center}
\end{figure}

Fixed-target data taken in 
the range $400 \leq E_{\rm lab} \leq 800$\gev\ 
have shown that the \jpsi\ and 
$\psip$ absorption cross sections are not identical, as the basic color-octet 
absorption mechanism would 
suggest~\cite{Alessandro:2003pc,Alessandro:2006jt,Leitch:1999ea}. 
The difference between the effective
$A$ dependence of \jpsi\ and $\psip$ seems to decrease with beam energy.
The \jpsi\ absorption cross section at $y \sim 0$ is seen to decrease
with energy, regardless of the chosen shadowing 
parametrization~\cite{Lourenco:2008sk}, as shown in \Fig{fig:media_fig2}.

Recent analyses of \jpsi\ production
in fixed-target interactions~\cite{Lourenco:2008sk} 
show that the effective absorption
cross section depends on the energy of the initial beam and the rapidity or
$x_F$ of the observed \jpsi.  One possible interpretation is that 
low-momentum color-singlet states can hadronize in the
target, resulting in larger effective absorption cross sections at lower
center-of-mass energies and backward $x_F$ (or center-of-mass rapidity).
At higher energies, the states traverse the target more rapidly so that
the $x_F$ values at which they can hadronize in the target move 
back from midrapidity toward more negative $x_F$.
Finally, at sufficiently high energies, the quarkonium states pass 
through the target before hadronizing, resulting in negligible absorption
effects.  Thus the {\it effective} absorption cross section decreases with 
increasing center-of-mass energy because faster states are less likely 
to hadronize inside the target.

\begin{figure}[tb]
   \begin{center}
      \includegraphics[width=\figwid]{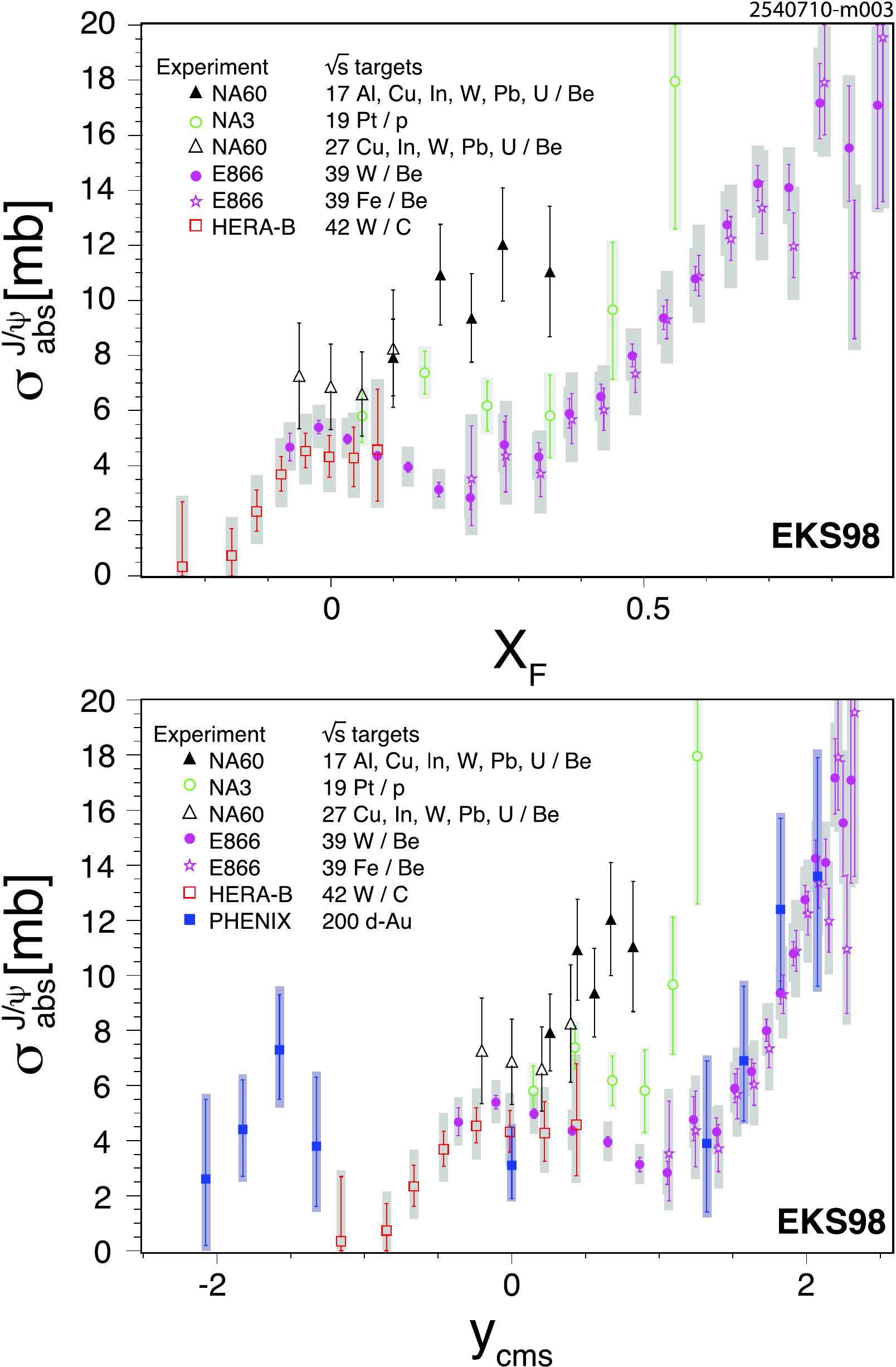} 
      \caption{Top: The $x_F$ dependence of \sigabsj\ for
               incident fixed-target energies from 
               158\,GeV\,\protect\cite{Scomparin:2009tg}, 
               200\,GeV\,\protect\cite{Badier:1983dg}, 
               400\,GeV\,\protect\cite{Alessandro:2006jt},
               450\,GeV\,\protect\cite{Alessandro:2003pc}, 
               800\,GeV\,\protect\cite{Leitch:1999ea}, and   
               920\,GeV\,\protect\cite{Abt:2008ya} obtained using the
               EKS98 shadowing 
               parametrization~\cite{Eskola:1998iy,Eskola:1998df}.  
               The E866~\cite{Leitch:1999ea} and 
               HERA-B~\cite{Abt:2008ya} results  were previously shown
               in \cite{Lourenco:2008sk}. 
               Bottom:  The same results as above
               but as a function of center-of-mass 
               rapidity $y_{\rm CMS}$.  The 
               absorption cross sections extracted from the preliminary
               PHENIX~\cite{daSilva:2009yy}  results 
               at $|y_{\rm CMS}| > 0$ and the central rapidity 
               result~\cite{Adare:2007gn} are also included.
               The {\it boxes} surrounding the data {\it points} 
               represent the statistical and total systematic uncertainties
               added in quadrature, except for PHENIX where the global term
               ($^{+5.0}_{-4.1}$~mb, common to all rapidity bins) is missing
                }
      \label{fig:media_fig3}
   \end{center}
\end{figure}

At higher $x_F$, away from midrapidity, the effective absorption cross
section 
becomes very
large, as shown in the top panel of \Fig{fig:media_fig3}.  The increase in
\sigabsj\ begins closer to midrapidity for lower incident
energies.

There appears to be some saturation of the effect since the 800\gev\ 
fixed-target data exhibit the same trend as
the most recent (preliminary) PHENIX data~\cite{tonyect} as a function of
center-of-mass rapidity, $y_{\rm cms}$, as seen in the
bottom panel of \Fig{fig:media_fig3}.
Model calculations including CGC effects can 
reproduce the general trend of the high-$x_F$ behavior of \jpsi\ production
at 800\gev\ without invoking energy loss~\cite{Kharzeev:2008nw}.  However,
the fact that the NA3 data at $\sqrtsNN = 19$\gev\ exhibit the same 
trend in $x_F$ as E866 calls the CGC explanation into question.

As previously discussed, such an increase in the apparent
absorption cannot be due to interactions with nucleons.  In addition, since
the large-$x_F$ dependence seems to be independent of the quarkonium 
state (\ie the same for \jpsi\ and $\psip$~\cite{Leitch:1999ea},
and also for \UnS{1}\ and 
$\UnS{2} + \UnS{3}$)~\cite{Alde:1991sw}, 
it likely cannot be attributed to the size of the final state
and should thus be an initial-state effect,
possibly energy loss.  (See \cite{Vogt:1999dw} for a discussion of
several types of energy-loss models and their effect on \jpsi\ production.)
Work is in progress to incorporate this
effect using a new approach, based on the number of 
soft collisions the projectile parton undergoes
before the hard scattering to produce the $Q \overline Q$ pair.

It is also well known that feeddown 
from $P$ and higher-$S$ states through radiative 
and hadronic transitions, respectively,
accounts for almost half of the observed \jpsi\ and \UnS{1}\ 
yields. The excited quarkonium states have very 
different sizes and formation times 
and should thus have different absorption cross sections.
For example, the absorption
cross section of quarkonium state $C$ may be proportional to its area,
$\sigma_C \propto r_C^2$~\cite{Povh:1987ju}. 

It should be noted, however, that the fitted absorption cross sections 
used for extracting the ``normal absorption'' baseline for Pb+Pb collisions
at the SPS have treated \jpsi\ and $\psip$ 
absorption independently, ignoring feeddown 
and formation times, and have not taken initial-state shadowing into 
account~\cite{Alessandro:2006jt,Alessandro:2003pc}. 
As discussed above, more detailed analyses show
that the quarkonium absorption cross section decreases with increasing 
energy~\cite{Lourenco:2008sk,Vogt:2004dh}. 
More recent fixed-target
analyses~\cite{Arnaldi:2009it,Scomparin:2009tg}, 
comparing measurements at 158 and 400\gev, 
have begun to address these issues (see \Sec{sec:media_sec4}).  
Indeed, the
extracted absorption cross section is found to be larger at 158\gev\ than
at 400\gev, contrary to previous analyses, which assumed a universal,
constant absorption cross section~\cite{Alessandro:2003pc,Alessandro:2006jt}.
When these latest results are extrapolated to nucleus-nucleus collisions
at the same energy, the anomalous suppression is significantly
decreased relative to the new baseline~\cite{Scomparin:2009tg}.

The cold-nuclear-matter effects suggested (initial-state energy loss, 
shadowing, final-state breakup, {\it etc.}) 
depend differently on the quarkonium 
kinematic variables and the collision energy. It is clearly unsatisfactory to 
combine all these mechanisms into an {\it effective} absorption cross section,
as employed in the Glauber formalism, 
that only evaluates final-state absorption. 
Simply taking the $\sigma_{\rm abs}$ obtained from 
the analysis of the $pA$ data 
and using it to define the Pb+Pb baseline may not be sufficient. 

A better understanding of absorption 
requires more detailed knowledge of the 
production mechanism. Most calculations of the $A$ dependence use the color 
evaporation model (CEM), in which all quarkonia are assumed to be 
produced with the same underlying kinematic distributions~\cite{Gavai:1994in}. 
This model works 
well for fixed-target energies and for RHIC~\cite{Frawley:2008kk}, as does
the LO color-singlet model (CSM)~\cite{Brodsky:2009cf}.
In the latter case, but contrary to the 
CEM at LO, \jpsi\ production is necessarily accompanied by the 
emission of a perturbative final-state gluon which can be seen as an 
extrinsic source of transverse momentum. This induces modifications in 
the relations between the initial-state gluon momentum fractions and the 
momentum of the \jpsi. In turn, this modifies~\cite{Ferreiro:2008wc} 
the gluon-shadowing corrections relative to those
expected from the LO CEM where the transverse momentum of the 
\jpsi\ is intrinsic to the initial-state gluons. Further studies are 
being carried out, including the impact of 
feeddown, the extraction of absorption cross sections for each of the
charmonium states, and the 
dependence on the partonic \jpsi\ 
production mechanism.  A high precision measurement of the $\psip$ and
\UnS{3}\ production ratios in $pA$ interactions as a function of rapidity
would be desirable since they are not affected by feeddown contributions.
In addition, measurements of the feeddown contributions to \jpsi\ and
\UnS{1}\ as a function of rapidity and $p_T$ would be very useful.

On the other hand, the higher-$p_T$ Tevatron predictions have 
been calculated within the nonrelativistic QCD (NRQCD) 
approach~\cite{Bodwin:1994jh}, which 
includes both singlet and octet matrix elements. These high $p_T$ calculations 
can be tuned to agree with the high $p_T$ data but cannot reproduce the 
measured quarkonium polarization at the same 
energy~\cite{Brambilla:2004wf}. If some fraction of the 
final-state quarkonium yields can be attributed to color-singlet production, 
then absorption need not be solely due to either
singlet or octet states but rather some 
mixture of the two, as dictated by 
NRQCD~\cite{Vogt:2001ky,Beneke:1996tk,Zhang:1997xq}. 
A measurement of the $A$ dependence of $\chi_c$ 
production would be particularly helpful to ensure
significant progress toward understanding the production mechanism.

\subsection{Quarkonium in hot medium}
\label{sec:media_sec3}

\subsubsection{Spectral properties at high temperature}
\label{sec:media_subsec31}

There has been considerable interest in studying quarkonia 
in hot media since publication of the famous Matsui and Satz 
paper~\cite{Matsui:1986dk}.  It has been argued that color screening 
in a deconfined QCD medium will destroy all $Q\overline Q$ bound states
at sufficiently high temperatures. Although 
this idea was proposed long ago, first principle QCD calculations, 
which go beyond qualitative arguments, have been performed only recently. 
Such calculations include lattice QCD determinations of quarkonium 
correlators~\cite{Umeda:2002vr,Asakawa:2003re,Datta:2003ww,Jakovac:2006sf,Aarts:2007pk},
potential model calculations 
of the quarkonium spectral functions with potentials based on lattice 
QCD~\cite{Digal:2001ue,Wong:2004zr,Mocsy:2005qw,Mocsy:2004bv,Alberico:2006vw,Cabrera:2006wh,Mocsy:2007yj,Mocsy:2007jz},
as well as effective 
field theory approaches that justify potential models and reveal new medium 
effects~\cite{Laine:2007qy,Laine:2007gj,Laine:2008cf,Brambilla:2008cx}.  
Furthermore, better modeling of 
quarkonium production in the medium created by heavy-ion collisions has 
been achieved.   These advancements make it possible to disentangle the cold- 
and hot-medium effects on the quarkonium states, crucial for the 
interpretation of heavy-ion data.

\subsubsection{Color screening and deconfinement}
\label{sec:media_subsec32}

At high temperatures, strongly-interacting matter undergoes a deconfining
phase transition to a quark-gluon plasma. 
This transition is triggered by a rapid increase of the energy and
entropy densities as well as the disappearance of hadronic states. (For a 
recent review, see \cite{DeTar:2009ef}.) 
According to current lattice 
calculations~\cite{Aoki:2006br,Cheng:2007jq,Aoki:2009sc,Bazavov:2009zn,Cheng:2009zi,Bazavov:2009mi,Bazavov:2010sb,Bazavov:2010bx,Borsanyi:2010bp,Borsanyi:2010cj} 
at zero net-baryon density, deconfinement occurs at $T\sim 165-195$~MeV.
The QGP is characterized by color screening: the range of interaction between
heavy quarks becomes inversely proportional to the temperature. Thus at
sufficiently high temperatures, it is impossible to produce a bound state 
between a heavy quark ($c$ 
or $b$) and its antiquark.

\begin{figure}[b]
   \begin{center}
      \includegraphics[width=\figwid]{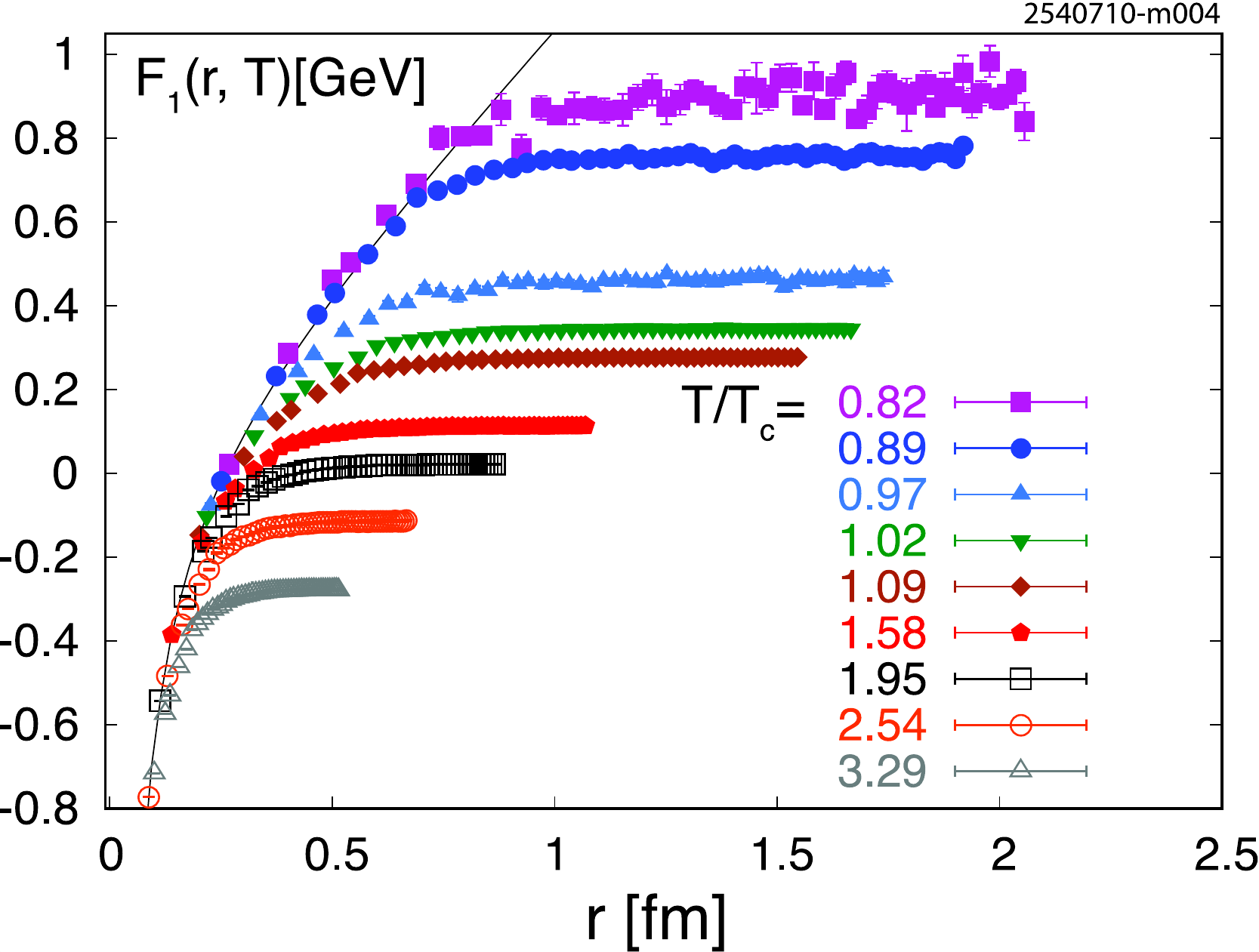}
      \caption{Heavy-quark-singlet free energy versus quark separation 
               calculated in 2+1 flavor QCD
               on $16^3 \times 4$ lattices at different 
               temperatures~\cite{Petreczky:2009ip,Petreczky:2010yn}  
               }
      \label{fig:media_fig4}
   \end{center}
\end{figure}

Color screening is studied on the lattice by 
calculating the spatial correlation function of a static quark and
antiquark in a color-singlet state which propagates in Euclidean time 
from $\tau=0$ to $\tau=1/T$, where $T$ is the temperature 
(see \cite{Bazavov:2009us,Petreczky:2005bd} for reviews). 
Lattice calculations of this quantity with dynamical quarks have been
reported 
\cite{Kaczmarek:2002mc,Digal:2003jc,Petreczky:2004pz,Kaczmarek:2005ui,Kaczmarek:2007pb,Petreczky:2010yn}.
The logarithm of the singlet
correlation function, also called the singlet free energy,
is shown in \Fig{fig:media_fig4}. 
As expected, in the zero-temperature limit the
singlet free energy coincides with the zero-temperature potential. 
\Figure{fig:media_fig4} also illustrates that,
at sufficiently short distances, the singlet free energy is
temperature independent and equal to the zero-temperature potential. 
The range of interaction decreases with increasing temperature.  For 
temperatures above the transition temperature, $T_c$, the heavy-quark 
interaction range becomes
comparable to the charmonium radius. Based on this general observation, 
one would expect that the charmonium
states, as well as the excited bottomonium states, do not remain bound at
temperatures just above the
deconfinement transition, often referred to as 
{\it dissociation} or {\it melting}.

\subsubsection{Quarkonium spectral functions and quarkonium potential}
\label{sec:media_subsec33}

In-medium quarkonium properties are encoded in the corresponding 
spectral functions, as is quarkonium dissolution
at high temperatures. Spectral functions are defined as
the imaginary part of the retarded correlation function of quarkonium
operators. Bound states appear as peaks in the spectral functions.
The peaks broaden and eventually disappear with
increasing temperature. The disappearance of a peak signals the melting of 
the given quarkonium state. 

In lattice QCD, the meson
correlation functions, $G(\tau,T)$, are calculated in Euclidean time. 
These correlation functions are related 
to the spectral functions $\sigma(\omega,T)$ by
\begin{equation}
G(\tau,T)=\int_0^{\infty} d \omega  \, \sigma(\omega,T) 
\frac{\cosh(\omega (\tau-1/(2 T)))}{\sinh(\omega/(2 T))}\, .
\label{eqn:media_eq1}
\end{equation}
Detailed information on $G(\tau,T)$ would allow
reconstruction of the spectral function from the lattice
data. In practice, however, this turns out to be a very difficult
task because the time extent is limited to $1/T$ (see the discussion
in \cite{Jakovac:2006sf} and references therein). Lattice artifacts
in the spectral functions at high energies $\omega$ are also a problem
when analyzing the correlation functions calculated
on the lattice \cite{Karsch:2003wy}.

\begin{figure}[tb]
   \begin{center}
      \includegraphics[width=\figwid]{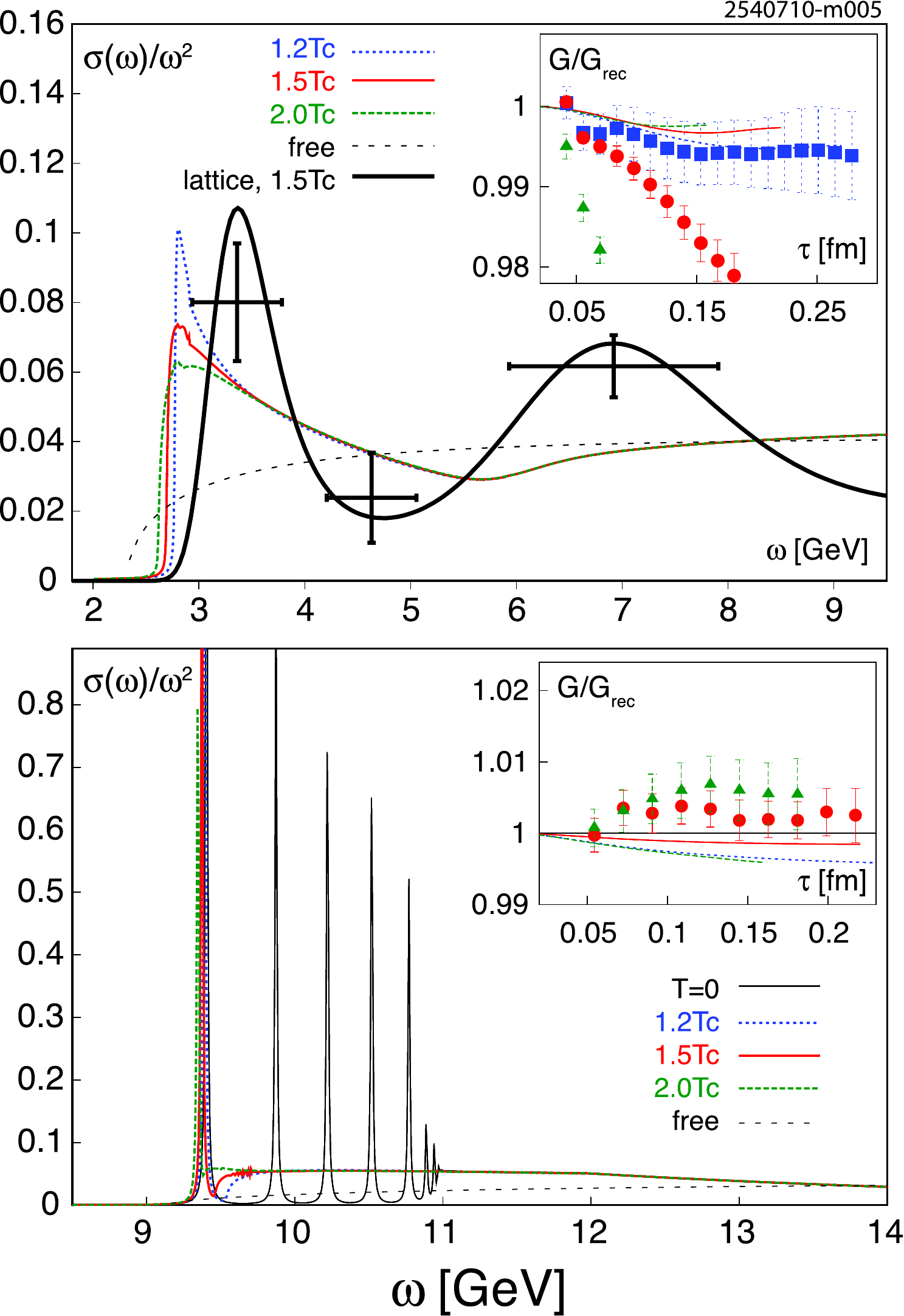}
      \caption{The $S$-wave charmonium (upper) and 
               bottomonium (lower) spectral 
               functions calculated in potential 
               models. 
               Insets: correlators compared to lattice data.  
               The {\it dotted curves} are the
               free spectral functions. 
               \AfigPermAPS{Mocsy:2007yj}{2008} }
      \label{fig:media_fig5} 
   \end{center}
\end{figure}

The quarkonium spectral functions can be calculated in potential models 
using the singlet free energy from \Fig{fig:media_fig4} or with different 
lattice-based potentials obtained using the singlet free energy
as an input~\cite{Mocsy:2007yj,Mocsy:2007jz} (see also \cite{Mocsy:2008eg}
for a review). The results for quenched QCD calculations are shown in 
\Fig{fig:media_fig5}
for $S$-wave charmonium (top) and bottomonium (bottom) 
spectral functions~\cite{Mocsy:2007yj}.
All charmonium states
are dissolved in the deconfined phase while the bottomonium $1S$
state may persist up to $T \sim 2T_c$. 
The temperature dependence of the Euclidean correlators can be predicted 
using \Eq{eqn:media_eq1}. 
Somewhat surprisingly, the Euclidean correlation functions in the pseudoscalar
channel show very little temperature dependence, 
irrespective of whether a state 
remains bound ($\eta_b$) or not ($\eta_c$). 
Note also that correlators from potential models are in accord with the 
lattice calculations (see insets in \Fig{fig:media_fig5}). Initially, 
the weak temperature dependence of the pseudoscalar correlators was 
considered to be evidence for the survival of $1S$ quarkonium
states~\cite{Datta:2003ww}. It is now clear that this conclusion was 
premature. 
In other channels one sees significant temperature dependence of the Euclidean
correlation functions, especially in the scalar and axial-vector channels where
it has been interpreted as evidence for dissolution of the quarkonium $1P$
states.  However, this temperature dependence is due to the zero-mode 
contribution, \ie a peak in the finite temperature spectral functions at 
$\omega \simeq 0$~\cite{Petreczky:2005nh,Umeda:2007hy}. After subtracting 
the zero-mode contribution, the 
Euclidean correlation functions show no temperature dependence within 
the uncertainties~\cite{Petreczky:2008px}.
Thus melting of the quarkonium states is not visible 
in the Euclidean correlation functions.

There is a large enhancement in the threshold region of the spectral functions
relative to the free spectral function, as shown in \Fig{fig:media_fig5}.
This threshold enhancement compensates for the absence of bound states
and leads to Euclidean correlation functions with very weak
temperature dependencies~\cite{Mocsy:2007yj}. 
It further indicates strong residual
correlations between the quark and antiquark, even in the 
absence of bound states.
Similar analyses were done for the $P$-wave charmonium and bottomonium 
spectral functions~\cite{Mocsy:2007yj,Mocsy:2007jz}. An upper bound on the
dissociation temperature (the temperatures above which no
bound states peaks can be seen in the spectral function and bound state 
formation is suppressed) can be obtained from the analysis of the spectral 
functions. Conservative upper limits on the dissociation
temperatures for the different quarkonium states obtained from 
a full QCD calculation~\cite{Mocsy:2007jz} are given in 
\Tab{tab:media_tab1}.

\begin{table}[tb]
   \caption{Upper bounds on the dissociation 
             temperatures.
             \AfigPermAPS{Mocsy:2007jz}{2007} }
   \label{tab:media_tab1}
   \setlength{\tabcolsep}{0.41pc}
   \begin{center}
      \begin{tabular}{ccccccc}
      \hline\hline
      \rule[10pt]{-1mm}{0mm}
      State & $\chi_{cJ}(1P)$ & $\psip$ & \jpsi\ & $\UnS{2}$ & $\chi_{bJ}(1P)$ &
      \Ups \\[1.0mm]
      \hline 
      \rule[10pt]{-1mm}{0mm}
      $T_{\rm diss}$ & $\le T_c$ & $\le T_c$ & $1.2T_c$ & $1.2T_c$
      & $1.3T_c$ & $2T_c$\\ 
\hline\hline
\end{tabular}
\end{center}
\end{table}

The problems with the potential model approach described above are
how to relate it to first principles QCD and how to
define the in-medium quark-antiquark potential. These problems can be 
addressed using an EFT approach to heavy-quark bound states in a real time 
formalism.  The first steps in this direction were taken in 
\cite{Laine:2006ns,Laine:2007qy}.  The static potential (for infinitely heavy
$Q \overline Q$ pairs)  was calculated 
in the regime $T \gg 1/r \simg m_D$, where $m_D$ is the Debye mass and 
$r$ the $Q \overline Q$ separation, by analytically continuing the 
Euclidean Wilson loop to real time. The calculation was done in 
weak-coupling resummed perturbation
theory. The imaginary part of the gluon self-energy gives an imaginary part to
the static potential and thus introduces a thermal width in
the $Q \overline Q$ bound state.  The heavy quarkonium
contribution to the spectral function of the electromagnetic current was
calculated in the same framework \cite{Laine:2007gj,Burnier:2007qm}.
Static $Q \overline Q$ pairs in a real-time 
formalism are considered in~\cite{Beraudo:2007ky}, where 
the potential for distances $1/r \sim m_D$ in a hot 
QED plasma is derived. 
The real part of the static potential was in agreement with the
singlet free energy while the damping factor (imaginary part of the potential)
at large distances agreed with the one found in \cite{Laine:2006ns}.  This
damping can be thought of as quarkonia scattering with light partons in a 
thermal bath where the collisional width increases with temperature.
The real part of the static potential was found to agree with the
singlet free energy and the damping factor with the one found in 
\cite{Laine:2006ns}.
In \cite{Escobedo:2008sy}, a study of bound states in a hot QED 
plasma was performed in a non-relativistic EFT framework. 
In particular, the hydrogen atom was studied for temperatures ranging from 
$T\ll m\alpha^2$ to $T\sim m$, where the imaginary part of the
potential becomes larger than the real part and the hydrogen ceases to
exist.  The same study has been extended to muonic hydrogen in 
\cite{Escobedo:2010tu}.
 
The EFT framework quarkonium at 
finite temperature for different distance regimes was developed in 
\cite{Brambilla:2008cx} using the real time formalism and weak coupling
techniques. In a zero-temperature medium, the behavior of quarkonia
is characterized by different energy and momentum scales related to 
non-relativistic bound states with heavy quark velocity $v$ and mass $m$.
The scale related to the inverse distance between the $Q$ and $\overline Q$
is $mv$ while $mv^2$ is the scale related to the binding energy of the state.
Finally, $\lamQ$ is related to
the nonperturbative features of the QCD vacuum. In the weak-coupling regime, 
$v \sim \als$ and the hierarchy of scales follows 
$m \gg m v \gg m v^2 \gg \lamQ$\footnote{A 
    hierarchy of bound state scales may exist beyond
    the weak-coupling regime, \eg $mv \simeq \lamQ$, but has not yet
    been considered at finite temperature.}.   
In addition, there are thermodynamical scales: the temperature 
$T$; the inverse of the screening 
length of chromo-electric interactions, the Debye mass ($m_D \sim g T$); and 
the static magnetic scale, $g^2 T$. In the weak-coupling regime at finite 
temperature, the ordering of these scales follows $T \gg g T \gg g^2 T$. 

If there exists such a hierarchy of scales, any quantity of interest 
maybe expanded in some ratio of the scales. 
If the contributions from the different scales are separated explicitly
at the Lagrangian level, this amounts to replacing QCD with a hierarchy of 
EFTs equivalent to QCD order by order in the expansion parameters.
As described in \ref{sec:SpecTh_eft}, at $T=0$ the EFTs
that follow from QCD by integrating over the scales $m$ and $mv$ are called 
nonrelativistic QCD (NRQCD) and potential NRQCD (pNRQCD) respectively.
This procedure for constructing EFTs can be generalized to finite temperatures.
The construction of EFTs for heavy-quark bound states and their resulting forms
depend on how the bound-state scales are related to the thermal scales. If all 
the bound state scales are larger than $T$, the relevant EFT is zero 
temperature pNRQCD. In this case, there are no corrections to the heavy 
$Q \overline Q$ potential even though there may be thermal corrections to the 
quarkonium binding energies and
widths. If $m v > T > mv^2$ the relevant EFT can be constructed similarly to 
the $T=0$ case but now the temperature scale is integrated over.
Now there are thermal corrections to the potential which 
have both real and imaginary parts 
\cite{Brambilla:2008cx,Vairo:2009ih}. Finally, when $T> mv$ the temperature 
scale must be integrated
out after the scale $m$ is integrated over but before the scale $mv$. This 
procedure leads to an EFT very similar to 
NRQCD but a modification of the Lagrangian corresponding to gluon and light 
quark fields is necessary.
This part of the Lagrangian is replaced by the hard thermal loop (HTL) 
Lagrangian  
\cite{Braaten:1989mz,Braaten:1990az,Braaten:1991gm,Frenkel:1989br}.
The resulting effective theory is referred to
as NRQCD$_{\rm HTL}$ \cite{Brambilla:2008cx,Vairo:2009ih}.  Subsequent 
integration over the scale $mv$ leads to a new
EFT called pNRQCD$_{\rm HTL}$ similar to pNRQCD. Now the heavy quark potential 
receives both real and imaginary thermal corrections. Furthermore, for 
$r \sim 1/m_D$, the real part of
the potential is exponentially screened and the imaginary part is much larger
than the real part \cite{Brambilla:2008cx,Vairo:2009ih}.
Below we summarize what has been learned from the EFT approach to heavy quark
bound states at finite temperature.

The thermal part of the potential has both a real and  an imaginary part. 
The imaginary part of the potential smears out the bound state 
peaks of the quarkonium spectral function, leading to their dissolution prior
to the onset of Debye screening in the real part 
of the potential (see, \eg the 
discussion in \cite{Laine:2008cf}).  Therefore, quarkonium dissociation 
appears to be a consequence of the thermal decay width rather than color 
screening of the real part of the potential.  This conclusion follows from the 
observation that the thermal decay width becomes comparable to the binding 
energy at temperatures below that required for the onset of color screening.

Two mechanisms contribute to the thermal decay width: the imaginary part of 
the gluon self-energy induced by Landau damping, as also observed in QED, see  
\cite{Laine:2006ns}  and the quark-antiquark color singlet to color 
octet thermal break-up 
(a new effect, specific to QCD)~\cite{Brambilla:2008cx}.
Parametrically, the first mechanism dominates at temperatures where the Debye 
mass $m_D$ is larger than the binding energy, while the latter 
effect dominates for temperatures where $m_D$ is smaller than the binding 
energy.  The dissociation temperature is related to the coupling by  
$\pi T_{\rm dissoc} \sim m g^{4\over 3}$ \cite{Escobedo:2008sy,Laine:2008cf}.

The derived color-singlet thermal potential, $V$, is neither the color-singlet 
$Q \overline Q$ free energy nor its internal energy. Instead it has an 
imaginary part and may contain divergences that eventually cancel in physical 
observables \cite{Brambilla:2008cx}.

Finally, there may be other finite-temperature effects other than screening.
These typically may take the form of power-law corrections or have a 
logarithmic temperature dependence \cite{Brambilla:2008cx,Escobedo:2008sy}.

The EFT framework thus provides a clear definition of the potential and a 
coherent and systematic approach for calculating quarkonium masses and widths 
at finite temperature.  In \cite{Brambilla:2010vq},
heavy quarkonium energy levels and decay widths in a quark-gluon plasma, 
below the quarkonium dissociation temperature where the
temperature and screening mass satisfy the hierarchy  
$m \als  \gg \pi T  \gg m \als^2 \gg m_D$, have been calculated to order 
$m \als^5$, relevant for bottomonium $1S$ states ($\Upsilon(1S)$, $\eta_b$) 
at the LHC.  At leading order the quarkonium masses increase quadratically 
with $T$, the same functional increase with energy as dileptons produced in 
electromagnetic decays \cite{Brambilla:2010vq}.  A thermal correction 
proportional to $T^2$ appears in the quarkonium electromagnetic decay rates.  
The leading-order decay width grows linearly with temperature, implying that
quarkonium dissociates by decaying to the color-octet continuum. 

This EFT approach was derived assuming weak coupling and neglecting 
nonperturbative effects.  In particular, the role of the color-octet degrees 
of freedom in pNRQCD beyond perturbation theory needs to
be better understood (see \eg \cite{Jahn:2004qr}). Comparison of certain
static $Q \overline Q$ correlators 
calculated in the EFT framework with results 
from lattice QCD, partly discussed in subsection \ref{sec:media_subsec32}, 
could prove useful in this respect. The correlation function of two
Polyakov loops, which is gauge invariant and corresponds to the free energy 
of a static $Q \overline Q$ pair could
be particularly suitable for this purpose. Therefore, in 
\cite{Brambilla:2010xn} the Polyakov loop and the correlator of two Polyakov 
loops have been calculated to next-to-next-to-leading order at finite
temperature in the weak-coupling regime and at $Q \overline Q$ separations 
shorter than the inverse of the temperature and for 
Debye masses larger than the Coulomb potential.  The relationship between the 
Polyakov loop correlator and the  singlet and octet 
$Q \overline Q$ correlator has been established in the EFT framework.
A related study of cyclic Wilson loops at finite temperature in perturbation 
theory was reported in \cite{Burnier:2009bk}. A further attempt to relate 
static $Q \overline Q$ correlation functions to the real-time potential was 
discussed in \cite{Rothkopf:2009pk}.
Very recently the first lattice NRQCD calculations of bottomoniun at
finite temperature have appeared~\cite{Aarts:2010ek}.
The initial discussion on the possibility of calculating quarkonium 
properties at finite temperature using NRQCD goes back to the first 
meeting of the QWG at CERN in 2002.

In addition, the effects of medium
anisotropies on the quarkonium states have been considered, 
both on the real~\cite{Dumitru:2007hy} and imaginary 
\cite{Dumitru:2009fy,Burnier:2009yu,Philipsen:2009wg} parts of the potential, 
as well as on bound-state production. 
Polarization of the $P$ states has been predicted to arise from 
the medium anisotropies, resulting in a significant ($\sim 30\%$) effect on 
the $\chi_b$ states~\cite{Dumitru:2009ni}. A weak medium anisotropy 
may also be related to the shear viscosity~\cite{Asakawa:2006tc}.  Thus the
polarization can directly probe the properties of the medium 
produced in heavy-ion collisions. 

\subsubsection{Dynamical production models}
\label{sec:media_subsec34}

While it is necessary to understand the quarkonium spectral functions in 
equilibrium QCD, this knowledge is insufficient for predicting 
effects on quarkonium production in 
heavy-ion collisions because, unlike the light degrees of freedom, heavy 
quarks are not fully thermalized in heavy-ion collisions. Therefore it is 
nontrivial to relate the finite temperature quarkonium spectral functions
to quarkonium production rates in heavy-ion collisions without further model
assumptions. The bridge between the two is provided by dynamical models of 
the matter produced in heavy-ion collisions. Some of the simple models 
currently available are based on statistical 
recombination~\cite{Andronic:2006ky}, 
statistical recombination and dissociation rates~\cite{Zhao:2007hh}, 
or sequential melting~\cite{Karsch:2005nk}. 
Here we highlight a more recent model, which makes closer contact with both
QCD and experimental observations~\cite{Young:2008he}.

The bulk evolution of the matter produced in heavy-ion collisions is 
well described by hydrodynamics (see \cite{Teaney:2009qa} for a recent 
review). The large heavy quark mass makes it possible to model its interaction 
with the medium by Langevin dynamics~\cite{Moore:2004tg}. Such an approach 
successfully describes the anisotropic flow of charm quarks observed at RHIC 
\cite{Moore:2004tg,vanHees:2004gq} (see also the review~\cite{Rapp:2009my} and
references therein).  Potential models have shown that, in the absence of 
bound states, the $Q \overline Q$ pairs are correlated in space 
\cite{Mocsy:2007yj,Mocsy:2007jz}. This correlation can be modeled classically
using Langevin dynamics, including a drag force and a random force between the 
$Q$ (or $\overline Q$) and the medium as well as the forces between the $Q$ 
and $\overline Q$ described by the potential. It was recently shown that a 
model combining an ideal hydrodynamic expansion of the medium with a
description of the correlated $Q \overline Q$ pair dynamics by the Langevin 
equation can describe charmonium suppression at RHIC
quite well~\cite{Young:2008he}. In particular, this model can explain
why, despite the fact that a deconfined medium is created at RHIC, there is
only a 40-50\% suppression in the charmonium yield. The attractive potential
and the finite lifetime of the system prevents the complete decorrelation of
some of the $Q \overline Q$ pairs~\cite{Young:2008he}. Once the matter has
cooled sufficiently, these residual correlations make it possible for 
the $Q$ and $\overline Q$ to form a bound state.  Charmonium production by
recombination can also be calculated in this approach \cite{Young:2009tj}.
Although recombination was found to be significant for the most central 
collisions, it is still subdominant \cite{Young:2009tj}.

The above approach, which neglects quantum effects, is applicable only if 
there are no bound states, as is likely to be the case for the \jpsi. If 
heavy quark bound states are present, as is probable for the \UnS{1}, 
the thermal dissociation rate will be most relevant for 
understanding the quarkonium yield. It is expected that the interaction of a 
color-singlet quarkonium state with the medium is much smaller than
that of heavy quarks. Thus, to first approximation, medium effects will only 
lead to quarkonium dissociation.

\subsubsection{Summary of hot-medium effects}
\label{sec:media_subsec35}

Potential model calculations based on lattice QCD, as well as resummed 
perturbative QCD calculations, indicate that all charmonium states and the
excited bottomonium states dissolve in the deconfined medium. This leads to 
the reduction of the quarkonium yields in heavy-ion collisions 
compared to the binary scaling of $pp$ collisions.   Recombination and edge
effects, however, guarantee a nonzero yield.

One of the great opportunities of the LHC and RHIC-II heavy-ion programs 
is the ability to study bottomonium yields. From a theoretical perspective, 
bottomonium is an important and clean probe for at least two reasons. 
First, the effective field theory approach, which provides a link to first 
principles QCD, is more applicable for bottomonium due to better separation of 
scales and higher dissociation temperatures. Second, the heavier bottom quark 
mass reduces the importance of statistical recombination effects, 
making bottomonium a good probe of dynamical models. 
 
\subsection{Recent results at SPS energies}
\label{sec:media_sec4}

Studies of charmonium production and suppression in cold 
and hot nuclear matter have been carried out by the NA60 
collaboration~\cite{Scomparin:2009tg,Arnaldi:2007zz,Arnaldi:2009ph}.  
In particular, data have been taken 
for In+In collisions at 158\gev/nucleon and for $pA$ collisions at 158 and 
400\gev.  In the following, the primary NA60 results and their impact on the 
understanding of the anomalous \jpsi\ suppression, first observed by the
NA50 collaboration in Pb+Pb collisions~\cite{Alessandro:2004ap}, 
are summarized. 
A preliminary comparison between the suppression patterns observed 
at the SPS and RHIC is discussed in \Sec{sec:media_sec6}.

\subsubsection{\jpsi\ production in $pA$ collisions}
\label{sec:media_subsec42}

One of the main results of the SPS heavy-ion program was the observation of 
anomalous \jpsi\ suppression.  Results obtained in Pb+Pb collisions at 
158\gev/nucleon by the NA50 collaboration showed that
the \jpsi\ yield was suppressed with respect to estimates that include only
cold-nuclear-matter effects~\cite{Alessandro:2004ap}.
The magnitude of the cold-nuclear-matter effects has typically been extracted 
by extrapolating the \jpsi\ production data obtained in $pA$ collisions. 
Until recently the reference SPS $pA$ data were based on samples 
collected at 400/450\gev\ by the NA50 collaboration, at higher energy than the 
nuclear collisions and in a slightly different rapidity 
domain~\cite{Alessandro:2003pc,Alessandro:2006jt,Alessandro:2003pi}.

The need for reference $pA$ data taken under the same conditions as the $AA$ 
data was a major motivation for the NA60 run with an SPS 
primary proton beam at 158\gev\ in 2004.
Seven nuclear targets (Be, Al, Cu, In, W, Pb, and U) were simultaneously 
exposed to the beam. The sophisticated NA60 experimental 
setup~\cite{Arnaldi:2008er},
based on a high-resolution vertex spectrometer coupled to the muon 
spectrometer inherited from NA50, made it possible to unambiguously identify 
the target in which the \jpsi\ was produced as well as measure muon pairs 
from its decay with a $\sim 70$\mev\ invariant mass resolution.  
During the same period, a 400\gev\ 
$pA$ data sample was taken with the same experimental setup.

\begin{figure}[t]
   \begin{center}
      \includegraphics[width=\figwid]{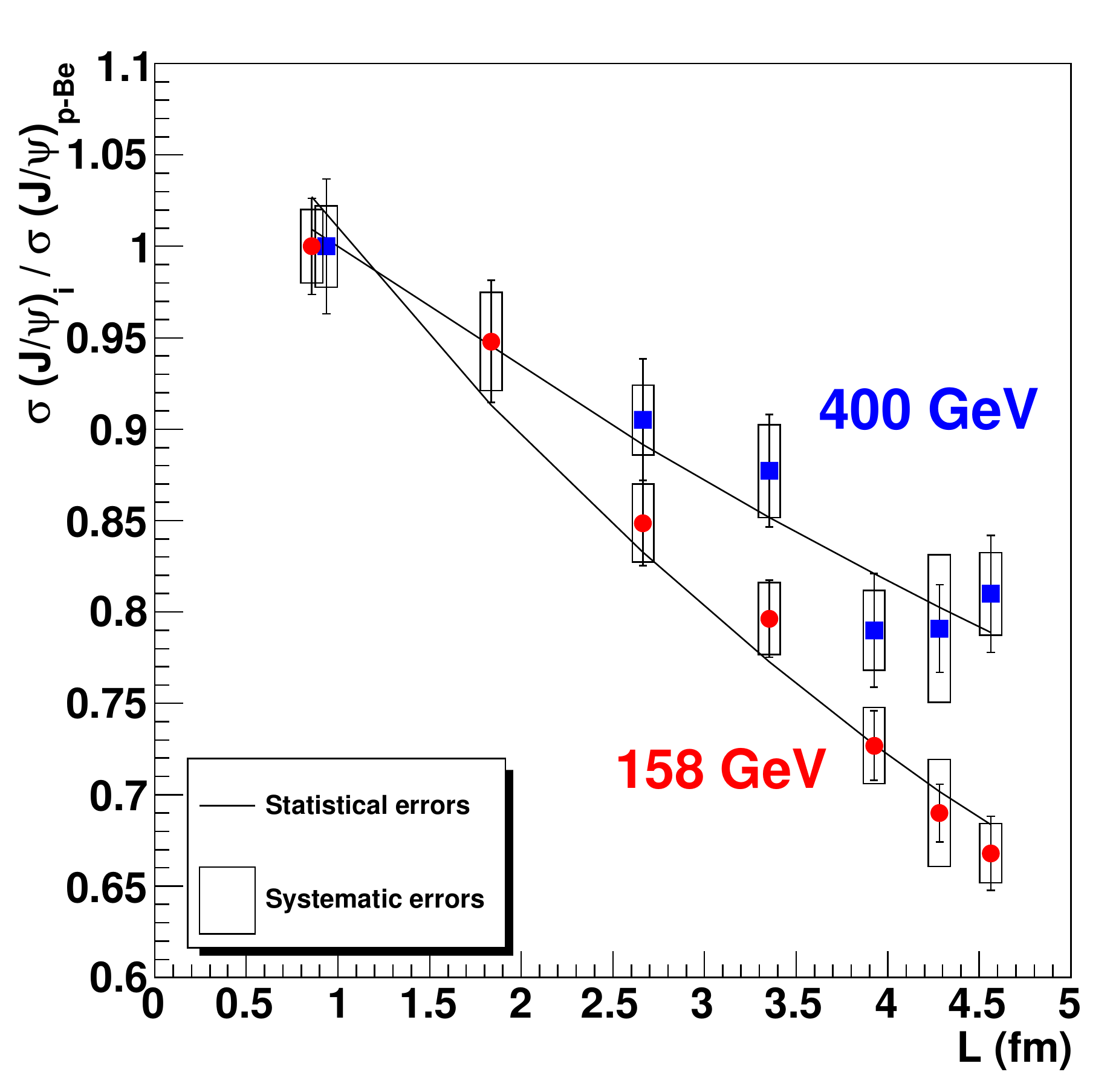}
      \caption{The \jpsi\ cross section ratios for $pA$ 
               collisions at 158\gev\ 
               {\it (circles)} and 400\gev\ {\it (squares)}, 
               as a function of $L$, the mean thickness 
               of nuclear matter traversed by the \jpsi.
               \figPermXNPA{Scomparin:2009tg}{2009} }
      \label{fig:media_fig6}
   \end{center}
\end{figure}

Cold-nuclear-matter effects were evaluated comparing 
the cross section ratio 
\beq
\frac{\sigma_{pA}^{\jpsi}}{ \sigma_{p {\rm Be}}^{\jpsi}}\, ,
\eeq 
for each nucleus with mass number $A$, relative to the lightest target (Be). 
The beam luminosity factors cancel out in the ratio, apart from a small 
beam-attenuation factor. However, 
since the sub-targets see the vertex telescope 
from slightly different angles, the track reconstruction 
efficiencies do not completely cancel out. 
Therefore an accurate evaluation of the time evolution of such quantities was  
performed target-by-target, with high granularity and on a run-by-run basis. 
The results~\cite{Scomparin:2009tg,Arnaldi:2009ph}, 
shown in \Fig{fig:media_fig6}, 
are integrated over $p_{T}$ and are given in the 
rapidity region covered by all the sub-targets, $0.28 < y_{\rm CMS} < 0.78$ 
for the 158\gev\ sample and $-0.17< y_{\rm CMS}< 0.33$ for the 400\gev\ 
sample. 
Systematic errors include uncertainties in the target thickness, 
the rapidity distribution used in the acceptance calculation, and the 
reconstruction efficiency. Only the fraction of systematic errors not common 
to all the points is shown since it affects the evaluation of nuclear
effects.

Nuclear effects have usually been parametrized by fitting the $A$ dependence 
of the \jpsi\ production cross section using the expression 
\beq
\sigma_{pA}^{\jpsi} = \sigma_{pp}^{\jpsi} A^\alpha\,\, .
\eeq
Alternatively, the effective absorption cross section, 
\sigabsj, can be extracted from the data using the Glauber
model.  Both $\alpha$ and \sigabsj\ are effective quantities
since they represent the strength of the cold-nuclear-matter effects that 
reduce the \jpsi\ yield.  However, they cannot distinguish among the 
different effects, \eg shadowing and nuclear absorption, contributing
to this reduction.  The results in \Fig{fig:media_fig6} were used to
extract 
\begin{eqnarray}
\sigabsj &=& 7.6 \pm 0.7 \, {\rm (stat.)} \, \pm 0.6 \,
{\rm (syst.)}~{\rm mb}\,;\nonumber\\
 \alpha &=& 0.882 \pm 0.009 \pm 0.008
\end{eqnarray} 
at 158\gev\ and 
\begin{eqnarray}
\sigabsj &=& 4.3 \pm 0.8 \, {\rm (stat.)} \pm 
0.6\, {\rm (syst.)}~{\rm mb}\,;\nonumber\\
\alpha &=& 0.927 \pm 0.013 \pm 0.009
\end{eqnarray}
at 400\gev. 
Thus \sigabsj\ is larger at 158\gev\ than at 400\gev\
by three standard deviations.
The 400\gev\ result is, on the other hand, in excellent agreement with the 
previous NA50 result obtained at the same energy~\cite{Alessandro:2006jt}.

\begin{figure}[t]
   \begin{center}
      \includegraphics[width=\figwid]{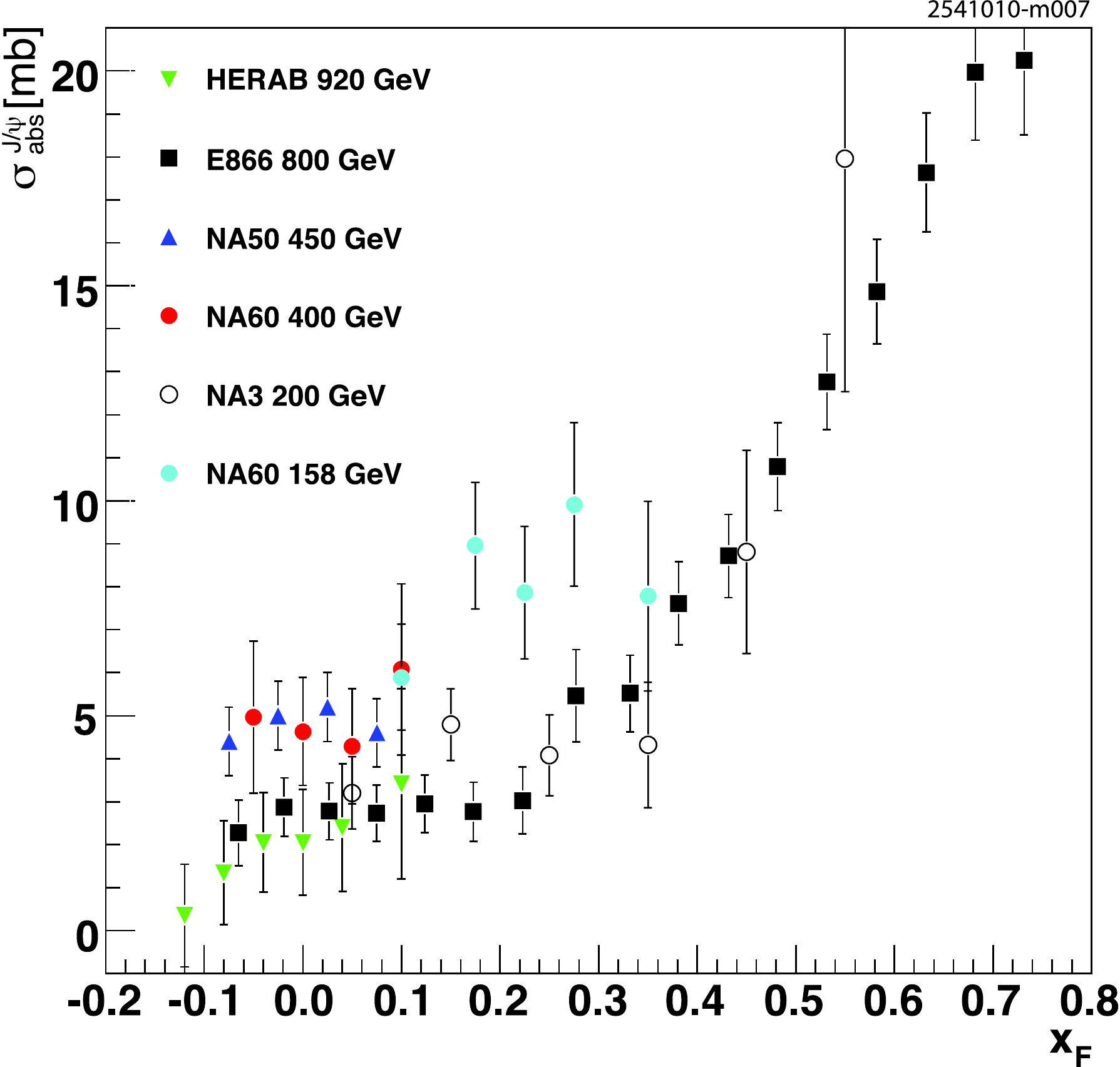}
      \caption{Compilation of \sigabsj\ 
               as a function of $x_F$
               with no additional cold-matter effects included.
               \figPermXNPA{Arnaldi:2009ph}{2009} }
      \label{fig:media_fig7}
   \end{center}
\end{figure}

The study of cold-nuclear-matter effects at fixed-target energies is a subject 
which has attracted considerable interest. In \Fig{fig:media_fig7}, a 
compilation of previous results for 
\sigabsj\ as a function of
$x_{F}$~\cite{Alessandro:2003pc,Leitch:1999ea,Badier:1983dg,Abt:2008ya} 
is presented, together with the new NA60 results~\cite{Arnaldi:2009ph}.  
Contrary to \Fig{fig:media_fig3}, the values of 
\sigabsj\ in \Fig{fig:media_fig7} do not 
include any shadowing contribution, only absorption.
There is a systematic increase in the nuclear effects 
going from low to high $x_F$ as well as when from high to low incident 
proton energies.  As shown in \Fig{fig:media_fig7}, the new NA60 results 
at 400\gev\ confirm the NA50 values obtained at a similar energy. On the other 
hand, the NA60 158\gev\ data suggest higher values of \sigabsj\ 
and hint at increased absorption over the $x_{F}$ range.  Note also
that the older NA3 \jpsi\ results are in partial contradiction with these 
observations, giving lower values of \sigabsj, similar to 
those obtained from the higher energy data samples. 
Such a complex pattern of nuclear effects results from a delicate interplay 
of various nuclear effects (final-state absorption, shadowing, 
initial-state energy loss, {\it etc.}) 
and has so far not been satisfactorily explained 
by theoretical models~\cite{Vogt:1999dw}.  A first attempt to 
disentangle the 
contribution of shadowing from 
\sigabsj\ (as extracted from the NA60 results) has been 
carried out 
using the EKS98~\cite{Eskola:1998df} parametrization of the 
nuclear PDFs. It was found that a larger 
\sigabsj\ is needed to describe the measured data:
\begin{eqnarray}
\sigabsj(158\gev) &=& 9.3 \pm 0.7 \pm 0.7~{\rm mb}\, ; \nonumber\\
\sigabsj(400\gev) &=& 6.0 \pm 0.9 \pm 0.7~{\rm mb}\, .
\end{eqnarray}  
The results thus depend on the parametrization of the nuclear modifications of the
PDFs. For example, slightly higher (5-10\%) values of \sigabsj\ 
are obtained if
the EPS08~\cite{Eskola:2008ca} parametrization is used.

\subsubsection{Anomalous \jpsi\ suppression}
\label{sec:media_subsec43}

The $pA$ results at 158\gev\ shown in the previous section have been collected 
at the same energy and in the same $x_{F}$ range as the SPS $AA$ data.  
It is therefore natural to use these results to calculate the expected 
magnitude
of cold-nuclear-matter effects on \jpsi\ production in nuclear collisions. 
In order to do so, the expected shape of the \jpsi\ 
distribution as a function of the forward energy 
in the zero degree calorimeter,
$dN^{\rm expect}_{\jpsi}/dE_{\rm ZDC}$, 
has been determined using the Glauber model.  The \jpsi\ yield is assumed to
scale with the number of $NN$ collisions.  The effective
\jpsi\ absorption cross
section in nuclear matter is assumed to be the same as the value at
158\gev\ deduced in the previous section.

\begin{figure}[b]
   \begin{center}
      \includegraphics[width=\figwid]{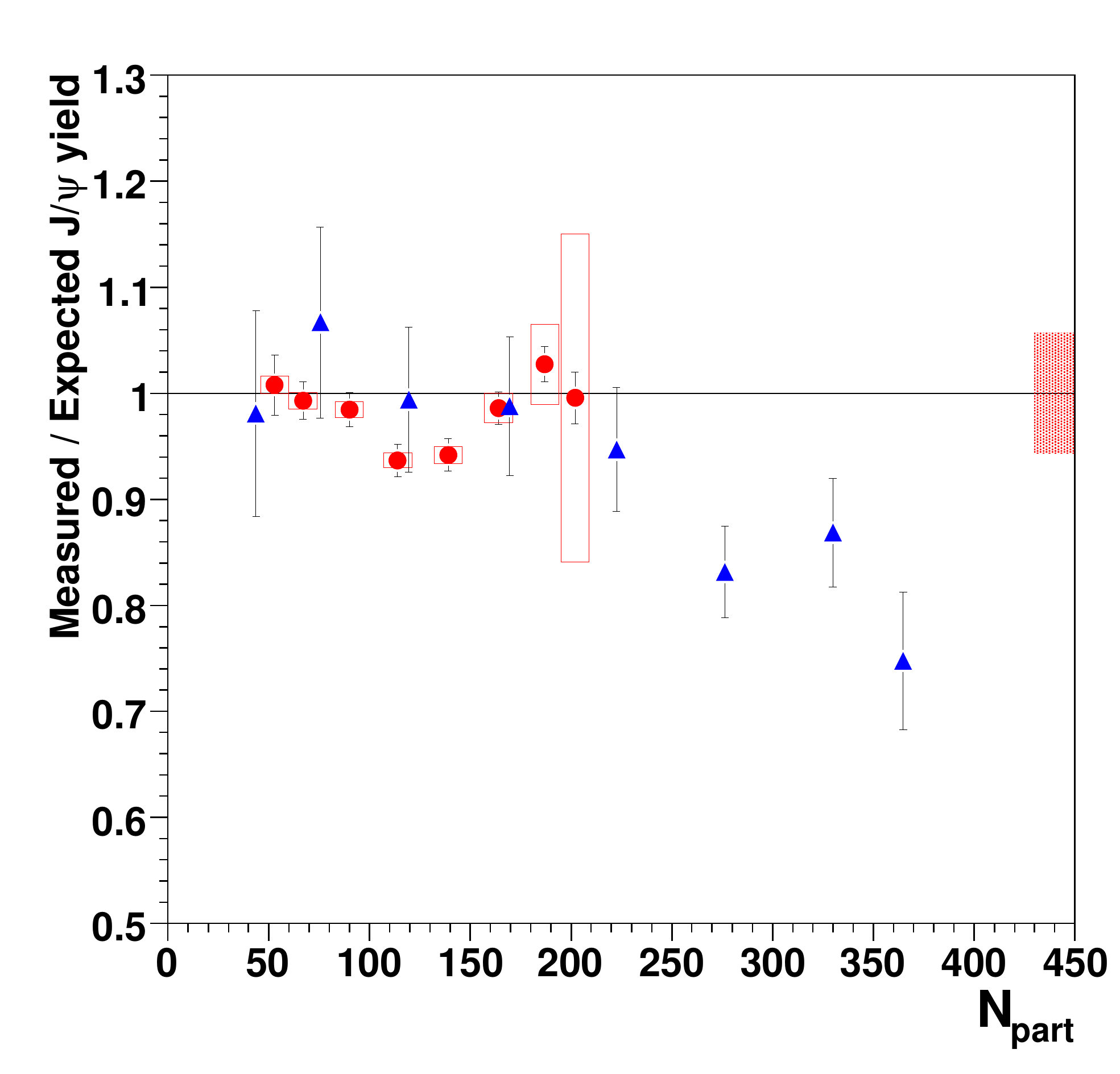}
      \caption{Anomalous \jpsi\ suppression in 
               In+In {\it (circles)} and Pb+Pb collisions
               {\it (triangles)} as a function of $N_{\rm part}$. 
               The {\it boxes} around the 
               In+In {\it points} represent correlated systematic 
               errors. The {\it filled box} on the
               right corresponds to the uncertainty 
               in the absolute normalization 
               of the In+In {\it points}. A 12\% global error, 
               due to the uncertainty on 
               \sigabsj\ at 158\gev\ is not shown.
               \figPermXNPA{Arnaldi:2009ph}{2009} }
      \label{fig:media_fig8}
   \end{center}
\end{figure}

The measured \jpsi\ yield, 
$dN_{\jpsi}/dE_{\rm ZDC}$, is
normalized to $dN^{\rm expect}_{\jpsi}/dE_{\rm ZDC}$ using the 
procedure detailed in \cite{Arnaldi:2007zz}.
This procedure previously did not take shadowing effects into account when
extrapolating from $pA$ to $AA$ interactions.  In $pA$ collisions, only the
target partons are affected by shadowing, while in $AA$ collisions, effects on 
both the projectile and target must be taken into account.  If shadowing is
neglected in the $pA$ to $AA$ extrapolation, a small bias is introduced,
resulting in an artificial $\sim 5$\% suppression of the \jpsi\ yield with
the EKS98 parametrization~\cite{Arnaldi:2009it}. 
Therefore, if shadowing is properly accounted for in the $pA$ to $AA$
extrapolation, the amount of the anomalous \jpsi\ suppression is reduced.
\Figure{fig:media_fig8} presents the new results for the 
anomalous \jpsi\ suppression in In+In and Pb+Pb 
collisions~\cite{Scomparin:2009tg,Arnaldi:2009ph}
as a function of $N_{\rm part}$, the number of participant nucleons.
Up to $N_{\rm part}\sim 200$ the \jpsi\ yield is, within errors, compatible 
with the extrapolation of cold-nuclear-matter effects.  When 
$N_{\rm part}> 200$, there is an anomalous 
suppression of up to $\sim20-30\%$ in the most 
central Pb+Pb collisions.  This new, smaller anomalous suppression is
primarily due to the larger \sigabsj\ extracted from the
evaluation of cold-nuclear-matter effects.

\subsection{Recent hadroproduction results from RHIC}
\label{sec:media_sec5}

The strategy of the RHIC \jpsi\ program has been to measure production cross 
sections in \sqrtsNN \, = 200\gev\ collisions for \ppcoll, \dAu, \AuAu\ and 
\CuCu\ collisions. 
RHIC  has also studied $J/\psi$ production in Cu+Cu
collisions at $\sqrtsNN = 62$\gev\ and will also study $J/\psi$
production in $pp$ collisions at $\sqrt{s} = 500$ GeV.
The \ppcoll\ collisions are studied both to learn about 
the \jpsi\ production mechanism and to provide baseline production cross 
sections needed for understanding the \dAu\ and \AAcoll\ data. Similarly, the 
\dAu\ measurements are inherently interesting because they study the 
physical processes that modify \jpsi\ production cross sections in nuclear 
targets and also provide the crucial cold-nuclear-matter 
baseline for understanding \jpsi\ production in \AAcoll\ collisions. 
Note that \dAu\ collisions are studied at RHIC instead of \pAu\ collisions 
for convenience -- \pAu\ collisions
are possible at RHIC, but would require a dedicated \pAu\ run.

The last few years of the RHIC program have produced \jpsi\ data from 
PHENIX for
 \ppcoll, \dAu\ and \AuAu\ collisions with sufficient statistical precision to 
establish the centrality dependence of both hot and cold-nuclear-matter 
effects 
at \sqrtsNN \, = 200\gev. The data cover the rapidity range $|y| < 2.4$.

We introduce some quantities that have been applied to d+Au and $AA$ collisions
at RHIC to describe the impact parameter, $b$ (also called {\it centrality}) ,
dependence of the quarkonium results.  While most of the data taken are at
large impact parameter (peripheral collisions), the small impact parameter
(central) collisions are more likely to produce a quark-gluon plasma.
Therefore it is important to study quantities over a range of centralities,
using impact-parameter dependent variables such as the number of participant
nucleons,  $N_{\rm part}$, and the number of collisions, $\Ncoll$.
The number of participants depends on $b$ as
\begin{eqnarray}
N_{\rm part}(b)  = \int \,d^2s\hspace{2.0in}\non\\ 
 \bigg[\, T_A(s) 
\left(\, 1 - \exp\left[ -\sigma_{\rm inel}(s_{_{NN}})
T_B\,(|\vec b - \vec s| )\,\right]\,\right)\, + \non\\
  T_B ( |\vec{b} - \vec{s}| ) \, (\, 1 - 
\exp\left[-\sigma_{\rm inel}(s_{_{NN}})
\, T_A(s)\,\right]\, )\,  \bigg] \, \, .~~~~~
\label{eqn:Med_npartdef}
\end{eqnarray}
Here $\sigma_{\rm inel}$ is the inelastic nucleon-nucleon cross section,
42~mb at RHIC, and $T_{A/B}(s) = \int dz \rho_{A/B}(s,z)$, the line integral
of the nuclear density, $\rho$, in the beam direction, is the nuclear
profile function.
Large values of $N_{\rm part}$ are obtained for small impact parameters with
$N_{\rm part}(b=0) = 2A$ for spherical nuclei.  Small values of $N_{\rm part}$
occur in very peripheral collisions.  
The number of collisions, $\Ncoll(s_{_{NN}};b) 
= \sigma_{\rm inel}(s_{_{NN}})
T_{AB}(b)$, depends on the nuclear overlap integral,
\begin{eqnarray}
T_{A B} (b) = \int d^2s \,dz\, dz'\, \rho_A(s,z)\, \rho_B(|\vec{b}-\vec{s}|,z') \, \, .
\label{eqn:Med_tabdef}
\end{eqnarray}
In $pA$ collisions, we assume that the proton has a negligible size,
$\rho_A(s,z) = \delta(s) \delta (z)$ so that $T_{AB}(b)$ 
collapses to the nuclear profile function.
The deuteron cannot be treated as a point particle since it is large
and diffuse.  Thus the H\'{u}lthen
wave function~\cite{Kharzeev:2002ei,hulthen} is used to calculate the deuteron density
distribution.  No shadowing effects are included on the deuteron.

The nuclear suppression factor, $R_{AB}$, for d$A$, and $AA$
collisions is defined as the ratio
\begin{equation}
R_{AB}(N_{\rm part};b)  = {d \sigma_{AB}/dy \over T_{AB}(b) \,d\sigma_{pp}/dy }
\label{eqn:Med_rab}
\end{equation}
where $d\sigma_{AB}/dy$ and $d\sigma_{pp}/dy$ are the quarkonium rapidity
distributions in $AB$ and $pp$
collisions and $T_{AB}$ is the nuclear overlap function, defined in 
\Eq{eqn:Med_tabdef}.  In $AA$ collisions, $R_{AA}$ is sometimes shown relative
to the extracted cold-nuclear-matter baseline, $R_{AA}^{\rm CNM}$.
PHENIX has also shown both d+Au and $AA$ data as a function of $R_{CP}$, 
the ratio of $AB$ cross sections
in central relative to peripheral collisions, 
\begin{eqnarray} 
R_{CP}(y) = \frac{T_{AB}(b_P)}{T_{AB}(b_C)} 
\frac{d\sigma_{AB}(b_C)/dy}{d\sigma_{AB}(b_P)/dy} \, \, ,
\label{eqn:Med_rcpdef}
\end{eqnarray} 
where $b_C$ and $b_P$ correspond to the central and peripheral values of the
impact parameter since systematic uncertainties cancel in the ratio.  
Another quantity of interest is $v_2$, the second harmonic of the azimuthal
Fourier decomposition of the momentum distribution, $dN/dp_T \propto
1 + 2v_2 \cos(2(\phi - \phi_r))$ where $\phi$ is the particle emission angle
and  $\phi_r$ is the reaction plane angle, known as the elliptic flow.  It
gives some indication of the particle response to the thermalization of the medium.
A finite $J/\psi$ $v_2$ would give some indication of whether the $J/\psi$
distribution becomes thermal.  The strength of $v_2$ depends on the proportion
of $J/\psi$ produced by coalescence.

In the next few years the increased RHIC luminosity and the commissioning of 
upgraded detectors and triggers for PHENIX and STAR will enable a next 
generation of RHIC measurements, extending the 
program to the \ups\ family, excited charmonium states, and 
\jpsi\ $v_2$ and 
high-\ptrans\ suppression measurements. There have already been low-precision, 
essentially proof-of-principle, measurements of most of those signals.
Very importantly, upgraded silicon vertex detectors for both PHENIX and STAR 
are expected to produce qualitatively better open charm measurements that 
will provide important inputs to models of \jpsi\
production in heavy-ion collisions.

In addition to the results discussed here, there have been PHENIX results on 
\jpsi\ photoproduction in peripheral \AuAu\ collisions~\cite{Afanasiev:2009hy} 
and a proof-of-principle measurement of the \jpsi\ $v_2$ in \AuAu\ 
collisions by PHENIX~\cite{Atomssa:2009ek} 
with insufficient precision for physics conclusions. 

\subsubsection{Charmonium from \ppcoll\ collisions}
\label{sec:media_subsec52}

\begin{figure}[b]
   \begin{center}
      \includegraphics[width=\figwid]{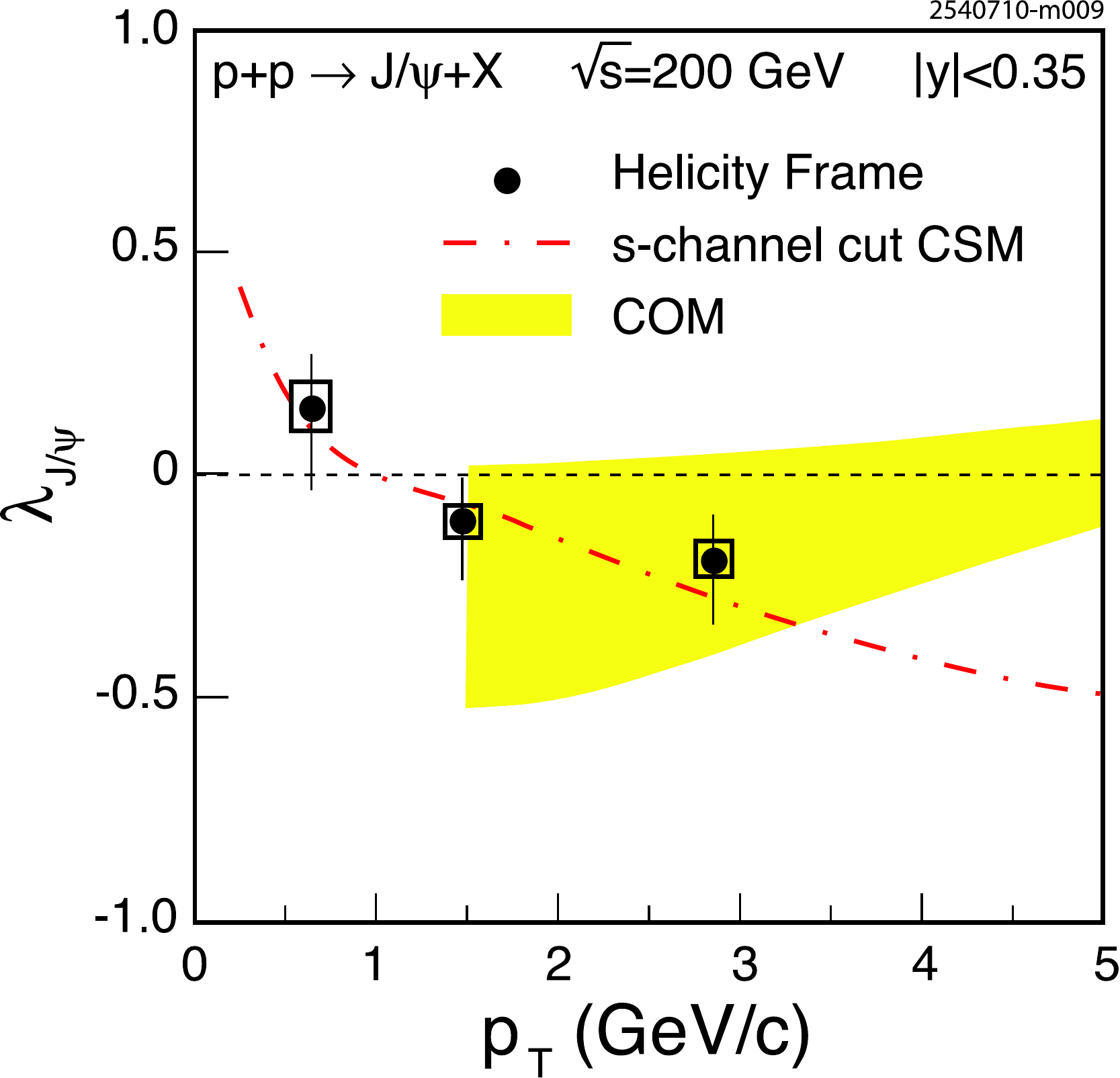}
      \caption{The polarization extracted from 200\gev\ PHENIX
               \ppcoll\ data at 
               midrapidity as a function of $p_T$.  
               The data are compared with the 
               $s$-channel-cut 
               CSM~\cite{Haberzettl:2007kj} and a 
               COM~\cite{Chung:2009xr} prediction.
               \AfigPermAPS{Adare:2009js}{2010}  }
      \label{fig:media_fig9}
   \end{center}
\end{figure}

PHENIX~\cite{Adare:2009js} has reported measurements of the inclusive 
\jpsi\ polarization in 200\gev\ \ppcoll\
collisions at midrapidity.  Results for the 
polarization parameter $\lambda$, defined in the Helicity frame, 
are shown in 
\Fig{fig:media_fig9} and
compared to COM~\cite{Chung:2009xr}
and $s$-channel-cut CSM~\cite{Haberzettl:2007kj} predictions. The latter
has been shown to 
describe the rapidity and \ptrans\ dependence of the PHENIX 200\gev\ 
\ppcoll\ \jpsi\ data~\cite{Lansberg:2008jn} using a two-parameter 
fit to CDF data at $\sqrts = 1.8\tev$.

At Quark~Matter~2009, PHENIX~\cite{daSilva:2009yy} 
showed preliminary measurements of the \ptrans\ 
dependence of the \psip\ cross section at 200\gev. 
This is the first measurement of the \ptrans-dependence 
of an excited charmonium state at RHIC. PHENIX measured the 
feeddown contribution of the \psip\
to the \jpsi\ to be $(8.6 \pm 2.3)\%$, in good agreement 
with the world average. 

STAR~\cite{Abelev:2009qaa} has published measurements of the \jpsi\ 
cross section in 200\gev\ \ppcoll\ collisions for 
$5 < \ptrans < 13\gevc$. 
This greatly extends the \ptrans\ range over which \jpsi\ data are 
available at 
RHIC. Although PHENIX can trigger at all \ptrans, it has so far
been limited to \ptrans\ below about 9\gevc~\cite{daSilva:2009yy} 
because of its much smaller acceptance.

\subsubsection{Charmonium from \CuCu\ collisions}
\label{sec:media_subsec53}

Measured quarkonium production rates from heavy-ion collisions are
commonly presented in terms of a nuclear modification factor, \RAA, 
defined in \Eq{eqn:Med_rab}.
PHENIX~\cite{Adare:2008sh} 
results on the rapidity and \ptrans\ dependence of \RAA\ 
values for \jpsi\ from 200\gev\ \CuCu\ collisions were published some time 
ago. However, those results were limited to $p_T < 5$\gevc, 
and do not address the high-\ptrans\ behavior of the 
measurements very well.
STAR~\cite{Abelev:2009qaa} has now published \CuCu\ \RAA\ data
for \jpsi\ at 
5.5 and 7\gevc\ that yield an average 
$\langle\RAA\rangle = 1.4\pm0.4({\rm stat})\pm0.2({\rm sys})$ 
above 5\gevc\ for the 0-20\% most central collisions. 
The \RAA\ data for the 0-60\% most central collisions have very similar values,
in contrast to the PHENIX data below 5\gevc\ that yield 
$\langle\RAA\rangle\approx 0.52$ for central \CuCu\ collisions. 

PHENIX~\cite{Atomssa:2009ek} has also 
released preliminary data on the \RAA\ for \jpsi\ from minimum 
bias (0-94\% centrality) 
\CuCu\ data at 7 and 9\gevc. The 
minimum bias PHENIX data should be comparable to the STAR 0-60\% data, but 
the PHENIX results are more consistent with a nearly \ptrans-independent \RAA. 
However, both measurements have large statistical uncertainties and a direct 
comparison~\cite{LindenLevy:2009td} 
of the STAR and PHENIX \CuCu\ \RAA\ data at high 
\ptrans\ suggests
that  more data will be required to definitively determine
the high-\ptrans\ behavior of \RAA\ in central collisions.

\subsubsection{Bottomonium production}
\label{sec:media_subsec54}

PHENIX~\cite{Leitch:2007wa} showed a preliminary result for the
\Usum\ cross section at forward and 
backward rapidity ($1.2 < |y| < 2.4$) at Quark Matter 2006.
More recently, PHENIX~\cite{daSilva:2009yy} 
showed a preliminary result at Quark Matter 2009 for 
\Usum\ production in 200\gev\ \ppcoll\ collisions at midrapidity 
($|y| < 0.35$). The measured cross 
section is $\Brat d\sigma/dy = 114^{+46}_{-45}$~pb at $y=0$,
where the presence of the $\Brat$ reflects that
the results have not been separated by individual
\UnS{n}\ resonance nor corrected for the
dilepton branching fractions $\Brat(\UnS{n}\to\epem)$.

STAR~\cite{:2010am} published a measurement of the 
$\Usum \to e^+e^-$ cross section at $|y| < 0.5$ for 
200\gev\ \ppcoll\ collisions. The measured value is
$\Brat d\sigma/dy = 114 \pm 38 \, ({\rm stat}) ^{+23}_{-24} \, ({\rm syst})$~pb
at $y=0$. 
STAR~\cite{Liu:2009wa} also has a preliminary result for the 
$\Usum \to e^+e^-$ 
cross section at midrapidity in \dAu\ collisions 
at 200\gevc. The cross section was
found to be 
$\Brat d\sigma/dy = 35 \pm 4\, ({\rm stat}) \pm 5\, ({\rm syst})$~nb. 
The midrapidity value of \RdAu\ was found to be 
$0.98 \pm 0.32 \, ({\rm stat}) \pm 0.28 \, ({\rm syst})$, consistent with 
binary scaling.

PHENIX has made a preliminary measurement of the dielectron yield in the 
\Usum\
mass range at midrapidity in \AuAu\ 
collisions~\cite{Atomssa:2009ek}. In combination with 
the PHENIX \Usum\ \ppcoll\ result 
at midrapity, a 90\% CL upper limit on \RdAu\
of 0.64 was found for the \Usum\ 
mass region.  The significance of this 
result is not yet very clear since the measurement is for all three \ups\ 
states combined. 

\subsubsection{\jpsi\ production from d+Au collisions}
\label{sec:media_subsec55}

As discussed previously, modification of the \jpsi\ production cross section 
due to the presence of a nuclear target is expected to 
be caused by shadowing, breakup of the precursor \jpsi\ state by collisions 
with nucleons, initial-state energy loss, and other possible effects.  
Parametrizing these effects by employing a Glauber model with a fitted 
effective \jpsi-absorption cross section, \sigabsj, 
results in an effective cross section with strong rapidity and 
$\sqrtsNN$ dependencies~\cite{Lourenco:2008sk} that are not well 
understood. A large increase in the 
effective absorption cross section is observed
by E866/NuSea~\cite{Leitch:1999ea}  at forward rapidity.
This increase cannot be explained by shadowing models 
alone, suggesting that there are 
important physics effects omitted from the Glauber 
absorption-plus-shadowing model.

The extraction of 
hot-matter effects in the \AuAu\ \jpsi\ data at RHIC has been seriously 
hampered by the poor understanding of \jpsi\ production in nuclear targets, 
including the underlying \jpsi\ production mechanism. Thus the 
cold-nuclear-matter baseline has to be obtained experimentally.

The PHENIX \jpsi\ data obtained in the 2003 RHIC \dAu\ run did not have 
sufficient statistical precision either for studies of cold-nuclear-matter 
effects or for setting a cold-nuclear-matter baseline for the \AuAu\ 
data~\cite{Adare:2007gn}. This low-statistics measurement has been augmented
by the large \jpsi\ data set obtained in the 2008 \dAu\ run. 
PHENIX~\cite{daSilva:2009yy} has 
released \dAu\ \RCP\ data for \jpsi\ production
in nine rapidity bins 
over $|y| < 2.4$.  Systematic uncertainties associated with the 
beam luminosity, detector acceptance, trigger efficiency, and tracking 
efficiency cancel in \RCP, defined in \Eq{eqn:Med_rcpdef}.
There is a remaining systematic uncertainty due to the centrality 
dependence of the tracking and particle identification efficiencies.

The use of a Glauber model also gives rise to 
significant systematic uncertainties in the centrality dependence of
\RCP. The model is used to calculate the
average number of nucleon-nucleon collisions as a means of estimating the 
relative normalization between different centrality bins. The systematic 
uncertainty due to this effect is independent of rapidity. 

\begin{figure}[t]
   \begin{center}
      \includegraphics[width=\figwid]{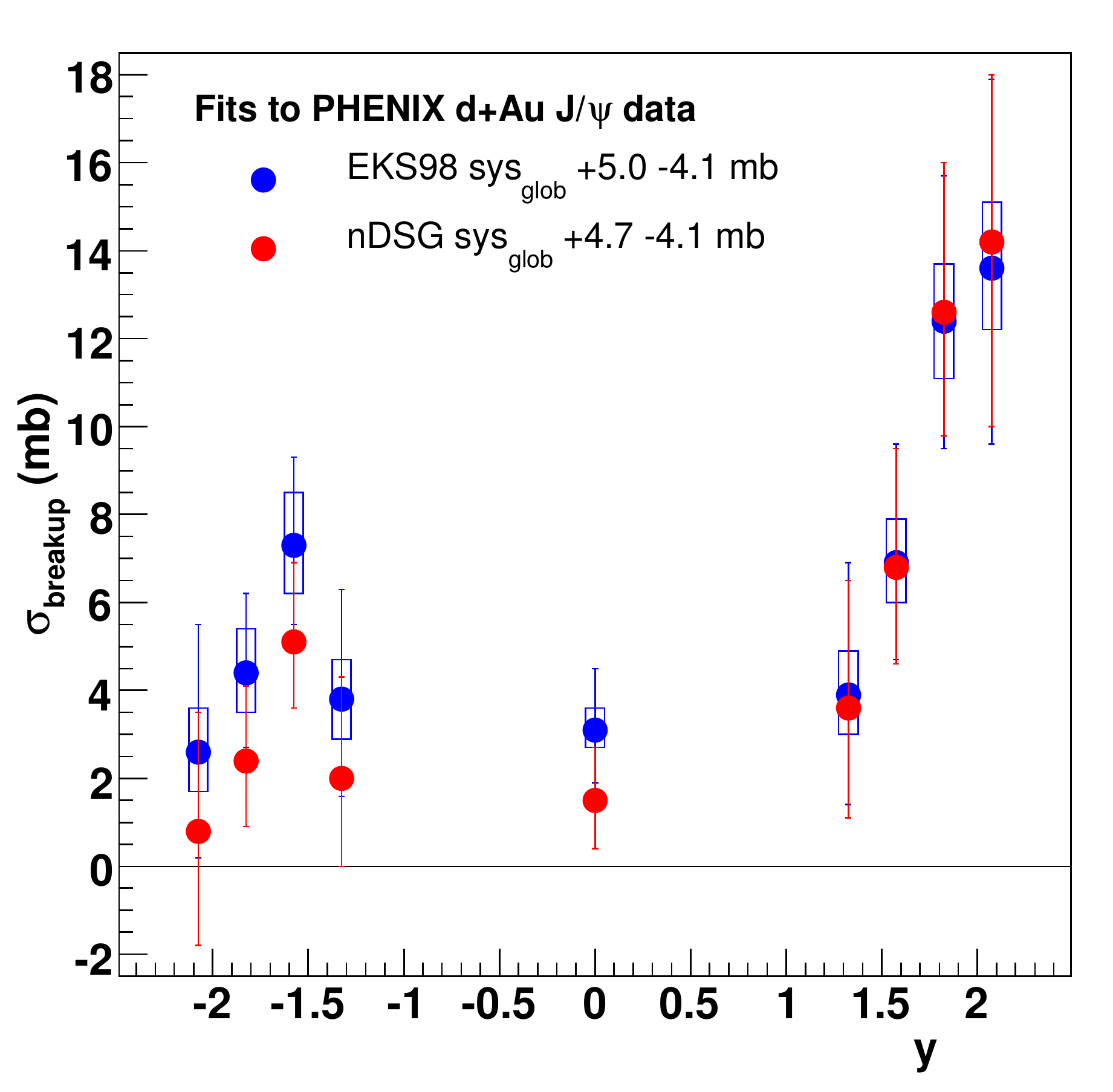}
      \caption{The effective absorption cross section as a 
               function of rapidity 
               extracted from PHENIX \dAu\ \RCP\ data 
               using the EKS98 and nDSg shadowing 
               parametrizations. The {\it vertical bars} show 
               uncorrelated point-to-point 
               uncertainties, the {\it boxes} show correlated 
               uncertainties, and the global 
               uncertainties are given in the legend }
      \label{fig:media_fig10}
   \end{center}
\end{figure}

The PHENIX \dAu\ \RCP\ data have been independently fitted at each of the nine 
rapidities~\cite{tonyect} employing a model including shadowing and 
\jpsi\ absorption.  The model calculations~\cite{Vogt:2004dh} use the EKS98
and nDSg shadowing parametrizations with $0 \leq \sigma_{\rm abs} \leq 15$~mb.
The best fit absorption cross section was determined at each rapidity, along 
with the $\pm 1\sigma$ uncertainties
associated with both rapidity-dependent and
rapidity-independent systematic effects. 
The results are shown in \Fig{fig:media_fig10}.
The most notable feature is the stronger effective absorption cross section at 
forward rapidity, similar to the behavior observed at lower energies 
\cite{Leitch:1999ea}. In fact, it is striking that 
the extracted cross sections at forward rapidity are very similar for PHENIX 
($\sqrtsNN = 200\gev$) and E866~\cite{Lourenco:2008sk}
($\sqrtsNN= 38.8\gev$) 
(see the lower panel of \Fig{fig:media_fig3}),
despite the large difference in center-of-mass energies.

Note the large global systematic uncertainty in $\sigma_{\rm abs}$ extracted 
from the PHENIX \RCP\ data, dominated by the uncertainty in the 
Glauber estimate of the average number of collisions at each centrality. 
Although it does not affect the shape of the rapidity dependence of \sigabsj, 
it results in considerable
uncertainty in the magnitude of the effective absorption cross section. 

It has been suggested~\cite{Ferreiro:2008wc} that the large increase in 
effective absorption cross section at forward rapidity obtained from a CEM
calculation~\cite{tonyect}
may be moderated significantly if the $2 \to 2$ kinematics of
the leading-order CSM is used.  This difference emphasizes the importance of
understanding the underlying production mechanism.

PHENIX has very recently released~\cite{Adare:2010fn} final 
\RdAu\ and \RCP\ data from the 2008 d+Au RHIC run. The final \RCP\
data are in good agreement with the preliminary data, 
discussed earlier, as well as in the next section. A 
comparison~\cite{Adare:2010fn} of the \RdAu\ data, which has not 
been shown before, with the \RCP\ data shows that a simultaneous 
description of the two observables will require a stronger than 
linear dependence of the $J/\psi$ suppression on the nuclear thickness 
function at forward rapidity.  The dependence of the suppression on 
nuclear thickness is at least quadratic, and is likely 
higher. The result has important implications for the understanding of
forward-rapidity d+Au physics.  Since the calculations of the cold matter
contributions to \RAA\ assumed that shadowing depends linearly on the nuclear
thickness, the calculations of \RAA\ shown in the next section should be 
be revisited.

\subsubsection{\jpsi\ production from Au+Au collisions}
\label{sec:media_subsec56}

\begin{figure}[tb]
   \begin{center}
      \includegraphics[width=\figwid]{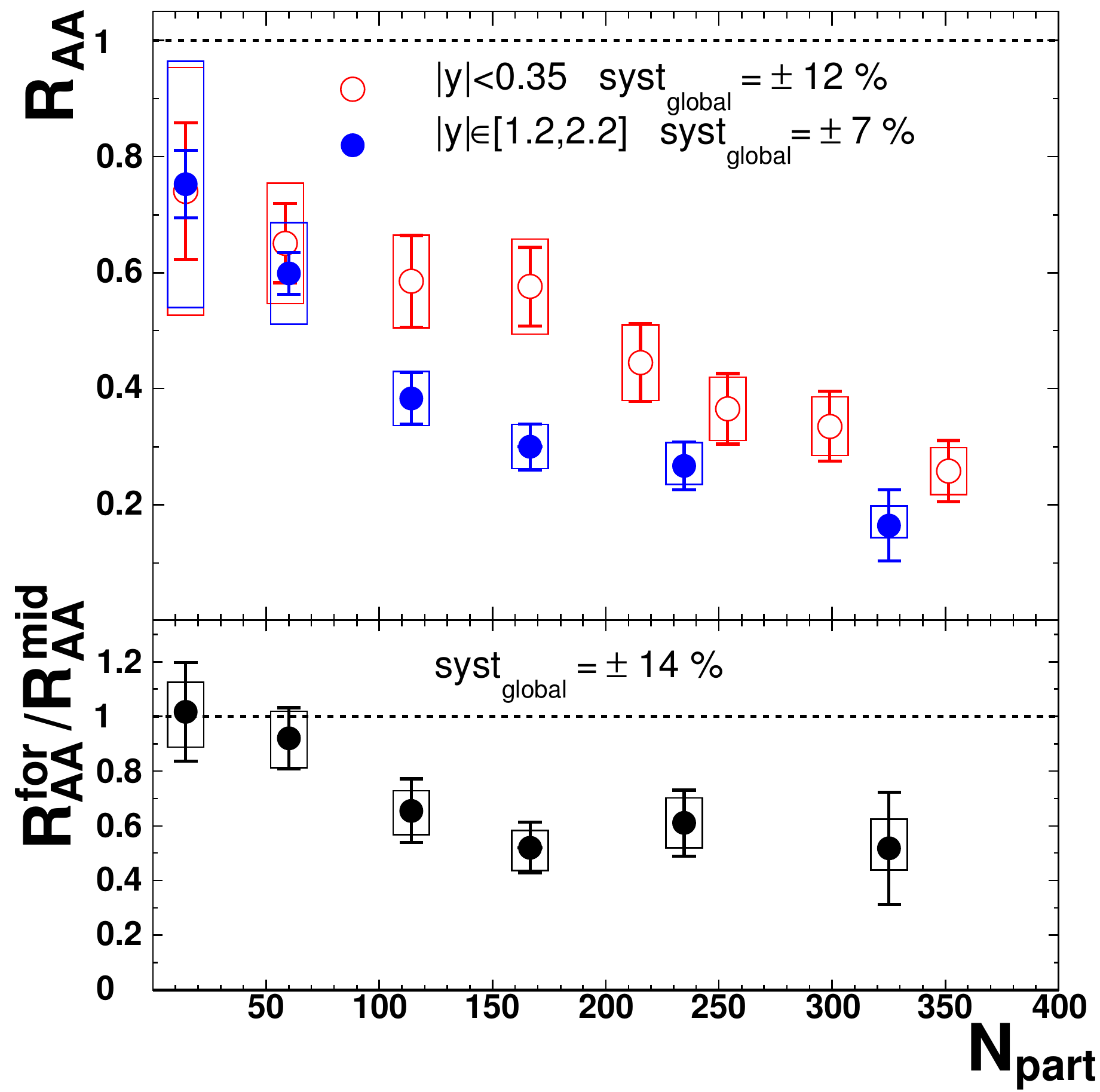}
      \caption{The PHENIX \AuAu\ \RAA\ as a function of centrality for 
               $|y| < 0.35$ and $1.2 < |y| < 2.2$ }
      \label{fig:media_fig11}
   \end{center}
\end{figure}

PHENIX~\cite{Adare:2006ns} has published the centrality dependence 
of \RAA\ for \AuAu\ collisions using \AuAu\ data from the 
2004 RHIC run and \ppcoll\ data from the 2005 run. The data are
shown in \Fig{fig:media_fig11}. The suppression is considerably 
stronger at forward rapidity than at midrapidity. The significance of 
this difference with respect to hot-matter effects will not be clear, however, 
until the suppression due to cold-nuclear-matter effects is
more accurately known.

To estimate the cold-nuclear-matter contribution to the \AuAu\ \jpsi\ \RAA\, 
the \dAu\ \jpsi\ \RCP\ data were extracted using the EKS98 and nDSg shadowing 
parametrizations, as described earlier, except that, in this case,
the \sigabsj\ values in \dAu\ collisions were fitted independently in three 
rapidity intervals: $-2.2 < y < -1.2$, $|y| < 0.35$, and $1.2 < y < 2.2$. 
In effect, this tunes the calculations to reproduce the \dAu\ \RCP\ 
independently in each of the three rapidity windows in which the \AuAu\ 
\RAA\ data were measured. The cold-nuclear-matter \RAA\ for Au+Au collisions
was then estimated in a Glauber calculation 
using the fitted absorption cross sections and the centrality-dependent \RpAu\ 
values
calculated using EKS98 and nDSg shadowing parametrizations~\cite{Vogt:2010aa}.
Each nucleon-nucleon collision contributes differently to the \RAA\  
in each rapidity window. To more directly simulate nucleon-nucleus 
interactions, the analysis assumes that \RAA\ can be treated
as a convolution of $p+$Au and Au$+p$ collisions in the 
three rapidity windows.  The impact-parameter
dependence of \RpAu\ is determined separately to infer the \RAA\ centrality
dependence for a rapidity-dependent absorption cross section.  Thus the value
of \RpAu\ at the impact parameter of nucleon 1 in the projectile is convoluted
with the value of $R_{{\rm Au}p}$ at the impact parameter of nucleon 2 in the
target.  Effectively, this means that to obtain \RAA\ for $1.2 < |y| < 2.2$, 
\RpAu\ for the forward-moving nucleon ($1.2 < y < 2.2$) is multiplied by 
\RpAu\ for the backward-moving nucleon ($-2.2 < y < -1.2$).
When $|y| < 0.35$, the \RpAu\ calculations at midrapidity are used.  
The number of participants, obtained from a Glauber calculation, is used to 
bin the collisions in centrality with a cut on peripheral events to mimic the 
effect of the PHENIX trigger efficiency at large impact parameter. The 
uncertainty in the calculated CNM \RAA\ was estimated by repeating 
the calculation with \sigabsj\ varied away from
best-fit values. This variation ranged over the
rapidity-dependent systematic uncertainty 
determined when fitting the \dAu\ \RCP. 

The global systematic uncertainty in \sigabsj\ was neglected in the 
calculation of the CNM \RAA. This was done because the same Glauber model 
was used to obtain both the number of nucleon-nucleon collisions,
\Ncoll, in \dAu\ and \AuAu\ interactions and the fitted \sigabsj\ values. 
Therefore, if, for example, \Ncoll\ is underestimated for the \dAu\ \RCP, 
the fitted absorption cross section will be overestimated.  However, 
this would be compensated in the calculated CNM \RAA\ by the underestimated 
\Ncoll\ value. Any possible differences in the details of the \dAu\ and \AuAu\ 
Glauber calculations would result in an imprecise cancellation of the
uncertainties.  This effect has not yet been studied.

Note that there is a significant difference between the impact-parameter 
dependence of the \RpAu\ and \RdAu\ calculations~\cite{tonyect},
primarily for peripheral collisions, due to the 
smearing caused by the finite size of the deuteron.   Since \RdAu\
and \RpAu\ are calculated using the same basic model, this smearing does not
present a problem in the present analysis.  However, if the measured \RdAu\ 
was used directly in a Glauber model, as was done with the RHIC 2003
data~\cite{Adare:2007gn}, a correction would be necessary.

\begin{figure}[t]
   \begin{center}
      \includegraphics[width=\figwid]{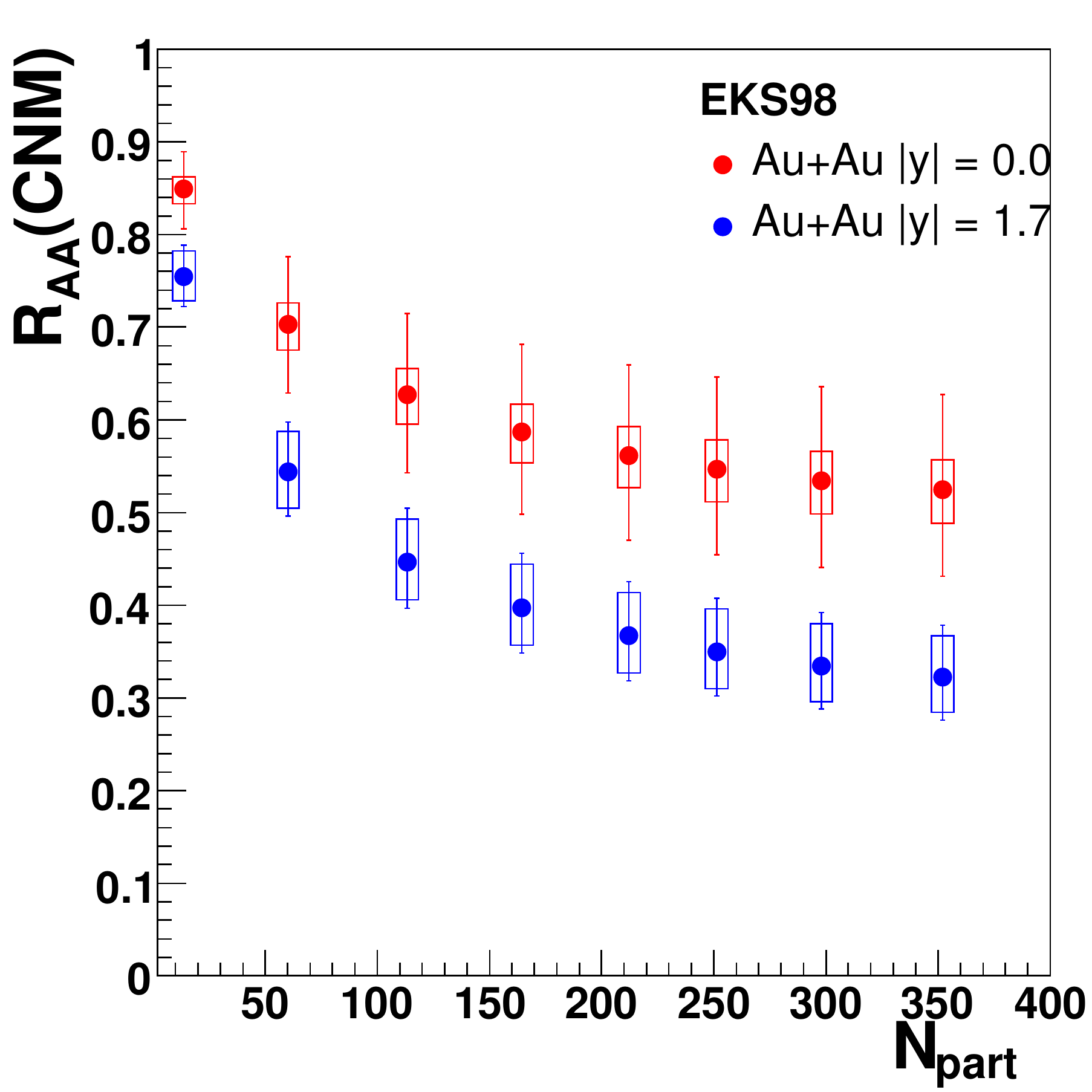}
      \caption{The estimated \AuAu\ cold-nuclear-matter \RAA\ as a function of 
               centrality for $|y| < 0.35$ and $1.2 < |y| < 2.2$. 
               The {\it vertical bar} represents 
               the rapidity-dependent systematic uncertainty in 
               the fitted \sigabsj }
      \label{fig:media_fig12}
   \end{center}
\end{figure}

The resulting Glauber calculations of the cold-nuclear-matter \RAA\ using
the EKS98 shadowing parametrization are shown in 
\Fig{fig:media_fig12}. The values obtained with nDSg are almost 
identical, as they should be since both methods parametrize the same data.

\begin{figure}[t]
   \begin{center}
      \includegraphics[width=\figwid]{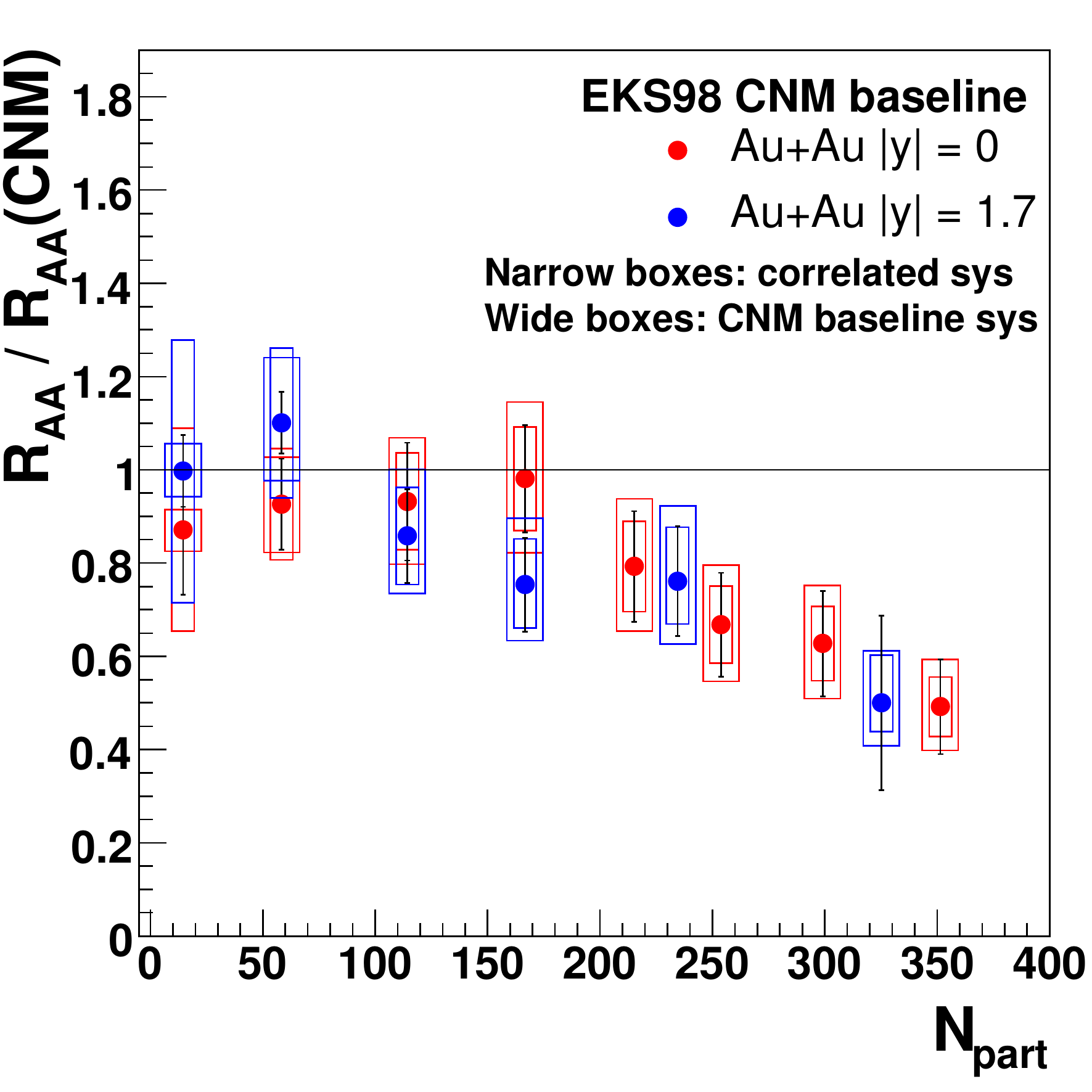}
      \caption{The estimated \AuAu\ suppression relative to the 
               cold-nuclear-matter \RAA\ as a 
               function of centrality for $|y| < 0.35$ and 
               $1.2 < |y| < 2.2$. The systematic uncertainty of the 
               baseline cold-nuclear-matter \RAA\ is depicted by the 
               {\it wide box} around each {\it point}. The 
               {\it narrow box} is the systematic uncertainty 
               in the \AuAu\ \RAA }
      \label{fig:media_fig13}
   \end{center}
\end{figure}

We emphasize that the kinematic-dependent differences in the effective 
absorption cross sections noted in the previous section do not 
affect the cold-nuclear-matter \RAA\ derived from the data.  As long as the 
method of fitting the \dAu\ data is consistent with the estimate of the cold 
nuclear matter \RAA, the result should be model independent.

The \jpsi\ suppression beyond CNM effects in \AuAu\ collisions 
can be estimated by dividing the measured \RAA\ by the estimates of the CNM 
\RAA. The result for EKS98  is shown in \Fig{fig:media_fig13}. The 
result for nDSg is nearly identical.

Assuming that the final PHENIX \RdAu\ confirms the 
strong suppression at forward rapidity seen in \RCP, it would suggest that 
the stronger suppression seen at forward/backward rapidity in 
the PHENIX \AuAu\ \RAA\ data is primarily due to cold-nuclear-matter effects. 
The suppression due to hot-matter effects seems to be comparable at midrapidity
and at forward/backward rapidity.

Finally, it is possible to use the effective absorption cross sections obtained
from the \dAu\ \jpsi\ \RCP\ data in a similar
Glauber calculation of \RpCu\ to estimate the cold-nuclear-matter \RAA\ for 
\CuCu\ collisions. However, the resulting 
CNM \RAA\ for \CuCu\ is significantly 
different for EKS98 and nDSg~\cite{tonyect}, most likely due to the
different $A$ dependences of EKS98 and nDSg. 
Measurements of \jpsi\ production in \pCu\ or \dCu\ collisions would be needed
to reduce the model dependence of the estimated CNM \RAA\ for \CuCu\ 
collisions. 

\subsection{Anomalous suppression: SPS vs RHIC}
\label{sec:media_sec6}

\begin{figure}[t]
   \begin{center}
      \includegraphics[width=\figwid]{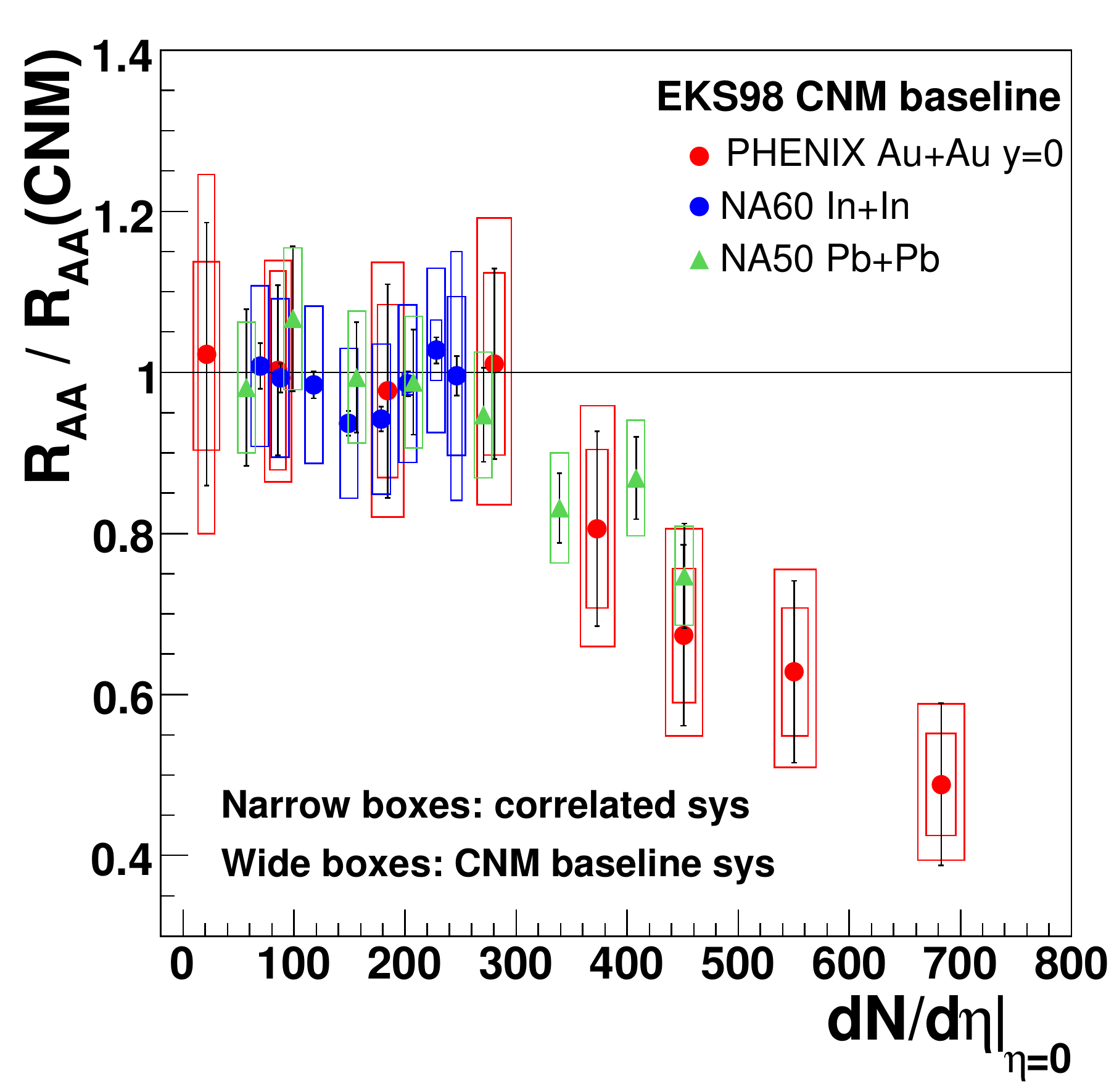}
      \caption{Comparison of the anomalous suppression at the SPS and RHIC 
               as a function of $dN_{\rm ch}/d\eta$ at $\eta=0$ }
      \label{fig:media_fig14}
   \end{center}
\end{figure}

The preliminary PHENIX d+Au results at $\sqrt{s}=200$\gev\ are, for the first 
time, based on a high-statistics sample~\cite{tonyect}.  Comparing these 
results with the previous Au+Au data gives an estimate of the magnitude 
of the anomalous \jpsi\ suppression at RHIC.
The newly-available NA60 $pA$ results at 158\gev, described in 
\Sec{sec:media_sec4}, allow significant comparisons of 
the centrality dependence of the anomalous suppression at the SPS and 
that obtained at RHIC. 
Work is in progress to make such a comparison as a function of several  
variables of interest, such as the charged particle multiplicity, 
$dN_{\rm ch}/d\eta$, and the Bjorken energy density reached in the 
collision.  The anomalous suppression patterns in In+In and Pb+Pb collisions
at the SPS and the midrapidity Au+Au results at RHIC are presented
as a function of $dN_{\rm ch}/d\eta$ in 
\Fig{fig:media_fig14}~\cite{robertaect}.  Note that the
magnitude of the anomalous \jpsi\ suppression is practically system- and 
$\sqrt{s}$-independent when expressed as a function of 
$dN_{\rm ch}/d\eta|_{\eta =0}$.

\subsection{Photoproduction in nuclear collisions}
\label{sec:media_subsec57}

In addition to in-medium hadroproduction, photoproduction of quarkonium may 
also occur in nucleus-nucleus collisions.  In this case, one nucleus acts as a 
photon source (the photon flux is given by the Weizsacker-Williams formalism). 
The photons fluctuate to virtual quark-antiquark pairs which interact with the 
opposite (target) nucleus \cite{Bertulani:2005ru,Baur:2001jj} 
and emerge as heavy quarkonia 
(\eg $\jpsi$ and $\Upsilon$) or other, light, vector mesons.   Such 
$\jpsi$ photoproduction has been observed in Au+Au collisions with 
PHENIX~\cite{Afanasiev:2009hy} and in $\overline p p$ collisions at the 
Tevatron~\cite{Aaltonen:2009kg}. CDF~\cite{Aaltonen:2009kg} 
has also observed $\psi(2S)$ 
photoproduction.   

At the LHC, photoproduction can be studied at far higher energies than 
available at fixed-target facilities or at HERA.  At the maximum $pp$ energy of
the LHC, $\gamma p$ collisions with center-of-mass energies up to 
$\sqrt{s_{\gamma p}} = 8.4$\tev\ are 
accessible, forty times the energy reached 
at HERA.  With Pb beams at maximum energy, the per-nucleon center-of-mass 
energy can reach $\sqrt{s_{\gamma N}} = 950$\gev~\cite{Baltz:2007kq}, 
equivalent to a 480\tev\ photon beam on a fixed target.

Photoproduction is of interest because it is sensitive to the gluon 
distribution in the target nucleus.  The cross section for 
$\gamma p\rightarrow V p$ scales as~\cite{Ryskin:1992ui}
$[x\,\,g(x,\,Q^2)]^2$,
where $x$ is the gluon momentum fraction, $Q^2=m_V^2/4$ 
is the photon virtuality, and 
$m_V$ is the vector meson mass.  For low $p_T$ vector mesons, the gluon 
momentum fraction, $x$, may be related to the final state rapidity, $y$, by 
\beq
y = -\frac{1}{2}\, \ln\left(\frac{2\,x\,\gamma\, m_p}{m_V}\right)\,,
\eeq
where $\gamma$ is the Lorentz boost of 
the nuclear beam and $m_p$ is the proton mass.  The higher energies available 
at the LHC allow studies at much lower $x$ than previously available,
possibly down to $10^{-6}$ \cite{Baltz:2007kq,Rebyakova:2009yg}.

\subthreesection{Photoproduction cross sections}

The cross section for vector meson production may be calculated by integrating 
over photon momentum $k$ (equivalent to integrating over rapidity $y$):
\beq
\sigma(AA\rightarrow AAV) = 2 \int dk\, \frac{dN_\gamma}{dk}\, 
\sigma(\gamma A\rightarrow VA)\,,
\eeq
where $dN_\gamma/dk$ is the photon flux, determined from the 
Weizsacker-Williams method, and $\sigma(\gamma A\rightarrow VA)$ is the 
photoproduction cross section.   This cross section may be extrapolated from 
HERA data.  A Glauber calculation is 
used to determine the cross sections for nuclear targets.  Two Glauber 
calculations of $\jpsi$ and $\Upsilon$ photoproduction are available 
\cite{Klein:2003vd,Baltz:2002pp,Klein:1999qj,Frankfurt:2002sv,Frankfurt:2002wc};
a third uses a color-glass condensate/saturation 
approach to describe the nuclear target \cite{Goncalves:2005yr}.  

\begin{figure}[t]
   \begin{center}
      \includegraphics[width=\figwid]{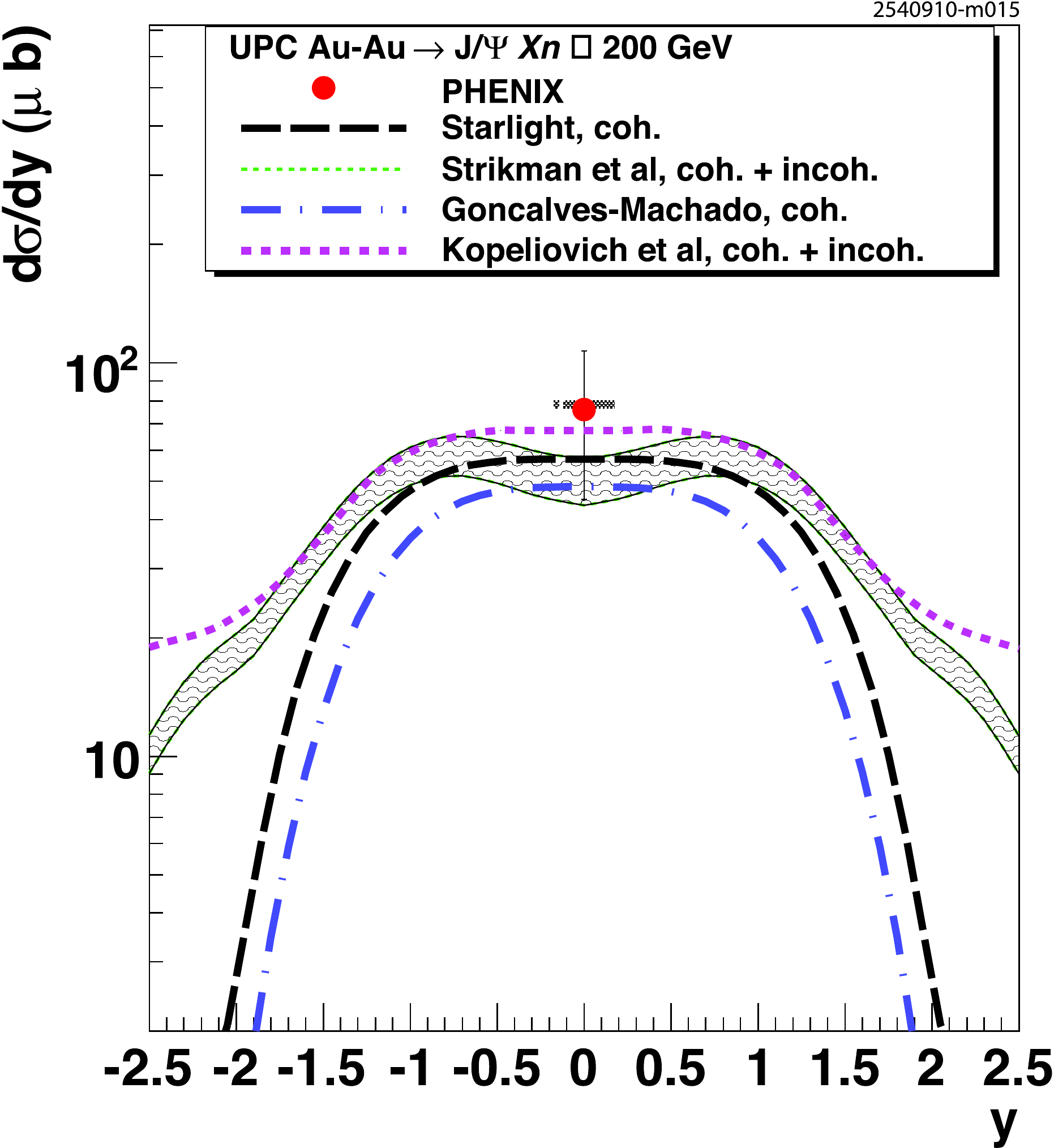}
      \caption{The rapidity distribution, $d\sigma/dy$, of $\jpsi$
               photoproduction measured by PHENIX, compared with three 
               calculations. Coherent and incoherent predictions are 
               summed. Coherent production dominates until $|y|>2$, 
               where the coherent cross section is kinematically 
               suppressed and incoherent production becomes important. 
               \figPermXPLB{Afanasiev:2009hy}{2009} }
      \label{fig:media_fig15}
   \end{center}
\end{figure}

The Glauber calculations successfully predict the rapidity distribution and 
cross section for $\rho^0$ photoproduction in Au+Au collisions 
\cite{Abelev:2007nb,Adler:2002sc},
while the saturation calculation predicts a somewhat higher cross section.  
Calculations have also provided a reasonable estimate of the 
cross sections of excited meson production, such as $\rho^*$ states 
\cite{:2009cz,Frankfurt:2008er,Frankfurt:2006tp}. 
The Tevatron $\jpsi$ and $\psi(2S)$ cross sections are compatible with 
expectations \cite{Aaltonen:2009kg}.  The $\jpsi$ 
photoproduction cross section
in Au+Au collsions is sensitive to nuclear shadowing. The uncertainty of the 
PHENIX~\cite{Afanasiev:2009hy} measurement  is still large, but, as 
\Fig{fig:media_fig15} shows, the central point indicates that shadowing 
is not large.  (At RHIC, midrapidity $\jpsi$ photoproduction corresponds to 
$x\approx 0.015$.) 

\subthreesection{Transverse momentum spectra}

The $p_T$ spectrum of quarkonium photoproduction is the sum of the photon and 
Pomeron $p_T$-dependent contributions.  Since the photon $p_T$ is small, the 
spectrum is dominated by the momentum transfer from the target nucleus.  
In $pp$ collisions, the typical momentum scale is $\sim 300$~MeV, set by the 
size of the nucleon, while for heavy-ion collisions, the momentum scale is 
$\sim \hbar c/R_A$, where $R_A$ is the nuclear radius.  

Photoproduction has a unique feature \cite{Klein:1999gv}:  either nucleus can 
emit the photon while the other serves as the target.  Because the two 
possibities are indistinguishable, their amplitudes add.  In $pp$ and $AA$
collisions, the possibilities are related by a parity transformation.  Since
vector mesons have negative parity, the two amplitude subtract, leading to a 
net amplitude $A\approx A_1-A_2\exp{(ip_T\cdot b)}$ where $b$ is the 
impact parameter.  The two amplitudes, 
$A_1$ and $A_2$, are equal at midrapidity, but may differ for $y\ne 0$ 
because the photon energies differ, depending on which proton or nucleus
emits the photon.  The exponential is a propagator from one nucleus to the 
other.  The cross section is suppressed for $p_T< \langle b\rangle$ with a 
suppression factor proportional to $p_T^2$.  Such suppression has been 
observed by STAR~\cite{Abelev:2008ew}.

The bulk of the cross section from a nuclear target is due to coherent 
production since the virtual $q\overline q$ pair interacts in-phase with the 
entire nucleus.  The $p_T$ transfer from the nucleus is small with a $p_T$ 
scale on the order of a few times $\hbar c/R_A$.
The cross section for coherent photoproduction scales as $Z^2$ 
(from the photon flux) times $A^\delta$, where $4/3 < \delta < 2$.  
Here, $\delta=2$ corresponds to small interaction probabilities, as expected
for heavy quarkonia.   Larger interaction probabilities lead to smaller values
of $\delta$.  Studies of $\rho^0$ photoproduction at RHIC suggest 
$\delta\approx 5/3$.

\begin{figure}[t]
   \begin{center}
      \includegraphics[width=\figwid]{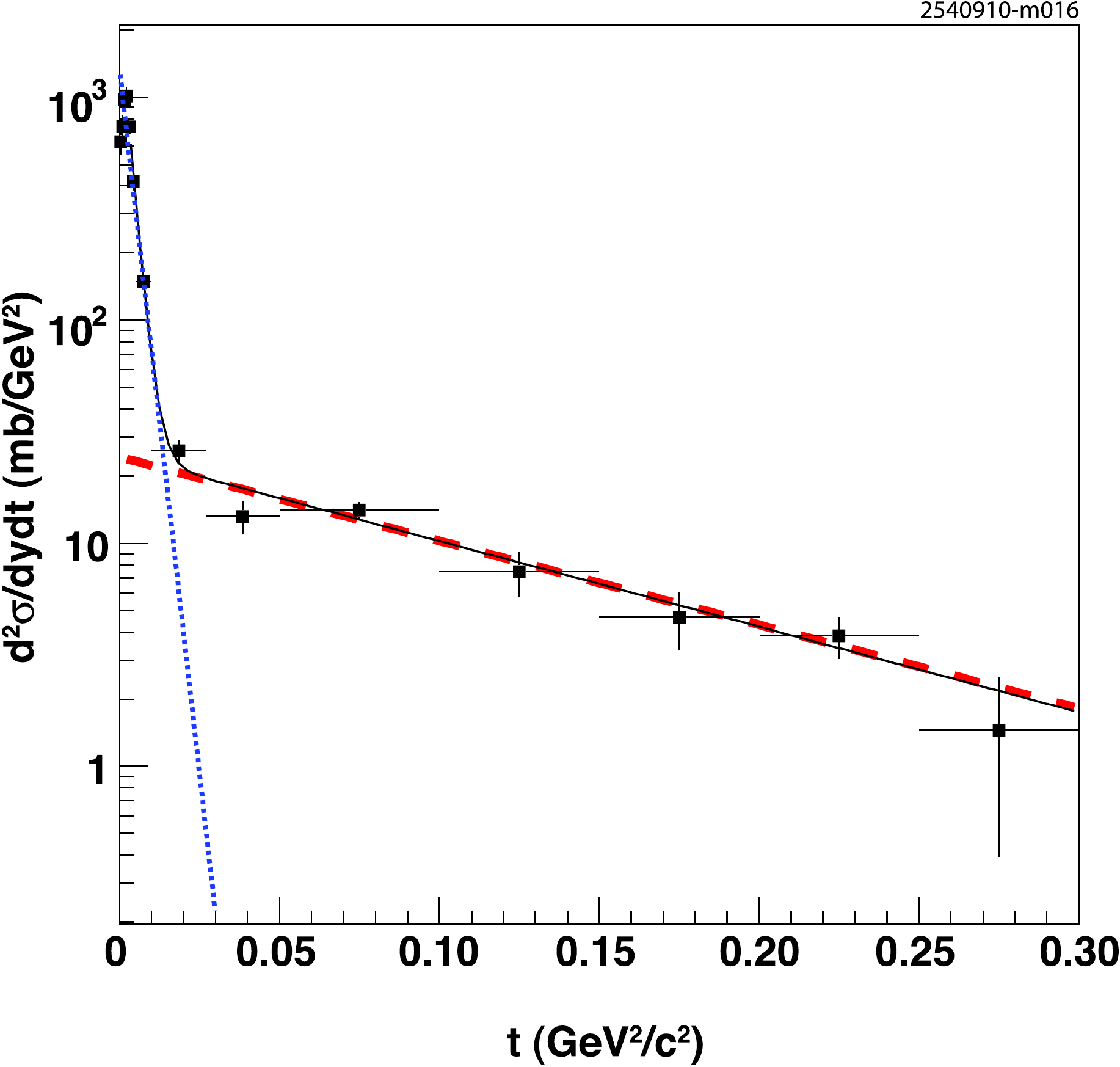}
      \caption{The $t = p_T^2$ spectrum for $\rho^0$ photoproduction 
               observed by STAR in 200\gev\ Au+Au collsions,
               averaged over $|y|<1$. The 
               data are shown by {\sl points with error bars},
               and the {\sl solid curve} is a fit of the data
               to a sum of two components, each exponential in $t$,
               representing that is coherent ({\sl dashed curve}) at low $t$ 
               and incoherent ({\sl dotted}) at  high $t$.  
               \AfigPermAPS{Abelev:2007nb}{2008} }
      \label{fig:media_fig16}
   \end{center}
\end{figure}

At larger $p_T$, the $q\overline q$ pair interactions are out of phase so that
the pair effectively interacts with a single nucleon.  This contribution thus
gives a harder slope in momentum transfer, $t$, corresponding to the size of 
a single nucleon, as can be seen in the STAR data in 
\Fig{fig:media_fig16}.  
At these higher $p_T$, the struck nucleon may be ejected from the nucleus, 
resulting in nuclear dissociation, making it possible to probe the dynamics of 
hard Pomerons \cite{Frankfurt:2008er,Frankfurt:2006tp}.  
Expected at higher $p_T$ are additional components
which probe the nucleon substructure.   In this 
regime, the cross sections become sensitive to the quark distributions 
\cite{Goncalves:2009gs}. Because of the higher momentum transfer from the 
target, incoherent interactions contribute at larger rapidities than coherent 
interactions, explaining the `wings' at large $|y|$ in 
\Fig{fig:media_fig15}.

\subthreesection{Plans for the LHC}

 ALICE, CMS, and ATLAS are all planning to study quarkonium photoproduction 
\cite{Baltz:2007kq}.  These events have a very clean topology: two nearly 
back-to-back electrons or muons, with almost nothing else in the detector. 
At RHIC, STAR and PHENIX found that such an analysis is relatively 
straightforward.  The most difficult part of the study is devising a trigger 
to select these events.  However, the LHC experiments will benefit from 
vastly more sophisticated triggers than are available at RHIC.  Indeed CMS, 
ATLAS, and the ALICE forward muon spectrometer have triggers primitives that 
can be employed for this purpose.  CMS, in particular, may be able to 
separately measure 
$\Upsilon(1S)$, $\Upsilon(2S)$ and $\Upsilon(3S)$ photoproduction.

The LHC energy and luminosity are sufficient for copious $\jpsi$ and 
significant $\psi(2S)$ and $\Upsilon$ signals.  The greatest physics interest 
may be in probing the gluon distributions, and, in particular, 
measuring nuclear 
shadowing.  At midrapidity, quarkonium production probes $x$ values between 
$2\times10^{-4}$ ($\jpsi$ in $pp$ collisions) and $1.7\times 10^{-3}$ 
($\Upsilon$ in Pb+Pb collisions).  Away from midrapidity it is possible to 
probe $x$ values as low as $10^{-6}$ \cite{Baltz:2007kq}.  

In the case of $AA$ collisions, since the photon can be emiitted from either
nucleus, ambiguities arise because the photon energies, and hence the $x$ 
values probed, are different for the two possibilities.  We mention two
possible ways to resolve this two-fold ambiguity.  
Conceptually, the easiest is to study $pA$ collisions (or $dA$ at RHIC).  
Here, the ion is usually the photon emitter.  In addition, it is possible to
employ the difference in the $p_T$ spectra for photons scattering on protons 
and ions to separate the two possibilities.  Unfortunately, $pA$ runs at 
the LHC are some years off.  
A second possiblity is to use bootstrapping, usually by comparing results 
at different beam energies.  At each beam energy, the midrapidity cross
section can be unambiguously determined, giving the cross section at a 
specific photon energy.  At a different beam energy, the same photon energy 
corresponds to a different rapidity.  By measuring the cross section at this 
rapidity and subtracting the known cross section determined previously,
one obtains the cross section at the new photon energy.  Unfortunately, the 
uncertainties add each time the cross sections are subtracted, increasing 
the relative error.   A similar procedure may also employed by using data 
taken under different running conditions, such as exclusive $\jpsi$ production
relative to $\jpsi$ production accompanied by mutual Coulomb excitation.  
At a given rapidity, these two processes contribute differently to the 
cross section, depending on the direction.

The LHC measurements allow for a relatively clean measurement of nuclear 
shadowing by taking the ratio of the $AA$ and $pp$ cross sections.  In this 
approach, many of the theoretical and experimental uncertainties cancel
in the ratio, resulting in a relatively clean determination.  Additional
$pA$ data would help this study by allowing cross checks between $pA$ and $pp$ 
interactions as well as between $AA$ and $pA$ interactions.
Another possibility, for $pp$ collisions, is to use Roman pots or other 
small-angle detectors to tag the outgoing protons \cite{Cox:2009ag}.  
The proton that emitted the photon will usually have lower $p_T$.
Of course, some of these techniques are also applicable at RHIC, where the 
experiments are collecting large data sets with improved triggers and 
particle identification. 

Such measurements of the nuclear gluon distributions will be important for 
understanding the properties of cold nuclear matter, which, in turn,
clarifies the 
interpretation of the quarkonium signals in central heavy-ion collisions. 

\section[Experimental outlook]{Experimental outlook$^{21}$}

\addtocounter{footnote}{1}
\footnotetext{Contributing authors:
S.~Eidelman$^\dag$, P.~Robbe$^\dag$,
A.~Andronic, D.~Bettoni, J.~Brodzicka, G.~E.~Bruno,
A.~Caldwell, J.~Catmore, E.~Chudakov, P.~Crochet, P.~Faccioli,
A.~D.~Frawley,
C.~Hanhart, F.~A.~Harris, D.~M.~Kaplan, 
H.~Kowalski, E.~Levichev, V.~Lombardo, C.~Louren\c{c}o, M.~Negrini, K.~Peters,
W.~Qian, E.~Scomparin, P.~Senger, F.~Simon, S.~Stracka, Y.~Sumino, C.~Weiss, H.~K.~W\"ohri, and C.-Z.~Yuan}
\label{sec:FutChapter}

Moving beyond the present status of heavy quarkonium 
physics described in the previous 
sections poses major challenges to the next generation of
accelerators and experiments. In 
this section the future ``players'' in the field will be described with 
special emphasis on the potential to resolve the important open questions. 
Here we will present the rationale for and status of the newer 
facilities and experiments, from those already 
running (BESIII at BEPCII; ALICE, ATLAS, CMS, and LHCb at the LHC) 
to those under 
construction or only planned (\PANDA and CBM at FAIR, 
SuperB and tau-charm 
factories, lepton-hadron colliders,
and high-energy linear \epem\ colliders).

\subsection{BESIII}

\label{sec:Future_besiiiintro}

For BESIII, the future is now. The Beijing Electron-Positron
Collider (BEPC) and the Beijing Spectrometer (BES) operated in the
tau-charm center-of-mass energy region between 2 and $5\gev$ from
1990 until 2003. Now, BEPC has been upgraded to a two-ring
collider (BEPCII), and a brand new detector (BESIII) has been
constructed.
Commissioning of the upgraded accelerator and new detector began
in spring 2008, and the first event was obtained on July 20, 2008.
Approximately $13\times 10^6$ $\psip$ events were accumulated in fall 2008,
which provided data for studies of the new detector and for
calibration.
In spring 2009 after running for about one month, \mbox{$106\times 10^6$} $\psip$
events were obtained, and in summer 2009 after running for six
weeks, about $226\times 10^6$ $\jpsi$ events were accumulated. These are the
world's largest such data sets and are approximately four times
larger than the CLEO-c $\psip$ sample and the BESII $\jpsi$
sample, respectively.  The new data will allow more detailed
studies of detector performance, and offers many physics
opportunities.

The peak design luminosity of BEPCII is 
$10^{33}\,{\rm cm}^{-2}{\rm s}^{-1}$ 
($1\,{\rm nb}^{-1}{\rm s}^{-1}$) at a beam energy of
$1.89\gev$, an improvement of a factor of 100 with respect to the
BEPC. It will operate at a center-of-mass energy between 2 and
$4.6\gev$, which allows production of almost all known charmonium and
charmonium-like states.
The detector performance is also greatly improved compared to
BESII. BESIII~\cite{Asner:2008nq} 
is a new, general-purpose detector. It features a beryllium beam pipe;
a small-cell, helium-based drift chamber (MDC); a Time-of-Flight
(TOF) system; a CsI(Tl) electromagnetic calorimeter; a $1\,\rm{T}$
superconducting solenoidal magnet; and a muon identifier using
the magnet yoke interleaved with Resistive Plate Chambers.

Running at design luminosity, BESIII will be able to accumulate
$10\times 10^9$ $\jpsi$ events or $3\times 10^9$ $\psip$ events in one
year's running. It will take around $20\,{\rm fb}^{-1}$ of data each at
$3.77\gev$ and $4.17\gev$ for charm physics. There is also the
possibility of a high-statistics fine scan between 2 and $4.6\gev$,
allowing the direct study of states with $J^{PC}=1^{--}$. 
States with even charge parity may be studied using radiative decays of high
mass excited $\psi$ states, such as $\psip$, $\psit$, $\psift$,
$\psifto$, and $\psiftf$.  All these data samples allow detailed
studies of charmonium physics, including the spectroscopy of
conventional charmonium (see \Secs{sec:SpecExp_ConVec}-\ref{sec:SpecExpNewCon})
and charmonium-like (see \Sec{sec:SpecExp_Unanticipated}) 
states, charmonium transitions 
(see \Secs{sec:Dec_radtrans} and \ref{sec:Dec_hadtrans}),
and charmonium decays
(see \Secs{sec:Dec_radlepdec} and \ref{sec:Dec_haddec}).
Charmonium hadronic decay
dynamics are especially interesting because of the 
$\rho\pi$ puzzle (see \Sec{sec:Dec_RhoPiPuzzle}).
The new datasets should also
enable a better understanding of the physics of the strong
interaction in the transition region between perturbative and
nonperturbative QCD.

\subsubsection{Spin singlets: $\hsubc$, $\eta_c(1S,2S)$}
\label{sec:Future_besiiicharmonium}

Below open charm threshold there are three spin-singlet states,
the $S$-wave spin-singlet, $\etac$, its radially
excited state, $\etacp$, and the $P$-wave spin-singlet, $\hsubc$.
All these may be reached from $\psip$ transitions. The $\etac$ can
also be studied in $\jpsi$ radiative decays. Their
properties are less well measured because of their low production
rates in previous $\epem$ experiments.

BESIII will measure the $\hsubc$ mass, width, spin-parity, production
rate via $\psip\to \piz\hsubc$, and its $E1$ transition rate $\hsubc\to
\gamma \etac$ (see \Secs{sec:SpecExp_hc} and \ref{sec:Dec_hcradtrans}). 
BESIII will also search for its hadronic decays
(see \Sec{sec:Dec_hchaddec}),
which are expected to be about 50\% of the total decay width, and
search for other transitions.

Extraction of the \etac\ mass and width from
radiative \jpsi\ or \psip\ radiative transitions 
is not straightforward due to the unexpected
lineshape observed in such transitions
(see \Sec{sec:Dec_1Sc}). With theoretical
guidance and more data, these transitions
may become competitive with other \etac\
production sources in determination of
its mass and width. In addition to
increased statistics, more decay
modes will be found and their
branching fractions measured.

Despite the passage of eight years since the
observation of the \etacp\ (see \Sec{sec:SpecExp_etac2s}),
the discovery mode $\etacp\to K\bar{K}\pi $ remained the
only mode observed until the summer of 2010,
at which time Belle~\cite{NakazawaICHEP2010} reported preliminary
observation of several hadronic \etacp\ decay modes
in two-photon production of \etacp.
BESIII will search for \etacp\ in $\psip$ radiative decays.
With much less data, CLEO-c sought 
11 exclusive hadronic decay modes in radiative
transitions but saw none
(see \Sec{sec:Dec_PsipToGEtac}),
even in the discovery mode and in the
three new modes found by Belle.
With more data, BESIII will have a better chance with
exclusive decay modes.
However, it will be a challenge to isolate 
the low-energy ($\simeq50\mev$) radiative photon
due to the many background
photon candidates, both genuine and fake. 
Observation of a signal in the inclusive photon
spectrum is even more challenging, but is the only way to
get the absolute $\psip\to \gamma \etacp$ transition rate. This
task will require a good understanding of
both backgrounds and the electromagnetic
calorimeter performance.
If the \etacp\ is found in radiative \psip\ decays, 
the photon energy lineshape can 
then be studied and compared to that of the \etac\ (see \Sec{sec:Dec_1Sc}).

\subsubsection{Vectors above \psit: $\psi$'s and $Y$'s }

There are many structures between $3.9\gev$ and $4.7\gev$, including
the excited $\psi$ (see \Sec{sec:SpecExp_ConVecCharm}) 
and the $Y$ (see \Sec{sec:SpecExp_UnconVector})
states~\cite{Swanson:2006st,Klempt:2007cp,Godfrey:2008nc}. 
By doing a fine
scan in this energy range, BESIII may study the inclusive cross
section, as well as the cross sections of many exclusive modes,
such as $\DDbar$, $D^*\bar{D}+c.c.$, $\DstDst+c.c.$,
$\DDbar\pi$, {\it etc.} (see \Sec{sec:SpecExp_ConVecCharm}). 
This will help in understanding the
structures, for instance, whether they are really resonances, due
to coupled-channel effects, final-state interactions, or even
threshold effects. BESIII will also measure the hadronic and
radiative transitions of these excited $\psi$ and the $Y$ states.
Other $XYZ$ particles can also be sought in these transitions.

\subsubsection{Hadronic decays}

As discussed in
\Sec{sec:Dec_RhoPiPuzzle}, the 12\% rule is
expected to hold for exclusive and inclusive decays,
but is violated by many such modes, including the
namesake mode of the $\rho\pi$ puzzle. A plethora
of experimental results exists (see \Sec{sec:Dec_RhoPiPuzzle}
and references therein). With much larger datasets, a
variety of theoretical explanations can be tested 
by BESIII at higher accuracy~\cite{Mo:2006cy}.
Moreover, studies should be made not only of ratios of
$\psip$ to $\jpsi$ decays, but also of other ratios such as those
between $\etacp$ and $\etac$~\cite{Feldmann:2000hs}, between $\psit$
and $\jpsi$~\cite{Yuan:2005es}, and other ratios between different
resonances for the same channel or between different channels from
the same resonance~\cite{Chernyak:1983ej} 
(\eg $\gamma\eta$ and $\gamma\etap$, as discussed in
\Sec{sec:Dec_gammaP}). All such studies are important
to our understanding of charmonium decays. 

BESIII also has an opportunity
to measure the direct photon spectrum in
both \jpsi\ and \psip\ decays and values
of $R_\gamma$ (see \Eq{eqn:SpecTh_Rgamma}) for both resonances
(see discussion in \Secs{sec:SpecTh_alphas}
and \ref{sec:Dec_Gammaglueglue} and in~\cite{Asner:2008nq}),
building on the work of CLEO for \jpsi~\cite{Besson:2008pr}
and \psip~\cite{Libby:2009qb}.

\subsubsection{Excited $C$-even charmonium states}

Above open charm threshold, there are still many $C$-even
charmonium states not yet observed, especially the excited
$P$-wave spin-triplet and the $S$-wave
spin-singlet~\cite{Li:2009zu,Barnes:2005pb}. In principle, these states can
be produced in the E1 or M1 transitions from excited $\psi$
states. As BESIII will accumulate much data at $4.17\gev$ for the
study of charm physics, the sample can be used for such a search.

\subsubsection{Decays of $\chicJ(1P)$}

Approximately 30\% of $\psip$ events decay radiatively to
$\chicJ$, which decay hadronically via two or more gluons
(see \Sec{sec:Dec_chicjhaddec}). 
These events and radiative $\jpsi$ decays are thought
to be important processes for the production of glueball, hybrid,
and other non-$q\bar{q}$ states. BESIII will study these processes
and also search for charmonium rare decays.
The decay $\chicOne\to \eta\pi\pi$ is a golden channel for the study of states
with exotic quantum numbers $I^G(J^{PC})=1^-(1^{-+})$, that is, the
$\pi_1$ states~\cite{Amsler:2008zzb}, since these states, can be produced in
$\chicOne$ $S$-wave decays. A detailed partial wave analysis with a
large $\chicOne$ sample will shed light on these exotic states.

\subsubsection{Prospects}

The present and future large BESIII data sets and excellent new
detector will allow extensive studies of charmonium states and
their decays.

\subsection{ALICE}
\label{sec:Future_aliceintro}

ALICE~\cite{Aamodt:2008zz} is the experiment dedicated 
to the study of nucleus-nucleus 
collisions at the Large Hadron Collider (LHC). 
The study of heavy quarkonium production in 
nuclear collisions is 
one of the most important sources of information on the
characteristics of the hadronic/partonic 
medium. (For a discussion of quarkonium physics
in this medium, see \Sec{sec:MedChapter}.)
ALICE will study Pb+Pb collisions at top LHC Pb energy
($\sqrt{s_{_{NN}}}=5.5$~TeV), at a nominal luminosity 
$L=5\times10^{26}\,{\rm cm}^{-2}{\rm s}^{-1}$. 

In the ALICE physics 
program~\cite{Carminati:2004fp,Alessandro:2006yt}, 
the study of $pp$ 
collisions is also essential, in order to provide reference data for the 
interpretation of nuclear collision results. In addition, many aspects 
of genuine $pp$ physics can be addressed. 
The $pp$ luminosity in ALICE will be restricted
so as to not exceed
$L = 3\times 10^{30}\,{\rm cm}^{-2}{\rm s}^{-1}$.
Despite this luminosity limitation, most physics topics related 
to charmonium and bottomonium production remain accessible.

Heavy quarkonia will be measured 
in the central barrel, covering
the pseudorapidity range $-0.9< \eta <0.9$, and in the forward muon arm, 
which has a coverage $2.5 < \eta < 4$. In the central barrel, heavy quarkonia 
will be detected through the $\epem$ decay. 
ALICE can push its transverse momentum ($\ptrans$) reach for 
charmonium down to $\ptrans$$\sim$0. 
Electron identification is performed jointly in the TPC through the $dE/dx$ 
measurement and in the Transition Radiation Detector (TRD). 
In the forward region, quarkonia will be studied via their decay 
into muon pairs. Muons with momenta larger than $4\gevc$
are detected by means of a spectrometer which 
includes a 3~Tm dipole magnet, 
a front absorber, a muon filter, tracking 
(Cathode Pad Chambers, CPC) and triggering (Resistive Plate
Chambers, RPC) devices.

In the following sections we will review 
the ALICE physics capabilities for 
heavy-quarkonium measurements at the top LHC energy
within the running conditions 
specified above.
A short overview of the measurements 
that could be performed 
in the first high-energy run of the LHC
will also be presented.
For the ALICE physics run in 2010, the forward muon spectrometer 
and most of the 
central barrel detectors have been installed and 
commissioned, including seven 
TRD supermodules (out of 18).

In the central barrel, the geometrical acceptance for $\jpsi$ produced 
at rapidity $|y|<0.9$ (with no $\ptrans$ cut on either the $\jpsi$ or 
the decay electrons) is 29\% for the complete TRD setup.
The electron reconstruction and identification efficiency in the TRD 
is between 80 and 90\% for $\ptrans>0.5\gevc$, while the probability of 
misidentifying a pion as an electron is $\sim$1\%. Below a few\gevc, particle 
identification in the TPC~\cite{Alme:2010ke} contributes substantially to hadron 
rejection, with an overall TPC+TRD electron reconstruction efficiency 
of $\sim$75\%.

The acceptance of the forward spectrometer, relative to the 
rapidity range $2.5<y<4$, is $\sim$35\% for the
$\jpsi$. Since most of the background is due to 
low transverse momentum muons,
a $\ptrans$ cut is applied to each muon at the trigger level.
With a $1\gevc$ $\ptrans$  cut, there is a $\sim$20\% 
reduction of the $\jpsi$ acceptance. 
The combined efficiency for $\jpsi$ detection in the
forward spectrometer acceptance, taking into account the 
efficiency of tracking and triggering detectors, 
is expected to be about 70\%.

\subsubsection{\jpsi\ production from Pb+Pb collisions}

Heavy quarkonium states probe the medium created in 
heavy-ion collisions. Color screening in a deconfined state
is expected to suppress the charmonium and bottomonium yields. 
In addition, at the LHC, 
a large multiplicity of heavy quarks (in particular, charm) 
may lead to significant regeneration of bound states 
in the dense medium during the hadronization phase. ALICE will 
investigate these topics through a study of the yields 
and differential distributions of various quarkonium states, 
performed as a function of the centrality of the collision.

A simulation 
has been performed~\cite{Grigoryan:2008zz}
for $\jpsi$ production in the forward muon arm,
using as an input a 
Color Evaporation Model (CEM) 
calculation, based on the MRST HO set of Parton Distribution
Functions (PDF), 
with $m_{\rm c}=1.2\gevcc$ and 
$\mu = 2 m_{\rm c}$~\cite{Bedjidian:2004gd}.
(For a discussion of the CEM, see \Sec{prod_sec:CEM}.)
 With such a choice of parameters, the total 
$pp$ $\jpsi$ cross section at 
$\sqrt{{s}_{_{NN}}}=5.5$~TeV, 
including the feeddown from 
 higher resonances, amounts to $31\,{\rm \mu b}$. The 
$pp$ cross section has been scaled to Pb+Pb 
assuming binary collision scaling and taking 
into account nuclear shadowing through the 
EKS98~\cite{Eskola:1998df} 
parametrization. The differential
 $\ptrans$ and $y$ shapes have been obtained 
via an extrapolation of
 the CDF measurements and via CEM predictions, 
respectively, and assuming
 that the $\jpsi$ are produced unpolarized. The
hadronic background was simulated using a parametrized 
HIJING generator tuned to $dN_{\rm ch}/dy=8000$ for
central events at midrapidity (such a high value, 3-4 times 
that realistically expected, represents a
rather extreme evaluation of this source). 
Open heavy quark production was simulated using PYTHIA, 
tuned to reproduce the single particle results 
of NLO pQCD calculations.

At nominal luminosity, the expected $\jpsi$ Pb+Pb statistics 
for a $10^6\,{\rm s}$ run, 
corresponding to the yearly running time with the Pb beam, 
are of the order of $7\times 10^5$~events. 
The mass resolution will be 
$\sim$70$\mevcc$~\cite{Carminati:2004fp,Alessandro:2006yt}. 
A simulation of
the various background sources to the muon pair invariant 
mass spectrum in the $\jpsi$  region 
(including combinatorial $\pi$ and $K$ decays, 
as well as semileptonic decays of open heavy flavors)
shows that the signal-to-background ratio, $S/B$, 
ranges from 0.13 to $\approx 7$ when moving from central to
peripheral collisions. With such statistics and $S/B$ values
it will be possible to study the proposed
theoretical scenarios for the modification of 
the $\jpsi$ yield in the hot medium.

The transverse momentum distributions can be addressed 
with reasonable statistics even for the relatively 
less populated peripheral Pb+Pb collisions. 
In particular, for collisions with an impact parameter 
$b>12$~fm, we expect having more than 
1000 events with $\ptrans > 8\gevc$.

Finally, a study of the $\jpsi$ polarization will be 
performed by measuring the angular distribution of
the decay products. With the expected statistics, 
the polarization parameter
$\lambda$ extracted from the fit 
$d\sigma/d\cos\theta = \sigma_{0}(1+\lambda\cos^2\theta)$ 
can be measured,
defining five impact parameter bins, with a statistical error 
$< 0.05$ for each bin.

Another simulation of $\jpsi$ production at central rapidity 
has been carried out using as input the rates 
obtained from the CEM 
calculation described above. For high-mass electron pairs, 
the main background sources 
are misidentified pions and electrons from 
semileptonic $B$ and $D$ decays. 
The value $dN_{\rm ch}/dy=3000$ 
for central events at $y=0$ was used 
for the simulation of the hadronic background. 
PYTHIA was used for open heavy quark production, 
with the same tuning used for the 
forward rapidity simulations. 

For Pb+Pb, the expected $\jpsi$ statistics, 
measured for $10^6\,{\rm s}$ 
running time at the nominal luminosity, are about 
$2\times 10^5$ candidates from the $10^8$
collisions passing the 10\% most central
impact parameter criteria. 
The mass resolution will be $\sim$$30\mevcc$~\cite{Sommer:2007nw}. 
The background under the $\jpsi$ peak,
dominated by misidentified pions, is 
at a rather comfortable level ($S/B=1.2$). 
As for the forward region, 
it will therefore be possible to test the 
proposed theoretical models.

The $S/B$ ratio is expected to increase as a 
function of $\ptrans$, 
reaching a value of $\sim 5$ at $10\gevc$. 
The expected statistics 
at that $\ptrans$ are still a few 
hundred counts, implying that 
differential $\jpsi$ spectra 
can also be studied.

\subsubsection{\jpsi\ production from $pp$ collisions}

Quarkonium hadroproduction is an issue which is not yet 
quantitatively understood theoretically. 
A study of $\jpsi$ production 
in $pp$ collisions at ALICE aims 
at a comprehensive measurement of interesting 
observables (production cross sections, 
$\ptrans$ spectra, polarization) useful 
to test theory in a still unexplored energy 
regime. Furthermore, the forward rapidity measurement 
offers a possibility to access the gluon 
PDFs at very low $x$ ($< 10^{-5}$).

\begin{figure}[b]
   \begin{center}
      \includegraphics[width=\figwid]{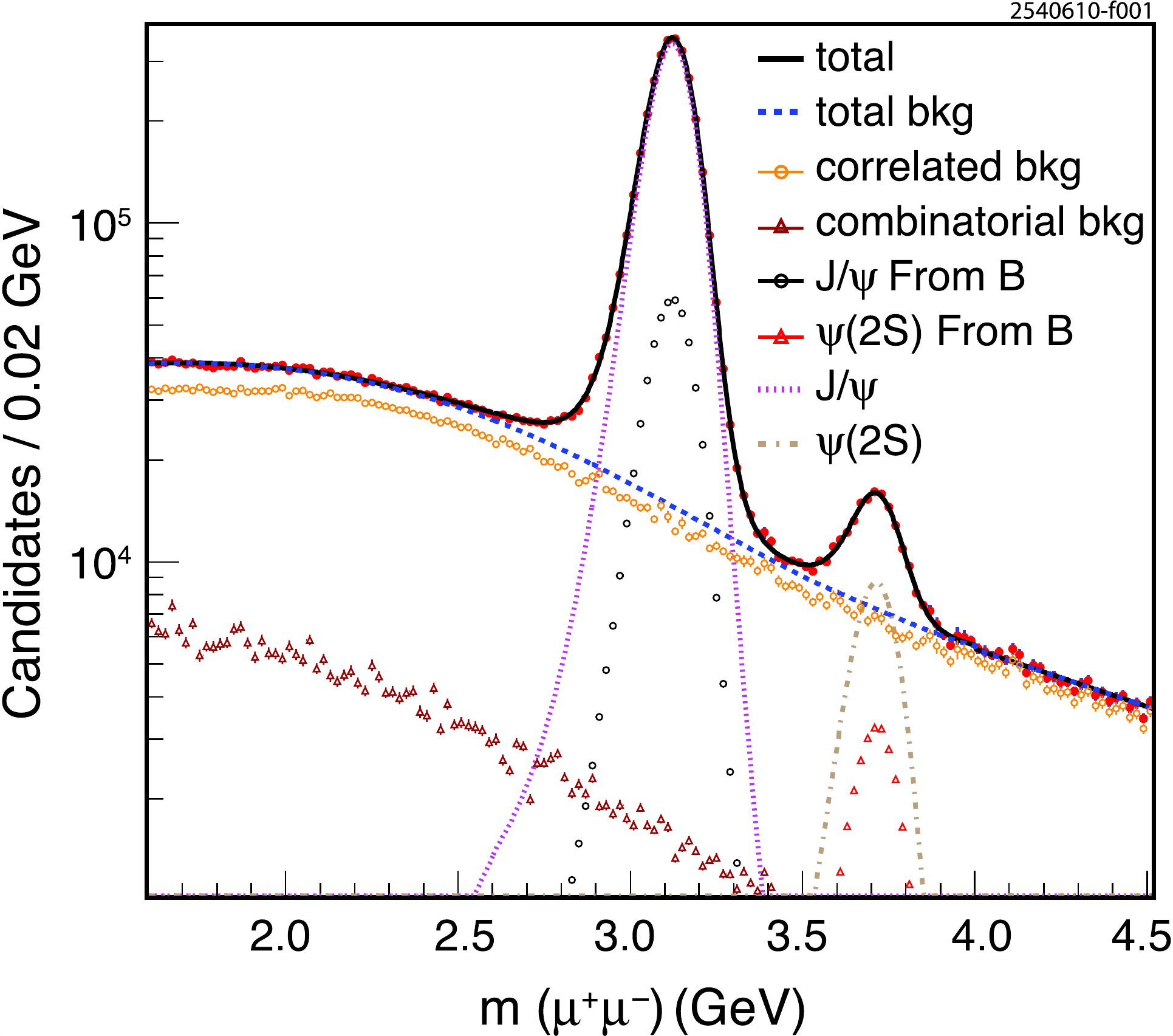}
      \caption{Opposite-sign dimuon mass spectrum in $pp$
               collisions at $\sqrt{s}=14\,{\rm TeV}$ 
               for a $10^7\,{\rm s}$ running time
               at $L = 3\times 10^{30}\,{\rm cm}^{-2}{\rm s}^{-1}$ }
      \label{fig:Future_invMassPsiDec08}
   \end{center}
\end{figure}

In the forward muon arm, $\jpsi$ 
production at $\sqrt{s}=14\,{\rm TeV}$ 
has been simulated using the CEM, with
parameters identical to those 
listed above for 
Pb+Pb collisions. 
The $\jpsi$ total cross
section turns out to be $53.9\,{\rm \mu b}$, 
including the feeddown from higher-mass resonances. 
A typical data taking
period of one year (assuming $10^7\,{\rm s}$ running time) 
at $L = 3 \times 10^{30}\,{\rm cm}^{-2}{\rm s}^{-1}$
gives an integrated luminosity of $30\,{\rm pb}^{-1}$. 
The corresponding dimuon invariant mass spectrum, 
for opposite-sign pairs, 
is shown in \Fig{fig:Future_invMassPsiDec08}. 

The expected $\jpsi$ statistics are 
$\sim 2.8\times 10^6$ events~\cite{Stocco:2006zz}. 
The background under the 
$\jpsi$ peak is dominated by correlated 
decays of heavy flavors 
but is anyway expected to be quite small
($S/B=12$). It will be possible to study 
the transverse momentum 
distribution of the $\jpsi$ with negligible
statistical errors up to at least $\ptrans = 20\gevc$. 
By studying the shape of the $\jpsi$ rapidity 
distribution in the region $2.5<y<4$ it will 
be possible to put 
strong constraints on the gluon PDFs and,
in particular, to discriminate between the 
currently available 
extrapolations in the region around $x=10^{-5}$.
With the expected statistics it will also 
be possible to carry out 
a detailed analysis of the
$\ptrans$ dependence of the $\jpsi$ polarization.

In a $pp$ run, a sample of a few thousand $\jpsi$ 
events is expected to be
acquired in minimum bias collisions. 
With such statistics it will be 
possible to measure $d\sigma/dy$ at 
midrapidity. 
In order to improve these statistics, 
the implementation of a Level-1 trigger for 
electrons is foreseen. Assuming a 
conservative trigger efficiency of 10\%, 
one would get about $7 \times 10^5$ $\jpsi$. 
Such a yield 
would open up the possibility
of measuring differential spectra up to 
high $\ptrans$ and polarization.

\subsubsection{\jpsi\ production from $b$-hadron decays}

 When measuring $\jpsi$ production at the LHC, 
a significant fraction of the 
measured yield comes from $b$-hadron decays. 
This $\jpsi$ source is a very interesting physics signal 
for the evaluation of the
open-bottom production cross section, nicely complementing 
measurements performed via single leptons. 
It is also an important component to be disentangled 
when one wants to identify prompt 
$\jpsi$ production, as it is the case
for studies of yield modifications in nuclear collisions.

At midrapidity, thanks to the excellent 
vertexing capabilities of ALICE, 
the secondary $b$-decay vertex can be separated. 
A good measure of the separation from the main vertex
is $L_{xy}$, the signed projection of the 
\jpsi\ flight distance, $\vec{L}$, on its transverse momentum,
$\vec{p}_T$, defined as
\beq
L_{xy}\equiv\frac{\vec{L}\times\vec{p}_{\rm T}}{\ptrans}\,.
\label{eqn:Fut_Lxy}
\eeq
To reduce the dependence on the $\jpsi$\ 
transverse momentum distribution, the variable 
$x$\ is used instead of $L_{xy}$, 
\beq
x\equiv L_{xy} \times \frac{m(\jpsi)}{\ptrans}\,,  
\eeq
where $m(\jpsi)$ is the known $\jpsi$ mass.
Studies based on Monte Carlo simulation have shown that the  
fractions of secondary $\jpsi$ as a 
function of $\ptrans$ can be extracted 
by a likelihood fit to the dielectron invariant mass and the $x$ 
variable defined above with  uncertainties smaller than 10\%.  
This approach will also provide a  
measurement of the open-bottom $\ptrans$-differential 
cross section down to $\ptrans\approx 0$.  

The situation is more difficult at forward rapidity. 
Due to the presence of a 
thick hadron absorber in the path of the muons, 
the accuracy on the position of 
the $\jpsi$ production vertex is not sufficient. 
Work is in progress 
in order to evaluate the secondary $\jpsi$  
yield starting from 
the study of events with 
3 muons detected in the muon spectrometer 
in $pp$ collisions.
Finally, the option of introducing 
a Si vertex tracker covering the $2.5<\eta <4$ 
rapidity domain is currently 
under study. 
It should be noted that open-bottom
production at forward rapidity 
will be estimated from 
the study of the single-$\mu$ $\ptrans$
distributions and from the 
contribution to the dimuon continuum 
of correlated semileptonic 
decays of $b$-hadrons. These measurements will allow us 
to estimate the fraction of the $\jpsi$  yield
coming from $b$ decays.  

\subsubsection{Production of $\chicJ(1P)$ and \psip}

It is well known that a significant 
fraction (up to $\approx 40$\%) 
of the measured $\jpsi$ yield comes
from $\chicJ$ and $\psip$ decays. 
An accurate measurement of the 
yield of these resonances is therefore an
important ingredient in the interpretation of the $\jpsi$ 
production data. At the same time, these 
higher-mass resonances suffer from a much smaller 
feeddown contribution 
than the $\jpsi$ and may represent cleaner 
signals for theoretical calculations.

The dilepton yield from $\psip$ is much 
smaller than that of $\jpsi$. At the nominal LHC energy and 
luminosity described above, one expects about
$7.5\times 10^4$ events in the forward 
muon arm for a standard 
$pp$ run, with $S/B\approx0.6$~\cite{Stocco:2006zz}. 
In Pb+Pb collisions, the
situation is not so favorable, due to the much larger 
combinatorial background. The expected statistics are 
about $1.5 \times 10^4$ events, but with 
a $S/B$ ratio ranging 
from 18\% to only 1\% from peripheral 
to central collisions.
The background levels at midrapidity
are prohibitive 
for Pb+Pb collisions;
a measurement in $pp$ collisions
also appears to be problematic.

Concerning $\chi_c$, a feasibility 
study has been performed 
on the detection of the radiative decay
$\chi_c\rightarrow \jpsi \gamma$ at midrapidity 
in $pp$ collisions~\cite{Gonzalez:2008nc}. 
The $\jpsi$ has been reconstructed via its
$\epem$ decay, while the photon 
conversion has been reconstructed
from opposite-sign tracks 
with opening angle $<0.1\,{\rm rad}$ and mass $<0.175\gevcc$.
The  $\chicOne$ and $\chi_{c2}$ states 
can be separated in the 
$\Delta m = m(\epem\gamma) - m(\epem)$ spectrum. 
The mean reconstruction efficiency is 0.9\%. 
As for $\jpsi$ 
production at midrapidity, triggering is crucial also
for this signal. With a 10\% trigger efficiency, 
several thousand events could be collected in a $pp$ run.

\subsubsection{\Ups\ production}

In nucleus-nucleus collisions, 
the yield of $\Upsilon(1S,2S,3S)$ 
states should exhibit various degree of 
suppression due to the screening of the color force 
in a Quark-Gluon Plasma. Results from $pp$
collisions will be essential as a normalization for 
Pb+Pb results and extremely interesting in 
order to understand the related QCD topics
(see \Sec{sec:MedChapter}).

In the forward rapidity region, where the muon-pair 
invariant-mass resolution is $\sim$100$\mevcc$, the
$\Ups$ states can be clearly separated. 
The expected yields 
are of the order of $7\times 10^3$ events for the   
$\UnS{1}$ in Pb+Pb collisions, and factors 
$\approx4$ and $\approx6.5$ smaller 
for the higher-mass resonances $\UnS{2}$ and $\UnS{3}$, 
respectively~\cite{Grigoryan:2008zz}.
The $S/B$ ratios will be more favorable than for the $\jpsi$ 
($\approx1.7$ for the $\UnS{1}$ in central
collisions). In $pp$ collisions, 
about $2.7\times 10^4$ $\UnS{1}$ events are expected for 
one run~\cite{Stocco:2006zz}. 
These statistics will allow, in addition to the integrated 
cross section measurement, a study of 
$\ptrans$ distributions and polarization.

At midrapidity, a possibility of measuring the $\Upsilon$ states 
is closely related to the implementation of a Level-1 trigger 
on electrons~\cite{Sommer:2007nw}. Assuming a 
conservative 10\% trigger efficiency, about 7000 $\Upsilon$ events 
could be collected 
in a $pp$ run. In a Pb+Pb 
run, a significant $\UnS{1}$ sample
(several thousand events) can be collected 
with a comfortable $S/B\approx1$. 
The statistics for the higher-mass resonances depend 
crucially on the production
mechanism. Assuming binary scaling, as for 
$\UnS{1}$,  a measurement of 
$\UnS{2}$ looks very promising 
($\approx1000$ events with $S/B=0.35$).

\subsubsection{First LHC high-energy running}

In 2010 the LHC has begun to deliver proton beams at 
$\sqrt{s}=7\,{\rm TeV}$. 
Under the present running conditions, during 2010 it is expected that
a few $10^4$ $\jpsi\to\dimu$ will be collected
in the forward spectrometer using a single muon trigger.
With these statistics a measurement of the $\ptrans$ distribution 
and a $\ptrans$-integrated 
polarization estimate could be within reach. 
Several hundred $\psip$ and $\UnS{1}$ events could be 
collected, 
enough for an estimate of the 
$\ptrans$-integrated cross sections.

Assuming a sample of $10^{9}$ minimum bias events, 
the expected $\jpsi$ statistics in the 
central barrel are a few hundred 
events (due to the reduced coverage provided
by the presently installed TRD supermodules). 
Employing the TRD trigger would 
enhance this sample significantly and 
would enable measurements of 
other charmonium states,
as well as of the \Ups.

\subsection{ATLAS}

ATLAS~\cite{ATLASTDR1:1999zz,ATLASTDR2:1999zzz}
is a general-purpose $4\pi$ detector at the LHC.
Although primarily designed for the discovery of 
physics beyond the Standard Model through the 
direct observation of new particles, indirect 
constraints through precise measurements of known 
phenomena are also an important avenue of activity. 
The quarkonium program of ATLAS falls into this category. 
Of particular importance to these studies are the 
tracking detector and muon spectrometer. The silicon 
pixels and strips close to the interaction point allow 
primary and secondary vertex reconstruction with 
good resolution. The vast muon spectrometer in the 
outer parts of the machine provides 
a flexible muon trigger scheme that can fire on 
pairs of low-momentum (4\gevc) muons as well as
efficient muon identification and 
reconstruction. Together these factors have allowed 
ATLAS to assemble a strong quarkonium physics program. 

We review here the ATLAS capability relevant for
prompt quarkonium production at the LHC, 
in particular the methods of separating promptly 
produced $\jpsi\to\dimu$ and $\Upsilon\to\dimu$ 
decays from the various backgrounds. The outlook 
for the first measurements at 7\tev\ is also discussed. 
All of the results shown here are taken from the 
Computing Services 
Commissioning~\cite[p.~1083-1110]{Aad:2009wy} 
exercises on 
Monte Carlo carried out in 2008.

\subsubsection{Trigger considerations}
\label{sec:Future_atlastrigger}

A detailed account of the ATLAS trigger can be 
found in \cite{Aad:2008zzm}, and the full details 
of the trigger scheme to be used in the ATLAS 
bottom and quarkonia program are available in 
\cite[p.~1044-1082]{Aad:2009wy}. 
The quarkonium program relies on two trigger 
methods in particular. The first requires the 
lowest-level trigger to fire on two over-threshold 
muons independently, forming two conical 
``Regions of Interest" (RoIs) around the muon 
candidate. Full track reconstruction on hits 
within the RoIs is then performed by 
higher-level trigger algorithms to confirm and 
refine the low-level signature. 
The second method requires only one muon at 
the lowest level; the RoI in this case encompasses 
a larger volume and the second muon is sought only 
in the higher-level algorithms. These methods 
allow thresholds in $\ptrans$ 
down to 4\gevc. An alternative approach requires 
only one muon, with a higher $\ptrans$ threshold 
of 10\gevc; in this case the other muon is sought offline.

Any determination of the quarkonia cross sections 
requires a detailed 
understanding of the trigger efficiencies. 
With around $10\,{\rm pb}^{-1}$ of data it 
will be possible to measure the efficiency 
maps directly from the data, using the 
narrow $\jpsi$ resonance in the so-called 
``Tag and Probe'' method  \cite[p.~1069-1081]{Aad:2009wy}.
With fewer data such maps will have to be 
made from Monte Carlo.

\begin{figure}[b]
   \begin{center}
      \includegraphics[width=\figwid]{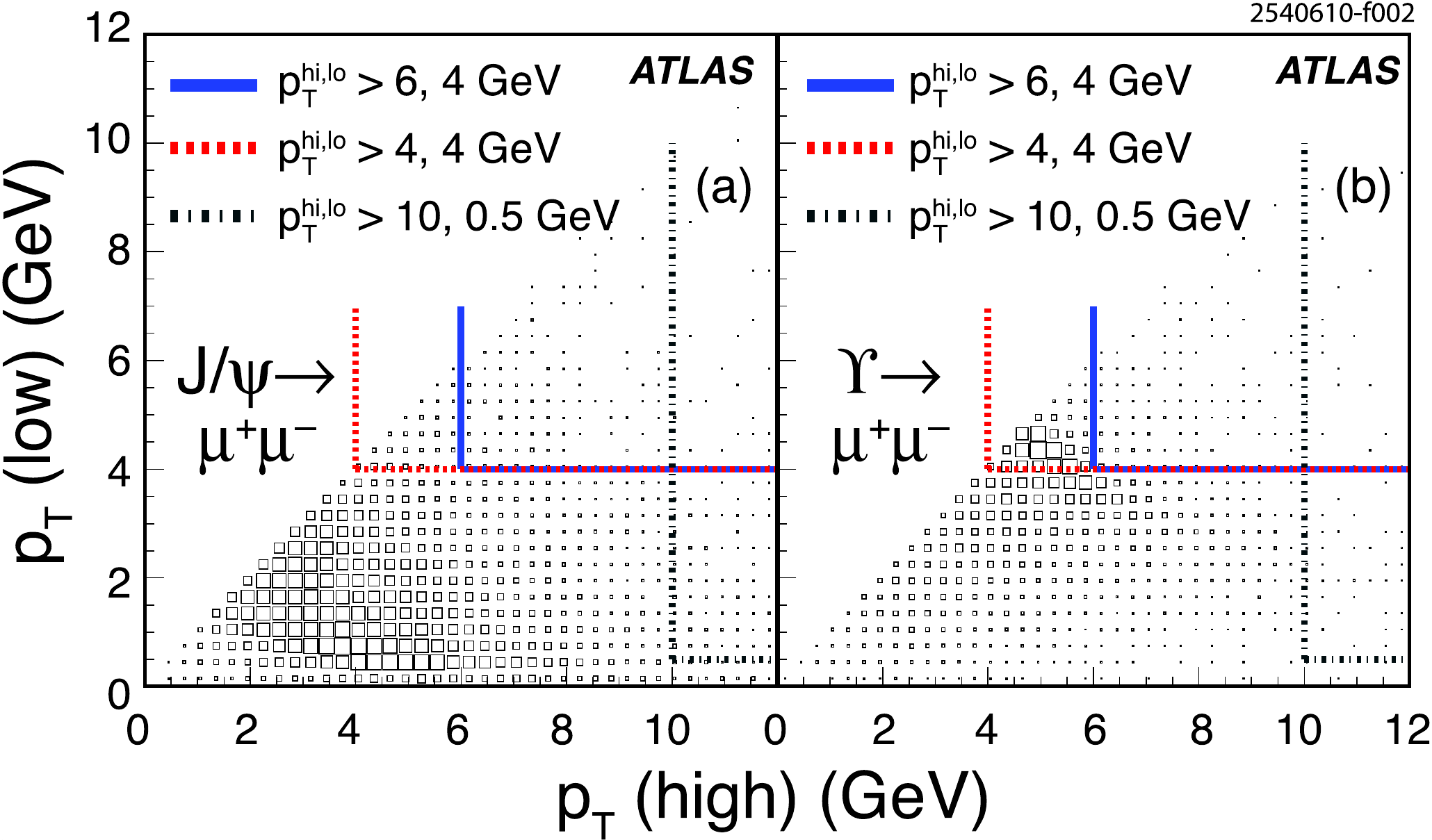}
      \caption{Density of production cross section for \jpsi\ and \Ups}
      \label{fig:Future_crossection}
   \end{center}
\end{figure}

\Figure{fig:Future_crossection} shows the 
density of the production cross section for 
$pp\to \jpsi\to\dimu$ and $pp\to\Upsilon\to\dimu$ 
as a function of the $\ptrans$ of the two muons, 
with cut lines representing dimuon triggers of 
$(4,4)$ and $(6,4)$\gevc\ and the single-muon 
trigger threshold of 10\gevc. It can be seen 
immediately that the situation for the two states 
is very different. In the case of $\jpsi$,
most of the decays produce muons 
with $\ptrans$ well below the $(4,4)$\gevc\ threshold, 
which is as low as the ATLAS muon triggers can go. 
Furthermore, it is clear that increasing the thresholds 
to $(6,4)$\gevc\ does not lose many additional events. 
On the other hand, the $\Upsilon$, which is three times 
as massive as the $\jpsi$, decays into muons with 
significantly higher $\ptrans$. In this case the 
difference between thresholds of $(4,4)$ and 
$(6,4)$\gevc\ is critical, with the lower cut 
capturing many more events and resulting in 
an order-of-magnitude increase in the accessible 
cross section. See \Tab{tab:Future_cs} 
for expected cross sections from a variety of 
quarkonium states for different trigger configurations. 
Although excited $\Ups$ states are included in the 
table, it is unlikely that ATLAS will have good enough 
mass resolution to be able to separate them. 
It should also be noted that the muon trigger 
configuration used early-on will have a nonzero 
efficiency below the $(4,4)$\gevc\ threshold, which 
will allow ATLAS to collect more events than suggested by
\Tab{tab:Future_cs}, which assumes hard cuts. 
Finally, the opening angle between 
the muons in $\Upsilon$ decay is typically 
much larger than for $\jpsi$, which presents a 
difficulty for RoI-guided triggers because the RoI 
is generally too narrow. Physics studies of $\Upsilon$ 
will therefore benefit from the ``full scan" 
dimuon triggers, which allow the whole tracking 
volume to be accessed by the higher-level 
trigger algorithms rather than just 
hits in the RoI. Full-scan triggers are 
CPU-intensive and will only be available 
at low luminosity.

\begin{table}[t]
   \caption{Predicted cross sections for various prompt 
            quarkonia production and decay into dimuons 
            for three trigger scenarios }
   \label{tab:Future_cs}
   \setlength{\tabcolsep}{0.95pc}
   \begin{center}
   \begin{tabular}{ccccc}
   \hline\hline
   \rule[10pt]{-1mm}{0mm}
State & \multicolumn{4}{c}{Cross section, nb} \\[0.7mm]
 & $\mu4\mu4$ & $\mu6\mu4$ & $\mu10$ & $\mu6\mu4\cap\mu10$ \\[0.7mm]
   \hline
   \rule[10pt]{-1mm}{0mm}
 \jpsi   & 28 & 23 & 23 & 5 \\[0.7mm]
 \psip   & 1.0 & 0.8 & 0.8 & 0.2 \\[0.7mm]
 \UnS{1} & 48 & 5.2 & 2.8 & 0.8 \\[0.7mm]
 \UnS{2} & 16 & 1.7 & 0.9 & 0.3 \\[0.7mm]
 \UnS{3} & 9.0 & 1.0 & 0.6 & 0.2 \\[0.7mm]
 \hline\hline
\end{tabular}
\end{center}
\end{table}

\begin{figure}[b]
   \begin{center}
      \includegraphics[width=\figwid]{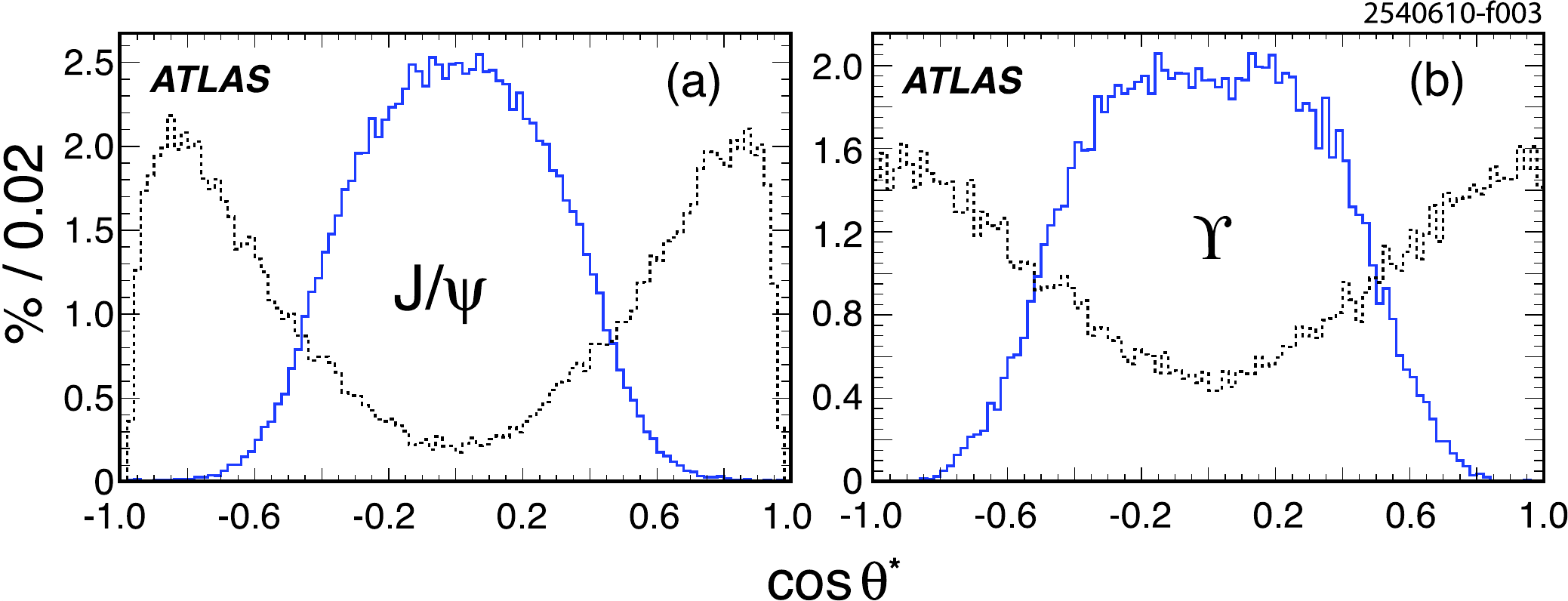}
      \caption{Polarization angle 
               $\cos \theta^{*}$ distributions
               for \jpsi\ and \Ups\ dimuon decays }
      \label{fig:Future_costhetastar}
   \end{center}
\end{figure}

The angle $\cos \theta^{*}$, used in quarkonium spin-alignment 
analyses, is defined (by convention) as the angle
in the quarkonium rest frame between 
the positive muon from the quarkonium 
decay and the flight 
direction of the quarkonium itself in the laboratory 
frame. The distribution of this angle may depend 
on the relative contributions of the different 
quarkonium production mechanisms that are not 
fully understood. Different angular distributions 
can have different trigger acceptances: 
until the spin alignment is properly 
understood, a proper determination of the 
trigger acceptance will not be possible. 
For quarkonium decays in which the two muons have roughly 
equal \ptrans, $\cos \theta^{*} \approx 0$;
such decays will have a high chance 
of being accepted by the trigger. Conversely, 
quarkonia decays with $|\cos\theta^{*}|\approx 1$ 
will have muons with very different \ptrans, 
and as the lower \ptrans\ muon is likely to fall 
below the trigger threshold, such events have 
a greater chance of being rejected. 
\Figure{fig:Future_costhetastar} shows the 
$\cos \theta^{*}$ distributions for $\jpsi$ and 
$\Upsilon$ after trigger cuts of $\ptrans>(6,4)\gev$ 
{\it (solid line)} and a single muon trigger cut of 
$\ptrans>10\gev$ {\it (dashed line)}. The samples were 
generated with zero spin alignment, so without 
trigger selection the $\cos \theta^{*}$ distribution 
would be flat across the range $-1$ to $+1$. 
The figures show, first, that a narrow acceptance 
in $\cos \theta^{*}$ would impair the spin-alignment 
measurements, and second, that the single-muon trigger 
has much better acceptance at the extreme ends of the 
$\cos \theta^{*}$ distribution, since it has a much 
better chance of picking up events with one low-\ptrans\ 
and one high-\ptrans\ muon. At low luminosity such a 
trigger will have an acceptable rate, and, used in 
conjunction with the dimuon triggers, will provide 
excellent coverage across the whole $\cos \theta^{*}$ range.

\subsubsection{Event selection}
\label{sec:Future_atlasselection}

Events passing the triggers are processed offline. 
Oppositely-charged pairs of tracks identified as muons 
by the offline reconstruction are fit to a common 
vertex, after which the invariant mass is calculated from the 
refitted track parameters. Candidates whose refitted 
mass is within $300\mevcc$ ($1\gevcc$) of the 
$\jpsi$ ($\Upsilon$) table mass of 
3097~(9460)\mevcc\ are regarded as quarkonia 
candidates and are accepted for further analysis. 
\Tab{tab:Future_massres} shows the mass resolution 
for $\jpsi$ and $\Upsilon$ candidates for three cases: 
both muon tracks reconstructed in the barrel 
($|\eta| < 1.05$), both in the endcaps 
($|\eta| > 1.05$), and one each in the barrel and an endcap.

 For prompt quarkonia candidates accepted by a dimuon 
trigger there are five major sources of background:
\begin{itemize}
  \item $\jpsi\to\dimu$ candidates from $b\bar{b}$ events
  \item nonresonant \dimu\ from $b\bar{b}$ events
  \item nonresonant \dimu\ from charm decays
  \item nonresonant \dimu\ from the Drell-Yan mechanism
  \item nonresonant \dimu\ from $\pi$ and $K$ decays-in-flight
  \end{itemize}
The first two in the list are the largest: decays of the form 
$b\to \jpsi\left(\to\dimu\right)X$ and dimuons 
from $b\bar{b}$ events. While the charm 
background may be higher in aggregate, the 
\ptrans\ spectrum of the muons falls off sharply 
and the probability of a dimuon having an invariant 
mass close to either of the quarkonia is much lower 
than for the $b\bar{b}$. Monte Carlo studies
indicate that the background from Drell-Yan 
is negligible
because only a tiny fraction passes the trigger thresholds. 
Muons from decays-in-flight also have a very sharply 
falling \ptrans\ spectrum and also need to be in 
coincidence with another muon, such that the 
two form an accepted quarkonium candidate,
and hence are not dominant background contributors.

\begin{table}[t]
   \caption{Mass peak positions and resolutions for prompt 
            quarkonia production in various pseudorapidity ranges }
   \label{tab:Future_massres}
   \setlength{\tabcolsep}{0.33pc}
   \begin{center}
   \begin{tabular}{cccccc}
   \hline\hline
   \rule[10pt]{-1mm}{0mm}
State & $m_{\rm rec}-m_{\rm PDG}$ & \multicolumn{4}{c}{Resolution $\sigma$~(MeV)} \\[0.7mm]
 & (MeV) & Average & Barrel & Mixed & Endcap \\[0.7mm]
   \hline
   \rule[10pt]{-1mm}{0mm}
 \jpsi & $+4\pm1$ & 53 & 42 & 54 & 75 \\[0.7mm]
 \Ups  & $+15 \pm 1$ & 161 & 129 & 170 & 225 \\[0.7mm]
 \hline\hline
\end{tabular}
\end{center}
\end{table}

\begin{figure}[t]
   \begin{center}
      \includegraphics[width=\figwid]{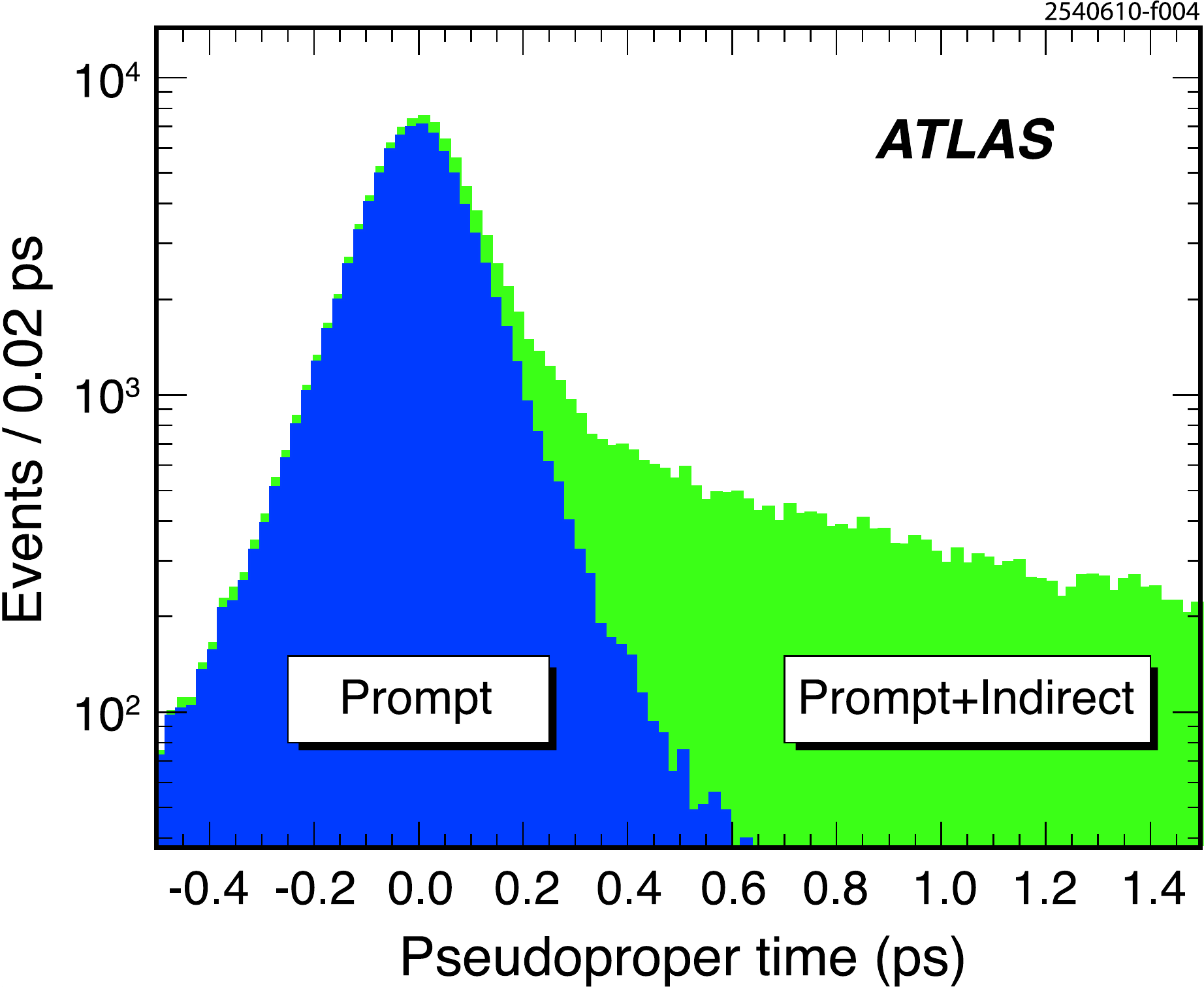}
      \caption{Pseudoproper decay-time distribution for 
               reconstructed prompt \jpsi\ {\it (dark shading)} 
               and the sum of direct and indirect 
               contributions {\it (lighter shading)} }
      \label{fig:Future_propertime}
   \end{center}
\end{figure}

Since all of the sources above (aside from Drell-Yan) 
produce muons which emerge from a secondary vertex, 
it is possible to suppress them by means of a 
secondary-vertex cut based on the pseudoproper time, defined as
\begin{equation}
\label{eqn:Fut_pseudoproper}
\tau=\frac{L_{xy}\times m}{\ptrans\times c}\,,
\end{equation} 
where $m$ and \ptrans\ are the invariant mass and 
transverse momentum of the quarkonium candidate, respectively,
and $L_{xy}$ is the measured radial displacement 
of the two-track vertex from the beamline
as in \Eq{eqn:Fut_Lxy}. 
A collection of prompt quarkonia will have a 
pseudoproper time distribution around zero, while 
distributions for nonprompt candidates will have 
an exponentially decaying tail on the positive side
due to the nonzero lifetime of the parent, as shown in
\Fig{fig:Future_propertime}. 
By making a cut on $\tau$, it is 
possible to exclude the nonprompt component by,
for instance, removing all prompt $\jpsi$ 
candidates with
$\tau>0.2\,{\rm ps}$, thereby obtaining a sample with an 
efficiency of $93\%$ and a purity of $92\%$. 
In the case of $\Upsilon$ there is no background 
from $b\to \jpsi\left(\dimu\right)X$ to address.
But $b\bar{b}\to\dimu$ is more problematic 
in this higher-mass region: the
two muons must have come from different decays, 
rendering the use of pseudoproper time less effective. 
However, it is possible, \eg to insist 
that both muon tracks in the candidate are used 
to build the same primary vertex: in this case 
the $b\bar{b}\to\dimu$ background under 
the $\Upsilon$ can be reduced by a factor of 
three or more while losing about 5\% of the signal. 

\begin{figure}[b]
   \begin{center}
      \includegraphics[width=\figwid]{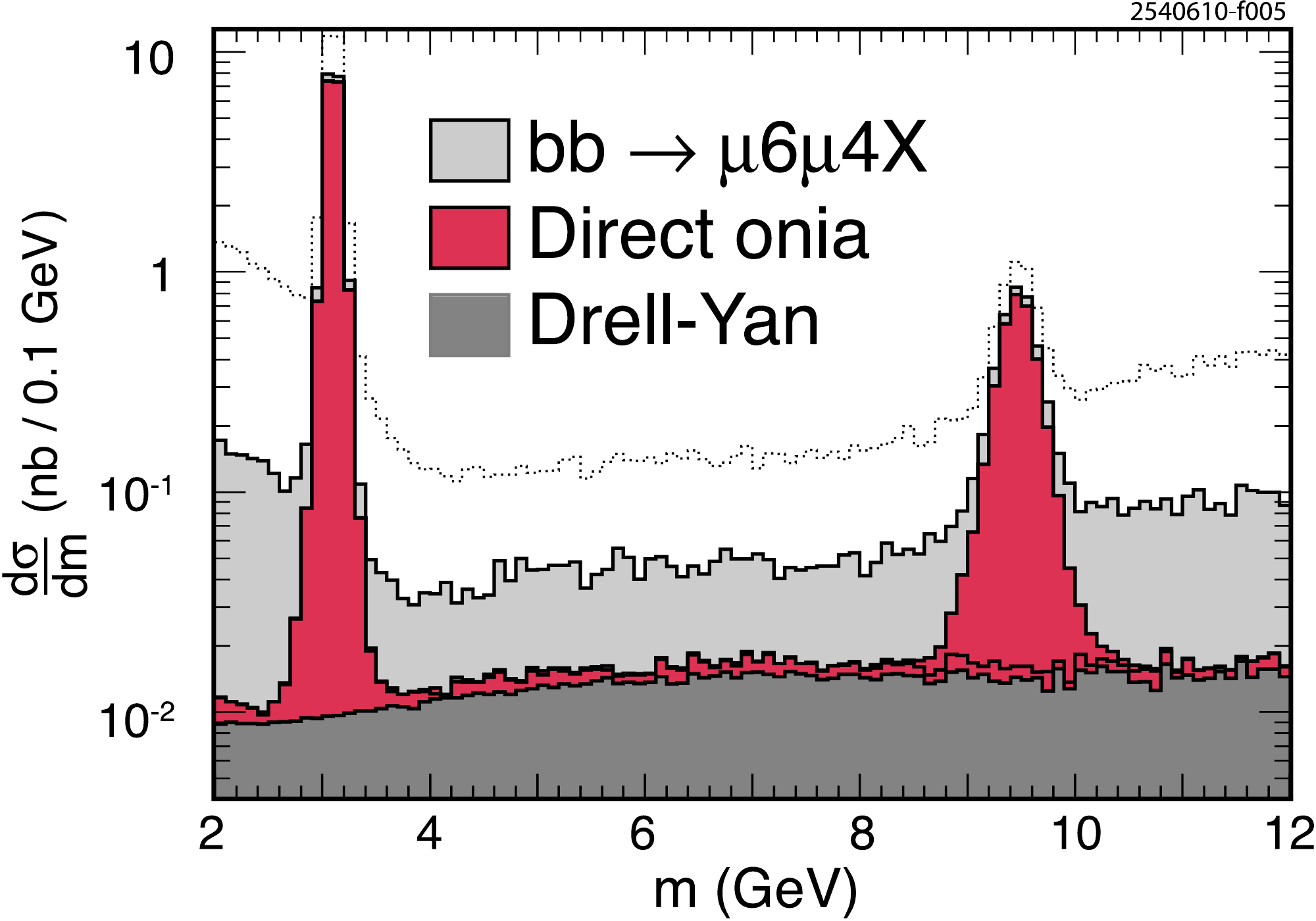}
      \caption{Cumulative plot of the invariant mass of 
               dimuons from various sources, reconstructed 
               with a dimuon trigger with thresholds of 
               $\ptrans>(6,4)\gev$, with the requirements 
               that both muons are identified as coming 
               from the primary vertex and with a 
               pseudoproper time cut of $0.2\,{\rm ps}$. 
               The {\it dotted line} shows the cumulative 
               distribution without vertex and pseudoproper time
               cuts }
      \label{fig:Future_resonances}
   \end{center}
\end{figure}

\Figure{fig:Future_resonances} shows the 
quarkonia signals and the principal backgrounds 
for the dimuon trigger with thresholds of 
$\ptrans>(6,4)\gev$. The higher resonances of 
the $\Upsilon$ were not included in the simulation;
hence their absence from the plot. A pseudoproper 
time cut of $0.2\,{\rm ps}$ has been applied as 
described above, and both muon tracks in a candidate 
are required to have been fitted to the same primary vertex. 
 \Tab{tab:Future_final} summarizes the 
reconstruction efficiencies of all of the cuts 
described above for the different trigger schemes. 

\begin{table}[tb]
   \caption{For prompt quarkonia with various selection and 
            background suppression cuts, predicted and observed cross sections,
            and efficiencies relative to 
            generator-level Monte Carlo }
   \label{tab:Future_final}
   \setlength{\tabcolsep}{0.28pc}
   \begin{center}
   \begin{tabular}{clcccc}
   \hline\hline
   \rule[10pt]{-1mm}{0mm}
   & Quantity & \jpsi & \jpsi & $\Upsilon$ & $\Upsilon$ \\[0.7mm]
   \hline
   \rule[10pt]{-1mm}{0mm}
   & Trigger type & $\mu6\mu4$ & $\mu10$ & $\mu6\mu4$ & $\mu10$ \\ [1.2mm]
 & MC cross sections & 23~nb & 23~nb & 5.2~nb & 2.8~nb \\[1.3mm]
 $\epsilon_{1}$ & Trigger, reconstruction,\\
                &\qquad and vertexing & $75\%$ & $90\%$ & $51\%$ & $90\%$ \\[0.7mm]
 $\epsilon_{2}$ & Offline cuts & $90\%$ & $76\%$ & $95\%$ & $75\%$ \\[0.7mm]
 $\epsilon$ & Overall efficiency $\epsilon_{1} \times \epsilon_{2}$ & $67\%$ & $69\%$ & $49\%$ & $68\%$ \\[1.2mm]
 & Observed signal $\sigma$ & 15~nb & 16~nb & 2.5~nb & 2.0~nb \\[0.7mm]
 & $N_{s}$ $(10\,{\rm pb}^{-1})$ & $150$K & $160$K & $25$K & $20$K \\[0.7mm]
 & $N_{b}$ $(10\,{\rm pb}^{-1})$ & $7$K & $700$K & $16$K & $2000$K \\[0.7mm]
 & Signal/bgd at peak & 60 & 1.2 & 10 & 0.05 \\[0.7mm]
 \hline\hline
\end{tabular}
\end{center}
\end{table}

\subsubsection{Prompt quarkonium polarization}

The Color Octet Model (COM) predicts that prompt quarkonia 
are transversely polarized, with the degree of 
polarization increasing as a function of the 
transverse momentum of the quarkonium. 
Other models predict different \ptrans-behaviors 
of the polarization, so this measurable quantity 
serves as an important discriminator of the various 
quarkonia production models. As discussed above,
the polarization can be accessed via the distribution 
of the angle $\cos\theta^{*}$. This measurement 
is challenging due to reduced acceptance at high 
$|\cos\theta^{*}|$ and the difficulty of disentangling 
acceptance corrections from the spin alignment. 
Additionally, feeddown from $\chi_c$ and $B$ mesons 
may act to reduce polarization in the final quarkonia sample.

The ATLAS program, in this respect, seeks to measure 
the polarization of prompt quarkonia states up to 
transverse momenta of
$\sim50\gevc$, with the coverage in 
$\cos\theta^{*}$ extended through the use of both 
single- and double-muon triggers. The high quarkonia 
rate at the LHC will allow ATLAS to obtain a 
high-purity prompt quarkonia sample through the 
use of the pseudoproper time cut, which
reduces the depolarization due to 
contamination from nonprompt quarkonia. 
Taken together, these techniques will allow 
ATLAS to control the systematics of the 
polarization measurement. Of the two main 
quarkonium states, the $\jpsi$ is easier to 
deal with than the $\Upsilon$ due to a 
higher production cross section and much lower 
backgrounds with the single muon trigger. 
Indeed, the background to the single-muon 
trigger for $\Upsilon$ renders the sample 
available at $10\,{\rm pb}^{-1}$ essentially 
unusable: the reduced 
acceptance in the high $|\cos\theta^{*}|$ 
part of the angular distribution cannot be 
offset with use of a single-muon trigger in the 
same way as for $\jpsi$. For this reason 
the uncertainties on the spin alignment 
for $\Upsilon$ are much higher than $\jpsi$.

The uncertainties from both integrated luminosity 
and spin alignment
need to be factored into
measurement errors for 
prompt quarkonia production cross sections.
Both are expected to be high in the early running of the new 
machine. However, the relative magnitudes of 
the cross sections measured in separate 
\ptrans\ slices will be unaffected by 
both luminosity and spin-alignment uncertainties. 

Summarizing the main conclusion of \cite[p.~1083-1110]{Aad:2009wy}:
after $10\,{\rm pb}^{-1}$ it should be possible to 
measure the spin alignment, $\alpha$, of prompt $\jpsi$  
with a precision of $\Delta\alpha=\pm0.02$-$0.06$ 
for $\ptrans>12\gevc$, depending 
on the level of polarization. 
For the reasons discussed above, in the case 
of $\Upsilon$, the precision is about 
ten times worse - of order $0.2$. 
With an integrated luminosity increased by a 
factor of 10, the uncertainties on $\Ups$ polarization could 
drop by a factor of around 5 because the
sample obtained with the $\mu10$ trigger will become more useful.
 
\subsubsection{Early 7~TeV LHC running}

The quarkonium program in ATLAS has begun with 
the first runs of the LHC. The first task is 
to observe the resonances in the data,
using the peaks as calibration points 
to assess the performance of the muon- and 
inner-detector track reconstruction and the muon 
triggers. These studies are being carried 
out in a rapidly changing luminosity and 
trigger environment as the LHC itself is commissioned. 

After about $1\,{\rm pb}^{-1}$ ATLAS should have 
collected some 15K $\jpsi$ and 
2.5K $\Upsilon$ candidates decaying to pairs 
of muons passing the dimuon trigger requiring 
both muons to have a \ptrans\ of 4\gevc\ and 
one having at 6\gevc. The single-muon trigger 
with a threshold of 10\gevc\ will provide largely 
independent additional samples of 16K $\jpsi$ 
and 2K $\Upsilon$ decays. Separately from these, some 
7K $\jpsi\to\dimu$ events are expected 
from $b$-hadron decays. All of these decays can be 
used for detector performance studies. 
Furthermore, a measurement of the  
fraction of $\jpsi$ arising from $B$ decays 
will be possible at this level, although the 
muon trigger and reconstruction efficiencies 
will have to be estimated with Monte Carlo at this stage.

After about $10\,{\rm pb}^{-1}$ there will be 
sufficient statistics to use the data-driven 
tag-and-probe method to calculate the efficiencies, 
leading to a reduction in the systematic 
uncertainties on the ratio measurement. 
The \ptrans-dependence of the production 
cross section for both \jpsi\ and \Ups\
should be fairly well measured by then, 
over a wide range of transverse momenta 
($10 \le \ptrans \le 50\gevc$). 

After around $100\,{\rm pb}^{-1}$ the \jpsi\ 
and \Ups\ differential cross sections 
will be measured up to transverse momenta 
around 100\gevc. With several million 
\jpsi\ and around 500K \Ups, 
and a good understanding of the efficiency 
and acceptance, polarization measurements 
should reach precisions of a few percent. 
Additional luminosity may allow the 
observation of resonant pairs of \jpsi\ 
in the \Ups\ mass region 
from the decays of $\eta_b$ and $\chi_{b}$ states.

\subsection{CMS}

The primary goal of the Compact Muon Solenoid (CMS)
experiment~\cite{Adolphi:2008zzk} is 
to explore particle physics at the TeV
energy scale exploiting the proton-proton 
collisions delivered by the
LHC.
The central feature of the CMS apparatus is a superconducting solenoid
of $6\,{\rm m}$ internal diameter which provides an axial magnetic field
of $3.8\,{\rm T}$.  Within the field volume are the silicon tracker,
the crystal electromagnetic calorimeter and the
brass/scintillator hadronic calorimeter in barrel and endcap
configurations. CMS also has extensive forward calorimetry, including a
steel/quartz-fibre forward calorimeter covering the $2.9<|\eta|<5.2$
region.  Four stations of muon detectors are embedded in the steel
return yoke, covering the $|\eta| < 2.4$ window.  Each station
consists of several layers of drift tubes in the barrel region and
cathode strip chambers in the endcap regions, both complemented by
resistive plate chambers.

Having a high-quality muon measurement was one of the basic pillars in
the design of CMS.  Around 44\% of the $\jpsi$ mesons produced in $pp$
collisions are emitted within the almost 5 units of pseudorapidity
covered by the muon stations, which cover an even larger phase-space fraction
for dimuons from $\Upsilon$ decays.  These detectors are crucial 
for triggering and for muon identification purposes; CMS
can easily select collisions
which produced one or more muons for writing on permanent storage.  
The good quality of the muon
measurement, however, is mostly due to the granularity of the silicon
tracker (1440 silicon-pixel and 15\,148 silicon-strip modules) and to
the very strong bending power of the magnetic field~\cite{Ragusa:2007zz}.
The silicon tracker also provides the vertex position with
$\sim$$15\,{\rm \mu m}$ accuracy~\cite{Adam:2008dia}.

The performance of muon reconstruction in CMS has been evaluated using
a large data sample of cosmic-ray muons recorded in
2008~\cite{Chatrchyan:2010zz}.  Various efficiencies, measured for a broad
range of muon momenta, were found to be in good agreement with
expectations from Monte Carlo simulation studies. The relative
momentum resolution for muons crossing the barrel part of the detector
is better than 1\% at $10\gevc$.

The CMS experiment, thanks to its good performance for the measurement
of dimuons, including the capability of distinguishing prompt dimuons
from dimuons produced in a displaced vertex, should be ideally placed
to study the production of several quarkonia, including the \jpsi,
\psip\ and \UnS{1S,2S,3}\ states.  Complementing the
dimuon measurements with the photon information provided by the
electromagnetic calorimeter should also allow reconstruction of the
\chic\ and \chib\ states.  Such measurements will lead to several
studies of quarkonium production. Some will be simple analyses
that will lead to the first CMS physics publications. 
Other rather complex ones will come later, 
such as the measurement of the polarization of
the directly-produced $\jpsi$ mesons as a function of their \ptrans, after
subtraction of feeddown contributions from \chic\ and $B$-meson
decays.

Here we do not describe an exhaustive description of all
the many interesting quarkonium physics analyses that can, in
principle, be performed by CMS.  Instead, we focus on only a few
representative studies.  We only mention measurements with dimuons in
proton-proton collisions, despite the fact that similar studies could
also be made with electron pairs, and/or in heavy-ion collisions, at
least to some extent.

\subsubsection{Quarkonium production}
\label{sec:Future_cms_xsec}

At midrapidity, the strong magnetic field imposes a minimum
transverse momentum of around $3\gevc$ for muons to reach the
muon stations.  At forward angles, the material thickness imposes a
minimum energy on the detected muons, rather than a minimum \ptrans.
In general, for a muon to trigger it needs to cross at least
two muon stations.  This requirement rejects a significant fraction of
the low \ptrans\ $\jpsi$ dimuons which could be reconstructed from a
data sample collected with a ``minimum bias'' trigger.  In the first
few months of LHC operation, while the instantaneous luminosity will
be low enough, less selective triggers can be used.  For instance, it
is possible to combine (in the ``high-level trigger'' online farm) a
single-muon trigger with a silicon track, such that their pair mass is
in a mass window surrounding the $\jpsi$ peak.  In this way, sizeable
samples of low \ptrans\ $\jpsi$ dimuons can be collected before the
trigger rates become too large.

\begin{figure}[b]
   \begin{center}
      \includegraphics[width=\figwid]{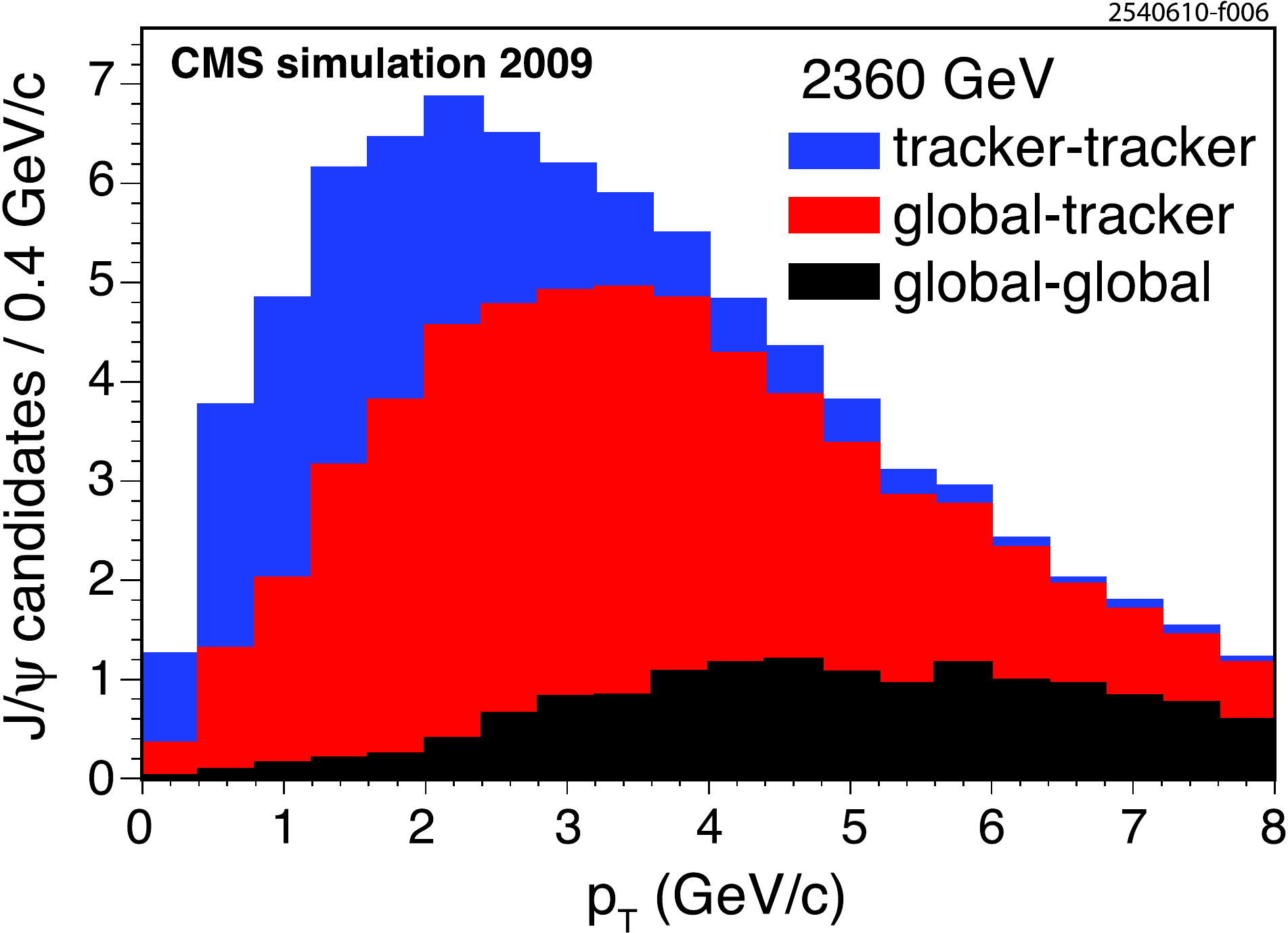}
      \caption{Transverse momentum distribution of dimuons from \jpsi,
               reconstructed by CMS in three different muon-pair categories,
               which depend upon the number of stations 
               crossed by the muons (MC
               study)~\cite{CMSDimuon:2010zz} }
      \label{fig:Future_cms_jpsi_pt_2360}
   \end{center}
\end{figure}

\Figure{fig:Future_cms_jpsi_pt_2360} shows a $\jpsi$ \ptrans\
distribution resulting from a Monte Carlo simulation study (based on a
tuned~\cite{Bargiotti:2007zz} version of the
PYTHIA~\cite{Sjostrand:2003wg} event generator).  This
study~\cite{CMSDimuon:2010zz} was made for $pp$ collisions at
2360\gev\ and corresponds to a minimum-bias event sample, collected
without any trigger selection of muon-station signals.
We see that CMS should have the capability of measuring very low \ptrans\
$\jpsi$ dimuons, especially if one of the two muons (or both) is
reconstructed as a ``tracker muon'', meaning that it only traverses
one muon station.  In fact, most of the yield that could be
reconstructed by CMS is contained in the muon-pair category where only
one of the muons crosses two or more muon stations (the
``global-tracker'' pairs).

\begin{figure}[b]
   \begin{center}
      \includegraphics[width=\figwid]{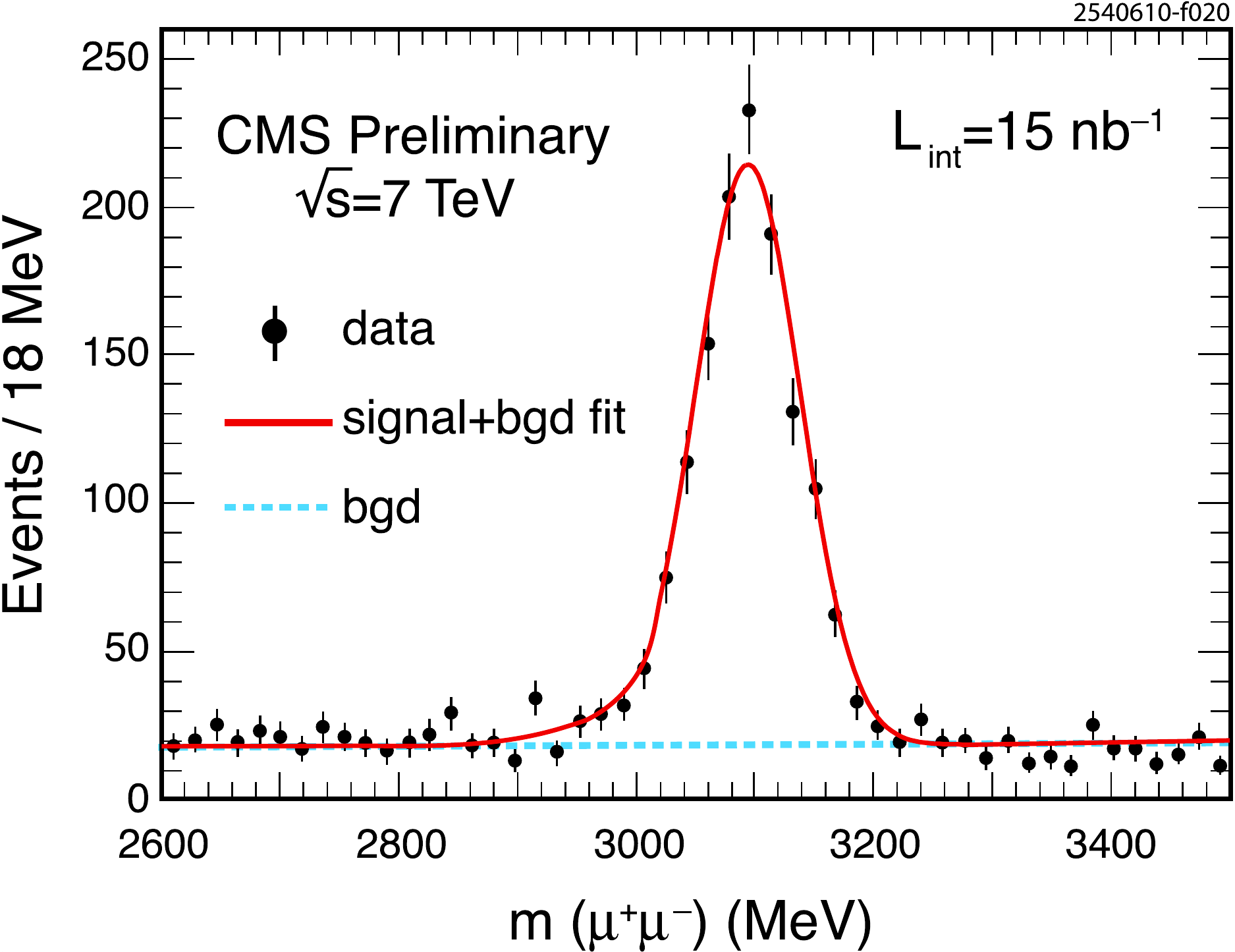}
      \caption{Dimuon mass distribution reconstructed for $pp$
               collisions at 7~TeV, for an integrated luminosity of
               $15~{\rm nb}^{-1}$~\cite{CMSJPsi:2010zz} }
      \label{fig:Future_cms_jpsi_mass_7TeV}
   \end{center}
\end{figure}

By accepting events with one of the muons measured only in one
station, the signal-to-background ratio in the $\jpsi$ dimuon mass
region becomes smaller than in the ``global-global'' category.
However, it remains rather good, as illustrated in
\Fig{fig:Future_cms_jpsi_mass_7TeV}, where we see the $\jpsi$
peak reconstructed from $pp$ collisions at 7~TeV after applying
certain selection cuts on the muons and requiring a minimum dimuon
vertex quality.

With $\sim$$10\,{\rm pb}^{-1}$ of integrated 
luminosity for $pp$ collisions at
$7\,\rm{TeV}$, CMS should collect a few hundred thousand $\jpsi$ dimuons and a
few tens of thousands of $\UnS{1}$ dimuons.  It is important to
note that CMS can measure muon pairs resulting from decays of zero
\ptrans\ $\Upsilon$s.  Indeed, the high mass of the $\Upsilon$ states
($\sim$$10\gevcc$) gives single muons enough energy to reach
the muon stations even when the $\Upsilon$ is produced at rest.

Given its very good muon momentum resolution, better than 1\% (2\%)
for the barrel (endcap) region for muon momenta up to
$100\gevc$~\cite{Bayatian:2006zz}, CMS will reconstruct the $\jpsi$
and $\Upsilon$ peaks with a dimuon mass resolution of around 30 and
80\mevcc, respectively, better at midrapidity than at forward
rapidity~\cite{CMSaafke:2007zz}.

The very good electromagnetic calorimeter of CMS, covering the range
$|\eta| < 3.0$, enables the study of \chic\ and \chib\
production through the measurement of their radiative decays.  Such
measurements are crucial to evaluate non-negligible feeddown
contributions to prompt $\jpsi$ and $\Upsilon$ production, a mandatory
ingredient to fully understand the physics of quarkonium production
from measurements of differential cross sections and polarization.
The contribution of \psip\ and \chic\ decays to prompt $\jpsi$
production has recently been evaluated to be $8.1\pm 0.3\%$ and
$25\pm5\%$, respectively~\cite{Faccioli:2008ir}, while around half of the
$\UnS{1}$ yield is due to decays of heavier bottomonium states, at
least for $\ptrans(\Upsilon) > 8\gevc$~\cite{Affolder:1999wm}.
The decays of $b$-hadrons also contribute to the observed \jpsi\ yield.  
This further complication can be kept under control
through the measurement of nonprompt $\jpsi$ production, which CMS can
do efficiently thanks to very good vertexing and $b$-tagging
capabilities, and profiting from the long $b$-hadron lifetimes.
In the $\Upsilon$ sector there are no feeddown decays from nonprompt
sources.

Given the performance capabilities of the CMS detector, which include
a good dimuon mass resolution, a broad rapidity coverage, acceptance
down to zero \ptrans\ for the $\Upsilon$ states (and also to
relatively low \ptrans\ for the \jpsi), CMS
is in a very good position to do detailed studies of the quarkonium
production mechanisms, hopefully answering some of the questions left
open by the lower energy experiments.  Naturally, the CMS quarkonium
physics program foresees measurements of the differential production
cross sections, versus \ptrans\ and rapidity, of many quarkonium
states.  Given the large charm and bottom production cross sections at
LHC energies, CMS should collect large $\jpsi$ and $\Upsilon$ event
samples in only a few months of LHC operation, leading to physics
publications on, in particular, their \ptrans\ distributions, very
competitive with respect to the presently available Tevatron results.

\subsubsection{Quarkonium polarization}
\label{sec:Future_cms_pol}

Polarization studies, particularly challenging because of their
multidimensional character, will exploit the full capabilities of the CMS
detector and the ongoing optimization of dedicated trigger
selections. CMS will study the complete 
dilepton decay distributions, including polar and azimuthal anisotropies 
and as functions of \ptrans\ and rapidity, in
the Collins-Soper (CS) and the helicity (HX) frames. These analyses 
will require
considerably larger event samples than the cross section measurements. The
acceptance in the lepton decay angles is drastically limited by the
minimum-\ptrans\ requirements on the accepted leptons (rather than reflecting
geometrical detector constraints). Polarization measurements will
therefore profit crucially from looser muon triggers.
Moreover, such trigger-specific acceptance
limitations determine a significant dependence of the global acceptances (in
different degrees for different quarkonium states) on the knowledge of the
polarization. The systematic contributions of the as-yet 
unknown polarizations to
early cross section measurements will be estimated, adopting the
same multidimensional approach of the polarization analyses. Plans for
high-statistics runs include separate determinations of the polarizations of
quarkonia produced directly and of those coming from the decays of heavier
states. Current studies indicate that CMS should be able to measure the
polarization of the \jpsi's that result from \chic\ decays, together with the
\ptrans-dependent $\jpsi$ feeddown contribution 
from \chic\ decays, from very low
to very high \ptrans.

All measurements will also be reported in terms of frame-invariant
quantities, which will be determined, for cross-checking purposes, in
more than one reference frame.  These plans reflect our conviction
that robust measurements of quarkonium polarization can only be
provided by fully taking into account the intrinsic
multidimensionality of the problem. As emphasized 
in~\cite{Faccioli:2008dx,Faccioli:2010kd}, the
measurements should report the full decay distribution in possibly more than
one frame and avoid kinematic averages (for example, over
the whole rapidity acceptance range) as much as possible.

\begin{figure}[tb]
   \begin{center}
      \includegraphics[width=\figwid]{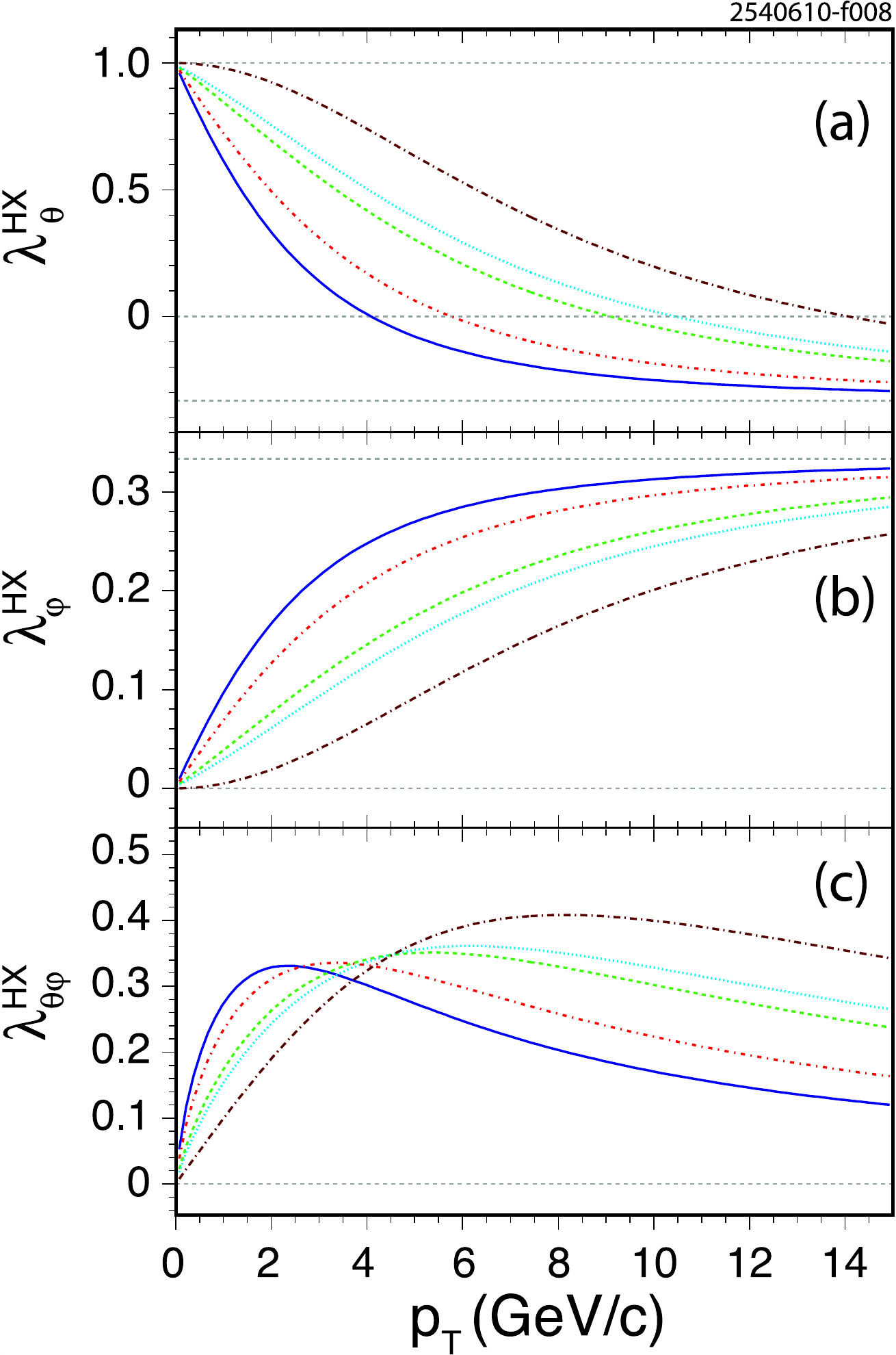}
      \caption{Anisotropy parameters in (a)~polar and (b)~azimuthal
               angle vs.~transverse momentum for 
               $\Ups\to\dilep$ decays in the HX frame
               for a natural polarization 
               $\lambda_{\vartheta} = +1$ in the CS frame. 
               The {\it curves} in each plot correspond to different rapidity
               intervals representative of different 
               experiments. Starting from the 
               {\it solid line}: $|y| < 0.6$ (CDF), $|y| < 0.9$ 
               (ALICE, $e^+e^-$ channel), $|y| < 1.8$
               (\DZero), $|y| < 2.5$ (ATLAS and CMS), 
               $2< |y| < 5$ (LHCb). For simplicity, the
               event populations have been assumed 
               to be flat in rapidity. The vertical axis of the
               polarization frame is here defined as 
               $\mathrm{sign}(p_\mathrm{L}) (
               \vec{P^\prime}_1 \times \vec{P^\prime}_2 ) 
               /|\vec{P^\prime}_1 \times
               \vec{P^\prime}_2|$, where $\vec{P^\prime}_1$ 
               and $\vec{P^\prime}_2$ are the
               momenta of the colliding protons in the 
               quarkonium rest frame (the sign of
               $\lambda_{\vartheta \varphi}$ depends on this definition) }
      \label{fig:Future_exp_kindep_lambda} 
   \end{center}
\end{figure}

\begin{figure}[t]
   \begin{center}
      \includegraphics[width=\figwid]{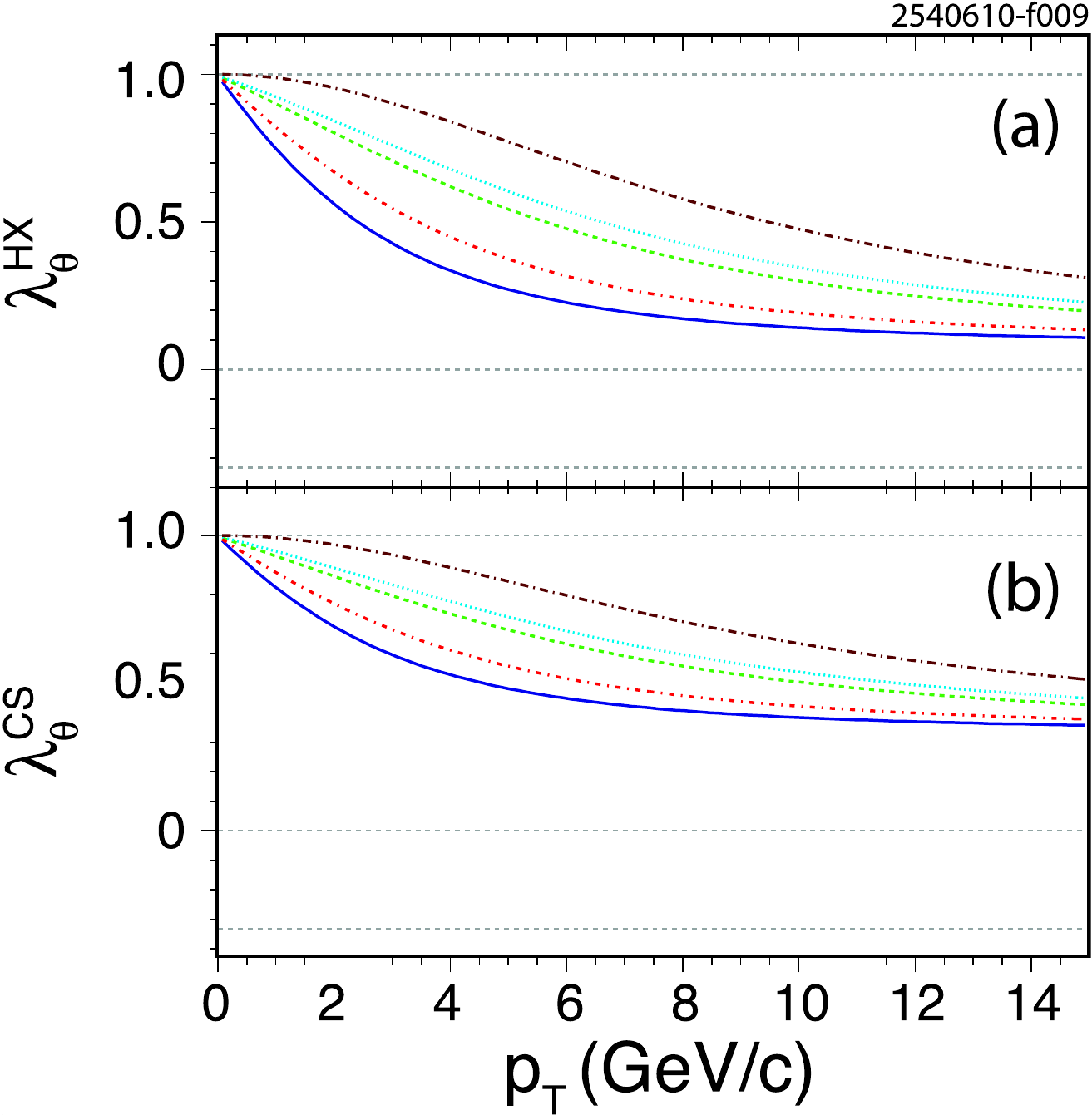}
      \caption{Polar angle anisotropy parameter
               vs.~transverse momentum for $\Ups\to\dilep$ 
               decay,
               observed in the (a)~HX and (b)~ CS frames
               when $40\%$ ($60\%$) of the 
               events have full transverse polarization
               in the HX (CS) frame. The {\it curves} in 
               each plot represent measurements performed
               in different experimental acceptance ranges, 
               as detailed in
               \Fig{fig:Future_exp_kindep_lambda}  }
      \label{fig:Future_mix_exp_kindep_lambda_th} 
   \end{center}
\end{figure}

\Figure{fig:Future_exp_kindep_lambda} shows, as
a simple example, how a hypothetical Drell-Yan-like polarization (fully
transverse and purely polar in the CS frame) in the $\Upsilon$ mass
region would translate into different \ptrans-dependent polarizations 
measured in
the HX frame by experiments with different rapidity acceptances. The anisotropy
parameters $\lambda_{\vartheta}$, $\lambda_{\varphi}$ and $\lambda_{\vartheta
\varphi}$ are defined as in~\cite{Faccioli:2008dx}. This example
illustrates the following general concepts:
\begin{itemize} 
\item{ The polarization depends very
strongly on the reference frame. The very concepts of ``transverse'' and
``longitudinal'' are frame-dependent.}
\item{The fundamental nature of the
polarization observed in one chosen frame can be 
correctly interpreted (without
relying on assumptions) only when the 
azimuthal anisotropy is measured together
with the polar anisotropy.}
\item{The measured polarization may be affected by
``extrinsic'' kinematic dependencies due to a 
nonoptimal choice of the observation frame. 
Such extrinsic dependencies can introduce artificial
differences among the results obtained by experiments performed in different
acceptance windows and give a misleading view of the polarization
scenario.}
\end{itemize}
On the other hand, these spurious effects cannot always be eliminated by a
suitable frame choice, as is shown by the further illustrative case represented
in \Fig{fig:Future_mix_exp_kindep_lambda_th}. 
Here it is assumed that 
60\% of the $\Upsilon$ events has natural polarization $\lambda_\vartheta =
+1$ in the CS frame and the remaining fraction has natural polarization
$\lambda_\vartheta = +1$ in the HX frame. While the polarizations of the two
event subsamples are intrinsically independent of the production kinematics, in
neither frame will measurements performed in different transverse and
longitudinal momentum windows find identical results for $\lambda_\vartheta$
(the same is true for the other two anisotropy parameters, not shown here).
This example provides a first motivation for the complementary use of a
frame-invariant approach~\cite{Faccioli:2010kd}, consisting of the
measurement of intrinsically rotation-invariant polarization parameters
like 
\beq
K= \frac{1 + \lambda_\theta + 2 \lambda_\phi}{
         3 + \lambda_\theta}\,.
\eeq 
In the example of \Fig{fig:Future_mix_exp_kindep_lambda_th}, 
all experiments would measure a
constant, frame-independent value $K=1/2$. This method facilitates the
comparison between different experiments, as well as between measurements and
theory. Furthermore, since the acceptance distributions for the polar and
azimuthal decay angles can be very different in different frames, checking
whether quantities like $K$ are, as they should be, numerically independent
of the reference frame provides a nontrivial systematic test of the
experimental analyses.

\subsection{LHCb}

LHCb~\cite{Alves:2008zz} is a dedicated experiment for 
$b$-physics at the LHC. 
Since $b$ production is peaked in the forward 
region at LHC energies, the LHCb detector has a 
forward spectrometer geometry
covering an angle between 15~mrad and 
300~mrad with respect to the beam axis. 
This corresponds to an $\eta$ range between
2 and 5, which will allow LHCb to have a unique 
acceptance coverage among the LHC experiments. 
Good vertex resolution and particle identification 
over a wide momentum range are key characteristics of LHCb. 
The trigger system retains muons with moderate $\ptrans$ 
as well as purely hadronic final states. 

\subsubsection{Charmonium physics}
\label{sec:Future_lhcbcharmonium}

The $\jpsi$ selection studies and in general 
all studies presented here have been 
performed using the full LHCb Monte Carlo 
simulation based on PYTHIA~\cite{Sjostrand:2003wg}, 
EvtGen \cite{Lange:2001uf}, and
GEANT4 \cite{Agostinelli:2002hh}. 
At the generation level, color-octet $\jpsi$ production 
models in PYTHIA have been tuned to reproduce the 
cross section and $p_T$ spectrum observed 
at Tevatron energies \cite{Bargiotti:2007zz}. 
A full-event reconstruction is applied to the
simulated events \cite{Needham:2009zz}. 
$\jpsi$ candidates are selected using track and vertex 
quality requirements, and also muon identification information. 
Since the first-level trigger (L0) requires at least one muon 
with a $\ptrans$ larger than $1\gevc$, 
a tighter selection is applied at reconstruction 
level to keep only candidates formed with at least 
one muon with a $\ptrans$ larger 
than $1.5\gevc$.
The $\jpsi$ selection yields an expected 
number of reconstructed events equal to 
$3.2\times 10^6$ at $\sqrt{s}=14$~\tev, 
with $S/B = 4$, for an integrated luminosity 
equal to 5~pb$^{-1}$. This number is obtained 
assuming a $\jpsi$ production cross section 
equal to $290\,\mu{\rm b}$. This amount of data
could be collected in a few days under nominal 
LHC running conditions. 
The mass resolution is $11.4\pm0.4\mevcc$.

\begin{figure}[t]
   \begin{center}
      \includegraphics[width=\figwid]{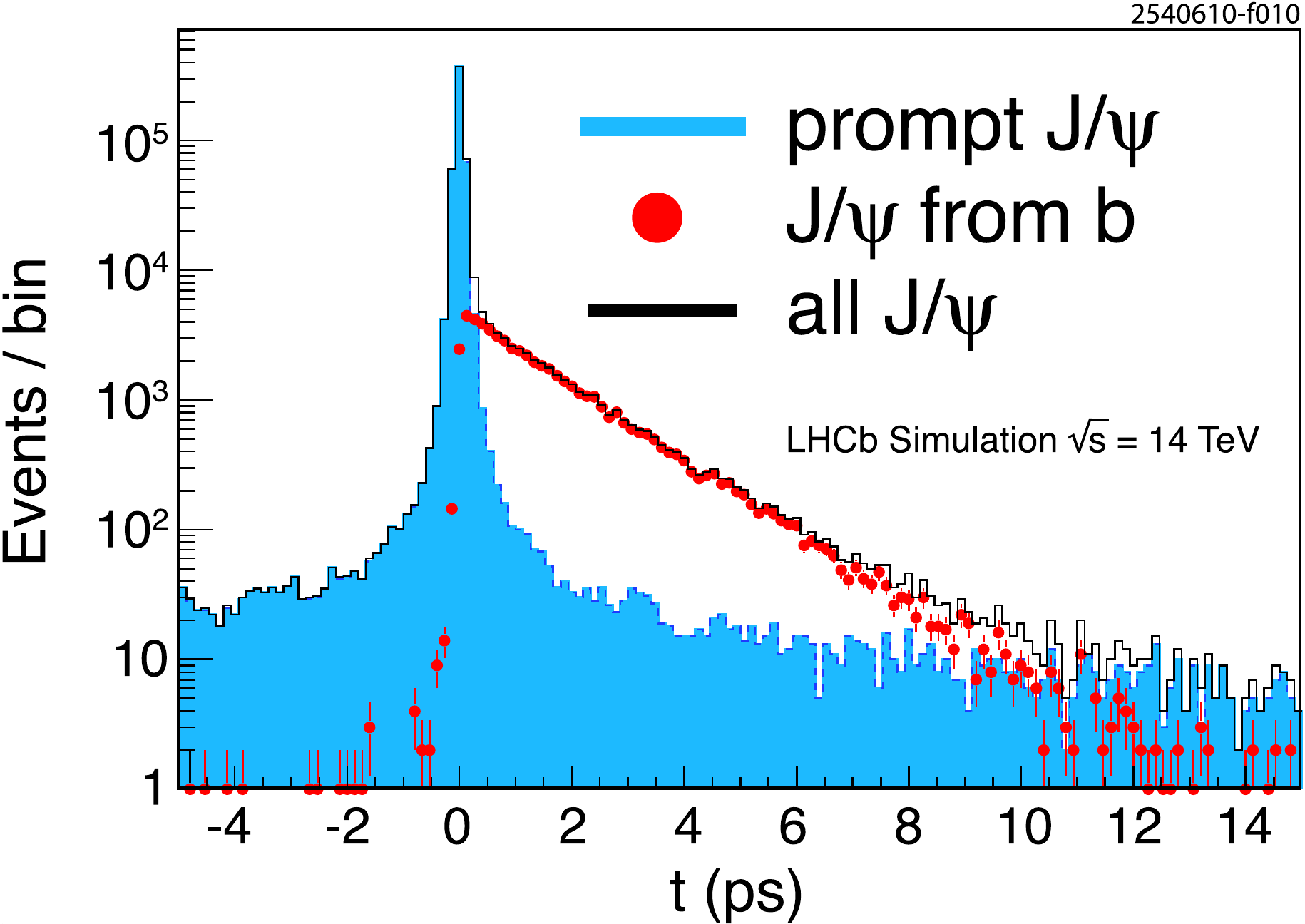}
      \caption{Time distribution of \jpsi\
               candidates obtained with LHCb Monte Carlo simulation }
      \label{fig:Future_notweighted}
   \end{center}
\end{figure}

One of the first goals of the LHCb experiment 
will be to measure the differential $\jpsi$ cross section in bins
of $\ptrans$ and $\eta$ in the range $0 < \ptrans < 7\gevc$ 
and $ 2 < \eta < 5$. Both the prompt-$\jpsi$ and the 
$b\to \jpsi X$ production
cross sections will be accessible, thereby measuring the 
total $b\bar{b}$-production cross section. 
The two contributions will be separated 
using a variable which approximates the $b$-hadron proper time 
along the beam axis 
\beq
t\equiv\frac{dz\times m({\jpsi})}{p_z^{\jpsi}\times c}\,,
\eeq
where $dz$ is the distance 
between the $\jpsi$ decay vertex and the primary vertex of
the event projected along the beam ($z$) axis, 
$p_z^{\jpsi}$ is the signed projection of the $\jpsi$ momentum 
along the $z$ axis, and 
$m({\jpsi})$ is the known \jpsi\ mass.
(Note that this is analogous to the ATLAS
pseudoproper time definition in \Eq{eqn:Fut_pseudoproper},
which uses the transverse decay length and momentum
instead of the longitudinal component employed here.)
The expected distribution of the $t$ variable is shown
in \Fig{fig:Future_notweighted}.
The distribution can be described by
\begin{itemize}
\item a prompt \jpsi\ component produced at the 
primary vertex of the event, represented by a Gaussian distribution to 
account for vertex resolution;
\item an exponential \jpsi\ component coming 
from $b$-hadron decays, convoluted with a Gaussian resolution function;
\item a combinatorial background component due to 
random combinations of tracks coming from the 
primary vertex (the form of this 
component will be extracted using events in the 
sidebands of the dilepton mass distribution);
\item a tail due to a wrong association of primary 
vertex when computing the $t$ variable
(the shape of this component will be
determined from data, associating the 
$\jpsi$ vertex with a different event's primary vertex).
\end{itemize}

A combined fit of the mass and $t$ distributions 
will extract the number of reconstructed 
$\jpsi$ in each 
$\ptrans$ and $\eta$ bin~\cite{Gao:2010zz}. 
The absolute cross section in each of the bins 
will be obtained from this measured yield,
efficiencies that will be computed from 
Monte Carlo simulations, and integrated luminosity. 
The measurement uncertainty will be dominated 
by systematic errors in the integrated 
luminosity, the resolution model, 
and the reconstruction and trigger 
efficiencies.

The unknown polarization at production of the 
$\jpsi$ will complicate the measurement. 
The acceptance of the LHCb
detector is not uniform as a function of the 
$\jpsi$ polarization angle, $\theta$, defined 
as the angle between the $\mu^+$
direction in the center-of-mass frame of the $\jpsi$ 
and the direction of the $\jpsi$ in the 
laboratory. 
Ignoring this effect adds a 25\% uncertainty 
to the cross section measurement. 
Performing the measurement in bins of $\theta$ 
will allow determination of the \jpsi\ production polarization.

Measurements of the rates and polarization 
of the $\jpsi$, and more generally of the charmonium and
bottomonium states, will be compared to predictions of 
different theoretical models.

\begin{figure}[t]
   \begin{center}
      \includegraphics[width=\figwid]{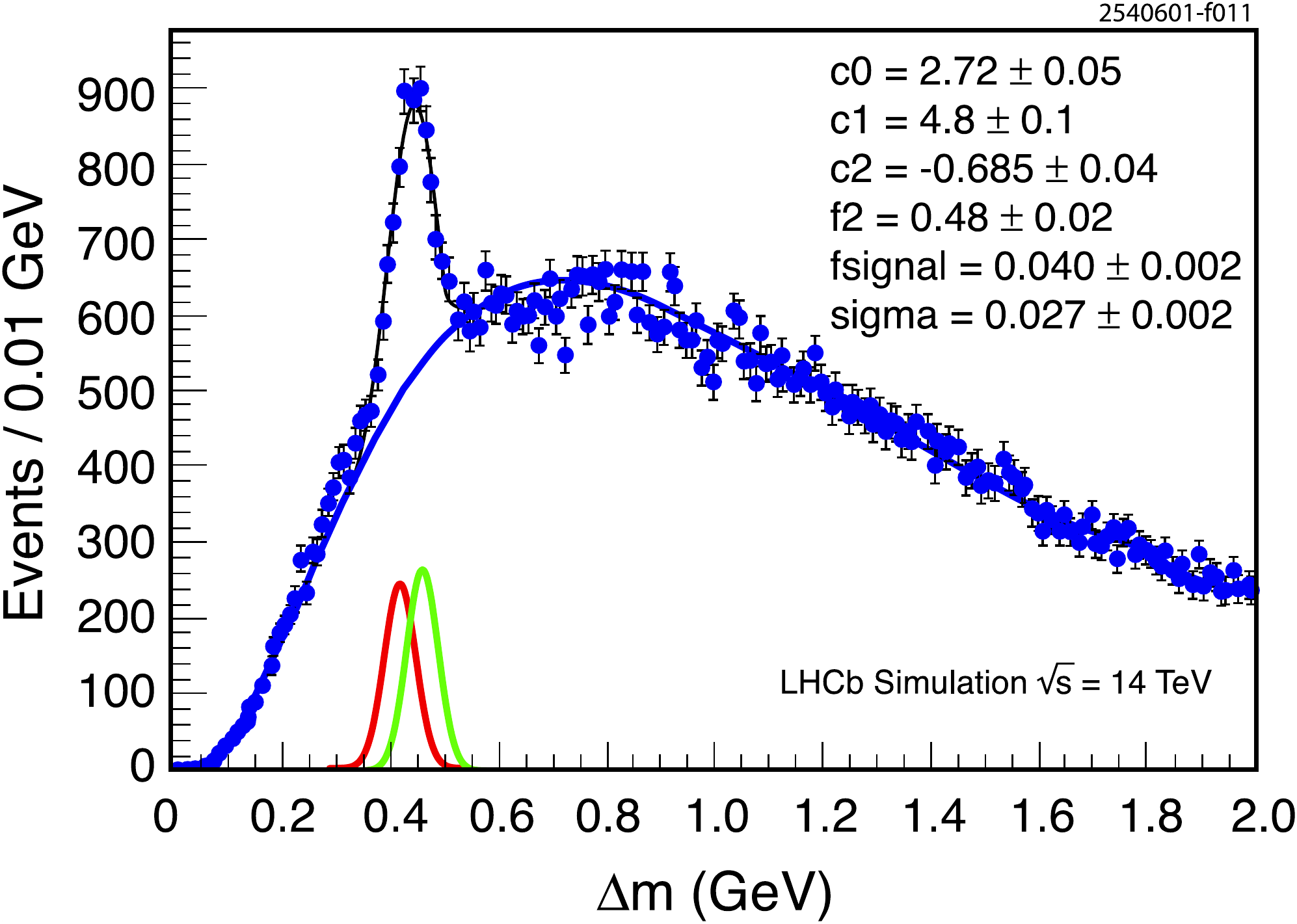}
      \caption{$\Delta m$ distribution of $\chi_c$ 
               candidates obtained with LHCb Monte Carlo simulation }
      \label{fig:Future_chic}
   \end{center}
\end{figure}

 Reconstruction of $\chicOne$ and $\chi_{c2}$ has also been 
studied~\cite{Needham:2009zz} using the decay
modes $\chi_{c1,2} \to \jpsi \gamma$. 
A photon candidate reconstructed in the electromagnetic calorimeter
with $\ptrans > 500\mev$ is associated with a $\jpsi$ 
candidate. \Figure{fig:Future_chic} shows the
$\Delta m = m({\dimu\gamma}) - m({\dimu})$ 
distribution obtained from a Monte Carlo simulation. 
A clear signal peak can be observed.
The $\Delta m$ resolution of 27\mevcc\ is dominated by 
the uncertainty in photon energy. Since the 
mass difference between $\chicOne$ and $\chi_{c2}$ 
is known ($55\mevcc$), imposing this constraint on the 
analysis should allow separation of the $\chicOne$ and 
$\chi_{c2}$ contributions.

A large number of $\psip$ mesons will also be 
collected at LHCb. It is expected that the number of
reconstructed $\psip \to \dimu$ decays will be 
equal to 2 to 4\% of the
number of reconstructed $\jpsi \to \dimu$, 
with $S/B$ between 1 and 2~\cite{Sabatino:2009zz}. 
Because the masses of the $\jpsi$ and $\psip$ are close, 
trigger and reconstruction efficiencies for the two states
are similar. A measurement of 
\beq
\frac{\sigma({\rm prompt}\, \jpsi)}{\sigma({\rm prompt}\,\psip)}\,,
\eeq 
where a number of systematic errors cancel, will be possible.

A large sample of $X(3872) \to \jpsi \pi^+ \pi^-$ decays 
will be reconstructed at LHCb, either 
prompt $X(3872)$, or $X(3872)$ from $b$-hadron decays. 
In particular, 
the decay channel\footnote{A charge-conjugate decay mode is implied in the
rest of the text} $B^+ \to X(3872) (\to \jpsi \rho^0) K^+$
will be studied because an angular analysis
of the decay products can lead to the determination of 
the now-ambiguous quantum numbers of the
$X(3872)$ (see \Sec{sec:SpecExpX3872}), 
allowing separation of the $1^{++}$ and 
the $2^{-+}$ hypotheses. 1850 reconstructed 
events are expected for $2\,\rm{fb}^{-1}$ of data 
at $\sqrt{s}=14\,\rm{TeV}$, with $B/S$ between 
0.3 and 3.4~\cite{Nicolas:2009zz}.
$Z(4430)$ will be sought in the decay 
$B^0\to Z(4430)^{\mp} K^{\pm}$, with 
\mbox{$Z(4430)^{\mp} \to \psip(\to \dimu) \pi^{\mp}$}. 
6200 events are expected with $B/S$ between 2.7 and 5.3, 
for $2\,\rm{fb}^{-1}$ of data at 
$\sqrt{s}=14\,\rm{TeV}$~\cite{Nicolas:2009zz}.

\subsubsection{$B_c$ physics}
\label{sec:Future_lhcbbc}

The expected $B_c^+$ cross section at the LHC 
is at the level of $1\,\mu\rm{b}$, so a very large number
of $B_c$ will be produced and recorded at LHCb. 
(For the present status of $B_c$ measurements,
see \Secs{sec:SpecExp_Bc} and \ref{sec:Prod_Bc}.)
First studies will use the reconstruction of the mode
with a large branching fraction 
$B_c^+\to \jpsi \mu^+ \nu$, and most promising 
results are expected
using the clean $B_c^+ \to \jpsi \pi^+$ decay mode. 
But the large number of $B_c^+$ produced will allow
a systematic study of the $B_c$ family at LHCb.

The selection of the decay channel $B_c^+ \to \jpsi \pi^+$, 
with $\jpsi \to \dimu$ has been studied using full 
Monte Carlo simulation
of events reconstructed by the LHCb 
detector~\cite{He:2009zz,Gao:2008zzj,Yushchenko:2003}. 
A specific generator, BCVEGPY~\cite{Chang:2003cq} 
has been 
used to generate $B_c$ events. 
Since the $B_c$ vertex is displaced with respect to 
the primary vertex, impact parameter selections are imposed
to the $\pi$ and $\jpsi$ candidates. 
Particle identification, quality of track and vertex fits, 
and minimum $\ptrans$ requirements
are applied to $B_c$ candidates in order to 
reduce the large background due to other $b$-hadron decays with a 
$\jpsi$ in the final state. The total 
reconstruction efficiency is estimated 
to be $(1.01 \pm 0.02)\%$, with
$1<B/S<2$ at 90\%~CL. 
Assuming a $B_c$ production cross section of 
$\sigma(B_c^+)=0.4 \,\mu \rm{b}$ for 
$\sqrt{s}=14\,\rm{TeV}$ and a branching fraction 
${\cal B}(B_c^+\to \jpsi \pi^+)=1.3\times 10^{-3}$, 
310 signal events
are expected with $1\,\rm{fb}^{-1}$ of data.

\begin{figure}[tb]
   \begin{center}
      \includegraphics[width=\figwid]{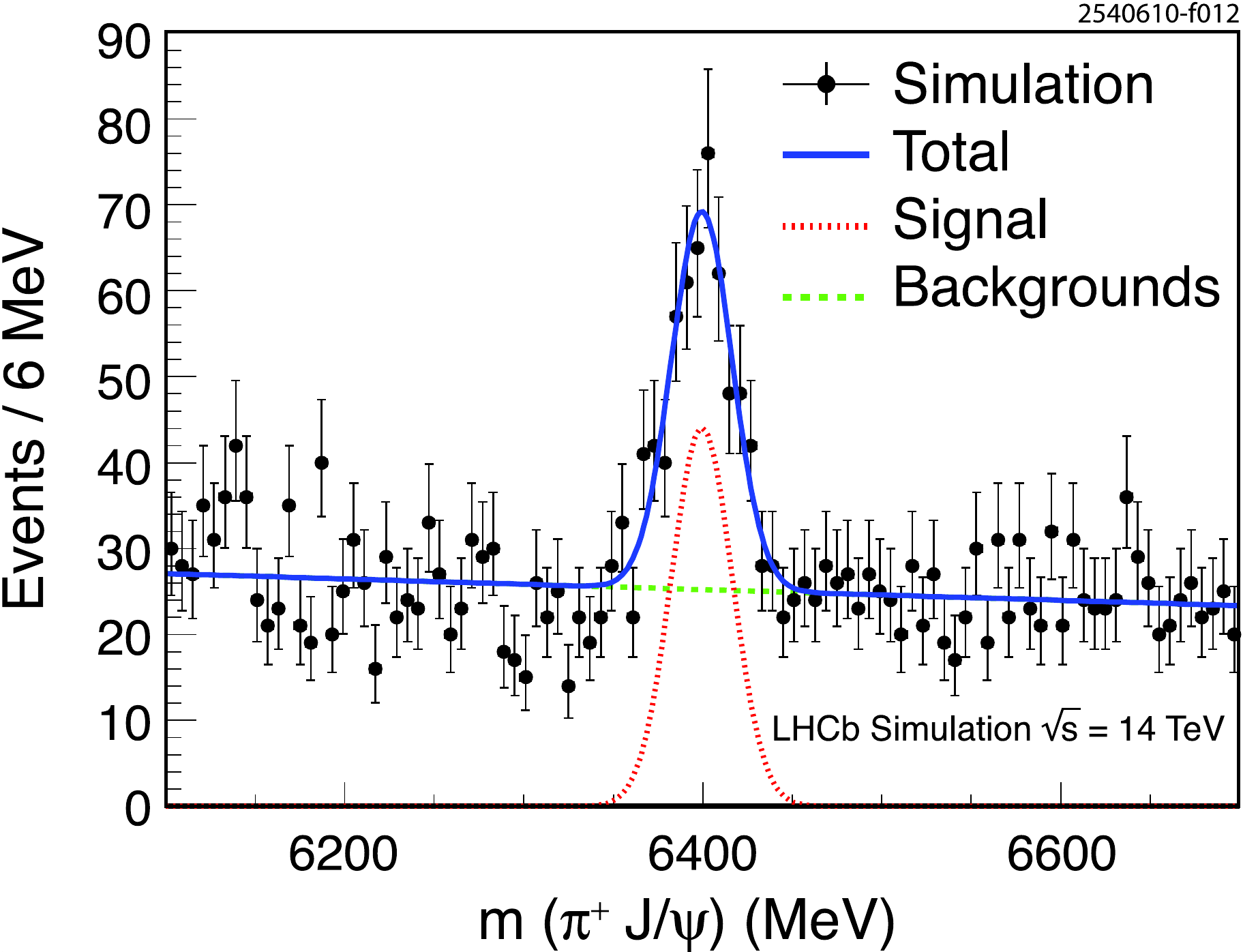}
      \caption{$B_c^+$ candidate mass distribution obtained 
               with LHCb Monte Carlo simulation }
      \label{fig:Future_bcmass}
   \end{center}
\end{figure}

 The potential of a mass measurement has been 
studied using an unbinned maximum likelihood
method to extract the $B_c^+$ mass from the 
reconstructed sample of $B_c^+ \to \jpsi \pi^+$ candidates.
Describing the invariant mass distribution of the 
signal by a single Gaussian and the combinatorial
background by a first-order polynomial, the fit procedure 
gives a $B_c^+$ mass of
$m(B_c^+)=6399.6 \pm 1.7\mevcc$, where the error is 
statistical only, consistent with the input value
of $6400\mevcc$. The size of the sample used
corresponds to the expected yield for 
$1\,\rm{fb}^{-1}$ of data. The result
of the fit and the mass distribution are 
shown in \Fig{fig:Future_bcmass}. 
The mass resolution
is  $\sigma=17.0 \pm 1.6\mevcc$.

The reconstructed $B_c^+$ candidates will also 
be used to measure the lifetime of the $B_c^+$. A combined
mass-lifetime fit is performed. The proper-time 
distribution is described by an exponential function 
convoluted with a resolution function and 
multiplied by an acceptance function $\epsilon(t)$ 
which describes
the distortion of the proper-time distribution 
due to the trigger and offline event selections
through impact parameter requirements. 
The form of these functions is determined 
from the full Monte Carlo simulation. 
Since the resolution of the impact parameter 
depends on the transverse momentum of the tracks, the
proper-time acceptance function $\epsilon(t)$ 
depends on the $\ptrans$ distribution of the 
$B_c^+$, and
then on the generation model used for 
the $B_c^+$ when determining $\epsilon(t)$. 
In order to evaluate
the systematics associated with this effect, 
a fit was performed on a $B_c^+$ sample generated with 
a $\ptrans$ spectrum identical to the $B^+$ 
spectrum observed in the simulation. 
A bias of $0.023\,\rm{ps}$ is then 
observed in the lifetime determination. 
In order to reduce this bias, the lifetime fit is performed
simultaneously on two samples with different $\ptrans$ 
ranges, $5<\ptrans<12\gevc$ and $\ptrans>12\gevc$. 
The resulting bias is then reduced to $0.004\,\rm{ps}$. 
The fit procedure applied to a sample corresponding to 
$1\,\rm{fb}^{-1}$ of data gives a $B_c^+$ lifetime
of $\tau(B_c^+)=0.438 \pm 0.027\,\rm{ps}$, 
where the error is statistical only, consistent with the input value
of $0.46\,\rm{ps}$ \cite{Gao:2008zzi}.

 LHCb capabilities should allow comprehensive 
studies of many other aspects of $B_c$ physics~\cite{Gouz:2002kk}.
Spectroscopy of the $B_c$ excited states, both below 
and above the $m(B^0)+m(D^0)$ threshold will be
performed. For example, searches for 
$B_c^{**} \to B_c^+ \pi^+ \pi^-$ have been envisaged~\cite{Ke:2009sk}.
Searches for new decay modes of the $B_c$ 
will also be made, \eg modes  
with a weak decay of the $c$ quark, such as 
$B_c^+ \to B_s^0 \pi^+$ or those with
a $\bar{b}c$ 
annihilation~\cite{DescotesGenon:2009ja}, such
as $B_c^+ \to \bar{K}^{*0} K^+$. 

\subsubsection{Bottomonium physics}

Analyses of bottomonium in LHCb have begun. 
First results show that the
reconstruction of $\UnS{1} \to \dimu$ is possible with a 
mass resolution of $37\mevcc$. 
Other $\Upsilon$ states will also be observed in the detector.
Using these candidates, measurements of the production cross section 
and of the 
production polarization will be performed 
as a function of $\ptrans$ and $\eta$. 
Similarly, the decays 
$\chi_{b1,2} \to \Upsilon \gamma$ are also
possible to reconstruct in the LHCb detector, 
due to low-\ptrans-photon reconstruction provided by the
electromagnetic
calorimeter~\cite{Needham:2009zz}.
In addition, the equivalent for the $b$ 
family of the exotic $X$, $Y$, and $Z$ states
(see \Sec{sec:SpecExp_Unanticipated})
will be sought
at LHCb, for example, in the decay mode $Y_b \to \UnS{1} \pi^+ \pi^-$
(for status of the $Y_b$,
see \Secs{sec:SpecExp_UnconVector} and \ref{sec:Dec_HT_Ups5Spipi}).

\subsection{RHIC}

The Relativistic Heavy-Ion Collider, RHIC, provides d+Au, Au+Au, and Cu+Cu 
collisions at $\sqrt{s_{_{NN}}} = 200$\gev, and 
polarized $pp$ collisions at both 200 and 500\gev. 
Polarized $pp$ collisions are used for spin studies, with those from
200\gev\ serving as reference data for the heavy-ion program. 
The d+Au collisions are used for 
studies of forward physics and to establish the cold nuclear matter baseline 
for heavy-ion collisions. The primary focus of the heavy-ion program is
to quantify the differences between the hot, dense final state and scaled
$pp$ and d+Au reference data.

In 200\gev\ $pp$ collisions, PHENIX~\cite{daSilva:2009yy,Leitch:2007wa} 
has measured cross sections for production 
of the (unresolved) $\Upsilon$ states and the \psip.  
A $\chi_{cJ}$ 
measurement will be published soon. There is also a preliminary 
PHENIX~\cite{Atomssa:2009ek} result 
showing that the $\Upsilon$ $R_{AA}$, defined in \Eq{eqn:Med_rab}), 
measured in Au+Au collisions
is below 0.64 at 90\%~CL. All of these measurements involve low yields and 
will benefit greatly from additional luminosity. PHENIX~\cite{Lansberg:2008jn}
has also measured the polarization of the $\jpsi$ in 200\gev\ $pp$ collisions 
and will also do so with existing 500\gev\ data. 

So far, the high-statistics heavy quarkonium data available from RHIC are
the $\jpsi$ data sets measured by PHENIX in $\sqrt{s_{_{NN}}}=200$\gev\ 
$pp$~\cite{Lansberg:2008jn}, 
d+Au~\cite{Adare:2007gn,daSilva:2009yy}, 
Cu+Cu~\cite{Adare:2008sh}, 
and Au+Au~\cite{Adare:2006ns} 
collisions. The data were measured in three rapidity 
ranges: $-2.2 < y < -1.2$; $-0.5 < y < 0.5$; and $1.2 < y < 2.2$. There are 
also more recent data sets with much higher yields in d+Au (2008) and Au+Au 
(2007 and 2010) that are still being analyzed.  While preliminary $\jpsi$ 
$R_{CP}$ (see \Eq{eqn:Med_rcpdef}) 
results were presented by PHENIX~\cite{daSilva:2009yy} 
at Quark~Matter~2009, the 
final data have not yet been released. \Section{sec:media_sec5} 
reviews the status of the $\jpsi$ program at RHIC.

STAR~\cite{Sakai:2009zza} 
has measured $\jpsi$-hadron azimuthal angular correlations which have 
been used to infer the $B$-meson feeddown contribution to the 
$\jpsi$. STAR
has also measured \RdAu\ for the combined $\Upsilon$ 
states using 2008 d+Au~\cite{Liu:2009wa} data and 
2006 $pp$~\cite{:2010am} data, and made measurements of 
the high $p_{T}$ $\jpsi$ suppression factor, $R_{AA}$, from 
Cu+Cu~\cite{Abelev:2009qaa} collisions.

The ongoing RHIC luminosity upgrades will be completed by the 2013 run. 
The introduction in 2011 and 2012 of silicon vertex detectors into PHENIX at 
mid- and forward-rapidity, respectively, and of the STAR Heavy Flavor Tracker 
in 2014, will enable open charm and open bottom to be measured independently 
with greatly improved precision.  The detector upgrades will also allow 
improved measurements of quarkonium states due to improved mass resolution 
and background rejection. 

The RHIC run plan for the next 5 years or so will be centered on exploiting 
the capabilities of the new silicon vertex detectors and other upgrades, 
combined with the increased RHIC luminosity.  Of greatest interest to 
in-medium heavy flavor physics, there will likely be long $pp$, d+Au, 
and Au+Au 
runs at 200\gev, plus shorter runs with the same species at 62\gev\ to explore 
the energy dependence of open and hidden heavy flavor production. 
The luminosity increase and the
PHENIX detector upgrades will enhance the PHENIX heavy quarkonium program in
several ways: increased $p_{T}$ reach for the $\jpsi$; new
studies of $\jpsi$ suppression with respect to the reaction plane; a first 
$\jpsi$ $v_{2}$ measurement; better understanding of cold nuclear matter 
effects on $\jpsi$ production; and low-statistics measurements of the 
modification of the combined $\Upsilon$ states. The increased luminosity will
enable the large-acceptance STAR detector to extend its $p_T$ reach to 
considerably higher values for $\Upsilon$ and $\jpsi$ measurements than 
previously possible. 

The RHIC schedule for the period beyond about 5 years is still under 
development.
RHIC experiments are presently engaged in preparing a decadal plan that 
will lay out their proposed science goals and detector upgrades for 2011 to 
2020. 
PHENIX is considering a conceptual plan that would keep the new silicon vertex 
detectors but completely replace the central magnet and the outer central 
arm detectors. The new magnet would be a 2T solenoid with an inner radius of 
70~cm.  Two new silicon tracking layers would be placed inside the solenoid
at 40 and 60~cm, followed by a compact electromagnetic calorimeter of 8~cm 
depth and a preshower layer. A hadronic calorimeter would be added outside 
the magnet with an acceptance of $|{\eta}| < 1$ and $2\pi$ in azimuth. 
The conceptual design is still being evaluated, but it promises to allow 
powerful measurements of light quark and gluon jets; 
dijets and $\gamma +$jet coincidences; 
charm and bottom jets; 
the $\jpsi$ modification factor over a range of energies; 
simultaneous studies of the modification of the three bound $\Upsilon$ states;
and direct $\gamma^{*}$ flow.  Removing the south muon 
spectrometer and replacing it with an electron/photon endcap spectrometer has
also been discussed. This replacement would be aimed at addressing spin 
physics questions and possibly providing electron-ion capabilities in PHENIX. 
The north muon spectrometer would be retained to provide forward rapidity 
heavy flavor measurements. 

In addition to the Heavy Flavor Tracker, STAR will add a new detector
that is of major 
importance to their quarkonium program. The Muon Telescope Detector will be a 
large acceptance muon detector located outside the STAR magnet at midrapidity, 
covering $|y| < 0.5$. It will provide a good signal-to-background ratio for 
measurements of the three $\UnS{n}$ states and add the capability of 
measuring $\jpsi$ elliptic flow and suppression at high $p_{T}$.
The upgrades being pursued by PHENIX and STAR, combined with very high RHIC 
luminosity, will provide the opportunity to compare, between RHIC and the LHC, 
in-medium quarkonium modification at energies of 62, 200, and 5500\gev.  This 
energy regime spans a wide range of medium temperatures and lifetimes, 
and also provides measurements with very different underlying heavy quark
rapidity densities and thus very different contributions to quarkonia from 
processes involving coalescence of heavy quarks from different hard 
collisions. 

\subsection{Super Flavor Factories}
\label{Sec:Future_superbintro}

At \epem\ machines, quarkonium can be produced through 
several processes: directly, {\it i.e.}, 
during energy scans for $J^{PC} = 1^{--}$ states in the $c\bar{c}$ and 
$b\bar{b}$ regions; 
through ISR (for $J^{PC} = 1^{--}$ states below the
\epem\ center-of-mass energy) or two-photon 
(for $C=+1$ states) processes; 
in $B$-meson decays through color-suppressed $b \to c$ transitions.
All these have been successfully employed 
for quarkonium studies in the CLEO, \babar, Belle, and/or BESIII experiments,
the first two of which permanently ceased taking data in early 2008.
Belle  
acquired data until the middle of 2010, and then shut down for 
a significant upgrade.
CLEO-c collected $48\,{\rm pb}^{-1}$ 
at the \psip\ (about 27 million \psip\ produced) 
and $1485\,{\rm pb}^{-1}$ in the CM energy range 
$3.67 - 4.26\,{\rm GeV}$; BESIII has already quadrupled
the CLEO-c \psip\ sample, acquired 200M \jpsi\ decays,
and will, in time, exceed the CLEO-c samples above
open-charm threshold as well (\Sec{sec:Future_besiiiintro}).
What will happen to
$e^+e^-$ quarkonium physics after BESIII and Belle
programs are complete?

A new generation of Super Flavor Factories 
has recently been proposed 
in order to perform precision 
measurements in the flavor sector and 
complement New Physics (NP) searches at 
hadronic machines~\cite{Hashimoto:2004sm,Bona:2007qt}. 
An increase in statistics by a factor of 50-100 with respect 
to the current generation of still-running experiments 
is essential to such physics program. 

Two different approaches have been devised to reach
a design peak luminosity of $10^{36}\,{\rm cm}^{-2}{\rm s}^{-1}$.
In the original SuperKEKB~\cite{Hashimoto:2004sm} 
design this was to be achieved by 
increasing the beam currents, and introducing crab crossing to maintain
large beam-beam parameters~\cite{Oide:2009zz}. 
In the SuperB design~\cite{Bona:2007qt} 
a similar luminosity goal is pursued
through the reduction of the interaction point size using very small
emittance beams, and with a ``crab'' of the focal plane to 
compensate for a large crossing angle and mantain optimal 
collisions~\cite{Raimondi:2007vi}. 
After KEK revised their design in favor of the nanobeam
collision option~\cite{Raimondi:2007vi}, 
the machine parameters for the SuperB factory and the 
KEKB upgrade are very similar~\cite{Bona:2007qt,Oide:2009zz}. 

An NP-oriented program suggests that the
machine be operated primarily at the \UnS{4}\ 
resonance~\cite{Browder:2008em}, with integrated 
luminosity of order $25-75\,{\rm ab}^{-1}$.
However, the ability to run at other $\Upsilon$ resonances 
or at energies in the $c\bar{c}$ region would substantially 
enhance the physics potential of the machine. 
In one month at design luminosities it would be possible
to collect about 150~fb$^{-1}$ at \DDbar\
threshold~\cite{Bona:2007qt}. In about the same time, 
an \UnS{5}\ run would integrate about $1\,{\rm ab}^{-1}$,
corresponding to a ``short run'' scenario. In a ``long run'' scenario
at \UnS{5}, about $30\,{\rm ab}^{-1}$ could be collected~\cite{Bona:2007qt}.

The detector design will primarily address the 
requirements of an NP search program at the \UnS{4}. Other 
constraints are posed by the possibility to re-use components from 
the \babar\ and Belle detectors.
Improvements in the vertex resolution, to compensate a reduced beam 
asymmetry, 
as well as increased hermeticity of the detector 
are foreseen.
Assuming the same magnetic field, a similar momentum resolution is expected.
In a simplified approach, one can assume similar backgrounds and 
detector performances as at \babar~\cite{Aubert:2001tu} 
and Belle~\cite{Abashian:2000cg}.

  A heavy-quarkonium to-do list in a Super Flavor Factory 
physics program can be 
found in the Spectroscopy (\Sec{sec:SpecChapter}), 
Decays (\Sec{sec:DecChapter}), and Summary (\Sec{sec:SumChapter})
sections of this review. Precision measurements
or simply observation of some conventional, expected
processes is warranted; the
as-yet-unexplained phenomena also demand attention
and offer great reward.
Highlights of such a program would include the following:
\begin{itemize}

\item{Precision measurements of \etac\ and \etacp\
masses and widths (probably in $\gamma\gamma$-fusion), 
understanding of their 
observed lineshapes in radiative \jpsi\ and \psip\ decays,
and a comprehensive inventory of radiative and
hadronic decay modes and respective branching fractions
(\Secs{sec:SpecExp_etac2s},~\ref{sec:Dec_1Sc} 
and \Tab{tab:Spec_ExpSumCon}).}

\item{First observation and study of $\eta_b(2S)$, $h_b(^1P_1)$, and
\UoneDJ~($J=1,\, 3$) (\Sec{sec:SpecExp_Ups1D} and 
Tables~\ref{tab:Spec_ExpSumCon} and \ref{tab:Spec_U1D_mass}).}

\item{For $X(3872)$ (see \Sec{sec:SpecExpX3872} and
Tables~\ref{tab:Spec_ExpSumUnc}, \ref{tab:Spec_XMass}, 
\ref{tab:Spec_DMass}, \ref{tab:Spec_RatBchgBneu}, and \ref{tab:Spec_DeltaMX} ), 
detailed study of the $\dipi\jpsi$ 
and $\DstnDn$ lineshapes,
high-statistics measurements of $\gamma\jpsi$ and $\gamma\psip$
branching fractions, and a search for decay to $\dipiz\jpsi$. 
The fruit of such an effort could be an answer to
how much tetraquark, molecular,
and/or conventional charmonium content this state contains.}

\item{For the $Y(4260)$ and nearby states (see
\Sec{sec:SpecExp_UnconVector}
and
Tables~\ref{tab:Spec_ExpSumUnc}, \ref{tab:Spec_Y4260},
\ref{tab:Spec_Y4360}, and \ref{tab:Spec_Y4260_opencharm}),
a comprehensive lineshape measurement
accompanied by precision branching fractions could
shed light on the molecular, tetraquark, or hybrid hypotheses.}

\item{Confirmation and study of the exotic charged $Z^+$ states
(\Sec{sec:SpecExpChargedExotic} and
\Tab{tab:Spec_ExpSumUnc}).}

\item{Confirmation and study of the $Y_b$ (\Sec{sec:SpecExp_UnconVector}
and
\Tab{tab:Spec_ExpSumUnc}).
A high-statistics scan of $20\,{\rm fb}^{-1}$ per point, necessary to 
reduce the relative error to the $10^{-3}$ level, 
might be needed~\cite{Hitlin:2008gf}.
If the $Y_b$ is below $10800\mev$, it could also be produced by ISR with 
\UnS{5}\ data collected by a Super $B$-factory. If it is above 
that energy, a direct scan will be necessary.
}
\end{itemize}

\subsection{\PANDA}
\label{sec:Future_pandapanda}

\PANDA is one of the major projects at
the future Facility for Antiproton
and Ion Research (FAIR)~\cite{FAIR:2006zz} at GSI in Darmstadt, 
which is expected to start its operations in 2014. 
\PANDA will use the  antiprotons circulating in the 
High Energy Storage Ring (HESR), 
to study their interactions with protons or nuclei 
on a fixed target. 
The antiproton momentum in the range 1.5 to $15\gevc$
corresponds to a center-of-mass energy in $\pbarp$ 
collisions in the range $2.5-5.5\gev$. 
The purpose of the \PANDA experiment is to investigate QCD in the 
nonperturbative regime. This is achieved through the study 
of several topics, like QCD bound states, 
nonperturbative QCD dynamics, study of hadrons in nuclear 
matter, hypernuclear physics, electromagnetic processes, and 
electroweak physics.
In particular, 
\PANDA has an extensive research program in charmonium physics. 
Experimentally, charmonium has been studied mainly in 
$\epem$ and $\pbarp$ experiments. 
While in $\epem$ collisions, only the $J^{PC} = 1^{--}$ states 
can be directly formed, in $\pbarp$ interactions, direct formation is 
possible for all the states with different quantum numbers, through 
coherent annihilation of the three quarks of the protons with the 
three antiquarks of the antiproton. 
An additional advantage of this technique is that the interaction energy 
can be precisely determined from the beam parameters and it 
is not limited by the detector resolution, allowing fine energy 
scans of narrow resonances. 
Historically, this experimental technique was successfully 
used at CERN and at Fermilab.
With a higher luminosity and a better beam energy resolution 
with respect to previous $\pbarp$ experiments, \PANDA 
will be able to obtain high-precision data on charmonia 
and measure the masses, widths, and excitation 
curves for the recently observed states with unprecedented precision.

\subsubsection{Experimental technique}

Quarkonium physics is one of the main research fields for \PANDA. 
The precision study of resonance parameters and excitation curves 
is an area where the close interplay between machine and 
detector is fundamental. 
For the \PANDA charmonium program, 
the antiproton beam collides with a fixed hydrogen target. 
In order to achieve the design luminosity 
($2 \times 10^{32}$~cm$^{-2}$~s$^{-1}$),
a high-density target thickness is needed. Two
solutions are under study: the cluster-jet target 
and the pellet target, 
with different implications for the beam quality and the 
definition of the interaction point. 

Thanks to the cooling of the antiproton beam, at the HESR 
the energy spread of the beam will be approximately $30\,{\rm keV}$, 
which is comparable with the width of the narrowest 
charmonia states. 
The knowledge of the total interaction energy from the beam parameters 
and the narrowness of the beam energy distribution 
allow the direct measurement of the mass and the 
width of narrow states through scans of the excitation curve.

The resonance parameters are determined from a 
maximum-likelihood fit to the number of observed 
events in a specific channel $N_i$, where the subscript $i$ 
refers to different center-of-mass energies:
\begin{equation}
N_i = \epsilon_i {\cal L} \int{\sigma_{BW}(E^\prime) B_i(E^\prime) dE^\prime}
\label{eqn:Future_obsevts}
\end{equation}
where $\sigma_{BW}$ is the Breit-Wigner resonance cross section 
to be measured and $B_i$ is the center-of-mass energy 
distribution from beam parameters. 

\begin{figure}[t]
   \begin{center}
      \includegraphics[width=\figwid]{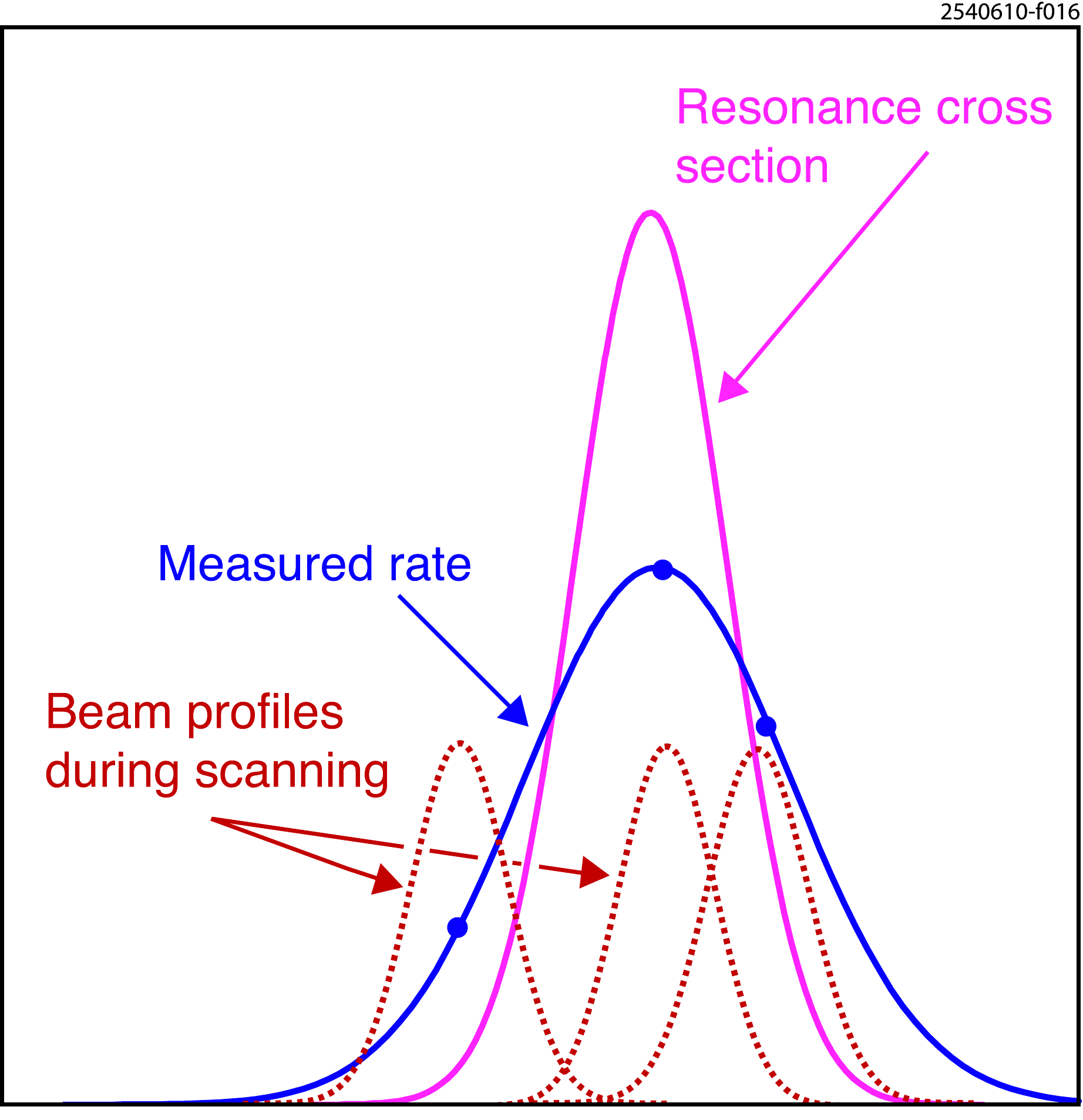}
      \caption{Schematic representation of the scan of a 
               resonance {\it (lighter solid line)}: 
               the center-of-mass energy is varied 
               with the HESR to several values 
               across the resonance (the energy distributions are shown by 
               {\it (dotted lines)}. The {\it solid circles} 
               represent the observed 
               cross section for each energy point, defining 
               the measured excitation curve {\it (darker solid line)}, 
               which is the convolution of the resonance curve 
               and the beam profile }
      \label{fig:Future_Scan}
   \end{center}
\end{figure}

The energy scan of a resonance is schematically represented 
in \Fig{fig:Future_Scan}: 
the energy of the interaction, shown on the 
horizontal axis, is obtained from 
the beam parameters, and is varied in the energy region 
to be explored. The detector is used as a simple 
event counter and the cross section values (vertical axis) 
can be obtained from the number of events observed at 
each energy point. 
Using \Eq{eqn:Future_obsevts}, it is then possible to 
determine the resonance parameters. 
By means of fine scans it will be possible to 
measure masses with accuracies of the order of 
100\kev\ and widths at the 10\% level. 
The entire region below and above the open charm 
threshold (from 2.5\gev\ to 5.5\gev\ in the 
center-of-mass energy) will be explored.

As described above,
antiproton-proton formation will become an important tool for decisive
studies of the natural widths and lineshapes of open and hidden charm states.
It will also have a powerful role in measuring decay modes
of very narrow states. Unfortunately, charm production cross sections 
in proton-antiproton annihilation in this energy regime are either
unknown or, at best, poorly measured. 
Thus predictions for sensitivities are not reliable. In order
to provide experimental input to event generators, and thereby
allow credible sensivity studies, it would be worthwhile to
measure inclusive open charm production
at the existing antiproton facility at Fermilab with a
relatively simple and small test apparatus equipped with a microvertex 
detector and a tracking device.
Such an experiment would likely require less than a month of datataking
to acquire a sufficient statistics.

\subsubsection{Detector}
\label{sec:Future_pandadetector}

The \PANDA detector is a multipurpose detector, 
designed to obtain the highest possible 
acceptance, good tracking resolution, particle identification 
and calorimetry. 
It is composed of two parts: a target spectrometer with an axial 
field generated by a superconducting solenoid for the measurement 
of the particles emitted at large angles with respect to the beam 
direction, and a forward spectrometer with a dipole magnet in the forward 
direction for the measurement of the particles leaving the interaction 
region at small angles with respect to the beam. 

 The target spectrometer is arranged in a barrel part, for angles 
larger than $22^\circ$, and an endcap part, covering the 
forward region down to $5^\circ$ in the vertical plane and 
$10^\circ$ in the horizontal plane. 
A silicon tracker is located close to the interaction region. 
It is composed of two inner layers of hybrid pixel detectors and 
two outer layers with silicon strip detectors, with cylindrical 
symmetry around the beam pipe. In addition, six silicon discs with 
both pixel and strip modules are located in the forward direction. 
The tracking system (solutions with straw tubes detector or TPC 
are under study) 
is required to provide momentum measurement 
with resolution $\delta p / p$ at the percent level, handling 
the high particle fluxes that are foreseen at the maximum luminosity. 
In the forward part of the target spectrometer, 
particles exiting the interaction region at polar angles below 22$^\circ$ 
are tracked with 3 GEM detectors placed downstream. 
Charged particle identification is obtained by collecting 
complementary information from different detectors. 
Most charged particles with momentum exceeding 0.8\gevc\ 
are identified using a Cherenkov detector. 
A detector of internally-reflected Cherenkov light based on the 
\babar~design \cite{Staengle:1997xp} will provide particle 
identification in the region $22^\circ$ to $140^\circ$. 
Time-of-flight will be also used in \PANDA for the 
identification of low-momentum particles. 
To obtain muon-pion separation, 
the yoke of the superconducting solenoid is segmented in 12 
layers and the gaps are instrumented for the measurement of 
the interaction length in iron. 
Electromagnetic calorimetry is required over a wide energy range, 
from the MeV to the GeV scale. 
A lead-tungstate ($\rm PbWO_4$) calorimeter, with crystal length 
corresponding to $22X_0$, will measure energy with a 
resolution better than 2\% at 2\gev\ and good time resolution. 

 A dipole magnet with 1~m gap and 2~m aperture is located 
between the target and the forward spectrometer. 
The measurement of the deflection of charged tracks 
is obtained by using a set of wire chambers 
placed before, within, and behind the dipole magnet. 
The expected $\delta p / p \sim 3\%$ for $3\gevc$ 
protons is limited by the scattering on gas and wires in 
the chambers. 
In the forward spectrometer, RICH or time-of-flight detectors
are proposed as particle identification systems. 
Electromagnetic calorimetry in the forward region 
will be performed by a shashlyk-type 
calorimeter with high resolution and efficiency. 
An energy resolution of $4\% / \sqrt{E}$ is foreseen. 
In the very forward part, a muon detector similar to 
the one employed in the target region will provide 
muon-pion separation. 

\subsubsection{Charmonium and open charm physics}
\label{sec:Future_pandacharmonium}

The study of charmonium and open charm 
are among the main topics 
in the \PANDA physics program. 
Compared to Fermilab E760/E835, 
a multipurpose detector with magnetic field, 
\PANDA has better momentum 
resolution and higher machine luminosity, 
which will allow the realization of an extensive 
research program in these fields. 

The existence of a large hadronic background in $\pbarp$ annihilation 
represents a challenge for the study of many final states and 
the capability of observing a particular final state 
depends on the signal-to-background ratio after the selection. 
A full detector simulation has been developed and used to 
test the separation of signal from background 
sources, with cross sections that are orders of 
magnitude larger than the channels of interest, and to prove 
the capability of background reduction at the level needed 
to perform charmonium studies in \PANDA. 
Here we will summarize the results,
which appear in more detail 
in the \PANDA Physics Performance Report~\cite{Lutz:2009ff},
of using the full simulation on some channels of interest. 
In all the following cases, a simple selection is adopted 
for the channel of interest: particles are reconstructed 
starting from detected tracks, using associated PID criteria, 
and a kinematic constraint 
to the energy and momentum of the beam is imposed
in order to improve on detector resolutions.

 The identification of charmonium states decaying 
into $\jpsi$ is relatively clean due to the presence 
of a pair of leptons \dilep\ in the final state. 
These channels can be used to study decays from 
$\psip$, $\chicJ$, $X(3872)$, and $Y(4260)$
that contain a $\jpsi$ in the final state.
These analyses rely on the positive identification 
of the two leptons in the final state for the 
reconstruction of $\jpsi \to\dilep$, where 
the main background is represented by pairs 
of tracks, like $\pi^+ \pi^-$, associated with large energy 
deposition in the electromagnetic calorimeter. 
The simulation shows that the resolution of the 
$\epem$ invariant mass for a 
reconstructed $\jpsi$ in the final state 
is in the range $4-8\mevcc$, depending 
on the total center-of-mass energy of the interaction. 

The $\jpsi\pi^+\pi^-$ final state 
is a key decay channel in charmonium studies. 
New states in the charmonia 
mass region, like the $X(3872)$ and the $Y(4260)$, have been 
discovered at the $B$-factories through this decay mode
(see \Sec{sec:SpecExp_Unanticipated}). 
The \PANDA performance on $\jpsi\pi^+\pi^-$ has been 
analyzed in detail through a complete simulation of 
this final state in the detector. 
A simple selection has been performed: adding two 
charged pions to a reconstructed $\jpsi$ and 
performing a kinematic fit with vertex constraint, 
the efficiency of the complete selection is approximately 
30\% over the energy region of interest (3.5 to $5.0\gev$).
The main source of hadronic background for this channel 
comes from $\pbarp \to \pi^+\pi^-\pi^+\pi^-$, 
where two pions may be erroneously identified as electrons and 
contaminate the signal. 
At a center-of-mass energy around $4.26\gev$, the cross section for 
this process is a few tens of $\rm \mu b$ 
\cite{Flaminio:1970zz}, which is $10^6$ times larger than the 
expected signal, estimated from previous results of Fermilab E835. 
With the selection described, the rejection power 
for this background source is of the order of $10^6$ and 
the signal-to-background ratio is about 2, 
which should provide well-identified and
relatively clean \jpsi\dipi final states in \PANDA.

The discovery mode (see \Sec{sec:SpecExp_hc})  for
$\hsubc$ is the electromagnetic transition to 
the ground state charmonia $\hsubc \to \etac \gamma$, 
where the $\etac$ can then be detected through many decay modes. 
The decay $\etac \to \gamma \gamma$ is characterized by 
a reasonably clean signature, due to the presence of 
two energetic photons in the detector with the $\etac$ 
mass, albeit with a small branching fraction (see \Sec{sec:Dec_twophoton}).
A study has been performed to assess the \PANDA capability 
to detect the $\hsubc \to \etac \gamma \to 3 \gamma$ decay, 
in presence of background sources 
due to final states such as $\pbarp \to \dipiz$, $\piz\eta$,
and $\eta\eta$. Each presents hard photons in the final 
state and no charged tracks, a signature 
that could mimic the channel of interest. 
(See \Sec{sec:SpecExp_hc} for a description of a previous
$p\bar{p}\to\hsubc\to\gamma\etacp\to 3\gamma$ analysis and 
\Sec{sec:Dec_jpsito3gamma} for a description of a $\jpsi\to 3\gamma$ 
observation in $\psip\to\dipi\jpsi$ decays.)
In order to improve the background rejection, additional 
cuts are applied after the event reconstruction:
\begin{itemize}
\item a cut on the center-of-mass energy of the energy of the 
$\gamma$ coming from the radiative transition: $0.4<E_\gamma<0.6\gev$;
\item an angular cut $|\cos\theta_{CM}| < 0.6$ allows 
rejection of a large fraction of backgrounds (like $\dipiz$)
that are strongly peaked in the forward direction;
\item to suppress \etap\ decays,
the invariant mass of the radiative $\gamma$ paired 
with either photon coming from the decay of the 
$\etac$ candidate is required to be larger than 1\gevcc. 
\end{itemize}
After these cuts, the efficiency on the signal is about 8\% 
and the background suppression of the order of $10^{-6}$ 
or larger on many background channels. 
The production cross section observed by E835~\cite{Andreotti:2005vu}, 
although with large uncertainties, 
can be combined with the present background suppression, 
to obtain an estimate of the order of 90 or more for 
the signal-to-background ratio.

\begin{figure}[b]
   \begin{center}
      \includegraphics[width=\figwid]{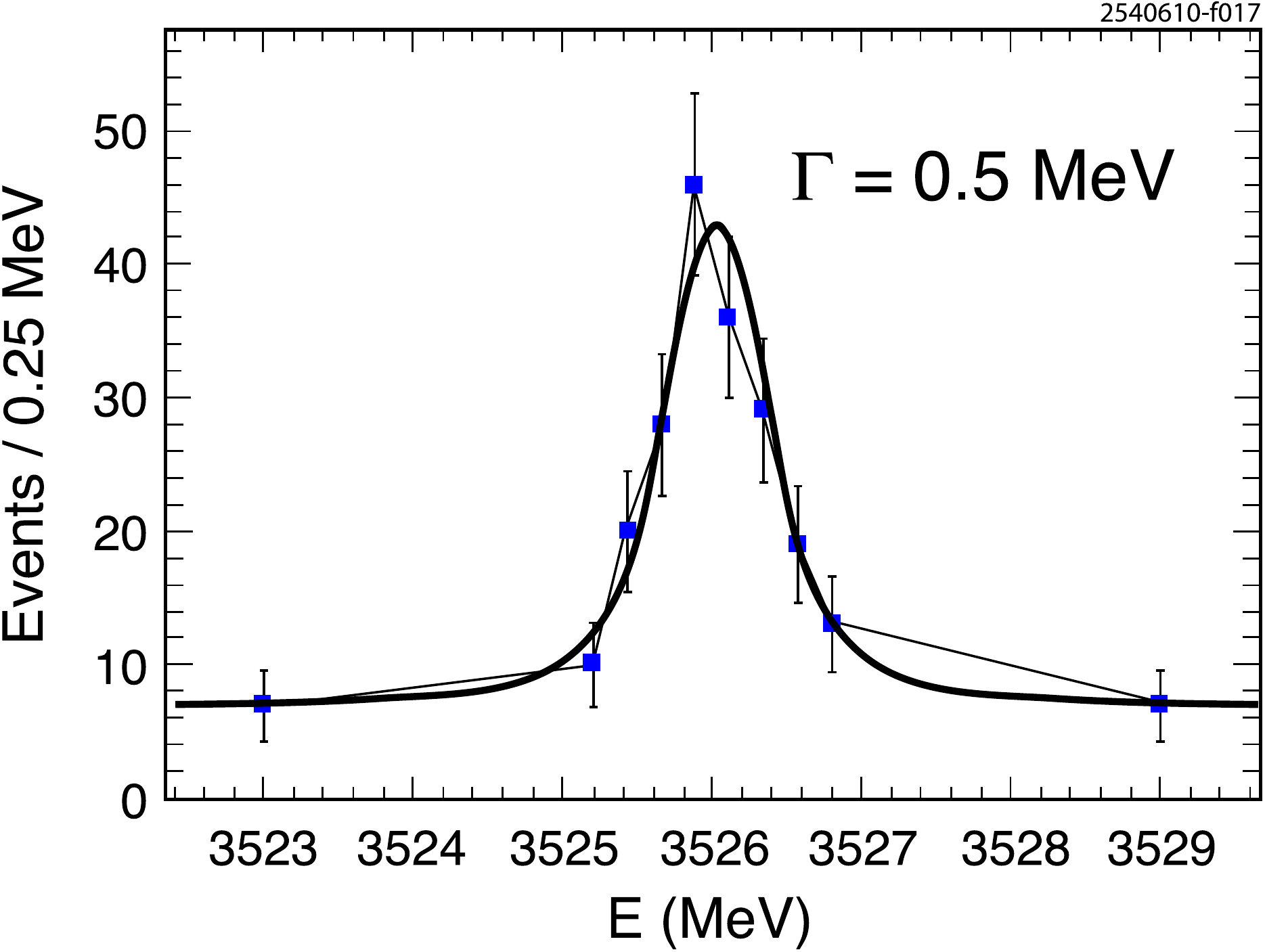}
      \caption{Simulation of the scan of the $\hsubc$ 
               (with $\Gamma(\hsubc) = 0.5\mevcc$) for the measurement 
               of the resonance width. Each point corresponds to 
               5 days of data-taking with \PANDA }
      \label{fig:Future_hc_width_05MeV}
   \end{center}
\end{figure}

 As a benchmark channel of hadronic decays, we 
consider $\hsubc \to \etac \gamma \to \phi\phi\gamma$, 
with $\phi \to K^+ K^-$. 
Three reactions are considered to be dominant contributions to the 
background: $\pbarp \to K^+ K^- K^+ K^- \piz$, 
$\pbarp \to \phi K^+ K^- \piz$, 
and $\pbarp \to \phi\phi \piz$, with one photon 
from the $\piz$ undetected.
To suppress such background,
it is additionally required that no 
$\piz$ candidates (photon pair with invariant mass 
in the 0.115-0.150\gevcc\ region) be present in the event. 
The overall efficiency for signal events is $\sim 25\%$. 
Since no experimental data is available for the three 
background cross sections, the only way to estimate the 
background contribution is to use the dual parton model (DPM); 
none out of $2 \times 10^7$ simulated events pass the selection. 
The three main background channels have been also simulated 
separately. With a total $\pbarp$ cross section of 60~mb, 
we estimate that
$\sigma(\pbarp \to K^+ K^- K^+ K^- \piz) = 345$~nb,
$\sigma(\pbarp \to \phi K^+ K^- \piz) = 60$~nb, and
$\sigma(\pbarp \to \phi\phi \piz) = 3$~nb.
Using these values,
a signal-to-background ratio of $\ge 8$ 
for each of the background channels is obtained. 
Using these signal-to-background values, it is possible to 
estimate the \PANDA sensitivity in the $\hsubc$ width 
measurement.  
A few scans of the $\hsubc$ have been simulated for different 
values of $\Gamma(\hsubc)$. The expected shape of the 
measured cross section is obtained from the 
convolution of the Breit-Wigner resonance curve 
with the normalised beam energy distribution 
plus a background term. 
The simulated $\hsubc$ resonance shape for the case 
$\Gamma(\hsubc) = 0.5\mevcc$, 
assuming 5 days of data taking per point in high-resolution mode, 
is shown in \Fig{fig:Future_hc_width_05MeV}. 
The accuracy on the width measurement is 
of the order of $0.2\mevcc$ for $\Gamma(\hsubc)$ values in 
the range $0.5-1.0\mevcc$. 

 The ability to study charmonium states above $\DbarD$ 
threshold is important for the major part of the 
\PANDA physics program, in topics like the study of open 
charm spectroscopy, the search for hybrids, and CP violation 
studies. 
The study of $\pbarp \to \DDbar$ as a benchmark channel
will also assess the capability to separate a hadronic decay 
channel from a large source of hadronic background. 
Two benchmark channels are studied in detail:
\begin{itemize}
\item $\pbarp \to D^+ D^-$ (with $D^+ \to K^- \pi^+ \pi^+$)
\item $\pbarp \to D^{*+} D^{*-}$ (with $D^{*+} \to D^0 \pi^+,~ D^0 \to K^-\pi^+$)
\end{itemize}
The first one is simulated at the $\psi(3770)$ and the second at 
the $\psi(4040)$ mass energies. 
We assume a conservative estimate for the charmonium production 
cross section above the open-charm threshold, on the order 
of 3~nb for $D^+ D^-$ and 0.9~nb for 
$D^{*+} D^{*-}$ production. The background is simulated using the DPM 
to produce inelastic reaction in $\pbarp$ annihilations. 
A background suppression of the order of $10^7$ is achieved with the 
previous selection. 
A detailed study of specific background reactions is also 
performed. In particular, nonresonant production of 
$K^+ K^- 2\pi^+ 2\pi^-$ has a cross section which is $10^6$ 
times larger than the $D^+ D^-$ signal. 
A cut on the longitudinal and transverse 
momentum of the $D^\pm$ can reduce the background 
by a factor $\sim$26, and the remaining events leave a 
nonpeaking background in the loose mass region defined in the 
preselection. The reconstructed decay vertex location will 
further improve the background rejection, reaching a 
signal-to-background ratio near unity with an 
efficiency for signal events of $\approx 8\%$. 
Under these assumptions, a conservative estimate of 
the number of reconstructed events per year of \PANDA 
operation is of the order of $10^4$ and $10^3$ 
for $D^+ D^-$ and $D^{*+} D^{*-}$, respectively. 

\begin{figure}[t]
   \begin{center}
      \includegraphics[width=\figwid]{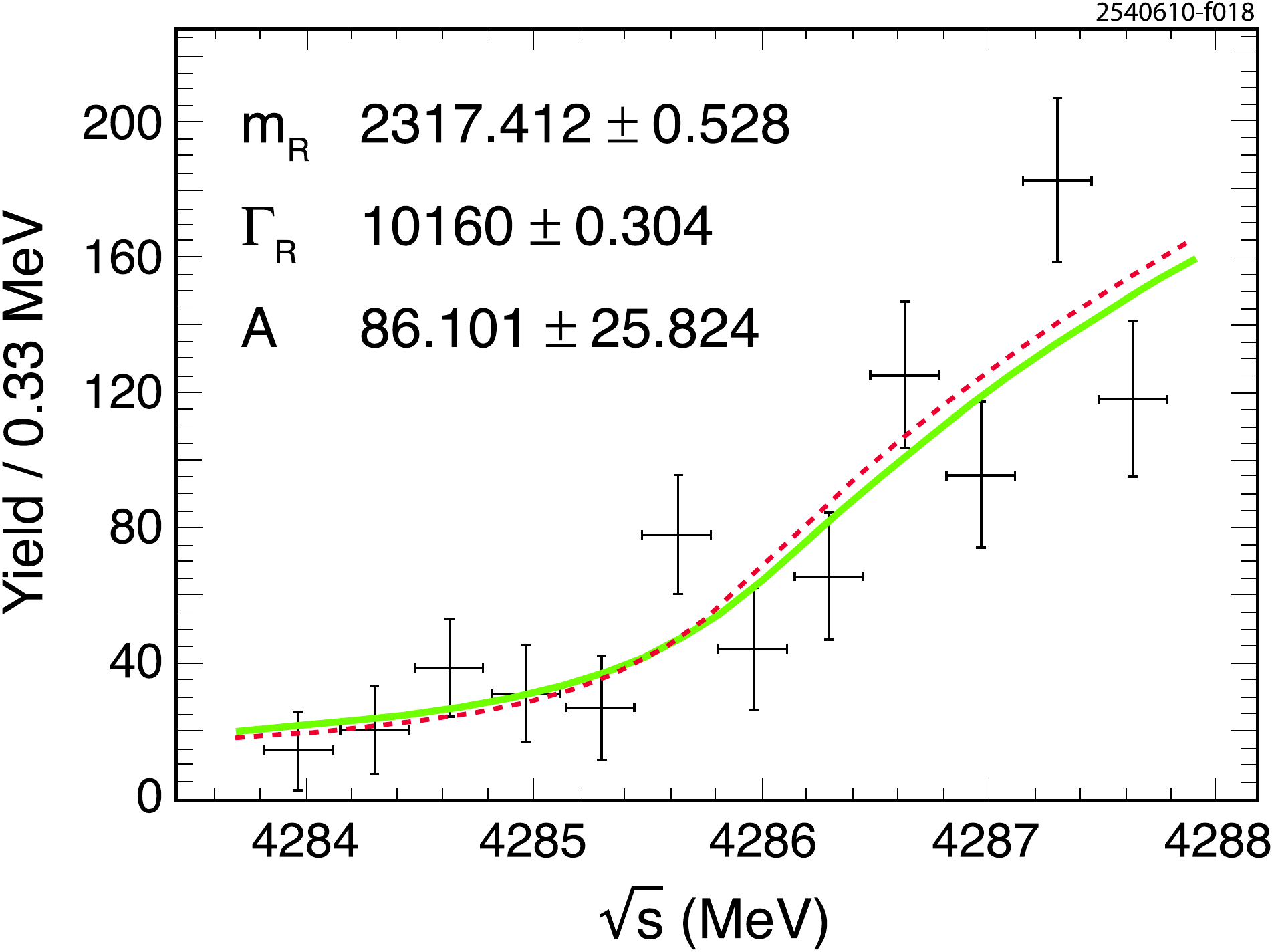}
      \caption{Fit of the simulated excitation function for 
               near-threshold $\pbarp \to D_s^\pm D_{s0}^*(2317)^\mp$. 
               The {\it dashed line} corresponds to the simulated function and 
               the {\it solid line} is the reconstructed curve }
      \label{fig:Future_thr_scan}
   \end{center}
\end{figure}

 Performing fine energy scans, \PANDA will be able to 
observe the energy dependence of a cross section 
in proximity to a threshold. 
Here we report the result obtained in the simulation 
of $\pbarp \to D_s^\pm D_{s0}^*(2317)^\mp$, 
to test the sensitivity to resonance parameters 
measurements with this technique. 
The assumptions used in this study are:
\begin{itemize}
\item a 12-point scan in a $4\mev$-wide region;
\item 14 days of data with a total integrated luminosity of 9~pb$^{-1}$/day;
\item a signal-to-background ratio of 1:3;
\item 1\mevcc\ total width for the $D_{s0}^*(2317)$.
\end{itemize}
The results of a fit to
the scan simulation are presented in \Fig{fig:Future_thr_scan}, 
where the mass and the width of the $D_{s0}^*(2317)$ are 
free parameters. 
The study yields the results:
\begin{eqnarray}
m & = & 2317.41 \pm 0.53\mevcc, \nonumber \\
\Gamma & = & 1.16 \pm 0.30\mevcc,
\end{eqnarray}
to be compared with the input values of the simulation: 
$m = 2317.30\mevcc$ and $\Gamma = 1.00\mevcc$, 
demonstrating that such a scan would yield accurate
values of mass and width
with the precisions shown.

 In conventional charmonia, the quantum numbers are derived 
directly from the excitation of the $\cbarc$ pair. 
The glue tube adds degrees of freedom that manifest themselves in 
unconventional quantum numbers; in the simplest $\cbarc g$ scenario 
this corresponds to the addition of a single gluon quantum number 
($J^{PC} = 1^-$ or $1^+$ for color-electric or color-magnetic 
excitation). 
This would result in charmonium hybrids with non-exotic and exotic 
quantum numbers which are expected in the 3-5\gevcc\ mass region. 
Here we will sketch out the strategy for hybrids studies in 
\PANDA, with more details available in~\cite{Lutz:2009ff}. 
Formation experiments would generate non-exotic charmonium 
hybrids with high cross sections while production 
experiments would yield a hybrid together with another particle 
like $\pi$ or $\eta$. 
In \PANDA both processes are possible; the strategy would be to start
searching for hybrids in production processes, fixing the $\pbarp$ 
center-of-mass energy at the 
highest possible value ($\sqrt{s} \simeq 5.5\gev$) 
and studying all the production channels. 
Then hybrid formation could be studied through energy scans 
over the regions where possible signals have been observed in 
production measurements.

Aside from the benchmark channels used for the detection of 
conventional charmonia, hybrids can be identified through reactions like:
\begin{itemize}
\item $\pbarp \to \widetilde{\eta}_{c0,1,2} \eta \to \chicOne \dipiz \eta$;
\item $\pbarp \to \widetilde{h}_{c0,1,2} \eta \to \chicOne \dipiz \eta$;
\item $\pbarp \to \widetilde{\psi} \eta \to \jpsi \omega \eta$;
\item $\pbarp \to [\widetilde{\eta}_{c0,1,2},\widetilde{h}_{c0,1,2},\widetilde{\chicOne}] \eta \to D D^* \eta$;
\end{itemize}
namely final states with charmonia accompanied by light hadrons 
or final states with a $DD^*$ pair.
As a case study, we will present the results 
obtained for the benchmark channel 
$\pbarp \to \widetilde{\eta}_{c1} \eta \to \chicOne \dipiz \eta$. 
It can be assumed that the 
$\pbarp \to \widetilde{\eta}_{c1} \eta$ production cross section 
is of the same order of $\pbarp \to \psip \eta$, 
which is estimated to be $33 \pm 8$~pb \cite{Lundborg:2005am}. 
As possible sources of background, several reactions with 
similar topology have been considered:
\begin{itemize}
\item $\pbarp \to  \chi_{c0}(1P)\, \dipiz \eta$;
\item $\pbarp \to  \chicOne(1P)\, \piz \eta \eta$;
\item $\pbarp \to  \chi_{c0}(1P)\, \dipiz \piz \eta$;
\item $\pbarp \to  \jpsi\, \dipiz \piz \eta$.
\end{itemize}
A simple analysis is carried out. 
Two photons are accepted as $\piz$ or $\eta$ candidates 
if their invariant mass is in the range $115-150\mevcc$ or 
$470-610\mevcc$, respectively. 
The $\chicOne$ is formed adding a radiative photon 
to a $\jpsi$ candidate, with total invariant mass within 
3.3-3.7\gevcc. 
From these, $\chicOne \dipiz \eta$ candidates are 
created and kinematically fit to the original beam 
energy-momentum, with an additional constraint for the $\jpsi$ 
mass. Additional cuts on the kinematic fit CL and on the 
invariant masses of the intermediate decay products are applied, 
obtaining a total efficiency around 7\% for this channel. 
The $\widetilde{\eta}_{c1}$ peak reconstructed in this way 
has a FWHM of $30\mevcc$.
The background suppression is estimated applying the 
same analysis to background events. The results are 
summarized in \Tab{tab:Future_1}. 

\begin{table}[b]
   \caption{Background suppression ($\eta$) for the 
            individual background reactions }
   \label{tab:Future_1} 
   \setlength{\tabcolsep}{1.88pc}
   \begin{tabular}{lc}
   \hline\hline
   \rule[10pt]{-1mm}{0mm}
Background channel & Suppression $(10^3)$  \\[0.7mm]
   \hline
   \rule[10pt]{-1mm}{0mm}
$\chi_{c0}(1P)\, \dipiz \eta$ & $5.3$ \\[0.7mm]
$\chi_{c1}(1P)\, \piz \eta \eta$ & $26$  \\[0.7mm]
$\chi_{c0}(1P)\, \dipiz \piz \eta$ & $> 80$  \\[0.7mm]
$\jpsi\, \dipiz \piz \eta$ & $10$  \\[0.7mm]
   \hline\hline
   \end{tabular}
\end{table}

 The $\jpsi\, N$ dissociation cross section is as yet
experimentally unknown, except for indirect information deduced
from high-energy $\jpsi$ production from nuclear targets.
Apart from being a quantity of its own interest, 
this cross section is closely related to the attempt
of identifying quark-gluon plasma (QGP) formation in 
ultra-relativistic nucleus-nucleus collisions: the interpretation
of the $\jpsi$ suppression observed at the CERN 
SPS~\cite{Alessandro:2004ap,Arnaldi:2005yk,Arnaldi:2007zz} as a signal
for QGP formation relies on the knowledge of the "normal" suppression
effect due to $\jpsi$ dissociation in a hadronic environment.
Nuclear $\jpsi$ absorption can only be deduced from models,
since the available data do not cover the kinematic regime
relevant for the interpretation of the SPS results.
In antiproton-nucleus collisions the $\jpsi\, N$ dissociation cross 
section can be determined for momenta around $4\gevc$ with very
little model dependence. 
The determination of the $\jpsi\,N$ dissociation cross section is
in principle straightforward: the $\jpsi$ production cross section is
measured for different target nuclei of mass number ranging from 
light (d) to heavy (Xe or Au) by scanning the $\pbar$ momentum across
the $\jpsi$ yield profile whose width is essentially given by the
known internal target-nucleon momentum distribution. The $\jpsi$
is identified by its decay to \epem\ or \dimu.
The attenuation of the $\jpsi$ yield per effective target proton is a
direct measure of the $\jpsi\, N$ dissociation cross section, which can
be deduced by a Glauber-type analysis.
These studies may be extended to higher charmonium
states like the $\psip$, which would allow determination of the
cross section for the inelastic process 
$\psip\, N \rightarrow \jpsi\, N$, which is also relevant for the
interpretation of the ultrarelativistic heavy-ion data. 
The benchmark channel studied in this context is the reaction:
\begin{equation}
\pbar\,\, {}^{40}\textrm{Ca} \rightarrow \jpsi X \rightarrow \epem X\,.
\end{equation}
The cross section for this process is estimated to be nine orders
of magnitude smaller than the total antiproton-nucleus cross section.
The results of the simulations show that it is possible to identify
the channel of interest with good efficiency and acceptable
signal-to-background ratio~\cite{Lutz:2009ff}.

\subsection{CBM at FAIR}

The Compressed Baryonic Matter (CBM) experiment will be one of the
major scientific activities at FAIR~\cite{FAIR:2006zz}.
The goal of the
CBM research program is to explore the QCD phase diagram in the
region of high baryon densities using high-energy nucleus-nucleus
collisions. This includes a study of the equation-of-state of
nuclear matter at high densities, and a search for the
deconfinement and chiral phase transitions. The CBM research
program comprises a comprehensive scan of observables, beam
energies, and collision systems. The observables include low mass
dileptons, charmonia and open charm, but also collective flow
of rare and bulk particles, correlations and fluctuations.

Particles with open and hidden charm are expected to provide
valuable information about the conditions inside the dense
fireball. For example, the excitation function of the charm
particle ratios such as the $\psip/(\jpsi)$ ratio and the
$(\jpsi)/D$ ratio may vary when passing the deconfinement phase
transition. In addition, the initial pressure of the partonic
phase influences the elliptic flow of charmonium. The transport
properties of open charm mesons in dense matter, which depend on
the interaction with the medium and hence on the structure of the
medium, can be studied via the yield, the elliptic flow, and the
momentum distributions of charmed particles. In a baryon-dominated
medium, these observables are expected to differ for $D$ and $\bar
D$ mesons.

The experimental goal is to measure these rare probes with
unprecedented precision. 
In order to compensate for the low yields, the measurements will be
performed at exceptionally high reaction rates (up to $10\,{\rm MHz}$ for
certain observables). These conditions require the development of
ultrafast and extremely radiation-hard detectors and electronics. A
particular challenge for the detectors, the front-end electronics,
and the data-acquisition system is the  online selection of displaced
vertices with the extraordinarily high speed and precision 
needed for open-charm measurements.

\begin{figure}[b]
   \begin{center}
      \includegraphics[width=\figwid]{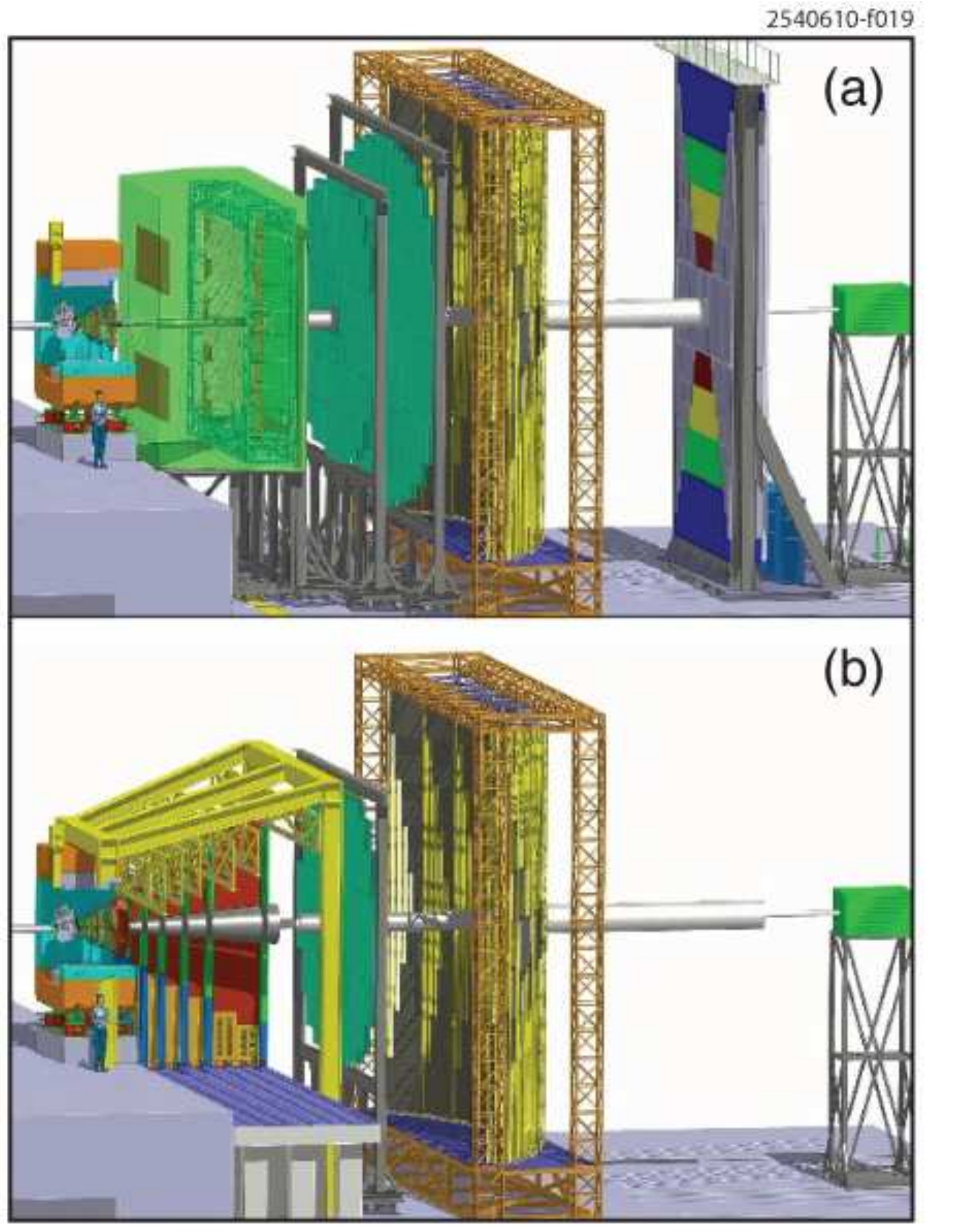}
      \caption{The CBM experimental facility with 
               (a) the RICH and TRD electron detectors, or
               (b) the muon detection system }
      \label{fig:Future_setup}
   \end{center}
\end{figure}

A schematic view of the proposed CBM experimental facility is
shown in \Fig{fig:Future_setup}. The core of the setup is a Silicon
Tracking and Vertexing System located inside a large aperture
dipole magnet. The vertex detector consists of 2 stations of
Monolithic Active Pixel Sensors, and the tracker comprises 8
stations of double-sided microstrip sensors. Particle
identification will be performed using the momentum information
from the silicon tracker and the time-of-flight measured with a
large area Resistive Plate Chamber (RPC) wall.
\Figure{fig:Future_setup}(a) depicts the setup with the Ring Imaging
Cherenkov (RICH) detector for the identification of electrons from
low-mass vector-meson decays. The Transition Radiation Detector
(TRD) will provide charged particle tracking and the
identification of high-energy electrons and positrons. The
Electromagnetic Calorimeter (ECAL) will be used for the
identification of electrons and photons. The muon detection/hadron
absorber system is shown in \Fig{fig:Future_setup}(b).
It consists of 5 double- or triple-stations of highly-granulated
gaseous micropattern chambers, \eg Gas-Electron
Multiplier (GEM) detectors, sandwiched by iron plates with a total
thickness equivalent to 13 absorption lengths. The status of
detector R\&D and recent results of detailed simulations are
documented in~\cite{Friese:2008zz}. The CBM collaboration consists of more
than 450 scientists from 55 institutions and 14 countries.

\subsection{Tau-charm factory in Novosibirsk}

A tau-charm factory can address various issues concerning $\tau$-leptons, 
charmonia, open-charm particles, and light-quark spectroscopy 
in a unique manner. Indeed, the $B$-factories have inadequate
sensitivity for some of these physics topics, leaving
a tau-charm factory as the only practical avenue for substantial progress.
A next-generation tau-charm factory is now under 
consideration in Novosibirsk. A novel approach of the Crab Waist 
collision scheme~\cite{Raimondi:2006zz,Raimondi:2007vi} 
allows reaching luminosity of
$(1\text{-}2) \times 10^{35}\,{\rm cm}^{-2} {\rm s}^{-1}$. 
Suggested priorities include
\begin{itemize}
\item $\DDbar$ mixing;
\item search for CP violation in charm decays;
\item study of rare and forbidden decays of open charm mesons;
\item high-precision study of regular charmonia and charmonium-like states;
\item tests of the Standard Model in $\tau$-lepton decays;
\item search for lepton flavor violation;
\item search for CP/T violation in $\tau$-lepton decays;
\item extensive study of light-quark ($u,~d,~s$) states between 1 and
  3\gev\ using ISR;
\item production of polarized antinucleons.
\end{itemize}

This experimental program would be carried out at a facility with 
the following basic features~\cite{Levichev:2008zz}:
\begin{itemize}
\item collision energy varying from 2-5\gev;
\item luminosity of $\sim$$5 \times 10^{34}\,{\rm cm}^{-2}{\rm s}^{-1}$ 
at 2\gev\ and more than $10^{35}\,{\rm cm}^{-2}{\rm s}^{-1}$ 
at the $\tau$-production threshold;
\item a longitudinally polarized electron beam at the interaction 
point (IP) extending the experimental possibilities of the facility;
\item extensive use of superconducting wigglers allowing 
control of damping parameters and tuning for optimal luminosity 
in the whole energy range;
\item \epem\ center-of-mass energy calibration with relative accuracy of 
$\approx 5 \times 10^{-4}$, achieved with the Compton backscattering technique.
\end{itemize}
The planned factory has separate rings for electrons and positrons 
and one interaction region. Each ring features 
$\sim$$800\,{\rm m}$ circumference and a racetrack shape, 
with two arcs and two long ($\sim$$100\,{\rm m}$) 
straight sections to accommodate the injection and radio-frequency 
(RF) equipment 
for the machine and the interaction region for the experiment. 
The injection facility includes a full-energy ($2.5\gev$) linear
accelerator equipped with the polarized-electron source and 
a positron complex with a $500\mev$ linac, converter, and accumulating 
ring. The injection facility operates at $50\,{\rm Hz}$ and can produce 
$\ge 10^{11}$ positrons per second.
High luminosity is provided by the Crab Waist collision approach, 
which assumes the beam intersection at large Piwinski angle 
and local focusing of the beams at the IP by means of two 
Crab Sextupoles with properly matched betatron phase advance in
between.  Local focusing rotates the vertical waist at IP 
according to the horizontal displacement of each individual particle, 
decouples the betatron oscillations, and therefore effectively reduces 
the beam-beam coupling betatron resonances.

\begin{figure*}
   \begin{center}
      \includegraphics[width=6.5in]{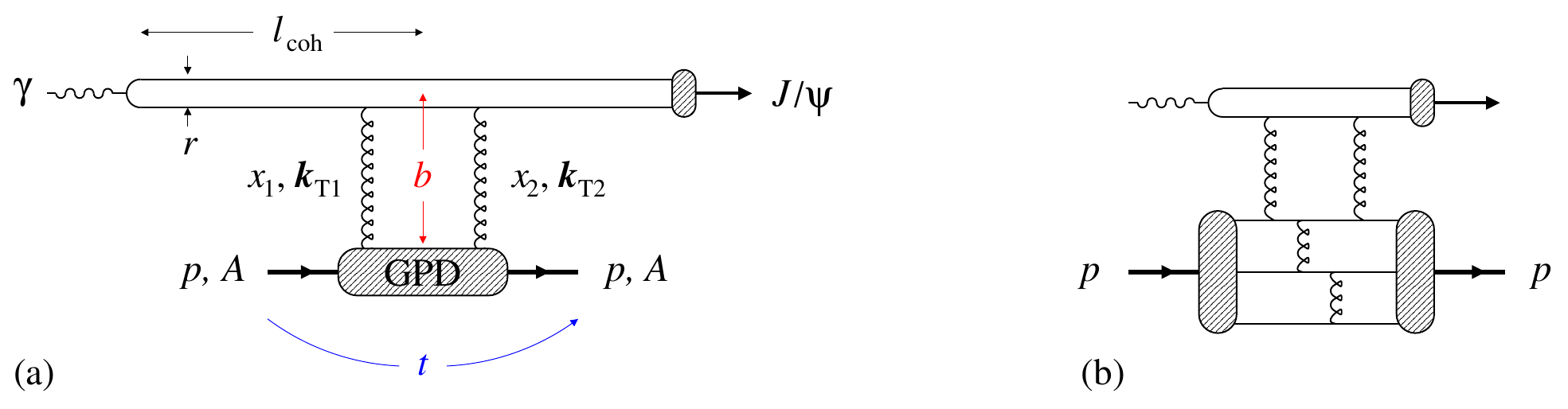}
      \caption{Dynamical mechanisms of \jpsi\ photoproduction.
               (a) Two-gluon exchange mechanism at high and 
                   intermediate energies, $W - W_{\rm th} > \textrm{few
               GeV}$, based on QCD factorization for hard exclusive 
               meson production. The space-time evolution in the target 
               rest frame corresponds to the dipole picture of 
               high-energy scattering~\cite{Frankfurt:1997fj,Caldwell:2010zza},
               where $l_{\rm coh}$ is the coherence length, $r$ the
               transverse size of the $c\bar c$ dipole and $b$ its
               impact parameter with the target. Here $x_{1,2}$ and 
               $\bm{k}_{T1,2}$ denote the longitudinal momentum
               fractions and transverse momenta of the exchanged gluons. 
               The invariant momentum transfer to the target proton or 
               nucleus is small and of the order of the inverse target 
               size, $|t| \sim |\bm{k}_{T1} - \bm{k}_{T2}|^2 \sim R_{\rm target}^{-2}$.
               (b) Coherent multi--gluon exchange proposed for 
               near-threshold production on the proton, analogous to the 
               hard scattering mechanism for high-$t$ elastic form
               factors. Near threshold the minimum value of the
               invariant momentum transfer $|t_{\rm min}|$ is large }
      \label{fig:Fut_jlab-jpsi-photopr} 
   \end{center}
\end{figure*}

\begin{figure}[t]
   \begin{center}
      \includegraphics[width=\figwid]{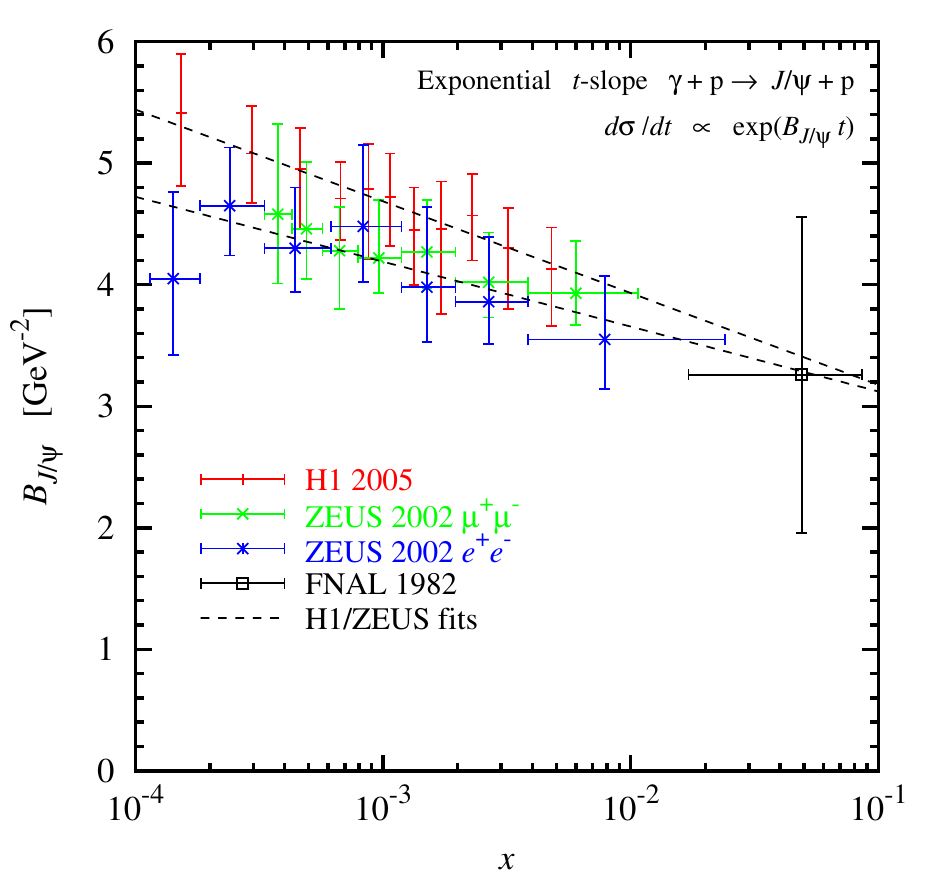}
      \caption{The exponential $t$-slope, $B_{\jpsi}$, 
               of the differential cross section of 
               exclusive \jpsi\ photoproduction measured 
               in the FNAL E401/E458~\cite{Binkley:1981kv}, 
               HERA H1~\cite{Aktas:2005xu}, and 
               ZEUS~\cite{Chekanov:2002xi} experiments, 
               as a function of 
               $x = m_{\jpsi}^2/W^2$~\cite{Frankfurt:2010ea}. 
               (In the H1 and ZEUS results, the quoted statistical 
               and systematic uncertainties were added
               linearly.) The {\it dashed lines} represent the 
               published two-dimensional fits to the H1 and ZEUS data. 
               The average squared transverse radius of gluons with 
               momentum fraction $x$ in the nucleon can be inferred 
               from the measured slope as 
               $\langle b^2 \rangle_{g} =2 (B_{\jpsi} - \Delta B)$, 
               where $\Delta B \approx 0.3-0.6 \, \textrm{GeV}^{-2}$ 
               accounts for the finite transverse size of the $c\bar c$ 
               pair in the production 
               amplitude~\cite{Frankfurt:1997fj,Caldwell:2010zza}. 
               The data show that the nucleon's gluonic transverse 
               radius at $x \sim 10^{-1}$ is smaller than the transverse 
               charge (Dirac) radius and increases slowly toward 
               small $x$; see~\cite{Frankfurt:2010ea,Caldwell:2010zza} for details }
      \label{fig:Future_kow}
   \end{center}
\end{figure}

\subsection{Charmonium photoproduction facilities}
\label{sec:Fut_photoprod}

The mechanism of charmonia photoproduction and their interactions with 
hadrons and nuclei have been the subject of much interest since their 
discovery (see \Sec{prod_sec:ep}). 
Generally, because of the small size of these heavy mesons 
on the hadronic scale of $\sim 1$~fm, it is expected that one can 
apply QCD to describe their interactions with hadronic matter. 
Heavy quarkonium production thus probes the local color fields
in the target and can reveal properties such as their response to 
momentum transfer and their spatial distribution, which are of
fundamental interest for understanding nucleon structure in QCD.
While this interpretation is valid at all energies, the details
(what mechanism produces the relevant color fields, which configurations
in the target are their main source) vary considerably between 
high energies and the near-threshold region, calling for detailed
experimental and theoretical study of this fascinating landscape.

\subsubsection{\jpsi\ photoproduction at high energy}
\label{sec:Fut_jpsiway}

The mechanism of exclusive \jpsi\ photoproduction is well-understood 
at high energies ($W > 10\gev$) and $|t| < 1\gevcc$, 
where the coherence length is large compared to the nucleon size, 
$l_{\rm coh} \gg 1 \, {\rm fm}$:
the process takes the form of the scattering of a small-size color 
dipole off the target (\Fig{fig:Fut_jlab-jpsi-photopr}(a)). 
The leading interaction in the small-size
expansion is via two-gluon exchange with the target. The nucleon 
structure probed in this case is the gluon generalized parton 
distribution (or GPD), which describes the two-gluon form factor 
of the target and is normalized to the usual gluon density in the 
zero momentum-transfer limit.
In the process shown in \Fig{fig:Fut_jlab-jpsi-photopr}(a), 
the transverse momenta of the exchanged gluons are large, 
$|\bm{k}_{T1, 2}| \sim m_{\jpsi}$, but their
difference can be small, resulting in a small invariant
momentum transfer to the target
$|t| \sim |\bm{k}_{T1} - \bm{k}_{T2}|^2$. Hence the reaction
can leave a proton or a nuclear target in its ground state or
a slightly excited state. 
Experiments in exclusive \jpsi\
photo- and electro-production at 
HERA~\cite{Aktas:2005xu,Chekanov:2004mw}
have confirmed this picture through detailed measurements of the 
$Q^2$-independence of $t$-slopes in electroproduction, 
energy dependence of the cross 
section, and comparison with other exclusive vector meson channels 
(universality of the gluon GPD); see \cite{Frankfurt:2005mc} 
for a review.
They have also measured the $t$-slope of the differential cross section 
and its change with energy, which allows one to infer the average 
transverse radius of gluons in the nucleon and its change with $x$ 
(\Fig{fig:Future_kow}). This 
information represents an essential 
input to small-$x$ physics (initial condition of evolution
equations) and the phenomenology of high-energy $pp$ collisions
with hard processes~\cite{Frankfurt:2006jp}.

Of particular interest is elastic or 
quasi-elastic scattering of a charmed dipole on nuclei
(\Fig{fig:Fut_jlab-jpsi-photopr}(a)), with 
a subsequent transformation of the $c\bar c$ into a $\jpsi$.  
The momentum transfer is measurable in this reaction because 
it is the difference between the transverse momentum of 
the incoming virtual photon and the final meson. 
Caldwell and Kowalski~\cite{Caldwell:2009ke}
propose investigation of the properties of nuclear matter 
by measuring the elastic scattering of $\jpsi$ on nuclei with 
high precision. 
The $\jpsi$ mesons are produced from photons 
emitted in high-energy electron-proton or electron-nucleus scattering 
in the low-$x$ region. 
Over the next few years, some relevant data should be available from 
RHIC, where STAR is collecting data on $J/\psi$ photoproduction.  
Based on the recent RHIC Au+Au luminosity, a sample of order 100 events 
might be expected.
Such measurements could be performed at the future 
ENC (Electron-Nucleon Collider, GSI),
Electron-Ion Collider (EIC, Brookhaven National Lab, or
Jefferson Lab (JLab)), or LHeC (Large Hadron-electron  
Collider, CERN) facilities.    
The advantage of $\jpsi$ photoproduction compared to the  
electroproduction of light vector mesons
is its high cross section and small 
dipole size, even at $Q^2 = 0$. 
In addition, the momenta of the decay products of 
$\jpsi\to \dimu$ or $\epem$ can be precisely measured.         
The smallness of the dipole in low-$x$ reactions assures that 
the interaction is mediated by gluon exchange only. 
Thus the deflection of the $\jpsi$ directly measures the 
intensity and the spatial distribution of the nuclear gluon field.     
  
\begin{figure}[t]
   \begin{center}
      \includegraphics[width=\figwid]{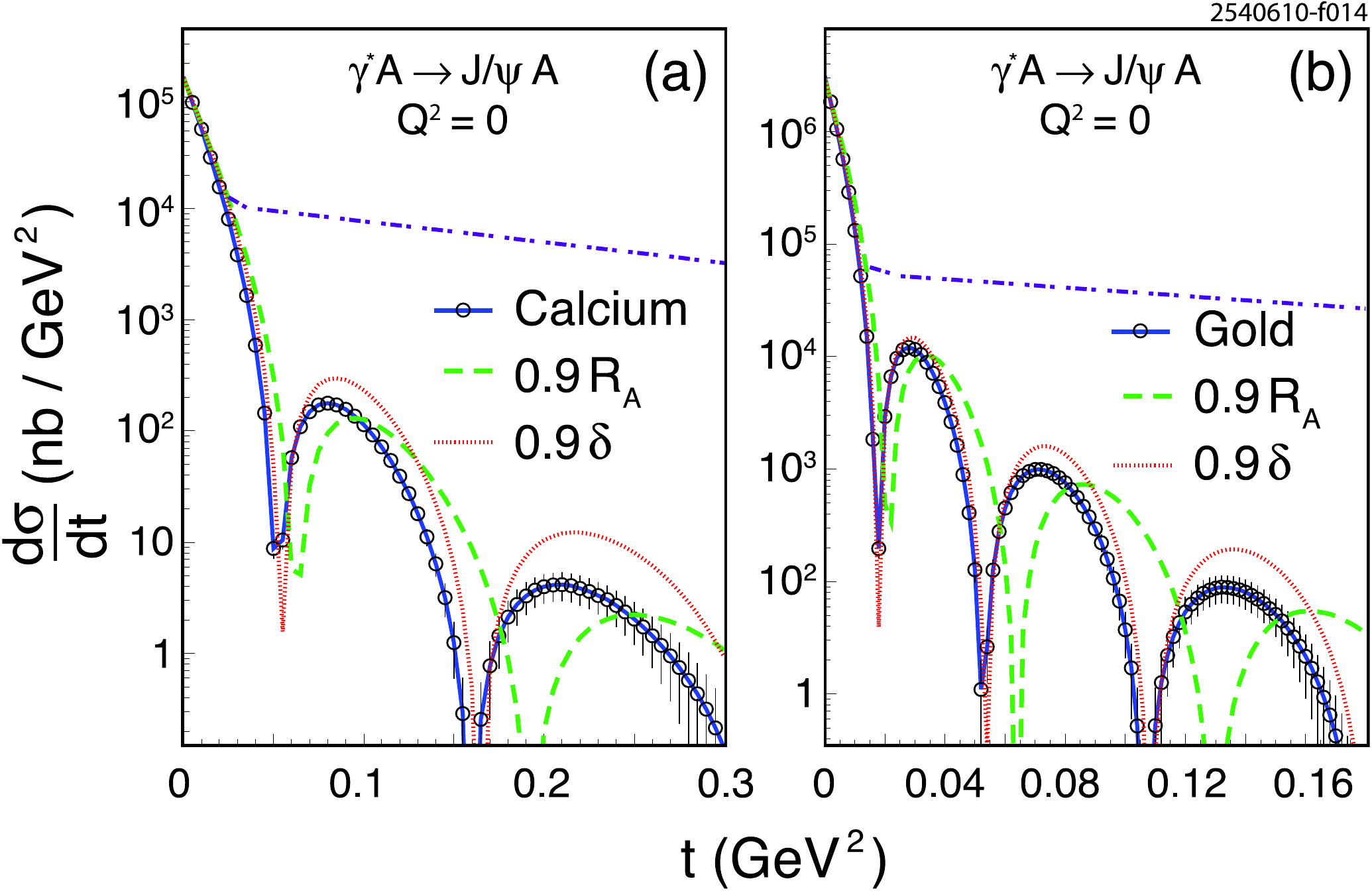}
      \caption{The prediction~\cite{Caldwell:2009ke} 
               of the dipole model for the $t$-distribution 
               of coherent \jpsi\ photoproduction 
               on (a) calcium and (b) gold nuclei, assuming 
               that the single nucleon distribution 
               can be identified with the 
               Woods-Saxon distribution, parametrized
               by the nuclear radius $R_A$ and skin depth $\delta$ 
               as in \Eq{eqn:Fut_WS}. 
               The simulated measurements with nominal $R_A$ and $\delta$
               {\it (open circles and solid curve)} 
               have statistical error bars based 
               on an assumed collected sample of $10^6$ events. 
               Somewhat smaller values of $R_A$ or $\delta$
               (by 10\% or 20\%, respectively)
               result in the {\it dashed} and {\it dotted}
               curves, respectively.
               The {\it dot-dashed curve} shows the 
               sum of the coherent and incoherent processes 
               in the case of no correlations }
      \label{fig:Future_dsca}
   \end{center}
\end{figure}

  The measurement of $\jpsi$ scattering on nuclei could become an 
important source of information on nuclear structure and 
high-density QCD. In the absence of nuclear shadowing, 
the interaction of  a dipole with a nucleus 
can be viewed as  a sum of dipole scatterings of the nucleons 
forming the nucleus. The size of the $c\bar c$ dipole in elastic  
$\jpsi$ scattering is around 0.15~fm, \ie it is much smaller 
than the nucleon radius. It is therefore possible that 
dipoles interact with smaller objects than nucleons; \eg
with constituent quarks or hot spots. 
The conventional assumption is that the nucleus 
consists of nucleons and that dipoles scatter on an ensemble 
of nucleons according to the Woods-Saxon~\cite{WoodsSaxon:1969} distribution,
\beqa
\rho_{\rm WS}(r) &=& \frac{N}{1 + \exp{\left(\dfrac{r-R_A}{\delta}\right)}}\,,\non\\
R_A  &\equiv& 1.12\,A^{1\over3} - 0.86\,A^{-{1\over3}}~{\rm fm}\,,
\label{eqn:Fut_WS}
\eeqa
where $R_A$ is the nuclear radius for atomic number $A$,
$\delta=0.54$~fm is the skin depth, and $N$ is chosen
to normalize $\int d^3\vec{r}\,\rho_{\rm WS}(r)\,=\,1$.
Under the foregoing assumption (or slight variations thereon),
the dipole model predicts the coherent and incoherent
nuclear cross sections 
shown in \Fig{fig:Future_dsca}.
Deviations from the predicted $|t|$-distribution
reveal the effects of nuclear shadowing, which depend on the
effective thickness of the target and thus change with the dipole
impact parameter, $b$. In the coherent process,  
the nucleus remains in its ground state. In the incoherent process,  
the nucleus gets excited and frequently breaks into nucleons or 
nucleonic fragments. 
Experimentally we expect to be able to distinguish cases 
where the nucleus remains intact and cases where the nucleus 
breaks up.  In the nuclear breakup process, there are several 
free neutrons and  protons in the final state, as well as other 
fragments. The number of free nucleons could depend on the 
value of the momentum  transfer.   The free nucleons and fragments 
have high momenta and different charge-to-mass ratios than the 
nuclear beam and should therefore be measurable in specialized detectors.  

The transverse momenta of the $\jpsi$ can be determined, 
in a TPC detector with $2\,{\rm m}$ radius  and $3.5\,{\rm T}$ magnetic field, 
with a precision of $O(1)\mev$. The momenta of the breakup protons 
can be precisely measured in the forward detector.  
Therefore a measurement of  $t$-distributions together with 
a measurement of nuclear debris could become a source of 
invaluable information about  the inner structure of gluonic 
fields of nuclei~\cite{Caldwell:2009ke}.

\subsubsection{\jpsi\ photoproduction at low energy}

Measurements of exclusive \jpsi\ photoproduction
at lower energies were performed in the Fermilab broadband beam 
experiment~\cite{Binkley:1981kv}, which detected the recoiling proton, 
as well as several other 
experiments~\cite{Gittelman:1975ix,Camerini:1975cy}. 
The few existing data suggest
that the two-gluon exchange mechanism of \Fig{fig:Fut_jlab-jpsi-photopr}(a)
continues to work in this region, 
and the gluonic size of the nucleon observed in these experiments 
consistently extrapolates to the HERA values (\Fig{fig:Future_kow}) \cite{Frankfurt:2002ka}.
However, no precise
differential measurements are available for detailed tests of the
reaction mechanism. New data in this region are expected from the 
COMPASS experiment at CERN and a future Electron-Ion Collider (EIC).

Charmonium photoproduction near threshold will be studied in 
experiments at JLab. The Continuous Electron Beam 
Accelerator Facility (CEBAF) at JLab delivers
500~MHz electron beams with energies up to 6\gev\ to 
three experimental halls. The ongoing upgrade will
increase the maximum energy to 12\gev. The first beam delivery
is planned for 2014. Halls~A and C are equipped with small-acceptance,
high-resolution
spectrometers and able to receive beam currents up to 100~$\mu$A.
Hall~B is equipped with a large acceptance toroidal spectrometer (CLAS)
and is able to receive up to 0.1~$\mu$A. A new Hall~D, being
built as a part of the upgrade, will use a tagged-photon beam
with a 12\gev\ endpoint and intensity up to 100~MHz/GeV.

In exclusive \jpsi\ photoproduction near threshold 
($E_{\gamma, {\rm thr}} = 8.21\gev$ for the free nucleon),
to be studied with JLab 12\gev, the reaction mechanism is expected 
to change in several important aspects. First, the minimum invariant
momentum transfer to the nucleon becomes large:
$|t_{\rm min}| = 2.23$\gev$^2$ at threshold, and
$|t_{\rm min}| = 1.3\,$-0.3\gev$^2$ at $E_{\gamma} = 8.5\,$-12\gev.
This requires
more exceptional high-momentum configurations in the target to 
bring about an exclusive transition, similar to elastic $eN$ 
scattering at large $t$. It is expected that, in this situation,
color correlations in the wave functions play an essential role,
suggesting a new avenue for their experimental study.
Second, while at high energies the production amplitude is mostly 
imaginary (absorptive), near threshold the real part plays an 
essential role, rendering the partonic interpretation of the
production process more complex. 

Two possible scenarios for near-threshold exclusive production on
the nucleon have been proposed. One scenario assumes that the two-gluon 
exchange mechanism of \Fig{fig:Fut_jlab-jpsi-photopr}(a) 
continues to apply, justified by the small size of 
the \jpsi, and extends this description to the 
near-threshold region~\cite{Frankfurt:2002ka}.
Here the challenge lies in modeling the gluon GPD in the ``extreme''
near-threshold region, characterized by large $t_{\rm min}$
and large ``skewness'' (difference in momentum gluon fractions 
$x_1 \neq x_2$). Some support for this picture comes from the fact
that the exclusive $\phi$ electroproduction data at JLab with 6 GeV
beam energy~\cite{Lukashin:2001sh,Santoro:2008ai}
are well described by a dynamical model based on
gluon GPDs~\cite{Goloskokov:2006hr}. 
The other scenario is based on analogy with the
hard scattering mechanism for high-$t$ elastic form factors.
It assumes that the production process happens predominantly
in the valence (3-quark) configuration of the nucleon and
that the momentum transfer is balanced via hard-gluon 
exchange~\cite{Brodsky:2000zc} as 
illustrated in \Fig{fig:Fut_jlab-jpsi-photopr}(b).
The two basic pictures make different predictions for the energy
dependence of the cross section and $t$-slope near threshold,
and can be tested with the expected JLab 12 GeV data. 
The quantitative implementation 
of the above scenarios, and a possible unified description, 
are the subjects of ongoing theoretical research. Independently of
the details, the expected JLab data will greatly advance our knowledge 
of color correlations and the gluonic response of the nucleon 
at low energies.

The cross section for exclusive \jpsi\ photoproduction at 11\gev\ 
is expected to be 0.2-0.5~nb, rapidly falling toward the threshold. 
The JLab experiment plans to map out the \jpsi\ differential
cross section in the region from 9 to 12\gev.

Another objective of \jpsi\ production experiments is
to study the interaction of the produced system with nuclear
matter at low energies.
The small coherence and formation lengths allow
extraction of the \jpsi-nucleon cross sections from the $A$-dependence of 
the \jpsi\ photoproduction cross section with minimal corrections to
the color-transparency effects. The only such experiment done at sufficiently
low energies~\cite{Anderson:1976hi} obtained $3.5\pm0.9$~mb; 
the signal was extracted from a 
single-muon transverse momentum spectrum
and the background level was not well understood. A new experiment
at JLab can reduce the statistical and systematic errors by a factor 
of three.

A proposal to study \jpsi\ photoproduction close to threshold in Hall~C
has been conditionally approved by the JLab Program Advisory Committee. 
The lepton
decay modes of \jpsi\ will be detected. In spite of a small acceptance of 
$\sim3\times10^{-4}$ to \jpsi\ decay products, the expected rate 
of detected \jpsi\ is 150-200 per hour in a 2.2\% radiation-length-thick liquid hydrogen
target. The effective photon flux of the $50\,\mu A$ electron beam will
be increased due to a radiator in front of the target 
which has a thickness of 7\% of a radiation length.
The high resolution of 
the Hall C spectrometers will provide strong background suppression.

\subsection{Proposed ${\bar p}p$ project at Fermilab}
\label{sec:Fut_Kaplan}

A uniquely capable and cost-effective multipurpose experiment could be 
mounted by adding a magnetic spectrometer to the existing Fermilab 
E760 lead-glass calorimeter~\cite{Bartoszek:1990ex} using an available 
BESS solenoid~\cite{Moiseev:1996xd}, fine-pitch scintillating fibers (SciFi), 
the \DZero\ SciFi readout system~\cite{Abazov:2005pn}, and hadron ID 
via fast timing~\cite{psec}. If the relevant cross sections are as 
large as expected, this apparatus could produce world-leading 
measurements of $X(3872)$ properties, along with those of other 
charmonium and nearby states, as is now 
proposed\footnote{The experiment would also 
address nonquarkonium  topics, such as charm mixing and {\em CP} 
violation.} to Fermilab~\cite{Asner:1900zz,New-pbar}. 
The Fermilab Antiproton Accumulator's 8\gev\ maximum 
kinetic energy and ability to decelerate down to $\approx3.5$\gev\ 
suit it well for studies in this mass region. If approved, the experiment 
could start about a year after completion of the Tevatron Collider run.

Antiproton Accumulator experiments E760 and E835 made the world's most 
precise ($\stackrel{<}{_\sim}$\,100\kev) measurements of charmonium 
masses and widths~\cite{Armstrong:1992wu,Andreotti:2007ur}, 
thanks to the precisely 
known collision energy of the stochastically cooled $\bar p$ beam 
(with its  $\approx0.02$\% energy spread) with a hydrogen 
cluster-jet target~\cite{Garzoglio:2004kw}. 
Significant charmonium-related questions remain, most notably the nature 
of the mysterious $X(3872)$ state~\cite{Eichten:2005ga} 
and improved measurements 
of the $\hsubc$ and $\etacp$~\cite{Brambilla:2004wf}. 
The width of the $X$ 
may well be $\ll 1$\mev~\cite{Stapleton:2009ey}. This unique 
${\bar p}p$ precision would have a crucial role in establishing whether 
the $X(3872)$ is a \DstnDn\ molecule~\cite{Tornqvist:2004qy}, 
a tetraquark state~\cite{Maiani:2004vq}, or something else entirely.

The ${\bar p}p\to X(3872)$ formation cross section may be 
similar to that of the $\chi_c$ states~\cite{Braaten:2004jg,Braaten:2007sh}. 
The E760 $\chi_{c1}$ and $\chi_{c2}$ detection rates of 
1~event/nb$^{-1}$ at the mass peak~\cite{Armstrong:1991yk} and the lower limit 
$\Brat(X(3872) \to\dipi\jpsi) > 0.042$ at 
90\%~CL~\cite{Aubert:2005vi} imply that, at the peak of the $X(3872)$, 
about 500 events/day can be observed. (Although CDF and \DZero\ could also
amass $\sim10^4$ $X(3872)$ decays, backgrounds and energy resolution 
limit their incisiveness.) 
Large samples will also be obtained in other modes besides 
$\dipi\jpsi$, increasing the statistics 
and improving knowledge of $X(3872)$ branching ratios.

While the above may be an under- or overestimate, perhaps by as much 
as an order of magnitude, it is likely that a new experiment at the 
Antiproton Accumulator could obtain 
the world's largest clean samples of $X(3872)$, in perhaps as little 
as a month of running. 
The high statistics, event cleanliness, and unique precision available 
in the ${\bar p}p$ formation 
technique could enable the world's smallest systematics. 
Such an experiment could 
provide a definitive test of the nature of the $X(3872)$. 
 
\subsection{Future linear collider}
\label{sec:Fut:ILC}

A high-energy \epem\ linear collider provides excellent possibilities 
for precision and discovery, within the Standard Model as well as for 
new physics. The International Linear Collider ILC~\cite{:2007sg} 
is a proposed machine based on superconducting RF cavities
that will provide center-of-mass energies of up to 500\gev, with the 
possibility for an upgrade to 1\tev. At the design energy of 500\gev, 
it will deliver a luminosity of $2 \times 10^{34}$~cm$^{-2}\,$s$^{-1}$, 
providing high-statistics datasets for precision studies. Two mature 
ILC detector concepts, ILD~\cite{:2010zzd} and SID~\cite{Aihara:2009ad} 
exist, both sophisticated general-purpose detectors with excellent 
tracking and vertexing capabilities and unprecedented jet energy 
resolution achieved with highly granular calorimeter systems and 
particle flow reconstruction algorithms. In parallel, the technology 
for the Compact Linear Collider CLIC~\cite{Assmann:2000hg}, using a 
two-beam acceleration scheme to reach center-of-mass energies up to
3\tev, is being developed. In a staged construction, such a machine 
would initially run at energies comparable to the design goals 
for the ILC before reaching the multi-TeV regime. The detector 
concepts for CLIC are based on the already-mature ILC detectors, 
with some modifications to account for the higher final collision energy. 
These planned colliders are excellent tools for top physics. 
The precise control of the beam energy at a linear \epem\ 
collider allows a scan of the $t\bar{t}$ production threshold 
to determine the top mass, width, and production cross 
section~\cite{Djouadi:2007ik}. 

There are several interesting observables in the $t\bar{t}$ 
threshold region that can be used for these measurements.
These are the total cross section, top momentum distribution,
top forward-backward asymmetry and the lepton angular
distribution in the decay of the top quark. The total 
cross section as a function of $\sqrt{s}$ rises sharply 
below the threshold and peaks roughly at the position
of the would-be $1S$ $t\bar{t}$ resonance mass \cite{Fadin:1988fn}.
The normalization at the peak is proportional to the
square of the resonance wave function at the origin and
inversely proportional to the top quark decay width.
Theoretically, the resonance mass (hence the peak position) can be
predicted very accurately as a function of $m_t$ and $\als$, so that
the top quark mass can be determined by measuring the peak position 
accurately. We may determine the top quark width and the top quark 
Yukawa coupling from the normalization, since the exchange of a
light Higgs boson between $t$ and $\bar{t}$ induces a Yukawa potential, 
which affects the resonance wave function at the origin~\cite{Jezabek:1993tj}.

The top quark momentum distribution is proportional to the square 
of the resonance wave function in momentum space
\cite{Sumino:1992ai,Jezabek:1992np}. The wave function of the 
would-be $1S$ toponium resonance can be measured, since the top 
quark momentum can be experimentally reconstructed from the final state, 
unlike in the bottomonium or charmonium cases. The wave function is also
predicted theoretically and is determined by $\als$ and $m_t$.
The top quark forward-backward asymmetry below threshold is generated 
as a result of interference between the $S$-wave and $P$-wave resonance states
\cite{Murayama:1992mg}. This interference is sizable, since the top quark width
is not very different from the level splitting between the $1S$ and $1P$
states, and since the width is much larger than the $S$-$P$
splittings of the excited states. Thus, by measuring the forward-backward 
asymmetry, one can obtain information on the level structure of the $S$ and
$P$-wave resonances in the threshold region. Both the top momentum
distribution and forward-backward asymmetry provide information 
on $t\bar{t}$ dynamics, which are otherwise 
concealed because of the smearing due to the large top decay width.

It is known that the charged lepton angular distribution is sensitive 
to the top quark spin. By measuring the top quark spin  
in the threshold region~\cite{Harlander:1994ac},
it is possible to extract the chromoelectric and
electric dipole moments of the top quark, which are
sensitive to $CP$-violations originating from 
beyond-the-standard-model physics \cite{Jezabek:2000gr}.

\begin{figure}[t]
   \begin{center}
      \includegraphics[width=\figwid]{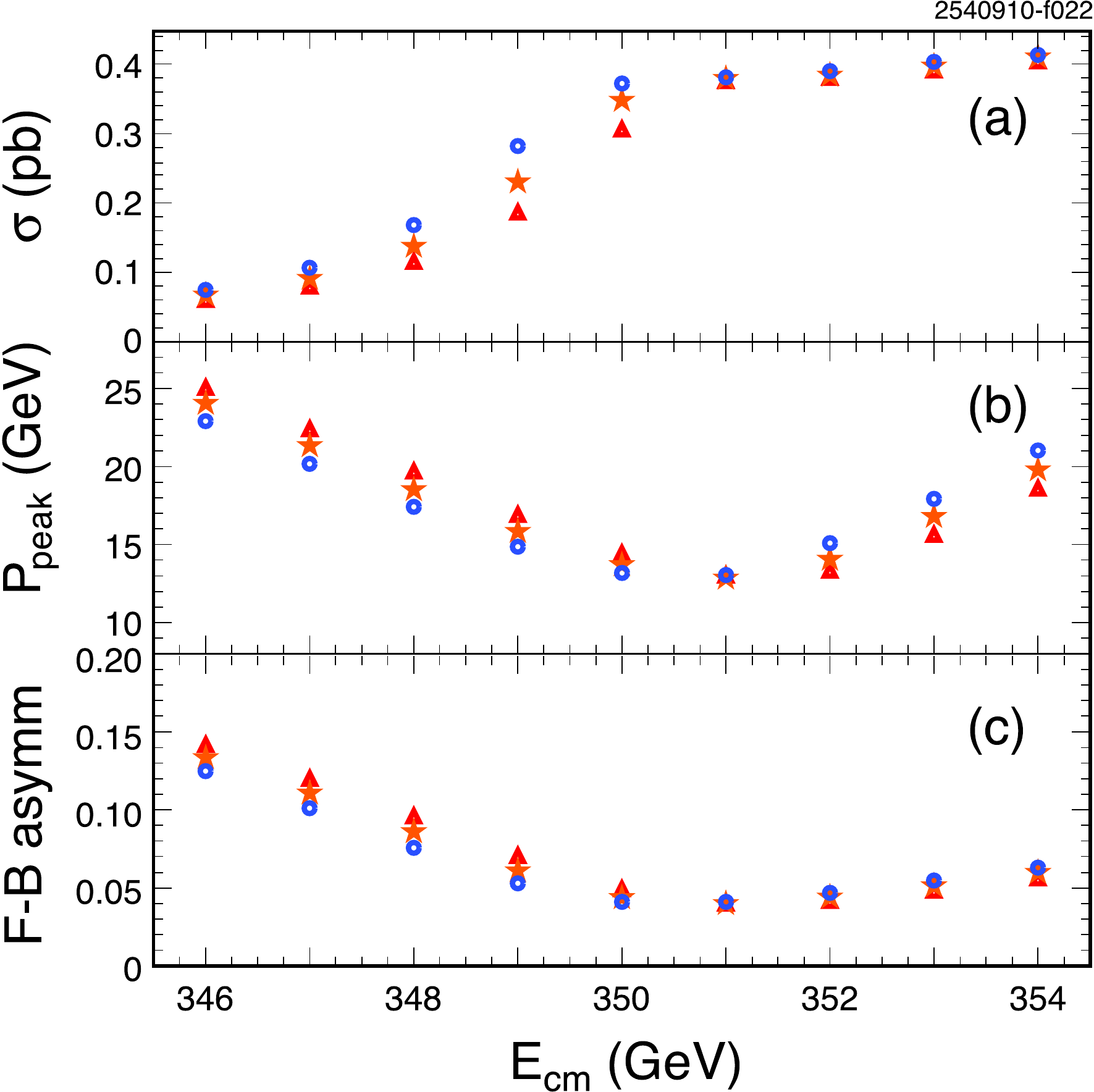}
      \caption{Sensitivity of three important observables 
               to the top mass from a simulated threshold scan 
               at a linear collider: (a) the top production 
               cross section, (b) the peak of the top
               momentum distribution $P_{\mathrm{peak}}$, and (c) the 
               forward-backward asymmetry.  The {\it solid triangles}
               {\it (open circles)} correspond to input MC generated
               with top mass that is 200\mev\ higher (lower) than
               that used for the MC sample with nominal top mass,
               represented by {\it stars}.
               \AfigPermSPV{Martinez:2002st}{2003} }
      \label{fig:Fut_ILC_TopScan}
   \end{center}
\end{figure}

By combining the collision energy dependence of these observables,
precise measurements of the top parameters are possible. 
\Figure{fig:Fut_ILC_TopScan} shows simulations of the sensitivity
of the cross section, the peak of the top momentum distribution and of
the forward-backward asymmetry to the top mass in a threshold scan at a
linear collider \cite{Martinez:2002st}. (See \cite{Fujii:1993mk} for an
earlier, similar analysis.) To illustrate the sensitivity, three
different input masses, spaced by 200\mev, are shown. From a
simultaneous fit to these three observables, a statistical uncertainty
of 19\mev\ on $m_t$ was obtained, neglecting sources of systematic errors
such as the determination of the luminosity spectrum of the collider or
theoretical uncertainties.  
To achieve the highest possible precision, detailed understanding of the
beam energy spectrum, of beamstrahlung and of initial state radiation is
crucial. The measured cross section in such a scan is theoretically well
described, as discussed in detail in Sect.~6 of
\cite{Brambilla:2004wf} and references therein. Using NNLO QCD
calculations, an overall experimental and theoretical error of 100\mev\ 
on the top mass is reachable with a threshold scan at a linear collider
\cite{Hoang:2000yr}. Further important improvements in precision 
arise from NNNLO QCD calculations,
the summation of large logarithms at NNLL order and a systematic
treatment of electroweak
and top quark finite-lifetime effects. 
Also, the determination of corrections beyond NLO
for differential observables such as the momentum distribution and the forward-backward
asymmetry used in the study described above would be required 
to fully exploit the precision of the expected data.

\begin{figure}[t]
   \begin{center}
      \includegraphics[width=\figwid]{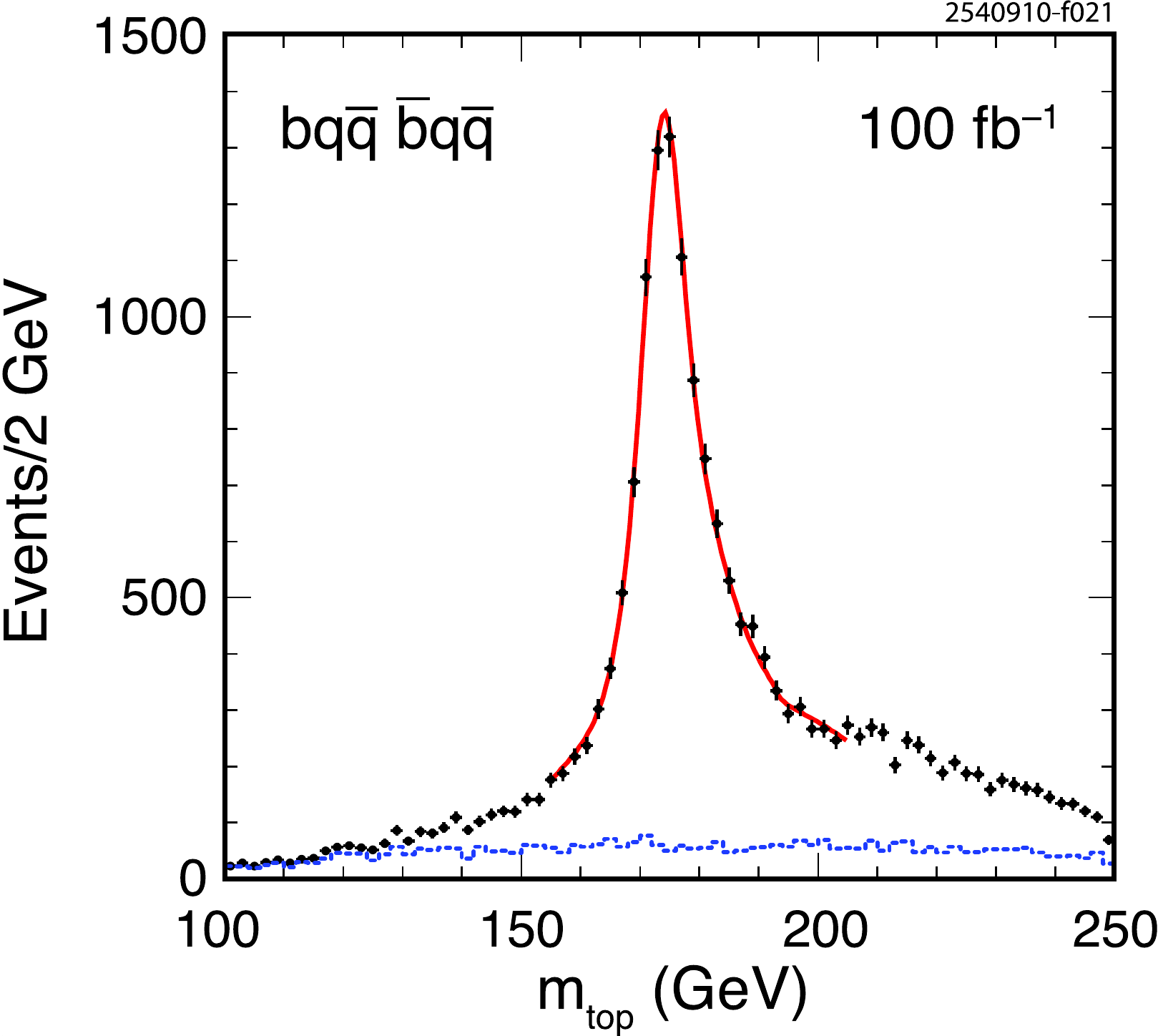}
      \caption{Distribution of the reconstructed top quark 
               invariant mass for the fully hadronic 
               $t\bar{t}\to bq\bar{q}\bar{b}q\bar{q}$ 
               decay channel in the ILD detector concept 
               for an integrated luminosity of 100~fb$^{-1}$, 
               including the full non-top background, indicated in
               the {\it dotted histogram}. The {\it solid curve}
               represents a fit for the top mass. From~\cite{:2010zzd} }
      \label{fig:Fut_ILC_TopMass}
   \end{center}
\end{figure}

A method has been developed to improve 
accuracies of kinematic variables in 
reconstructed top-quark events in the threshold 
region~\cite{Ikematsu:2003pg}.
This method would improve measurements
of the top quark observables mentioned above.

An alternative to the threshold-region measurements described above is
reconstruction of decay products in $t\bar{t}$ pair production far above
threshold. This approach has the advantage that the top measurement is
possible at full collider energy in parallel with other measurements,
resulting in much higher integrated luminosity. The top quarks decay
essentially exclusively into a $b$ quark and a $W$ boson, giving rise to
two $b$ jets and additional jets or leptons from the $W$ decays in the
final state. The expected performance of the ILC detectors leads to very
high precision in the invariant mass determination. Due to the clean
environment in $e^+e^-$ collisions, a measurement in the full-hadronic
six-jet mode, which has the highest branching fraction, is
possible. \Figure{fig:Fut_ILC_TopMass} shows the reconstructed top quark
invariant mass in the fully hadronic decay mode obtained from
a simulation for the ILD detector concept~\cite{:2010zzd} 
at an energy of 500\gev\ and corresponding to an integrated
luminosity of 100~fb$^{-1}$. The simulation uses a
complete detector modeling and standard model backgrounds. 
The statistical error on the
mass reconstruction is 90\mev\ in this channel alone. With a higher
integrated luminosity of 500~fb$^{-1}$ and by combining measurements in
fully hadronic and in semileptonic top decays, a statistical precision
of 30\mev\ seems reachable, comparable to the experimental precision of a
threshold scan. However, the theoretical interpretation of the invariant
mass is considerably more complicated than in the case of threshold
measurements. 
Recently, a systematic formalism to relate the measured invariant mass to 
theoretically meaningful parameters has been established
\cite{Fleming:2007qr, Fleming:2007xt}. This formalism has the potential
to compete with the threshold-scan calculations in terms of
achievable precision. The computations were carried out to NLL.
Further higher-order calculations are needed 
to reduce the present theoretical uncertainties and fully exploit the
experimental precision achievable at a future linear collider.
See also \Sec{sec:SpecTh_topmass}.

Beyond precision measurements of the top mass and width, a Linear Collider 
would also provide excellent conditions for a wealth of other top-related studies, 
such as the measurement of the top-Yukawa coupling in the process 
$e^+e^- \rightarrow t\bar{t}H$ and the search for anomalous couplings in decays.

\clearpage
\section{Conclusions and priorities}
\label{sec:SumChapter}

  Below we present a summary of the most crucial developments
in each of the major topics and suggested directions
for further advancement.\\

\noindent{\bf \underline{Spectroscopy:}}
An overview of the last decade's progress 
in heavy quarkonium spectroscopy
was given in \Sec{sec:SpecChapter}.

With regard to experimental progress, we conclude:

\begin{enumerate}

\item New measurements of inclusive hadronic cross sections (\ie $R$) 
for $e^+e^-$ collisions just above open $c\bar{c}$ and $b\bar{b}$
flavor thresholds have enabled improved determinations of some resonance 
parameters but more precision and fine-grained studies are needed 
to resolve puzzles and ambiguities. Likewise, progress has been 
made studying exclusive open-flavor two-body and multibody composition
in these regions, but further data are needed to clarify the details. 
Theory has not yet been able to explain the measured exclusive 
two-body cross sections.

\item Successful observations were made (\Tab{tab:Spec_ExpSumCon}) 
of 6 new conventional heavy quarkonium states (4 $c\bar{c}$, 2 $b\bar{b}$); 
of these, only the $\eta_b(1S)$ lacks a second, independent
5$\sigma$ confirmation.
Improved measurement of \etac\ and \etacp\ masses and widths 
would be quite valuable. Unambiguous observations and precise mass
and width measurements are needed for \etabp, $h_b(^1P_1)$, 
\UoneDO, and \UoneDH\ in order to constrain theoretical descriptions.

\item Experimental evidence has been gathered 
(\Tab{tab:Spec_ExpSumUnc}) for 
up to 17 
unconventional 
heavy quarkonium-like states. All but $Y_b(10888)$ are in the 
charmonium mass region,
and all but 5 remain unconfirmed at the 5$\sigma$ level. Confirmation or 
refutation of the remaining 12 is a high priority.

\item Theoretical interpretations for the unconventional states 
(\Tab{tab:Spec_ExpHyp}) range from coupled-channel effects to 
quark-gluon hybrids, mesonic molecules, and tetraquarks. More 
measurements and theoretical investigations 
are necessary to narrow the possibilities.
In particular, high-resolution measurements of lineshapes promise
deeper insights into the nature of various of those states.

\item The $X(3872)$ was the first unexpected state to be observed 
and has generated the most experimental and theoretical attention. 
Its sub-MeV proximity to \DstnDn-threshold
(\Tabs{tab:Spec_XMass}-\ref{tab:Spec_DMass}) and dominant $\DzDz\pi^0$
branching fraction suggest a \DstnDn-molecular component, although 
this interpretation is not universally shared. The $X(3872)$ has been 
confirmed in four decay modes
(\Tab{tab:Spec_RatBchgBneu}). The discovery mode, \ppjp, is still 
the best measured, and has a branching fraction comparable in size to 
that of $\omega\jpsi$; $\DzDz\pi^0$ is ten times more common and 
$\gamma\jpsi$ three times less.
The $X(3872)$ quantum numbers have been narrowed to $1^{++}$ or $2^{-+}$.

\item The charged $Z$ states observed in $Z^-\to\pi^-\psip$ and 
$\pi^-\chi_{c1}$ would be, if confirmed, manifestly exotic. Hence their
confirmation or refutation is of the utmost importance.

\end{enumerate}

  With regard to lattice QCD calculations:

\begin{enumerate}[resume]

\item Lattice QCD technology has progressed to the point that it  may
provide accurate calculations of the energies of quarkonium states below
the open flavor threshold, and also provide information about higher
states. 

\item Precise and definitive calculations of the $c\bar c$ and $b\bar b$ meson
spectra below threshold are needed.  Unquenching effects,
valence quark annihilation channels and spin contributions should be fully included.

\item Unquenched calculations of states above the open-flavor thresholds 
are needed. These would provide 
invaluable clues to the nature of these states.

\item The complete set of Wilson loop 
      field strength averages entering the
      definition of the nonperturbative $Q \bar{Q}$ potentials 
      must be calculated on the lattice.

\item Calculations of local and nonlocal gluon condensates on the lattice
      are needed as inputs to weakly-coupled pNRQCD spectra and 
      decay calculations.

\item NRQCD matching coefficients in the lattice 
      scheme at one loop (or more) are needed.

\item Higher-order calculations of all the relevant quantities 
      due to the lattice-to-$\overline{MS}$ scheme change
      are required in order to relate lattice 
      and continuum results in the EFT.

\item Lattice calculations of the overlap between quarkonia and
      heavy-light states in the threshold region, as well as with 
      hybrids or exotic states, should be performed.

\item A better determination of the $r_0$ lattice scale and a
      nonperturbative determination of $\LMSb$
      with 2+1 or 2+1+1 sea quarks is needed.

\end{enumerate}

  With regard to effective field theories (EFTs),

\begin{enumerate}[resume]

\item Higher-order perturbative EFT 
      calculations of static energies, static potentials, and 
      relativistic corrections to the potentials and energy levels have
      appeared recently for different heavy-quark/antiquark
      configurations. Further efforts in this direction are needed,
      and the emerging patterns of renormalons should be studied in relation
      to the behavior of the bound states perturbative series.

\item We have described simulation studies for a future linear collider 
      which demonstrate that precise determinations of the top mass 
      and the top Yukawa coupling can be obtained from a $t\bar t$ 
      production scan near threshold. To at least match
      this expected precision, a complete NNLL computation is necessary
      to obtain a firmer grasp of the theoretical uncertainties. 
      The complete NNNLO computation is also desirable to assess the 
      importance of the resummation of logarithms. 
      The electroweak and non-factorizable corrections may be calculated
      by developing an effective theory description for unstable
      particles.
      \label{enum:Summ_ttbar}

\item Better experimental data for $b\bar{b}$ production above 
      threshold would diminish the impact of 
      the experimental error in nonrelativistic sum rules determinations 
      of the bottom quark mass.

\item The $2\sigma$ discrepancy between the EFT calculation and measurements 
      of the \etab\ mass needs to be resolved.

\end{enumerate}

\noindent{\bf \underline{Decay:}}
Section~\ref{sec:DecChapter} described the enormous
progress on heavy quarkonium decays, showing
that many branching fraction, width, and spectra
measurements have attained high
precision. Some of these results provide
crucial anchors for theoretical approaches
while others have just scratched the surface
of what may be attainable with more
data and improved techniques. Accomplishments
and priorities are:

\begin{enumerate}[resume]

\item Precise measurements of the
dileptonic and total widths in charmonium
(\Tabs{tab:Dec_JPsi_ll}-\ref{tab:Dec_Psi_ll})
and bottomonium 
(\Tabs{tab:Dec_Ups_MM}-\ref{tab:Dec_Ups_Tot})
below the respective
open-flavor thresholds have been performed.

\item A novel and empirical measurement of
the nontrivial radiative photon lineshape
in \jpsi\ and \psip\ decays to \etac, which
in turn enabled determination of much-improved branching
fractions, has appeared. This has stimulated theoretical
activity to explain the photon spectral shape
and raised the importance of measuring
the corresponding spectra for $\psip\to\gamma\etacp$,
$\psip\to\gamma\chi_{cJ}(1P)$,
$\UnS{n}\to\gamma\eta_b(1S,2S)$, and
$\UnS{n}\to\gamma\chi_{bJ}(nP)$.
Precise measurement of the
$\psip\to\gamma\chi_{cJ}(1P)$ lineshapes
will be essential in quantifying
the branching fraction for
the nonresonant $\psip\to\gamma\gamma\jpsi$ decay.
\label{enum:radlineshapes}

\item There are new branching fraction measurements
for decays of \jpsi, \psip, and \Ups\
to $\gamma\eta(^\prime)$ that present
a puzzling and unexplained pattern (\Tab{tab:Dec_gampseudo}).

\item Measurements of the $\gamma gg$ spectra
and branching fractions (\Tab{tab:DecGammaGG}) 
for \UnS{1S,2S,3}, \jpsi, and \psip\ all have been published.
Considerably more attention to 
measurements, in which background-subtraction uncertainties have limited
the precision of charmonium results much more 
severely than for bottomonium, and predictions,
which have had mixed success on predicting
the radiative photon spectra, is needed.

\item First measurements of the two- and three-photon
partial widths of $\chi_{c0,2}$ (\Tab{tab:Dec_CC_2Gamma}) 
and \jpsi, respectively, have been performed. Although the experimental 
relative uncertainties on these widths are not yet below
the 10\% and 30\%, respectively, they already
present challenges to the theory.
Better experimental precision for the two-photon
couplings of \etac\ and \etacp\ would be
very useful to constrain cross-particle
branching fraction measurements as well as relevant theoretical
descriptions.

\item Detailed and provocative measurements of dipion and $\eta$
transitions for the $\psi$
(\Tab{tab:Dec_CLEO_BXJpsi})
and $\Ups$ (\Tabs{tab:Dec_CLEOMpipiUps}-\ref{tab:Dec_CLEOBRpipiUps})
systems below the respective open-flavor thresholds
challenge theoretical rate and dipion-mass
spectra predictions, while the surprisingly high
rates observed above these thresholds remain a mystery.

\item Measurements of $\psit\to\nonDDbar$
(\Tab{tab:Dec_BF_3770}) conflict with one another. Definitive measurements
of {\it exclusive decays} can best supply confidence in 
such a rate being more than a few percent,
because indirect or aggregate (\Tab{tab:Dec_XS_3770}) comparisons
appear to be quite challenging.

\item A multitude of measurements has been accumulated
in the realm of exclusive hadronic decays of heavy quarkonia.
These have deepened theoretical mysteries (\eg $\rho\pi$ puzzle,
conflicting measurements of \nonDDbar\ decays of \psit) but whetted the
community's appetite for more information on such decays
of $\chicJ$, $\chi_{bJ}$, \etac, \etacp, 
\etab, $h_c(^1P_1)$, and  $h_b(^1P_1)$.

\item Initial but nevertheless intriguing measurements of the rate
for deuterons to appear in \Ups\ decays and
the mechanism of baryon-number compensation
therein have been reported. Further experimental and theoretical attention
in \Ups\ decays, the LHC, and future facilities
are warranted. This information is useful for tuning
MC generators and may be relevant for the molecular
interpretation of $X(3872)$ (\Sec{sec:SpecTh_molec})
and other loosely-bound states.

\item A new measurement (\Tab{tab:Dec_tabmamp})
has resolved the longstanding discrepancy between experiment
and theory on multipole amplitudes in $\psip\to\gamma\chi_{cJ}$,
$\chi_{cJ}\to\gamma\jpsi$.

\item A theoretical understanding of the photon energy spectrum
from $\jpsi \to \gamma \etac$ and $\psip\to \gamma \etac$
(see item~\#\ref{enum:radlineshapes})
is required (\Secs{sec:Dec_EMT1} and \ref{sec:Dec_1Sc}).

\item It would be important to have a coherent EFT
treatment for all magnetic and electric transitions.
In particular, a rigorous treatment of the relativistic corrections
contributing to the E1 transitions and a nonperturbative analysis of the 
M1 transitions is missing. The first is relevant for transitions involving
$P$ states, the second for any transition from above the ground state.

\item New resummation schemes for the perturbative expressions of
the quarkonium decay widths should be developed.
At the moment, this is the major obstacle to precise theoretical
determinations of the \UnS{1}\ and \etab\ inclusive and electromagnetic 
decays (\Sec{sec:Dec_RadLepTheory}).

\item More rigorous techniques to describe above-threshold quarkonium decays 
and transitions, whose descriptions still rely upon models, should 
be developed (\Secs{sec:Dec_HadTranTheory} and \ref{sec:Dec_haddec}).

\end{enumerate}

\noindent{\bf \underline{Production:}}
The theoretical and experimental status of
production of heavy quarkonia was given in \Sec{sec:ProdChapter}.
Conclusions and priorities are as follows:

\begin{enumerate}[resume]

\item It is very important either to establish that the NRQCD 
factorization formula is valid to all orders in perturbation theory or
to demonstrate that it breaks down at some fixed order.

\item A more accurate treatment of higher-order corrections to the
color-singlet contributions at the Tevatron and the LHC is urgently
needed. The re-organization of the perturbation series that is
provided by the fragmentation-function approach
(\Sec{prod_sec:fragmentation}) may be an important tool.

\item An outstanding theoretical challenge is the development of methods 
to compute color-octet long-distance NRQCD production matrix elements on 
the lattice.

\item If NRQCD factorization is valid, it likely holds only for values
of $p_T$ that are much greater than the heavy-quark mass. Therefore, it
is important for experiments to make measurements of quarkonium
production, differentially in $p_T$, at the highest possible values of
$p_T$.

\item Further light could be shed on the NRQCD velocity expansion and its
implications for low-energy dynamics by comparing studies of charmonium
production and bottomonium production. The higher $p_T$ reach of the
LHC may be particularly important for studying bottomonium production at 
values of $p_T$ that are much greater than the bottomonium mass.

\item It would be of considerable help in disentangling the theoretical
issues in production of the $J/\psi$ and $\Upsilon$ if experimental
measurements could separately quantify direct and feeddown contributions.
Ideally, the direct production cross sections and
polarizations would both be measured differentially in $p_T$.

\item It is important to resolve the apparent discrepancy 
between the CDF and \DZero\ measurements of the $\Upsilon$ 
polarization, which were performed for different
rapidity ranges, $|y|<0.6$ (CDF) and $|y|<1.8$ (\DZero).
A useful first step would be for the two experiments to
provide polarization measurements that cover the same rapidity range.

\item It would be advantageous to measure complete quarkonium
polarization information in a variety of spin-quantization frames and
to make use of frame-invariant quantities to cross-check measurements in
different frames \cite{Faccioli:2008dx,Faccioli:2010ji,Faccioli:2010kd}.
Care should be taken in comparing different polarization measurements to
insure that dependences on the choices of frame and the kinematic ranges
of the experiments have been taken into account.

\item Measurements of inclusive cross sections, charmonium angular
distributions, and polarization parameters for $P$-wave charmonium states
would provide further important information about quarkonium production
mechanisms.

\item Studies of quarkonium production at different values of $\sqrt{s}$
at the Tevatron and the LHC, studies of hadronic energy near to and away
from the quarkonium direction at the Tevatron and the LHC, and studies
of the production of heavy-flavor mesons in association with a
quarkonium at $e^+e^-$, $ep$, $p\bar{p}$, and $pp$ machines could give
information that is complementary to that provided by traditional
observations of quarkonium production rates and polarizations.

\item Theoretical uncertainties in the region near the kinematic
endpoint of maximum quarkonium energy might be reduced through a
systematic study of resummations of the perturbative and velocity
expansions in both $ep$ and $e^+e^-$ quarkonium production.

\item In predictions for exclusive and inclusive quarkonium production
in $e^+e^-$ annihilation, large corrections appear at NLO. An important
step would be to identify the origins of these large corrections.
It might then be possible to improve the convergence of perturbation
series by resumming specific large contributions to all orders in
$\als$.

\item The central values of the Belle and \babar\ measurements of
$\sigma(e^+e^-\to  J/\psi +\etac)\times \Brat_{>2}$, 
where $\Brat_{>2}$ is the branching
fraction for the \etac\ to decay into a final state with
more than two charged particles, differ by about twice
the uncertainty of either measurement, suggesting that further
experimental attention would be valuable. Comparisons with theory
would be more informative if the uncertainty from 
the unmeasured branching fraction
$\Brat_{>2}$ 
were reduced or eliminated.

\item The central values for the prompt $J/\psi$ inclusive production cross
section that were obtained by \babar\ and Belle  
differ by more than a factor of two. It would be very
desirable to clear up this discrepancy.

\item Belle has presented results for
$\sigma(e^+e^-\to J/\psi+c\bar c)$ and 
$\sigma(e^+e^-\to J/\psi+X_{{\rm non}-c\bar c})$. 
It would be beneficial to have similar results from \babar.

\item Measurements of quarkonium production in $\gamma p$ and 
$\gamma\gamma$ collisions could provide additional information about production
mechanisms and should be carried out at the next opportunity at an
$e^+e^-$ or $ep$ collider. It may also be possible to measure quarkonium
production in $\gamma\gamma$ and $\gamma p$ collisions at the LHC
\cite{deFavereaudeJeneret:2009db}.

\item The observation and study of the $B_c$ mesons and their
excitations are new and exciting components of the quarkonium-physics
plans for both the Tevatron and the LHC.

\item Theoretical progress should always be mirrored by the development
of simulation tools for experiment. A particularly important goal would
be to develop experiment-friendly simulation tools that incorporate the
state-of-the-art theory for inclusive and associative quarkonium
production.

\end{enumerate}

\noindent{\bf \underline{In Medium:}}
The status of heavy quarkonium production
in cold and hot matter was presented in \Sec{sec:MedChapter}.
Conclusions are, for cold matter:

\begin{enumerate}[resume]

\item Studies are now attempting to place a limit on the 
      allowed level of quark energy loss in 
      next-to-leading order Drell-Yan dilepton production.
      These will determine the maximum amount of 
      gluon energy loss that can be applied to $J/\psi$ 
      production and suppression models of $pA$
      interactions.

\item A limit on the level of energy loss apparent in 
      $J/\psi$ production as a function of nuclear mass, 
      $A$, and longitudinal momentum, expressed either
      as a function of $x_F$ or rapidity, $y$, will put 
      constraints on the nuclear absorption cross section. 
      These constraints will help determine the 
      importance of formation-time effects and feeddown on the
      quarkonium absorption cross section.  Ultimately, the 
      cold-nuclear-matter baseline should include different asymptotic 
      absorption cross sections based
      on their final-state radii and formation times.

\end{enumerate}

  For hot matter:

\begin{enumerate}[resume]

\item It is important to calculate the quarkonium spectral functions
      at nonzero temperature using an effective field theory approach with
      different hierarchies of relevant scales amended with lattice QCD
      calculations of the relevant correlation functions.

\item We would like to be able to compare the hot-matter effects 
      more directly to heavy-ion data.  To do this, more realistic 
      dynamical models of quarkonium production in heavy-ion 
      collisions must be developed that rely on 
      state-of-the-art calculations of the quarkonium spectral functions.

\end{enumerate}

  Important future RHIC measurements include:

\begin{enumerate}[resume]

\item The open charm and 
      open bottom cross sections as a function of rapidity and 
      centrality in d+Au and Au+Au collisions, as well as a 
      measurement of the rapidity distribution in $pp$ collisions.  
      Such a measurement is complementary to quarkonium 
      production and necessary to establish whether the 
      reduced $J/\psi$ production cross section at forward 
      rapidity, manifested in the larger effective absorption cross
      shown in \Figs{fig:media_fig3}, \ref{fig:media_fig7} and 
      \ref{fig:media_fig10}, is associated with reduced charm 
      production or is particular to bound-state formation. 
      This should be feasible with the upgraded vertex detectors
      being installed in 2011 and 2012, in conjunction with 
      another d+Au run.

\item The $J/\psi$ elliptic flow, $v_2$, in Au+Au collisions.
      Statistical recombination models predict strong secondary 
      $J/\psi$ production in heavy-ion collisions. This measurement  
      will be an important test of this
      recombination picture. 

\item Higher-statistics measurements of $\psi$ and $\Upsilon$ 
      production in d+Au and Au+Au collisions, which can
      be expected with increasing RHIC luminosity.
      These will provide
      further tests of the quarkonium production and suppression 
      mechanisms in cold and hot matter.

\end{enumerate}

\noindent{\bf \underline{Experimental outlook:}}
Section~\ref{sec:FutChapter} gives an overview
of newly-commissioned, under-construction, and
only-planned experimental facilities and what their
activities relating to heavy quarkonium will be:

\begin{enumerate}[resume]

\item BESIII operating at BEPCII will continue
where BESII and CLEO-c left off, with a robust
program of charmonium spectroscopy and decay
investigations. Initial datasets at the \jpsi\
and \psip\ already exceed previous accumulations,
with additional data acquisition at \psit\ and the
\DDst\ peak at $\sqrt{s}=4170\mev$ planned for the near future. Fine
scans above open-charm threshold are likely.
At its highest energy, BEPCII will directly produce $Y(4260)$
for much-needed further study.

\item ALICE, ATLAS, CMS, and LHCb are being 
commissioned along with the LHC, and have
potent heavy quarkonium programs underway,
planning important 
production and polarization measurements
in both $pp$ and heavy-ion collisions.
The four experiments have 
distinct and complementary 
experimental strengths and specialties, but
also will have significant overlaps in 
many measurements for cross-checking of results.

\item The \PANDA ($p\bar p$) and CBM (nucleus-nucleus)
experiments at the FAIR facility at GSI
will complement the activities at other laboratories.

\item The knowledge of open-charm cross section in
proton-antiproton annihilations is extremely important
to shape the initial physics program of \PANDA at FAIR.
A collaboration should be formed to prepare a proposal
to perform inclusive measurements at Fermilab 
with $\bar{p}p$ collisions in the charmonium energy region
using existing detector elements.

\item Lepton-hadron colliders have significant
role in advancing the heavy-quarkonium-physics
agenda, with the energy, intensity, and
experimental upgrades at JLab likely to
contribute before other similar proposed facilities.

\item A tau-charm factory and/or a more flexible
super-flavor factory have been proposed to 
continue the giant strides in heavy quarkonium
physics taken at \epem\ machines, from
Mark-I at SPEAR to the very recent landmark
results from both $B$-factories, CLEO-c, 
and BESIII. The factory-level luminosities
combined with sophisticated detectors
and well-defined intitial-state
energy-momentum and quantum numbers give
\epem\ collisions many important advantages.

\item A future linear collider (CLIC or ILC) would
offer important opportunities to measure
top quark properties in the $t\bar t$ threshold region
(see item~\#\ref{enum:Summ_ttbar} above).

\end{enumerate}

\section*{Acknowledgments}

{\small
The authors appreciate and acknowledge support for work on this
document provided, in part or in whole, by
\begin{itemize}
\item the U.S. Department of Energy (DOE), under contracts 
      DE-FG02-91-ER40690~(P.~Artoisenet),\\
      DE-AC02-06-CH11357~(G.~T.~Bodwin),\\
      DE-AC05-06-OR23177~(E.~Chudakov and C.~Weiss)\\
      DE-AC02-07-CH11359, through FNAL,
         which is operated for DOE by the Fermi Research Alliance, LLC, under
         Grant No. DE-FG02-91-ER40676~(E.~Eichten and V.~Papadimitriou),\\
      DE-AC02-76-SF00515~(A.~Gabareen~Mokhtar and J.~P.~Lansberg),\\
      DE-AC02-05-CH11231~(S.~R.~Klein),\\
      DE-AC02-98-CH10886~(P.~Petreczky and J.~W.~Qiu),\\
      DE-FG02-96-ER41005~(A.~A.~Petrov),\\
      DE-AC52-07-NA27344f~(R.~Vogt), and\\
      DE-FG02-94-ER40823~(M.~Voloshin);
\item the German Research Foundation (DFG) 
      Collaborative Research Center 55 (SFB)
      and the 
      European Union Research Executive Agency (REA)
      Marie Curie Initial Training Network 
      (\href{http://www.physik.uni-regensburg.de/STRONGnet/}
      {www.physik.uni-regensburg.de/STRONGnet}), 
      under Grant Agreement 
      PITN-GA-2009-238353~(G.~Bali);
\item the European Union Marie Curie Research Training Network (RTN) 
      Flavianet, under Contract MRTN-CT-2006-035482, 
      and the German Research Foundation (DFG) 
      Cluster of Excellence {\it Origin and Structure of the 
      Universe}
      (\href{http://www.universe-cluster.de}{www.universe-cluster.de})\\
      (N.~Brambilla and A.~Vairo);
\item the Polish Ministry of Science and Higher Education~(J.~Brodzicka);
\item the National Natural Science Foundation of China (NSFC) 
      under Grants\\
      10875155 and 10847001 (C.-H.~Chang),\\
      10721063~(K.-T.~Chao),\\
      10920101072 and 10845003 (W.~Qian),\\
      and 10775412, 10825524, and 10935008 (C.-Z.~Yuan);
\item the Ministry of Science and Technology of China,
      under Grant 2009CB825200~(K.-T.~Chao);
\item The German Research Foundation (DFG) under grant 
      GZ~436~RUS~113/769/0-3 and the Russian 
      Foundation for Basic Research (RFBR) under grants
      08-02-13516 and 08-02-91969~(S. Eidelman);
\item the U.S. National Science Foundation (NSF), under contracts 
      PHY-07-56474~(A.~D.~Frawley),\\
      PHY-07-58312 and PHY-09-70024~(B.~K.~Heltsley),\\
      CAREER Award PHY-05-47794~(A.~Petrov), and\\
      PHY-05-55660~(R.~Vogt),;
\item Science and Engineering Research Canada (NSERC) (X.~Garcia~i~Tormo);
\item the Helmholtz Association, through funds provided to the 
      virtual institute {\it Spin and strong QCD} (VH-VI-231),
      the German Research Foundation (DFG) (under grants 
      SFB/TR~16 and 436~RUS~113/991/0-1) and the European
      Community-Research Infrastructure Integrating Activity 
      {\it Study of Strongly Interacting Matter} 
      (acronym HadronPhysics2, Grant Agreement 227431)
      under the European Union Seventh Framework Programme (C.~Hanhart);
\item the Belgian American Educational Foundation and 
      the Francqui Foundation (J.~P.~Lansberg);
\item the Belgian Federal Science Policy (IAP~6/11)\\
      (F.~Maltoni);  
\item the Brazil National Council for Scientific and Technological 
      Development (CNPq) and Foundation for Research Support of the 
      State of S\~ao Paulo (FAPESP) (F.~S.~Navarra and M.~Nielson);
\item the World Class University (WCU) project
      of the National Research Foundation of Korea,
      under contract R32-2008-000-10155-0~(S.~Olsen);
\item the Ministry of Education and Science of the Russian Federation
      and the State Atomic Energy Corporation ``Rosatom''~(P.~Pakhlov and
      G.~Pakhlova);
\item the France-China Particle Physics Laboratory (FCPPL)~(W.~Qian);
\item the French National Research Agency (ANR)
      under Contract ``BcLHCb ANR-07-JCJC-0146''~(P.~Robbe);
\item the Spanish Ministry of Science and Innovation (MICNN),
      under grant
      FPA2008-02878 and
      Generalitat Valenciana under grant
      GVPROMETEO2010-056~(M.~A.~Sanchis-Lozano);
\item the Portuguese Foundation for Science and Technology (FCT),
      under contracts SFRH/BPD/42343/2007 and 
      SFRH/BPD/42138/2007~(P.~Faccioli and H.~K.~W\"ohri)
\end{itemize}

N.~Brambilla, A.~D.~Frawley, C.~Louren\c{c}o, A.~Mocsy, P.~Petreczky, 
H.~K.~W\"ohri,  A.~Vairo, and R.~Vogt acknowledge the hospitality of 
Institute for Nuclear Theory at the University of Washington and the 
U.S.~Department of Energy for partial support during their attendance at
the CATHIE-INT mini-program on heavy quarkonium in Seattle.

We thank Hee Sok Chung for preparing 
Fig.~\ref{prod_fig:comp_polarization_D0_CDF}, and Roman Mizuk for
his perspective on the $Z^+$ analyses.
We thank Eric Braaten for his feedback on the manuscript,
and Dave Besson, Mathias Butensch\"on, 
and Bernd Kniehl for useful discussions.
We thank Jeanne Butler for assistance with
preparation of the figures.
}

\end{document}